\documentclass{aa}

\usepackage{graphicx}

\usepackage{txfonts}

\usepackage{lscape}
\usepackage{caption}
\usepackage{longtable}
\usepackage{xcolor}
\usepackage{rotating}
\usepackage{float}
\usepackage{subfig}
\usepackage{multicol}

\usepackage{natbib}
\usepackage[titletoc,toc,page]{appendix}
\usepackage{amstext}

\begin{document} 

   \title{Homogeneous study of Herbig Ae/Be stars from spectral energy distributions and Gaia EDR3}

   \author{J. Guzmán-Díaz\inst{1}\and I. Mendigutía\inst{1}\and B. Montesinos\inst{1}\and R.D. Oudmaijer\inst{2}\and M. Vioque\inst{2,3,4}\and C. Rodrigo\inst{1,5}\and E. Solano\inst{1,5}\and G. Meeus\inst{6}\and P. Marcos-Arenal\inst{1} 
          }

   \institute{$^{1}$Centro de Astrobiología (CSIC-INTA), ESA-ESAC Campus, 28692, Villanueva de la Cañada, Madrid, Spain\\$^{2}$School of Physics and Astronomy, University of Leeds, Leeds LS2 9JT, UK\\$^{3}$Joint ALMA Observatory, Alonso de Córdova 3107, Vitacura, Santiago 763-0355, Chile\\$^{4}$National Radio Astronomy Observatory, 520 Edgemont Road, Charlottesville, VA 22903, USA\\$^{5}$Spanish Virtual Observatory\\$^{6}$Departamento Física Teórica, Facultad de Ciencias, Universidad Autónoma de Madrid, Campus de Cantoblanco, 28049, Madrid, Spain}
   \date{Received September 24, 2020; accepted April 9, 2021}
 
  \abstract
   {Herbig Ae/Be stars (HAeBes) have so far been studied based on relatively small samples that are scattered throughout the sky. Their fundamental stellar and circumstellar parameters and statistical properties were derived with heterogeneous approaches before Gaia.}
   {Our main goal is to contribute to the study of HAeBes from the largest sample of such sources to date, for which stellar and circumstellar properties have been determined homogeneously from the analysis of the spectral energy distributions (SEDs) and Gaia EDR3 parallaxes and photometry.}
   {Multiwavelength photometry was compiled for 209 bona fide HAeBes for which Gaia EDR3 distances were estimated. Using the Virtual Observatory SED Analyser (VOSA), photospheric models were fit to the optical SEDs to derive stellar parameters, and the excesses at infrared (IR) and longer wavelengths were characterized to derive several circumstellar properties. A statistical analysis was carried out to show the potential use of such a large dataset.}
   {The stellar temperature, luminosity, radius, mass, and age were derived for each star based on optical photometry. In addition, their IR SEDs were classified according to two different schemes, and their mass accretion rates, disk masses, and the sizes of the inner dust holes were also estimated uniformly. The initial mass function fits the stellar mass distribution of the sample within 2 $<$ M$_*$/M$_{\odot}$ $<$ 12. In this aspect, the sample is therefore representative of the HAeBe regime and can be used for statistical purposes when it is taken into account that the boundaries are not well probed. Our statistical study does not reveal any connection between the SED shape from the \citet{Meeus_2001} classification and the presence of transitional disks, which are identified here based on the SEDs that show an IR excess starting at the K band or longer wavelengths. In contrast, only $\sim$28$\%$ of the HAeBes have transitional disks, and the related dust disk holes are more frequent in HBes than in HAes ($\sim$ 34$\%$ vs 15$\%$). The relatively small inner disk holes and old stellar ages estimated for most transitional HAes indicate that photoevaporation cannot be the main mechanism driving disk dissipation in these sources. In contrast, the inner disk holes and ages of most transitional HBes are consistent with the photoevaporation scenario, although these results alone do not unambiguously discard other disk dissipation mechanisms.}
   {The complete dataset is available online through a Virtual Observatory-compliant archive, representing the most recent reference for statistical studies on the HAeBe regime. VOSA is a complementary tool for the future characterization of newly identified HAeBes.}

\titlerunning{Homogeneous study of Herbig Ae/Be stars based on spectral energy distributions}
   \maketitle

\section{Introduction}\label{sect:intro}

The seminal work by \citet{Herbig_1960} defined the massive counterparts of T Tauri stars (TTs) as emission line objects with spectral type A or earlier that are located in obscured regions and illuminate bright and close nebulosities. This definition was subsequently nuanced. Currently, Herbig Ae/Be stars (HAeBes) are known as young ($\lesssim$ 10 Myr), optically visible pre-main-sequence (PMS) stars with emission lines in their spectra, typical spectral types A and B, stellar masses that typically range between $\sim$ 2 and $\sim$ 12 M$_{\odot}$, and infrared (IR) excesses associated with circumstellar disks. The initial list with dozens of sources \citep{Herbig_1960} was extended with new catalogs \citep[e.g.,][]{Finkenzeller_84,Herbig_88,The_1994,Carmona_2010,Chen_2016}, and more than 200 HAeBes are known today. This is far fewer than the thousands of TTs that are known, and this discrepancy can be partially explained by the shape of the initial mass function (IMF), which favors the formation of less massive objects and the faster evolution of massive stars. Statistical studies of HAeBes are generally less reliable than those for TTs because they are not based on complete samples in different star-forming regions but on small, scattered subsamples within the lists mentioned above and assume different stellar and circumstellar characterizations.

Many works have studied the stellar characterization of HAeBes \citep[see, e.g.,][and references therein]{Montesinos09}. However, before Gaia, most referred to relatively small samples and used different approaches to derive the stellar parameters. A major step toward a uniform characterization of HAeBes was made by \citet{Vioque_2018}, who placed 252 HAeBes in the HR diagram based on parallaxes from Gaia DR2 (\citealt{Gaia2016}; \citealt{Gaia2018}; \citealt{Lindegren18}). The stellar temperatures, gravities, extinctions, and other parameters were compiled from previous works in the literature, and the subsequent stellar luminosities, masses, and ages were then homogeneously inferred from the same atmospheric models and evolutionary tracks. Similarly, \citet{Arun_2019} derived a homogeneous set of stellar parameters for 131 HAeBes, this time based on Gaia DR2 parallaxes and magnitudes alone (\citealt{Lindegren18}; \citealt{Riello2018}). The most relevant reference for self-consistently obtained stellar parameters of HAeBes is the work by \citet{Wichittanakom20} (W2020 hereafter). This paper did not only update most of the spectroscopically determined stellar masses, luminosities, surface gravities, and ages of southern HAeBes derived by \citet{Fairlamb_2015} using Gaia DR2 distances, but extended the same spectroscopic analysis to additional HAeBes in the north. Thus, W2020 represents the most reliable and homogeneous stellar characterization of a wide sample of HAeBes to date, including spectroscopically determined data for 121 such stars mostly based on Gaia DR2 distances. 

An important caveat that was soon recognized as a result of the stellar characterization of HAeBes is that they do not constitute a homogeneous group. The group consists of two subsamples, HAes and HBes, that have specific ranges in stellar parameter space. In particular, HAes have relatively old ages and low masses (typically $>$ 3 Myr and 2-3 M$_{\odot}$), while HBes stars are younger and more massive (typically $<$ 3 Myr and $>$ 3 M$_{\odot}$). While the scarcity of old HBes results from the faster stellar evolution toward the main sequence (MS) with increasing stellar mass, the lack of very young HAes probably results from two main reasons. First, HAes become optically visible later in their evolution, and second, the younger population of intermediate-mass T Tauri stars (IMTTs) that will evolve into HAes is not well probed. For a related discussion, we refer to  \citet{vanvoekel2005,Calvet_2004} and \citet{Mendigutia_2011b}, among others. 

Regarding the circumstellar properties, disk-to-star accretion rates of relatively wide samples of HAeBes have been estimated either directly by modeling the near-ultraviolet (UV) excess \citep{Blondel06,Mendigutia_2011b,Fairlamb_2015} or indirectly through correlations with spectroscopic emission lines (\citealt{GarciaLopez06}; \citealt{Arun_2019}; W2020). Concerning the characterization of disk regions farther away from the star, the relative brightness and proximity of HAeBes make them the ideal targets for observations with high spatial resolution with the most advanced instrumentation \citep[e.g.,][and references therein]{Dong18}. However, these techniques are still affected by significant observational biases \citep{Garufi18}, and for statistical purposes, we must still rely mainly on the analysis of the spectral energy distributions (SEDs) in the IR and longer wavelengths \citep[e.g.,][]{Strom89,Hillenbrand_1992}. A commonly adopted scheme to classify the SEDs of HAeBes was proposed by \citet{Meeus_2001}, who divided 14 such sources into two groups depending on the shape of their SEDs: Group I, in which the continuum from the IR to the submillimeter region could be fit by a power-law component plus a cool blackbody, and group II, in which only the power-law component was necessary to make the fit. It has been shown that the Meeus groups might be related to the disk geometry \citep{Meeus_2001,Dullemond02,Dullemond04,Maaskant13}, the UXOr-type variability \citep{Dullemond03}, the presence of organic molecules \citep{Meeus_2001,AckeAncker04}, or the dust grain growth \citep{Acke_2004,Meijer08}, among others. While most previous works initially indicated an evolution from group I to group II, a more complex view is emerging, and there seems to be no evolutionary trend at the moment \citep{Mendigutia_2012,Maaskant13,Garufi17}. The difference between the two groups is mainly related to the presence of gaps and cavities, as shown in high-resolution imaging \citep{Maaskant13,Honda_2015,Garufi17}, which in turn may be connected with the potential presence of giant planets \citep{Kama15}. Other works characterizing the circumstellar properties of HAeBes include the determination of their disk masses \citep[e.g.,][]{Alonso_2009,Mendigutia_2012,Dong18} or their K band inner dust disk radii from interferometry \citep{Monnier_2002,Eisner03,Eisner04,Monnier05,Perraut19}. Taken together, previous works suggest that HBes tend to have lower dust disk masses from millimeter-continuum fluxes \citep{Alonso_2009,Vioque_2018} and dust inner radii from interferometric near- and mid-IR observations \citep[e.g.,][and references therein]{Monnier05} than expected from the corresponding trends followed by TTs and HAes. These add to the possible differences between the physical mechanisms driving accretion in the two types of sources (see, e.g., the recent review in \citealt{Mendi20}; W2020, and references therein). A major caveat is the relatively small samples and heterogeneous approaches from which many of the previous conclusions related to the circumstellar properties of HAeBes were inferred. This situation is changing because more complete samples are analyzed now \citep{Vioque_2018,Vioque20}. 

Here we contribute to the study of the HAeBe regime by providing a homogeneous characterization of the stellar and circumstellar properties of most of these confirmed sources, based on their complete SEDs from multiwavelength photometry. For all stars we make use of the recent Gaia Early Data Release 3 \citep[EDR3,][]{Gaia2016,Gaia2020} parallaxes and photometry \citep{Lindegren20,Riello2020}. Sect. \ref{results} describes how the stellar and disk parameters were obtained using the same procedure for all stars, as well as the online archive in which these data are publicly available. Sect. \ref{analysis} serves to exemplify the potential use of our dataset for the study of the HAeBe regime. It includes some general statistics that is mainly focused on the analysis of the different circumstellar properties of HAes and HBes as inferred from their SEDs. Finally, Sect. \ref{conclusions} includes a brief summary and conclusions.

\section{Sample and results} \label{results}

The initial sample we selected included the 252 HAeBes studied in \citet{Vioque_2018}. However, some sources were excluded for reasons related to strong variability or the lack of enough photometric points, which altogether prevented us from extracting the full set of parameters from an SED analysis. In addition, stars that were recently discarded as HAeBes in Appendix A of \citet{Vioque20} were not included here. Finally, some other sources were excluded because they lie below the MS in the HR diagram according to \citet{Vioque_2018} or W2020, and the new Gaia EDR3 distances do not change this situation (see also Sect. \ref{stellar_charac}). The SED-based stellar and circumstellar characterization was finally carried out for the 209 bona fide HAeBes with  coordinates (RA, DEC) listed in Cols. 1, 2, and 3 of Table \ref{Table:stellar_parameters}. Distance-independent stellar temperatures and visual extinctions inferred from them were derived by W2020 based on optical spectroscopy for 93 of the stars in our sample. These stars are indicated with italics, and their corresponding temperature (and extinction) values are included in Table \ref{Table:stellar_parameters} in addition to our estimates because they serve as a main reference (see Sects. \ref{stellar_charac} and \ref{analysis} and also Appendix \ref{appA}).

The SED analysis relies on the following distances and photometry. The distances were estimated based on the parallaxes provided in Gaia EDR3 (\citealt{Lindegren20}). Following \citet{Bailer21}, the inverse of the parallax was used to estimate the distances when the fractional parallax error (\textit{fpe}) was smaller than 0.1. A Bayesian method with a geometric prior from \citet{Bailer21} was used to derive the distances in case \textit{fpe} > 0.1, or when the parallax was negative. The distances for the 209 HAeBes in the sample are listed in Col. 4 of Table \ref{Table:stellar_parameters}. In addition, we assumed that a low-quality parallax and thus a potentially spurious distance can be obtained when any of the following criteria based on Gaia EDR3 parallaxes, uncertainties and quality flags are met: First, negative parallax or \textit{fpe} > 1. Second, renormalised unit weight error (RUWE) > 3. Third, astrometric{\_}excess{\_}noise > 0.5 and astrometric{\_}excess{\_}noise sig > 2. Finally, RUWE > 1.4 and ipd{\_}gof{\_}harmonic{\_}amplitude > 0.1. The last three criteria were defined according to Table 4 of \citet{Lindegren20} and Sect. 3.2 of \citet{Fabricius20}. The 42 objects indicated with the dagger in Table \ref{Table:stellar_parameters} have low-quality parallaxes according to the criteria listed above. Although we also derived the stellar and circumstellar properties of these objects, they were not considered in the analysis (Sect. \ref{analysis}). The potential spuriousness of Gaia EDR3 astrometric solutions is subject of ongoing work, for which we adopted conservative criteria. For instance, essentially all stars with low-quality parallaxes based on the very recent work by \citet{Rybizki21} (not yet submitted, according to the astro-ph notes) are also identified here, although roughly half of the sources with low-quality parallaxes according to our criteria are not classified as such based on \citet{Rybizki21}. 

We used the online tool Virtual Observatory SED Analyser (VOSA)\footnote{\url{http://svo2.cab.inta-csic.es/theory/vosa.}}, developed by the Spanish Virtual Observatory, to automatically compile photometry from the UV to the IR from the available catalogs in the Virtual Observatory\footnote{\url{http://svo2.cab.inta-csic.es/theory/vosa/help.php?what=credits#credits:vophot}}. In addition to the new Gaia EDR3 photometry \citep{Riello2020}, we included data from GALEX\footnote{Galaxy Evolution Explorer} \citep{Bianchi_2000}, APASS\footnote{AAVSO Photometric All-Sky Survey} 9 \citep{Evans_2002}, the Stroemgren-Crawford uvby$\beta$ photometry catalog \citep{Pauzen_2015}, 2MASS\footnote{Two Micron All-Sky Survey} \citep{Skrutskie_2006}, WISE\footnote{Wide-Field Infrared Survey Explore} \citep{Wright_2010}; IRAS\footnote{Infrared Astronomical Satellite} \citep{Beichman_1988}, Spitzer \citep{Evans_2009}, and AKARI (\citealt{Kawada_2007}; \citealt{Ishihara_2010}), among others. Information about the quality of the photometry is available in the catalogs, and the photometric points with poor-quality flags were discarded. Additional UBVRI and L,M photometry was extracted when necessary (\citet{Reed_2003, Vieira_2003, Kun_2009}; \citet{Alfonso12}; \citet{Mendigutia_2012, Zacharias_2012} and \citet{Fairlamb_2015}). Fluxes at millimeter wavelengths were also used in this work, as indicated in Sect. \ref{Sect:disk_mass}.

\subsection{Stellar characterization}\label{stellar_charac}

VOSA allows us to determine the stellar parameters comparing the observed SED with the synthetic photometry from photospheric theoretical models using a ${\chi}^2$ test \citep[see][for details]{Bayo_2008}. This tool enables us to restrict the models to those within a given range of stellar temperatures (T$_{\rm *}$). We used this option by using the models with T$_{\rm *}$ values that were constrained by the error bars provided by \citet{Vioque_2018}, except for the HAeBes that are listed in W2020, for which their more recent and homogeneous estimates have been considered (see Col. 5 of Table \ref{Table:stellar_parameters}). It is also possible to constrain the visual extinction, A$_{\rm v}$, but we let it be virtually free, providing a wide range of A$_{\rm v}$ from 0 to 10 mag for all HAeBes. VOSA then provides the combination of A$_{\rm v}$ and T$_{\rm *}$ that best fits Kurucz ODFNEW/NOVER models \citep{Castelli_1997} to the dereddened
SEDs using the interstellar extinction law by \citet{Fitzpatrick_1999} that was improved by
\citet{Indebetouw_2005}. The final A$_{\rm v}$ and T$_{\rm *}$ values resulting from these fits are listed in Cols. 6 and 7 of Table \ref{Table:stellar_parameters}. Uncertainties in both parameters were estimated by performing a 100-iteration \textit{\textup{Monte Carlo}} simulation. In the case of T$_*$, if the standard deviation is larger than half the grid step for the temperature, VOSA reports it as the uncertainty. Otherwise, half the grid step is the uncertainty of this parameter. Details about the procedure can be obtained from the online help provided in the VOSA web page. The uncertainties for the remaining stellar parameters (see below) depend on the previous uncertainties, which only reflect error bars derived from the SED fitting. A comparison with earlier determinations is included in Appendix \ref{appA}. The Kurucz models also depend on the surface gravities and metallicities, which were varied within the range of $\log g$ between 2.0 and 4.5 ($g$ in cm s$^{-2}$), and [Fe/H] between -0.5 and +0.5 (where [Fe/H] = ($\log N_{\rm Fe}/\log N_{\rm H})_\odot - (\log N_{\rm Fe}/log N_{\rm H})_*$). However, the corresponding best-fit results are not tabulated because they do not produce significant changes in the shape of the SEDs, and thus they may not be representative.  

Photometry at wavelengths shorter than U band or longer than J band was not generally considered to fit the photospheres because an excess may be present due to accretion and dust emission, respectively. However, after visual inspection, several sources without excesses show photospheric emission extending over a wider wavelength range. In these cases, the fitting procedure was repeated considering that range. This resulted in a lower ${\chi}^{2}$ value. In addition, only the photometric points with the highest fluxes were fit when multiple, scattered data were available for the same wavelengths. This serves to reflect the brightest (less extincted) state in variable sources. A minimum of six photometric points in the optical were used to carry out the fit for each star, which also served to minimize uncertainties. The dereddened SEDs and their photospheric fits are plotted in Appendix \ref{appC}. 

After we fit the photospheres, spectral types were associated with the resulting stellar temperatures using the relation in \citet{Kenyon_Hartmann_1995}. They are listed in Col. 8 of Table \ref{Table:stellar_parameters}. In addition, VOSA derives the stellar luminosity as L$_*$ = 4$\pi$d$^{2}$F$_*$, where F$_*$ is the dereddened flux derived from the integration of the best-fitting model \citep[see Appendix A in][]{Bayo_2008}. The stellar radius was then estimated as R$_*$ = (L$_*$/4$\pi$ $\sigma_{\rm SB}$ T$^4_*$)$^{1/2}$, where $\sigma_{\rm SB}$ is the Stefan-Boltzmann constant. The stellar radius and luminosity for each star, as well as their propagated errors, are listed in Cols. 9 and 10 of Table \ref{Table:stellar_parameters}.

Finally, VOSA allows us to estimate the stellar masses (M$_*$) and ages (t$_*$) through the comparison with several isochrones and evolutionary tracks. In our case, PARSEC V2.1s\footnote{\url{https://people.sissa.it/~sbressan/parsec.html}} stellar tracks and isochrones from \citet{Bressan_2012} were used. Using the error bars in T$_{*}$ and L$_*$ obtained in the fit process, we derived the minimum and maximum possible values for M$_*$ and t$_*$, which were used as an estimate of the corresponding error bars. Stellar ages, masses, and their corresponding errors are included in Cols. 11 and 12 of Table \ref{Table:stellar_parameters}.

Three objects for which different initial constraints of temperature were used with respect to those described previously deserve special mention. A wider initial temperature range was given for HD 290500 and HD 288012 (8500-10500 K and 7250-12250 K, respectively), whereas the relatively narrow range around the temperature value determined in \citet{Meeus_2010} was fixed for HD 169142 (7000-8000 K). We followed a different strategy for these HAeBes because under the initial approach, they appear to be located below the MS line in the HR diagram. The new T$_{\ast}$ constraints provide fits that are consistent with the PMS zone and allow VOSA to derive reasonable values for M$_*$ and t$_*$. 

\subsection{Circumstellar characterization}
The previous SEDs and stellar parameters allow us to characterize circumstellar properties of the sample of HAeBes in terms of IR-SED classification, disk-to-star accretion rates, disk masses, and inner disk sizes. We describe these next.

\subsubsection{SED classification}\label{sect:SEDs}

Infrared-SEDs were classified into groups I and II from the \citet{Meeus_2001} scheme, which we call ``M01'' classification hereafter. The ratio between the near-IR and the mid-IR luminosity, $L_{NIR}/L_{IR}$, and the non-color-corrected IRAS color, $[12]-[60]$, were used to carry out this classification following the procedure in \citet{Boekel_2003} and \citet{Acke_2004}. When the required J, H, K, L, M, or IRAS photometry was not available, the corresponding fluxes were estimated by interpolation from adjacent data. A few HAeBes that lie close to the limit of groups I or II or have IRAS fluxes that are upper limits were classified from direct visual inspection of their SEDs and comparing these classifications with those in other works \citep{Acke_2005, Acke_2006,Acke_2010, Mendigutia_2012}. As a result, four stars in the sample have a dubious group I or II classification. In addition, 23 sources cannot be classified because photometry was not available in the relevant ranges. It is also noted that the IRAS bands may be contaminated by the environment close to some stars (\citealt{Verhoeff_2012}; \citealt{Jimenez_2017}), which may affect the M01 group assignment.

\begin{figure}[h]
   \centering
   \includegraphics[width=\hsize]{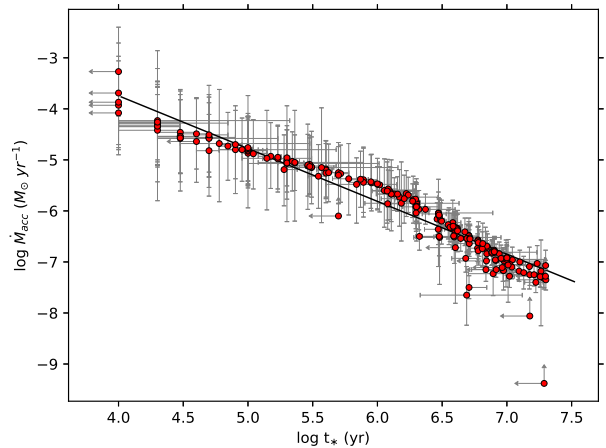}
      \caption{Mass accretion rate vs. stellar age. The best power-law fit ($\eta = 1.03 \pm 0.02$) is plotted with a solid line. Upper and lower limits are indicated by arrows.
              }
      \label{eta}
\end{figure}

In addition, the SEDs were also classified in a classical approach similar to that in \citet{Strom89}. In particular, we followed the criteria adopted in \citet{Mendigutia_2012} based on the shortest wavelength from which the infrared excess starts. This wavelength corresponds to the first photometric point that is not fit by the photosperic model from VOSA, which in turn is related to the sizes of the inner dust disk holes (see Sect. \ref{innerdisks}). In this way, sources belong to groups J, H, or K if the shortest wavelength at which the IR excess is apparent corresponds to the these near-IR bands (1.24, 1.66, and 2.16 $\mu$m, respectively), and they belong to group {$>$ K} if the IR starts at $\lambda$ $>$ 2.16 $\mu$m. This scheme is called JHK classification hereafter. The results of the SED classification from the M01 and JHK criteria are listed in Cols. 2 and 3 of Table \ref{Table:SEDs}.

\subsubsection{Disk-to-star accretion rates}\label{Sect:arates}

Accretion luminosities (L$_{\rm acc}$) were derived from the new empirical correlations with L$_*$ quantified in W2020, which depend on the stellar mass: log (L$_{\rm acc}$/L$_{\odot}$) = (-0.87 $\pm$ 0.11) + (1.03 $\pm$ 0.08)$\times$log(L$_*$/L$_{\odot}$) for M$_*$ $<$ 4 M$_{\odot}$, and log (L$_{\rm acc}$/L$_{\odot}$) = (0.19 $\pm$ 0.27) + (0.60 $\pm$ 0.08)$\times$log(L$_*$/L$_{\odot}$) for M$_*$ $>$ 4 M$_{\odot}$. Mass accretion rates were then estimated from the usual expression,

\begin{equation}
\dot{M}_{acc}=\frac{L_{acc}R_{*}}{GM_{*}}\Bigg(1-\frac{R_{*}}{R_{i}}\Bigg)^{-1}\label{macc},
\end{equation}
where R$_{i}$ is the disk truncation radius from which gas is channeled onto the star. For simplicity, we assumed that R$_{i}$ = 2.5 R$_{*}$ for all sources \citep{Mendigutia_2011a}. Mass accretion rates and accretion luminosities are included in Cols. 4 and 5 of Table \ref{Table:SEDs}. 

\begin{figure}[h]
   \centering
   \includegraphics[width=\hsize]{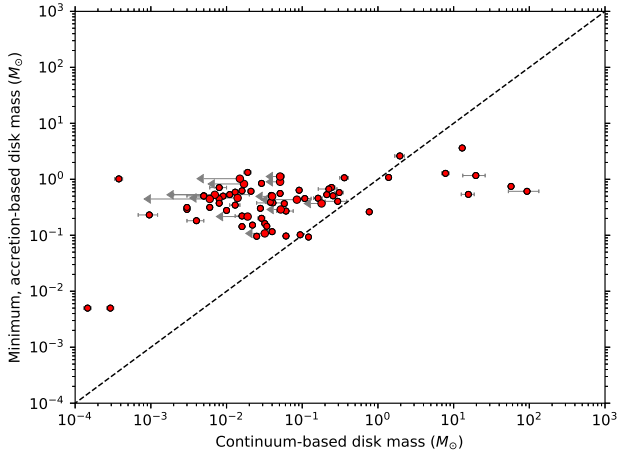}
      \caption{Disk masses from accretion and age vs. those from dust emission at millimeter wavelengths. Upper limits are represented by arrows. The dashed black line indicates equal values. 
              }
      \label{gas_dust}
\end{figure}

Although accretion estimates from emission line luminosities and from L$_*$ are roughly equivalent and accurate for most stars \citep{Mendigutia_2015}, the latter was chosen because it allowed us to derive accretion rates based on self-consistent data derived in this work using a homogeneous criterion for the whole sample. Emission line luminosities (in H$\alpha$ or any other line) in the literature come from relatively heterogenous measurements, and more importantly, they are not available for all stars. However, accretion luminosities and mass accretion rates based on empirical correlations with the stellar (or emission line) luminosities may be incorrectly high in a few sources that do not show evidence of accretion based on a direct probe such as the near-UV excess \citep[see][for the corresponding measurements in some HAeBes]{Mendigutia_2011b,Fairlamb_2015}. Moreover, it must be taken into account that mass accretion rates for HBes have strong associated uncertainties because accretion rates are poorly known in this regime (for details, see, e.g., \citealt{Mendi20}; W2020, and references therein).

\subsubsection{Disk masses} \label{Sect:disk_mass}

Continuum fluxes at millimeter and submillimeter wavelengths are commonly used to derive total (dust + gas) disk masses (M$_{\rm disk}$), making the assumption that the emission at these wavelengths is optically thin and that the gas-to-dust mass ratio is equal to 100 \citep[e.g.,][]{Beckwith_1990}. In the Rayleigh-Jeans limit,

\begin{equation}
M_{disk}=F_{\nu}\frac{d^{2}}{2kT_{D}}\frac{c^{2}}{\nu^{2}\kappa_{\nu}} \label{Disk_Mass_flux},
\end{equation}

where F$_{\nu}$ is the measured flux (normally at 1.3 mm), T$_{\rm D}$ is the dust temperature, $k$ is the Boltzmann constant, and $\kappa_{\nu}$ is the opacity. Total disk masses are determined from the previous expression for $\sim$ 1/3 of the stars in the sample (those for which previous F$_{\nu}$ measurements are available). We adopted $\kappa_{\rm 1.3mm}$ = 0.02 cm$^{-2}$gr$^{-1}$ , which already includes a gas-to-dust ratio = 100, and a wavelength dependence $\kappa_{\rm mm}$ $\propto$ $\lambda^{-\beta}$ \citep{Beckwith_1990}\footnote{i.e., opacities at different wavelengths are given by $\kappa_{\lambda} = \kappa_{\rm 1.3mm}/(\lambda/1.3)^{-\beta}$}, where 0 $<$ $\beta$ $\leq$ 2 is the dust opacity index associated with the typical dust grain size probed at (sub-) millimeter wavelengths (1.5-2 for the interstellar medium and smaller for larger grains). The values for T$_{\rm D}$ and $\beta$ are inferred from the best graybody model that fits the observed millimeter photometry using a ${\chi}^2$ criterion \citep[see, e.g.,][]{Andre_1993,Sandell_2000,Mendigutia_2012}. For the HAeBes with one single measurement of the millimeter flux, $\beta$ was fixed to 1 and T$_{\rm D}$ to the value in Table II of \citet{Natta_2000} associated with the corresponding spectral type. The values of $\beta$, T$_{\rm D}$ and M$_{\rm disk}$ are listed in Cols 2, 3, and 4 of Table \ref{Table:Disks_mm} whenever these estimates have been possible, and Cols 5 and 6 show the number of photometric points and the corresponding references that we used. The errors in M$_{\rm disk}$ were determined by propagation, considering only the uncertainties in the (sub-) millimeter fluxes. For the few flux values without associated error bars in the literature, typical $\sim$ 10$\%$ uncertainties were assigned.

Nevertheless, several studies have pointed out that disk masses derived from dust continuum emission could be underestimated \citep[see, e.g., ][and references therein]{Zhu19,Ballering_Eisner_2019}. An alternative method for estimating the disk mass is followed in this paper. It is based on previous works that inferred that parameter from measurements of the stellar mass accretion rates and ages \citep{Hartmann_1998,Andrews_Williams07,Mendigutia_2012,Dong18}. Given that a significant fraction of the disk is dissipated through accretion onto the star, a lower limit for the current gas disk mass is given by 

\begin{equation}
M_{disk}^{min}(t_{*}) \equiv M_{disk}^{min} = \int_{t_{*}}^{t_{MS}}\dot{M}_{acc}(t)dt \label{Disk_Mass_acc},
\end{equation} 

$t_{\rm MS}$ being the zero-age main sequence, when the gas disk mass is negligible. A relation between $\dot{\rm M}_{\rm acc}$ and age with the shape $\dot{\rm M}_{\rm acc} = {\rm A}{\rm t}^{\rm -\eta}$ is commonly considered \citep[e.g.,][]{Hartmann_1998,Sicilia-Aguilar10,Mendigutia_2012,DeMarchi17,Arun_2019}, where the constant $A$ is determined under the condition $\dot{\rm M}_{\rm acc}(\rm t_{*}) = \dot{\rm M}_{\rm acc}$. We replaced $\dot{\rm M}_{\rm acc}$ by the power law in equation \ref{Disk_Mass_acc} and resolved the integral,

\begin{equation}
M_{disk}^{min} = \frac{\dot{M}_{acc}(t_{\ast})t_{\ast}}{\eta -1} \times \left\lbrace 1 - \Bigg(\frac{t_{\ast}}{t_{MS}}\Bigg)^{\eta-1}\right\rbrace \label{Mass_macc}
.\end{equation} 

In order to derive M$_{\rm disk}^{\rm min}$ values from the previous expression, the value of $\eta$ characterizing the accretion evolution of the HAeBes is needed. Figure \ref{eta} shows the accretion rates and stellar ages that were previously derived for the stars in our sample. The best power-law fit without considering upper and lower limits has an exponent $\eta$ = 1.03 $\pm$ 0.02. This value indicates a slightly shallower decline than previously estimated by \citet{Mendigutia_2012} ($\eta$ = 1.8$^{+1.4}_{-0.7}$) and \citet{Arun_2019} ($\eta$ = 1.2 $\pm$ 0.1). M$_{\rm disk}^{\rm min}$ was then estimated for each star in our sample from Eq. \ref{Mass_macc}, using t$_{\rm MS}$ values from \citet{Tayler_1994}. Negative disk masses were obtained for 31 stars with t$_{*}$ > t$_{\rm MS}$, as indicated in Table \ref{Table:SEDs}. In these cases, M$_{\rm disk}^{\rm min}$ was derived again assuming either the minimum values of t$_{*}$ and M$_{*}$ from this work or the values of the same stellar parameters plus the $\dot{\rm M}_{\rm acc}$ values from the literature (\citealt{Vioque_2018}, W2020). Furthermore, for HAeBes with upper limits on t$_{*}$ and M$_{*}$, the values from \citet{Vioque_2018} and W2020 were also  used (for the specific case of HD 169142, the age was obtained from \citealt{Grady_2007}). M$_{\rm disk}^{\rm min}$ values for all stars in the sample are included in Col. 6 of Table \ref{Table:SEDs}.

Figure \ref{gas_dust} compares the disk mass estimates from dust continuum emission and from accretion rates. Six sources have extremely high continuum-based disk masses $>$ 2M$_{\odot}$ that may reflect contamination of the millimeter fluxes from envelopes or the surroundings. For the others, our data support previous claims indicating that continuum-based disk masses are typically lower than those inferred from accretion by $\text{about}$ one order of magnitude \citep{Hartmann_1998,Andrews_Williams07,Mendigutia_2012,Dong18}, probably suggesting that dust emission does not reflect the whole solid population in a protoplanetary disk. However, we note that accretion-based disk masses can be unrealistically high in several sources because the accretion rate decline commonly inferred through the comparison between accretion rates and ages (like in Fig. \ref{eta}) does not consider the corresponding nonaccreting fractions, which can be significant at least in well-studied samples of TTs \citep[e.g.,][]{Fedele_2010}. Moreover, the accretion rate decline shown in Fig. \ref{eta} also reflects the underlying dependence of $\dot{\rm M}_{\rm acc}$ and M$_*$ \citep[as described in Sect. \ref{sect:intro}, more massive objects are also younger and thus show higher accretion rates; see, e.g.,][]{Mendigutia_2011b}. W2020 studied the accretion rate decline by dividing the sample of HAeBes into different stellar mass bins, showing that the decay of accretion with age persists for the mass range 2.0-2.5 M$_{\odot}$ but perhaps not for more massive stars that were less well sampled. Accretion-based disk masses are provided here for the whole sample and not just for a fraction like continuum-based disk masses (Tables \ref{Table:SEDs} and \ref{Table:Disks_mm}, respectively), but the the previous caveats concerning both types of estimates must be taken into account.  

\subsubsection{Inner dust holes}
\label{innerdisks}
The sizes of the inner dust holes (r$_{\rm in}$) for the HAeBes of our sample were estimated from their SEDs as follows. First, we assumed that the disk temperature decreases with the distance as T$_{\rm disk}$ = $K$ $\times$ r$^{-3/4}$ \citep[e.g.,][]{Armitage_2009}. For each star, the constant $K$ was estimated by replacing T$_{\rm disk}$ by the dust sublimation temperature T$_{\rm sub}$, assumed to be 2000 K, and $r$ by the dust sublimation radius, r$_{\rm sub}$ = R$_*$(T$_{\rm sub}$/T$_*$)$^{-2.1}$ \citep{Robitaille_2006}. When the value of the constant is known, the size of the inner dust holes corresponding to the disk temperatures where the IR excess starts, determined from the Wien law and the SEDs, were derived from the above expression for T$_{\rm disk}$. The size of the inner dust holes is listed in Col. 8 of Table \ref{Table:SEDs} along with its corresponding errors, which were determined by propagation considering the effective width of the photometric filter where the IR excess starts as the only uncertainty. As described in Sect. \ref{sect:SEDs}, the SED classification based on the JHK groups is directly related to the size of the inner disk dust holes estimated here, with the relative sizes increasing from group J to group $>$ K. 

\subsection{Online archive of HAeBes}
All data derived in the previous sections are collected in an online archive of HAeBes\footnote{``HArchiBe'': http://svo2.cab.inta-csic.es/projects/harchibe/}, including the already introduced tables associated with this paper. This is a Virtual Observatory-compliant archive built in the framework of the Spanish VO using the SVOCat \footnote{http://svo2.cab.inta-csic.es/vocats/SVOCat-doc/} publishing tool. Additional features of the archive include individual figures showing the SEDs, and a tool for visualizing the on-sky position of the stars. The archive will be updated to include new HAeBes that are confirmed and characterized in the future, constituting a unique tool for the study of this type of objects from a homogeneous characterization of their properties.

\begin{figure}[h]
   \centering
   \includegraphics[width=\hsize]{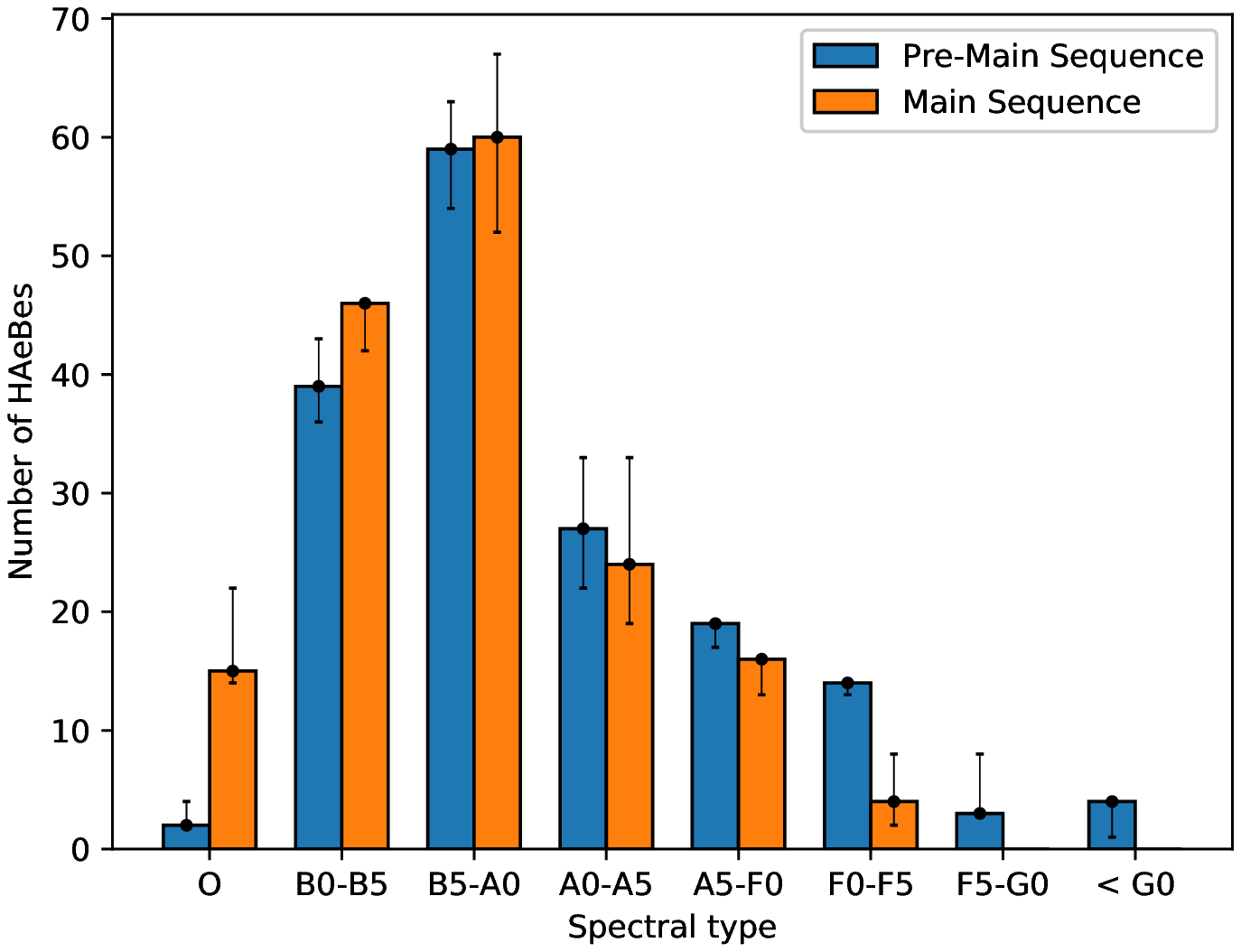}
   \includegraphics[width=\hsize]{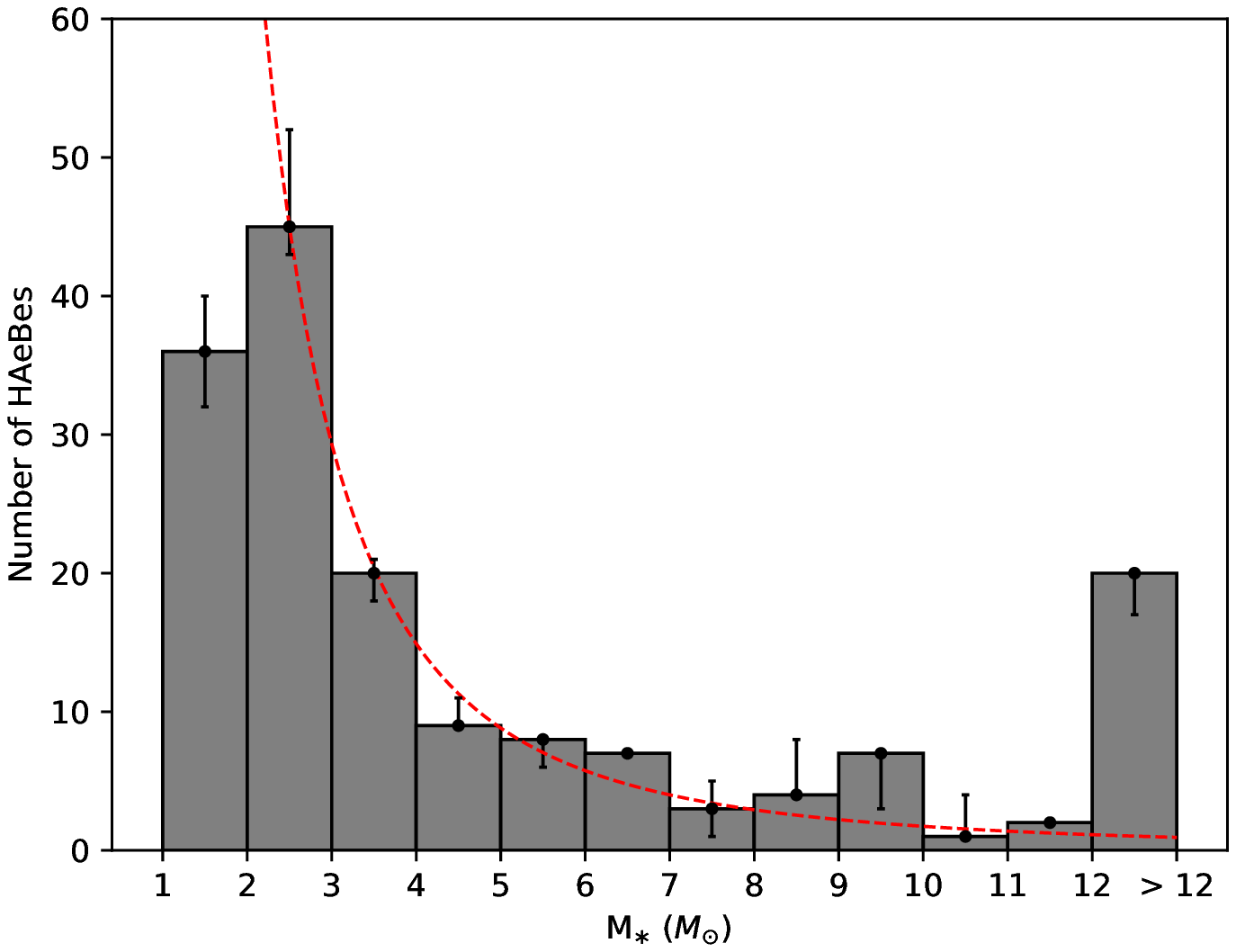}
         \caption{\textbf{Top panel}: Current distribution of spectral types in the PMS and expected distribution when the stars reach the MS, as indicated in the legend. The error bars reflect the uncertainties on the spectral types. \textbf{Bottom panel}: Number of HAeBes vs. stellar mass. The error bars for the number of stars per bin considering the uncertainties in M$_*$ are indicated. The dashed red line corresponds to the IMF \citep{Salpeter_1955} normalized to the bin 2-3 M$_\odot$. Stars with masses > 12 M$_\odot$ have been grouped together for simplicity.}
         \label{Sp_IMF}
\end{figure}

\section{Analysis and discussion}\label{analysis}

In this section we carry out a statistical analysis and discuss part of the previous results, focusing on different relations in the M01 and JHK SED classifications, stellar parameter accretion rates, and the mechanisms dissipating the inner part of the disks.
Stellar and circumstellar parameters derived in previous sections and based on up to date Gaia EDR3 distances are used in the analysis, with the only exception of the stellar effective temperature (i.e., spectral type). Although our own estimates are used for the majority of the stars, the effective temperatures in W2020 \citep[and the corresponding spectral types based on the relation in][]{Kenyon_Hartmann_1995} are used for the subsample included in both works. The reason is that T$_*$ is a distance-independent parameter that was derived homogeneously by W2020 based on optical spectra instead of photometry, thus using a more accurate method. However, because we used the error bars provided by W2020 to define the initial ranges of T$_*$ for the SED fitting (Sect. \ref{stellar_charac}), the overall results remain essentially the same when our T$_*$ estimates are used for the whole sample. Finally, we recall that the 42 objects whose parallaxes might be spurious according to the discussion at the beginning of Sect. \ref{results} are not considered in this statistical study. 

\begin{figure}[h]
   \centering
   \includegraphics[width=\hsize]{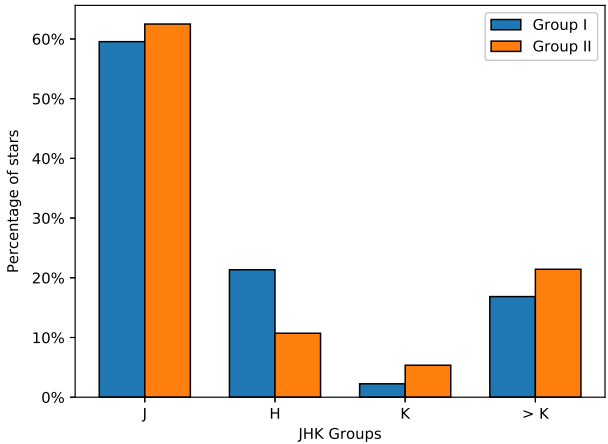}
      \caption{Histogram comparing the distributions of the JHK groups (based on the wavelength at which the infrared excess starts) and M01 groups (based on the shape of the SEDs), as indicated in the x-axis and the legend.
              }
      \label{meeus_jhk}
\end{figure}

\subsection{Representativeness of the sample}\label{sect: representativeness}

This work provides the largest dataset to date of stellar and circumstellar parameters of HAeBes homogeneously derived from SEDs. However, this sample was compiled from catalogs that do not take into account completeness criteria concerning specific star-forming regions or brightness limits \citep[see the detailed discussion in][]{Vioque_2018}, as is common for lower-mass TTs, for instance. Before we carry out any statistical analysis, it is therefore worth asking what the sample represents. As we mentioned in the introduction, HAeBes are heterogeneous by definition, and the specific distributions of HAes and HBes concerning stellar mass and age must be taken into account when statistical results are interpreted (see Sects. \ref{Sect:disk_mass}, \ref{Sect: M01 and SEDs}, \ref{Sect:stellar parameters and JHK}). In addition to the well-known caveats, we discuss below whether our sample represents HAes and HBes well, or if, on the contrary, the relative number of stars included in the subsamples is physically unrealistic. 

Figure \ref{Sp_IMF} (top) shows in blue the distribution of the sample as a function of the spectral type. The sample contains 46 HAes and 98 HBes (mostly late type), the remaining being within the boundaries, 21 IMTTs, defined here as the stars with F and later spectral types, and 2 O-type stars. These numbers imply proportions of $\sim$ 28$\%$ and $\sim$ 59$\%$ of HAes and HBes, respectively. That the number of HBes is larger than the number of HAes was discussed in \citet{Vioque_2018} (see also \citealt{The_1994}), who pointed out that it probably reflects an observational bias resulting from the fact that HBes are brighter and extend over a larger volume in the sky. As we show next, this bias is even more significant when we consider the fractions of A and B stars after the HAeBes reach the MS. 

\begin{figure}[h]
   \centering
    \includegraphics[width=\hsize]{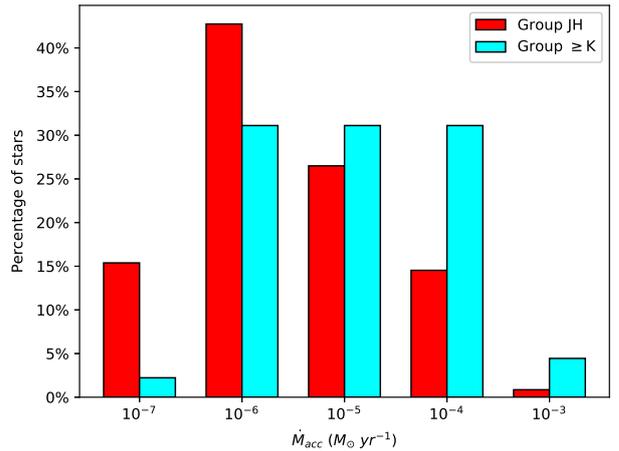}
         \caption{Histogram comparing the distributions of mass accretion rates and JHK groups, as indicated in the x-axis and the legend.}
         \label{Hist_disk}
\end{figure} 

The orange histogram in Figure \ref{Sp_IMF} (top) represents the future distribution of spectral types that the objects would have when they settle into the MS. These predicted spectral types were derived from the stellar temperatures corresponding to the points at which the MS starts, as inferred from the evolutionary tracks of each star. Although IMTTs will transform into A-stars (\citealt{Calvet_2004}), most of these objects are not considered in the original HAeBe catalogs from which our sample has been compiled. In turn, many HAes will transform into B sources, and only a few early-type HBes will end up as O-sources when they reach the MS. As a result, the ratio between the number of B and A in the MS (2.7) is even higher than the corresponding ratio between HBes and HAes in the PMS (2.1). This is a consequence of our own definition of HAeBes as ``the massive counterparts of classical TTs'' , that is, emission line PMS stars with spectral types A and B, instead of ``the precursors of A and B MS stars''. If the latter definition were adopted, our sample of HAeBes would be biased in the sense that it lacks IMTTs and contains an excess of hot sources that will transform into O-stars. 

Figure \ref{Sp_IMF} (bottom) shows the stellar mass distribution for the HAeBes in our sample, which has to be very similar to the mass distribution when the objects reach the MS. The expected distribution of spectral types (orange bars in Figure \ref{Sp_IMF}, top) was derived under the approximation that the mass gained by the stars due to accretion during the optically visible PMS evolution is negligible compared with the stellar mass in this evolutionary stage (i.e., most stellar mass is accreted during previous, embedded phases). Therefore it is useful to compare the histogram in Fig. \ref{Sp_IMF} (bottom) with the IMF, which by definition reflects the expected mass distribution when the stars reach the 
MS. The comparison with the IMF from Salpeter (1955)\footnote{All inferred IMFs have virtually the same shape for the stellar mass range we analyzed.} confirms that our sample is indeed representative in this sense (\citealt{Vioque_2018}), at least within the stellar mass range between 2 and 12 M$_{\odot}$ that is normally associated with the HAeBe regime. As discussed above, the first bin in the figure is smaller than expected mainly because of the lack of IMTTs. In turn, the last bin is larger than expected because the stars with M$_*$ >12 M$_\odot$ were all grouped together.

New searches based on machine-learning algorithms and making use of Gaia data (e.g., \citealt{Vioque20}) will give much more complete sets of stars that are even better accommodated to the IMF and provide volume-limited samples that reflect more physically realistic fractions of HBes versus HAes. Still, the statistical results presented in the following sections are based on a sample of HAeBes that is representative in the sense that it roughly follows the IMF in the mass range $\sim$ 2-12M$_\odot$. 

\subsection{M01 and JHK SED classifications}\label{Sect: M01 and SEDs}

Figure \ref{meeus_jhk} shows the distribution of the stars according to the M01 and JHK SED classification schemes. No trend is apparent, and the SED shape according to the M01 scheme is not related to the wavelength at which the IR excess departs from the photosphere. A two-sample Kolmogorov-Smirnov (K-S) test rejects the null hypothesis that both samples are drawn from the same parent distributions at a 1$\%$ significance level. Group I sources were associated to the presence of imaged dust gaps and cavities (see references in the introduction), but these are difficult to trace based on the SEDs alone. In turn, \citet{Perraut19} hypothesized that group I sources may also have larger inner disk sizes than group II, although no definitive trend based on interferometric data was found in that work. Our result above does not support the hypothesis that the M01 group I stars also tend to be group K and group $>$ K sources, whose inner dust holes are comparatively larger than for the stars in groups J and H. Moreover, the larger number of group I stars compared to group II sources (89 vs. 56; the 21 remaining objects have a doubtful or unknown classification) sharply contrasts with the smaller fraction of transitional disks compared to full disks in the lower-mass regime, which roughly ranges between 20$\%$ and 50$\%$ depending on the age \citep{Currie11}. However, given that IMTTs are not well probed in the sample (Sect. \ref{sect: representativeness}) and they are the precursors of HAes (thus in principle hosting less evolved disks), the currently observed proportion of sources in groups J and H is potentially underestimated. In turn, if group II stars probe disks without holes, the inclusion of additional IMTTs in the sample would potentially increase the observed proportion of group II sources.

\begin{figure}[h]
   \centering
   \includegraphics[width=\hsize]{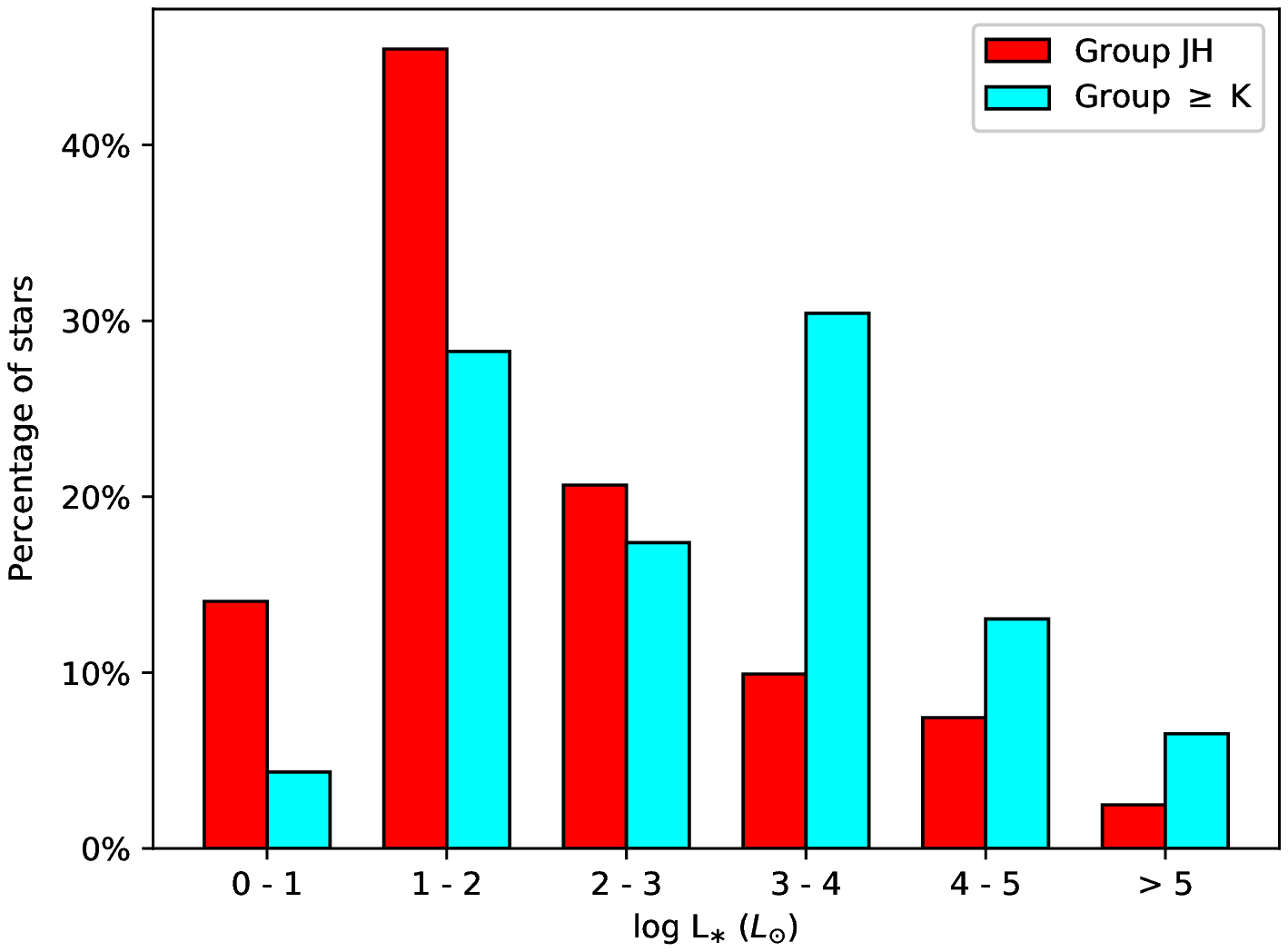}
   \includegraphics[width=\hsize]{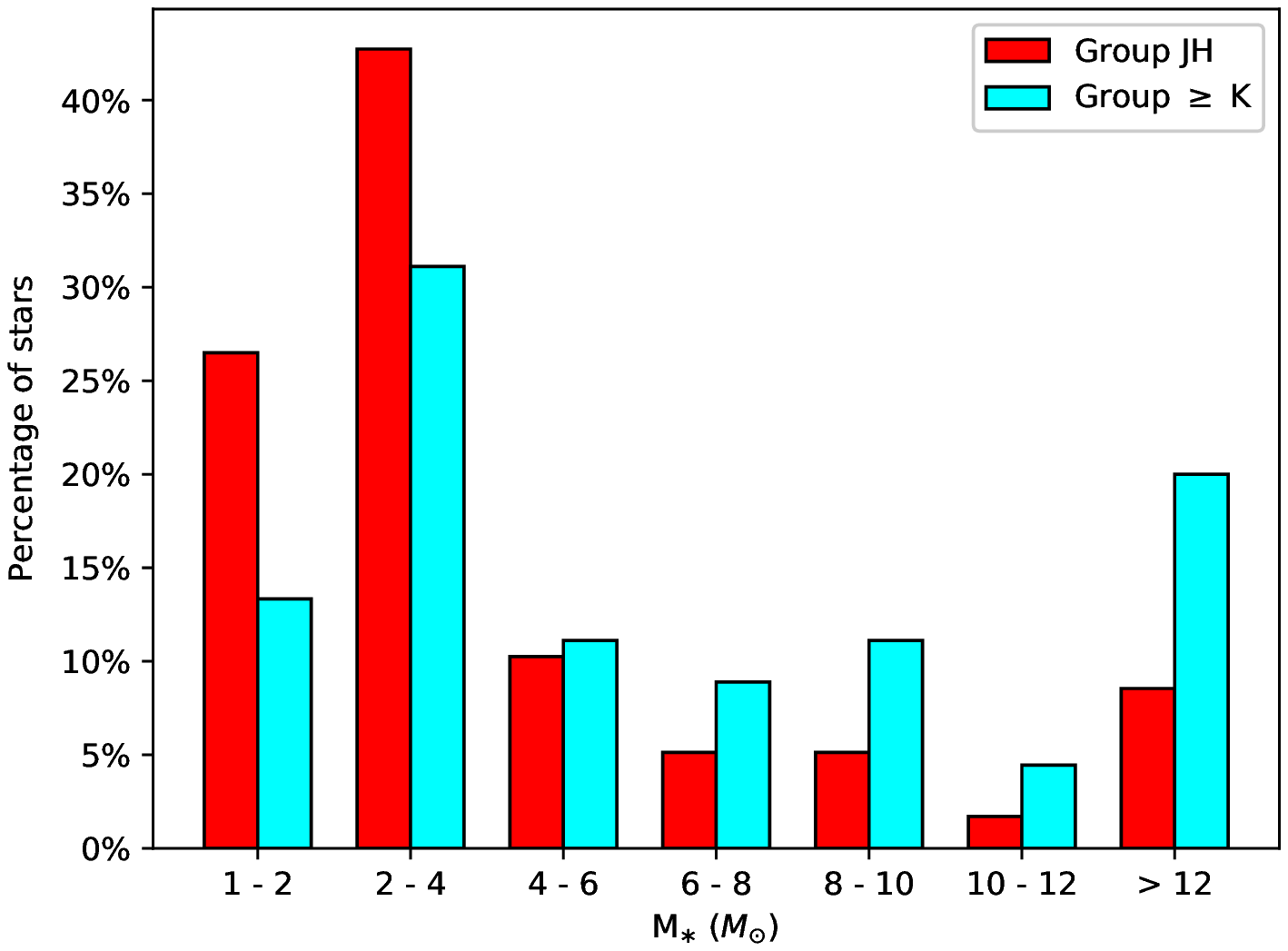}
   \includegraphics[width=\hsize]{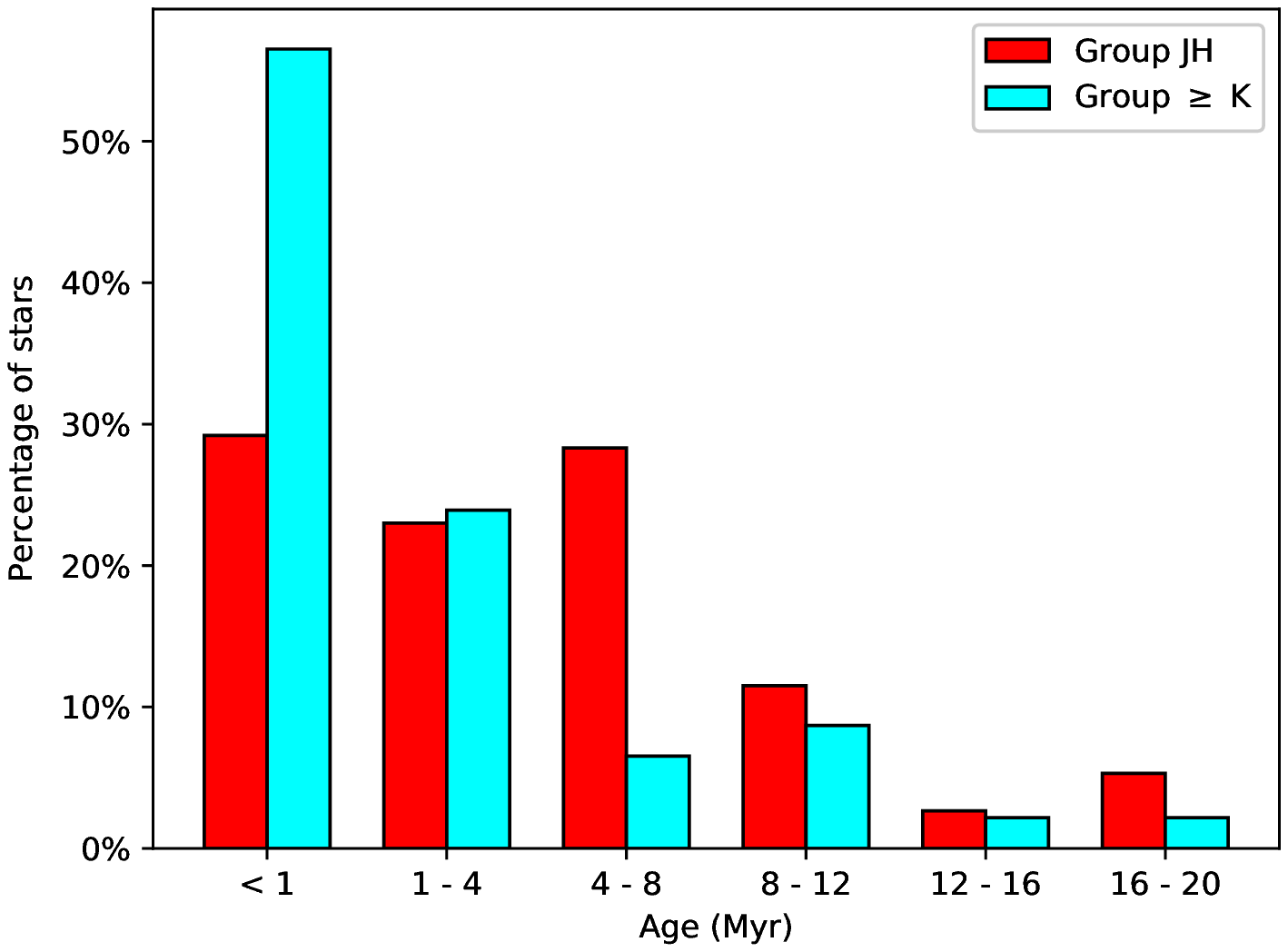}
         \caption{Histogram comparing the distributions of stellar parameters (luminosity, stellar mass, and age) and JHK groups, as indicated in the x-axis and the legend.}
         \label{Hist_stellar}
\end{figure}

The JHK classification can be roughly translated into the ``full'' and ``transitional'' disk nomenclature similar to that used in the lower-mass regime. We considered that HAeBes have ``full disks'' if their IR excess starts at the J or H bands (group J and group H stars), while HAeBes with ``transitional'' disks show IR excess only at longer wavelengths (group K and group $>$ K stars). We note that although the definition of transitional disk in TTs is usually based on an IR excess starting at wavelengths ($\sim$ 10 $\mu$m) longer than adopted here (K band), the size of the associated inner disk holes are roughly the same because HAeBes are hotter (see Sect. \ref{innerdisks} and the discussion in \citealt{Mendigutia_2012}). In this respect, the two definitions are therefore physically equivalent. However, our definition of a transitional disk for HAeBes does not consider the stars that have relatively weak emission at short, IR wavelengths (sometimes called ``pretransitional'' disks), but only the sources that are completely devoid of dust in their cavities according to their SEDs. Under the previous definitions and caveats, we found that 121 HAeBes have full disks and only 46 can be considered transitionals. This represents $\sim$ 28$\%$ of HAeBes, which is significantly smaller than the fraction of Class I HAeBes, but more similar to the fraction of transitional disks in TTs. We repeat that the inclusion of additional IMTTs in the sample would in principle decrease the fraction of transitional disks compared to what is currently observed.

On the other hand, our empirically based accretion rates do not show any apparent relation with the M01 groups, supporting previous claims in this respect \citep{Mendigutia_2012}. Moreover, our data also support previous findings indicating that group II sources tend to have larger dust grains than group I \citep{Acke_2004}, and that the former tend to show UXOr-like variability as well \citep{Dullemond03}. Concerning the two latter statements, the typical dust opacity index $\beta$ reflecting the submillimeter and millimeter dust grain size (Sect. \ref{Sect:disk_mass} and Table \ref{Table:Disks_mm}) changes from $\sim$ 1.15 (group I) to $\sim$ 0.72 (group II), and the sources classified as UXOrs in \citet{Vioque_2018} mostly belong to group II ($\sim$ 78 $\%$, vs. $\sim$ 22 $\%$ in group I).

\subsection{Accretion rates and JHK SED classification}

Figure \ref{Hist_disk} shows the distributions of the stars according to their $\dot{\rm M}_{\rm acc}$ values and the JHK SED classification scheme. HAeBes belonging to Group $\geq$K tend to show higher accretion rates, which is confirmed by a K-S test. This result disagree with a previous similar analysis in \citet{Mendigutia_2012}. The opposite result was found in that work, where HAeBes belonging to the JH group tend to be the strongest accretors. However, the sample analyzed in \citet{Mendigutia_2012} was comparatively smaller and mainly dominated by HAes, whereas our current result is driven mainly by the HBes in our sample, as we show below.

\subsection{Stellar parameters and JHK SED classifications}\label{Sect:stellar parameters and JHK}

\begin{figure}[h]
   \centering
   \includegraphics[width=\hsize]{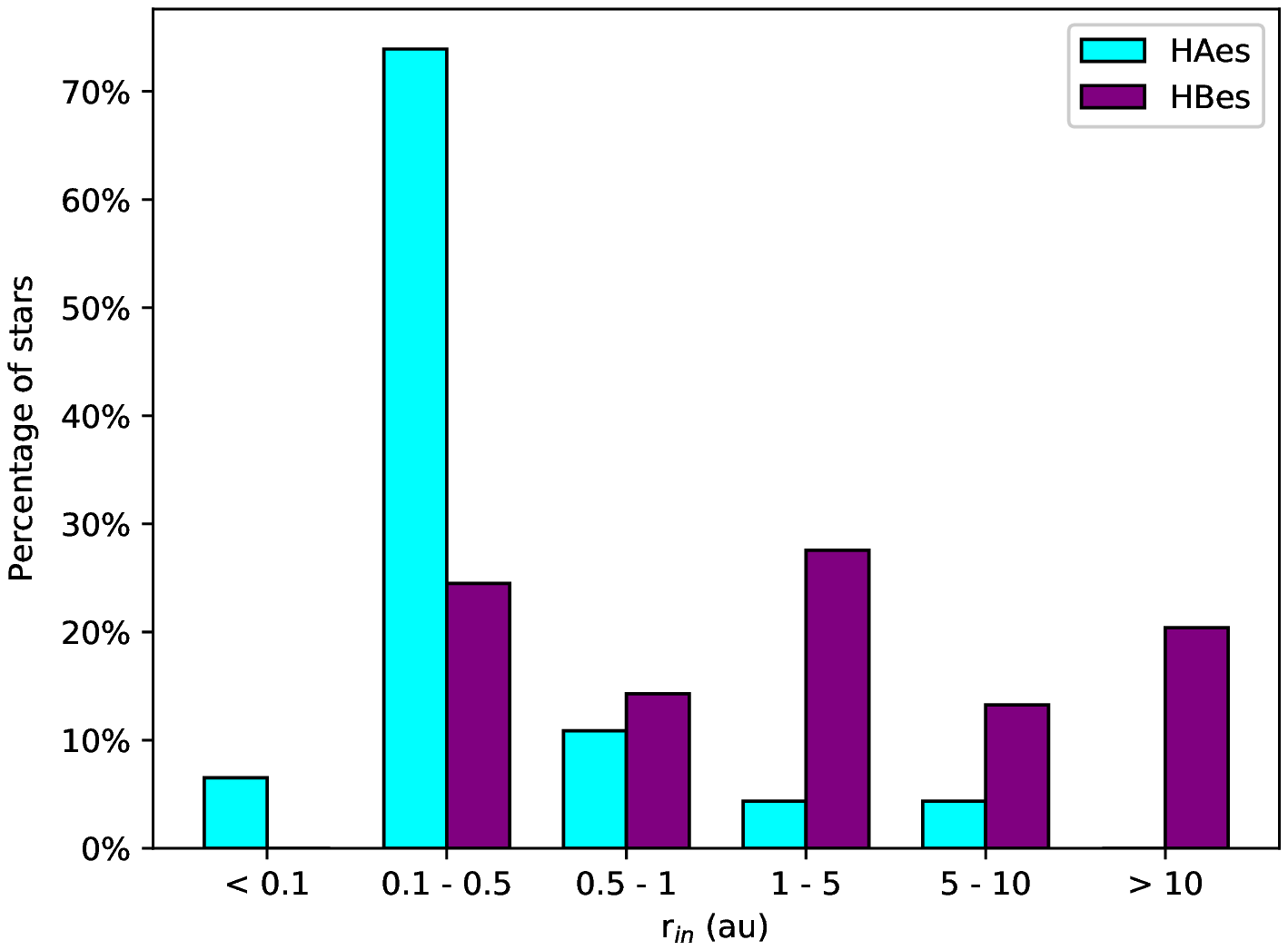}
   \includegraphics[width=\hsize]{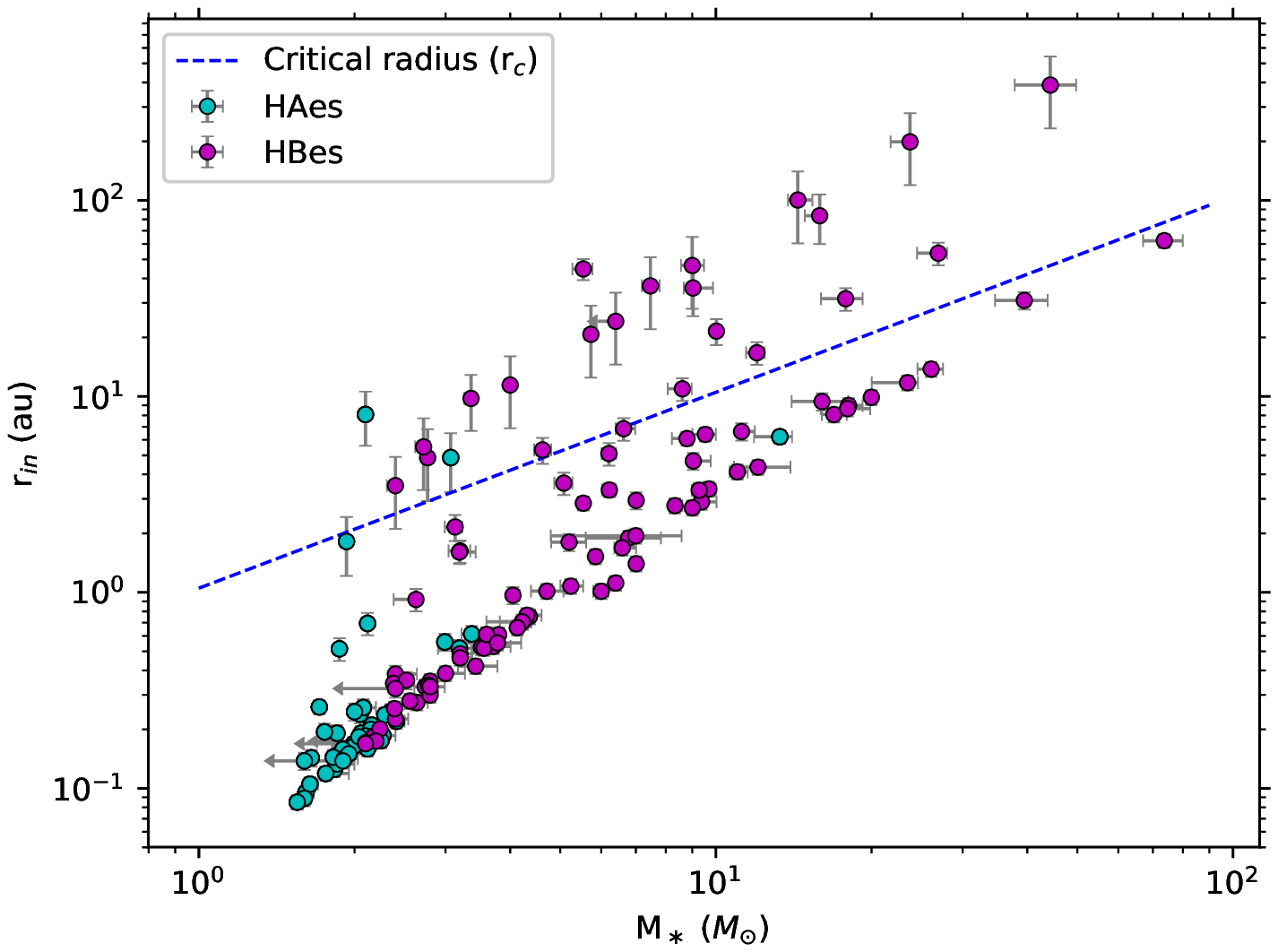}
         \caption{\textbf{Top panel}: Histogram comparing the distributions of sizes of the inner disk holes for HAes and HBes, as indicated in the x-axis and the legend. \textbf{Bottom panel}: Inner disk holes as a function of the stellar mass for HAes and HBes. The dashed blue line represents the critical radius above which the stars are consistent with the photoevaporation scenario (see text). Upper limits are indicated by arrows.}
         \label{inner_size}
\end{figure}

Figure \ref{Hist_stellar} shows three histograms representing the percentage of HAeBes classified in JHK groups as a function of the stellar luminosity, mass, and age. Two-sample K-S tests indicate that there are low probabilities that JH and $\geq$K groups are drawn from the same parent distribution when these groups are related to the stellar parameters. Group $\geq$K HAeBes are typically younger, brighter, and more massive than group JH HAeBes, which mainly corresponds to the early-type HBes in our sample. A similar analysis does not reveal any trend relating the M01 groups and the stellar parameters. In turn, the JHK groups are associated with the size of the inner dust holes (see Sect. \ref{innerdisks} and below). The top panel of Fig. \ref{inner_size} shows the distributions of r$_{\rm in}$ for HAes and HBes separately. HBes tend to have larger inner disk holes than HAes, which is confirmed by a K-S test. This is better seen in the bottom panel of Fig. \ref{inner_size}, where r$_{\rm in}$ is plotted against the stellar mass considering both subsamples. Associating groups K and $>$ K sources again to ``transitional'' disks, we found that $\sim$ 34$\%$ of the HBes in our sample have such disks, which is significantly higher than the $\sim$ 15$\%$ ratio found for the HAes. If less evolved IMTTs were well probed in the sample (Sect. \ref{sect: representativeness}), the difference between the fraction of transitional disks in HBes and lower-mass stars would in principle be even larger.

\subsection{Inner disk holes and associated physical processes}

The inner disk holes associated with excesses starting at relatively long wavelengths can be explained mainly by four different processes \citep[see, e.g.,][and references therein]{Espaillat14}: viscous evolution, substantial grain growth causing a depletion of the small dust particles in the inner regions, companions that sweep the material in their orbits, and/or photoevaporation. Viscous evolution alone is not expected to produce a preferential depletion of dust at small radii, especially considering that HBes are typically young ($\leq$ 1 Myr; see below) and the timescales involved would be too short. The dust opacity index $\beta$ indicates that although disks around HAeBes show grain growth compared to the interstellar medium, the median value is roughly the same in group JH as in group $\geq$K (1.2 and 1.1, respectively). Our data therefore contain no evidence to support the idea that dust grain growth in HBes with
IR excesses starting at long wavelengths is different from that in the rest of the stars. Similarly, the fraction of HAeBes in group JH with stellar companions \citep[as tabulated in][]{Vioque_2018} is larger than the same fraction in group $\geq$K ($\sim$ 33$\%$ vs. $\sim$ 22$\%$), which contradicts the hypothesis that these companions causing larger inner disk holes in the second group. Still, a population of (undetected) substellar companions and planets might explain the presence of inner disk holes. However, the previous arguments and the available data leave photoevaporation as the most plausible mechanism that dominates the formation of inner holes in many HBes.

\begin{figure}[h]
   \centering
   \includegraphics[width=\hsize]{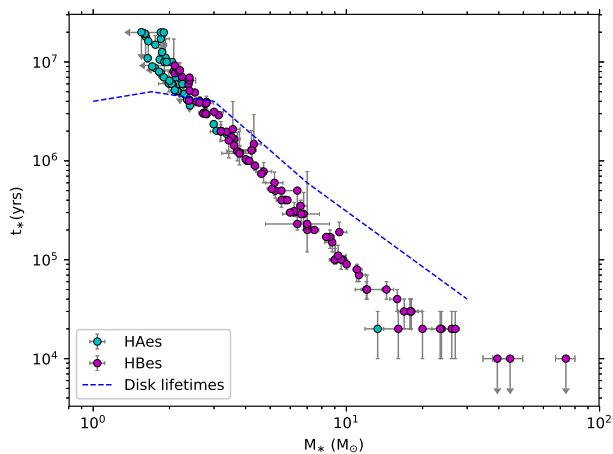}
       \caption{Stellar age vs. stellar mass for HAes and HBes. The dashed blue line represents the disk lifetime below which the stars are consistent with the photoevaporation scenario (see text). Upper limits are indicated by arrows.}
         \label{disk_life}
\end{figure}

In the photoevaporation scenario \citep[see, e.g.,][and references therein]{Alexander14} UV and/or X-ray photons cause photoevaporative winds limiting the resupply of inner disk material from accretion flows and producing a relatively fast inside-out disk dispersal. HBes have associated intrinsically high UV luminosities and excesses \citep[e.g.,][]{Fairlamb_2015}, and wind signatures in their spectra are significantly stronger than for HAes \citep[e.g.,][]{Mendigutia_2011a}. The fact that HBes show lower IR excess, disk masses, and accretion rates than expected from the trend followed by TTs and HAes (\citealt{Alonso_2009}; \citealt{Vioque_2018}; W2020) also supports photoevaporation as the main mechanism that drives the dissipation of their disks \citep[see also][]{Fuentes98,Mendigutia_2012}, although strong uncertainties are still involved in the determination of disk masses and accretion rates in HBes (Sects. \ref{Sect:arates} and \ref{Sect:disk_mass}). Two additional lines of evidence based on comparatively better known SEDs and stellar parameters are provided here to support photoevaporation as the dominant disk dissipation mechanism in HBes. 

On the one hand, the bottom panel of Fig. \ref{inner_size} overplots the critical radius (dashed line) representing the distance from the star from which a gap opens due to photoevaporation, r$_{\rm c}$ $\sim$ 1.05(M$_*$/M$_{\odot}$) for a typical temperature of the UV/X-ray heated gas $\sim$ 10$^4$ K \citep{Gorti09}. Although the majority of HAeBes show inner disk holes below the critical radius in agreement with interferometric measurements \citep{Perraut19}, the inferred inner holes for more than 60$\%$ of HBes with transitional disks (i.e., groups K and $>$ K) are equal to or above this limit. This proportion is significantly smaller for transitional HAes (30$\%$). On the other hand, Fig. \ref{disk_life} shows the stellar ages and masses of the HAeBes. The expected disk lifetime assuming photoevaporation as the main mechanism driving disk dissipation from the model by \citet{Gorti09} (see their Fig. 12) is overplotted. Almost all HAes are older than the photoevaporative lifetime. However, the trend is the opposite for the HBes because most of these sources are younger and do not survive for timescales longer than expected from photoevaporation. These two previous results indicate that while photoevaporation cannot be the main disk dissipation mechanism for most HAes, the inferred inner disk holes and ages of most HBes are consistent with this scenario.

\section{Conclusions} \label{conclusions}

We presented the largest database of stellar and circumstellar parameters that are uniformly derived from an SED analysis for 209 HAeBes. Stellar temperatures and spectral types, luminosities, radii, masses, and ages, as well as SED shape classifications, accretion rates, disk masses, and sizes of dust inner holes have been homogeneously derived based on multiwavelength photometry and Gaia EDR3 distances. This database is stored in an online archive of HAeBes (http://svo2.cab.inta-csic.es/projects/harchibe/) and constitutes a unique tool for the study of this type of sources based on the most recent information available for them.

This sample reproduces the IMF, with the exception of the lower- and higher- mass ends that belong to the IMTT and MYSO limits, respectively. In this sense, the stars characterized in this work represents the HAeBe regime within the stellar mass range 2 $<$ M$_*$/M$_{\odot}$ $<$ 12 well, but when the associated data are used for statistical purposes, the lack of sources in the boundaries, especially IMTTS, must be taken into account. A statistical analysis was carried out in order to exemplify the potential use of these data. We derived the following two main conclusions: 

\begin{itemize}
\item The distributions of the SEDs according to the M01 groups and the wavelength at which the IR excess starts (JHK) are not related to each other, implying that they reflect different properties. In particular, our more complete dataset confirms previous claims suggesting that the M01 groups are not connected to accretion rates but to dust grain growth or the UXOr-type variability. However, when the presence of inner dust disk holes is inferred from the JHK classification, no statistical evidence supports a relation between M01 group I sources and stars with transitional disks (defined here as those with IR excesses starting at wavelengths $\geq$ 2.16 $\mu$m).     
\item In turn, we found that relatively wide inner dust disk holes inferred from SEDs are present in $\sim$ 28 $\%$ of HAeBes, similar to the fraction of transitional disks in TTs, and are $\text{about twice}$ as frequent in early-type HBes than in HAes. The relatively small inner disk holes and old ages inferred for most transitional HAes cannot be explained from photoevaporation. In contrast, the inner holes and ages of most transitional HBes are compatible with the critical radii and lifetimes predicted by this scenario. This evidence supports photoevaporation as the main process driving disk dissipation in HBes, although we do not rule out other potential mechanims. 
\end{itemize}

Finally, our results show the potential use of SED analysis with VOSA for characterizing the stellar and circumstellar properties of newly discovered HAeBe candidates (see Appendix \ref{appA}). In particular, \citet{Vioque20} have published a new catalog with thousands of these objects, and methods complementary to spectroscopy could be of great help to study this large number of new stars that might belong to the HAeBe regime.

\begin{acknowledgements}
JGD, IM and PMA acknowledge the Government of Comunidad Aut\'onoma de Madrid (Spain) for funding this research through a `Talento' Fellowship (2016-T1/TIC-1890 PI I. Mendigutía). The research of IM, JGD, and BM is also partially funded by the Spanish "Ministerio de Ciencia, Innovación y Universidades" through the national project "On the Rocks II" (PGC2018-101950-B-100; PI E. Villaver).

MV acknowledges the STARRY project, which has received funding from the European Union’s Horizon 2020 research and innovation programme
under MSCA ITN-EID grant agreement No 676036.
 
This publication makes use of VOSA, developed under the Spanish Virtual Observatory project supported by the Spanish MINECO through grant AyA2017-84089.
VOSA has been partially updated by using funding from the European Union's Horizon 2020 Research and Innovation Programme, under Grant Agreement nº 776403 (EXOPLANETS-A).

This work has made use of data from the European Space Agency (ESA) mission
{\it Gaia} (\url{https://www.cosmos.esa.int/gaia}), processed by the {\it Gaia}
Data Processing and Analysis Consortium (DPAC,
\url{https://www.cosmos.esa.int/web/gaia/dpac/consortium}). Funding for the DPAC
has been provided by national institutions, in particular the institutions
participating in the {\it Gaia} Multilateral Agreement.

The authors also acknowledge the referee for her/his useful comments, which have served to improve the original manuscript.

\end{acknowledgements}

\bibliographystyle{aa}
\bibliography{ms.bib}
\newpage
\begin{appendix} 

\onecolumn

\section{Comparison with previous stellar parameters}\label{appA}

\begin{figure*}[h]
   \centering
   \includegraphics[width=0.45\textwidth]{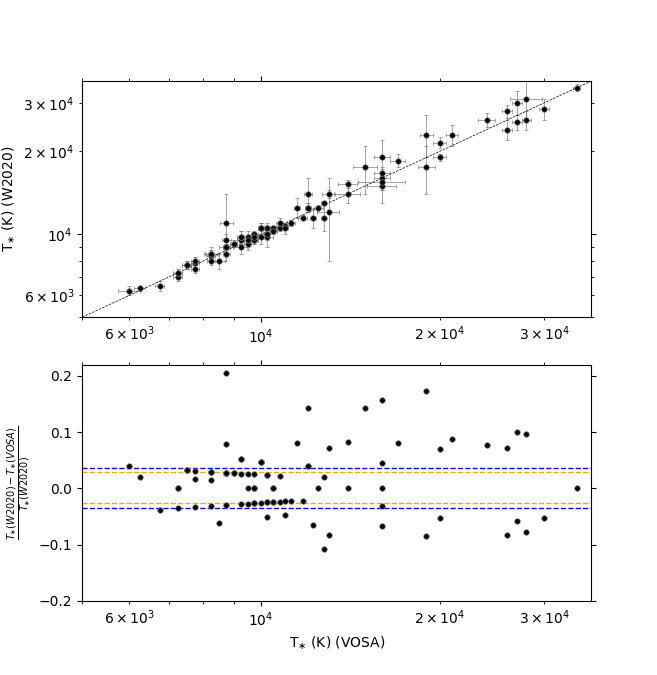}
    \includegraphics[width=0.45\textwidth]{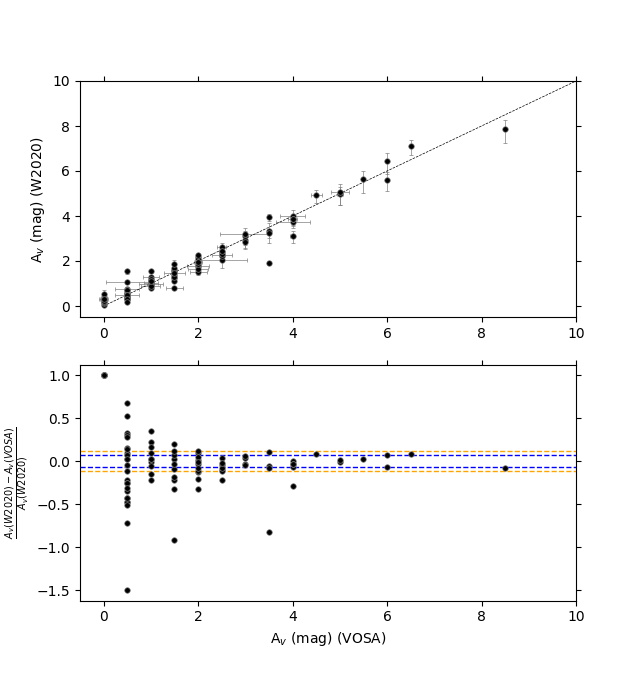}
        \caption{\textbf{Top:} Comparison between T$_{\ast}$ and A$_{v}$ estimated from VOSA and from W2020. The dashed black line indicates equal values. \textbf{Bottom:} Relative errors resulting from the comparison of the corresponding values from VOSA and W2020. The orange line indicates the typical median relative ($\pm$) error of this comparison. The median relative ($\pm$) error provided in W2020 is also plotted with the blue lines for reference.}
\label{Fig:T and Av}
\end{figure*}

In this appendix the stellar parameters (T$_*$, L$_*$, R$_*$, M$_*$, and t$_*$) and extinction (A$_{\rm v}$) derived from SED fitting with VOSA are compared with the corresponding values estimated by W2020 for the 93 stars included in both samples.
The T$_*$ values were inferred in W2020 from spectral typing in the optical, and the A$_{\rm v}$ values directly resulted from the comparison of the corresponding Kurucz synthetic models with optical photometry. In our case, the extinction was set virtually free (0 $<$ A$_{\rm v}$ $<$ 10) and the temperature range of the Kurucz fitting models was fixed by the error bars $\Delta$T$_*$ provided in W2020 for each star (Sect. \ref{stellar_charac}). Figure \ref{Fig:T and Av} compares the T$_*$ and A$_{\rm v}$ values obtained in both works. The bottom panels show that the typical (median) relative error from the comparison between the SED-based T$_*$ values and those spectroscopically determined collapses to $\pm$ 3$\%$ from the typical (median) relative errors of $\pm$ 4$\%$ reported in W2020. The difference between the T$_*$ values from VOSA and from W2020 increases for T$_*$ > 10000 K because of the comparatively larger error bars reported in W2020 for hotter stars. Concerning the optical extinction, the typical (median) relative error is $\pm$ 12$\%$. For reference, the corresponding median error reported in W2020 is $\pm$ 7$\%$.

\begin{figure}[h]
   \centering
   \includegraphics[width=0.42\hsize]{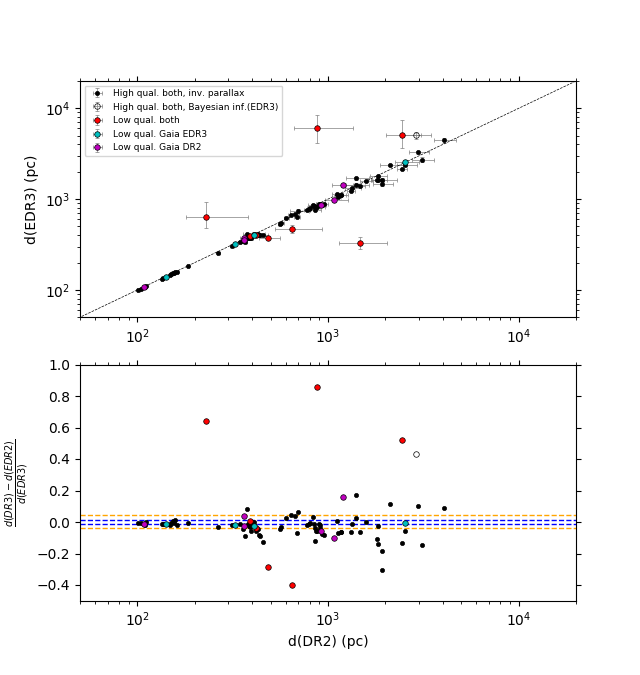}
        \caption{\textbf{Top:} Comparison of Gaia DR2 and Gaia EDR3 distances. Black, red, cyan, and purple circles correspond to objects that have high-quality parallaxes in both releases, low quality in both releases, low quality in Gaia EDR3, and low quality in Gaia DR2, respectively. In addition, objects with high quality in both releases whose EDR3 distances have been estimated using the inverse of parallax (filled black circles) or Bayesian inference (open black circles) are indicated \citep[note that all DR2 distances were estimated using Bayesian inference;][]{Vioque_2018}. \textbf{Bottom:} Relative errors resulting from the comparison of Gaia DR2 and Gaia EDR3 distances. The orange line indicates the typical median relative ($\pm$) error of such a comparison. The median relative ($\pm$) error in Gaia EDR3 distances is also plotted with the blue lines for reference. The relative error of VY Mon is $<$ - 1 and is not plotted for clarity. }
    \label{Fig:distance}
  
\end{figure}

The remaining stellar parameters determined in this work and in W2020 depend on the assumed distances to the sources. Figure \ref{Fig:distance} shows the distances used in W2020, based on Gaia DR2, versus the distances used in this work, based on Gaia EDR3. Using the code indicated in the legend of that figure, we find a good agreement within the error bars between Gaia DR2 and EDR3 for most stars with high-quality parallaxes in both releases. The uncertainties in EDR3 are smaller than in DR2. However, one of these sources (PDS 241) has a current Gaia EDR3 distance, $\sim$ 5000 pc, that almost doubles the previous Gaia DR2 distance. The reason for this discrepancy is the use of Bayesian inference under different priors to derive the distances \citep[see][and Sect. \ref{results} for details]{Vioque_2018}. A few stars are classified with low-quality parallaxes in DR2 or in EDR3 whose corresponding distances show a reasonable agreement as well. This is probably indicative of the conservative approaches used to make such classifications. Still, the most significant differences are found for the stars with low-quality parallaxes both in Gaia DR2 and in EDR3, indicating that the distances to these sources are basically unknown.  

\begin{figure*} [h]
 \centering
   \includegraphics[width=0.45\textwidth]{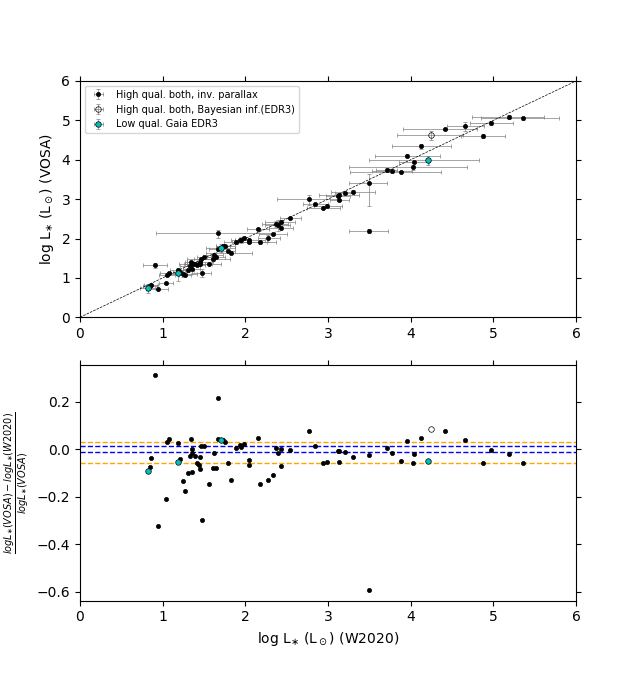}
    \includegraphics[width=0.45\textwidth]{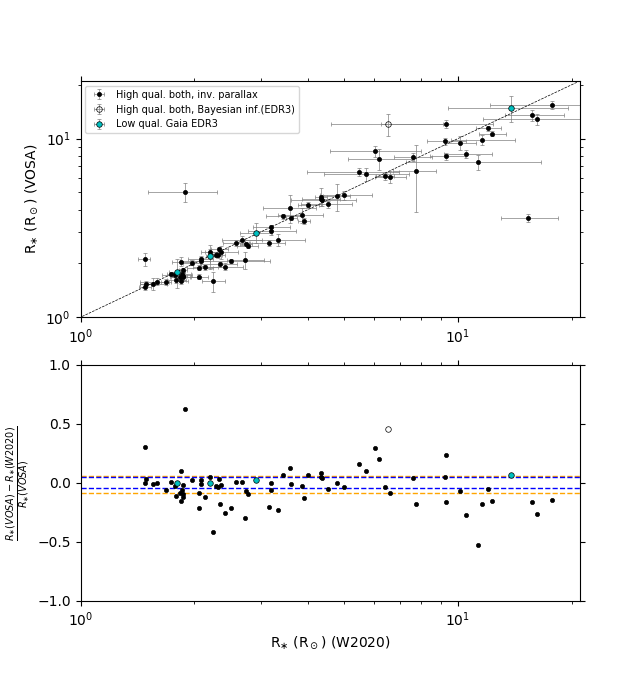}
    \caption{\textbf{Top:} Comparison of L$_{\ast}$ and R$_{\ast}$ estimated from VOSA and from W2020. The dashed black line indicates equal values. \textbf{Bottom:} Relative errors resulting from the comparison of the corresponding values from VOSA and W2020. The orange line indicates the typical median relative (+-) error of such a comparison. The median relative ($\pm$) error estimated in VOSA is also plotted with the blue lines for reference. The relative error of HD 250550 in R$_*$ is $<$ - 1 and is not plotted for clarity.}
    \label{Fig:L and R}
\end{figure*}

The remaining figures compare the stellar parameters in W2020 and in this work that are distance dependent, using the same code as in Fig. \ref{Fig:distance}, stellar luminosities and radii in Fig. \ref{Fig:L and R}, and stellar masses and ages in Fig. \ref{Fig:M and t}. We note that the previous parameters were not determined by W2020 for the stars with low-quality parallaxes in Gaia DR2 (red and purple circles in Fig. \ref{Fig:distance}), and thus they are not included in the analysis. For the remaining stars, the typical (median) relative errors resulting from the comparison are ${_{-6}^{+3}}$ $\%$, ${_{-9}^{+6}}$ $\%$, ${_{-7}^{+6}}$ $\%$, and ${_{-70}^{+28}}$ $\%$ for L$_*$, R$_*$, M$_*$, and t$_*$, respectively. For reference, the corresponding median errors obtained here and based on Gaia EDR3 are $\pm$ 1$\%$, $\pm$ 5$\%$, $\pm$ 3$\%$, and $\pm$ 9$\%$.

Second, Gaia allows us to identify thousands of new HAeBe candidates (\citealt{Vioque20}) that will need to be characterized in the future. Although an analysis of optical spectra is the best way for a proper stellar characterization, SED fitting with VOSA could be a complementary tool for specific subsamples of stars. Photometry of many candidates is available, which facilitates an SED analysis in a comparatively faster and cheaper way in terms of observational and data reduction time. In turn, a priori information on the stellar temperatures of the candidates is in principle very rough, and some independent knowledge of extinction could be available toward several star-forming regions. The previous comparisons were therefore carried out again assuming that we have very uncertain preliminary information on T$_*$ and A$_{\rm v}$. In particular, the VOSA models were run within a range of T$_*$ and A$_{\rm v}$ given by the corresponding values listed in W2020 $\sim$ $\pm$ 30\% and $\pm$ 50\% of these values, simulating the large uncertainties that will in principle be associated with most HAeBe candidates. Because L$_*$, R$_*$, M$_*$, and t$_*$ are distance dependent and W2020 worked with Gaia DR2 data, these parameters were derived again using VOSA and the same Gaia DR2 distances as in \citet{Vioque_2018} and W2020, allowing a direct comparison. As a result, the typical (median) relative errors are ${_{-12}^{+8}}$ \%, ${_{-20}^{+25}}$ \%, ${_{-5}^{+8}}$ \%, ${_{-6}^{+13}}$ \%, ${_{-12}^{+8}}$ \% and ${_{-13}^{+29}}$ \% for T$_{*}$, A$_{\rm v}$, L$_*$, R$_*$, M$_*$, and t$_*$, respectively. These numbers roughly quantify the accuracy that will be obtained with VOSA in future characterizations of newly discovered HAeBe candidates, where the values in W2020 have been taken as reference.

Finally, following this second case scenario, we also tested whether there is a parameter that primarily affects the accuracy obtained in the estimation of the stellar parameters with VOSA. Specifically, we analyzed whether there is any dependence on the extinction or on the number of photometric points used in the fit. Apart from the fact that the sources with relative errors in T$_*$ > 20\% have extinctions A$_{\rm v}$ > 1, there is no particular trend at least for the corresponding ranges explored here. In summary, the main requirement for a reasonable stellar characterization with VOSA is a proper photometric coverage (at least five  photometric points) in the optical range where HAeBes peak, the tool tending to perform slightly better for low-extincted (A$_{\rm v}$ < 1) sources.

\begin{figure*} [h]
 \centering
   \includegraphics[width=0.45\textwidth]{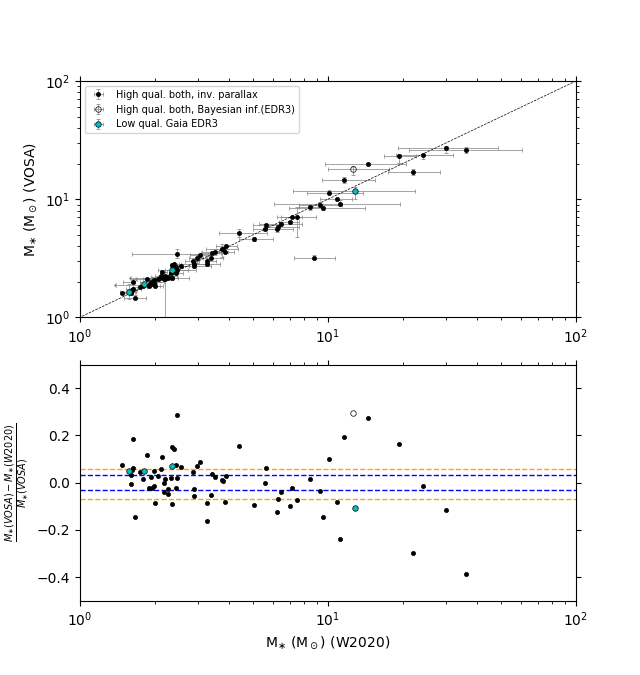}
    \includegraphics[width=0.45\textwidth]{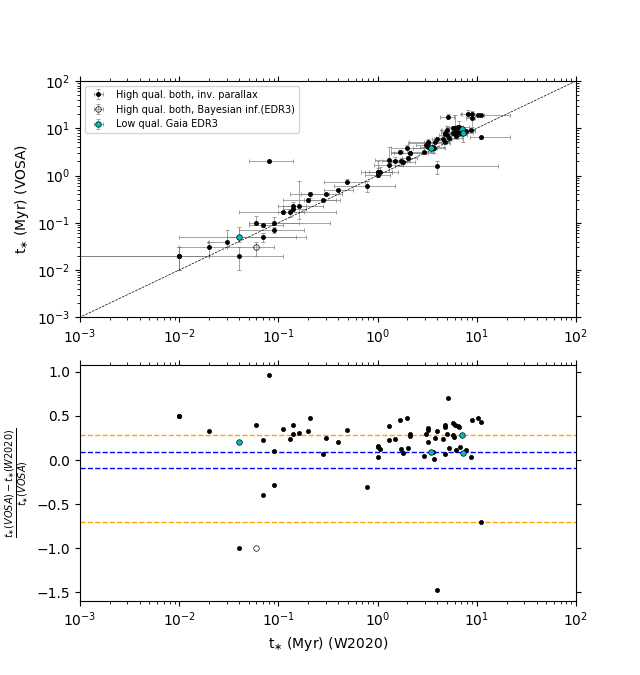}
    \label{Fig:M and t}
    \caption{\textbf{Top:} Comparison of M$_{\ast}$ and R$_{\ast}$ estimated from VOSA and from W2020. The dashed black line indicates equal values. \textbf{Bottom:} Relative errors resulting from the comparison of the corresponding values from VOSA and W2020. The orange line indicates the typical median relative (+-) error of such a comparison. The median relative ($\pm$) error estimated in VOSA is also plotted with the blue lines for reference. The relative error of HD 250550 in M$_*$ is $<$ -0.5 and is not plotted for clarity.}
        \label{Fig:M and t}
\end{figure*}

\newpage

\onecolumn

\section{Tables with results}\label{appB}

\begin{table}[h]
\centering
\begin{turn}{90}
\renewcommand{\arraystretch}{1.6}
\resizebox{21cm}{!} {
\begin{tabular}{l c c c c c c c c c c c}
\hline
\hline
  \multicolumn{1}{c}{Object} &
  \multicolumn{1}{c}{RA} &
  \multicolumn{1}{c}{DEC} &
  \multicolumn{1}{c}{d} &
  \multicolumn{1}{c}{$\Delta$ T$_{\rm \ast}$} &
  \multicolumn{1}{c}{A$_{\rm v}$} &
  \multicolumn{1}{c}{T$_{\rm \ast}$} &
  \multicolumn{1}{c}{Sp. Type} &
  \multicolumn{1}{c}{R$_{\rm \ast}$} &
  \multicolumn{1}{c}{Log(L$_{\rm \ast}$)} &
  \multicolumn{1}{c}{t$_{\rm \ast}$} &
  \multicolumn{1}{c}{M$_{\rm \ast}$} \\
  \multicolumn{1}{c}{} &
  \multicolumn{1}{c}{(h:m:s)} &
  \multicolumn{1}{c}{(d:m:s)} &
  \multicolumn{1}{c}{(pc)} &
  \multicolumn{1}{c}{(K)} &
  \multicolumn{1}{c}{(mag)} &
  \multicolumn{1}{c}{(K)} &
  \multicolumn{1}{c}{} &
  \multicolumn{1}{c}{(R$_\odot$)} &
  \multicolumn{1}{c}{(L$_\odot$)} &
  \multicolumn{1}{c}{(Myr)} &
  \multicolumn{1}{c}{(M$_\odot$)} \\
\hline
  HBC 1 $\dagger$ & 00:07:02.6 & +65:38:38 & 5043.53$_{-1565.26}^{+2276.09}$ & 8000-8500 & 5.50$\pm$0.00 & 8000$\pm$125 & A5-A6 & 13.24$\pm$5.15 & 2.81$_{-0.65}^{+0.25}$ & 0.20$_{-0.10}^{+0.83}$ & 6.42$_{-2.63}^{+1.57}$ \\
  MQ Cas $\dagger$ & 00:09:37.6 & +58:13:11 & 745.96$_{-42.26}^{+47.66}$ & 9500-10750 & 3.00$\pm$0.00 & 10250$\pm$125 & B9 & 2.51$\pm$0.16 & 1.80$_{-0.06}^{+0.05}$ & 3.57$_{-0.34}^{+0.37}$ & 2.58$_{-0.00}^{+0.26}$ \\
  VX Cas & 00:31:30.7 & +61:58:51 & 524.76$_{-5.47}^{+5.58}$ & 9750-10250 & 1.50$\pm$0.11 & 10000$\pm$125 & B9 & 1.68$\pm$0.10 & 1.40$_{-0.05}^{+0.05}$ & < 7.99 & 2.18$_{-0.00}^{+0.02}$ \\
  \textit{V594 Cas} & 00:43:18.3 & +61:54:40 & 551.16$_{-4.13}^{+4.20}$ & 11250-11750 & 2.50$\pm$0.21 (\textit{2.27$_{-0.23}^{+0.18}$}) & 11750$\pm$215 (\textit{11500$\pm$250}) & B7-B8 & 3.74$\pm$0.37 & 2.38$_{-0.09}^{+0.07}$ & 1.67$_{-0.46}^{+0.31}$ & 3.60$_{-0.21}^{+0.20}$ \\
  HBC 7 & 00:43:25.3 & +61:38:23 & 2512.43$_{-79.51}^{+84.88}$ & 16000-25000 & 4.50$\pm$0.00 & 19000$\pm$500 & B2-B3 & 6.81$\pm$0.48 & 3.74$_{-0.04}^{+0.04}$ & 0.15$_{-0.03}^{+0.03}$ & 8.79$_{-0.57}^{+0.21}$ \\
\hline
\end{tabular}
}
\end{turn}
\caption{Stellar parameters}
\label{Table:stellar_parameters}
\end{table}

\onecolumn

\begin{table}[h]
\centering
\begin{turn}{90}
\renewcommand{\arraystretch}{1.6}
\resizebox{21.5cm}{!} {
\begin{tabular}{l c c c c c c c c c c c}
\hline
\hline
  \multicolumn{1}{c}{Object} &
  \multicolumn{1}{c}{RA} &
  \multicolumn{1}{c}{DEC} &
  \multicolumn{1}{c}{d} &
  \multicolumn{1}{c}{$\Delta$ T$_{\rm \ast}$} &
  \multicolumn{1}{c}{A$_{\rm v}$} &
  \multicolumn{1}{c}{T$_{\rm \ast}$} &
  \multicolumn{1}{c}{Sp. Type} &
  \multicolumn{1}{c}{R$_{\rm \ast}$} &
  \multicolumn{1}{c}{Log(L$_{\rm \ast}$)} &
  \multicolumn{1}{c}{t$_{\rm \ast}$} &
  \multicolumn{1}{c}{M$_{\rm \ast}$} \\
  \multicolumn{1}{c}{} &
  \multicolumn{1}{c}{(h:m:s)} &
  \multicolumn{1}{c}{(d:m:s)} &
  \multicolumn{1}{c}{(pc)} &
  \multicolumn{1}{c}{(K)} &
  \multicolumn{1}{c}{(mag)} &
  \multicolumn{1}{c}{(K)} &
  \multicolumn{1}{c}{} &
  \multicolumn{1}{c}{(R$_\odot$)} &
  \multicolumn{1}{c}{(L$_\odot$)} &
  \multicolumn{1}{c}{(Myr)} &
  \multicolumn{1}{c}{(M$_\odot$)} \\
\hline
  CQ Tau $\dagger$ & 05:35:58.5 & +24:44:54 & 148.60$_{-1.31}^{+1.34}$ & 6000-7000 & 0.50$\pm$0.00 & 6750$\pm$125 & F2-F3 & 1.88$\pm$0.07 & 0.82$_{-0.01}^{+0.01}$ & 12.42$_{-1.43}^{+2.40}$ & 1.50$_{-0.01}^{+0.01}$ \\
  \textit{V380 Ori} $\dagger$ & 05:36:25.4 & -06:42:58 & 374.32$_{-15.02}^{+16.33}$ & 9000-10500 & 2.50$\pm$0.54 (\textit{2.05$_{-0.35}^{+0.26}$})& 10250$\pm$246 (\textit{9750$\pm$750}) & B9 & 3.11$\pm$1.10 & 1.98$_{-0.53}^{+0.23}$ & 2.96$_{-0.00}^{+4.82}$ & 2.80$_{-0.00}^{+0.60}$ \\
  \textit{HD 37258} & 05:36:59.3 & -06:09:16 & 376.55$_{-3.63}^{+3.70}$ & 9250-10250 & 0.50$\pm$0.00 (\textit{0.54$_{-0.04}^{+0.03}$}) & 9250$\pm$125 (\textit{9750$\pm$500}) & A0-A1 & 1.98$\pm$0.07 & 1.41$_{-0.02}^{+0.02}$ & 5.94$_{-0.78}^{+1.01}$ & 2.27$_{-0.08}^{+0.03}$ \\
  \textit{HD 290770} & 05:37:02.4 & -01:37:21 & 392.85$_{-3.80}^{+3.87}$ & 10250-10750 & 0.50$\pm$0.00 (\textit{0.34$_{-0.05}^{+0.05}$}) & 10750$\pm$125 (\textit{10500$\pm$250}) & B8 & 2.13$\pm$0.05 & 1.74$_{-0.01}^{+0.01}$ & 4.01$_{-0.04}^{+0.39}$ & 2.64$_{-0.04}^{+0.13}$ \\
\hline\end{tabular}
}
\end{turn}
\end{table}

\onecolumn

\begin{table}[h]
\centering
\begin{turn}{90}
\renewcommand{\arraystretch}{1.6}
\resizebox{21.5cm}{!} {
\begin{tabular}{l c c c c c c c c c c c}
\hline
\hline
  \multicolumn{1}{c}{Object} &
  \multicolumn{1}{c}{RA} &
  \multicolumn{1}{c}{DEC} &
  \multicolumn{1}{c}{d} &
  \multicolumn{1}{c}{$\Delta$ T$_{\rm \ast}$} &
  \multicolumn{1}{c}{A$_{\rm v}$} &
  \multicolumn{1}{c}{T$_{\rm \ast}$} &
  \multicolumn{1}{c}{Sp. Type} &
  \multicolumn{1}{c}{R$_{\rm \ast}$} &
  \multicolumn{1}{c}{Log(L$_{\rm \ast}$)} &
  \multicolumn{1}{c}{t$_{\rm \ast}$} &
  \multicolumn{1}{c}{M$_{\rm \ast}$} \\
  \multicolumn{1}{c}{} &
  \multicolumn{1}{c}{(h:m:s)} &
  \multicolumn{1}{c}{(d:m:s)} &
  \multicolumn{1}{c}{(pc)} &
  \multicolumn{1}{c}{(K)} &
  \multicolumn{1}{c}{(mag)} &
  \multicolumn{1}{c}{(K)} &
  \multicolumn{1}{c}{} &
  \multicolumn{1}{c}{(R$_\odot$)} &
  \multicolumn{1}{c}{(L$_\odot$)} &
  \multicolumn{1}{c}{(Myr)} &
  \multicolumn{1}{c}{(M$_\odot$)} \\
\hline 
  HD 259431 & 06:33:05.2 & +10:19:20 & 640.04$_{-10.98}^{+11.37}$ & 11000-16000 & 1.50$\pm$0.00 & 12500$\pm$332 & B7 & 6.10$\pm$0.51 & 2.91$_{-0.06}^{+0.05}$ & 0.50$_{-0.08}^{+0.10}$ & 5.24$_{-0.24}^{+0.30}$ \\
  HBC 217 & 06:40:42.2 & +09:33:37 & 704.61$_{-6.72}^{+6.85}$ & 6000-6500 & 0.50$\pm$0.00 & 6000$\pm$125 & F8-G1 & 2.38$\pm$0.10 & 0.82$_{-0.01}^{+0.01}$ & 6.89$_{-1.55}^{+1.13}$ & 1.75$_{-0.10}^{+0.10}$ \\
  HBC 222 & 06:40:51.2 & +09:44:46 & 701.14$_{-6.83}^{+6.97}$ & 6000-6500 & 0.50$\pm$0.17 & 6500$\pm$174 & F3-F6 & 2.42$\pm$0.22 & 0.97$_{-0.07}^{+0.07}$ & 8.24$_{-0.00}^{+1.77}$ & 1.70$_{-0.11}^{+0.12}$ \\
  \textit{PDS 24} & 06:48:41.7 & -16:48:06 & 1055.57$_{-16.07}^{+16.07}$ & 10000-11000 & 1.50$\pm$0.05 (\textit{1.54$_{-0.12}^{+0.10}$}) & 10000$\pm$125 (\textit{10500$\pm$500}) & B9 & 1.58$\pm$0.07 & 1.35$_{-0.03}^{+0.03}$ & < 8.27 & < 2.20 \\
  \textit{PDS 130} & 06:49:58.6 & -07:38:52 & 1237.04$_{-19.66}^{+20.30}$ & 10250-10750 & 2.50$\pm$0.00 (\textit{2.38$_{-0.15}^{+0.13}$}) & 10500$\pm$225 (\textit{10500$\pm$250}) & B8-B9 & 2.70$\pm$0.15 & 1.90$_{-0.03}^{+0.03}$ & 3.08$_{-0.08}^{+0.22}$ & 2.74$_{-0.13}^{+0.10}$ \\
\hline\end{tabular}
}
\end{turn}
\end{table}

\onecolumn

\begin{table}[h]
\centering
\begin{turn}{90}
\renewcommand{\arraystretch}{1.6}
\resizebox{21.5cm}{!} {
\begin{tabular}{l c c c c c c c c c c c}
\hline
\hline
  \multicolumn{1}{c}{Object} &
  \multicolumn{1}{c}{RA} &
  \multicolumn{1}{c}{DEC} &
  \multicolumn{1}{c}{d} &
  \multicolumn{1}{c}{$\Delta$ T$_{\rm \ast}$} &
  \multicolumn{1}{c}{A$_{\rm v}$} &
  \multicolumn{1}{c}{T$_{\rm \ast}$} &
  \multicolumn{1}{c}{Sp. Type} &
  \multicolumn{1}{c}{R$_{\rm \ast}$} &
  \multicolumn{1}{c}{Log(L$_{\rm \ast}$)} &
  \multicolumn{1}{c}{t$_{\rm \ast}$} &
  \multicolumn{1}{c}{M$_{\rm \ast}$} \\
  \multicolumn{1}{c}{} &
  \multicolumn{1}{c}{(h:m:s)} &
  \multicolumn{1}{c}{(d:m:s)} &
  \multicolumn{1}{c}{(pc)} &
  \multicolumn{1}{c}{(K)} &
  \multicolumn{1}{c}{(mag)} &
  \multicolumn{1}{c}{(K)} &
  \multicolumn{1}{c}{} &
  \multicolumn{1}{c}{(R$_\odot$)} &
  \multicolumn{1}{c}{(L$_\odot$)} &
  \multicolumn{1}{c}{(Myr)} &
  \multicolumn{1}{c}{(M$_\odot$)} \\
\hline
  PDS 324 & 10:57:24.2 & -62:53:13 & 3440.69$_{-227.77}^{+262.53}$ & 19000-29000 & 3.50$\pm$0.00 & 19000$\pm$500 & B2-B3 & 3.97$\pm$0.35 & 3.27$_{-0.07}^{+0.06}$ & 0.35$_{-0.06}^{+0.06}$ & 6.59$_{-0.23}^{+0.42}$ \\
  \textit{HD 95881} & 11:01:57.6 & -71:30:48 & 1098.03$_{-23.30}^{+24.34}$ & 9750-10250 & 0.50$\pm$0.00 (\textit{0.72$_{-0.04}^{+0.03}$}) & 9750$\pm$125 (\textit{10000$\pm$250}) & B9 & 10.66$\pm$0.36 & 2.97$_{-0.02}^{+0.02}$ & 0.23$_{-0.03}^{+0.04}$ & 6.40$_{-0.19}^{+0.07}$ \\
  \textit{HD 96042} & 11:03:40.5 & -59:25:59 & 2702.01$_{-165.08}^{+188.06}$ & 24000-27000 & 1.50$\pm$0.00 (\textit{1.13$_{-0.08}^{+0.08}$}) & 27000$\pm$500 (\textit{25500$\pm$1500}) & B0 & 13.51$\pm$1.02 & 4.94$_{-0.06}^{+0.05}$ & 0.02$_{-0.01}^{+0.01}$ & 23.74$_{-1.98}^{+0.30}$ \\
  \textit{HD 97048} & 11:08:03.2 & -77:39:17 & 184.11$_{-0.84}^{+0.85}$ & 10000-11000 & 1.50$\pm$0.00 (\textit{1.37$_{-0.06}^{+0.05}$}) & 11000$\pm$125 (\textit{10500$\pm$500}) & B8 & 2.22$\pm$0.06 & 1.81$_{-0.01}^{+0.01}$ & 3.90$_{-0.60}^{+0.07}$ & 2.80$_{-0.03}^{+0.03}$ \\
\hline\end{tabular}
}
\end{turn}
\end{table}

\onecolumn

\begin{table}[h]
\centering
\begin{turn}{90}
\renewcommand{\arraystretch}{1.6}
\resizebox{21.5cm}{!} {
\begin{tabular}{l c c c c c c c c c c c}
\hline
\hline
  \multicolumn{1}{c}{Object} &
  \multicolumn{1}{c}{RA} &
  \multicolumn{1}{c}{DEC} &
  \multicolumn{1}{c}{d} &
  \multicolumn{1}{c}{$\Delta$ T$_{\rm \ast}$} &
  \multicolumn{1}{c}{A$_{\rm v}$} &
  \multicolumn{1}{c}{T$_{\rm \ast}$} &
  \multicolumn{1}{c}{Sp. Type} &
  \multicolumn{1}{c}{R$_{\rm \ast}$} &
  \multicolumn{1}{c}{Log(L$_{\rm \ast}$)} &
  \multicolumn{1}{c}{t$_{\rm \ast}$} &
  \multicolumn{1}{c}{M$_{\rm \ast}$} \\
  \multicolumn{1}{c}{} &
  \multicolumn{1}{c}{(h:m:s)} &
  \multicolumn{1}{c}{(d:m:s)} &
  \multicolumn{1}{c}{(pc)} &
  \multicolumn{1}{c}{(K)} &
  \multicolumn{1}{c}{(mag)} &
  \multicolumn{1}{c}{(K)} &
  \multicolumn{1}{c}{} &
  \multicolumn{1}{c}{(R$_\odot$)} &
  \multicolumn{1}{c}{(L$_\odot$)} &
  \multicolumn{1}{c}{(Myr)} &
  \multicolumn{1}{c}{(M$_\odot$)} \\
\hline
  Hen 3-1191 & 16:27:15.1 & -48:39:27 & 1687.14$_{-119.46}^{+139.17}$ & 24000-33000 & 4.50$\pm$0.12 & 24000$\pm$500 & B1 & 4.17$\pm$0.50 & 3.72$_{-0.11}^{+0.09}$ & 0.19$_{-0.00}^{+0.05}$ & 9.38$_{-0.00}^{+0.65}$ \\
  HD 149914 & 16:38:28.6 & -18:13:14 & 153.45$_{-0.62}^{+0.63}$ & 9500-11250 & 1.00$\pm$0.00 & 9500$\pm$125 & B9-A0 & 3.92$\pm$0.11 & 2.05$_{-0.01}^{+0.01}$ & 2.02$_{-0.02}^{+0.05}$ & 3.07$_{-0.05}^{+0.07}$ \\
  \textit{HD 150193} & 16:40:17.9 & -23:53:45 & 150.02$_{-0.47}^{+0.48}$ & 9000-9500 & 1.50$\pm$0.00 (\textit{1.88$_{-0.20}^{+0.16}$}) & 9000$\pm$125 (\textit{9250$\pm$250}) & A1-A2 & 1.98$\pm$0.06 & 1.36$_{-0.01}^{+0.01}$ & 6.00$_{-0.02}^{+1.00}$ & 2.25$_{-0.08}^{+0.00}$ \\
  \textit{AK Sco} $\dagger$ & 16:54:44.8 & -36:53:19 & 139.15$_{-0.60}^{+0.61}$ & 6000-6500 & 0.50$\pm$0.25 (\textit{0.51$_{-0.02}^{+0.03}$}) & 6000$\pm$246 (\textit{6250$\pm$250}) & F7-G5 & 2.20$\pm$0.34 & 0.75$_{-0.13}^{+0.10}$ & 7.83$_{-2.78}^{+2.92}$ & 1.65$_{-0.22}^{+0.25}$ \\
\hline\end{tabular}
}
\end{turn}
\end{table}

\onecolumn

\begin{sidewaystable}[h]
\centering
\renewcommand{\arraystretch}{1.6}
\resizebox{21.5cm}{!} {
\begin{tabular}{l c c c c c c c c c c c}
\hline
\hline
  \multicolumn{1}{c}{Object} &
  \multicolumn{1}{c}{RA} &
  \multicolumn{1}{c}{DEC} &
  \multicolumn{1}{c}{d} &
  \multicolumn{1}{c}{$\Delta$ T$_{\rm \ast}$} &
  \multicolumn{1}{c}{A$_{\rm v}$} &
  \multicolumn{1}{c}{T$_{\rm \ast}$} &
  \multicolumn{1}{c}{Sp. Type} &
  \multicolumn{1}{c}{R$_{\rm \ast}$} &
  \multicolumn{1}{c}{Log(L$_{\rm \ast}$)} &
  \multicolumn{1}{c}{t$_{\rm \ast}$} &
  \multicolumn{1}{c}{M$_{\rm \ast}$} \\
  \multicolumn{1}{c}{} &
  \multicolumn{1}{c}{(h:m:s)} &
  \multicolumn{1}{c}{(d:m:s)} &
  \multicolumn{1}{c}{(pc)} &
  \multicolumn{1}{c}{(K)} &
  \multicolumn{1}{c}{(mag)} &
  \multicolumn{1}{c}{(K)} &
  \multicolumn{1}{c}{} &
  \multicolumn{1}{c}{(R$_\odot$)} &
  \multicolumn{1}{c}{(L$_\odot$)} &
  \multicolumn{1}{c}{(Myr)} &
  \multicolumn{1}{c}{(M$_\odot$)} \\
\hline
  \textit{V1685 Cyg} & 20:20:28.2 & +41:21:51 & 893.41$_{-15.10}^{+15.63}$ & 19000-27000 & 3.50$\pm$0.00 (\textit{3.33$_{-0.51}^{+0.34}$}) & 19000$\pm$500 (\textit{23000$\pm$4000}) & B2-B3 & 6.52$\pm$0.36 & 3.70$_{-0.02}^{+0.02}$ & 0.17$_{-0.02}^{+0.01}$ & 8.33$_{-0.20}^{+0.28}$ \\
  V1686 Cyg $\dagger$ & 20:20:29.3 & +41:21:28 & 665.82$_{-83.28}^{+78.36}$ & 5750-6250 & 2.50$\pm$0.69 & 6250$\pm$237 & F4-G0 & 4.00$\pm$1.33 & 1.34$_{-0.45}^{+0.22}$ & 3.02$_{-1.11}^{+5.98}$ & 2.40$_{-0.77}^{+0.60}$ \\
  MWC 1021 & 20:29:26.9 & +41:40:44 & 3090.47$_{-221.60}^{+258.70}$ & 10000-30000 & 9.00$\pm$0.00 & 13000$\pm$313 & B6-B7 & 154.47$\pm$17.18 & 5.79$_{-0.10}^{+0.08}$ & < 0.01 & 73.64$_{-6.63}^{+6.30}$ \\
  V1478 Cyg $\dagger$ & 20:32:45.5 & +40:39:37 & 1669.83$_{-176.25}^{+201.74}$ & 10000-30000 & 9.00$\pm$0.00 & 14000$\pm$900 & B5-B6 & 66.87$\pm$15.68 & 5.19$_{-0.22}^{+0.14}$ & < 0.01 & 36.19$_{-8.05}^{+8.78}$ \\
  V1977 Cyg & 20:47:37.5 & +43:47:25 & 820.69$_{-7.92}^{+8.08}$ & 10750-11250 & 2.50$\pm$0.00 & 10750$\pm$125 & B8 & 5.14$\pm$0.13 & 2.50$_{-0.01}^{+0.01}$ & 1.00$_{-0.07}^{+0.01}$ & 4.13$_{-0.10}^{+0.07}$ \\
\hline\end{tabular}
}
\tablefoot{Column 5 lists the input ranges for the stellar temperatures assumed to carry out SED fitting, as taken from \citet{Vioque_2018}. The stars indicated in italics in Col. 1 have stellar parameters determined homogeneously from spectroscopy in W2020, which has been followed to set the input ranges for T$_*$ in these cases. The rest of the columns list our results concerning the stellar characterization, as well as the distance-independent results from W2020 for the stars in common (T$_*$ and A$_{v}$, between parentheses). Our error bars refer to SED fitting (Sect. \ref{stellar_charac}; see also Appendix \ref{appA}). Stars indicated with the $\text{dagger}$ have low-quality Gaia EDR3 parallaxes according to the criteria described in Sect. \ref{results}, and their distance-dependent stellar parameters should be taken with caution. These stars have not been considered in the analysis.
}
\end{sidewaystable}

\newpage

\begin{table}[h]
\caption{SED classifications and disk parameters}
\label{Table:SEDs}
\renewcommand{\arraystretch}{1.4}
\center
\begin{tabular}{l c c c c c c c c}
\hline
  \multicolumn{1}{c}{Object} &
  \multicolumn{1}{c}{Meeus Group} &
  \multicolumn{1}{c}{JH-K Group} &
  \multicolumn{1}{c}{Log($\dot{\rm M}_{\rm acc}$)} &
  \multicolumn{1}{c}{Log(L$_{\rm acc}$)} &
  \multicolumn{1}{c}{M$_{\rm disk}^{\rm min}$} &
  \multicolumn{1}{c}{IR$_{\rm start}$} &
  \multicolumn{1}{c}{r$_{\rm in}$} &
  \multicolumn{1}{c}{Notes} \\
  \multicolumn{1}{c}{} &
  \multicolumn{1}{c}{} &
  \multicolumn{1}{c}{} &
  \multicolumn{1}{c}{(M$_\odot yr^{-1}$)} &
  \multicolumn{1}{c}{(L$_\odot$)} &
  \multicolumn{1}{c}{(M$_\odot$)} &
  \multicolumn{1}{c}{($\mu$m)} &
  \multicolumn{1}{c}{(au)} &
  \multicolumn{1}{c}{} \\
\hline
\hline
  HBC 1 & I & J & -5.07$_{-1.14}^{+1.13}$ & 1.88$_{-0.83}^{+0.76}$ & 1.278 & 1.24 & 0.92$\pm$0.08 &\\
  MQ Cas & II & J & -6.28$_{-0.38}^{+0.34}$ & 0.98$_{-0.31}^{+0.31}$ & 0.392 & 1.24 & 0.29$\pm$0.03 &\\
  VX Cas & II & J & -6.80$_{-0.30}^{+0.30}$ & 0.57$_{-0.27}^{+0.28}$ & 0.529 & 1.24 & 0.19$\pm$0.02 & $\ast$\\
  V594 Cas & II & J & -5.66$_{-0.45}^{+0.45}$ & 1.58$_{-0.39}^{+0.38}$ & 0.516 & 1.24 & 0.58$\pm$0.05 &\\
  HBC 7 & & Ks & -4.93$_{-0.63}^{+0.66}$ & 2.43$_{-0.59}^{+0.60}$ & 0.247 & 2.16 & 6.09$\pm$0.49 &\\
  PDS 2 & I & J & -7.25$_{-0.21}^{+0.20}$ & -0.08$_{-0.18}^{+0.03}$ & 0.230 & 1.24 & 0.09$\pm$0.01 &\\
\hline\end{tabular}
\end{table}

\onecolumn

\begin{table}[h]
\renewcommand{\arraystretch}{1.4}
\center
\begin{tabular}{l c c c c c c c c}
\hline
  \multicolumn{1}{c}{Object} &
  \multicolumn{1}{c}{Meeus Group} &
  \multicolumn{1}{c}{JH-K Group} &
  \multicolumn{1}{c}{Log($\dot{\rm M}_{\rm acc}$)} &
  \multicolumn{1}{c}{Log(L$_{\rm acc}$)} &
  \multicolumn{1}{c}{M$_{\rm disk}^{\rm min}$} &
  \multicolumn{1}{c}{IR$_{\rm start}$} &
  \multicolumn{1}{c}{r$_{\rm in}$} &
  \multicolumn{1}{c}{Notes} \\
  \multicolumn{1}{c}{} &
  \multicolumn{1}{c}{} &
  \multicolumn{1}{c}{} &
  \multicolumn{1}{c}{(M$_\odot yr^{-1}$)} &
  \multicolumn{1}{c}{(L$_\odot$)} &
  \multicolumn{1}{c}{(M$_\odot$)} &
  \multicolumn{1}{c}{($\mu$m)} &
  \multicolumn{1}{c}{(au)} &
  \multicolumn{1}{c}{} \\
\hline
\hline
  HD 290770 & II & J & -6.43$_{-0.29}^{+0.28}$ & 0.92$_{-0.26}^{+0.26}$ & 0.045 & 1.24 & 0.27$\pm$0.02 &\\
  BF Ori & II & J & -7.03$_{-0.41}^{+0.35}$ & 0.29$_{-0.27}^{+0.30}$ & 0.507 & 1.24 & 0.13$\pm$0.01 & $\ast$\\
  HD 37357 & II & J & -6.06$_{-0.44}^{+0.43}$ & 1.13$_{-0.36}^{+0.35}$ & 0.362 & 1.24 & 0.34$\pm$0.03 &\\
  HD 290764 & I & J & -6.63$_{-0.26}^{+0.26}$ & 0.51$_{-0.24}^{+0.24}$ & 0.715 & 1.24 & 0.17$\pm$0.02 &\\
  V1787 Ori & I & J & -6.57$_{-0.26}^{+0.25}$ & 0.61$_{-0.23}^{+0.24}$ & 0.648 & 1.24 & 0.19$\pm$0.02 &\\
  HD 37371 &  & > Ks & -6.16$_{-0.31}^{+0.32}$ & 1.19$_{-0.29}^{+0.29}$ & 0.627 & 4.60 & 2.16$\pm$0.33 & $\ast$\\
\hline\end{tabular}
\end{table}

\onecolumn

\begin{table}[h]
\renewcommand{\arraystretch}{1.4}
\center
\begin{tabular}{l c c c c c c c c}
\hline
  \multicolumn{1}{c}{Object} &
  \multicolumn{1}{c}{Meeus Group} &
  \multicolumn{1}{c}{JH-K Group} &
  \multicolumn{1}{c}{Log($\dot{\rm M}_{\rm acc}$)} &
  \multicolumn{1}{c}{Log(L$_{\rm acc}$)} &
  \multicolumn{1}{c}{M$_{\rm disk}^{\rm min}$} &
  \multicolumn{1}{c}{IR$_{\rm start}$} &
  \multicolumn{1}{c}{r$_{\rm in}$} &
  \multicolumn{1}{c}{Notes} \\
  \multicolumn{1}{c}{} &
  \multicolumn{1}{c}{} &
  \multicolumn{1}{c}{} &
  \multicolumn{1}{c}{(M$_\odot yr^{-1}$)} &
  \multicolumn{1}{c}{(L$_\odot$)} &
  \multicolumn{1}{c}{(M$_\odot$)} &
  \multicolumn{1}{c}{($\mu$m)} &
  \multicolumn{1}{c}{(au)} &
  \multicolumn{1}{c}{} \\
\hline
\hline
  HU CMa & I & J & -6.20$_{-0.45}^{+0.41}$ & 1.21$_{-0.37}^{+0.36}$ & 0.655 & 1.24 & 0.39$\pm$0.03 & $\ast$\\
  PDS 241 & I & > Ks & -4.46$_{-0.80}^{+0.81}$ & 2.97$_{-0.70}^{+0.70}$ & 0.747 & 3.35 & 31.51$\pm$4.16 &\\
  HD 56895 & II & > Ks & -6.98$_{-0.22}^{+0.23}$ & 0.14$_{-0.08}^{+0.20}$ & 0.499 & 10.15 & 1.80$\pm$0.72 &\\
  PDS 27 & I & J & -4.56$_{-0.78}^{+0.79}$ & 2.59$_{-0.66}^{+0.66}$ & 1.087 & 1.24 & 3.84$\pm$0.34 & \\
  GSC 6546-3156 & I & H & -6.81$_{-0.32}^{+0.31}$ & 0.52$_{-0.27}^{+0.27}$ & 0.093 & 1.66 & 0.26$\pm$0.03 &\\
  PDS 133 & I & J & -6.30$_{-0.37}^{+0.33}$ & 1.09$_{-0.31}^{+0.30}$ & 0.164 & 1.24 & 0.33$\pm$0.03 & $\ast$\\
\hline\end{tabular}
\end{table}

\onecolumn

\begin{table}[h]
\renewcommand{\arraystretch}{1.4}
\center
\begin{tabular}{l c c c c c c c c}
\hline
  \multicolumn{1}{c}{Object} &
  \multicolumn{1}{c}{Meeus Group} &
  \multicolumn{1}{c}{JH-K Group} &
  \multicolumn{1}{c}{Log($\dot{\rm M}_{\rm acc}$)} &
  \multicolumn{1}{c}{Log(L$_{\rm acc}$)} &
  \multicolumn{1}{c}{M$_{\rm disk}^{\rm min}$} &
  \multicolumn{1}{c}{IR$_{\rm start}$} &
  \multicolumn{1}{c}{r$_{\rm in}$} &
  \multicolumn{1}{c}{Notes} \\
  \multicolumn{1}{c}{} &
  \multicolumn{1}{c}{} &
  \multicolumn{1}{c}{} &
  \multicolumn{1}{c}{(M$_\odot yr^{-1}$)} &
  \multicolumn{1}{c}{(L$_\odot$)} &
  \multicolumn{1}{c}{(M$_\odot$)} &
  \multicolumn{1}{c}{($\mu$m)} &
  \multicolumn{1}{c}{(au)} &
  \multicolumn{1}{c}{} \\
\hline
\hline
  Hen 3-938 & II & J & -4.08$_{-0.82}^{+0.84}$ & 3.51$_{-0.75}^{+0.76}$ & 1.851 & 1.24 & 24.04$\pm$2.11 & \\
  PDS 69 & I & H & -5.37$_{-0.71}^{+0.69}$ & 1.99$_{-0.59}^{+0.58}$ & 1.028 & 1.66 & 1.80$\pm$0.18 & $\ast$\\
  DG Cir & I & J & -5.76$_{-0.48}^{+0.46}$ & 1.32$_{-0.38}^{+0.38}$ & 0.723 & 1.24 & 0.42$\pm$0.04 &\\
  HD 132947 & II & > Ks & -6.33$_{-0.28}^{+0.28}$ & 0.98$_{-0.26}^{+0.27}$ & 0.273 & 10.15 & 4.87$\pm$1.94 & $\ast$\\
  PDS 389 & I & H & -5.81$_{-032}^{+0.34}$ & 1.30$_{-0.30}^{+0.30}$ & 0.396 & 1.66 & 0.61$\pm$0.06 &\\
  HD 135344B & I & J & -7.28$_{-0.22}^{+0.21}$ & -0.14$_{-0.18}^{+0.10}$ & 0.346 & 1.24 & 0.08$\pm$0.01 &\\
\hline\end{tabular}
\end{table}

\onecolumn

\begin{table}[h]
\renewcommand{\arraystretch}{1.4}
\center
\begin{tabular}{l c c c c c c c c}
\hline
  \multicolumn{1}{c}{Object} &
  \multicolumn{1}{c}{Meeus Group} &
  \multicolumn{1}{c}{JH-K Group} &
  \multicolumn{1}{c}{Log($\dot{\rm M}_{\rm acc}$)} &
  \multicolumn{1}{c}{Log(L$_{\rm acc}$)} &
  \multicolumn{1}{c}{M$_{\rm disk}^{\rm min}$} &
  \multicolumn{1}{c}{IR$_{\rm start}$} &
  \multicolumn{1}{c}{r$_{\rm in}$} &
  \multicolumn{1}{c}{Notes} \\
  \multicolumn{1}{c}{} &
  \multicolumn{1}{c}{} &
  \multicolumn{1}{c}{} &
  \multicolumn{1}{c}{(M$_\odot yr^{-1}$)} &
  \multicolumn{1}{c}{(L$_\odot$)} &
  \multicolumn{1}{c}{(M$_\odot$)} &
  \multicolumn{1}{c}{($\mu$m)} &
  \multicolumn{1}{c}{(au)} &
  \multicolumn{1}{c}{} \\
\hline
\hline
  PDS 530 & I & H & -7.18$_{-1.07}^{+0.61}$ & 0.11$_{-0.59}^{+0.46}$ & 0.275 & 1.66 & 0.16$\pm$0.02 & $\ast$\\
  MWC 953 &  & J & -4.51$_{-0.82}^{+0.79}$ & 2.66$_{-0.68}^{+0.67}$ & 1.135 & 1.24 & 4.35$\pm$0.38 & \\
  PDS 543 & I & > Ks & -4.34$_{-0.75}^{+0.78}$ & 3.24$_{-0.70}^{+0.71}$ & 1.017 & 3.35 & 53.79$\pm$7.09 & \\
  HD 174571 & II & > Ks & -4.49$_{-0.69}^{+0.71}$ & 2.88$_{-0.66}^{+0.66}$ & 0.564 & 7.95 & 83.57$\pm$23.61 & \\
  HD 176386 &  & > Ks & -6.49$_{-0.26}^{+0.31}$ & 0.82$_{-0.25}^{+0.25}$ & 0.093 & 3.35 & 0.92$\pm$0.12 &\\
  TY CrA & I & J & -6.72$_{-1.09}^{+0.66}$ & 075$_{-0.65}^{+0.51}$ &0.826 & 1.24 & 0.23$\pm$0.02 & $\ast$\\
\hline\end{tabular}
\tablefoot{For the stars with $\text{an asterisk}$ in the last column, accretion-based disk masses have been reestimated because t$_{*}$ > t$_{\rm MS}$ (Sect. \ref{Sect:disk_mass}).
}
\end{table}

\begin{table}[h]
\caption{Disk masses derived by millimeter fluxes}
\label{Table:Disks_mm}
\renewcommand{\arraystretch}{1}
\center
\begin{tabular}{l c c c c c}
\hline
  \multicolumn{1}{c}{Object} &
  \multicolumn{1}{c}{$\beta$} &
  \multicolumn{1}{c}{T$_{\rm D}$} &
  \multicolumn{1}{c}{M$_{\rm disk}$} &
  \multicolumn{1}{c}{Nº photometric points} &
  \multicolumn{1}{c}{References} \\
  \multicolumn{1}{c}{} &
  \multicolumn{1}{c}{} &
  \multicolumn{1}{c}{(K)} &
  \multicolumn{1}{c}{(M$_\odot$)} &
  \multicolumn{1}{c}{} &
  \multicolumn{1}{c}{} \\
\hline
\hline
  HBC 1 & 1.0 & 21 & 7.808$\pm$0.887 & 1 & (16)\\
  VX Cas & 1.0 & 33 & < 0.007 & 1 & (28)\\
  V594 Cas & 1.1 & 32 & 0.039$\pm$0.001 & 2 & (7) (32)\\
  HD 9672 & 0.6 & 39 & 2.920E-4$\pm$3.060E-5 & 2 & (25)\\
  AB Aur & 1.2 & 40 & 0.009$\pm$0.002 & 5 & (1) (2) (3)\\
  HD 31648 & 0.9 & 20 & 0.051$\pm$4.410E-4 & 6 & (1) (2) (5) (20)\\
\hline
\end{tabular}
\end{table}

\newpage

\begin{table}[h]
\renewcommand{\arraystretch}{1}
\center
\begin{tabular}{l c c c c c}
\hline
  \multicolumn{1}{c}{Object} &
  \multicolumn{1}{c}{$\beta$} &
  \multicolumn{1}{c}{T$_{\rm D}$} &
  \multicolumn{1}{c}{M$_{\rm disk}$} &
  \multicolumn{1}{c}{Nº photometric points} &
  \multicolumn{1}{c}{References} \\
  \multicolumn{1}{c}{} &
  \multicolumn{1}{c}{} &
  \multicolumn{1}{c}{(K)} &
  \multicolumn{1}{c}{(M$_\odot$)} &
  \multicolumn{1}{c}{} &
  \multicolumn{1}{c}{} \\
\hline
\hline
  PDS 520 & 1.0 & 16 & 0.162$\pm$0.021 & 1 & (31)\\
  AS 310 & 1.0 & 215 & 0.768$\pm$0.050 & 1 & (6)\\
  HD 176386 & 1.0 & 33 & 0.121$\pm$0.010 & 1 & (15)\\
  TY CrA & 1.0 & 38 & < 0.017 & 1 & (17)\\
  HD 179218 & 1.4 & 24 & 0.029$\pm$0.003 & 2 & (16)\\
  WW Vul & 1.0 & 23 & 0.016$\pm$0.001 & 1 & (4)\\
  V1685 Cyg & 1.0 & 101 & 0.255$\pm$0.033 & 1 & (7)\\
  V1686 Cyg & 1.0 & 16 & < 0.040 & 1 & (8)\\
  V1478 Cyg & 0.0 & 35 & 12.973$\pm$0.349 & 2 & (1)\\
  HD 200775 & 1.0 & 71 & < 0.015 & 1 & (17)\\
  HD 235495 & 1.0 & 33 & 0.038$\pm$0.005 & 1 & (18)\\
  V361 Cep & 1.0 & 71 & < 0.032 & 1 & (7)\\
  V373 Cep & 1.0 & 44 & < 0.051 & 1 & (34)\\
  AS 477 & 1.0 & 33 & < 0.051 & 1 & (7)\\
  SV Cep & 1.0 & 23 & 0.005$\pm$0.001 & 1 & (28)\\
  MWC 1080 & 1.6 & 229 & 0.308$\pm$0.039 & 7 & (14)\\
  LkHa 259 & 1.0 & 18 & 93.196$\pm$40.276 & 1 & (10)\\
\hline\end{tabular}
\tablebib{mm flux measurements: (1) \citet{Sandell_2011}; (2) \citet{Mannings_1997}; (3) \citet{Guedel_1989}; (4) \citet{Acke_2004}; (5) \citet{Pericaud_2017}; (6) \citet{Henning_1994}; (7) \citet{Hillenbrand_1992}; (8) \citet{Mendigutia_2012}; (9) \citet{Ginsburg_2013}; (10) \citet{Planck_2013}; (11) \citet{Sylvester_1996}; (12) \citet{Barenfeld_2016}; (13) \citet{Meeus_2012}; (14) \citet{Mannings_1994}; (15) \citet{Francesco_2008}; (16) \citet{Mannings_2000}; (17) \citet{Pezzuto_1997}; (18) \citet{Kauffmann_2008}; (19) \citet{Kraus_2017}; (20) \citet{Pietu_2006}; (21) \citet{Pietu_2003};
(22) \citet{Sylvester_2001}; (23) \citet{Sheret_2004}; (24) \citet{Henning_1994_2}; (25) \citet{Cotten_2016}; (26) \citet{Henning_1993}; (27) \citet{Ribas_2017}; (28) \citet{Natta_1997}; (29) \citet{Planck_2018}; (30) \citet{Urquhart_2014}; (31) \citet{Enoch_2008}; (32) \citet{Boissier_2011}; (33) \citet{Natta_2000}; (34) \citet{Alonso_2009}; (35) \citet{Reipurth_1993}; (36) \citet{Giannini_1996}; (37) \citet{Mairs_2016};
}
\end{table}

\newpage

\onecolumn

\section{SEDs}\label{appC}

\begin{figure} [h]
 \centering
    \includegraphics[width=0.33\textwidth]{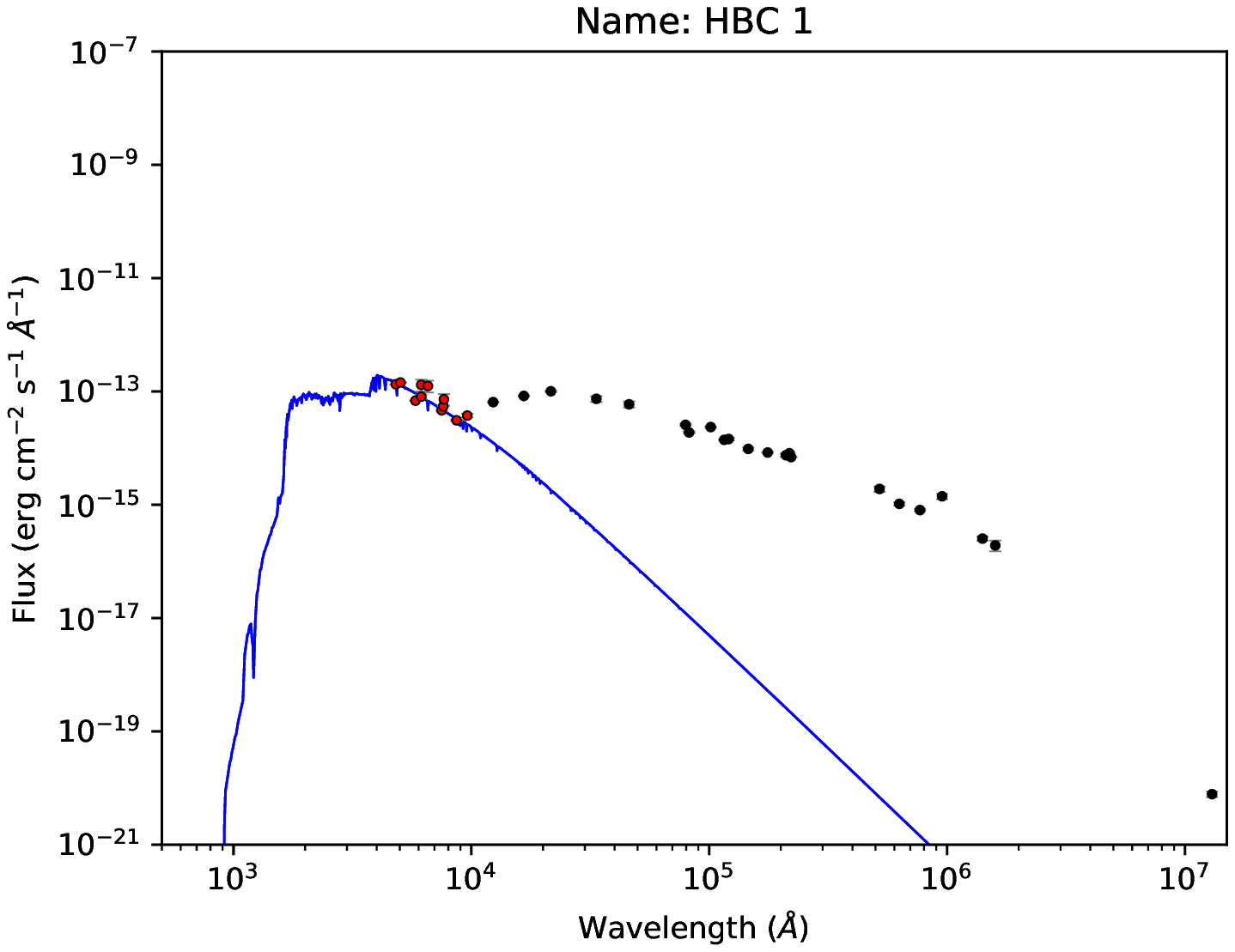}
    \includegraphics[width=0.33\textwidth]{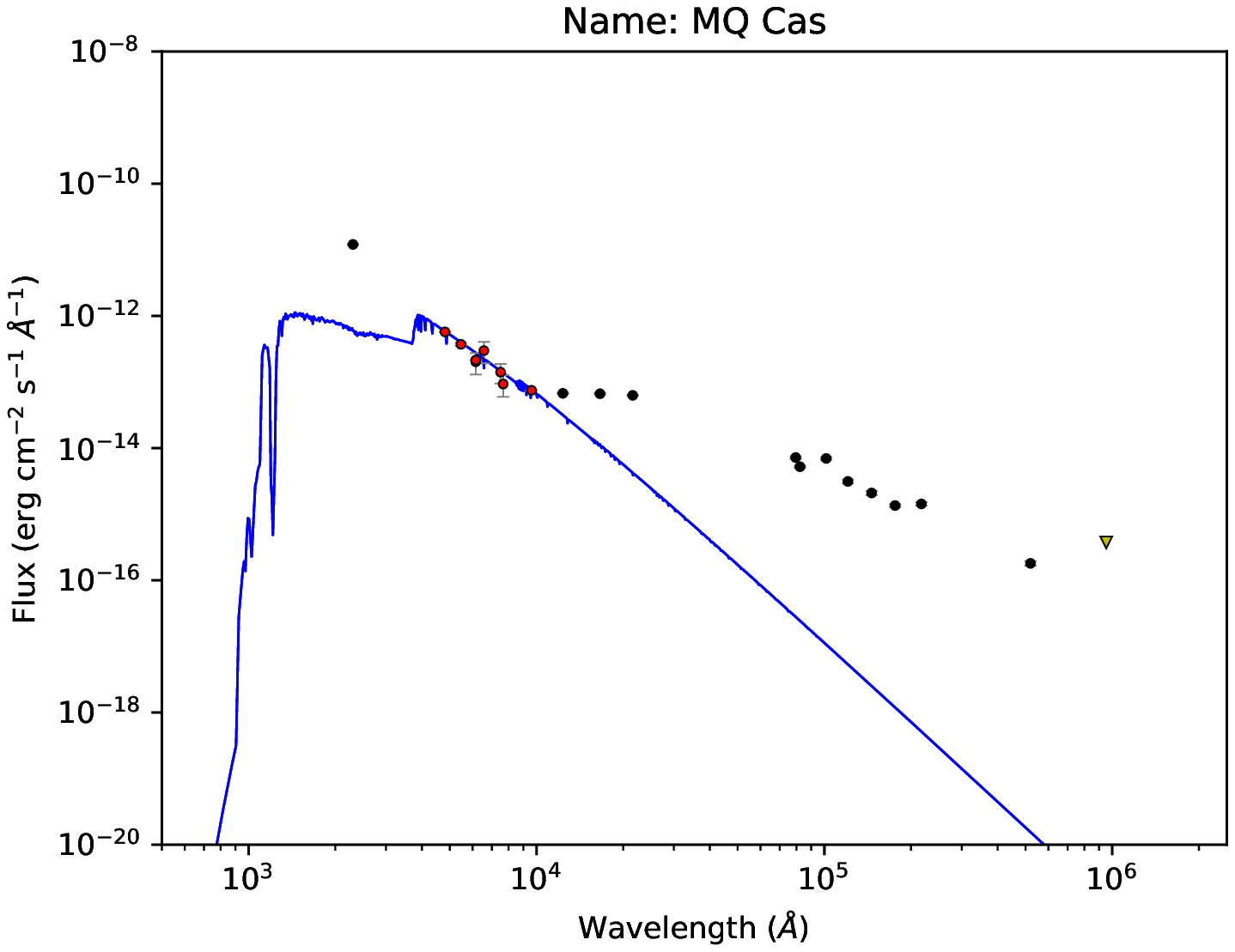}
    \includegraphics[width=0.33\textwidth]{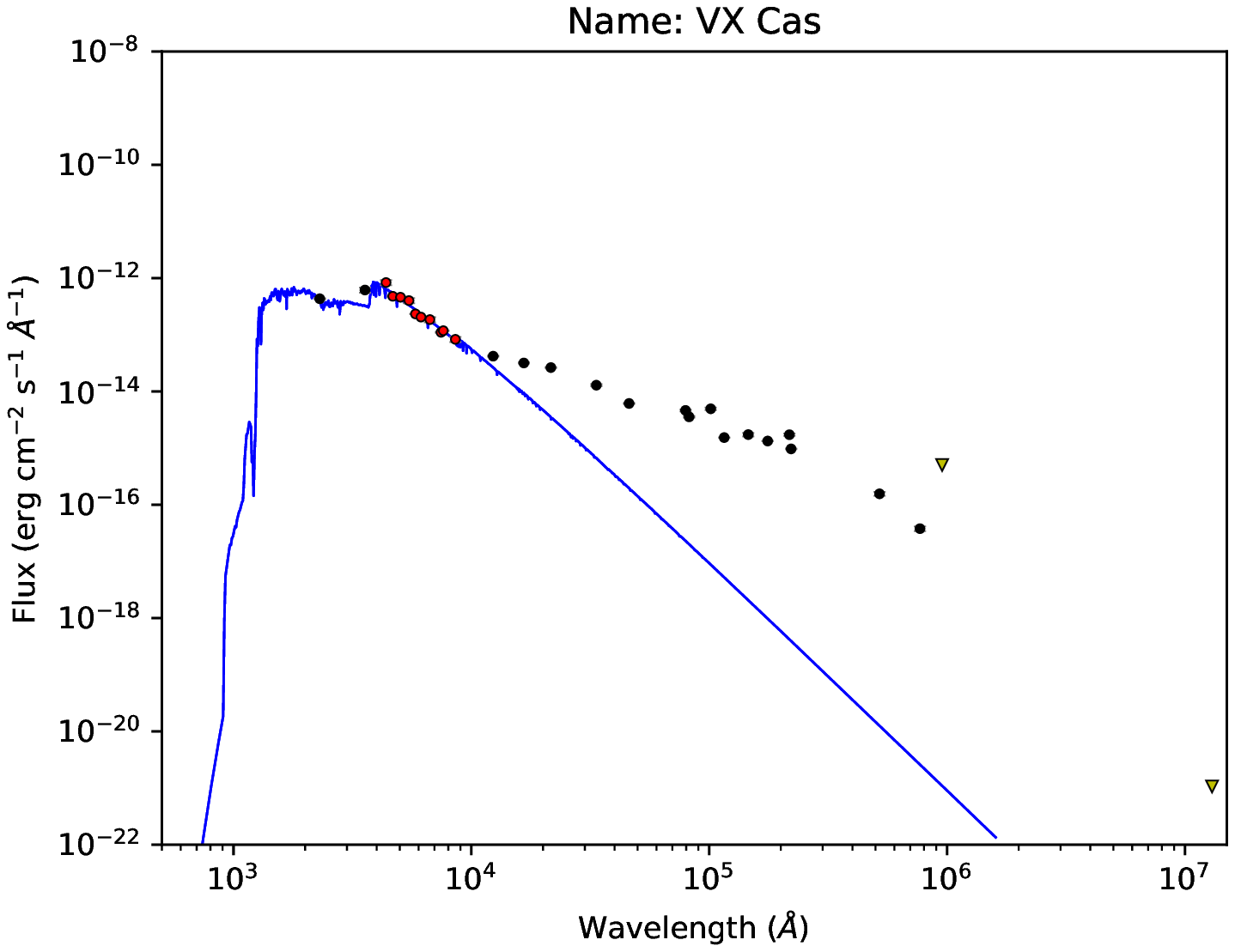}    
\end{figure}

\begin{figure} [h]
 \centering
    \includegraphics[width=0.33\textwidth]{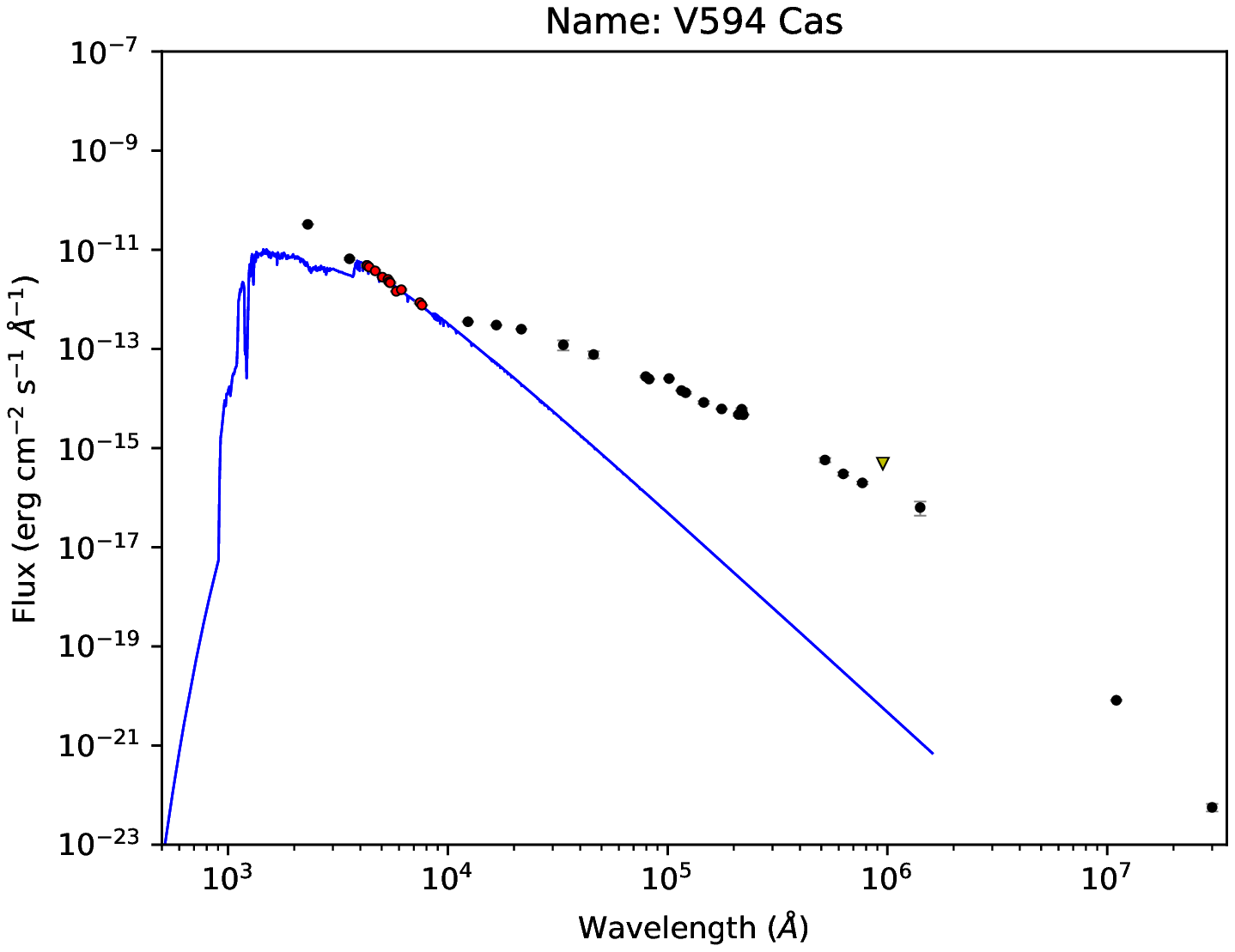}
    \includegraphics[width=0.33\textwidth]{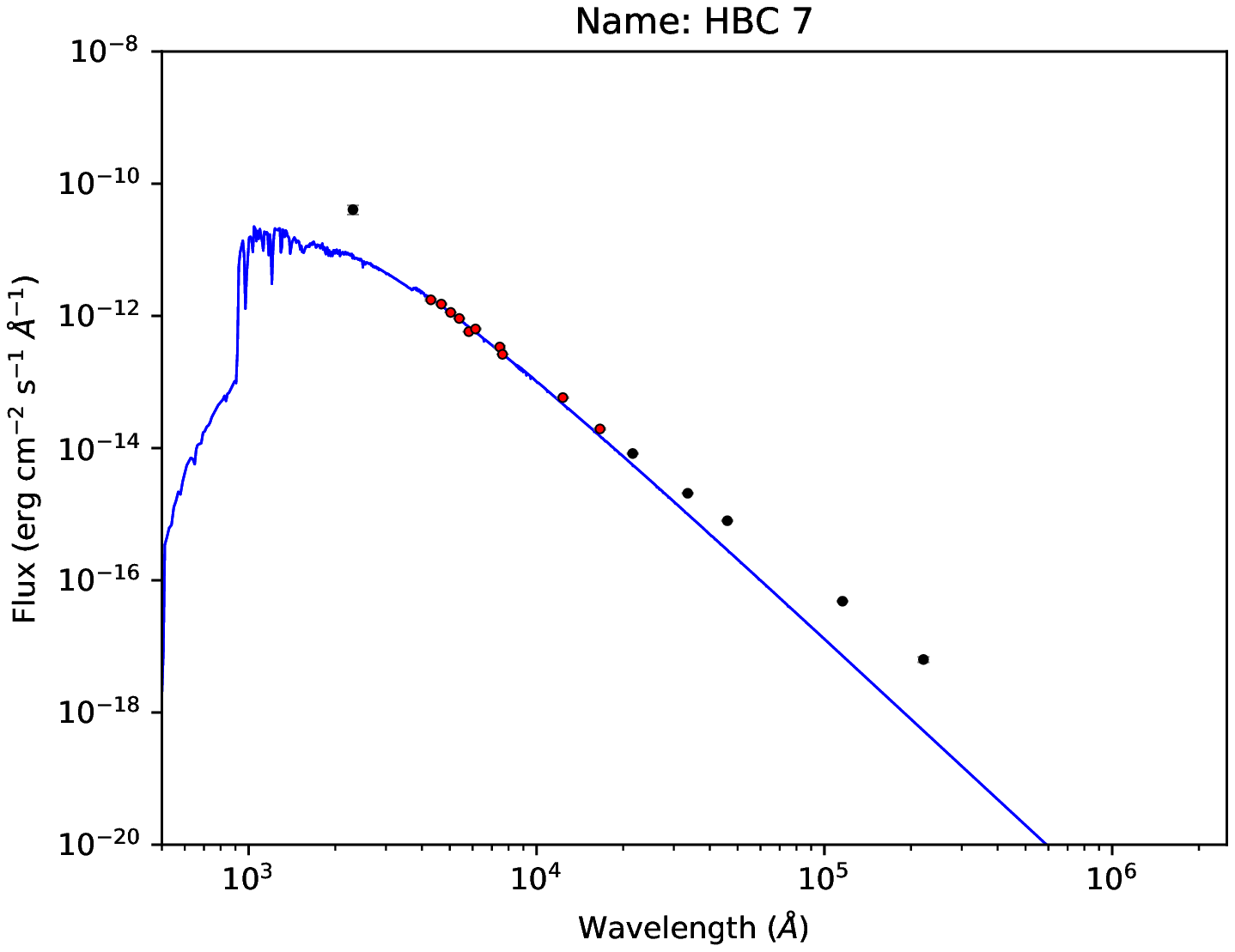}
    \includegraphics[width=0.33\textwidth]{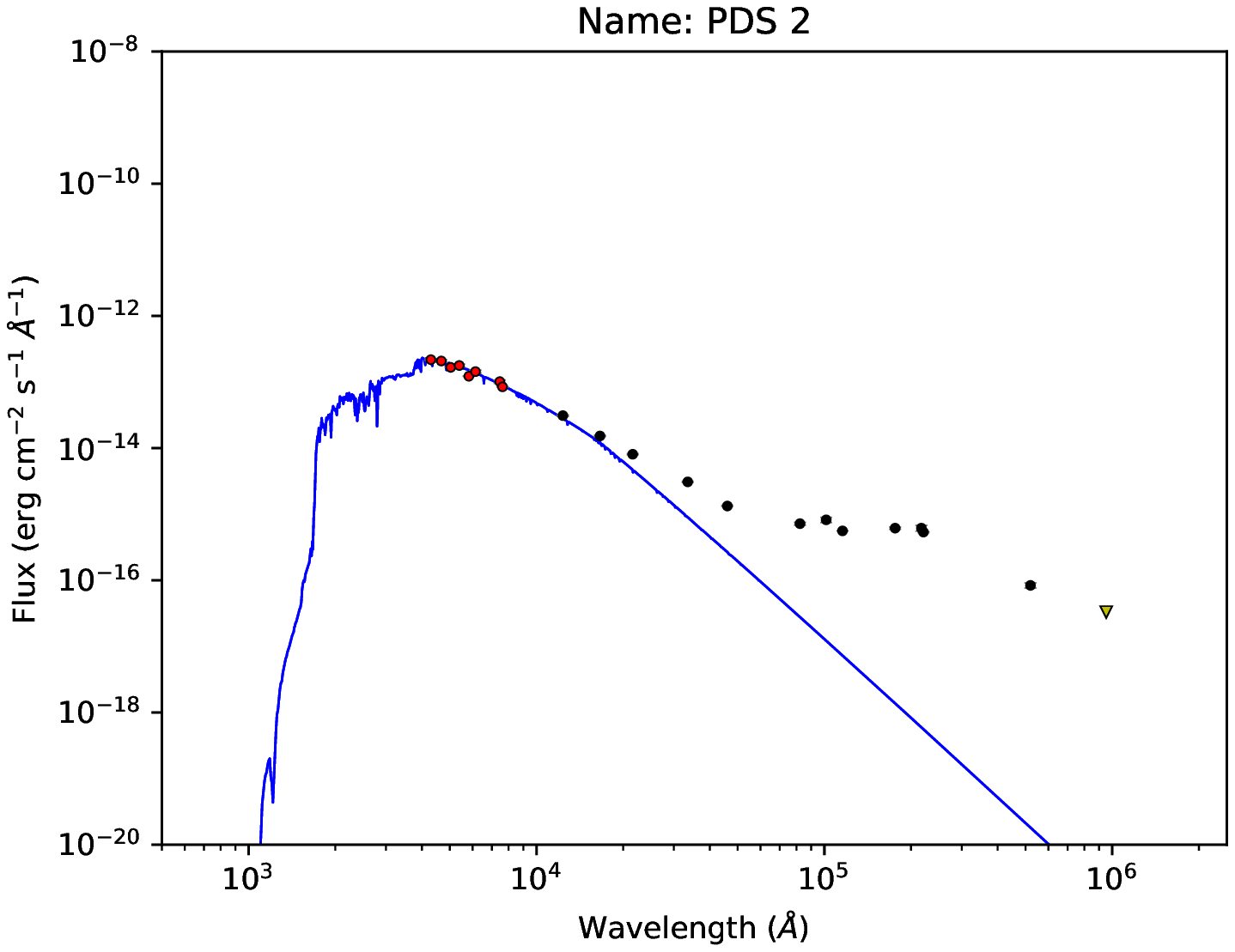}    
\end{figure}

\begin{figure} [h]
 \centering
    \includegraphics[width=0.33\textwidth]{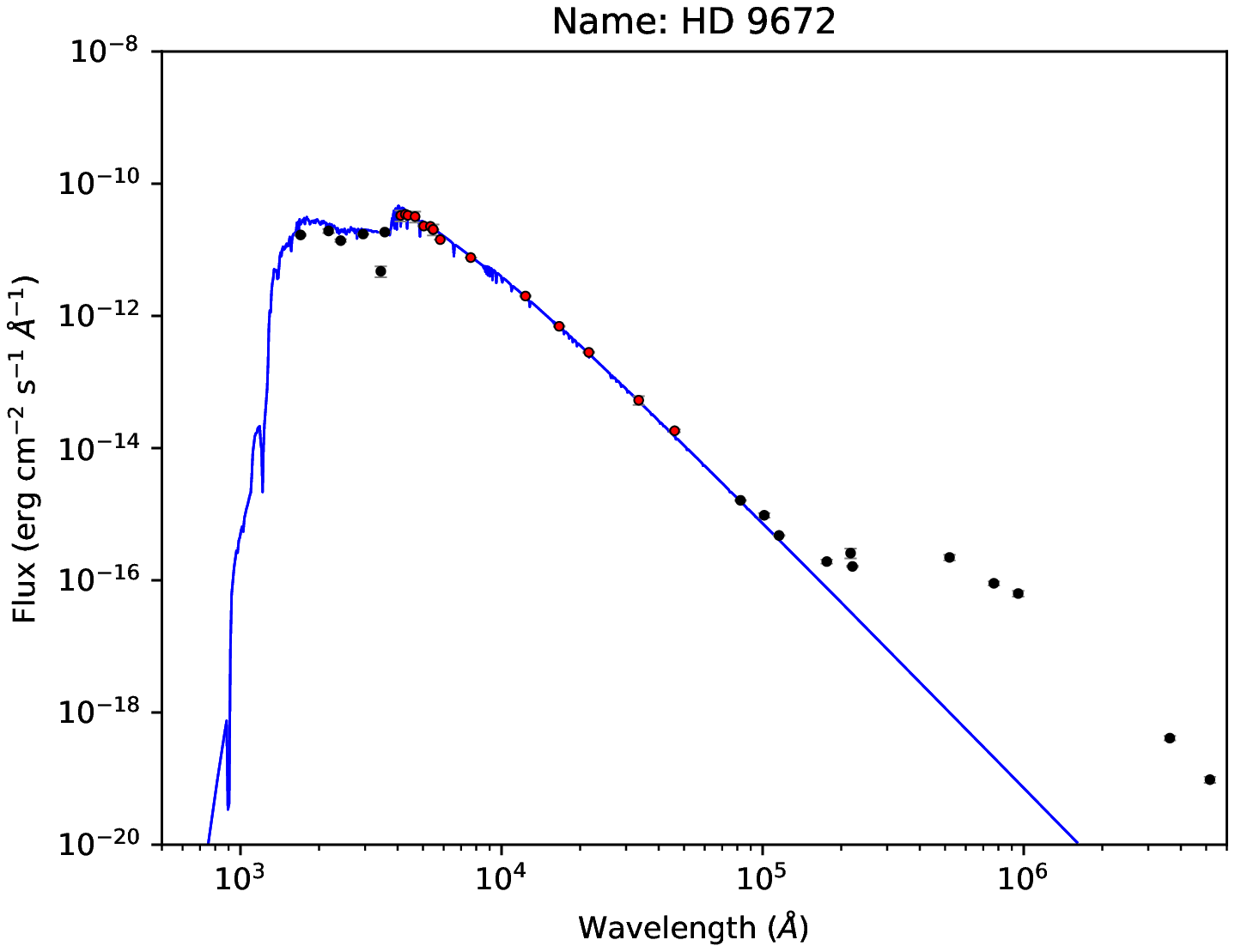}  
    \includegraphics[width=0.33\textwidth]{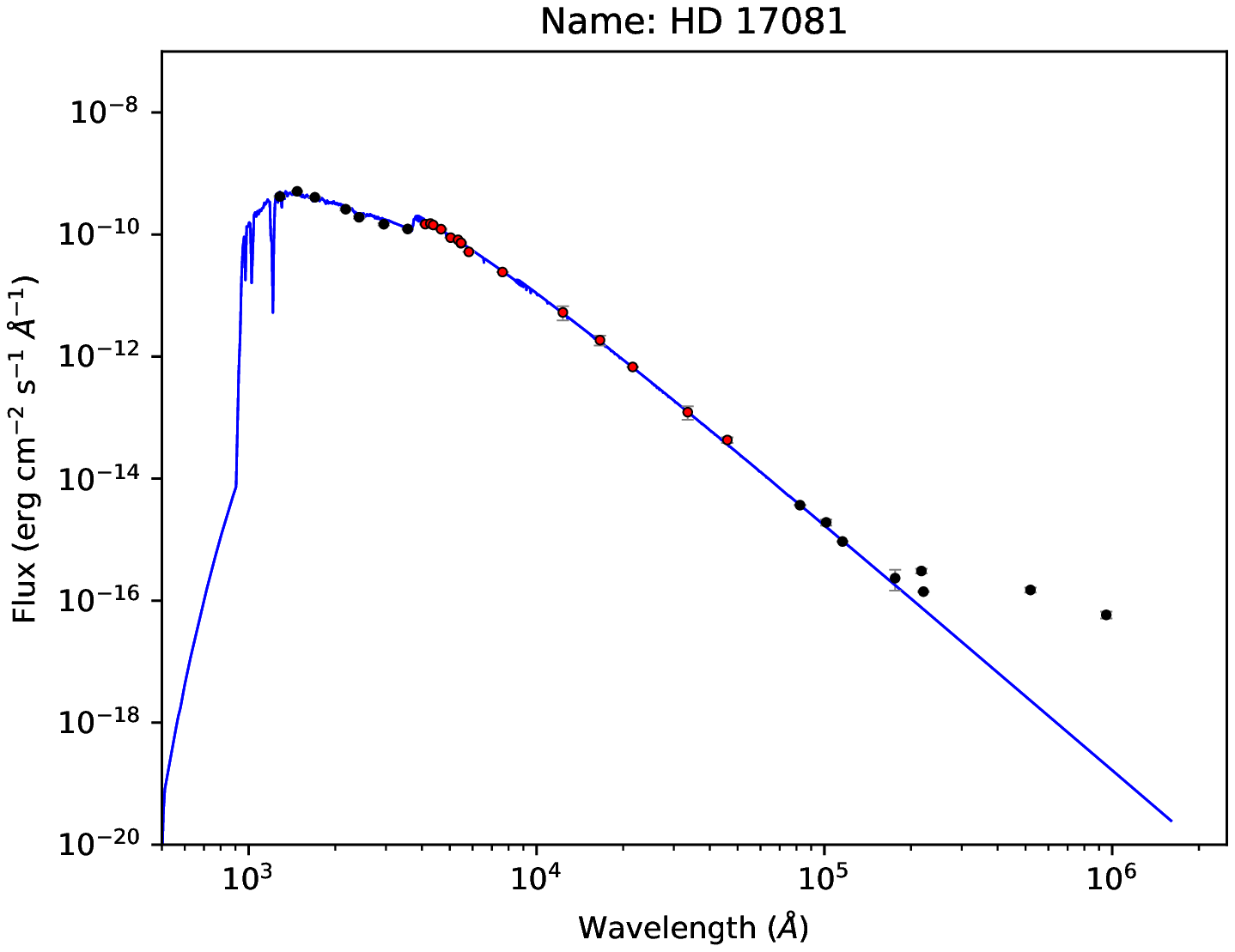}
    \includegraphics[width=0.33\textwidth]{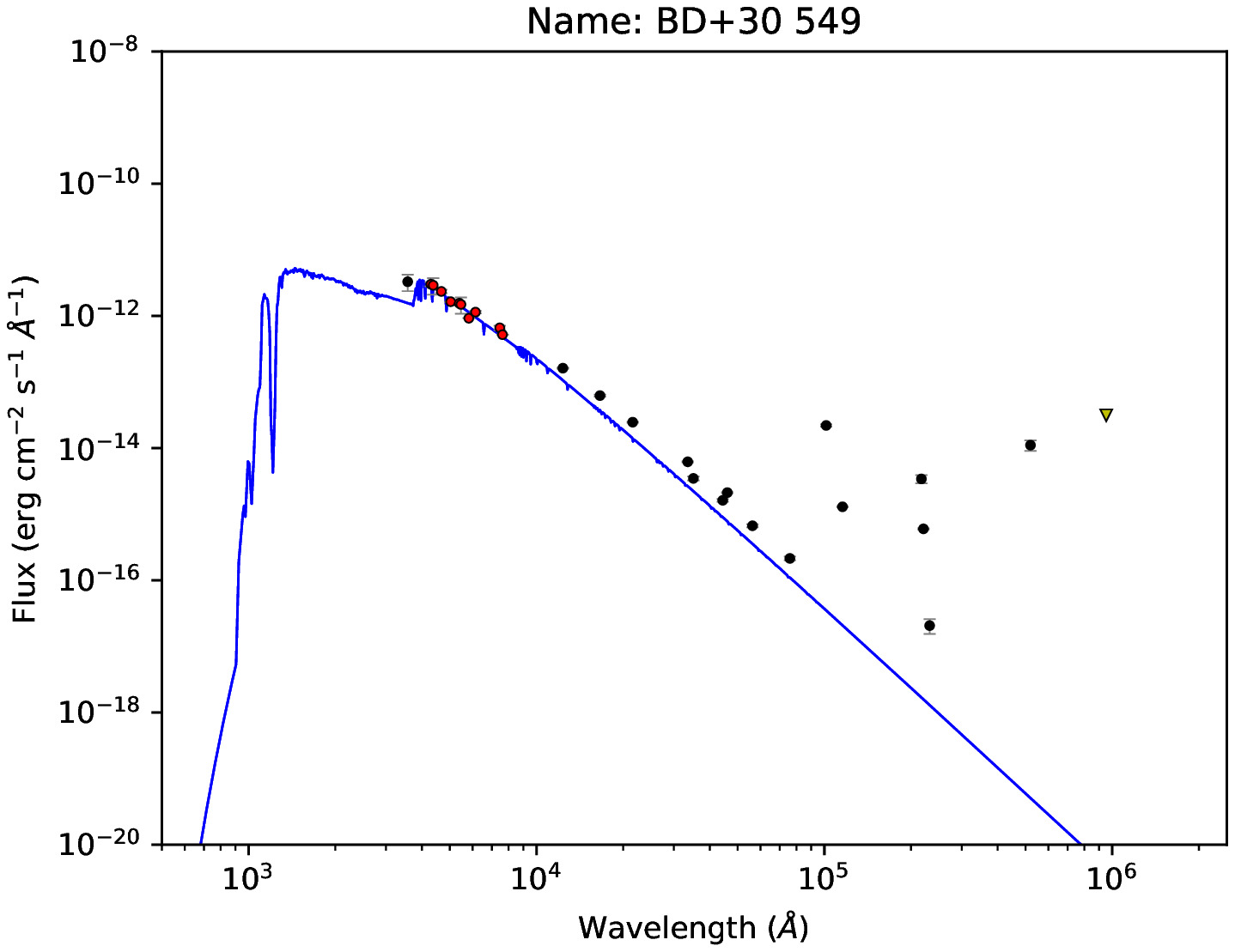}      
\end{figure}

\begin{figure} [h]
 \centering
    \includegraphics[width=0.33\textwidth]{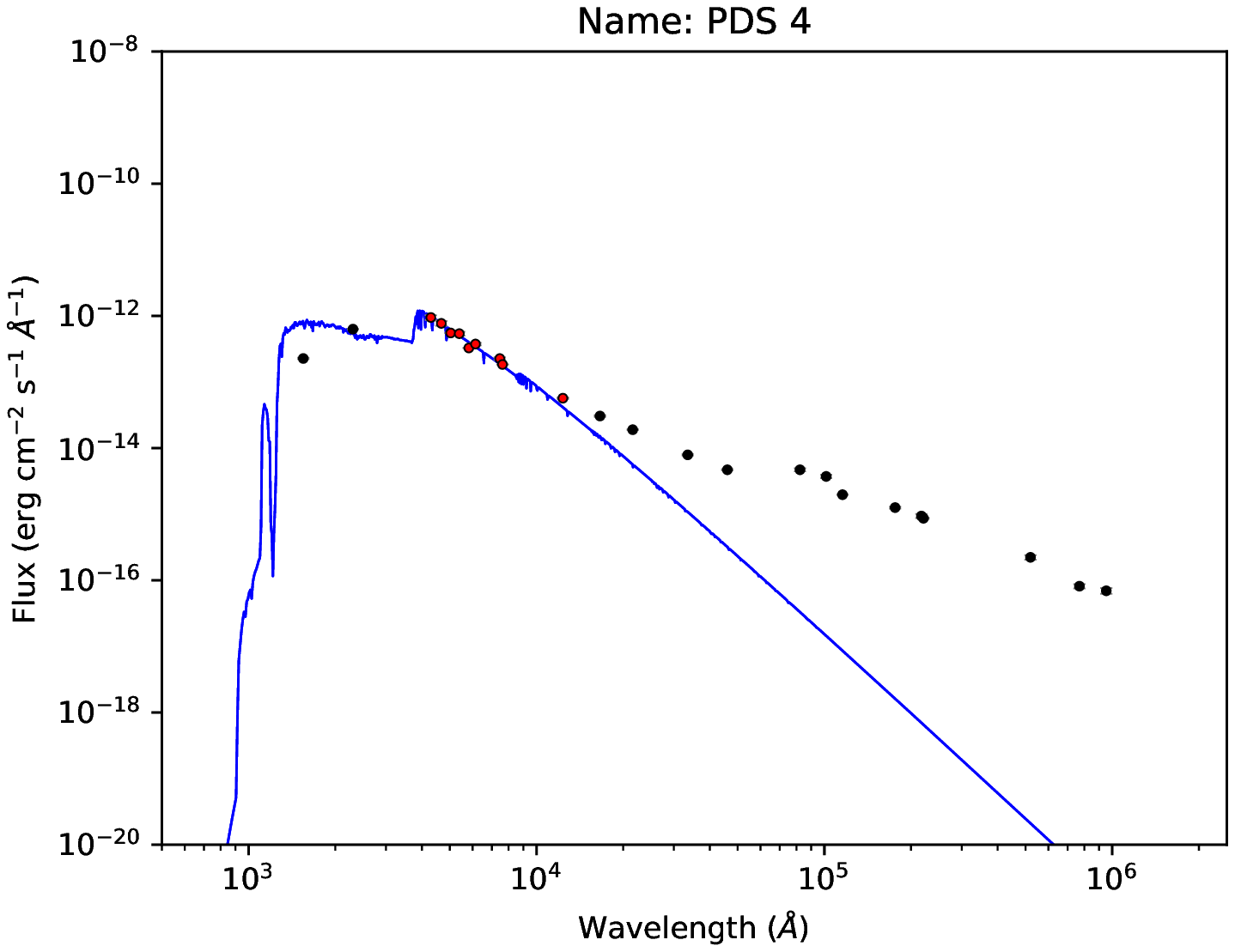}
    \includegraphics[width=0.33\textwidth]{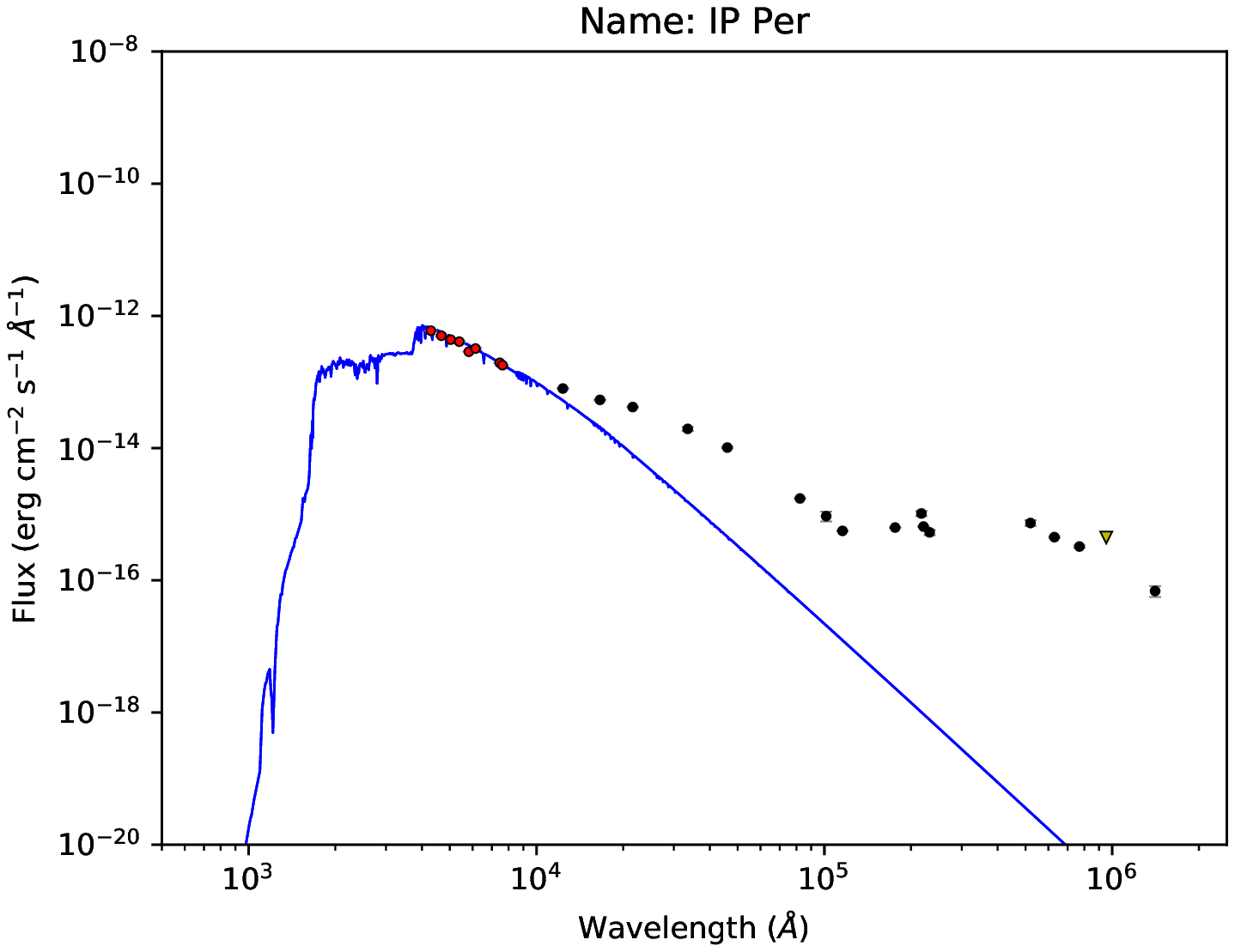}
    \includegraphics[width=0.33\textwidth]{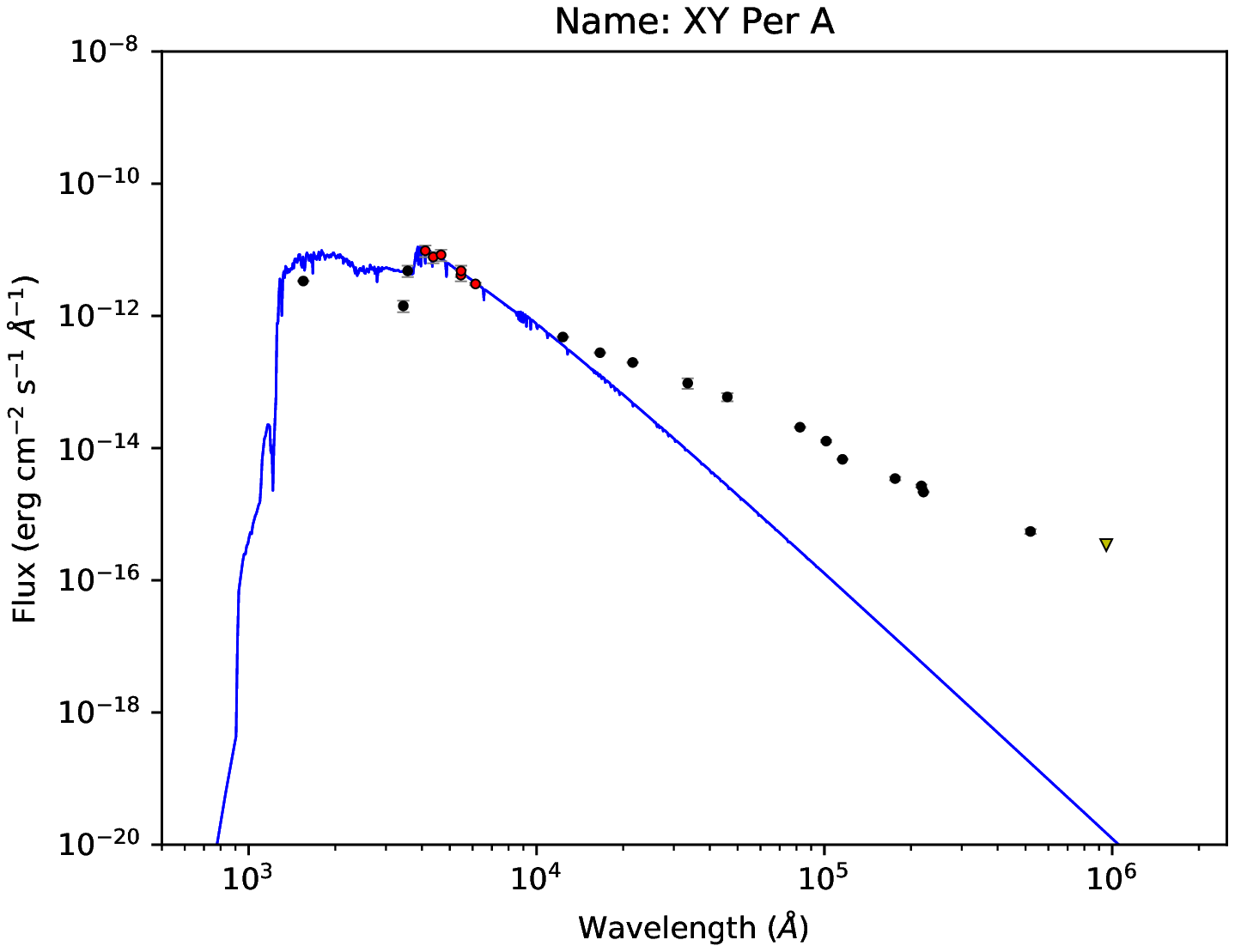}    
\end{figure}

\newpage

\onecolumn

\begin{figure} [h]
 \centering
    \includegraphics[width=0.33\textwidth]{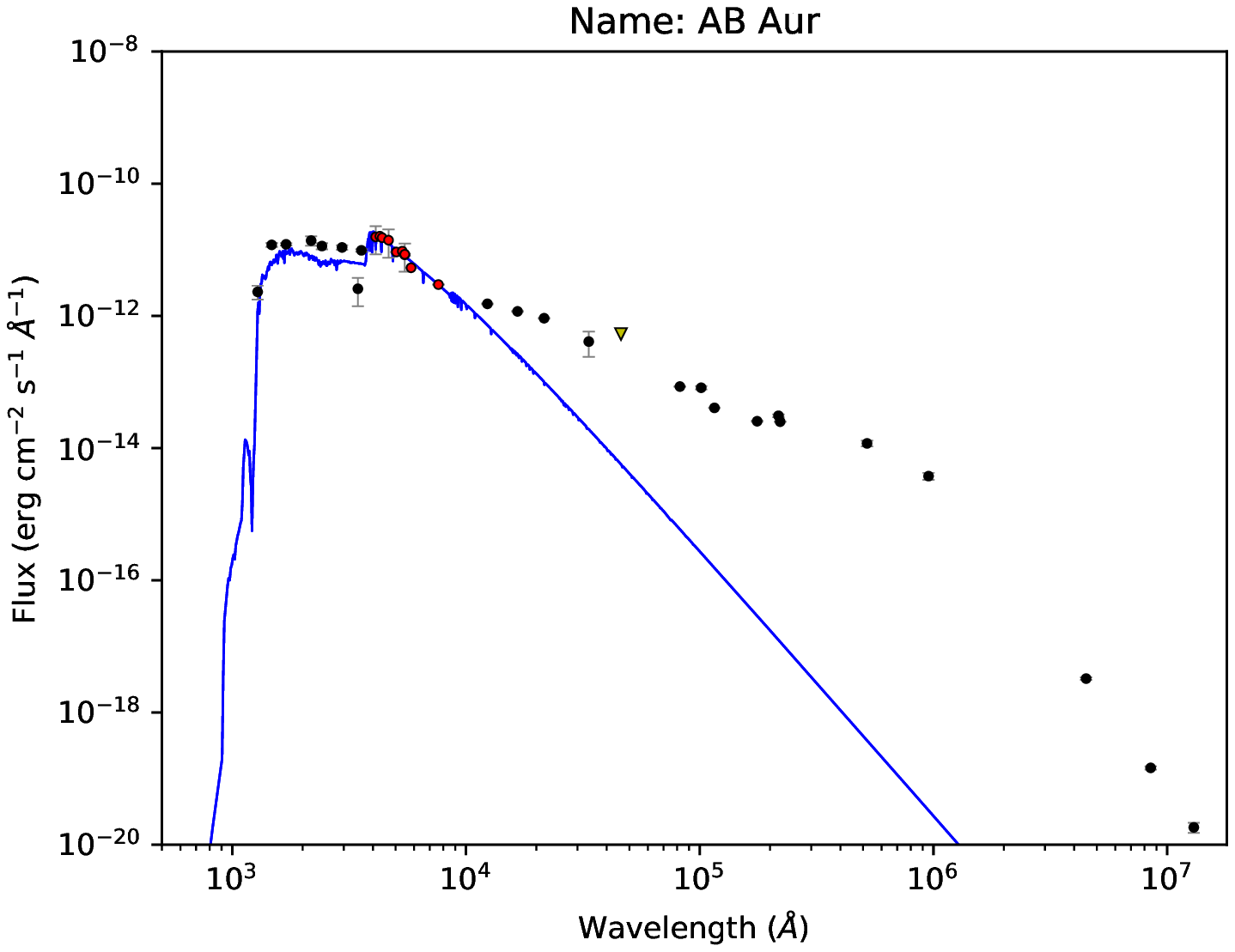}
    \includegraphics[width=0.33\textwidth]{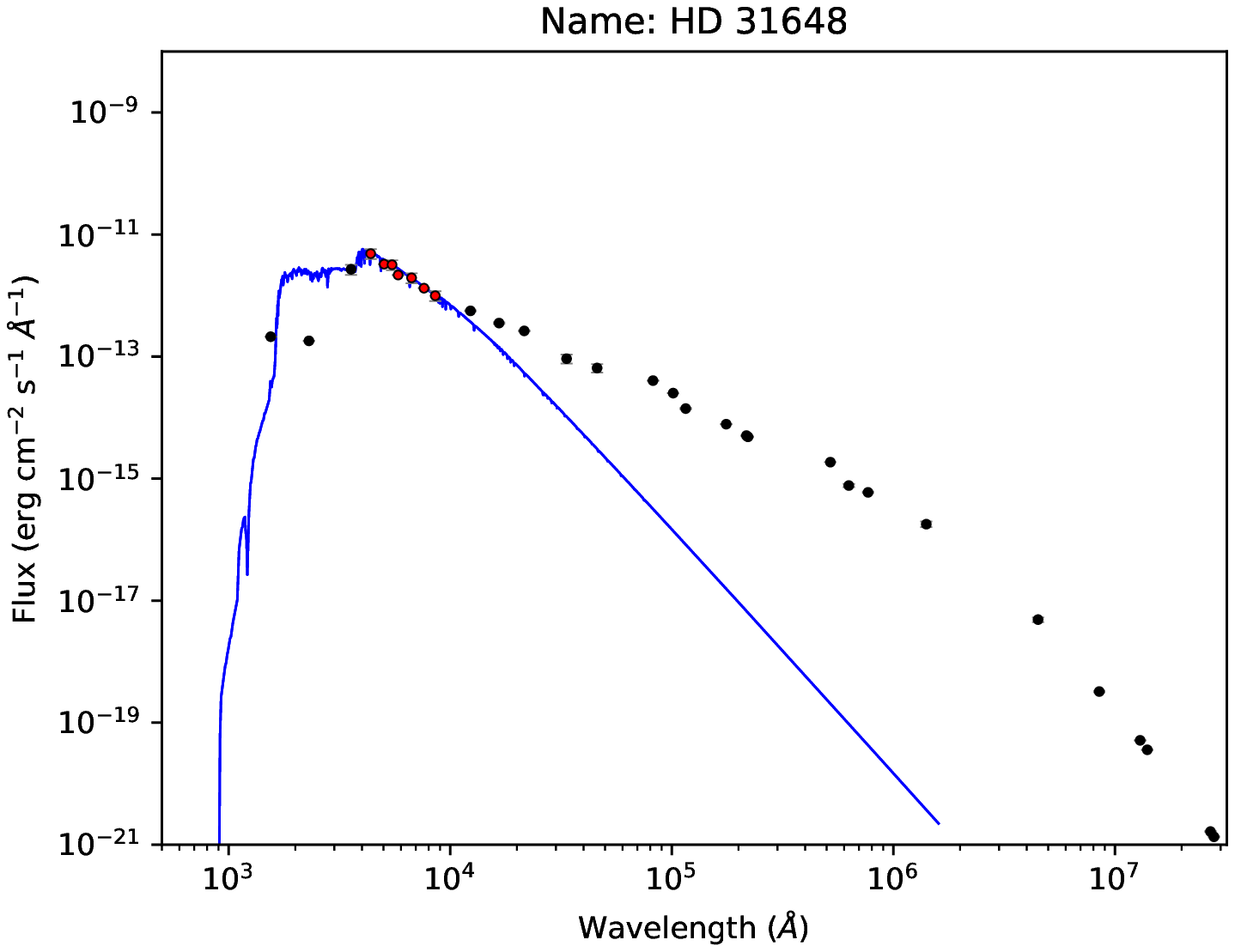}
    \includegraphics[width=0.33\textwidth]{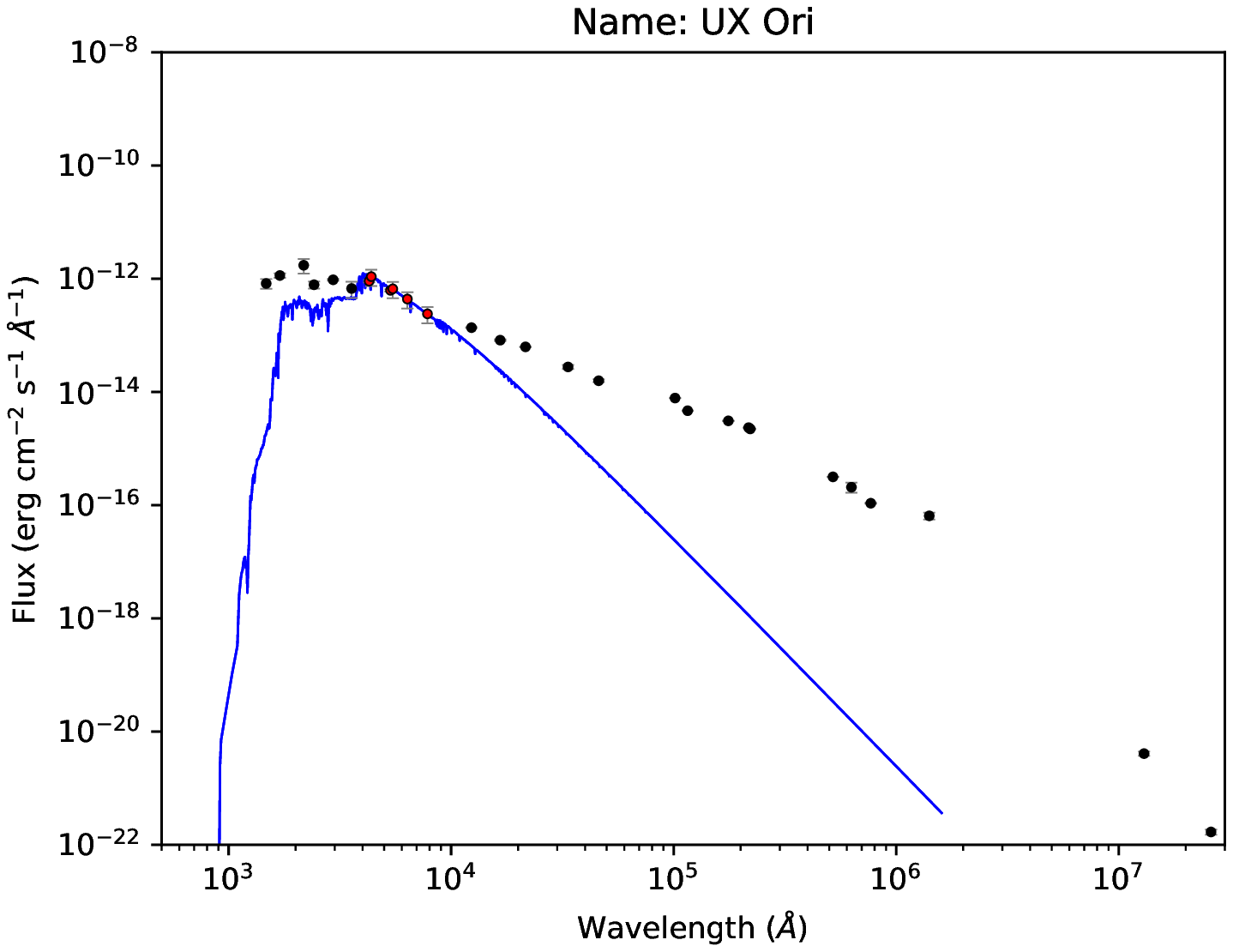}    
\end{figure}

\begin{figure} [h]
 \centering
    \includegraphics[width=0.33\textwidth]{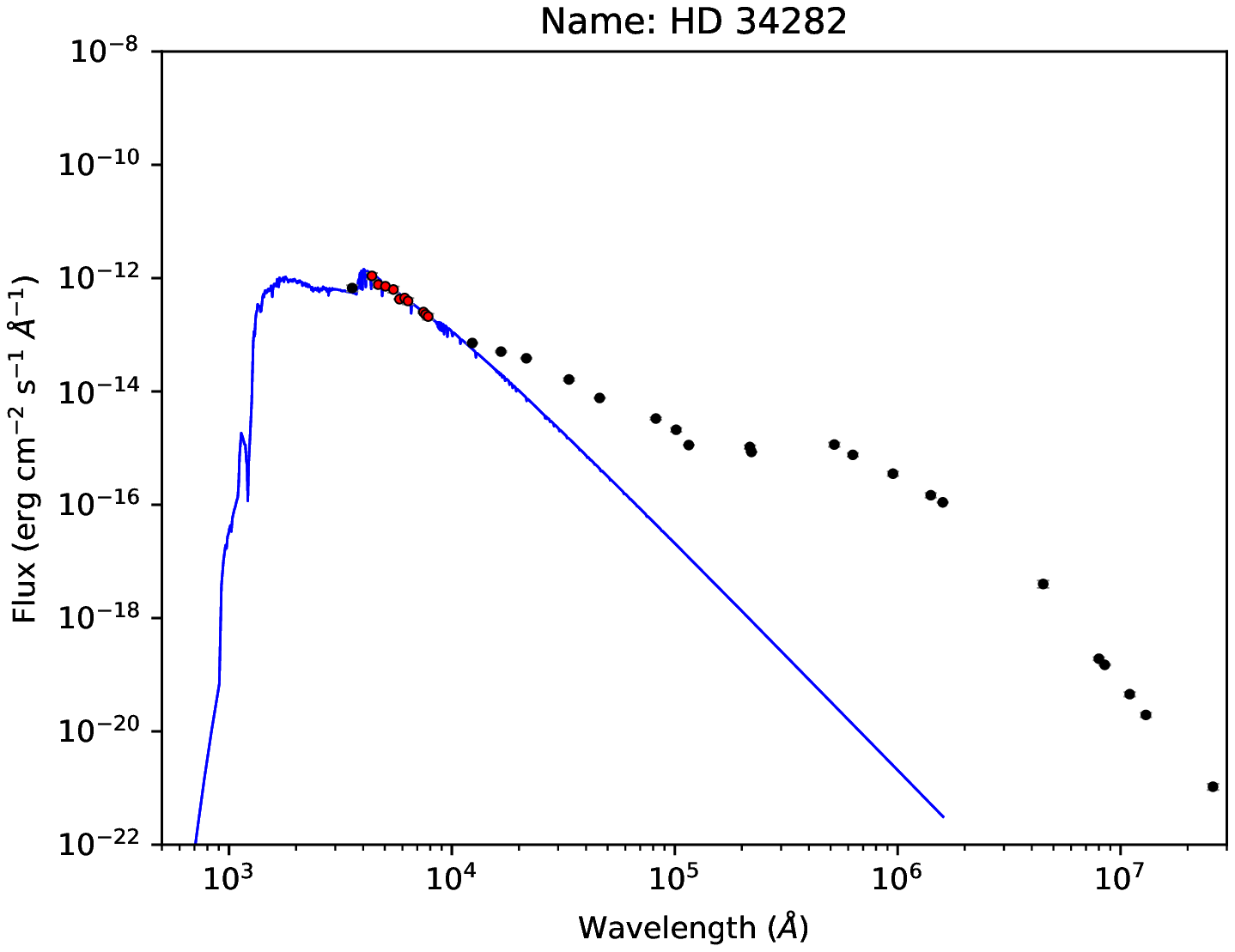}
    \includegraphics[width=0.33\textwidth]{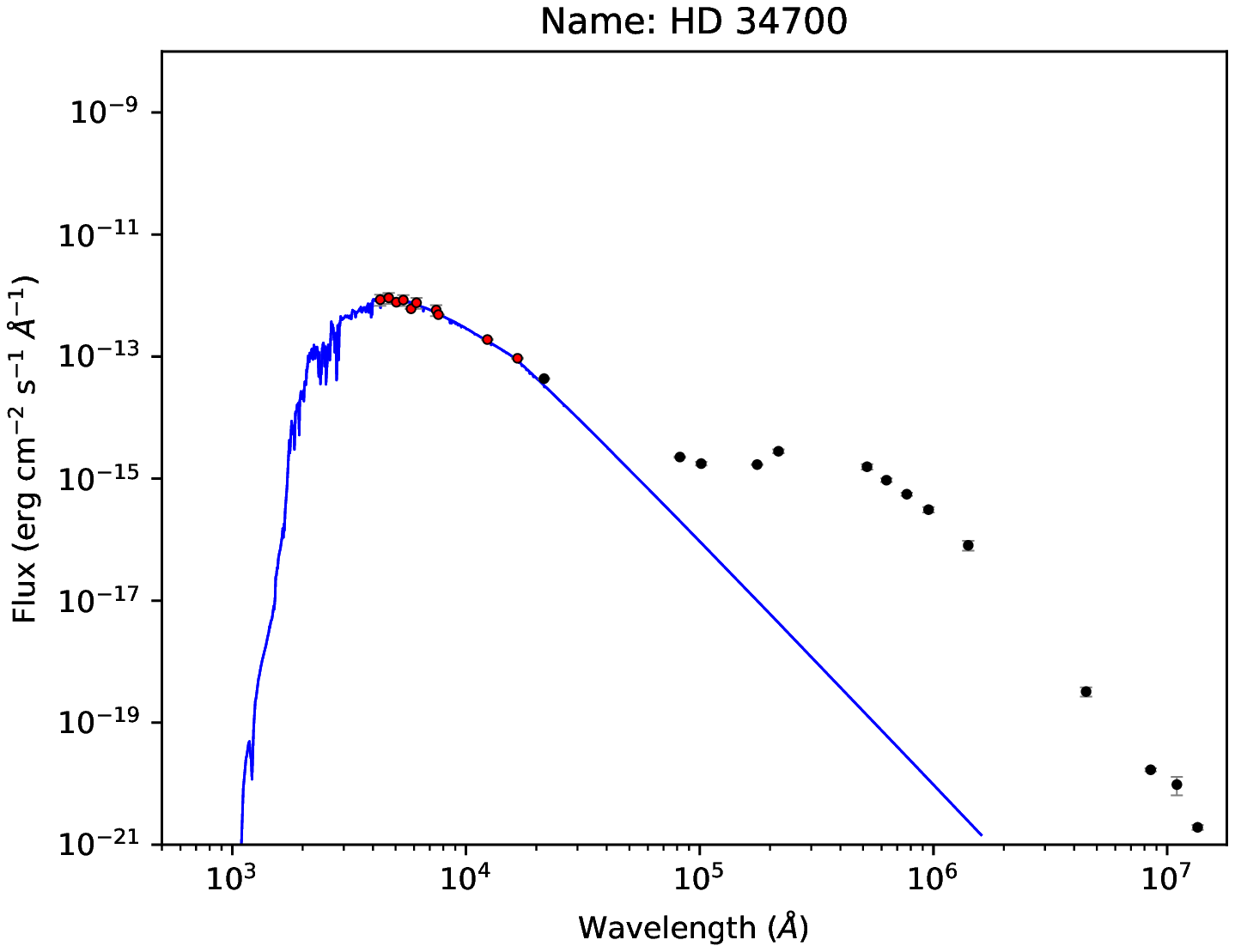}
    \includegraphics[width=0.33\textwidth]{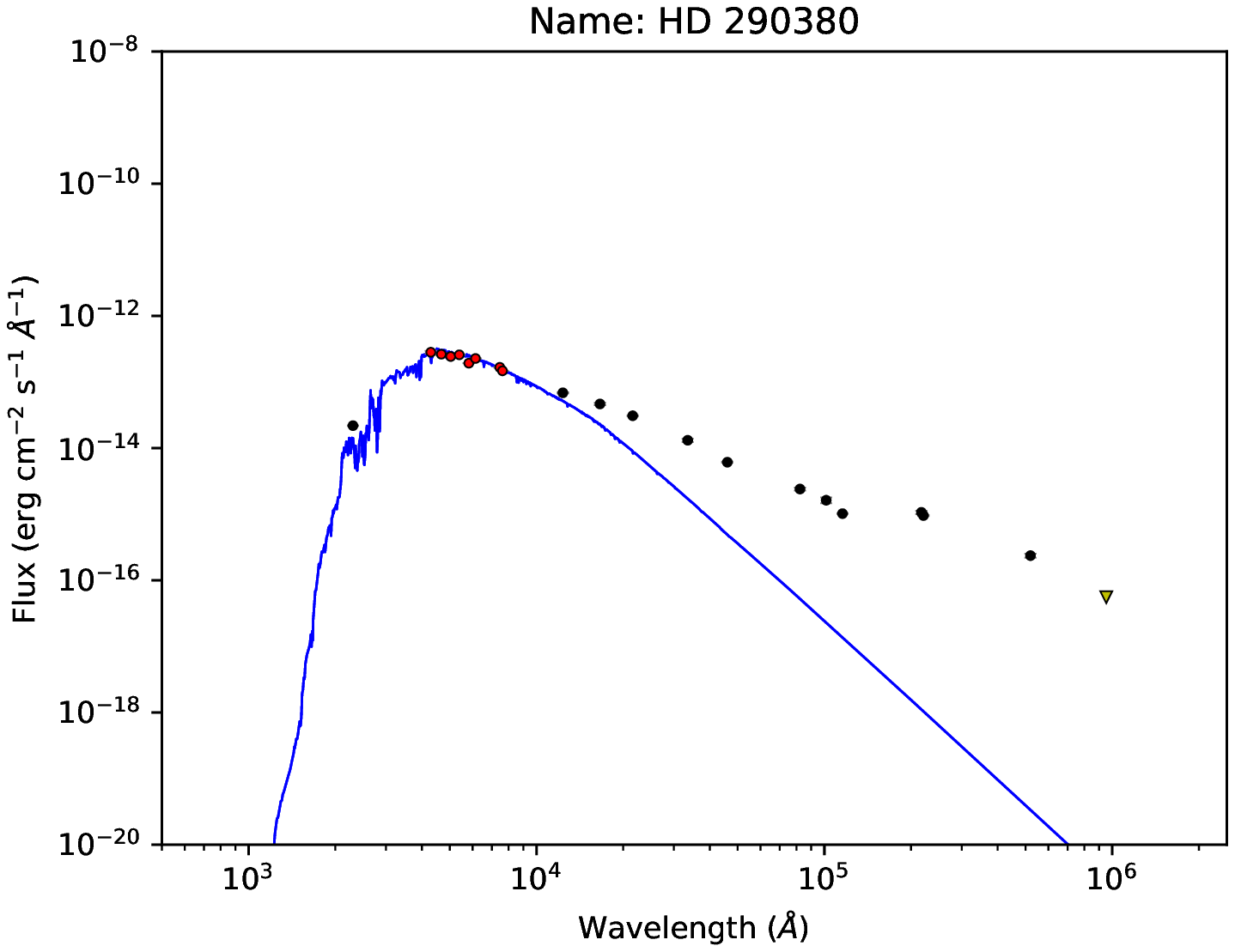}    
\end{figure}

\begin{figure} [h]
 \centering
    \includegraphics[width=0.33\textwidth]{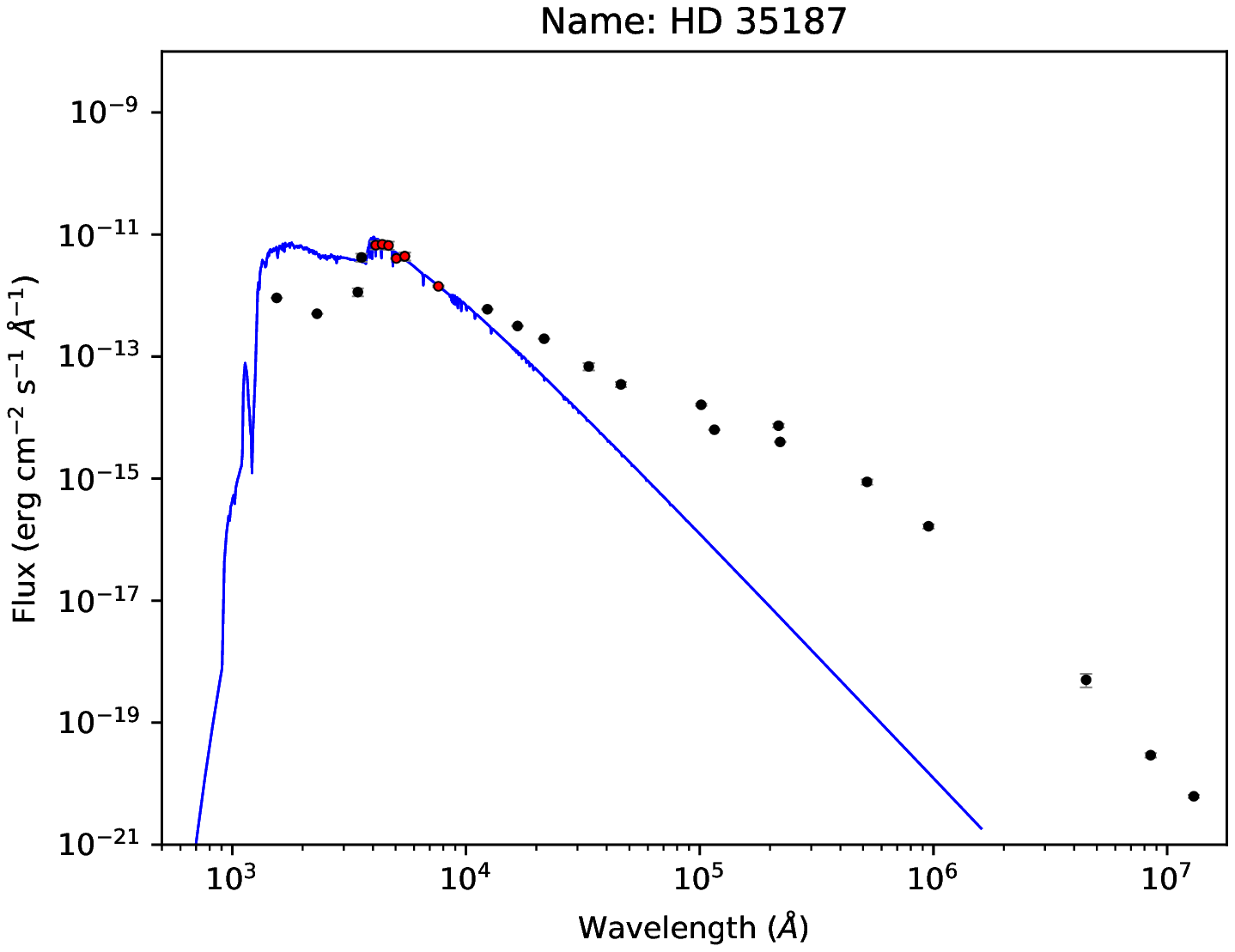}
    \includegraphics[width=0.33\textwidth]{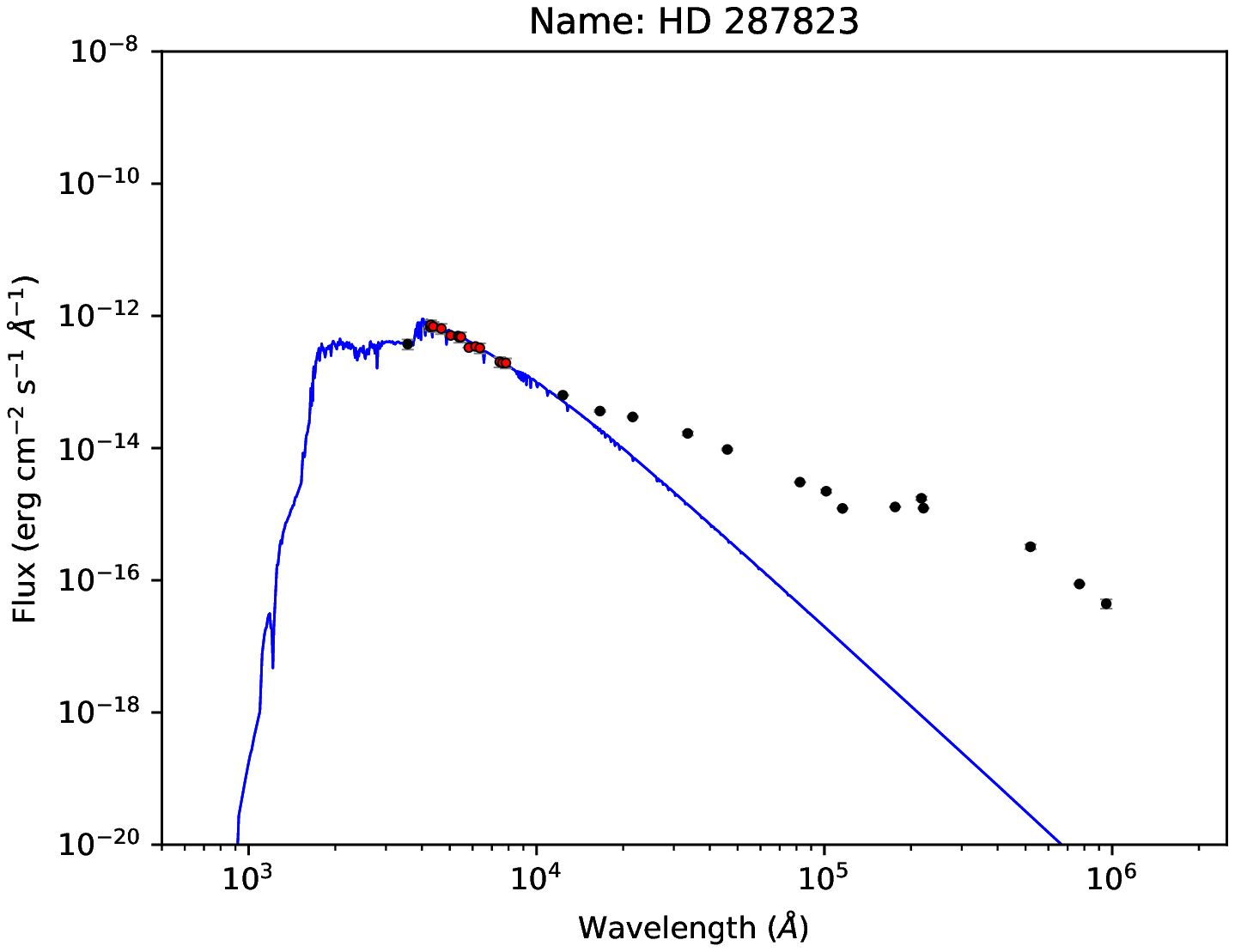}
    \includegraphics[width=0.33\textwidth]{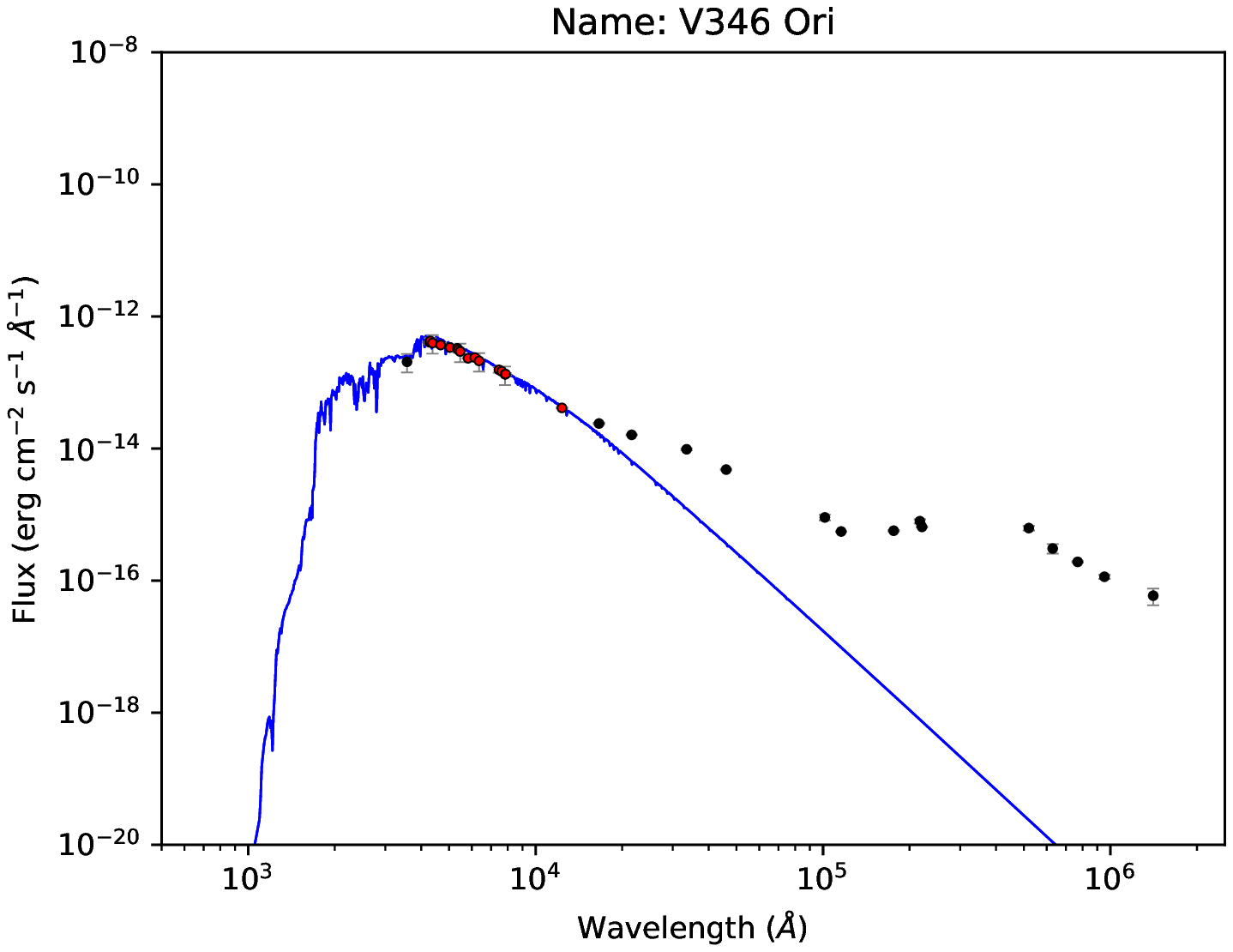}    

\end{figure}

\begin{figure} [h]
 \centering
    \includegraphics[width=0.33\textwidth]{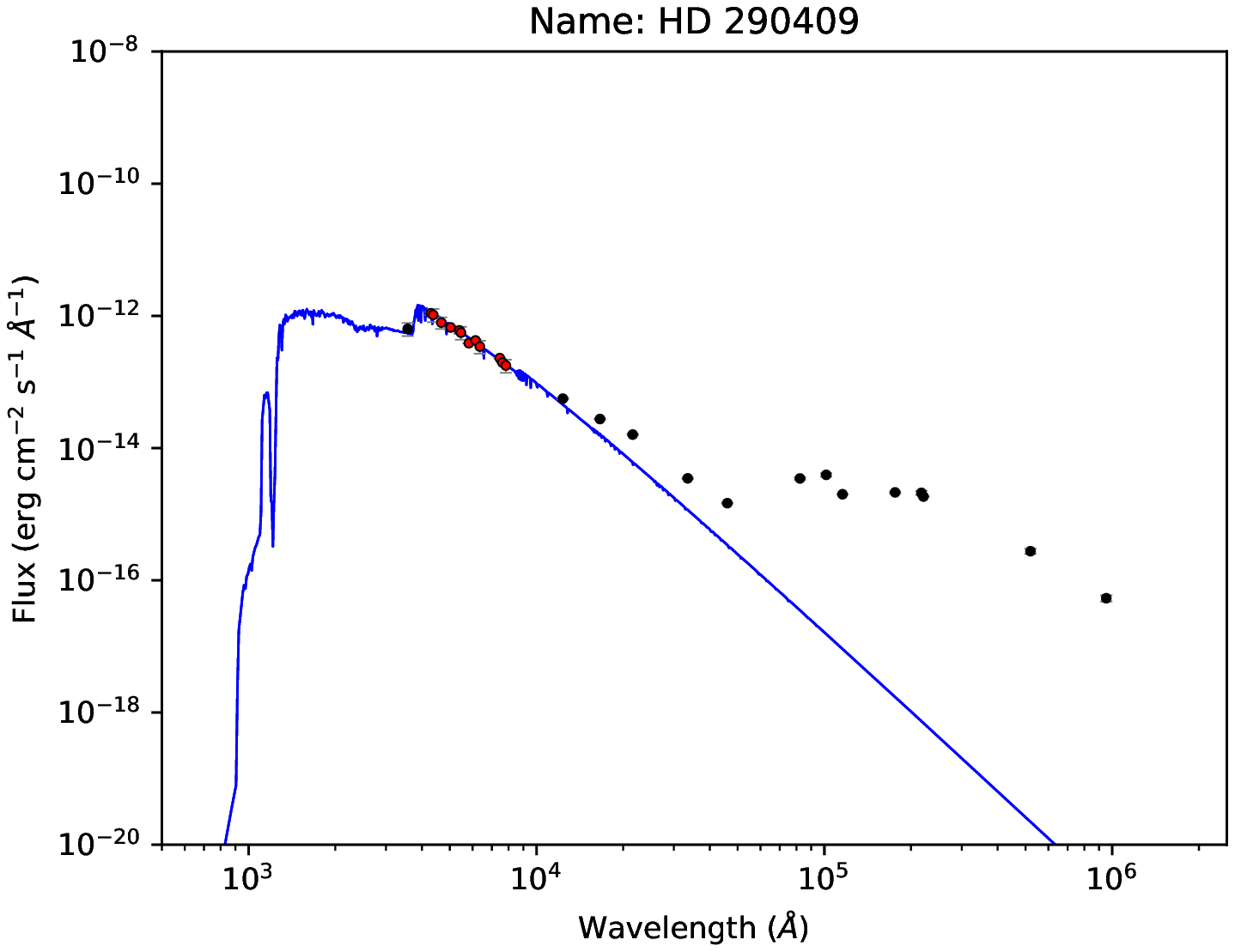}
    \includegraphics[width=0.33\textwidth]{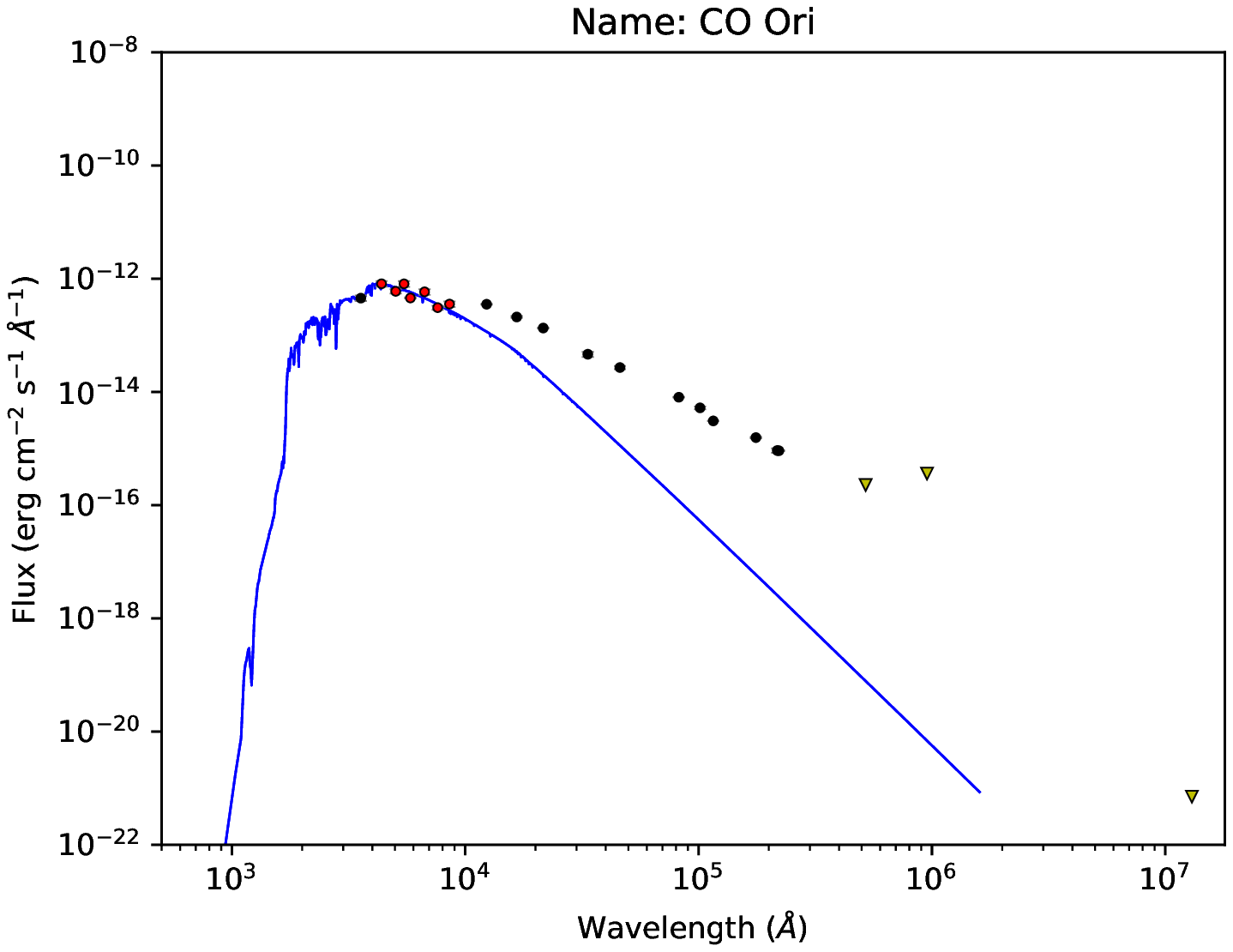}
    \includegraphics[width=0.33\textwidth]{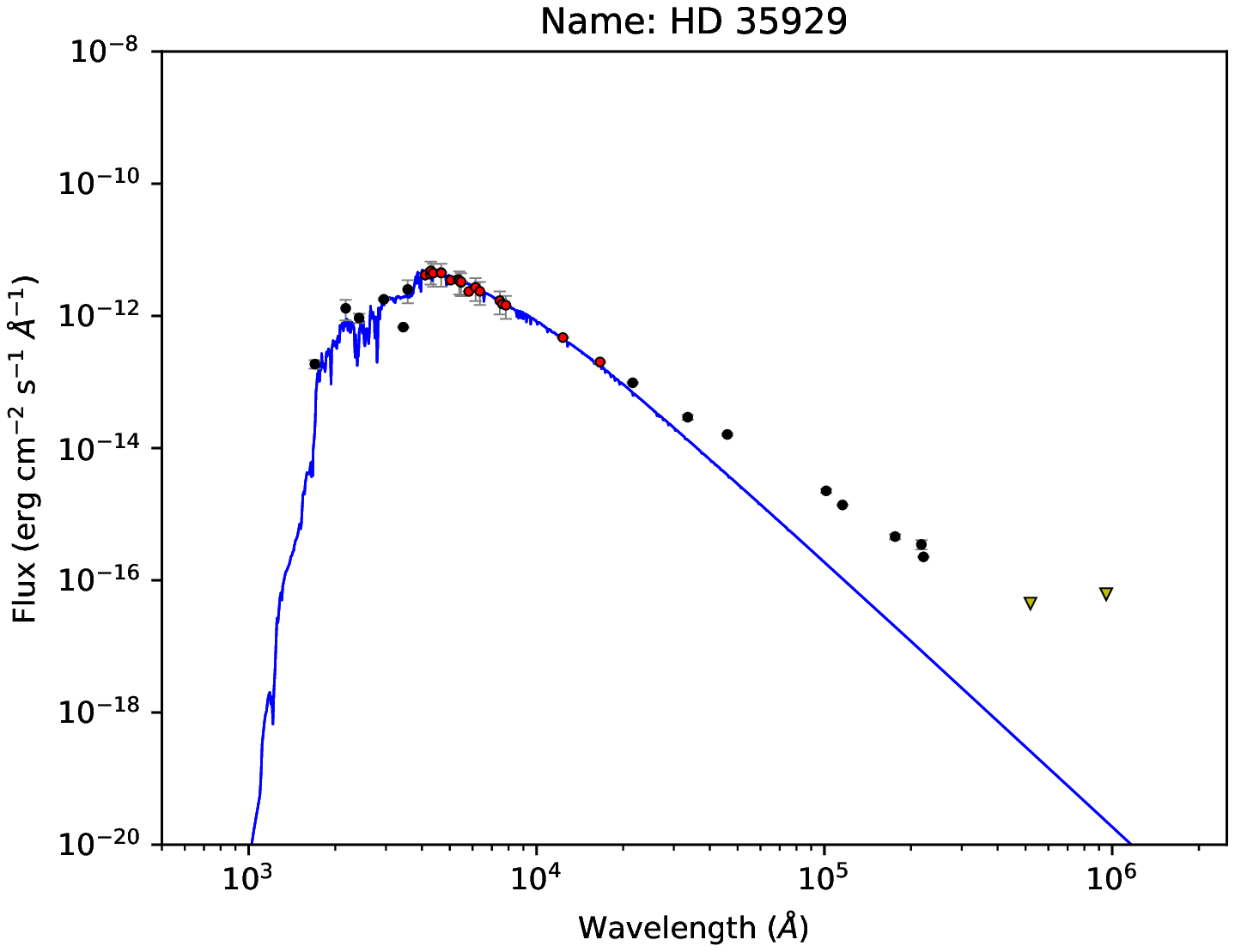}    
\end{figure}

\newpage

\onecolumn

\begin{figure} [h]
 \centering
    \includegraphics[width=0.33\textwidth]{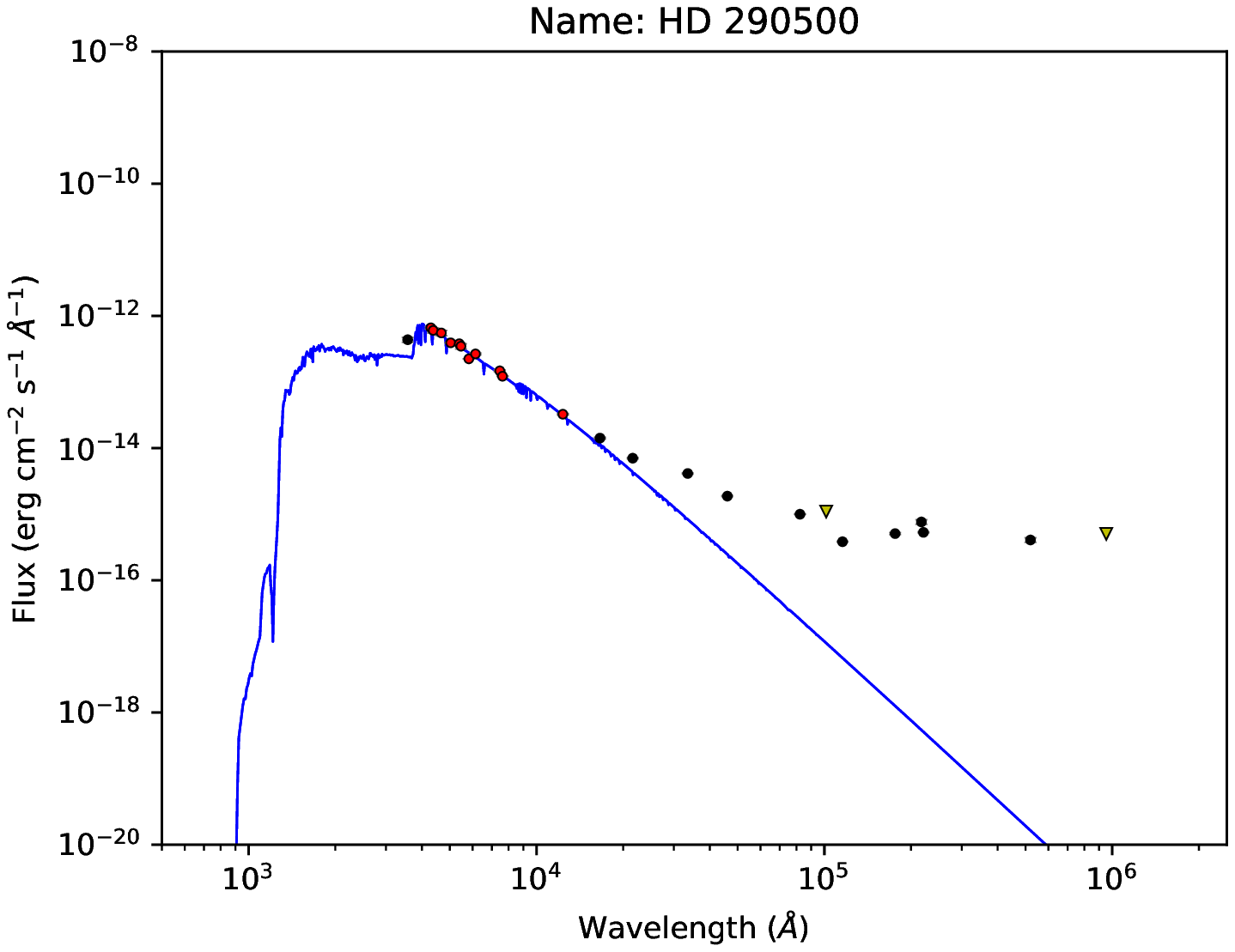}
    \includegraphics[width=0.33\textwidth]{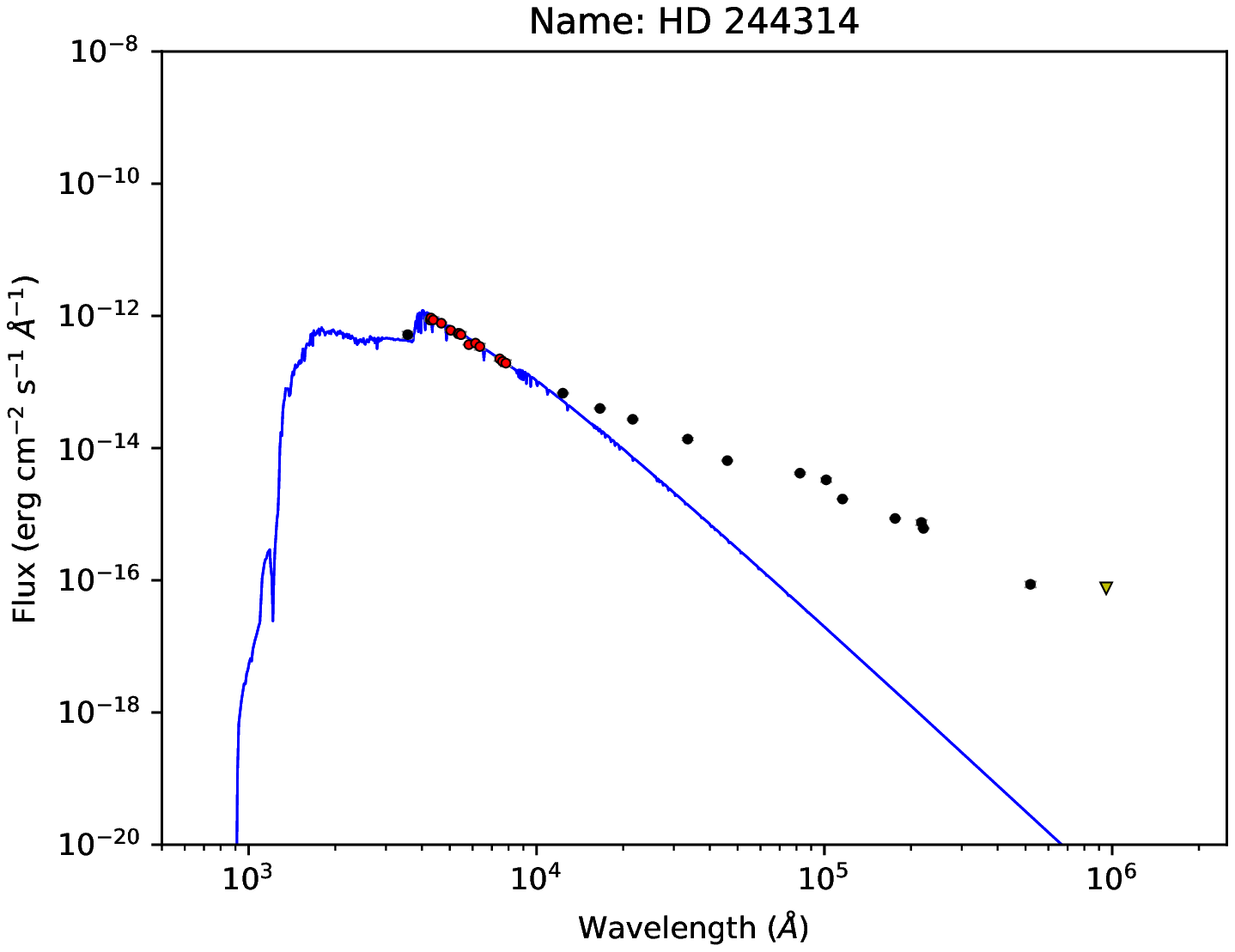}
    \includegraphics[width=0.33\textwidth]{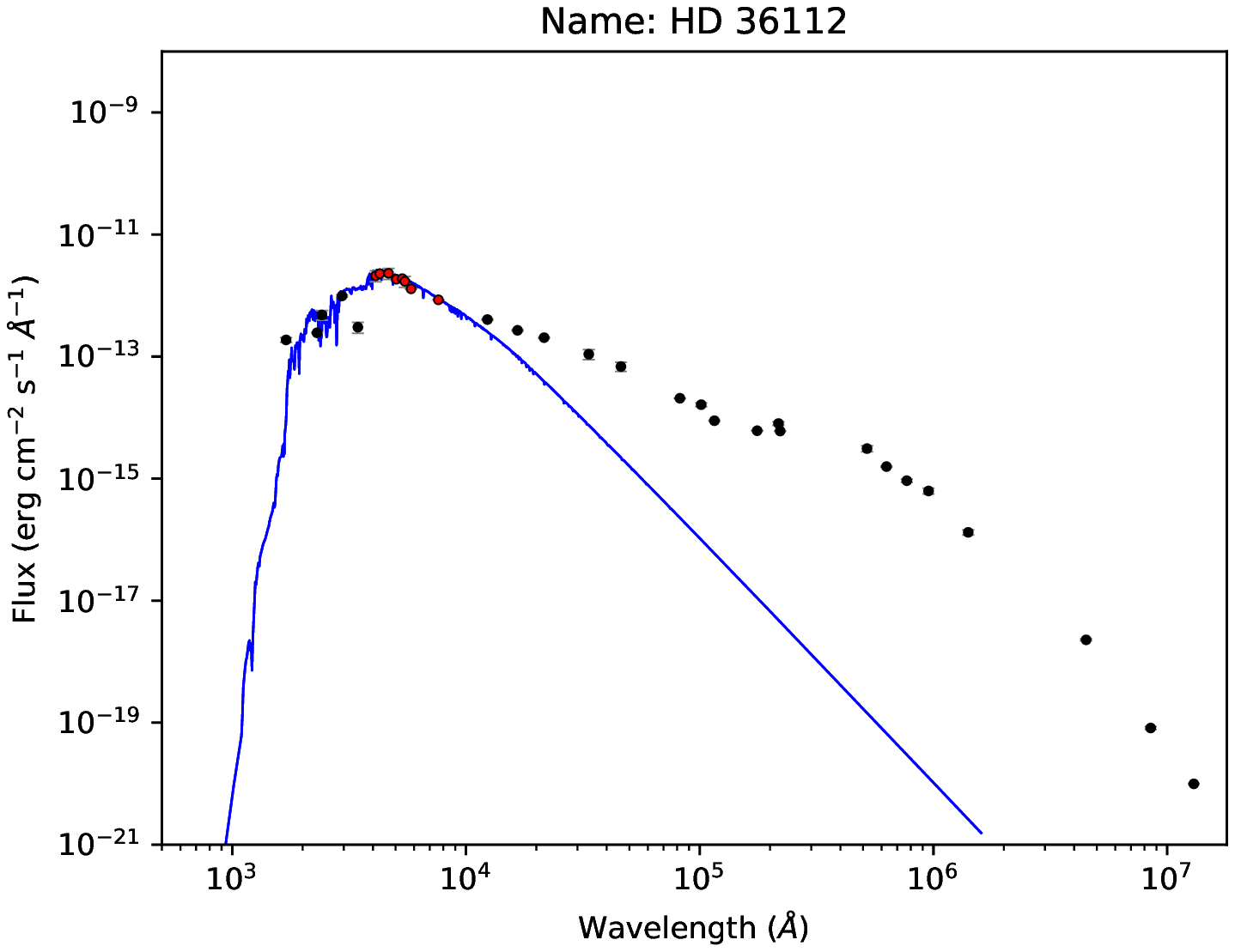}    
\end{figure}

\begin{figure} [h]
 \centering
    \includegraphics[width=0.33\textwidth]{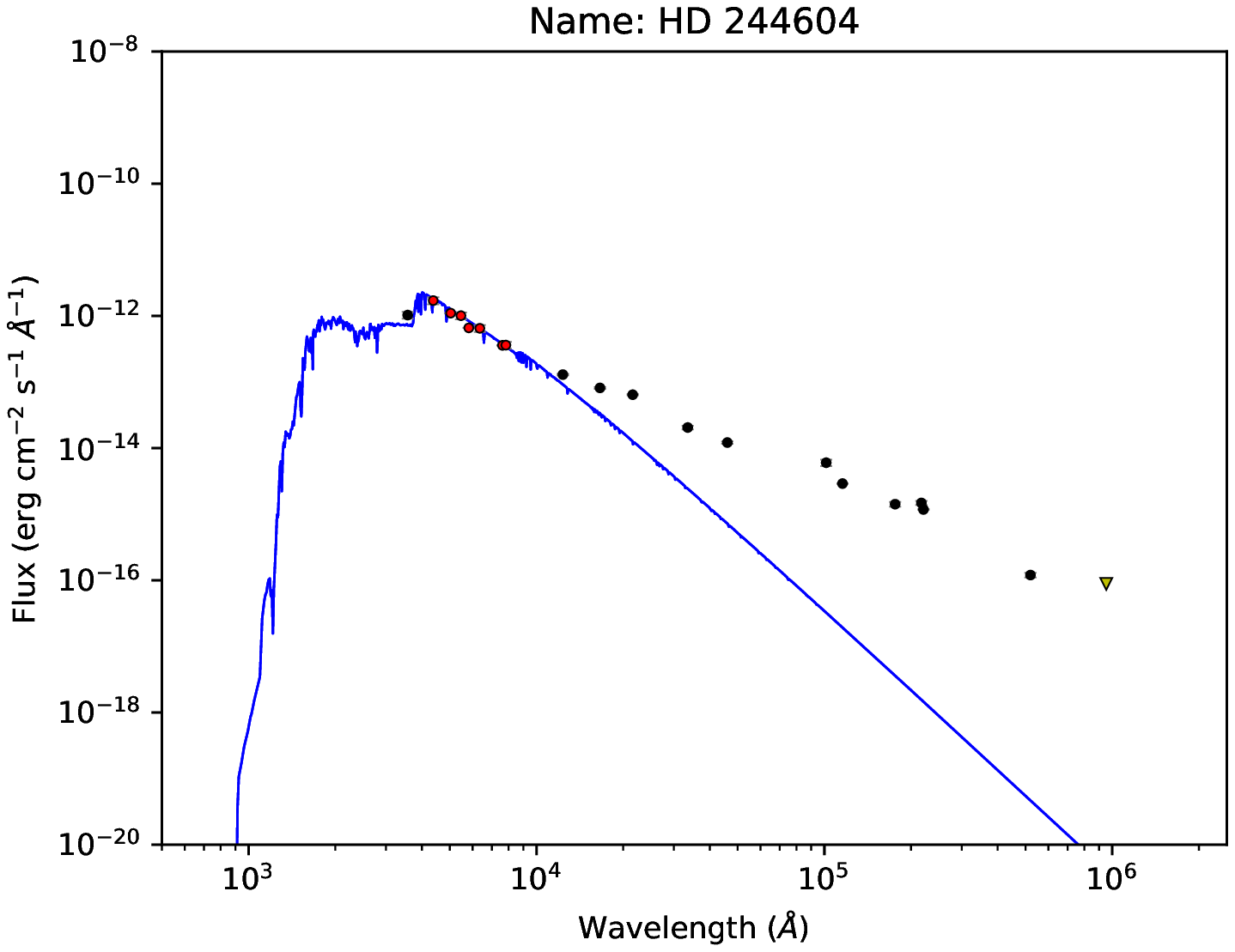}
    \includegraphics[width=0.33\textwidth]{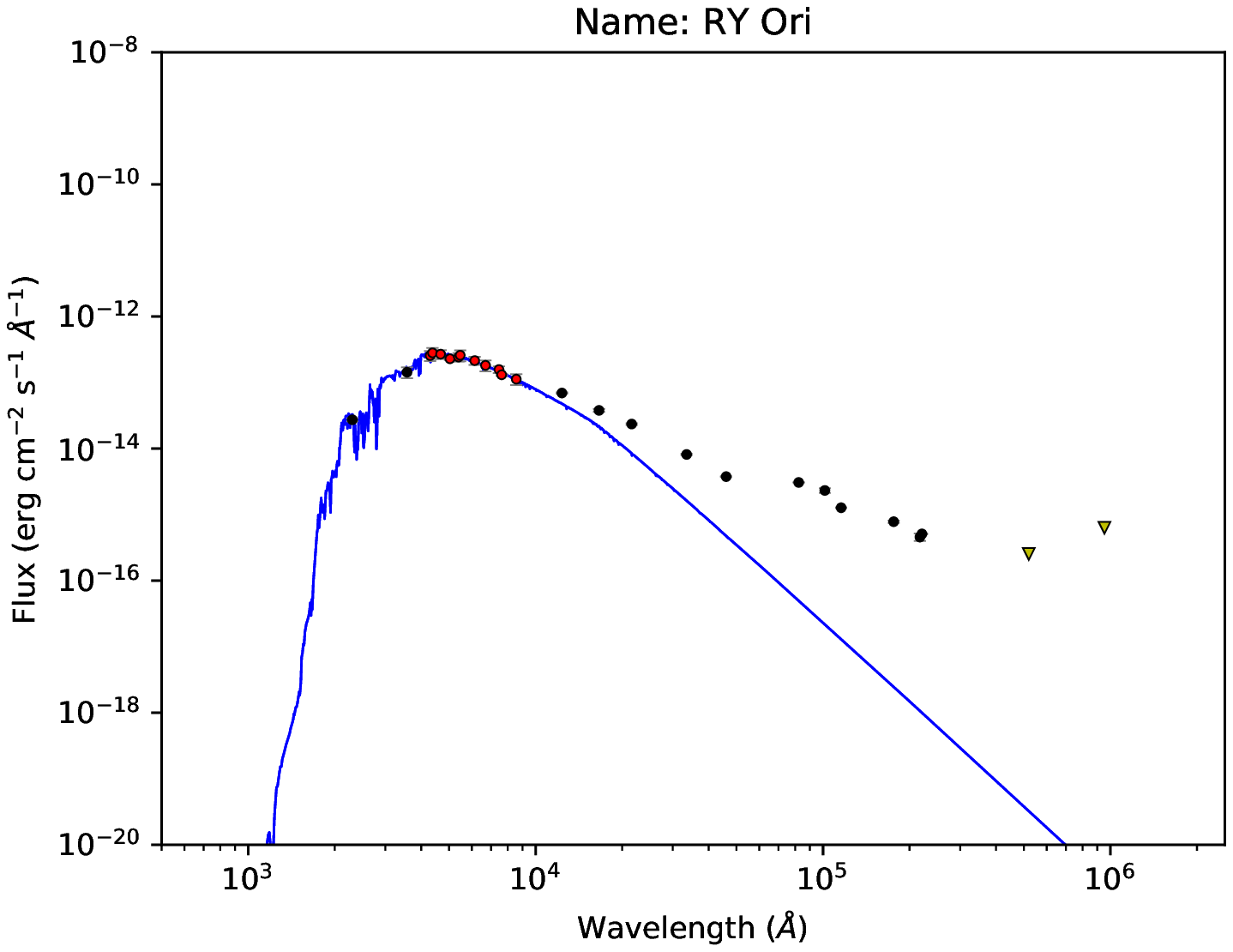}
    \includegraphics[width=0.33\textwidth]{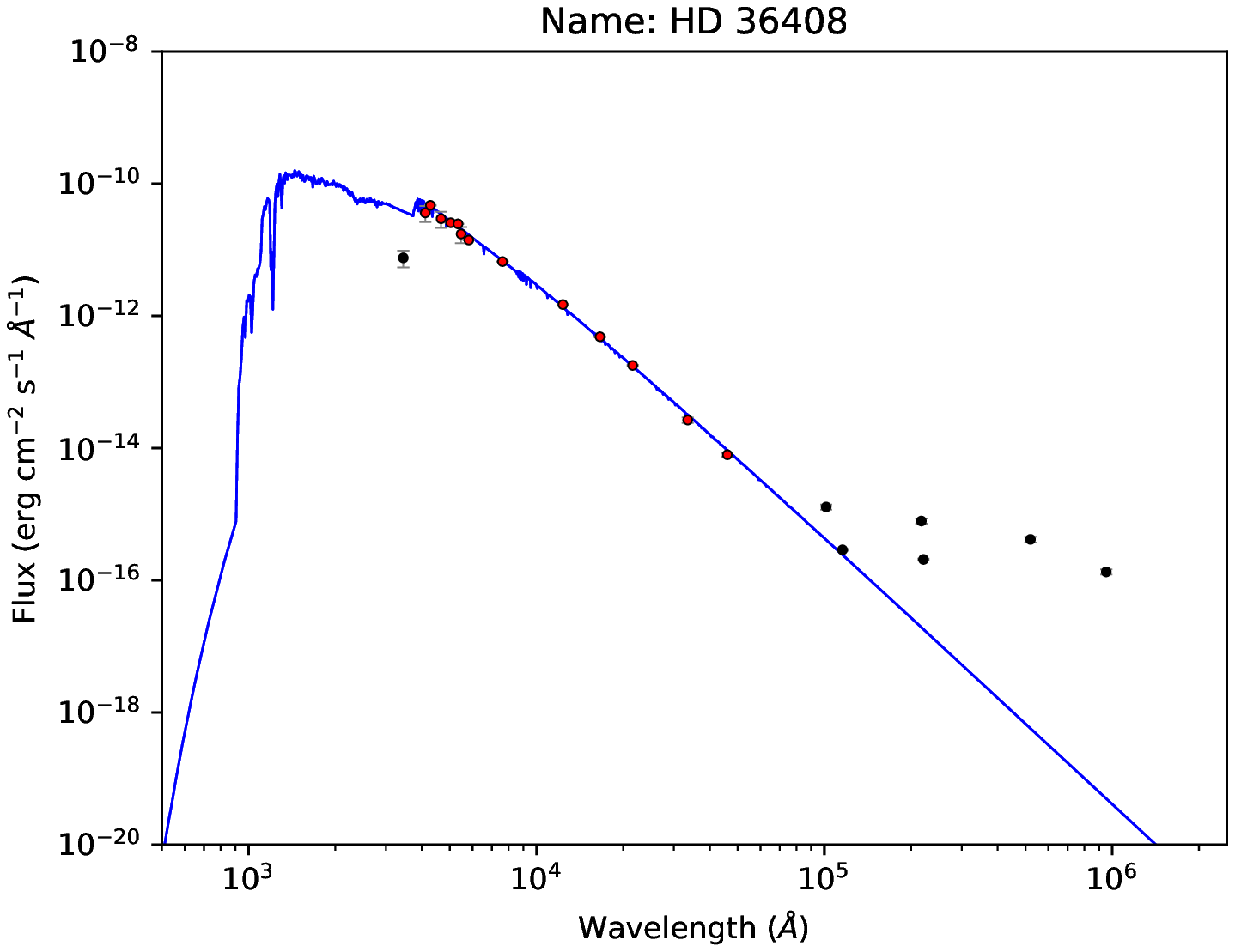}
\end{figure}

\begin{figure} [h]
 \centering
    \includegraphics[width=0.33\textwidth]{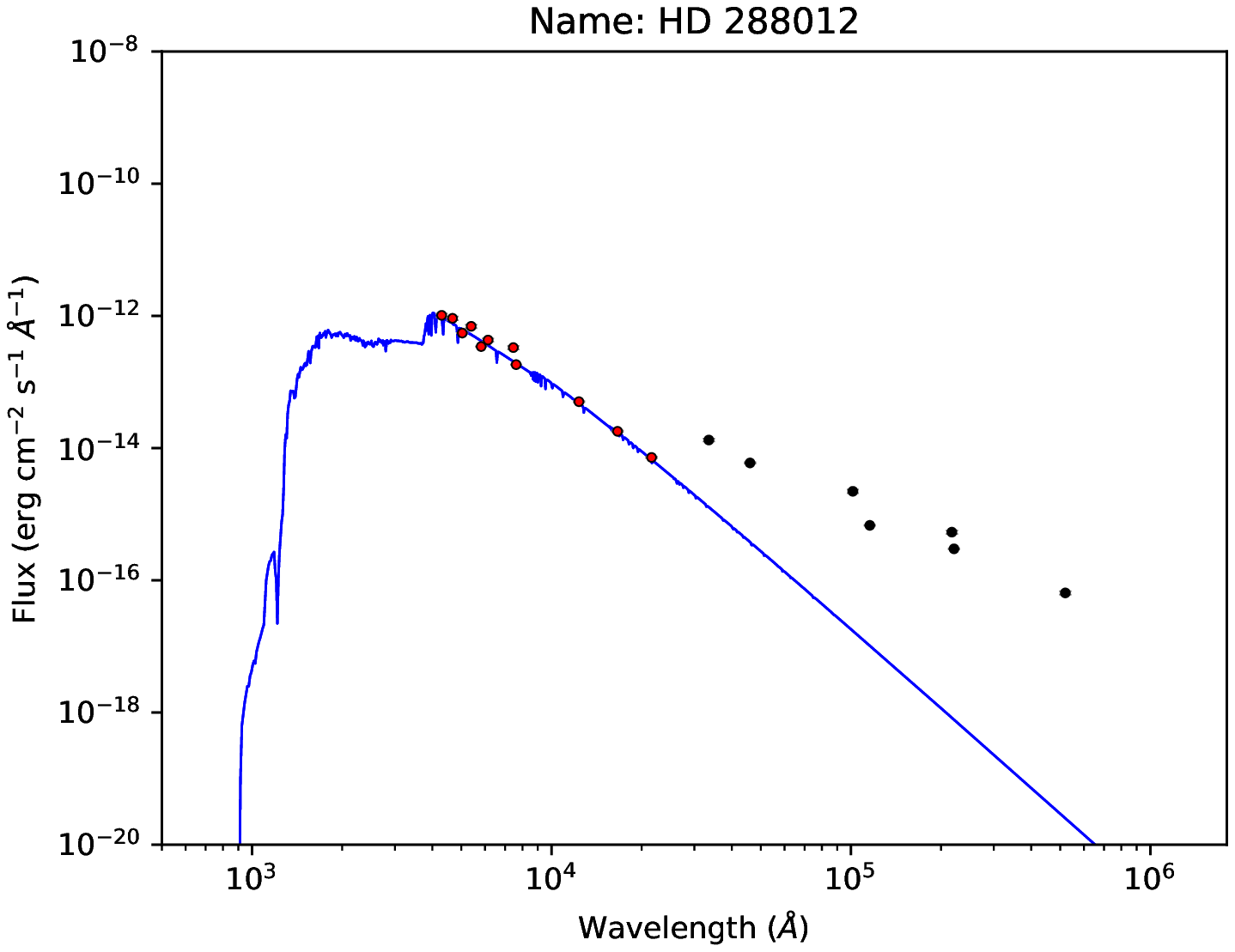}
    \includegraphics[width=0.33\textwidth]{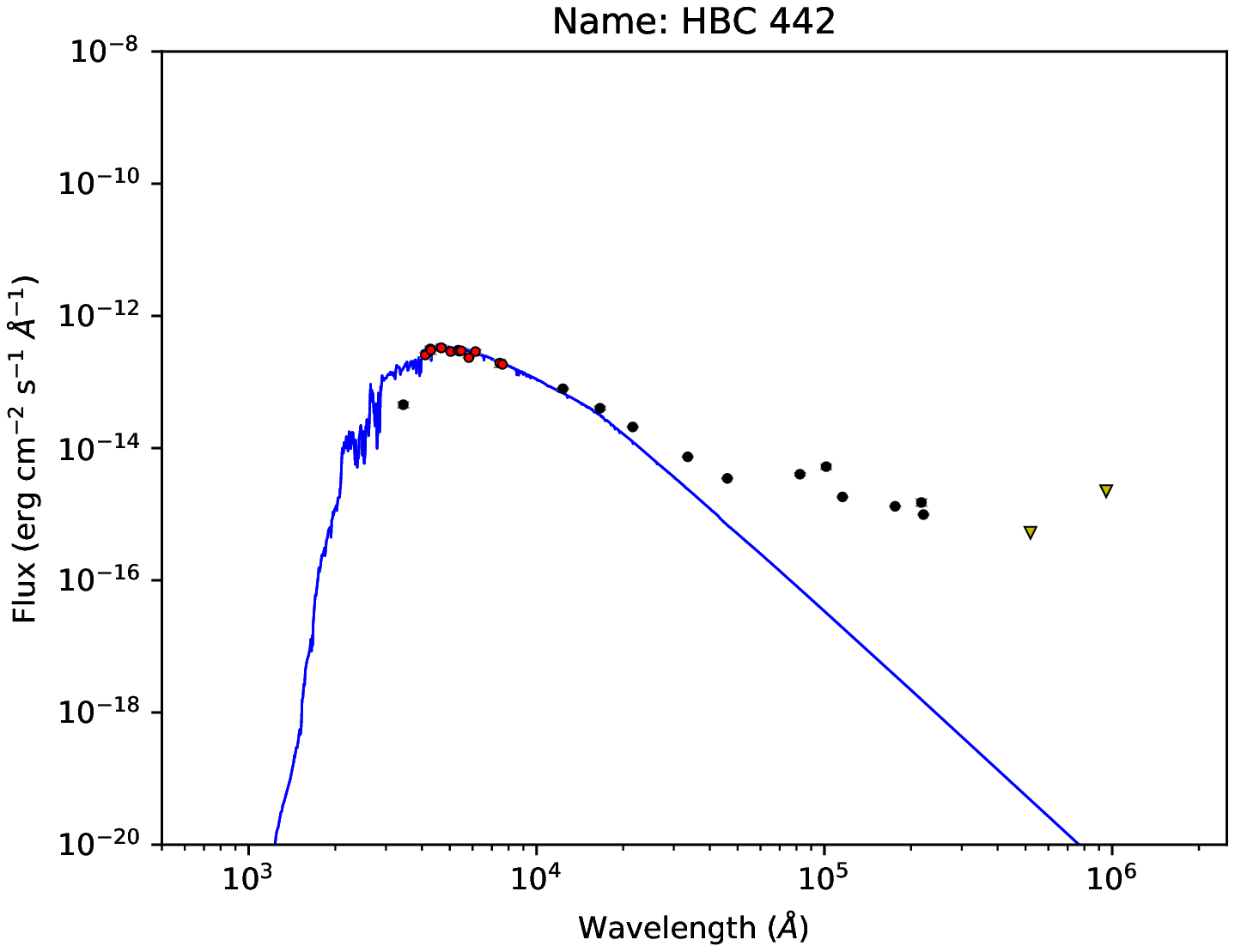}
    \includegraphics[width=0.33\textwidth]{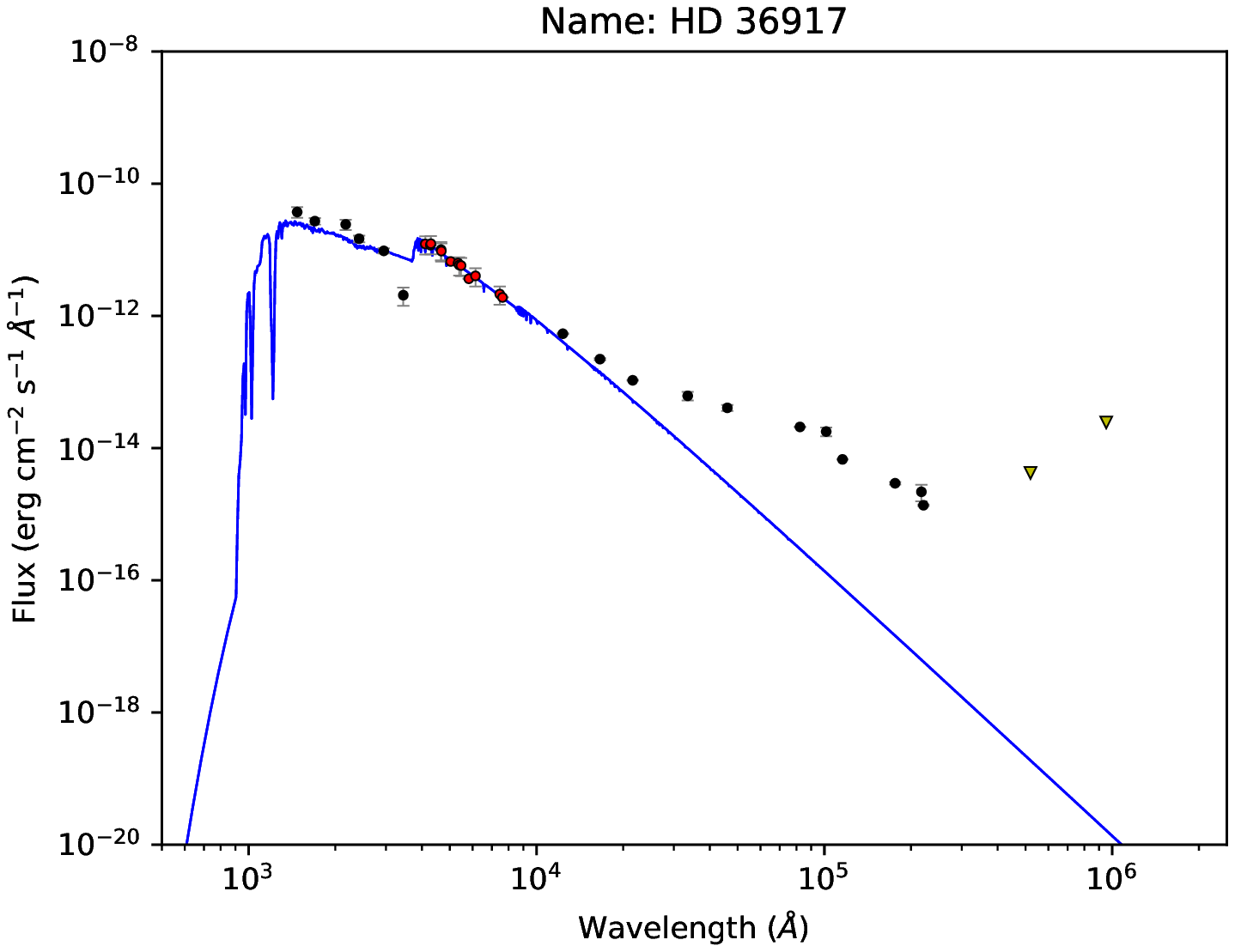}    
\end{figure}

\begin{figure} [h]
 \centering
    \includegraphics[width=0.33\textwidth]{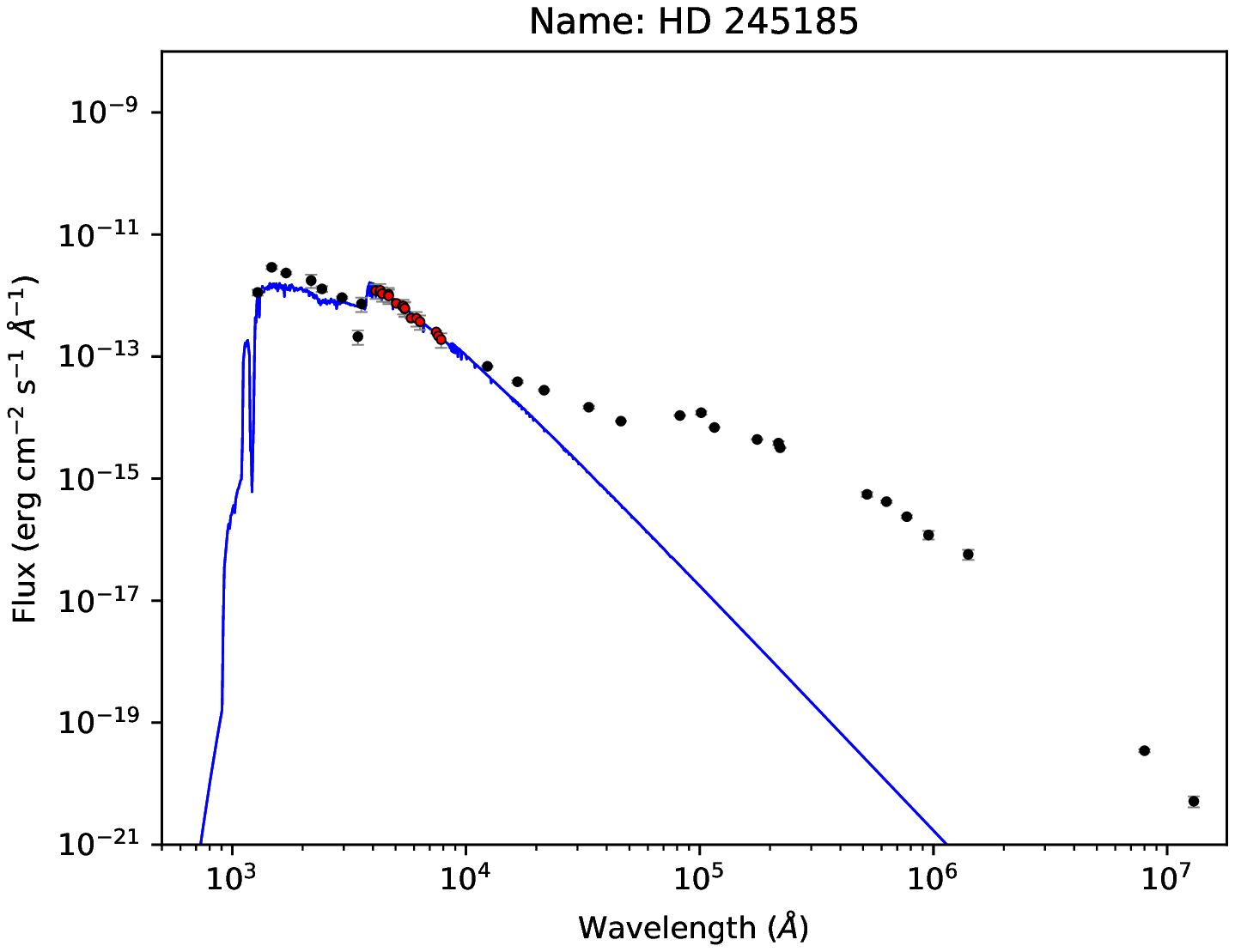}
    \includegraphics[width=0.33\textwidth]{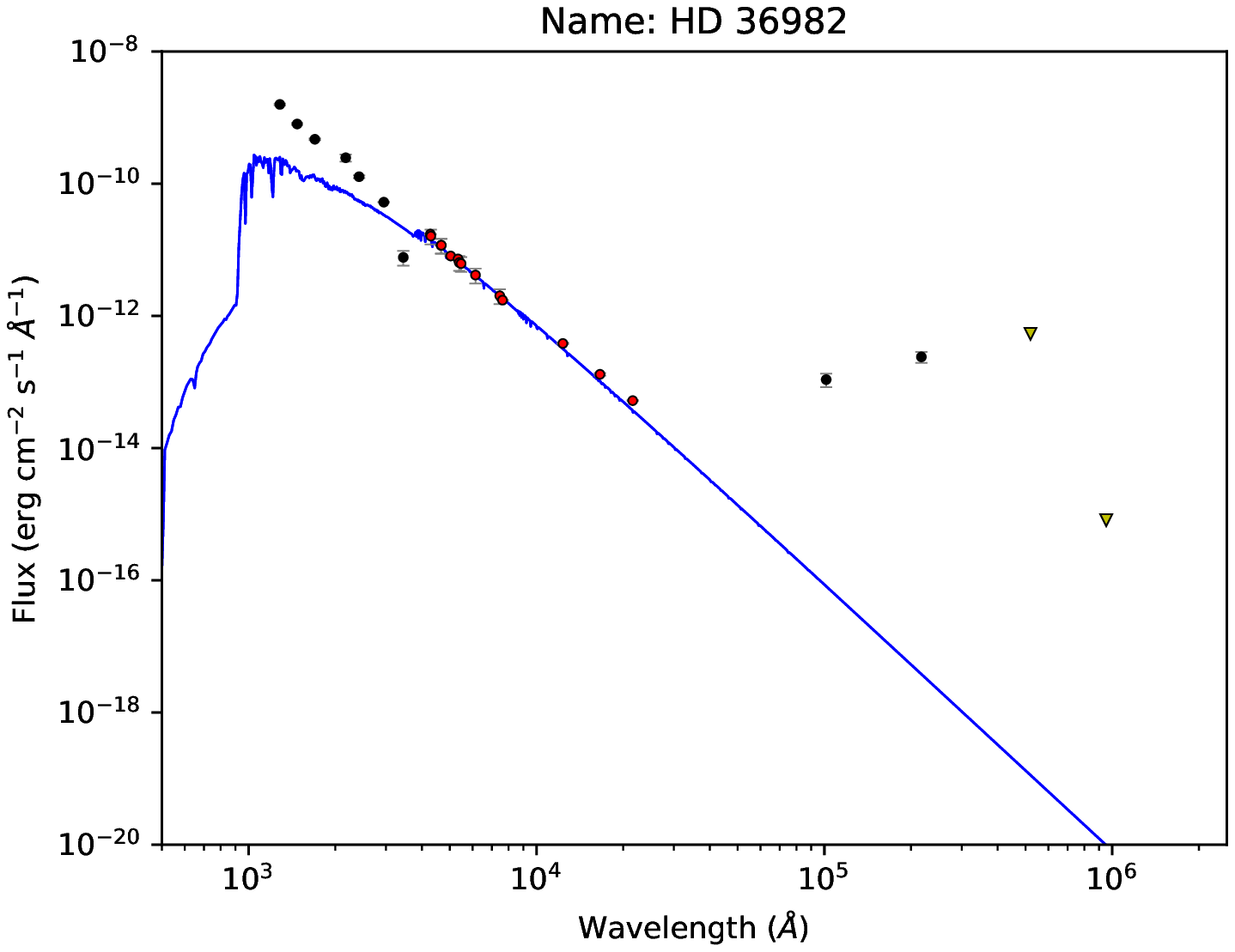}
    \includegraphics[width=0.33\textwidth]{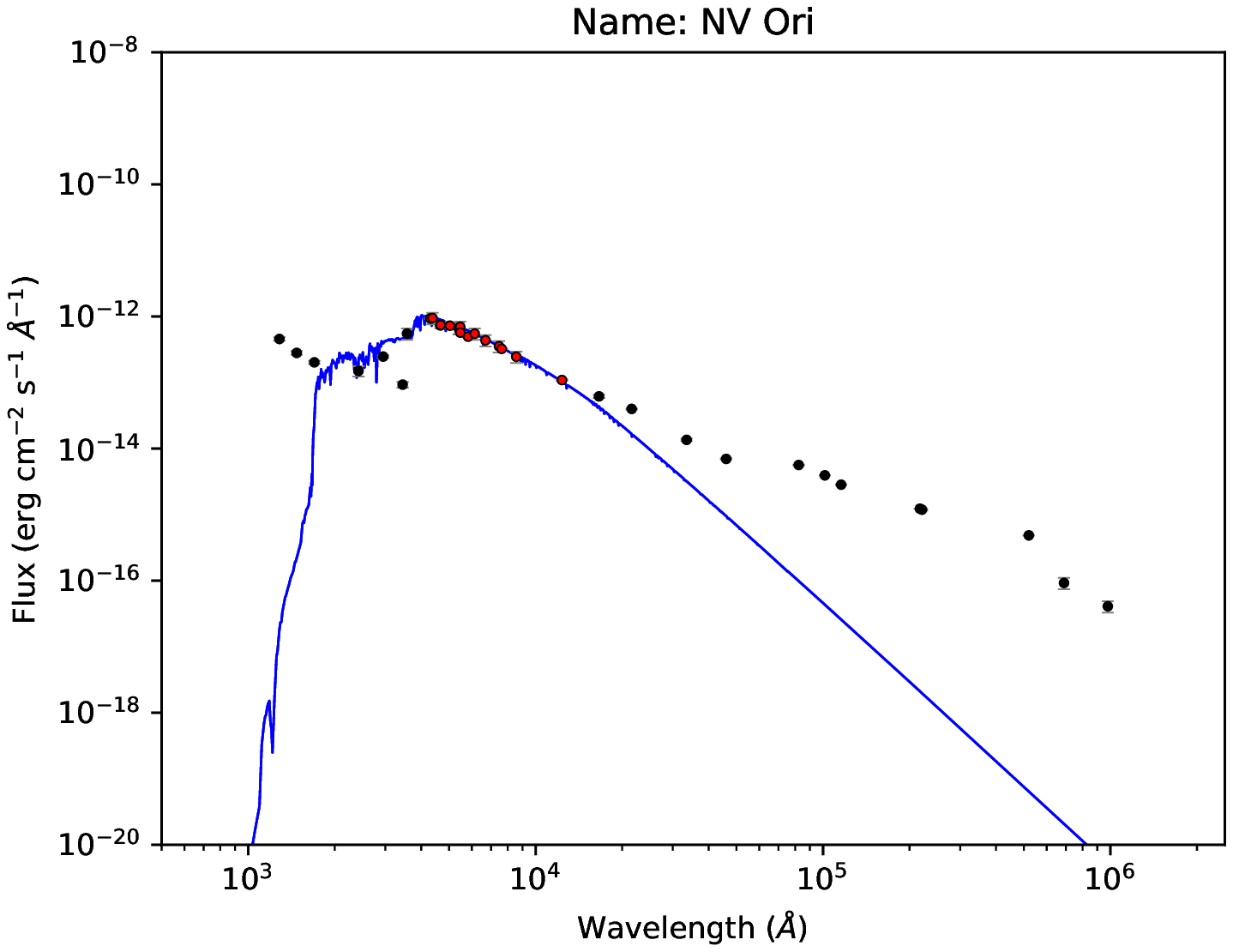}    
\end{figure}

\newpage

\onecolumn

\begin{figure} [h]
 \centering
    \includegraphics[width=0.33\textwidth]{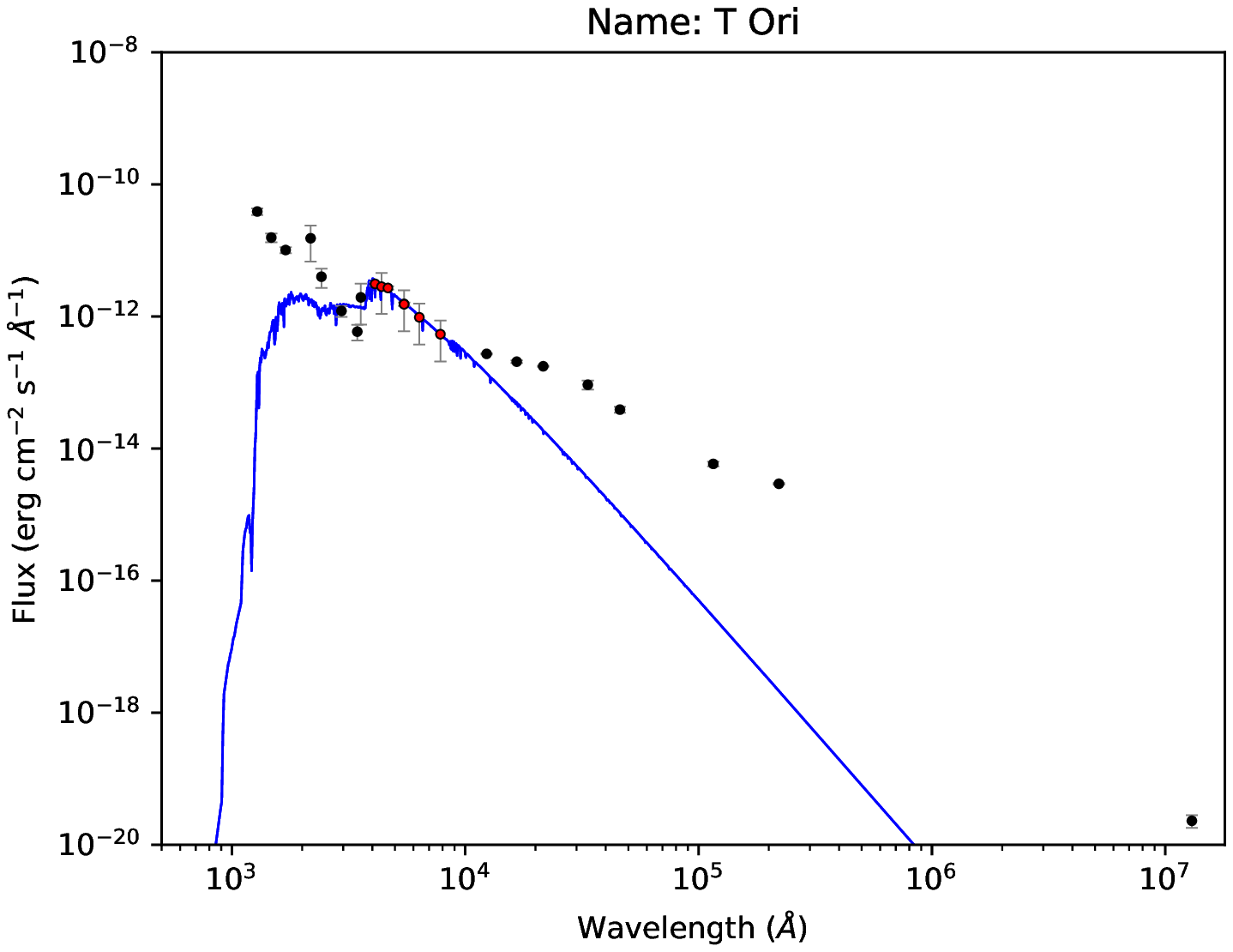}
    \includegraphics[width=0.33\textwidth]{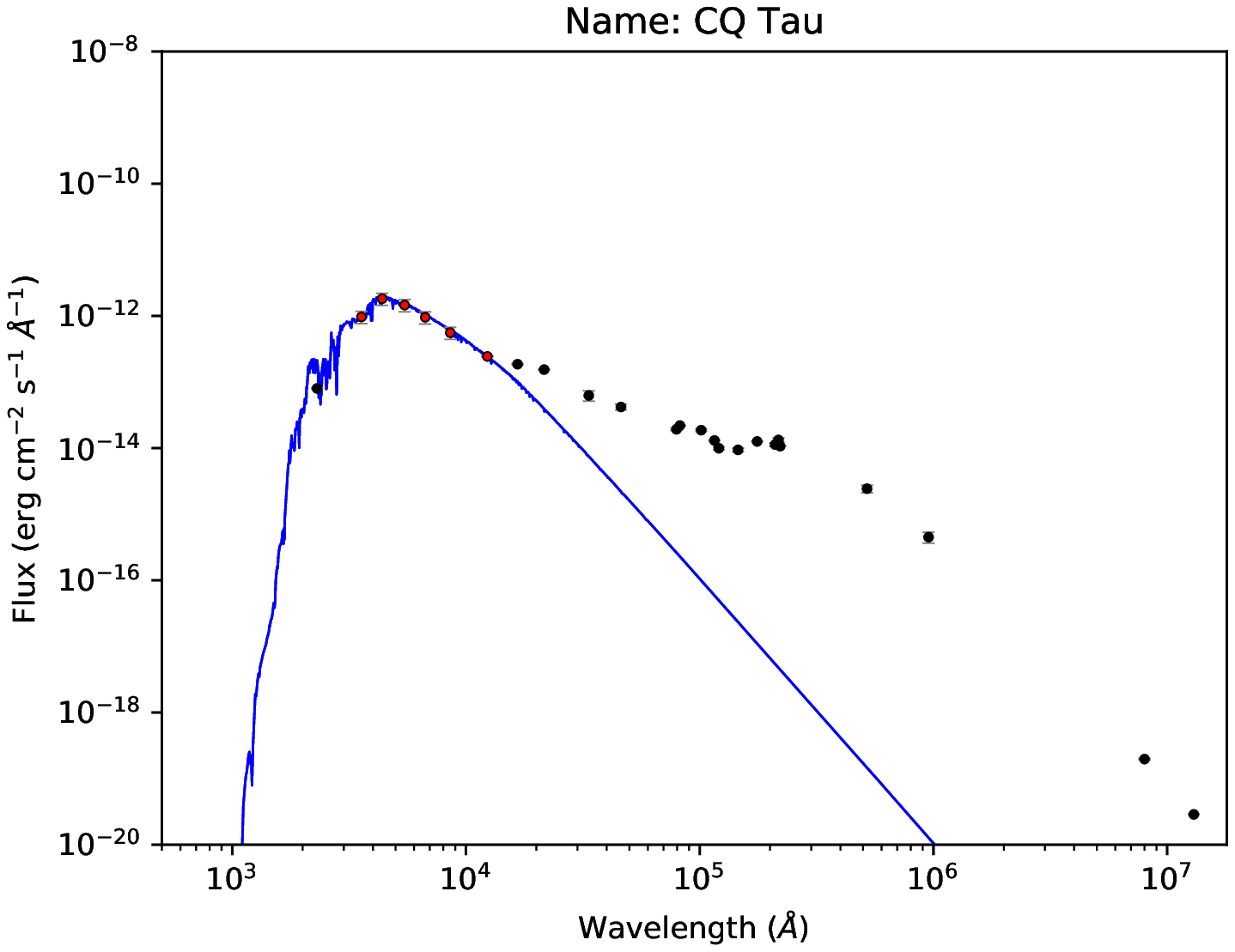}
    \includegraphics[width=0.33\textwidth]{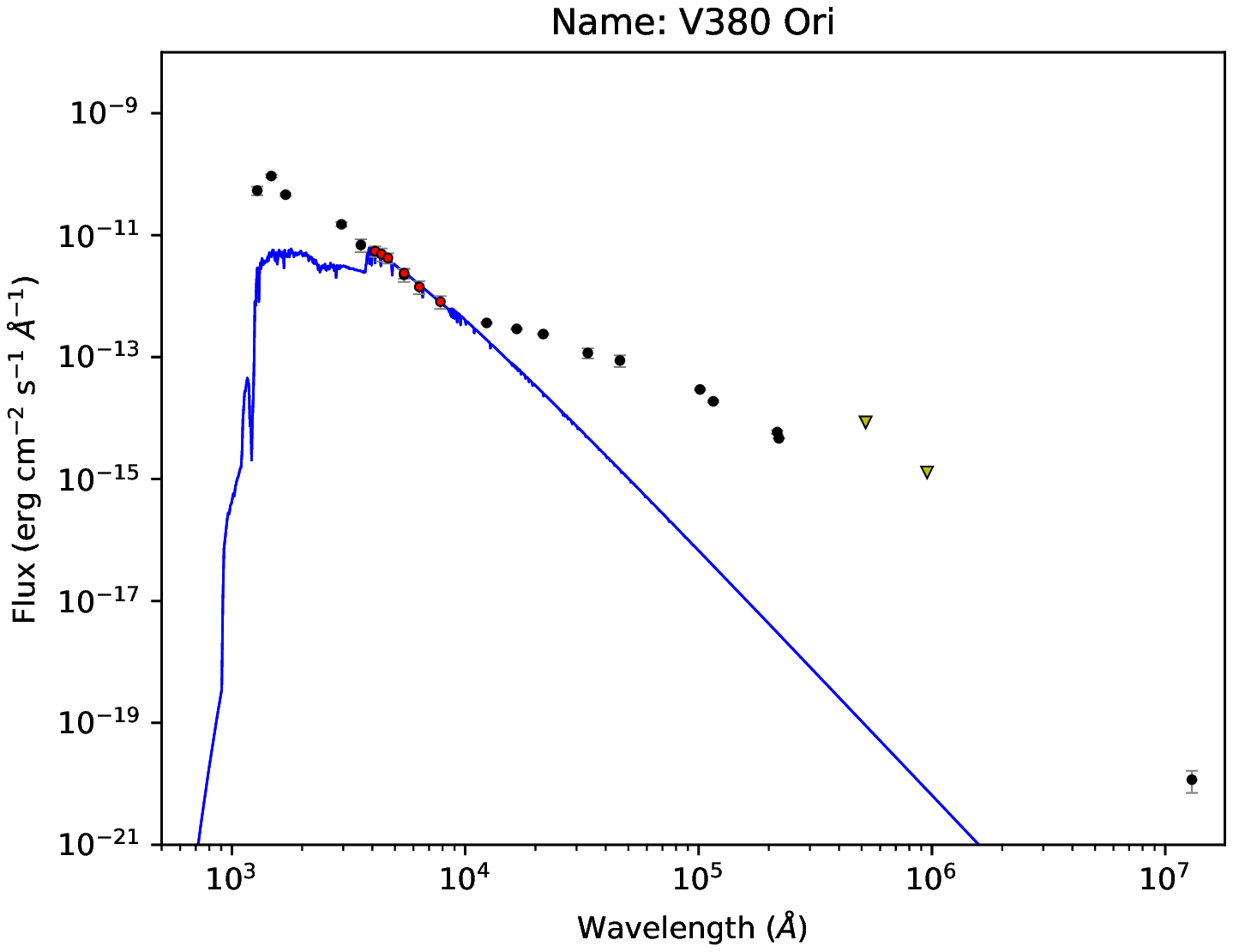}    
\end{figure}

\begin{figure} [h]
 \centering
    \includegraphics[width=0.33\textwidth]{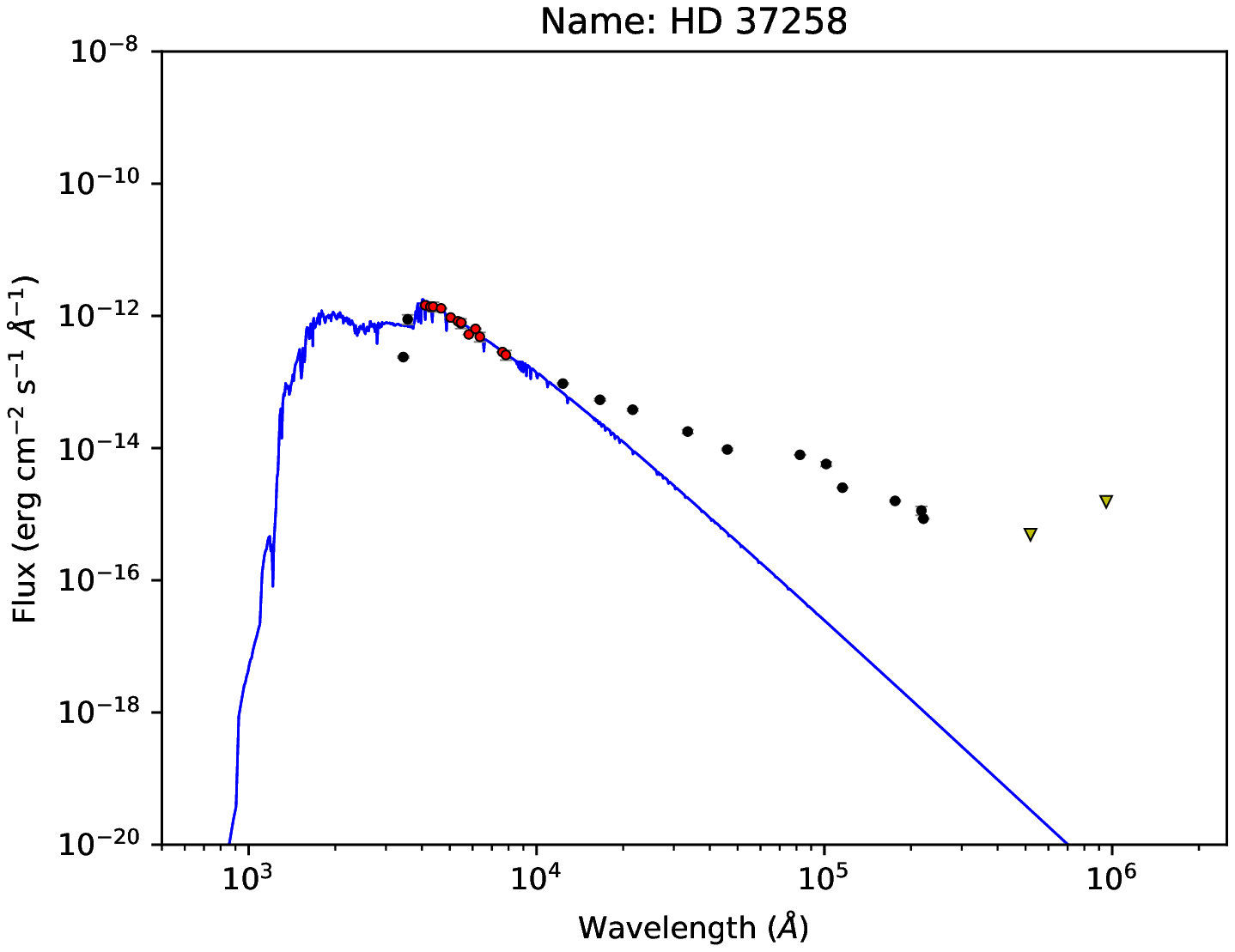}
    \includegraphics[width=0.33\textwidth]{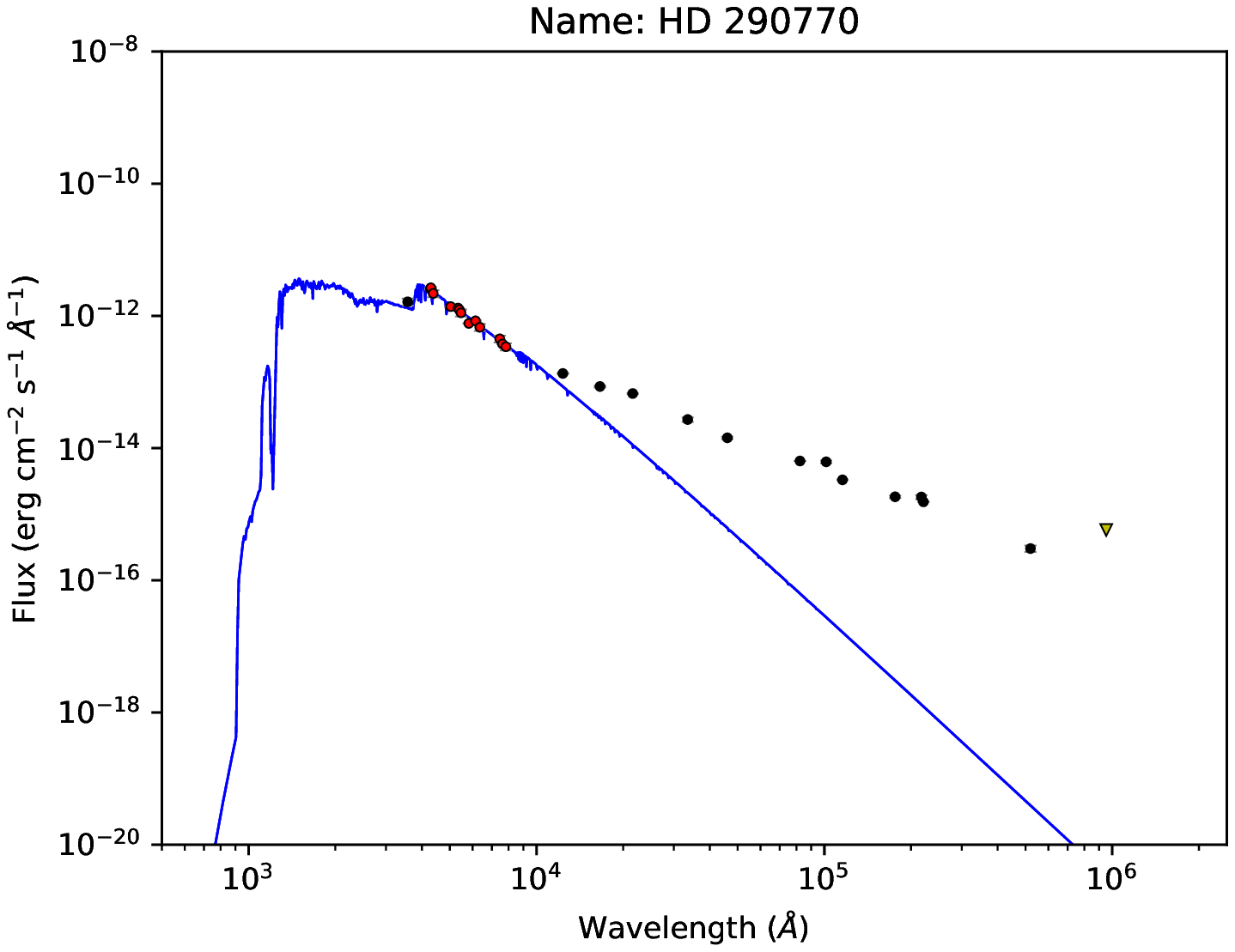}
    \includegraphics[width=0.33\textwidth]{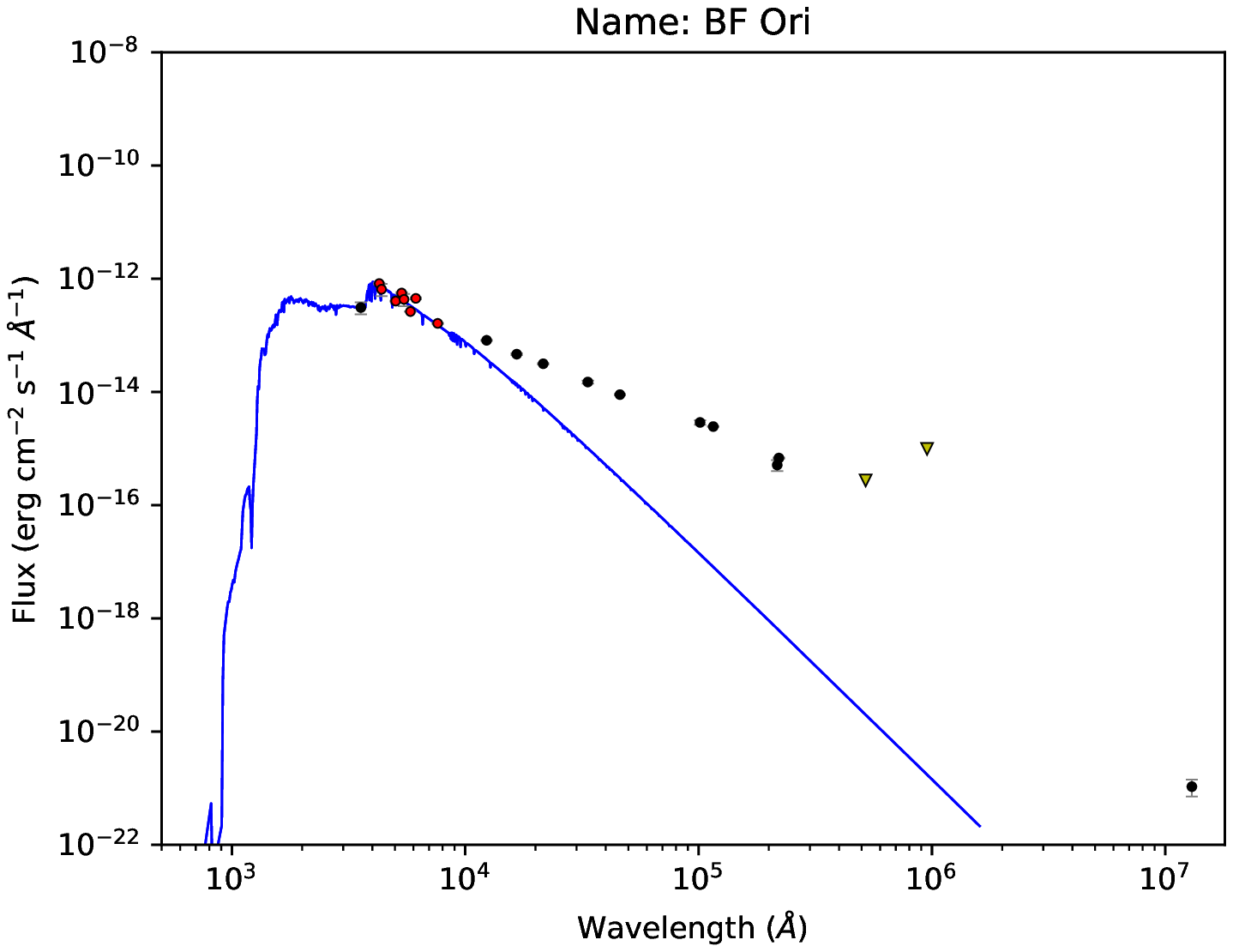}    
\end{figure}

\begin{figure} [h]
 \centering
    \includegraphics[width=0.33\textwidth]{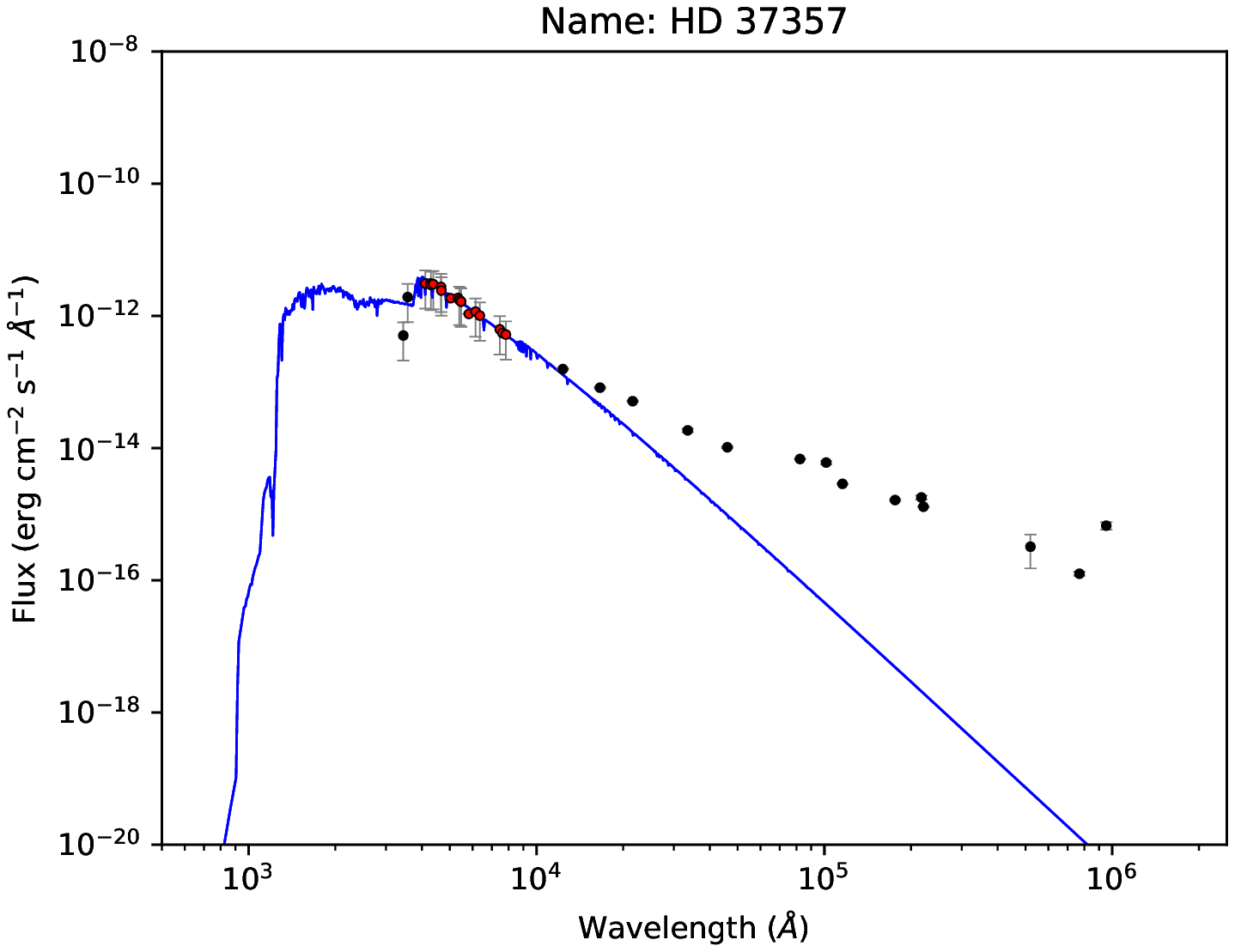}
    \includegraphics[width=0.33\textwidth]{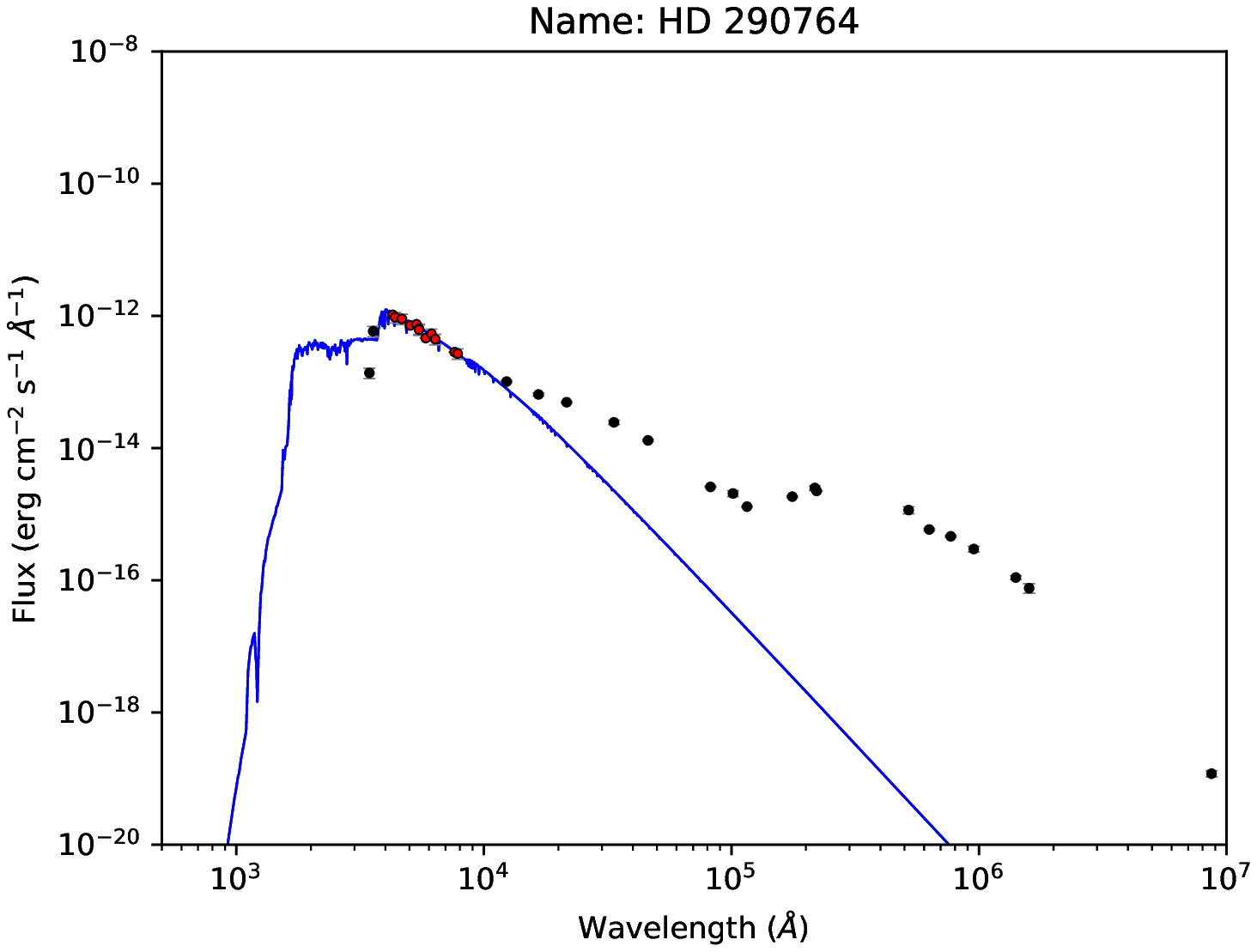}
    \includegraphics[width=0.33\textwidth]{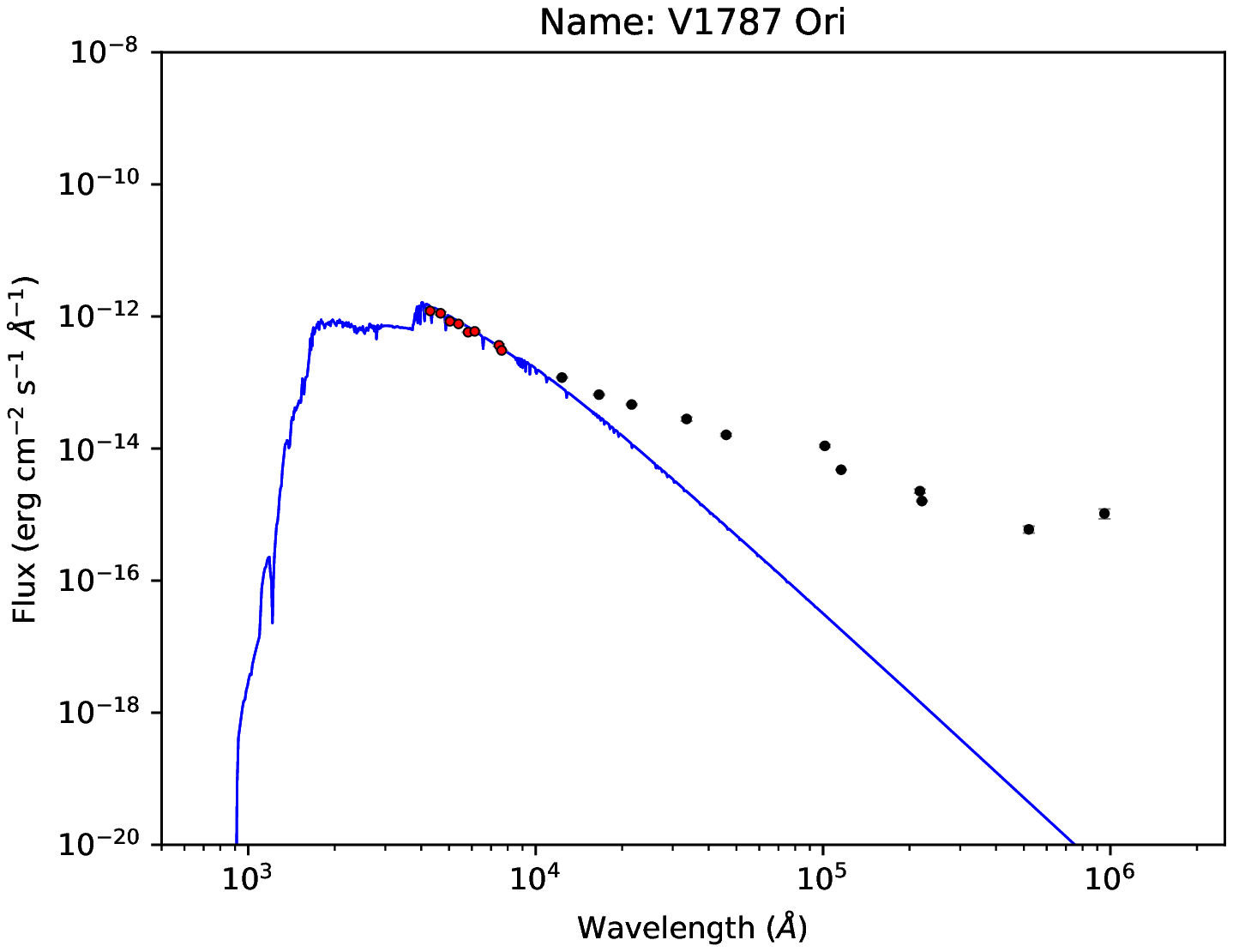}    
\end{figure}

\begin{figure} [h]
 \centering
    \includegraphics[width=0.33\textwidth]{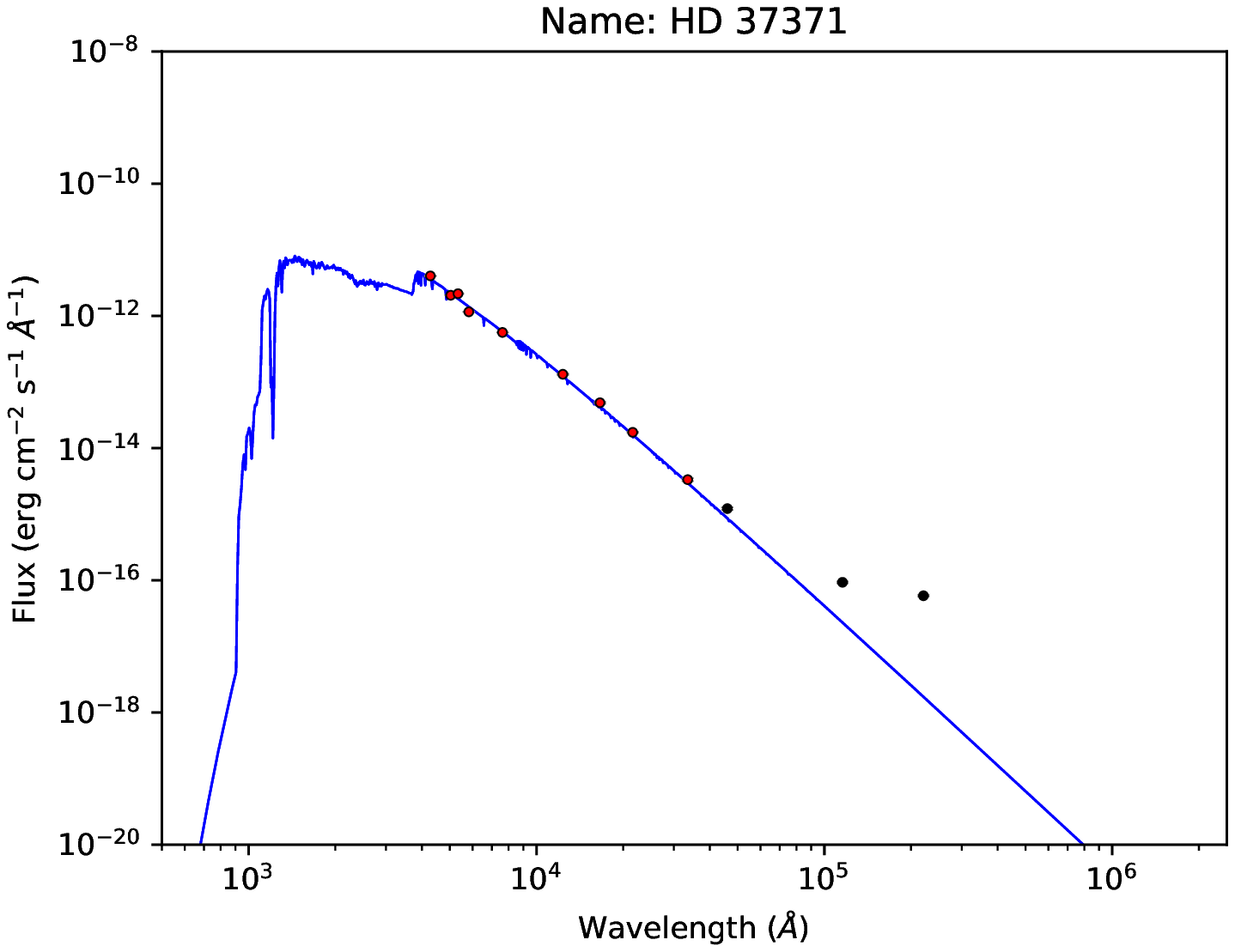}
    \includegraphics[width=0.33\textwidth]{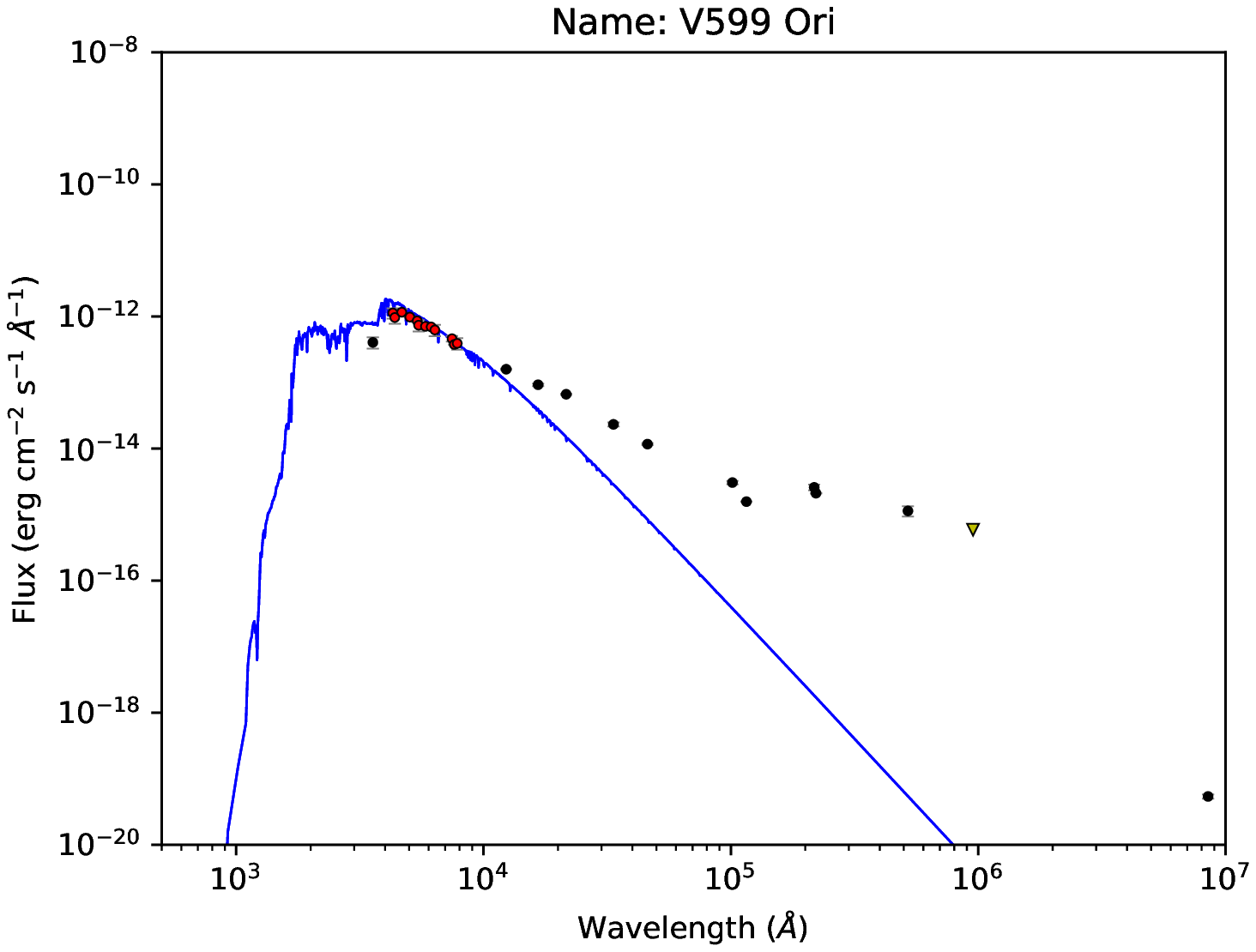}
    \includegraphics[width=0.33\textwidth]{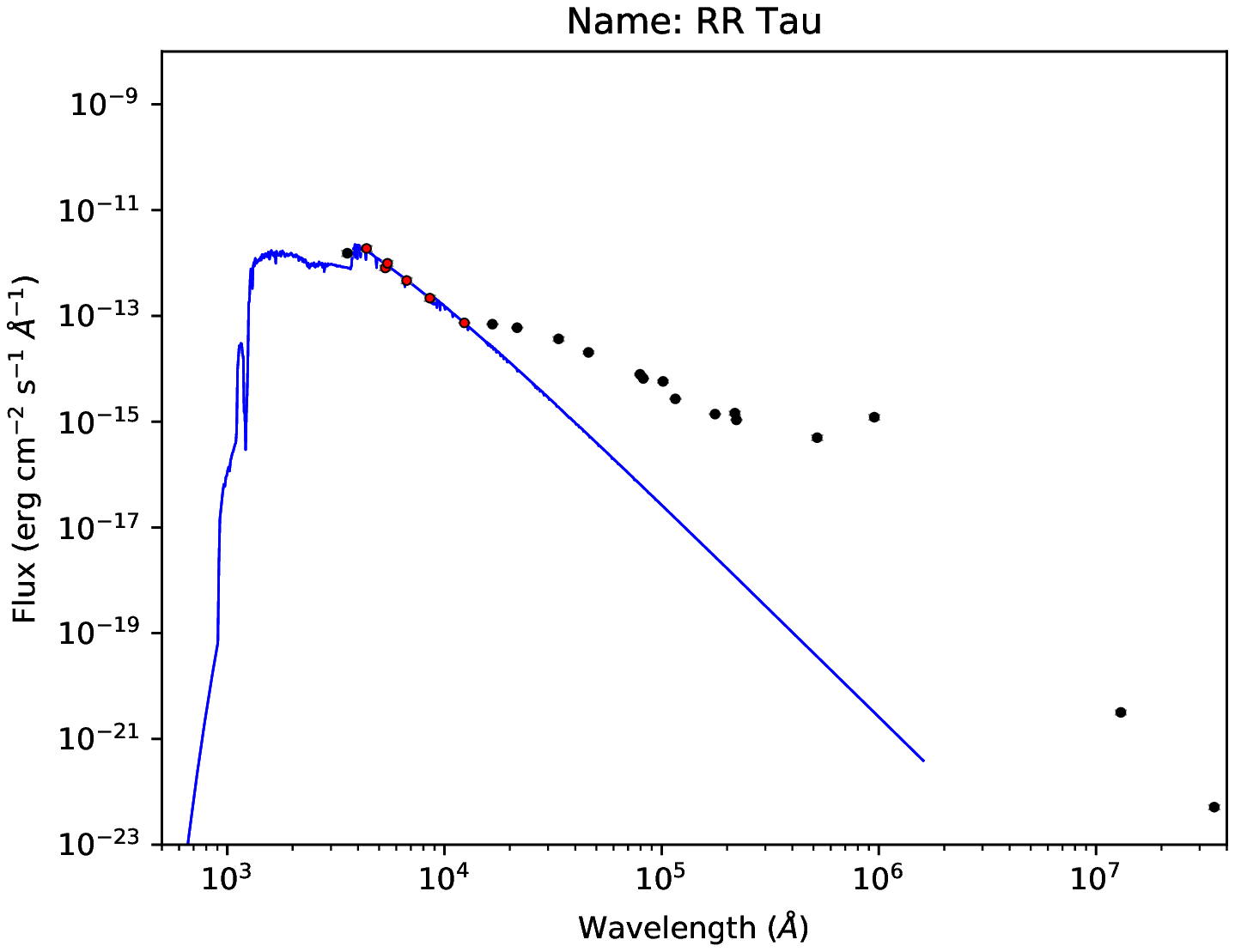}    
\end{figure}

\newpage

\onecolumn

\begin{figure} [h]
 \centering
    \includegraphics[width=0.33\textwidth]{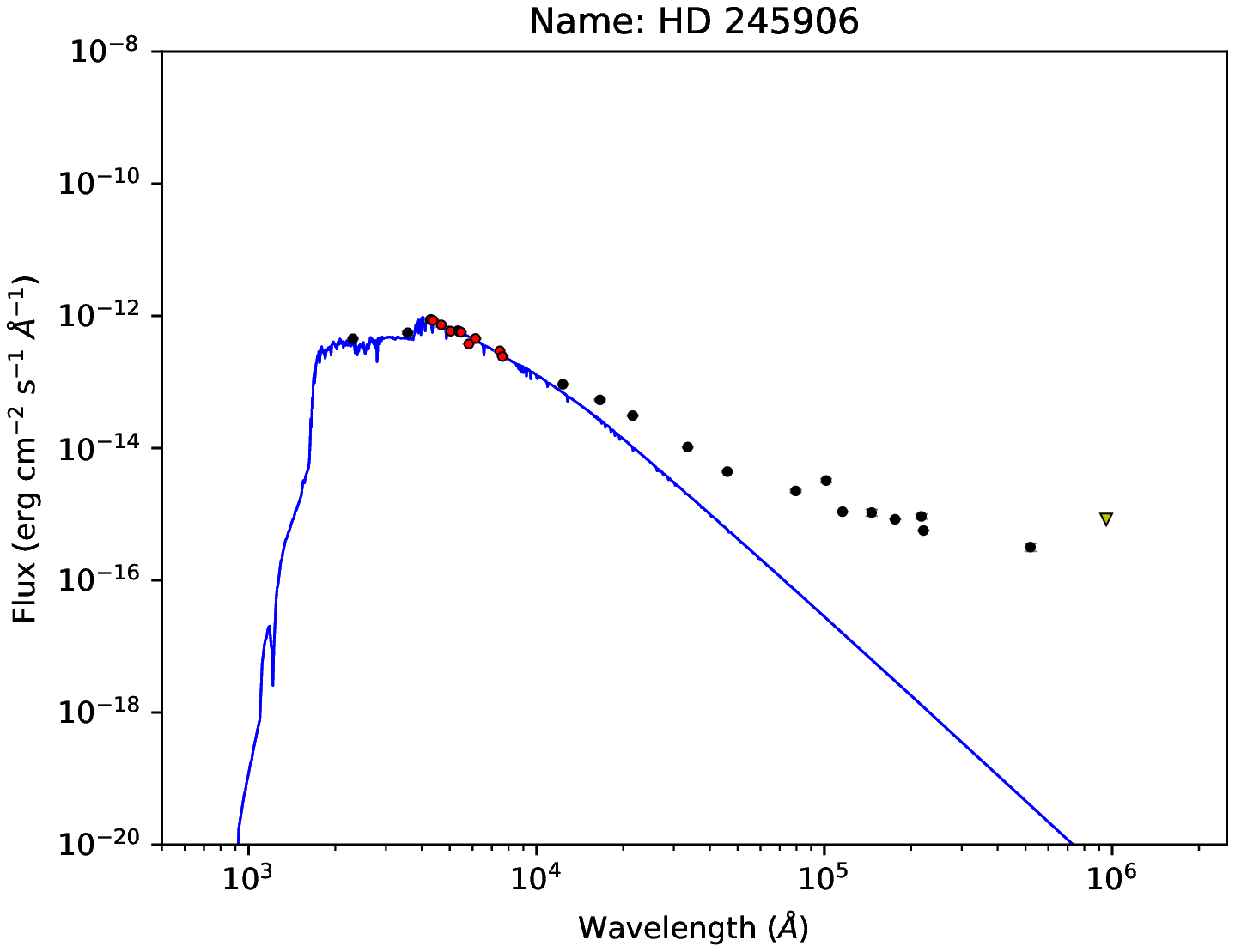}
    \includegraphics[width=0.33\textwidth]{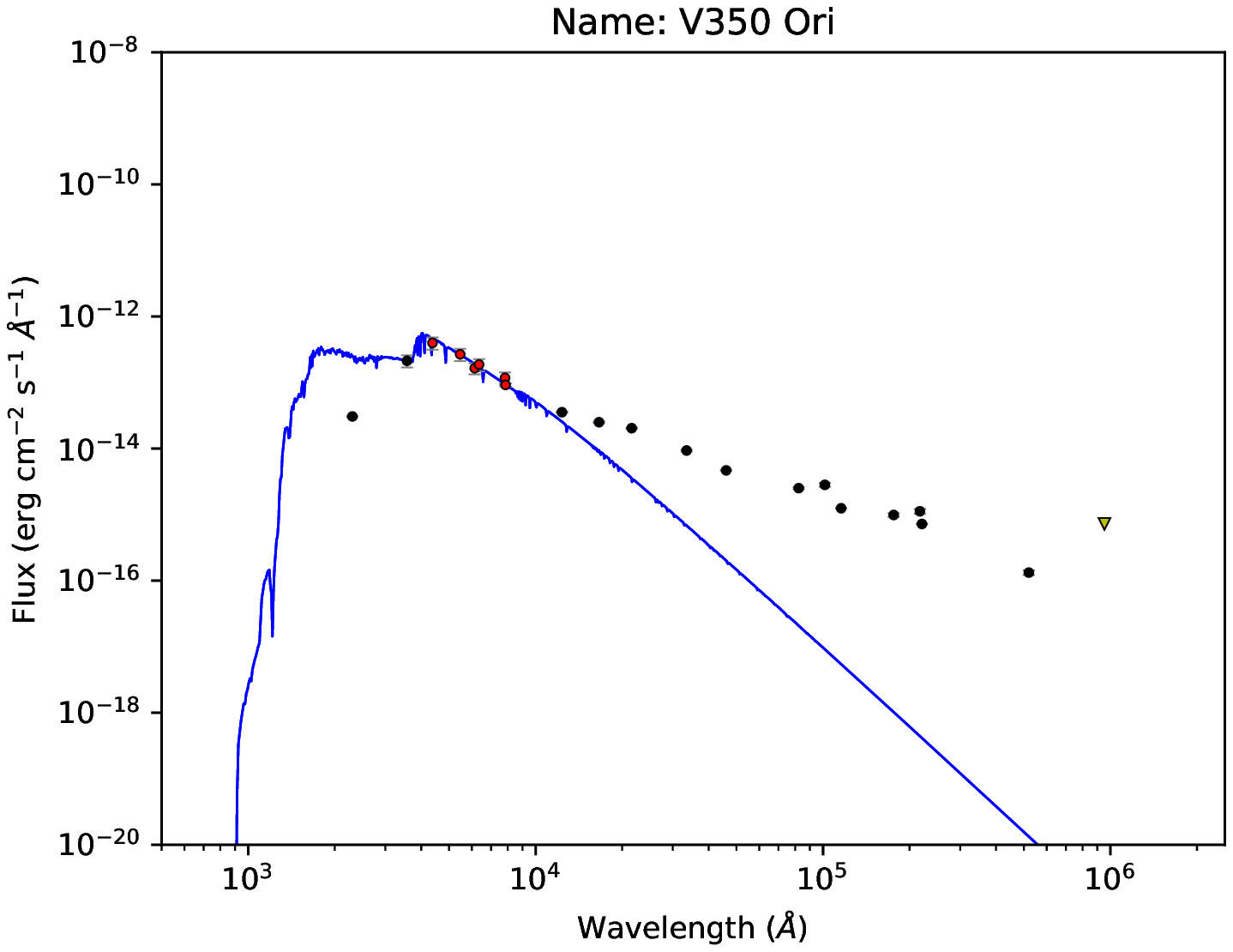}
    \includegraphics[width=0.33\textwidth]{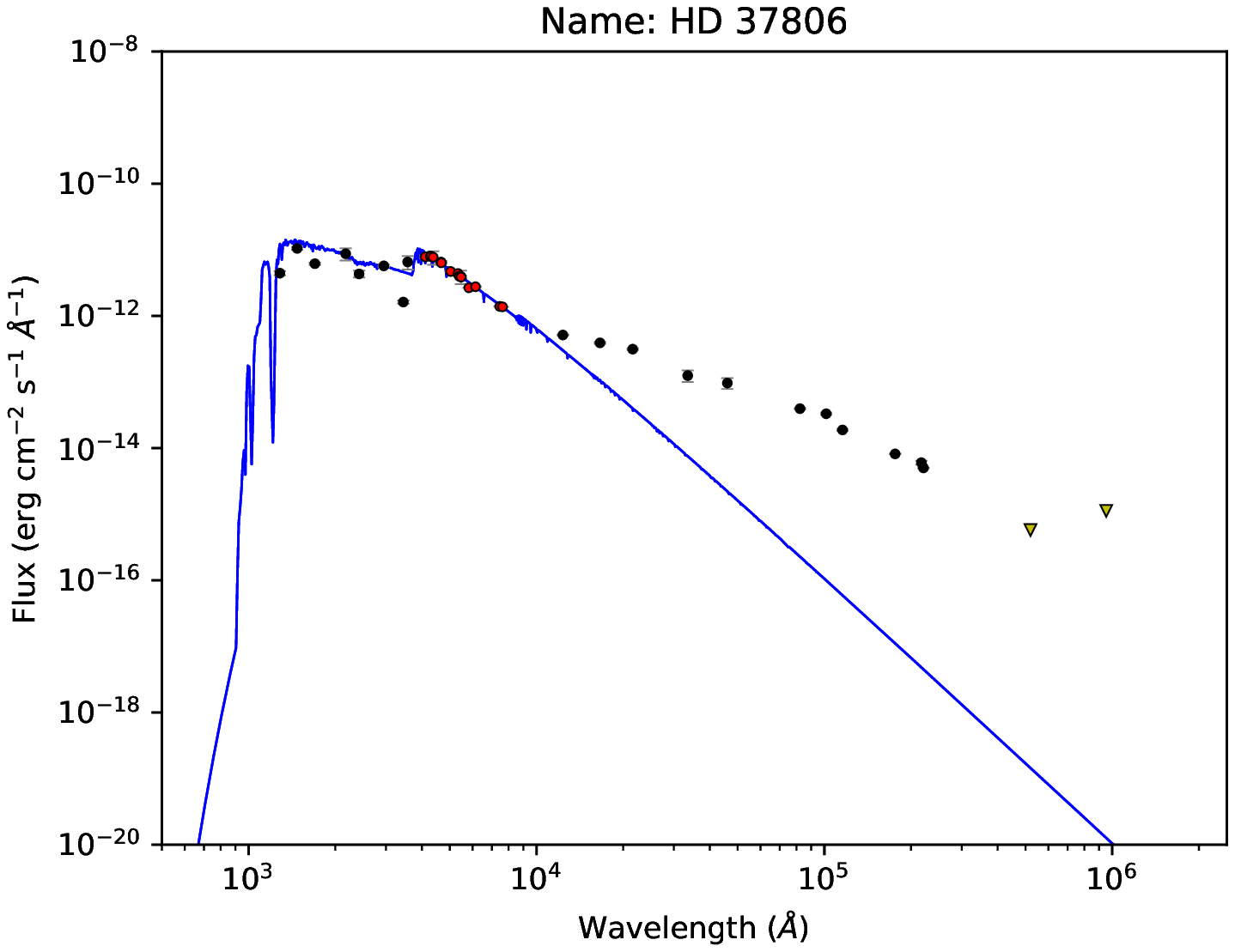}    
\end{figure}

\begin{figure} [h]
 \centering
    \includegraphics[width=0.33\textwidth]{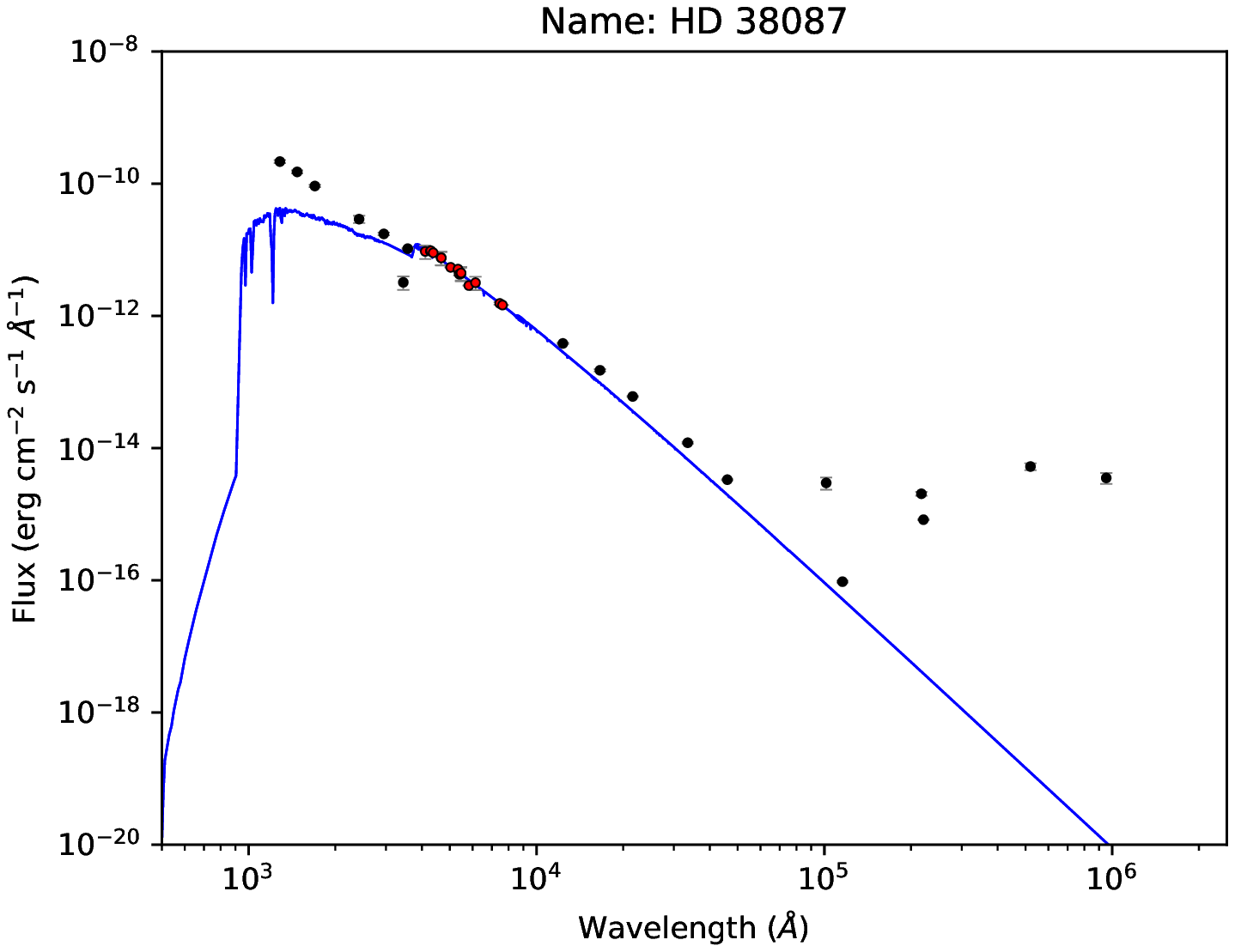}
    \includegraphics[width=0.33\textwidth]{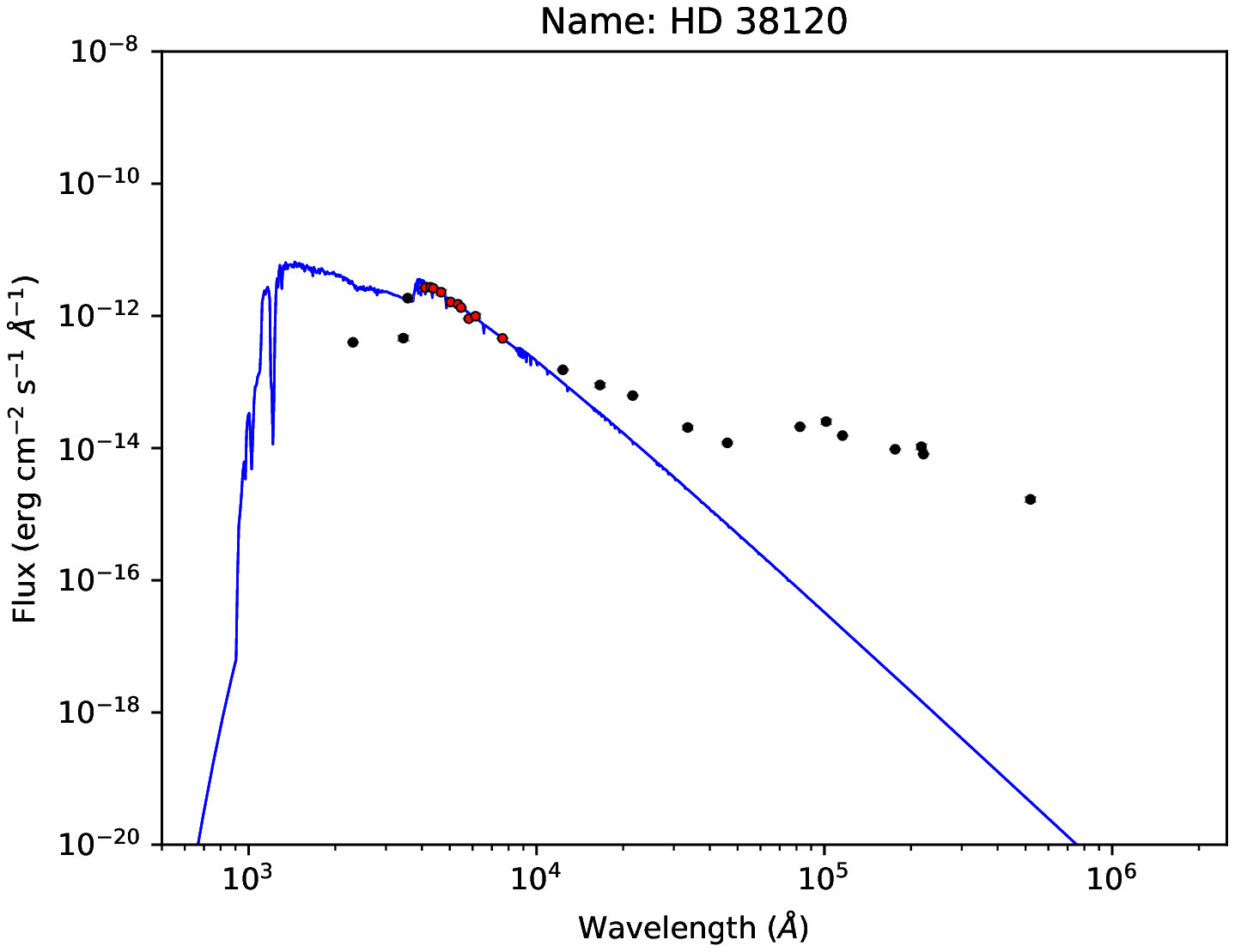}
    \includegraphics[width=0.33\textwidth]{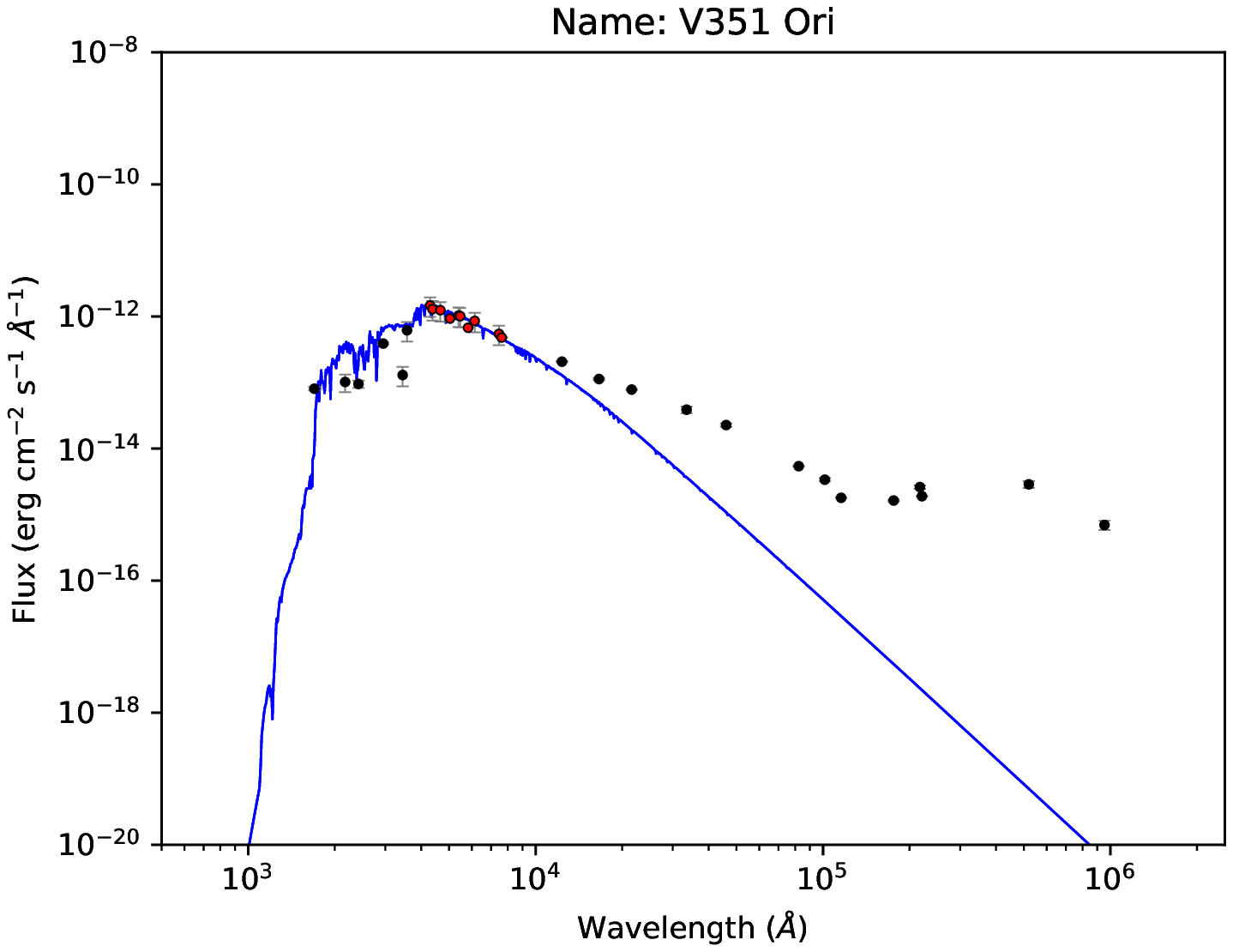}    
\end{figure}

\begin{figure} [h]
 \centering
    \includegraphics[width=0.33\textwidth]{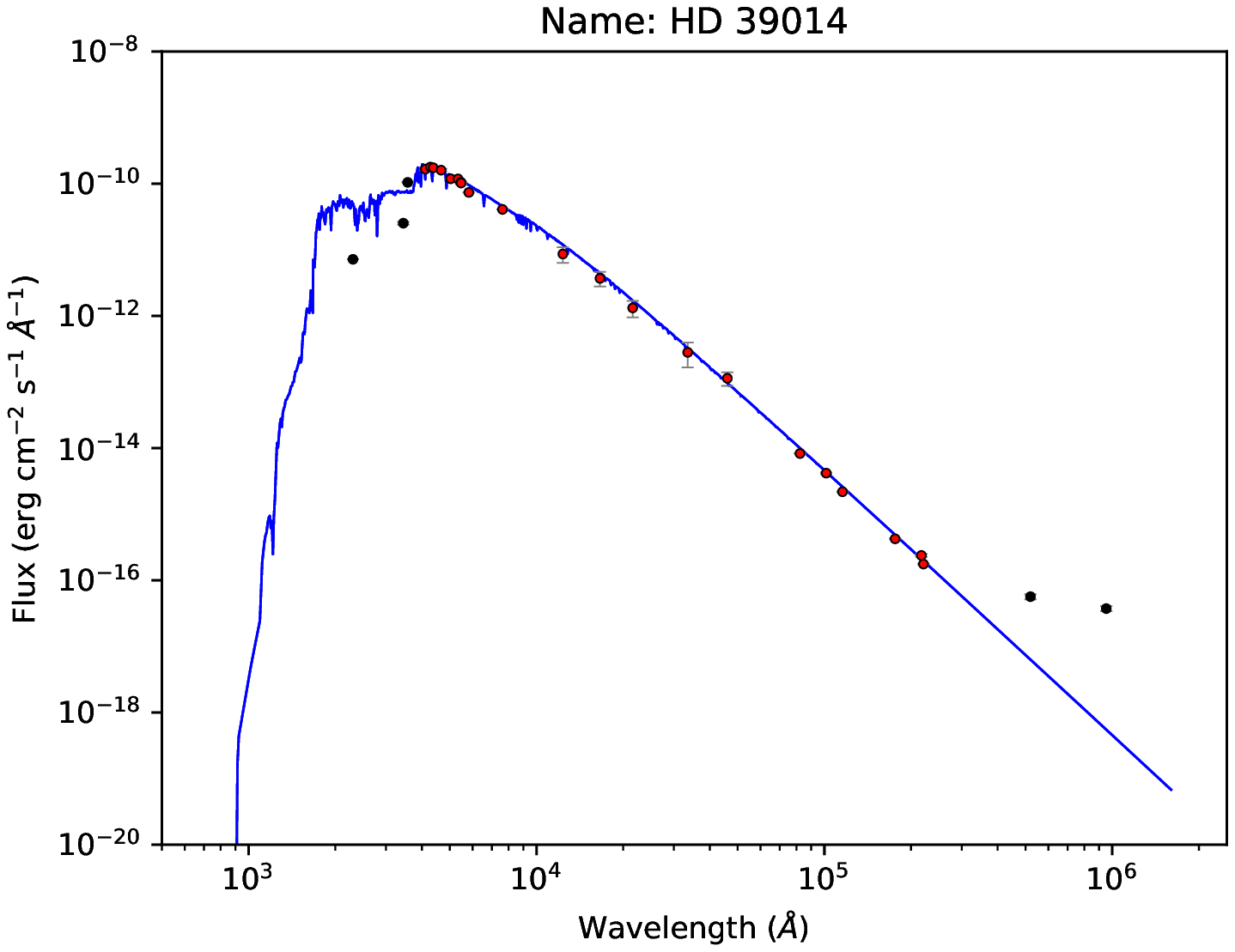}
    \includegraphics[width=0.33\textwidth]{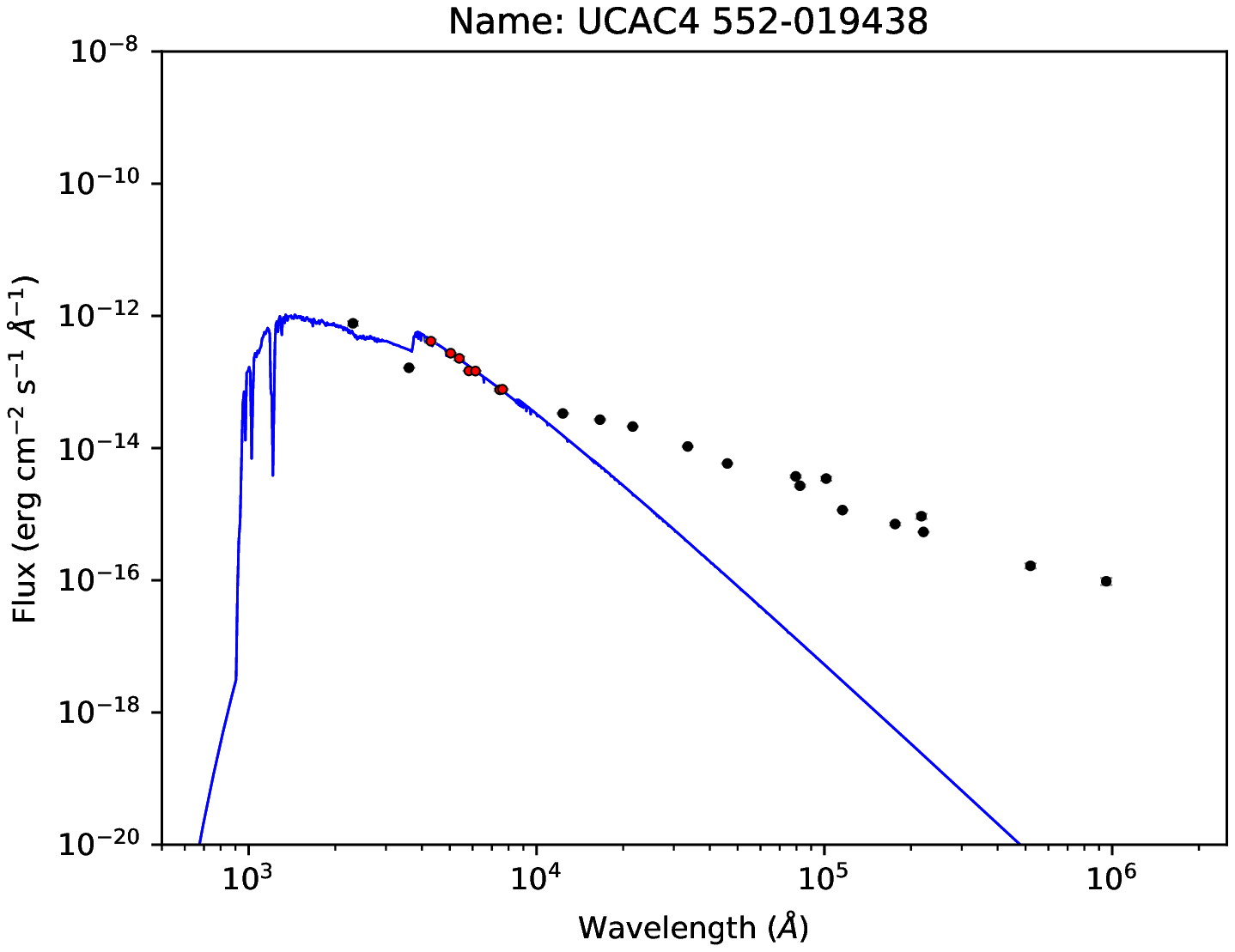}
    \includegraphics[width=0.33\textwidth]{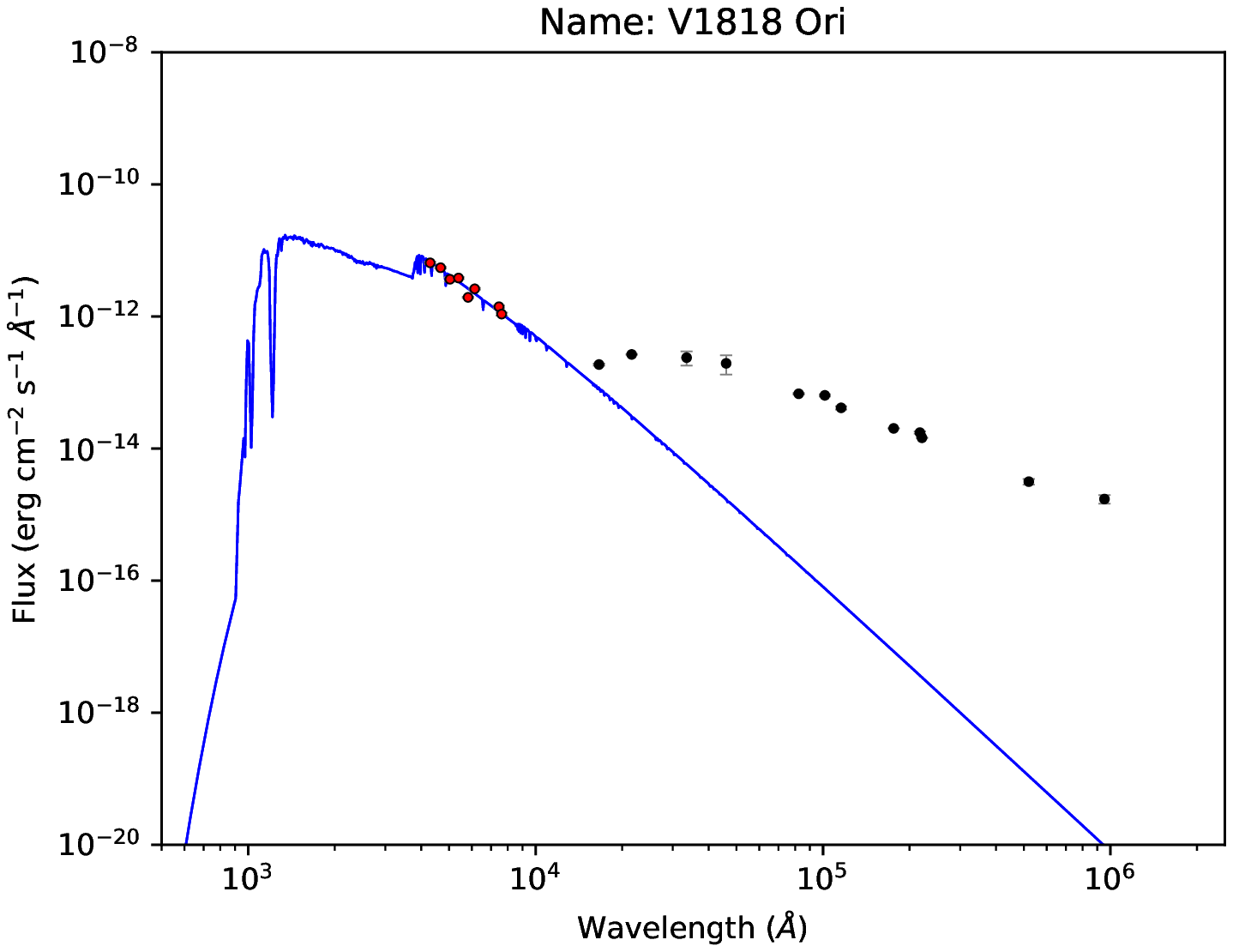}    
\end{figure}

\begin{figure} [h]
 \centering
    \includegraphics[width=0.33\textwidth]{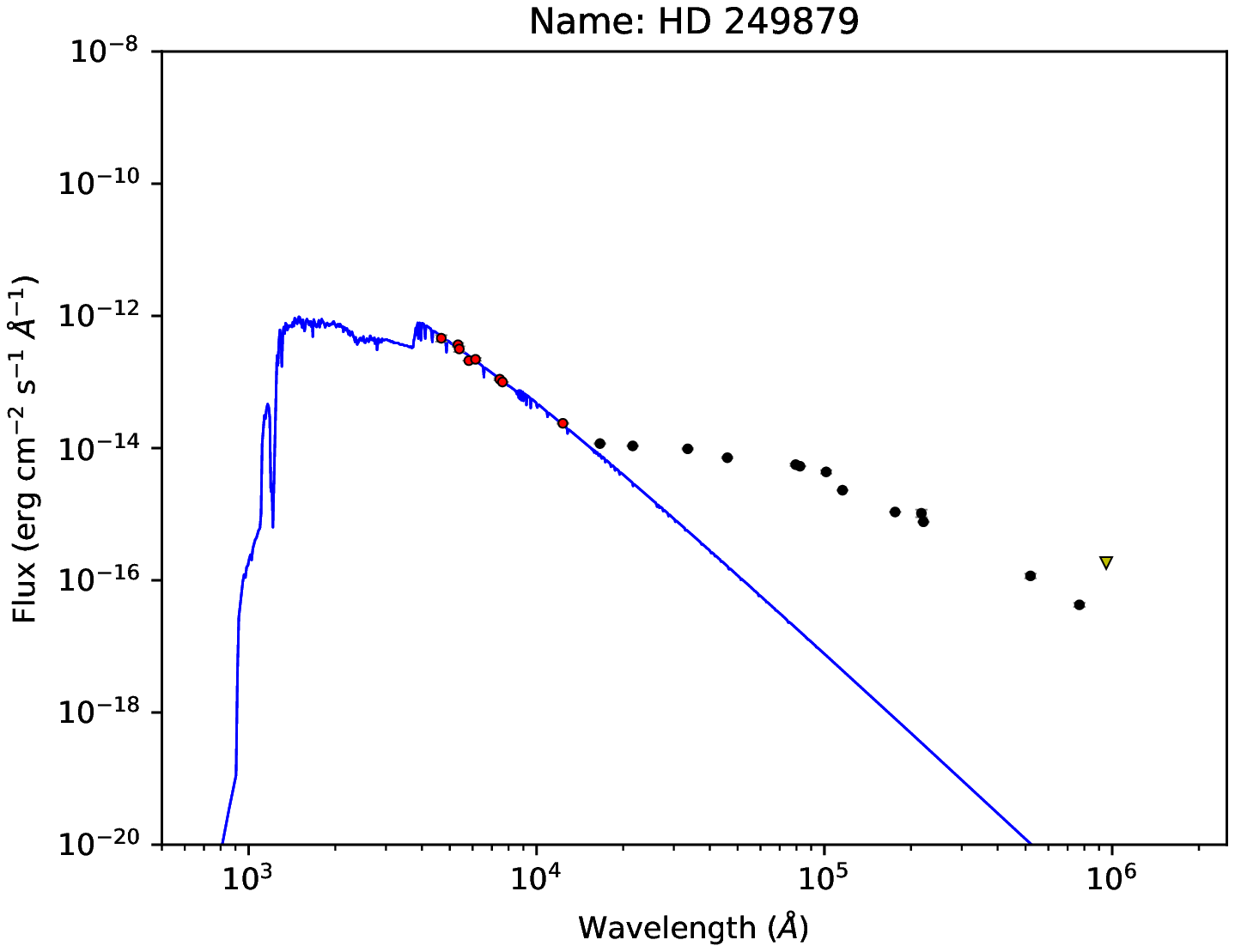}
    \includegraphics[width=0.33\textwidth]{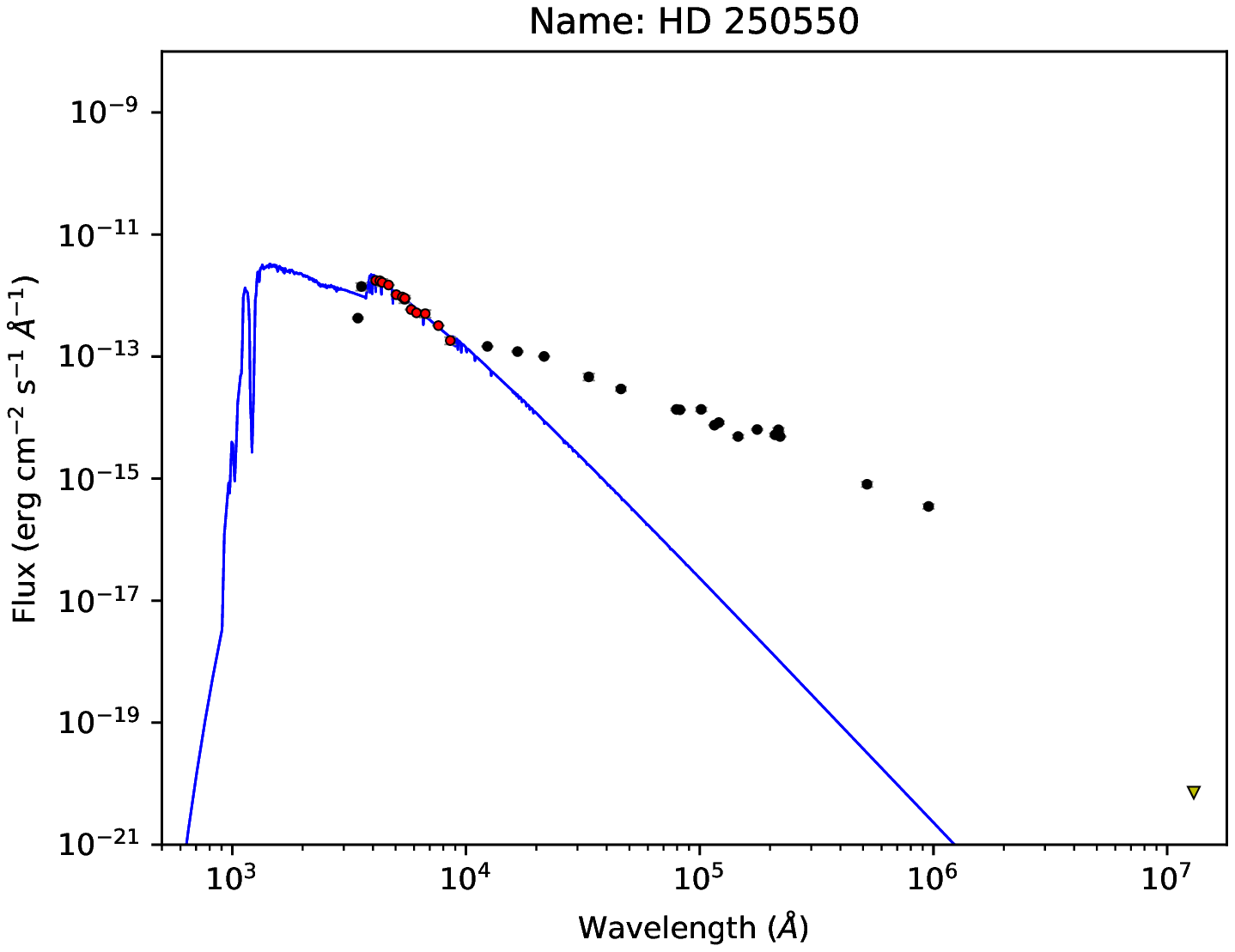}
    \includegraphics[width=0.33\textwidth]{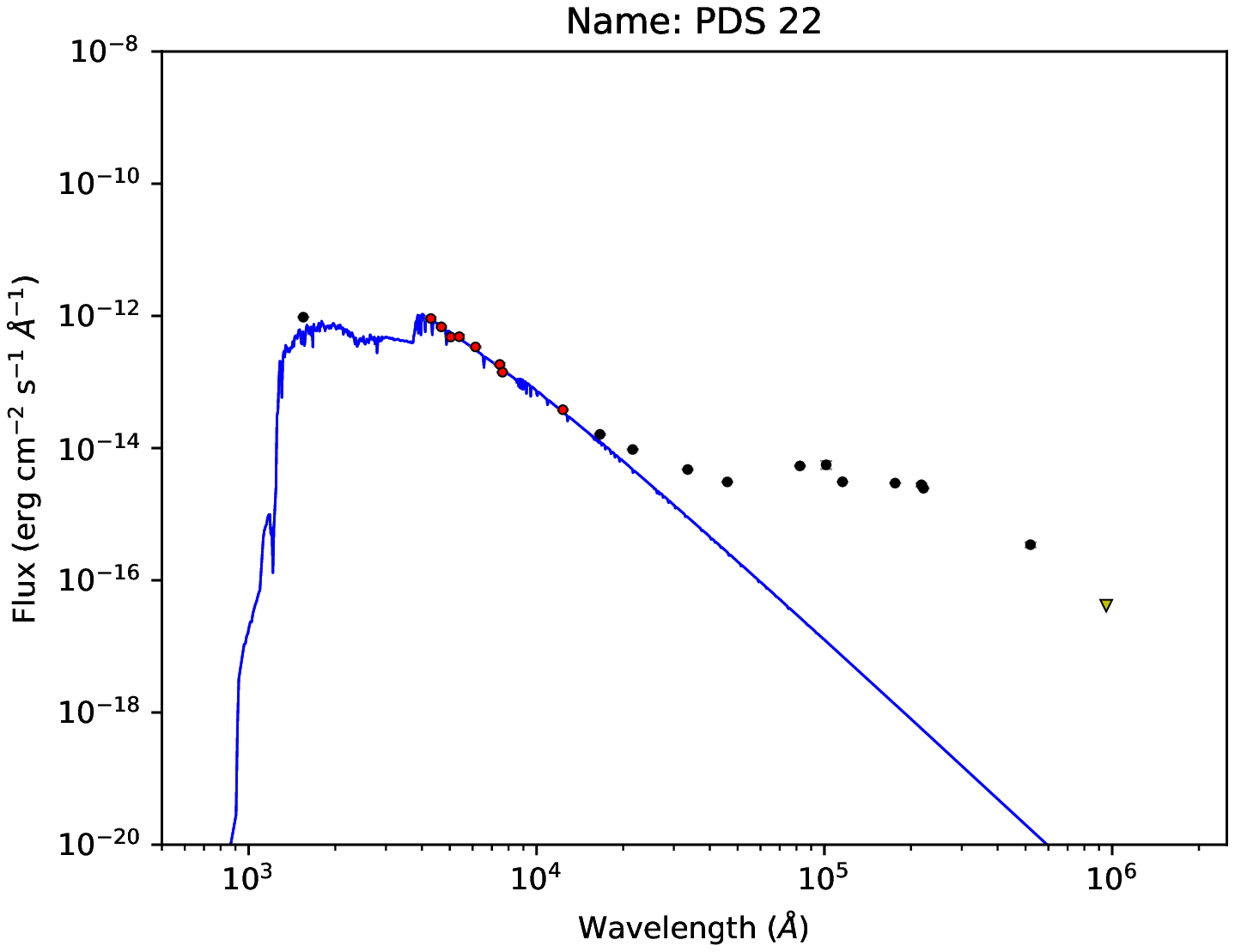}    
\end{figure}

\newpage

\onecolumn

\begin{figure} [h]
 \centering
    \includegraphics[width=0.33\textwidth]{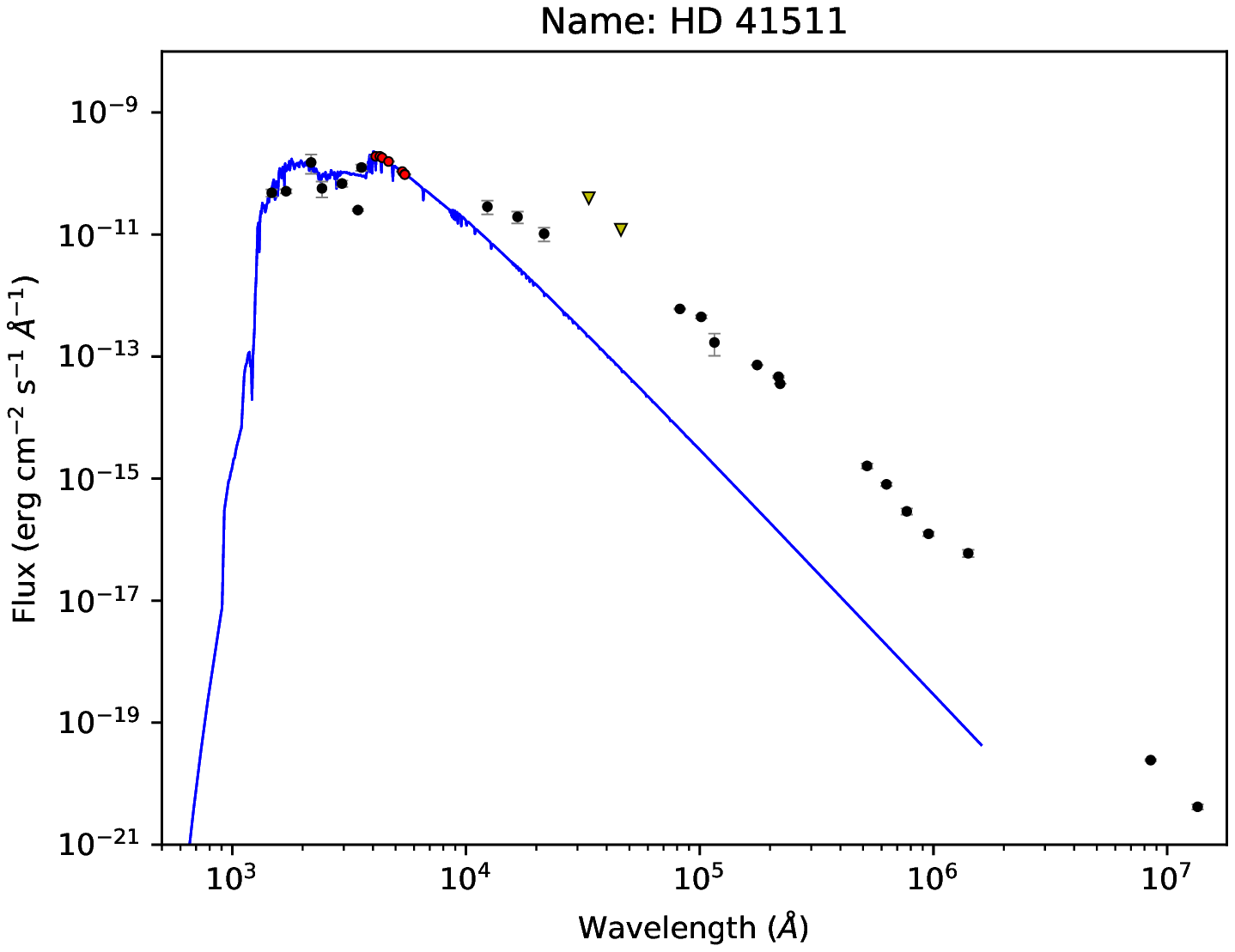}
    \includegraphics[width=0.33\textwidth]{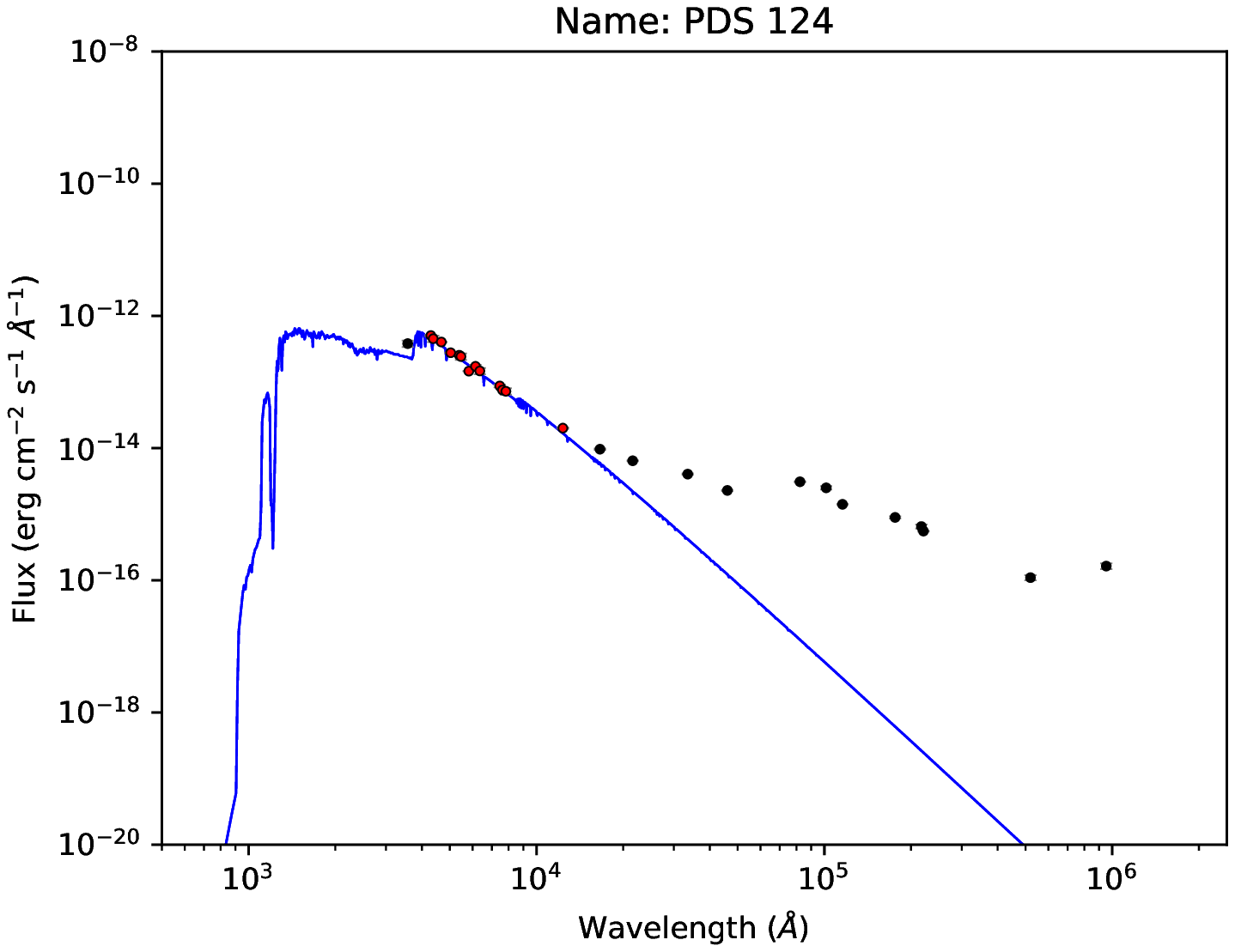}
    \includegraphics[width=0.33\textwidth]{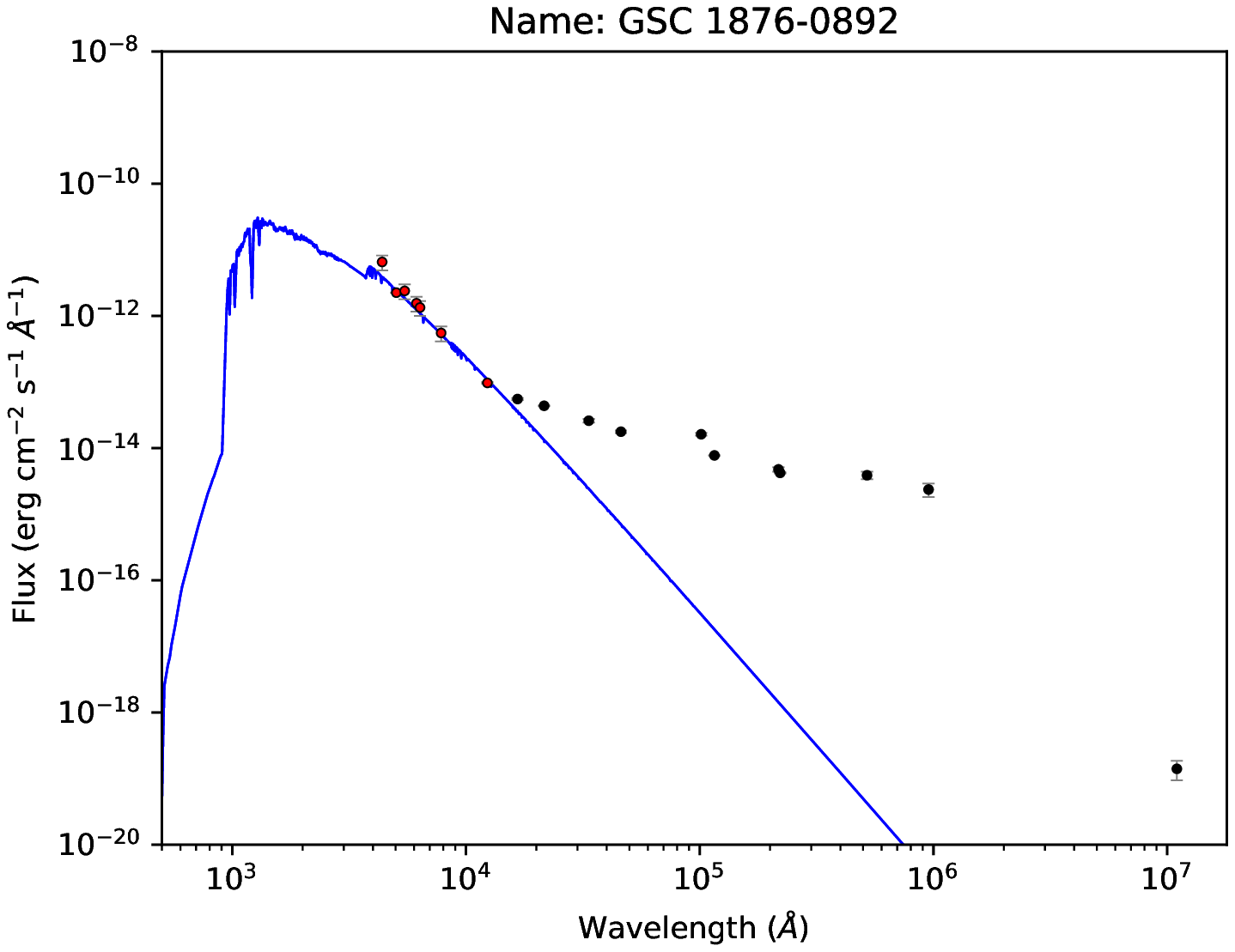}    
\end{figure}

\begin{figure} [h]
 \centering
    \includegraphics[width=0.33\textwidth]{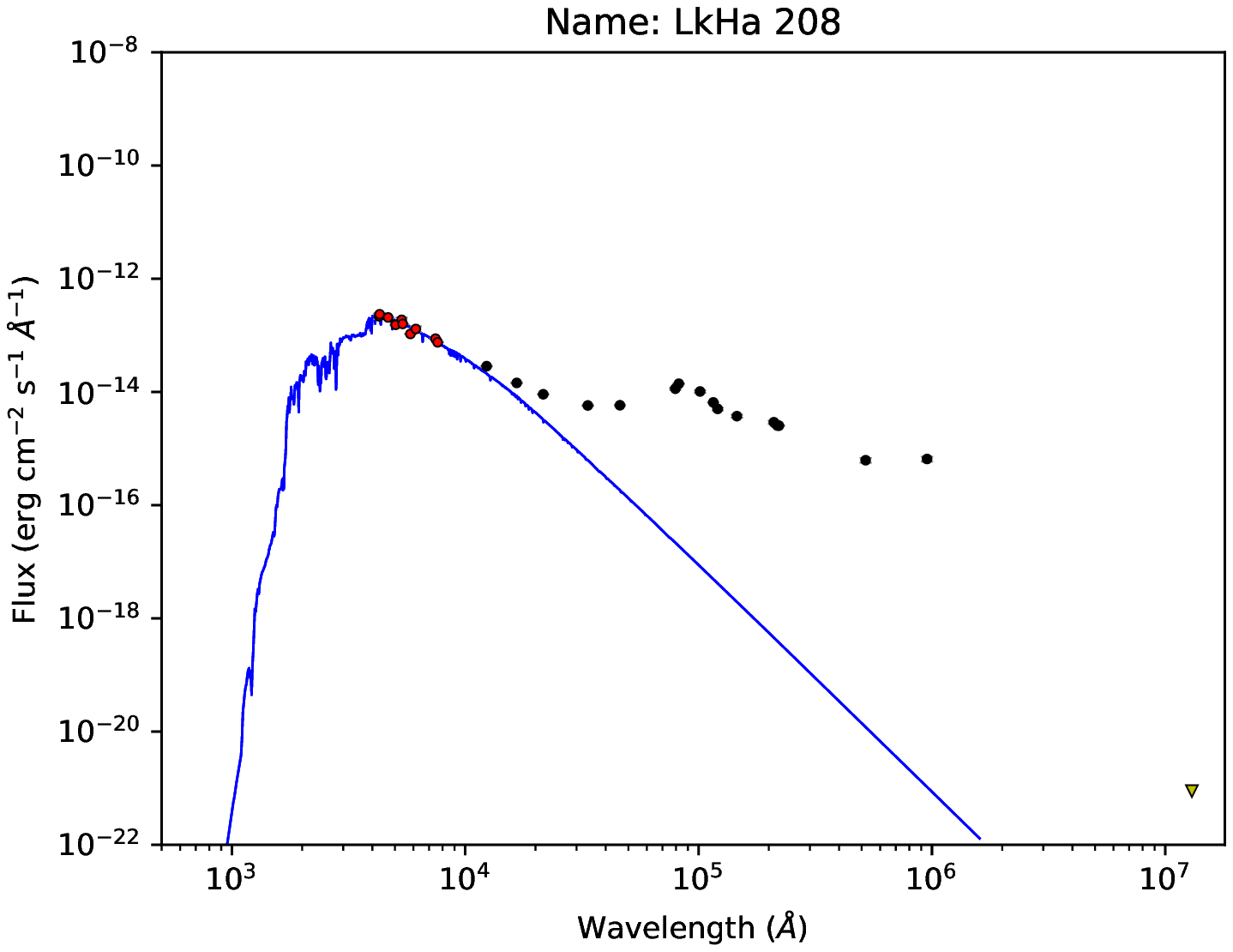}
    \includegraphics[width=0.33\textwidth]{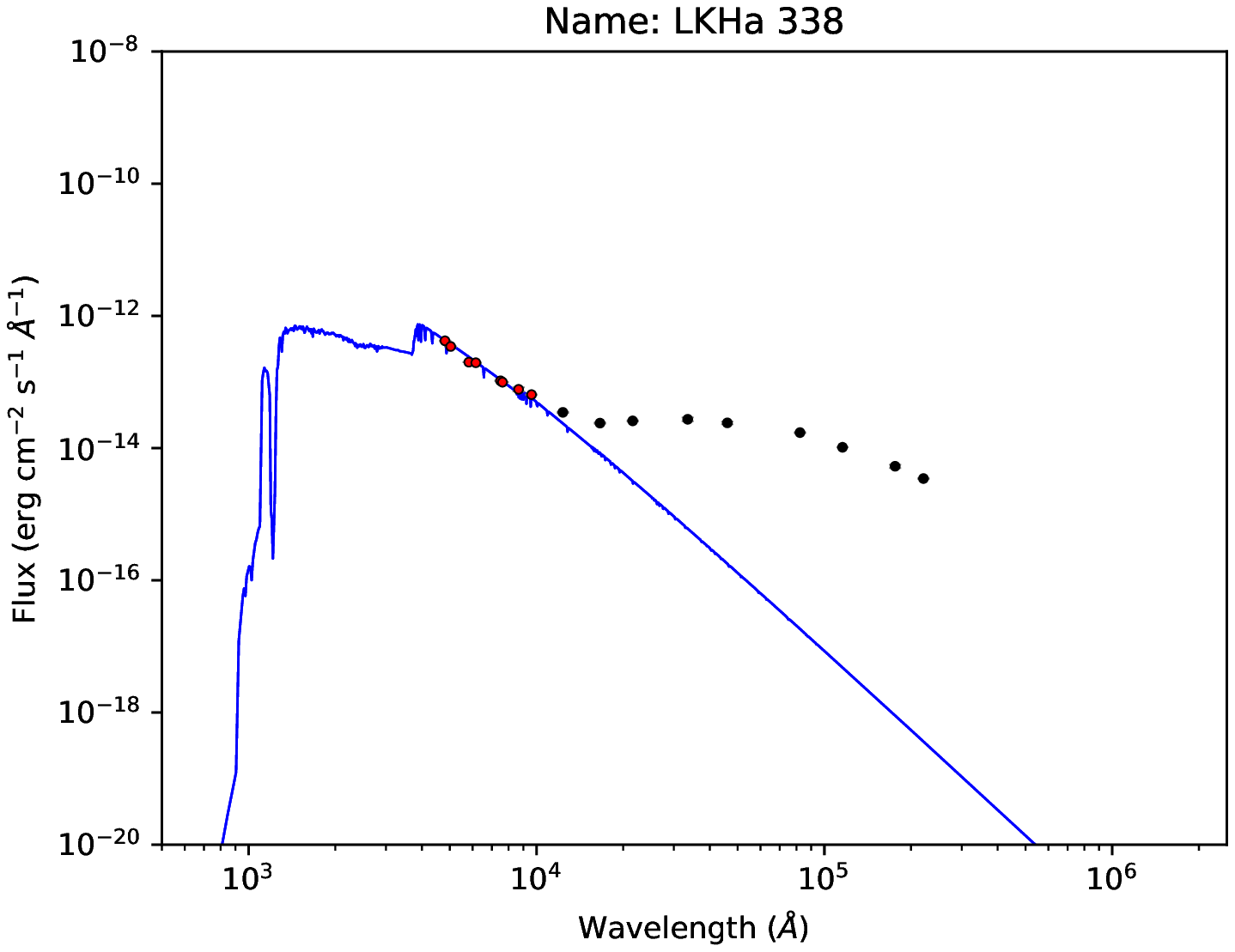}
    \includegraphics[width=0.33\textwidth]{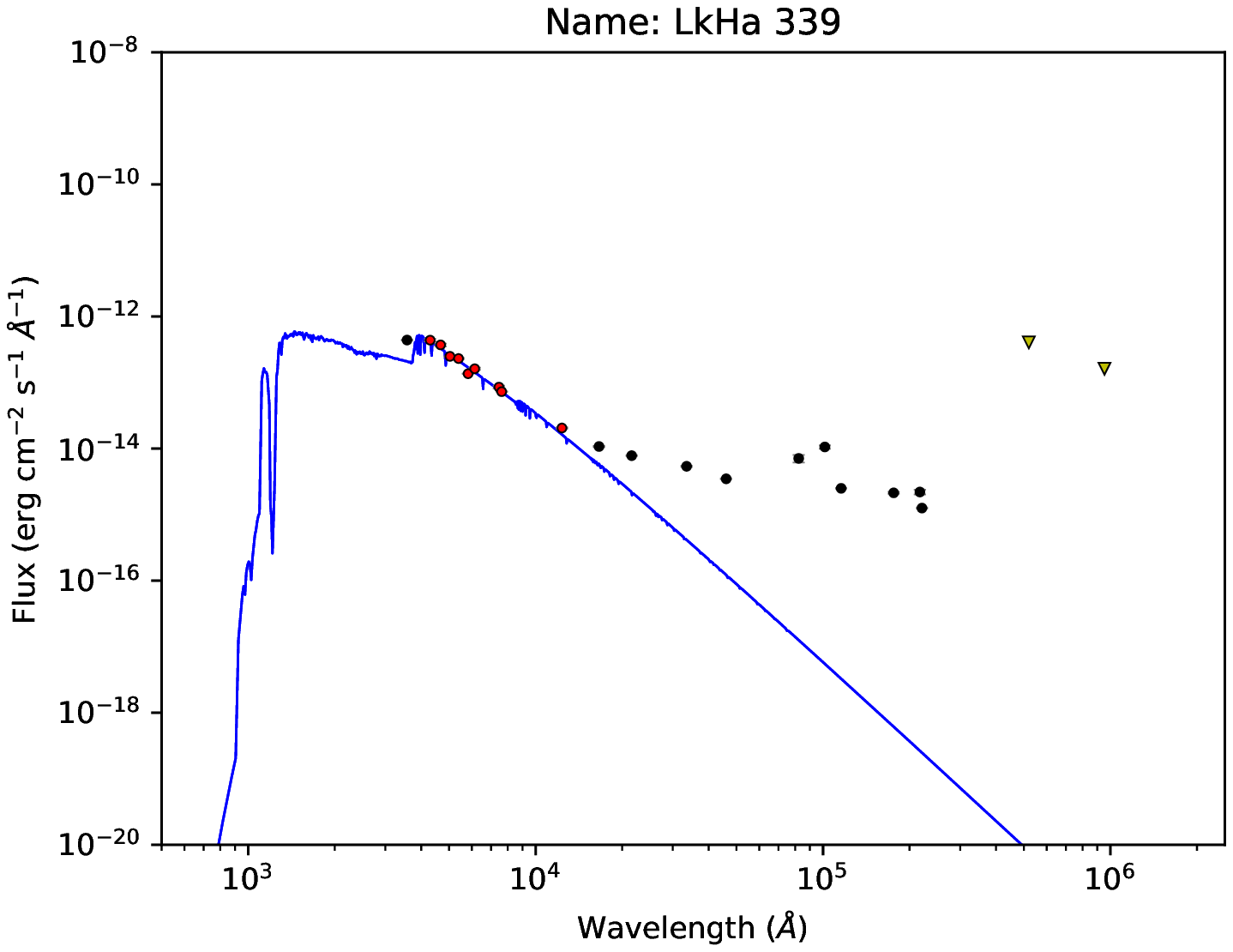}    
\end{figure}

\begin{figure} [h]
 \centering
    \includegraphics[width=0.33\textwidth]{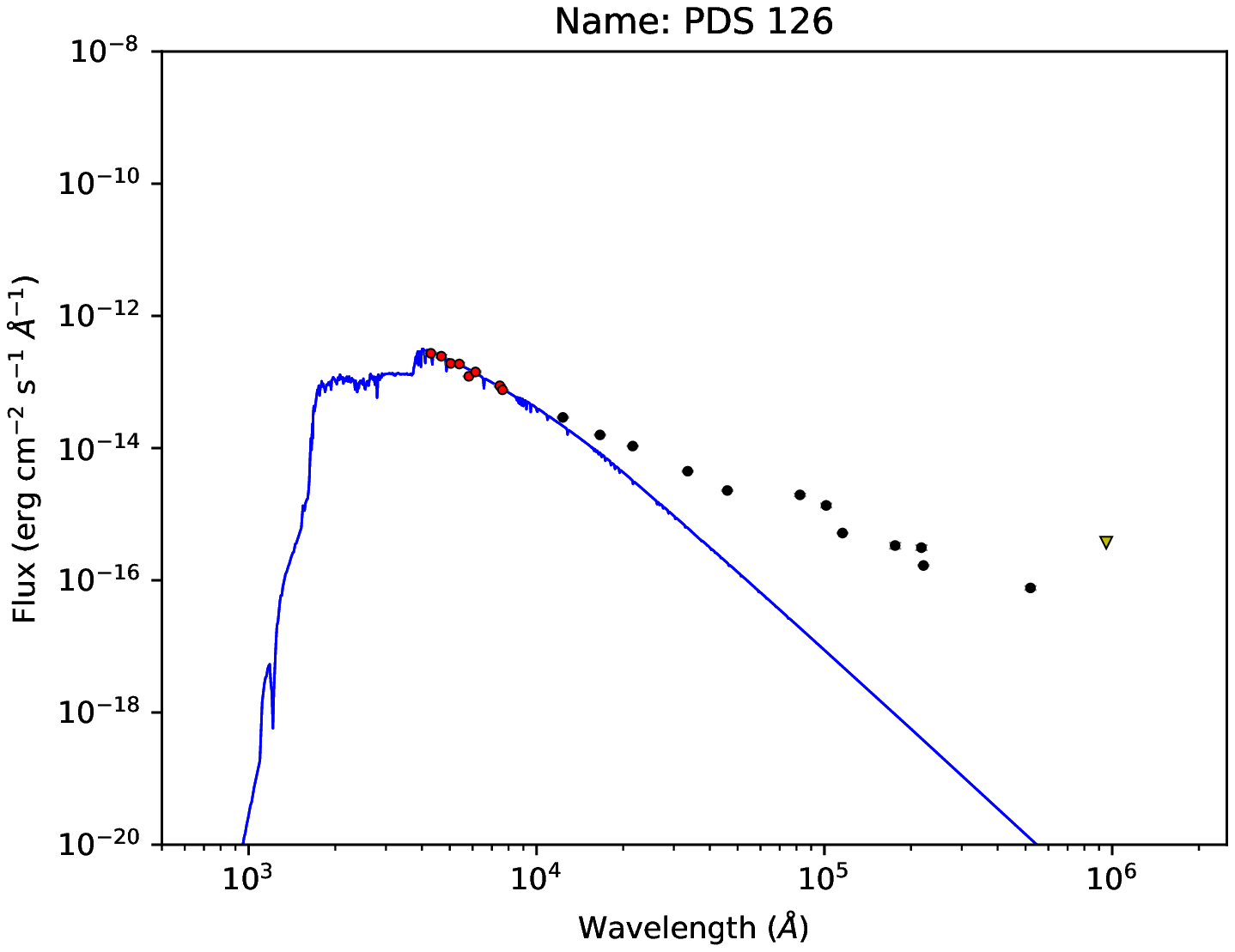}
    \includegraphics[width=0.33\textwidth]{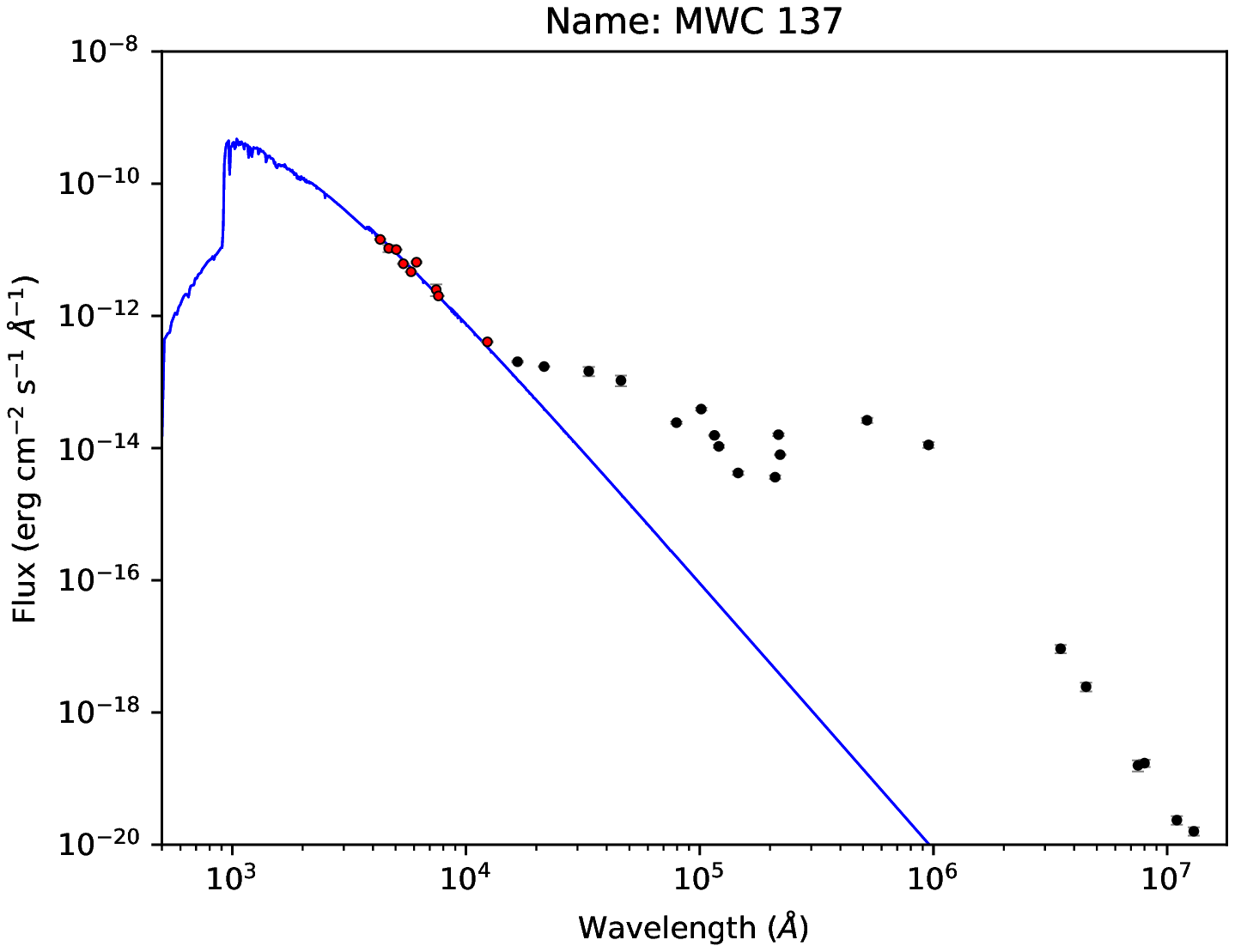}
    \includegraphics[width=0.33\textwidth]{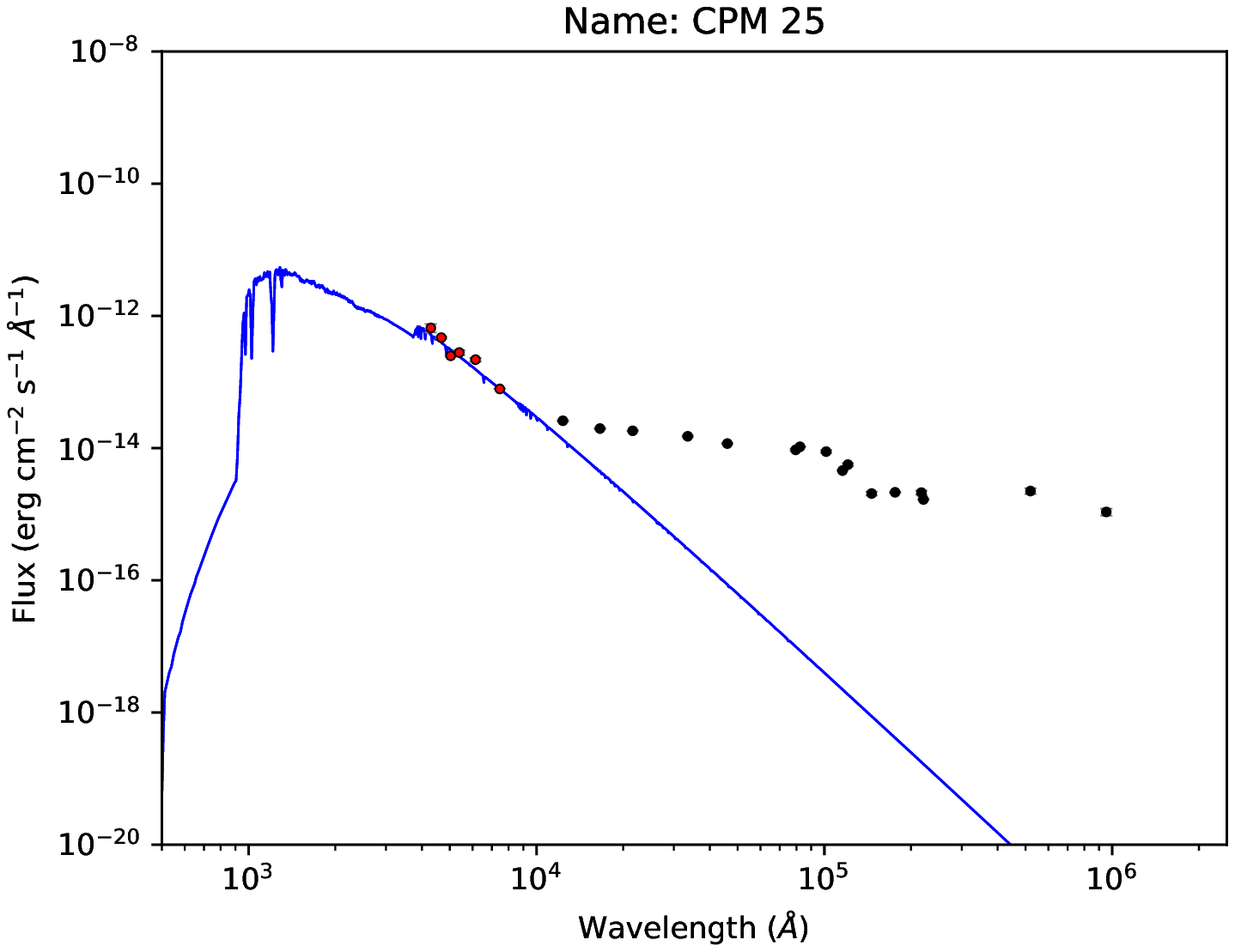}    
\end{figure}

\begin{figure} [h]
 \centering
    \includegraphics[width=0.33\textwidth]{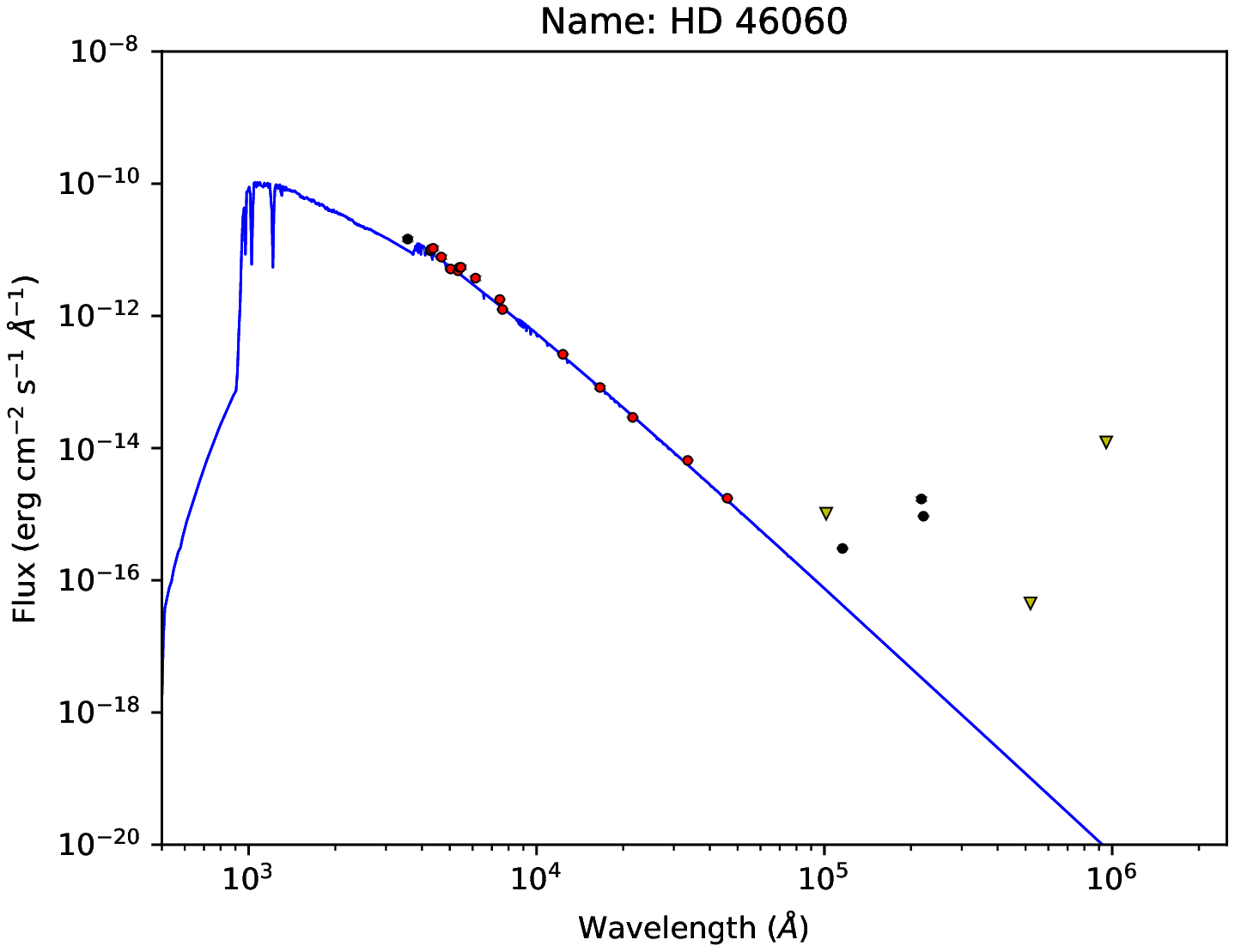}
    \includegraphics[width=0.33\textwidth]{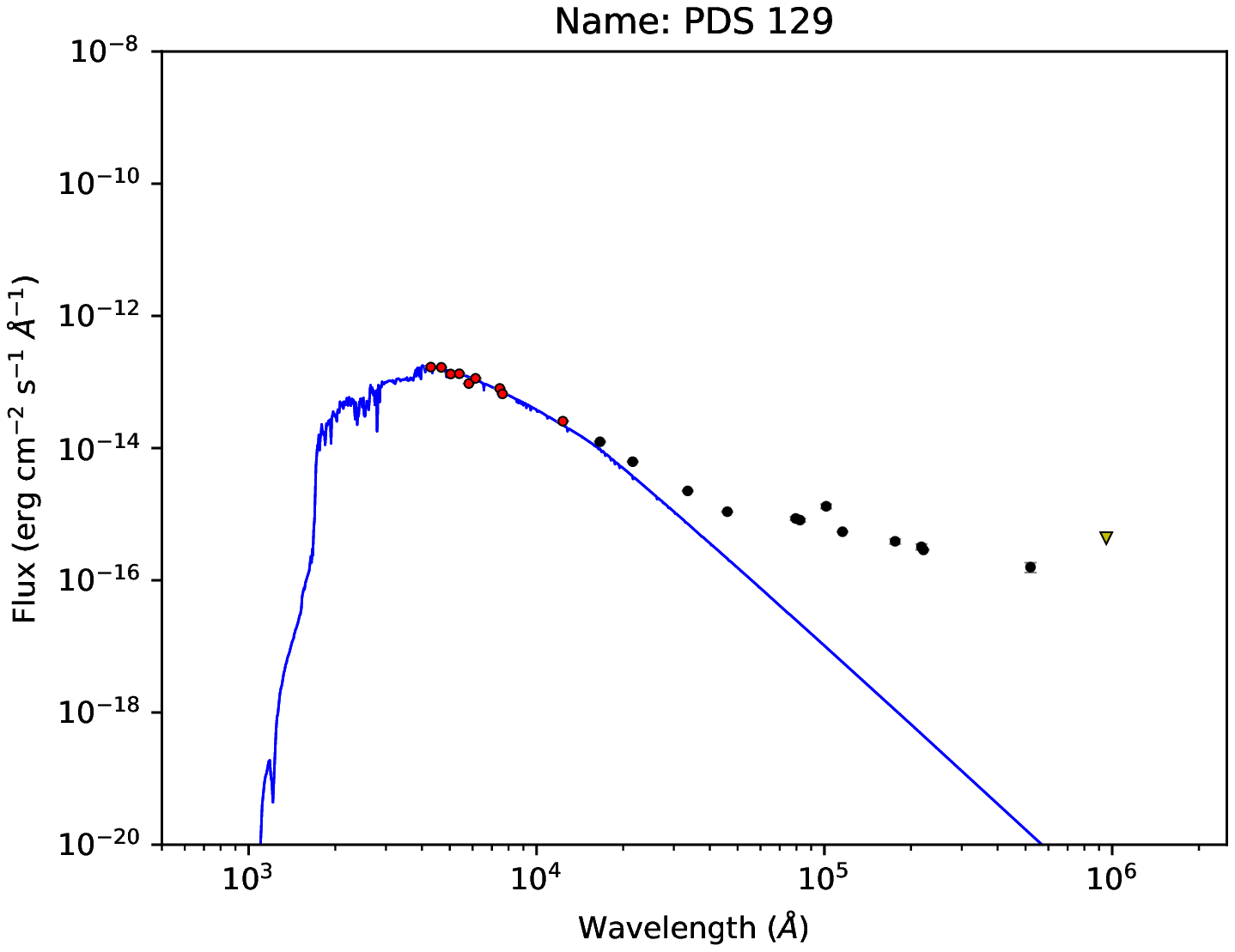}
    \includegraphics[width=0.33\textwidth]{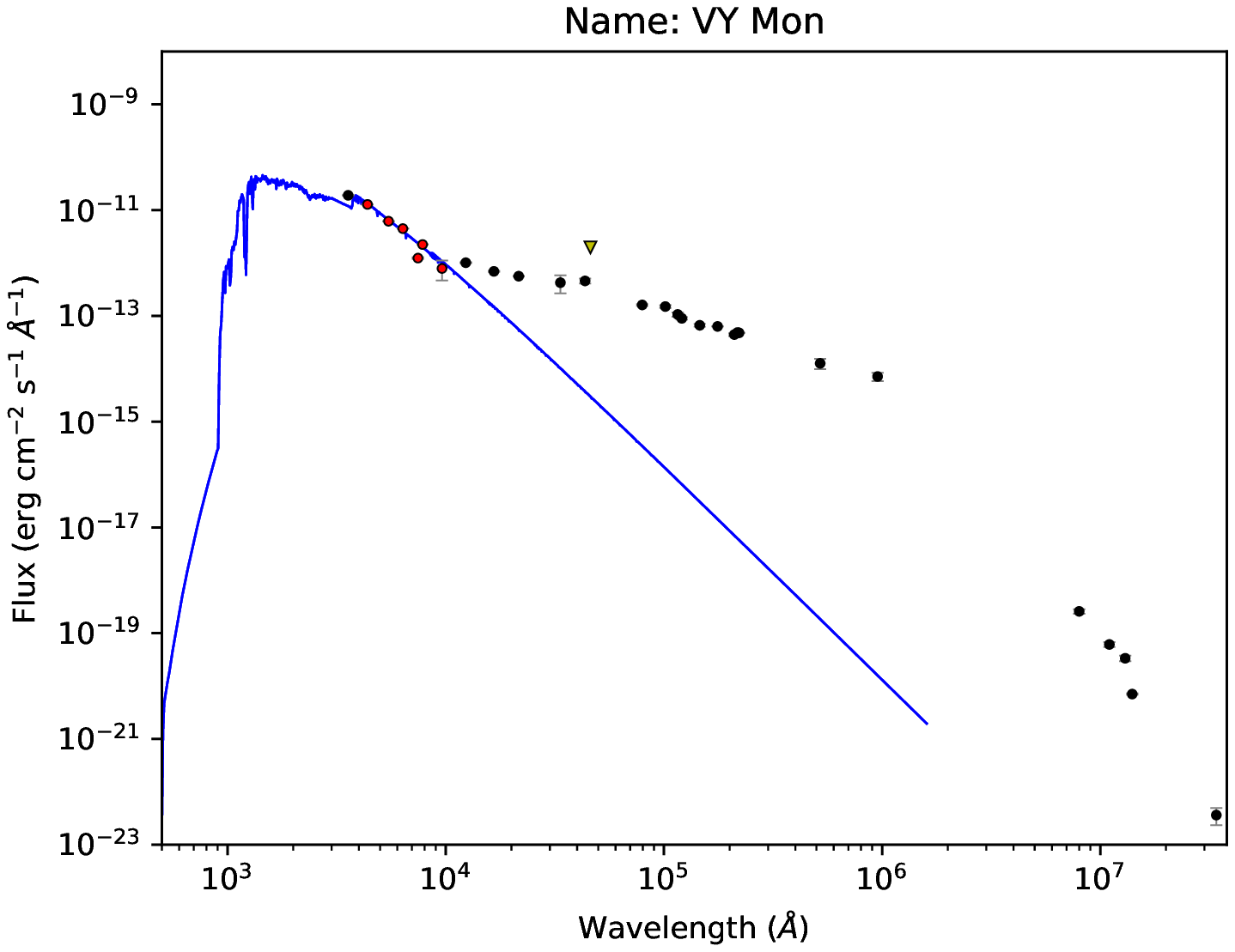}    
\end{figure}

\newpage

\onecolumn

\begin{figure} [h]
 \centering
    \includegraphics[width=0.33\textwidth]{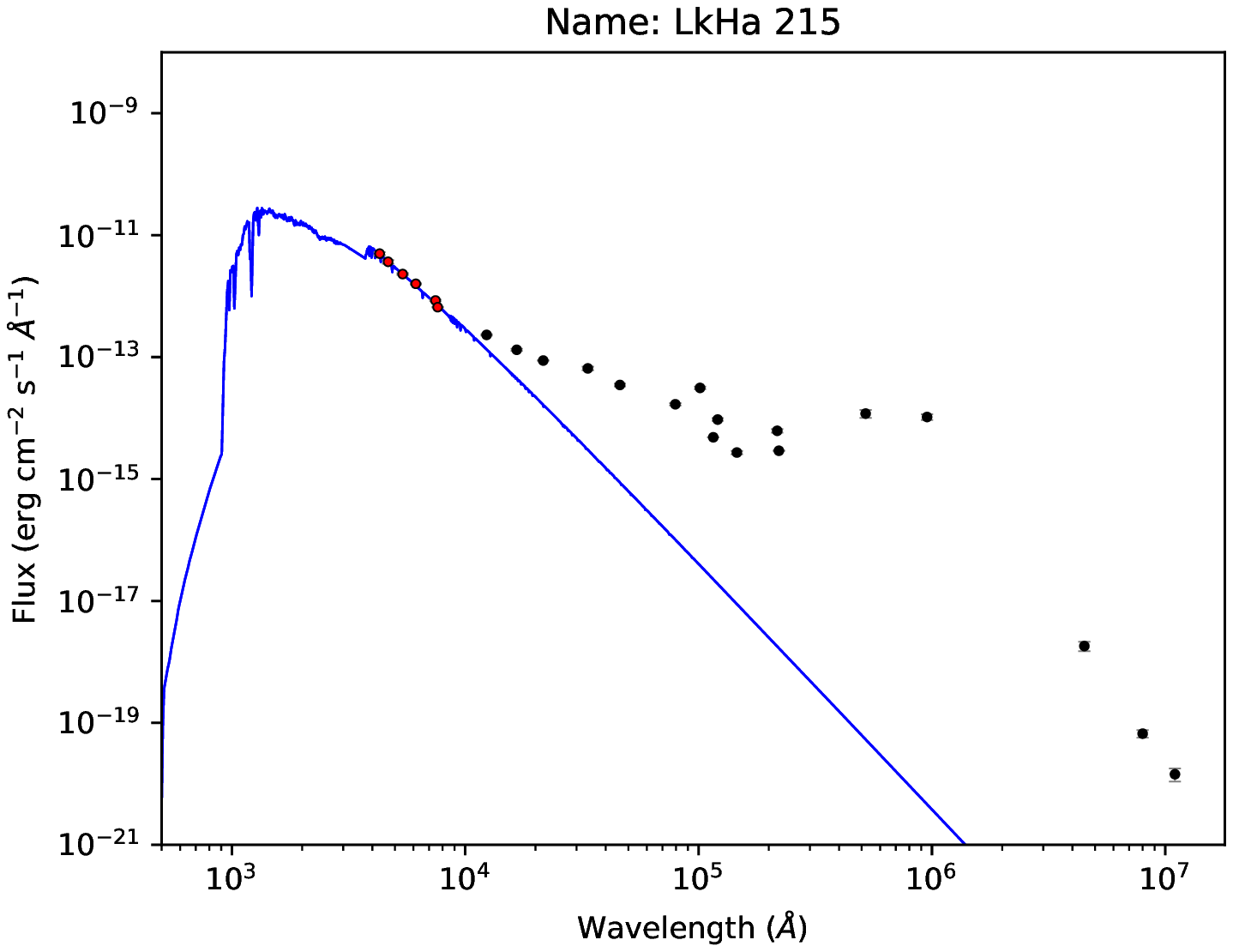}
    \includegraphics[width=0.33\textwidth]{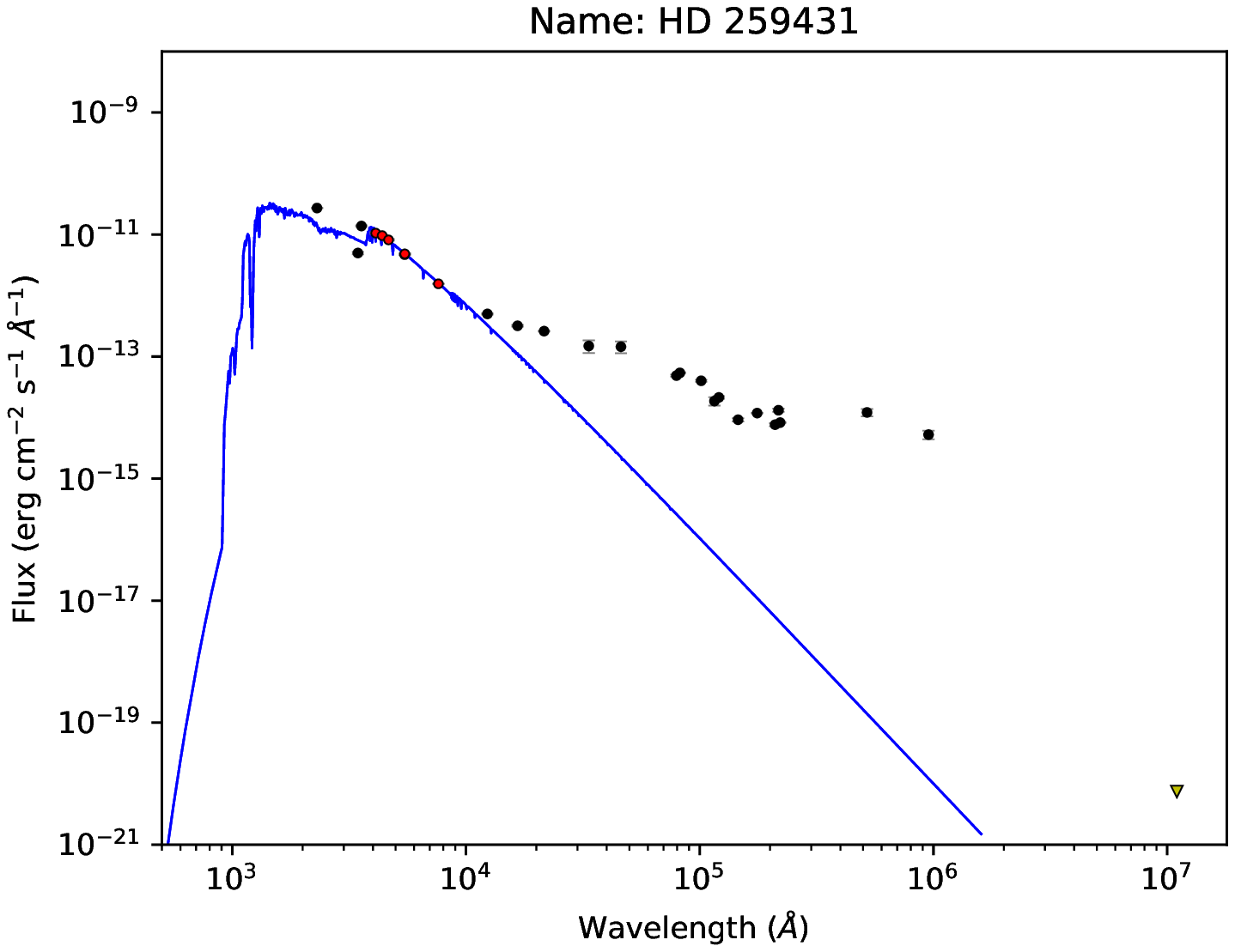}
    \includegraphics[width=0.33\textwidth]{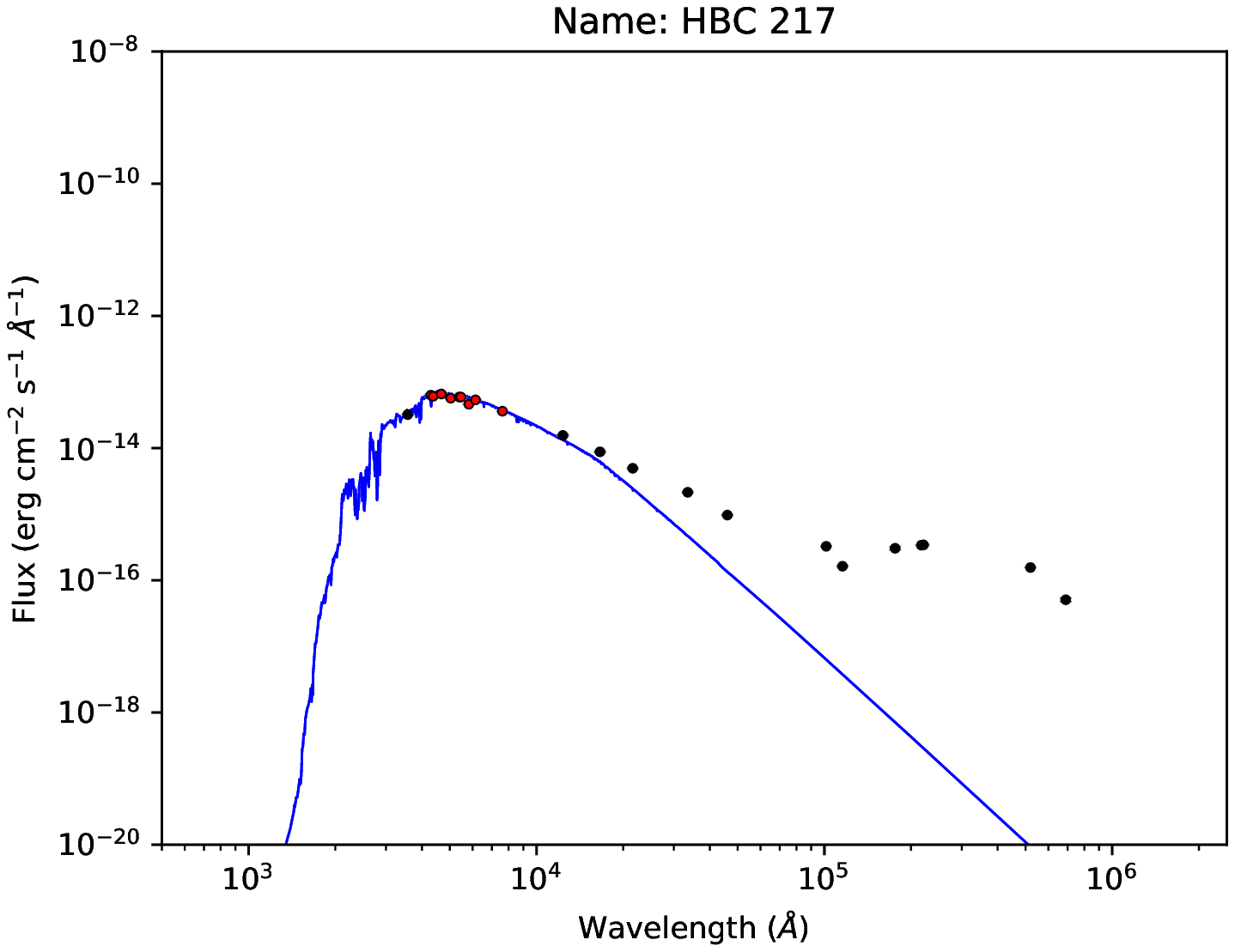}    
\end{figure}

\begin{figure} [h]
 \centering
    \includegraphics[width=0.33\textwidth]{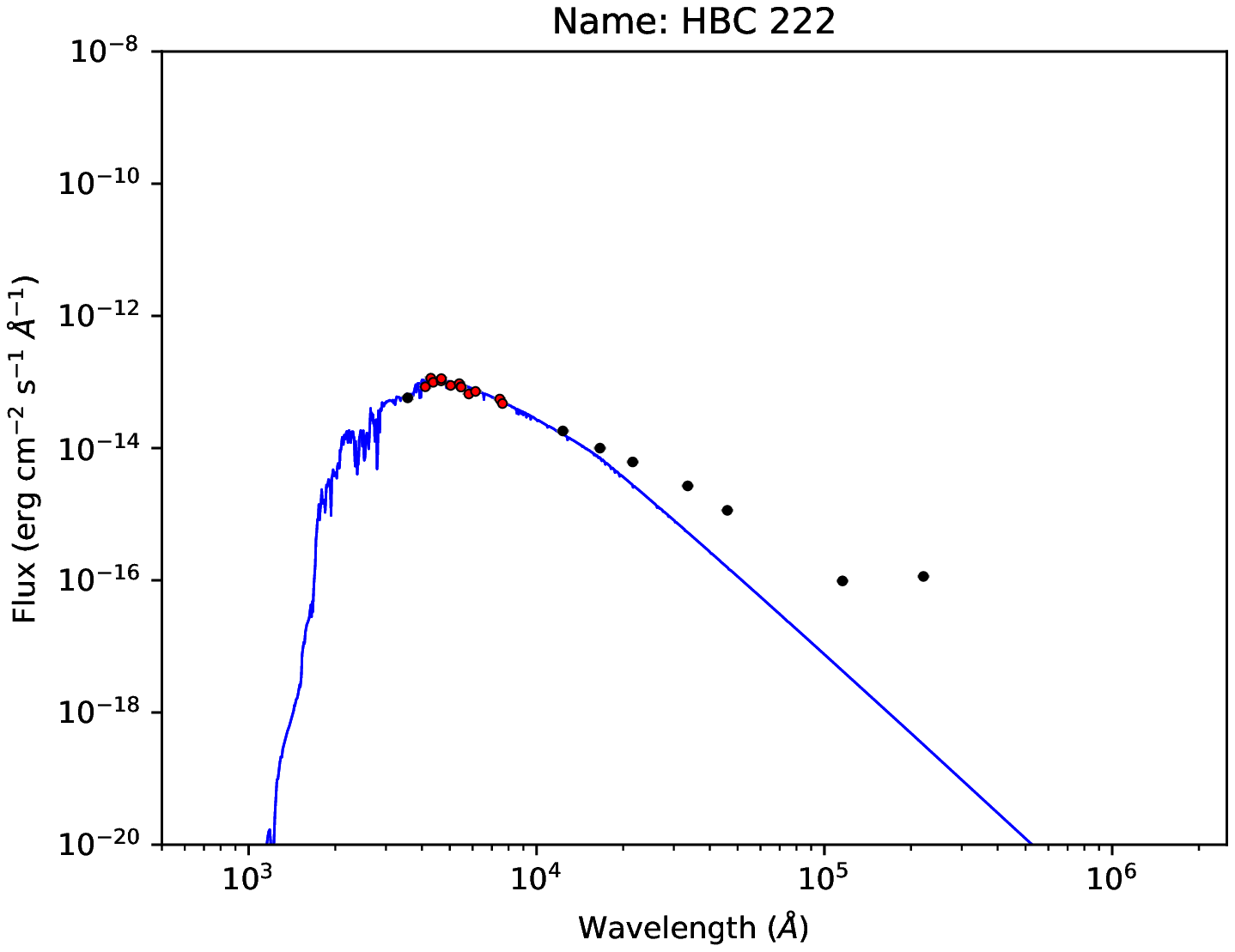}
    \includegraphics[width=0.33\textwidth]{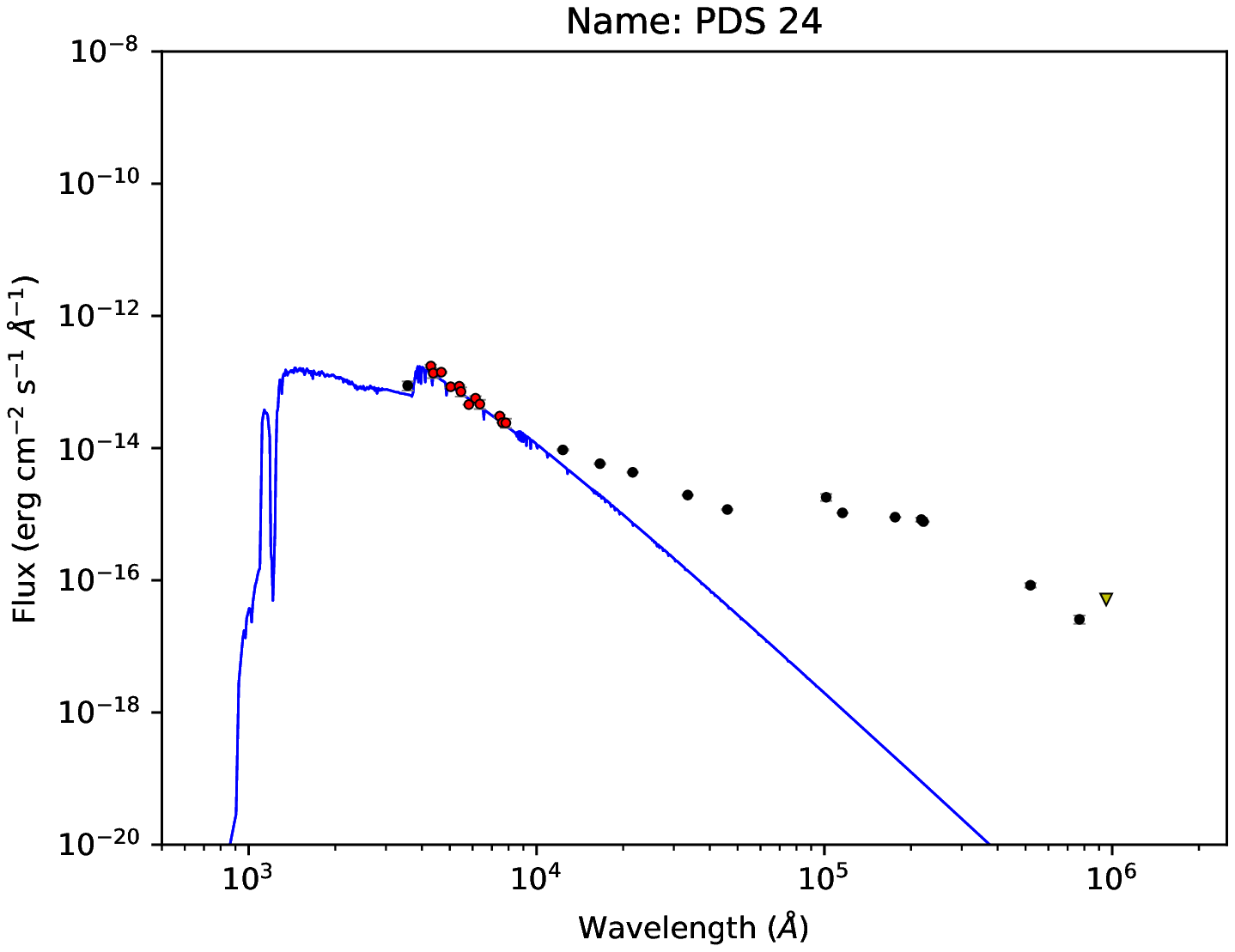}
    \includegraphics[width=0.33\textwidth]{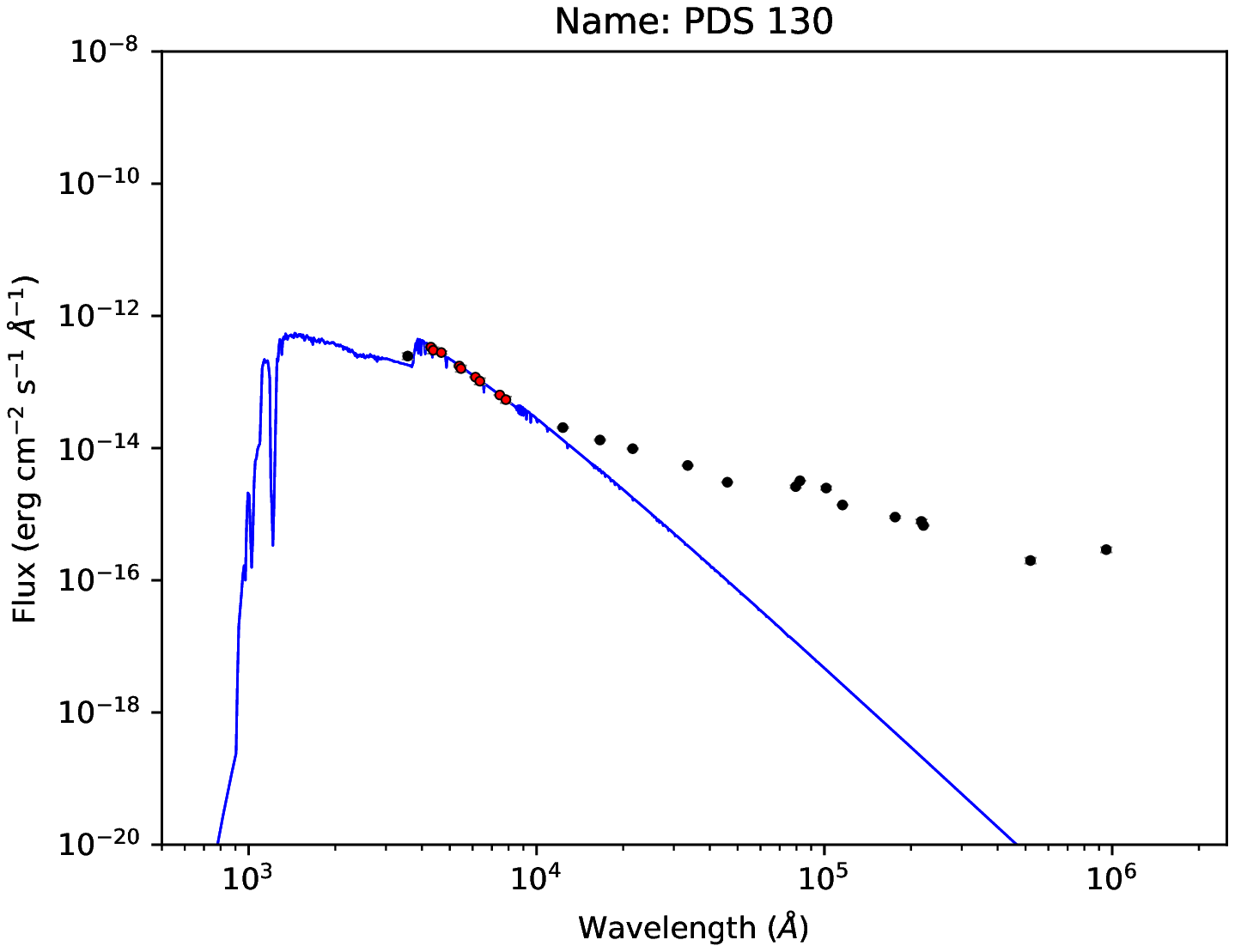}    
\end{figure}

\begin{figure} [h]
 \centering
    \includegraphics[width=0.33\textwidth]{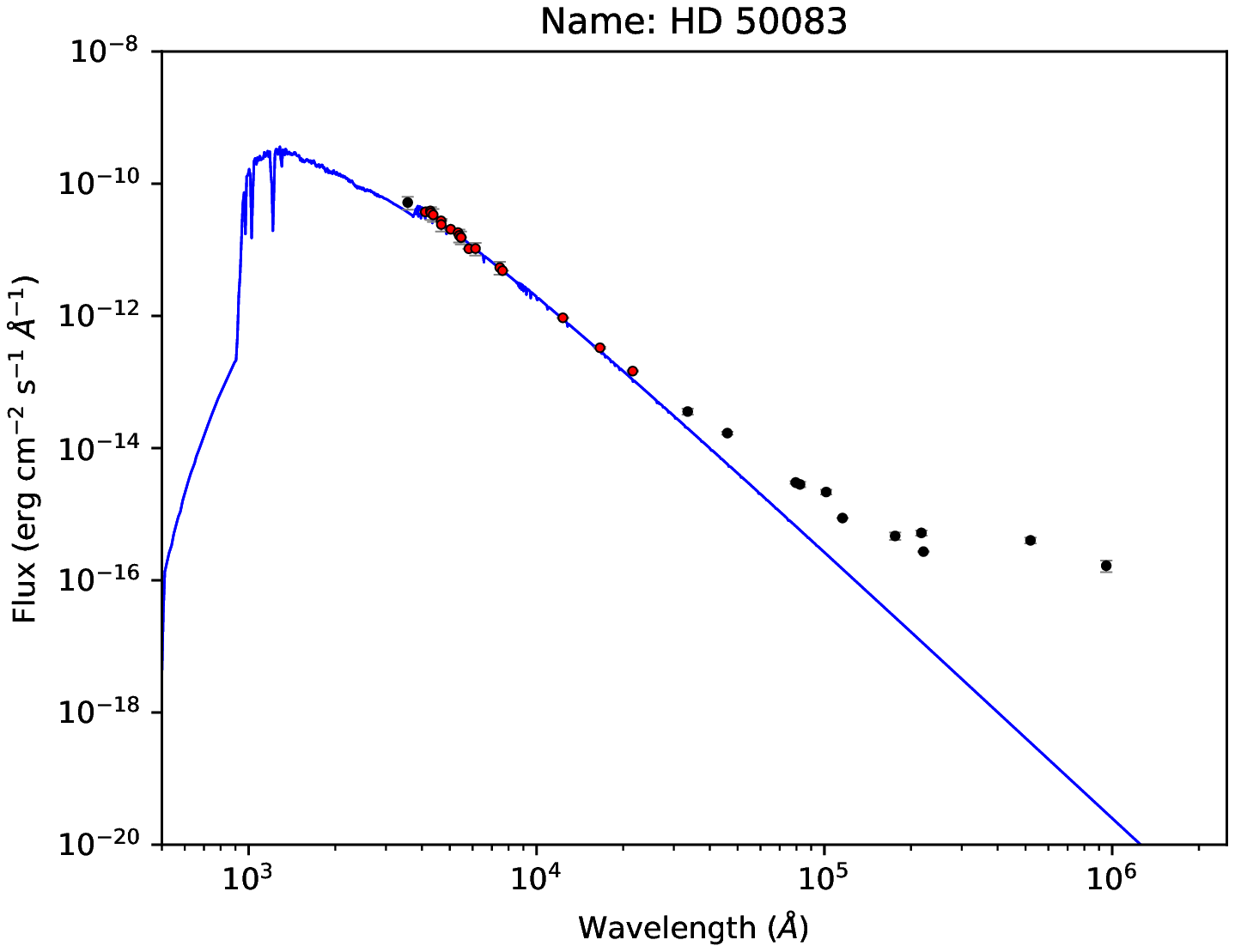}
    \includegraphics[width=0.33\textwidth]{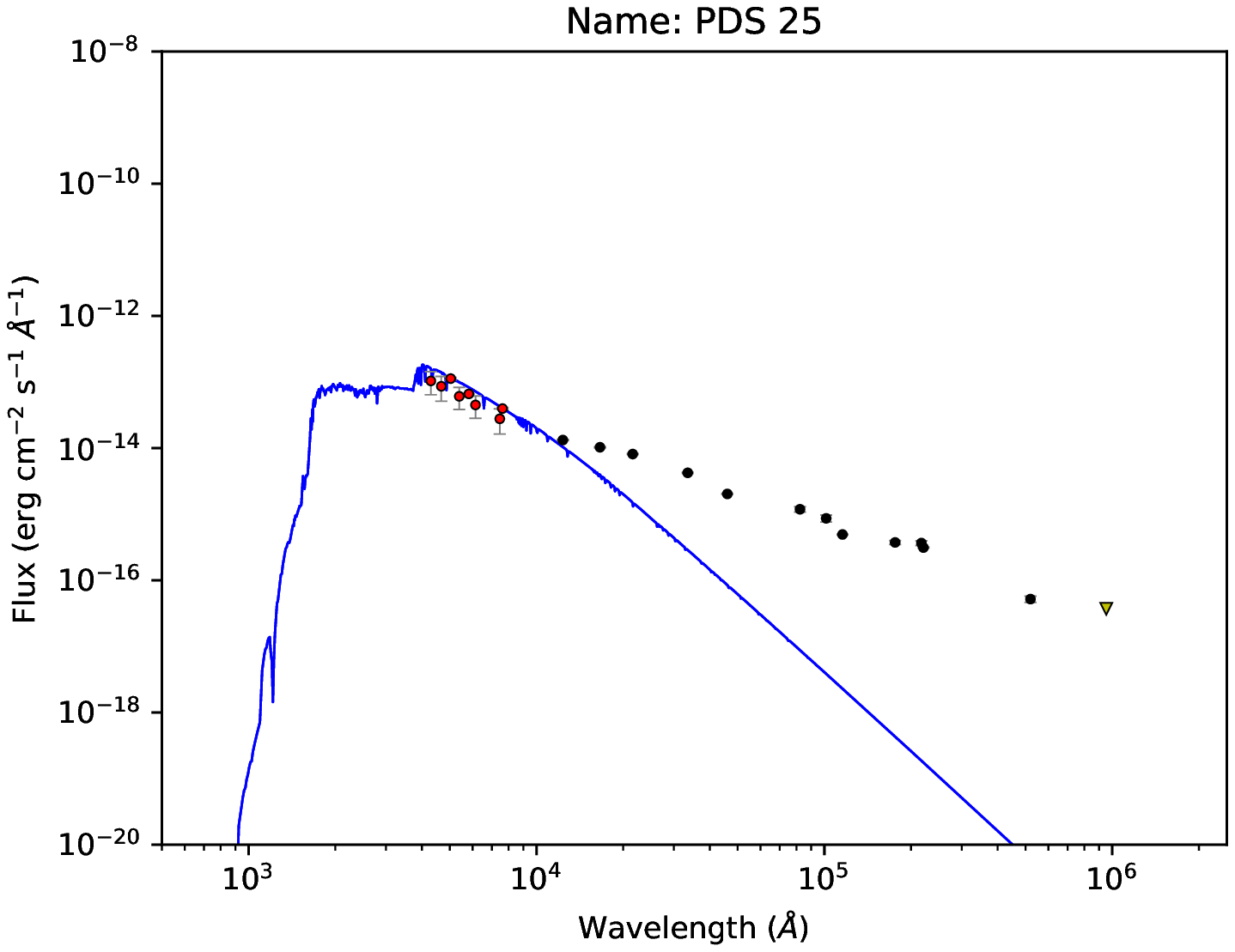}
    \includegraphics[width=0.33\textwidth]{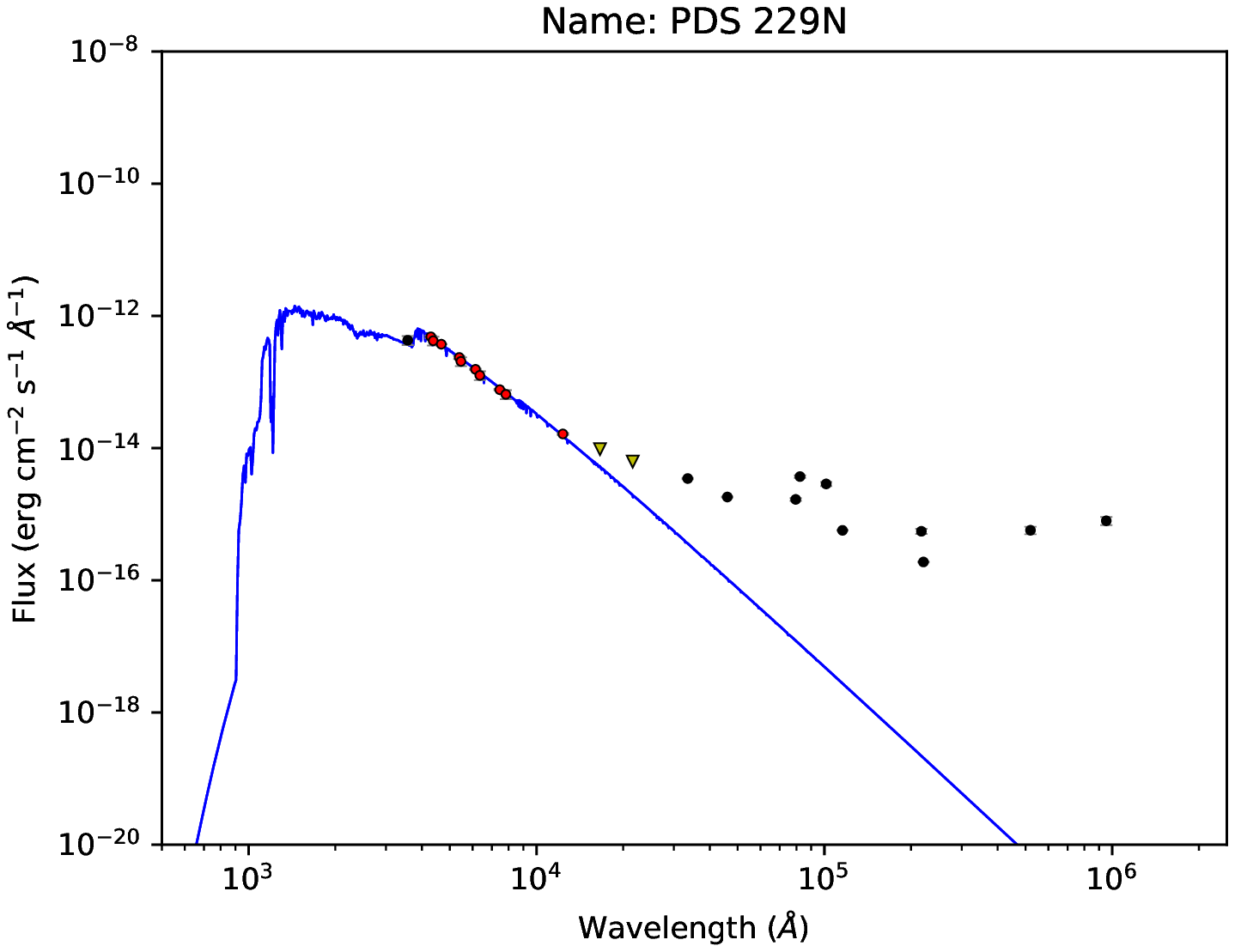}    
\end{figure}

\begin{figure} [h]
 \centering
    \includegraphics[width=0.33\textwidth]{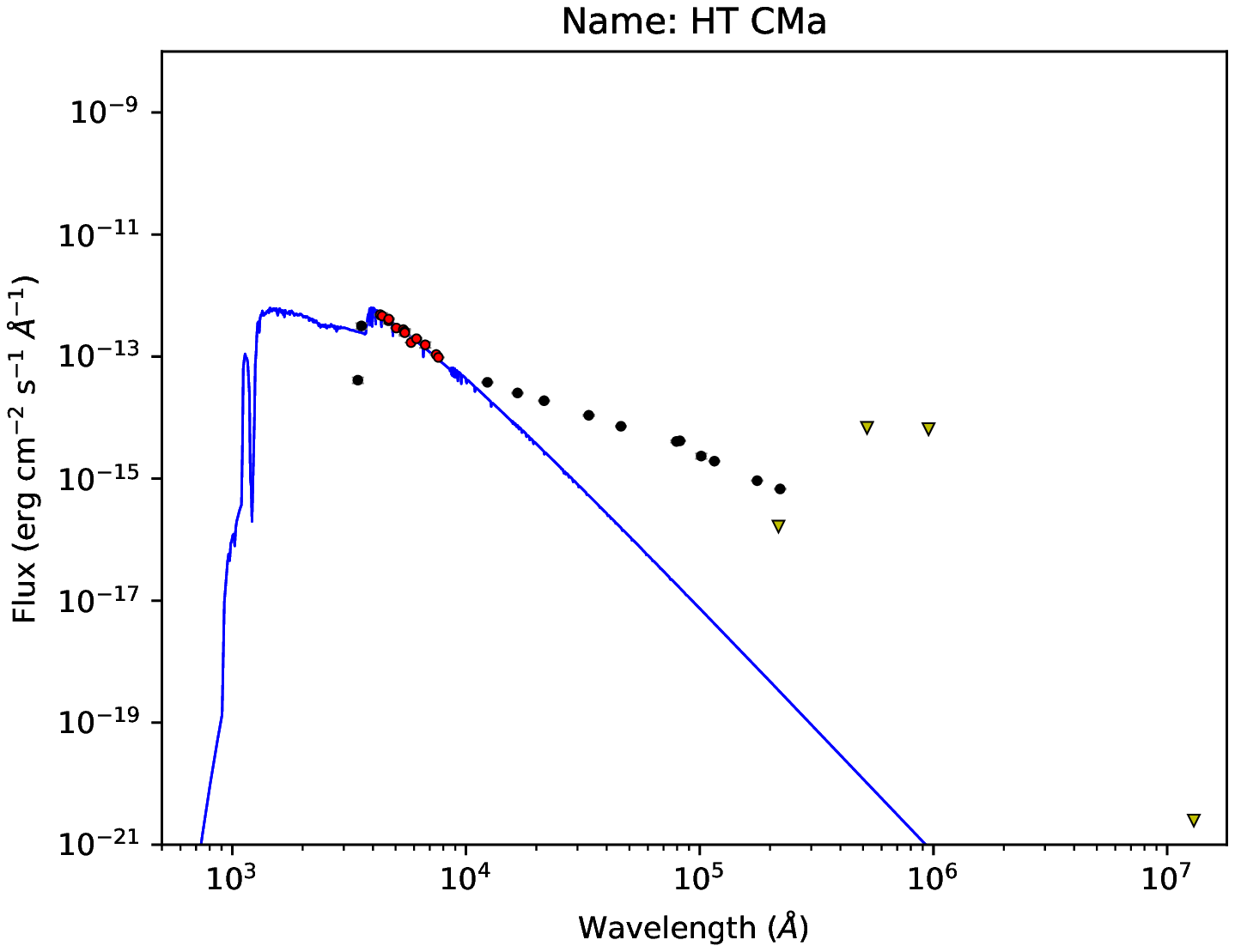}
    \includegraphics[width=0.33\textwidth]{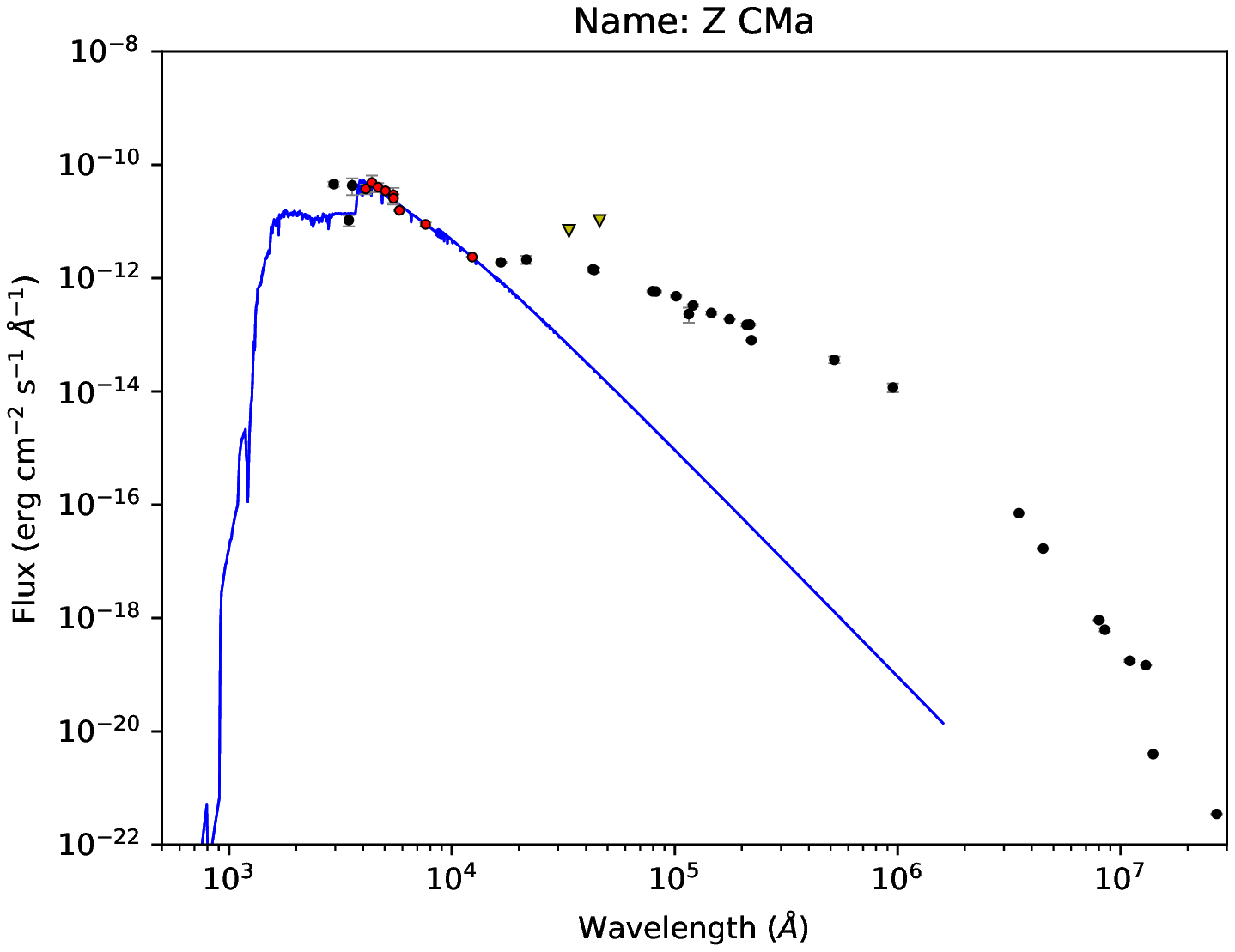}
    \includegraphics[width=0.33\textwidth]{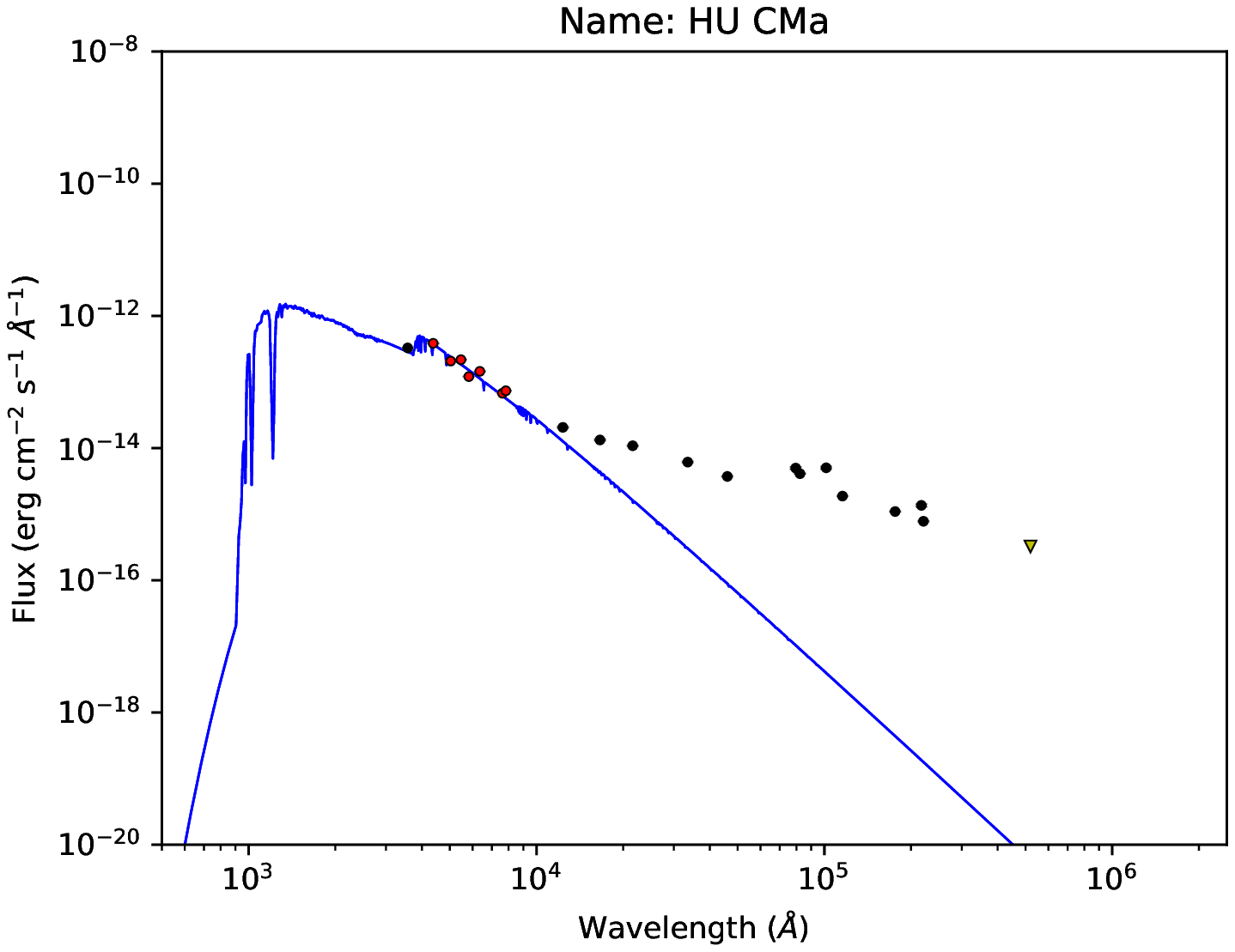}    
\end{figure}

\newpage

\onecolumn

\begin{figure} [h]
 \centering
    \includegraphics[width=0.33\textwidth]{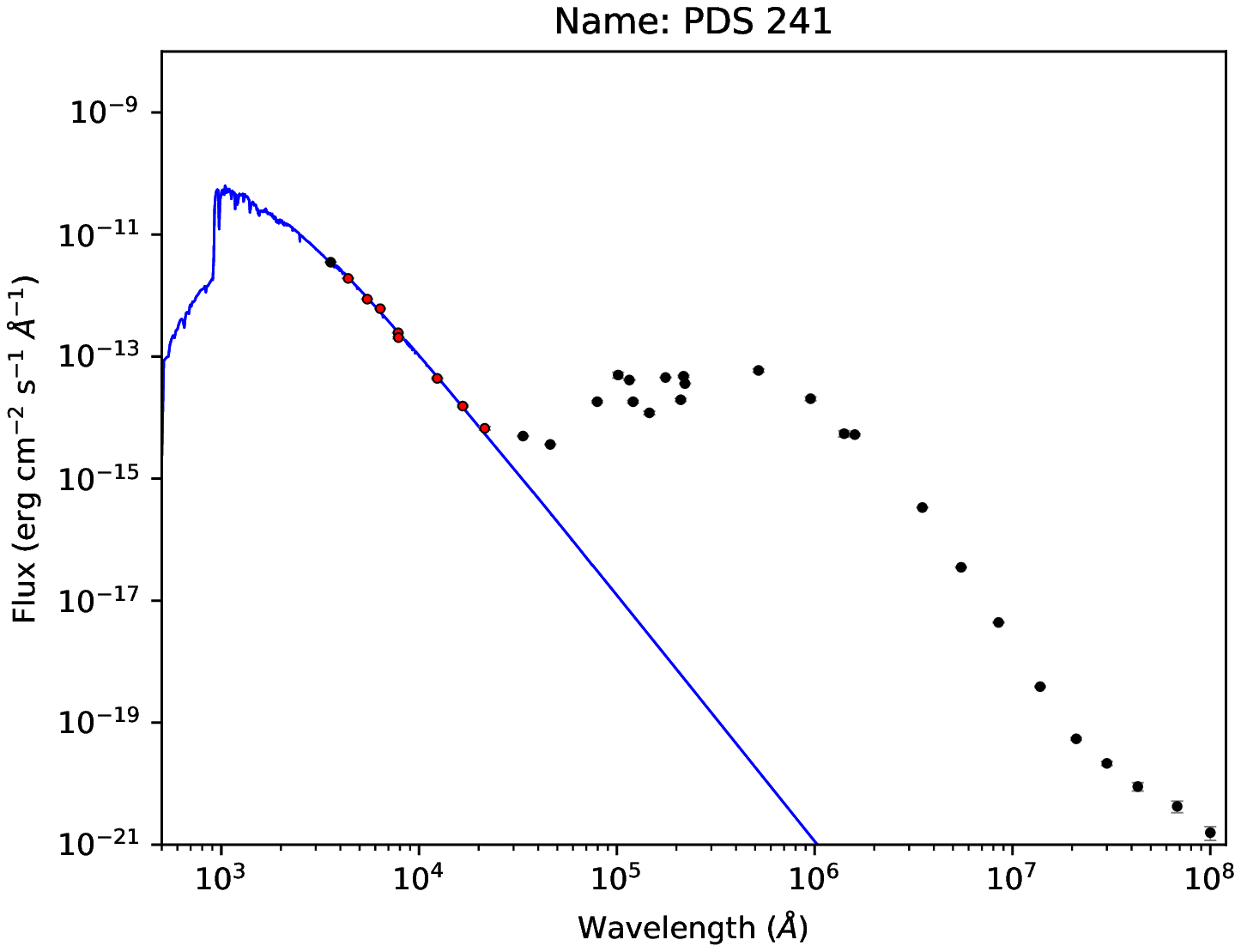}
    \includegraphics[width=0.33\textwidth]{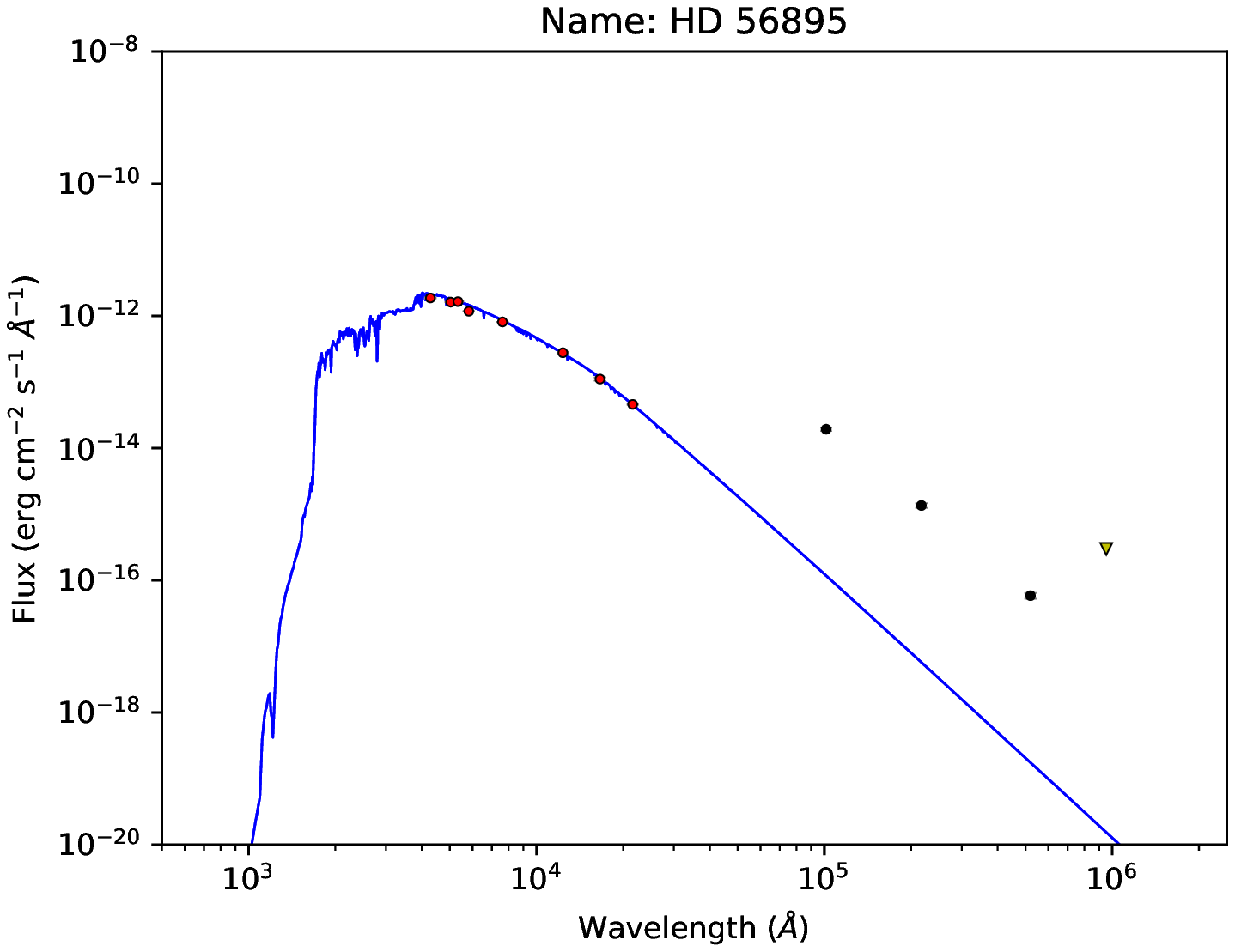}
    \includegraphics[width=0.33\textwidth]{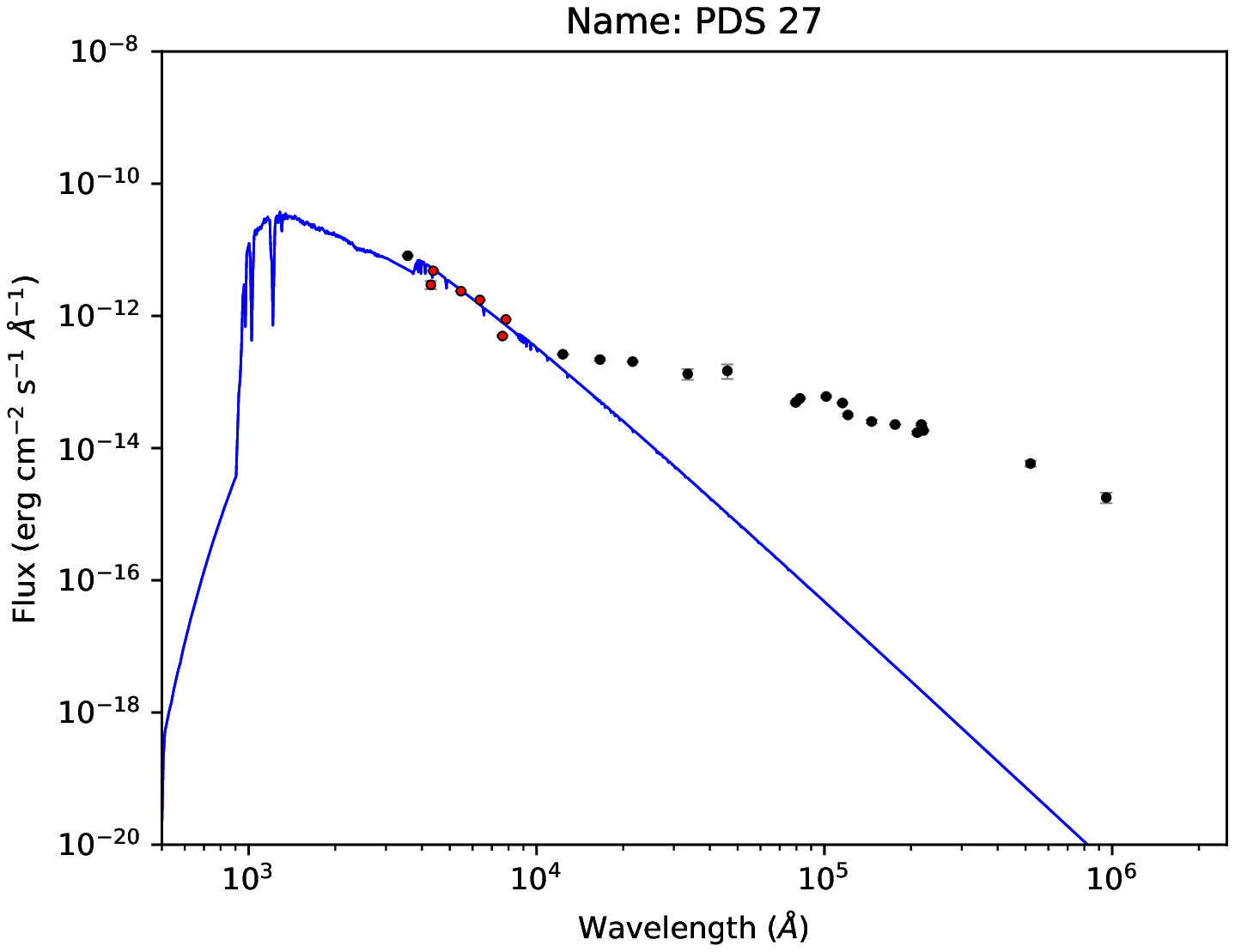}    
\end{figure}

\begin{figure} [h]
 \centering
    \includegraphics[width=0.33\textwidth]{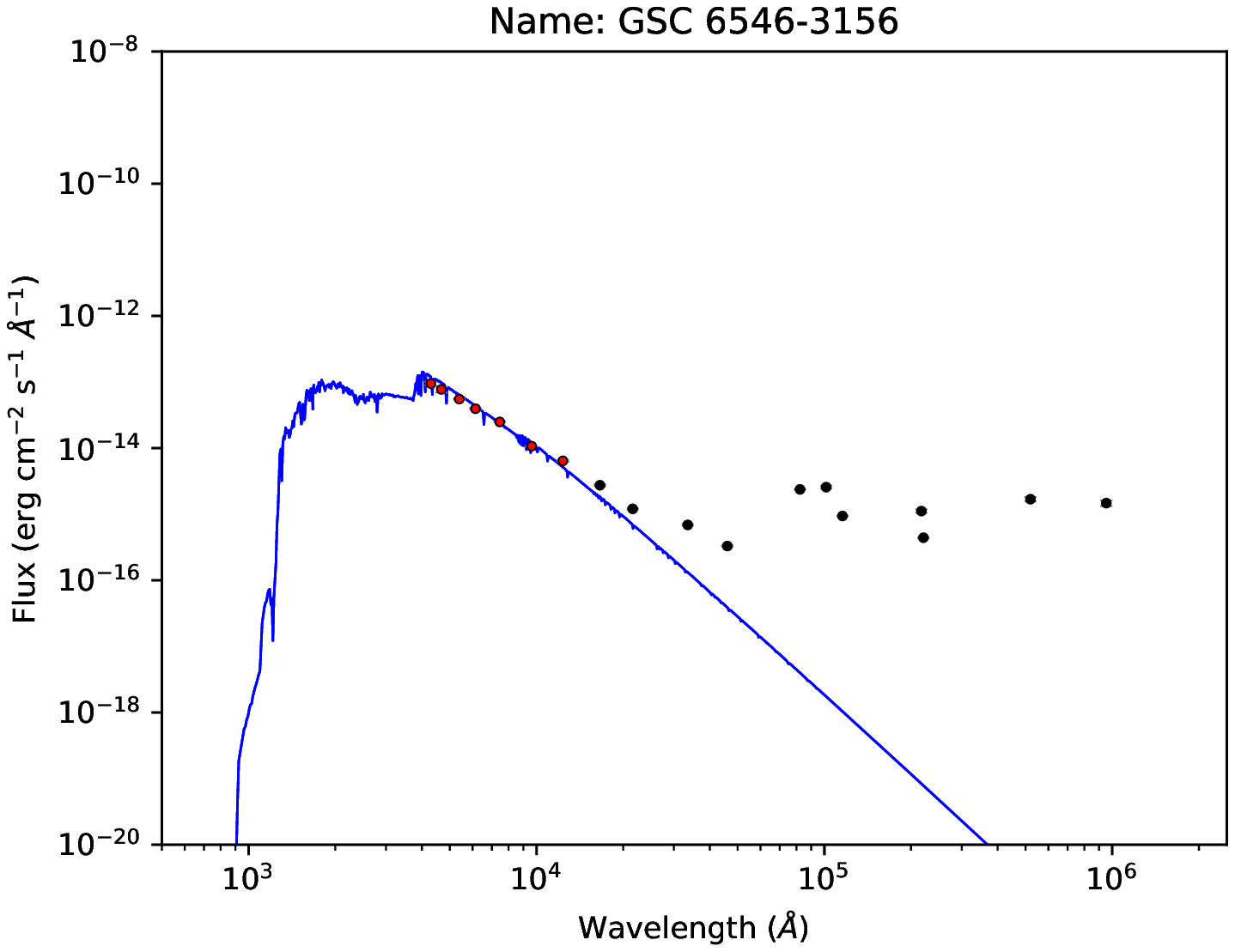}
    \includegraphics[width=0.33\textwidth]{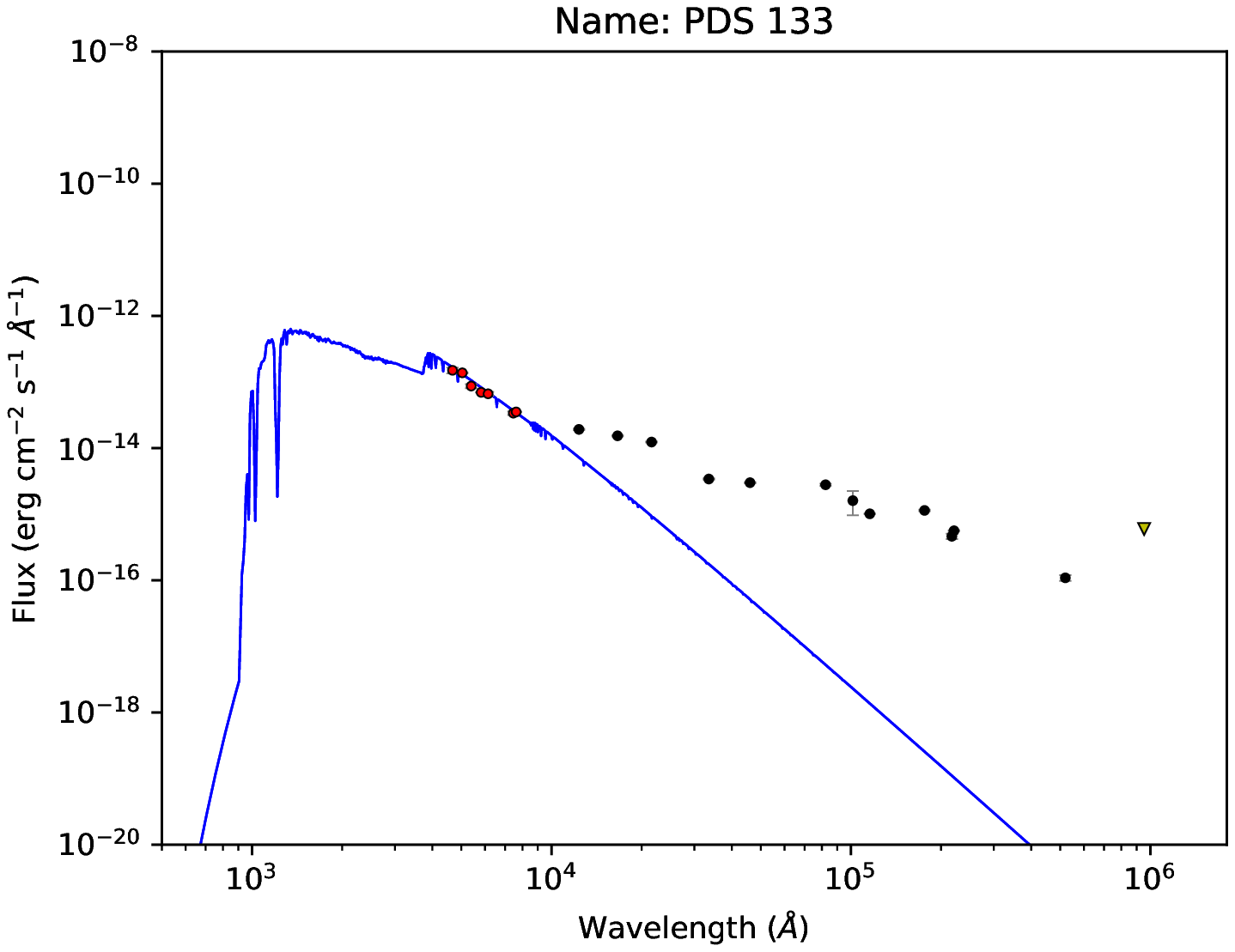}    
    \includegraphics[width=0.33\textwidth]{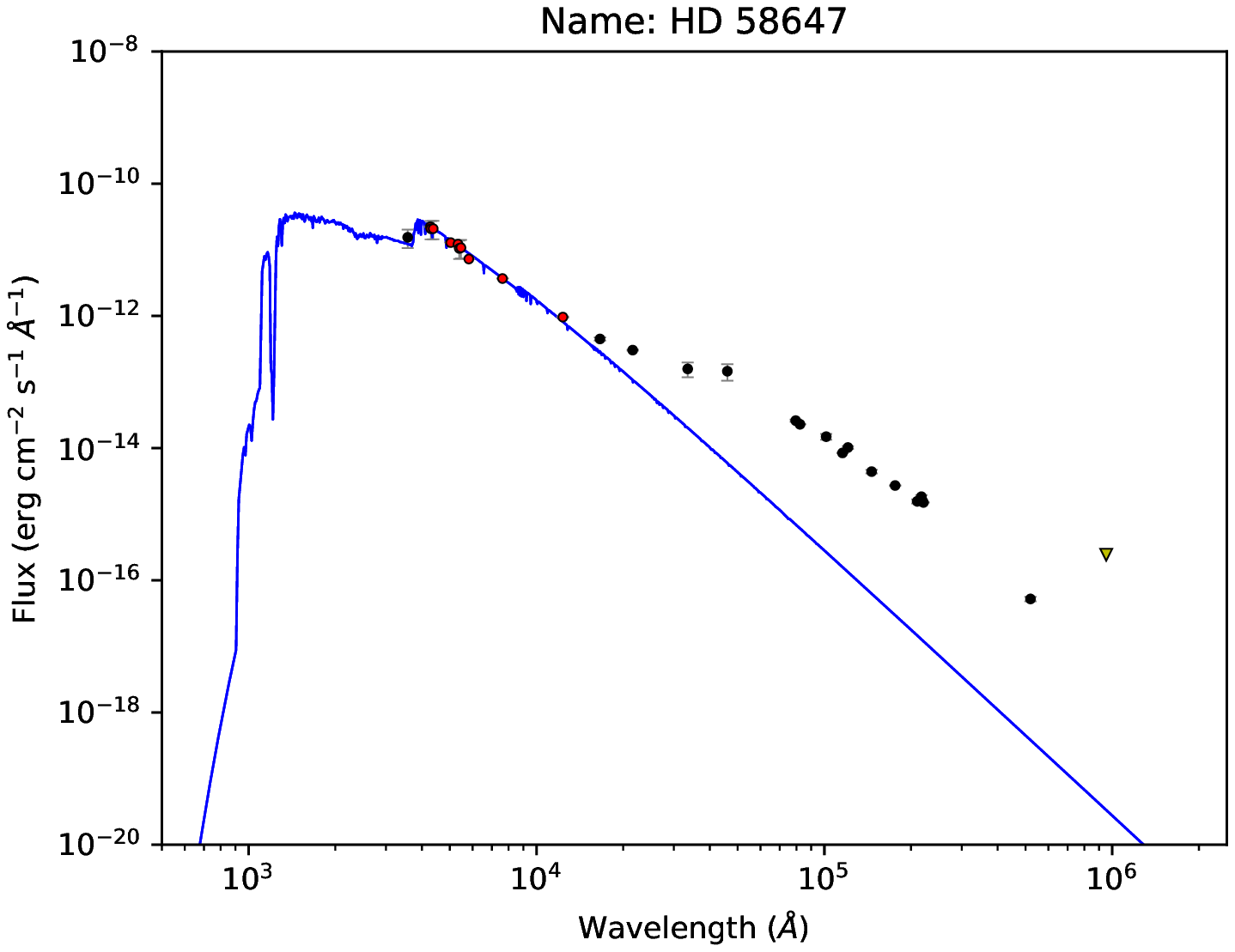}
\end{figure}

\begin{figure} [h]
 \centering
    \includegraphics[width=0.33\textwidth]{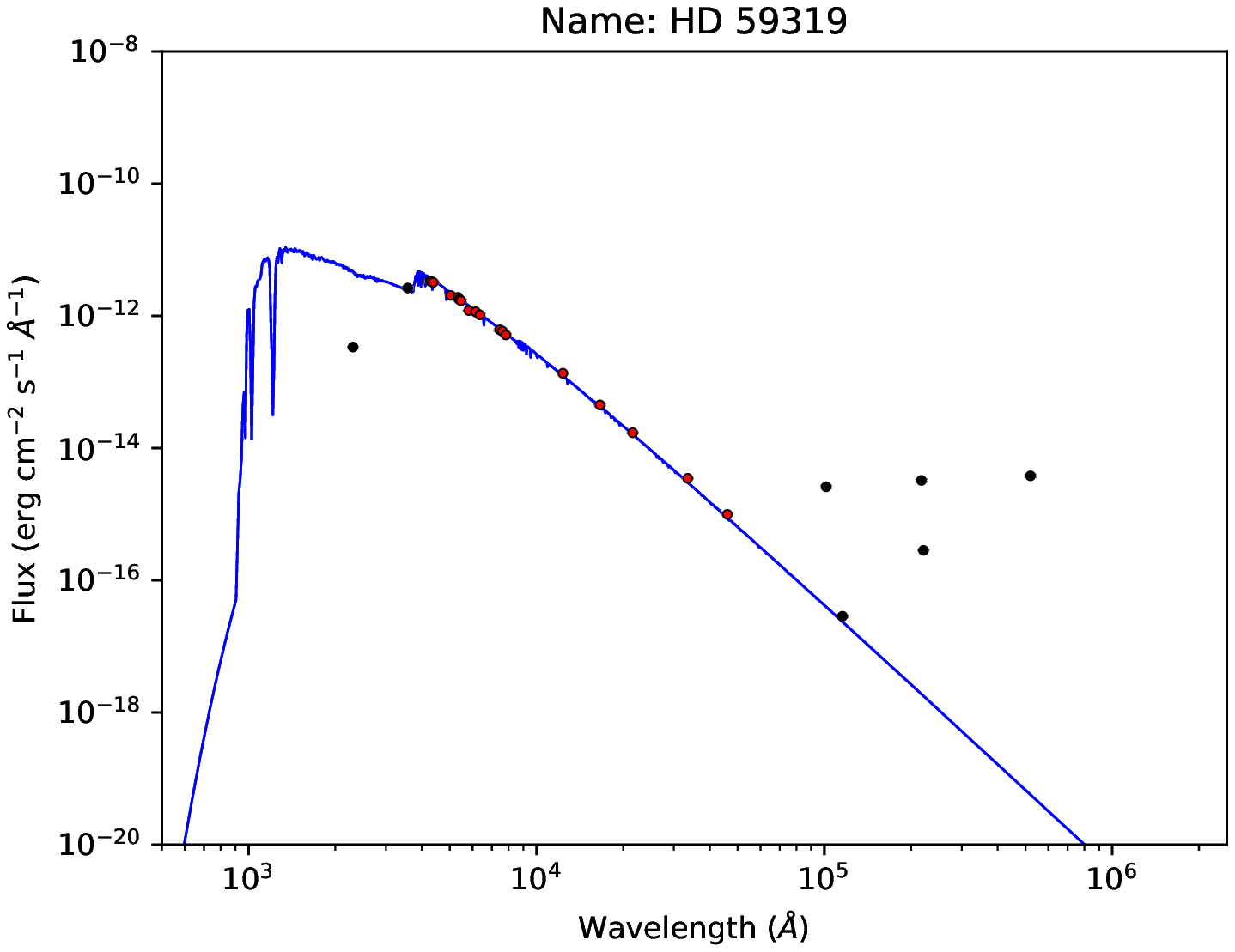} 
    \includegraphics[width=0.33\textwidth]{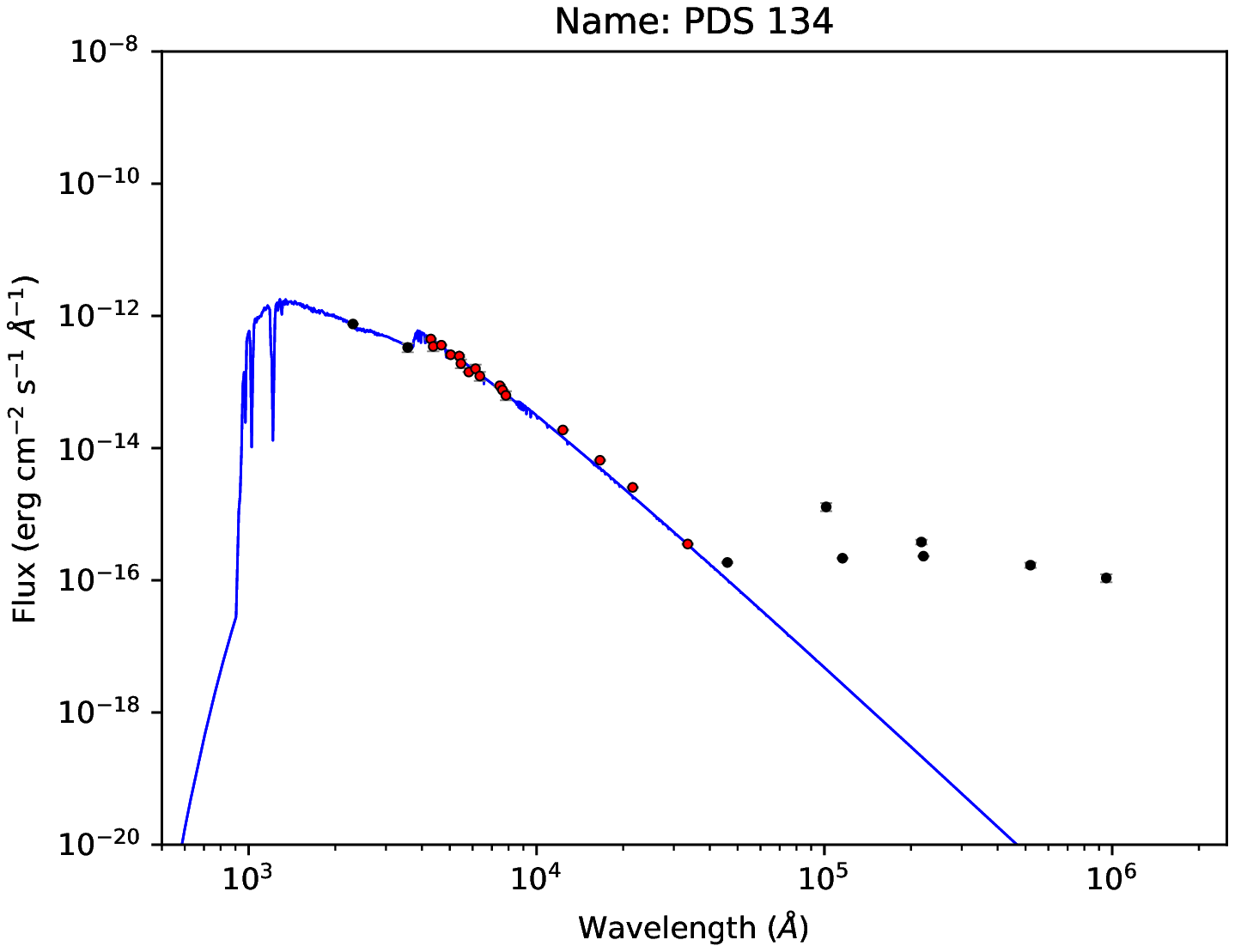}
    \includegraphics[width=0.33\textwidth]{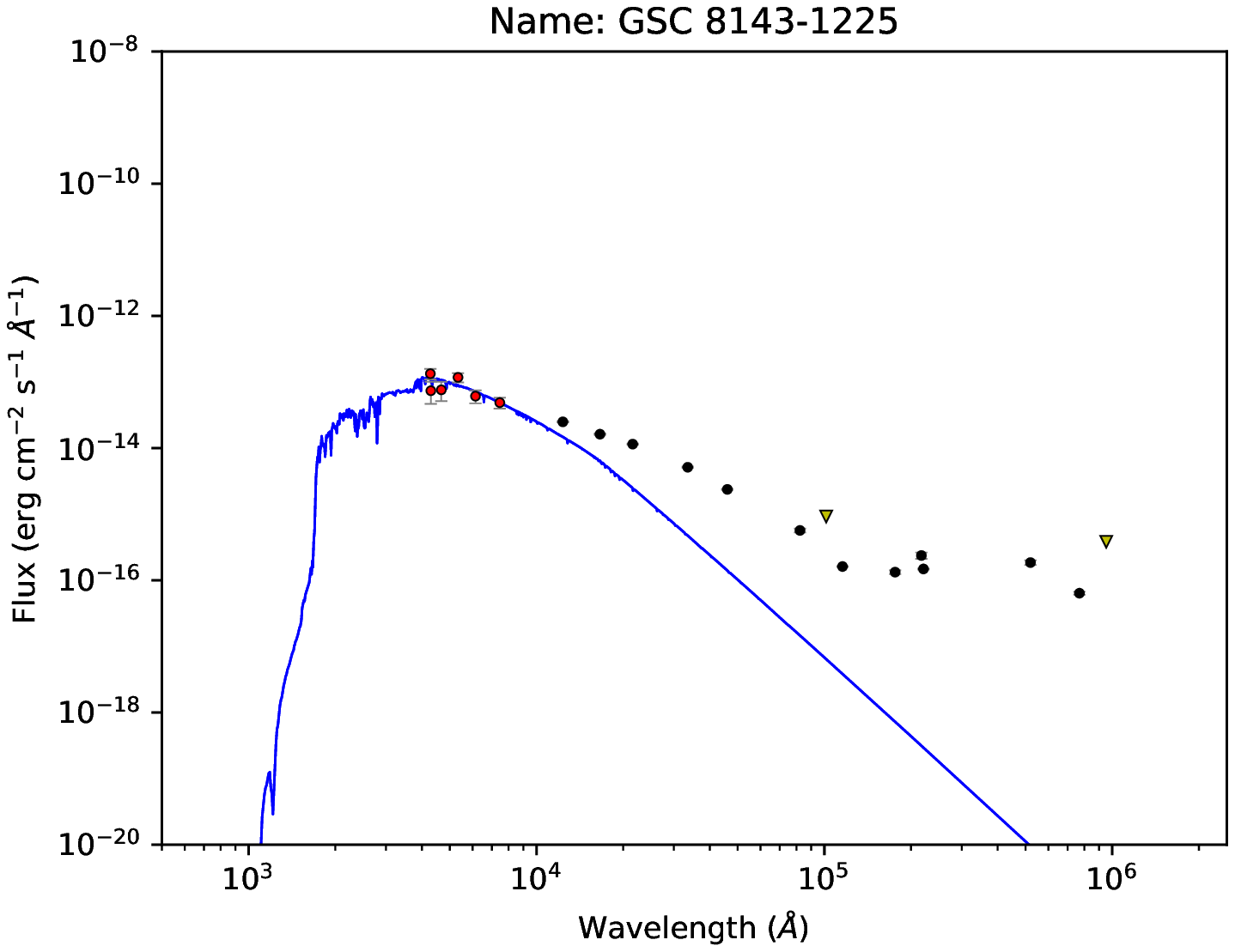}
\end{figure}

\begin{figure} [h]
 \centering
    \includegraphics[width=0.33\textwidth]{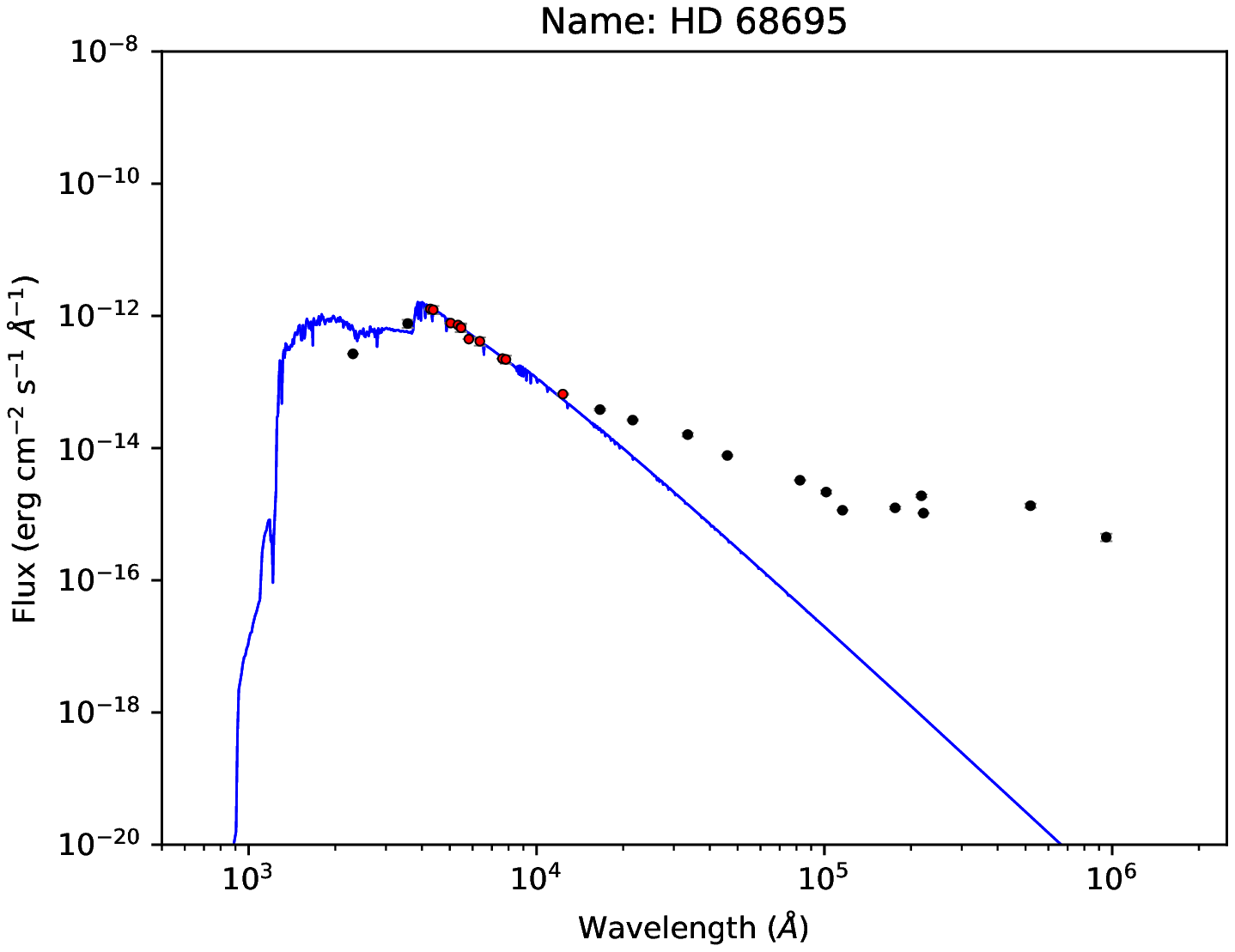}
    \includegraphics[width=0.33\textwidth]{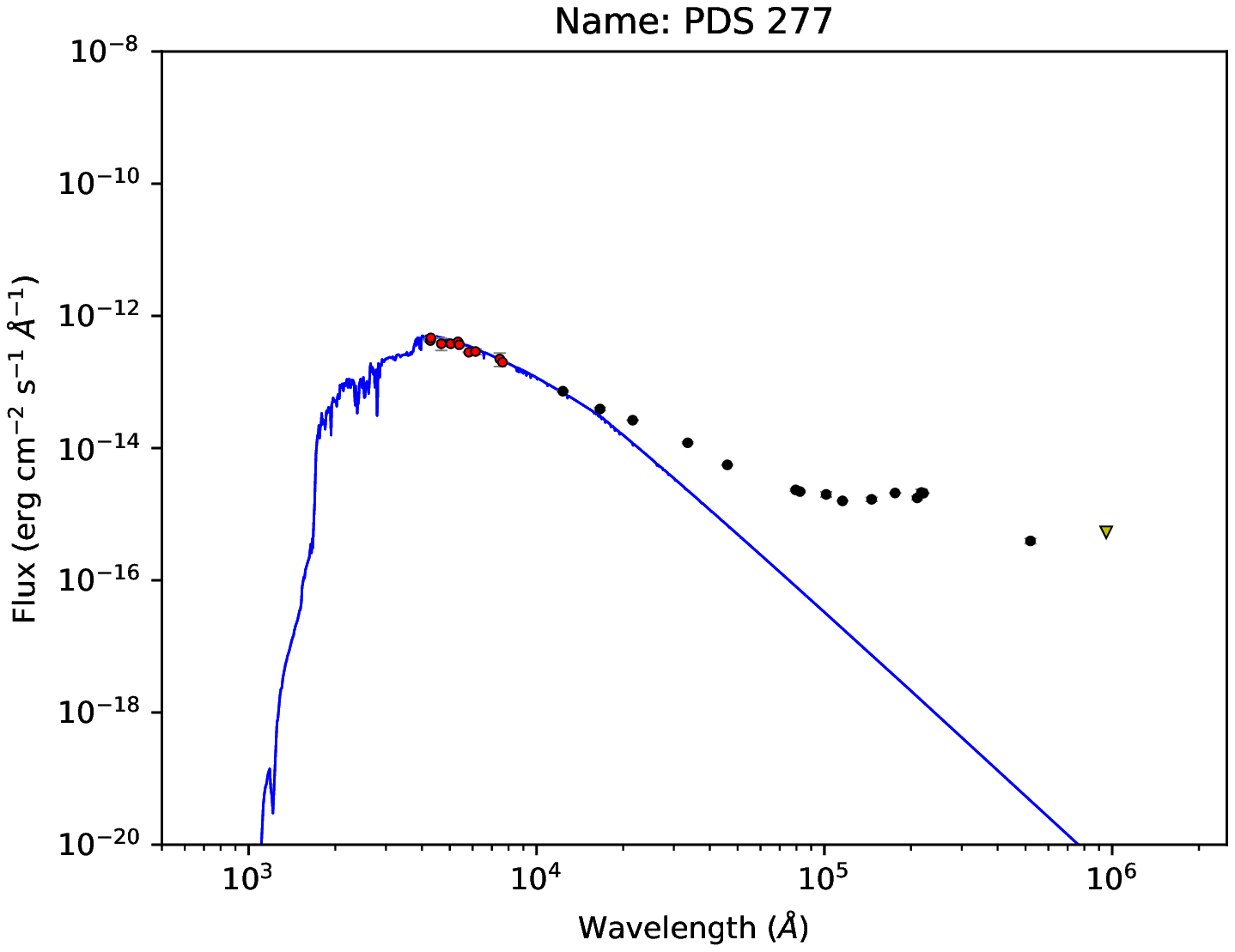}
    \includegraphics[width=0.33\textwidth]{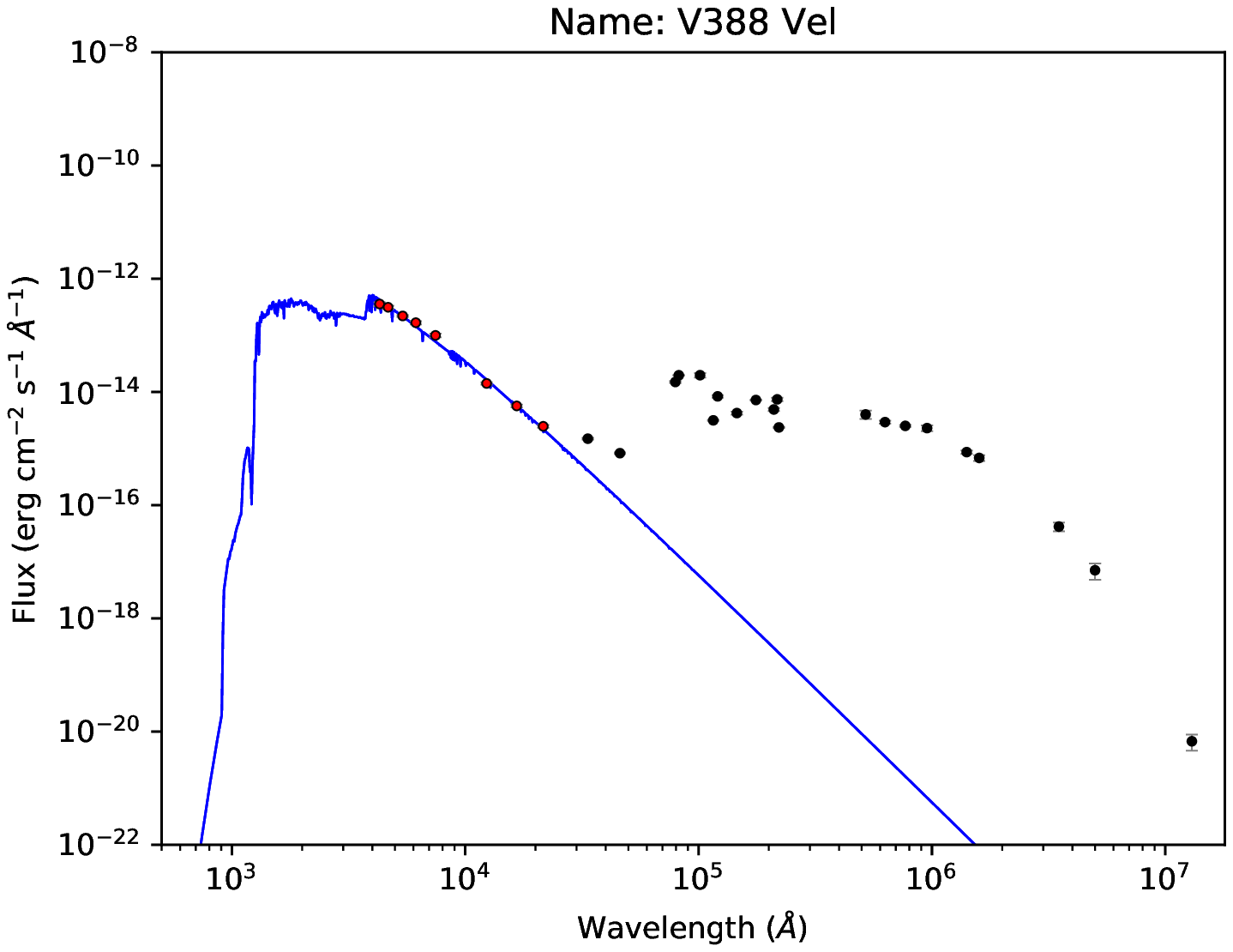}
\end{figure}

\newpage

\onecolumn

\begin{figure} [h]
 \centering
    \includegraphics[width=0.33\textwidth]{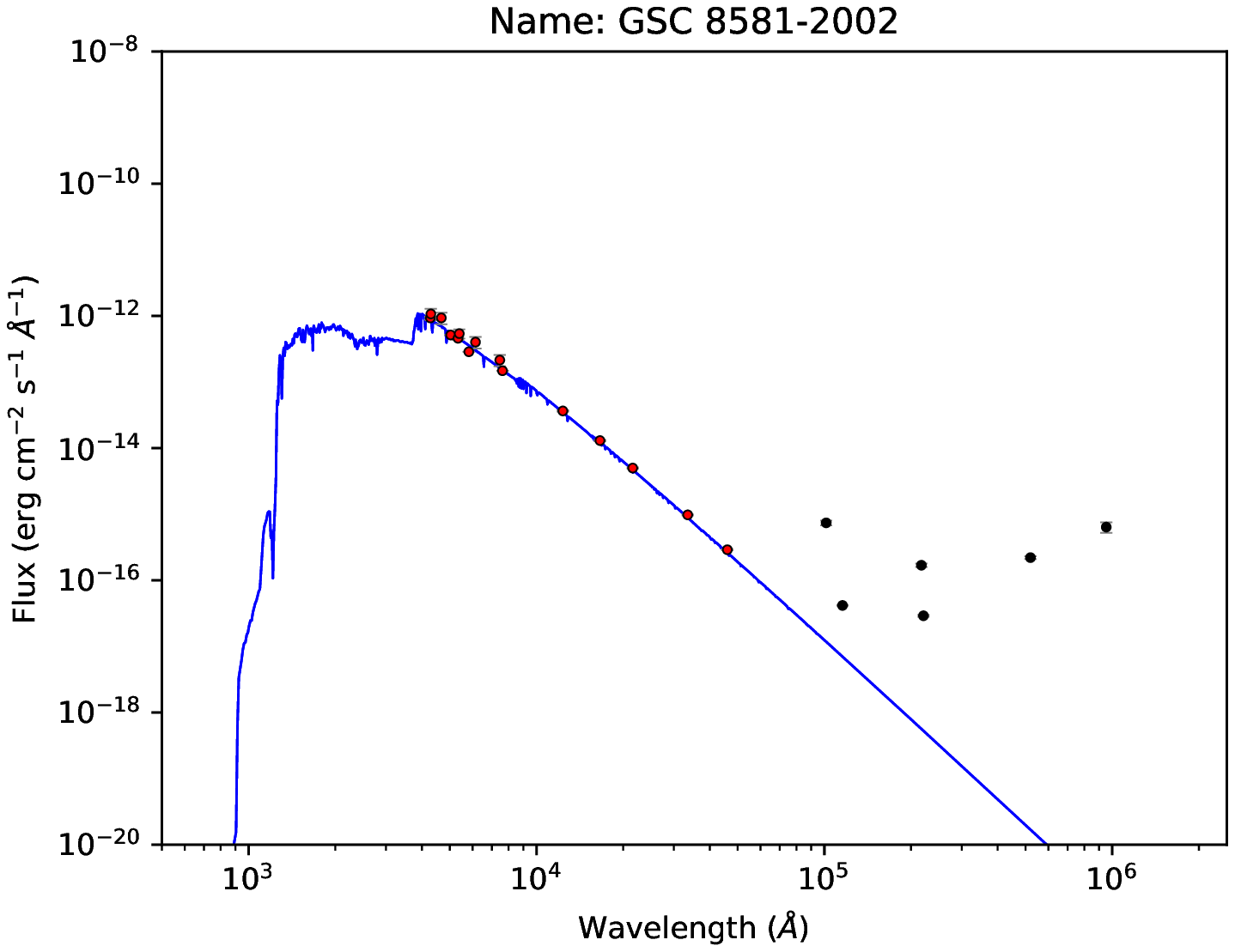}
    \includegraphics[width=0.33\textwidth]{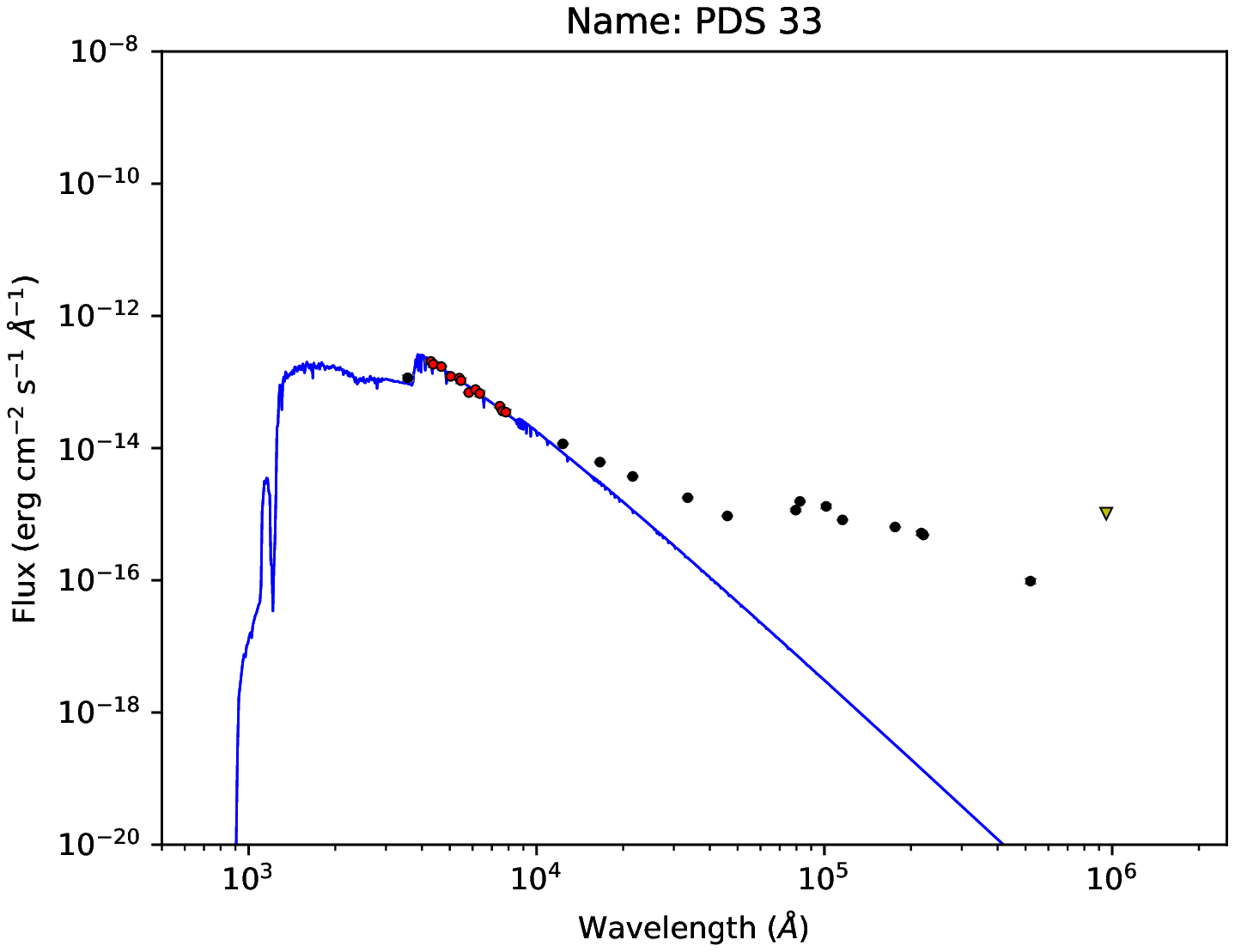}
    \includegraphics[width=0.33\textwidth]{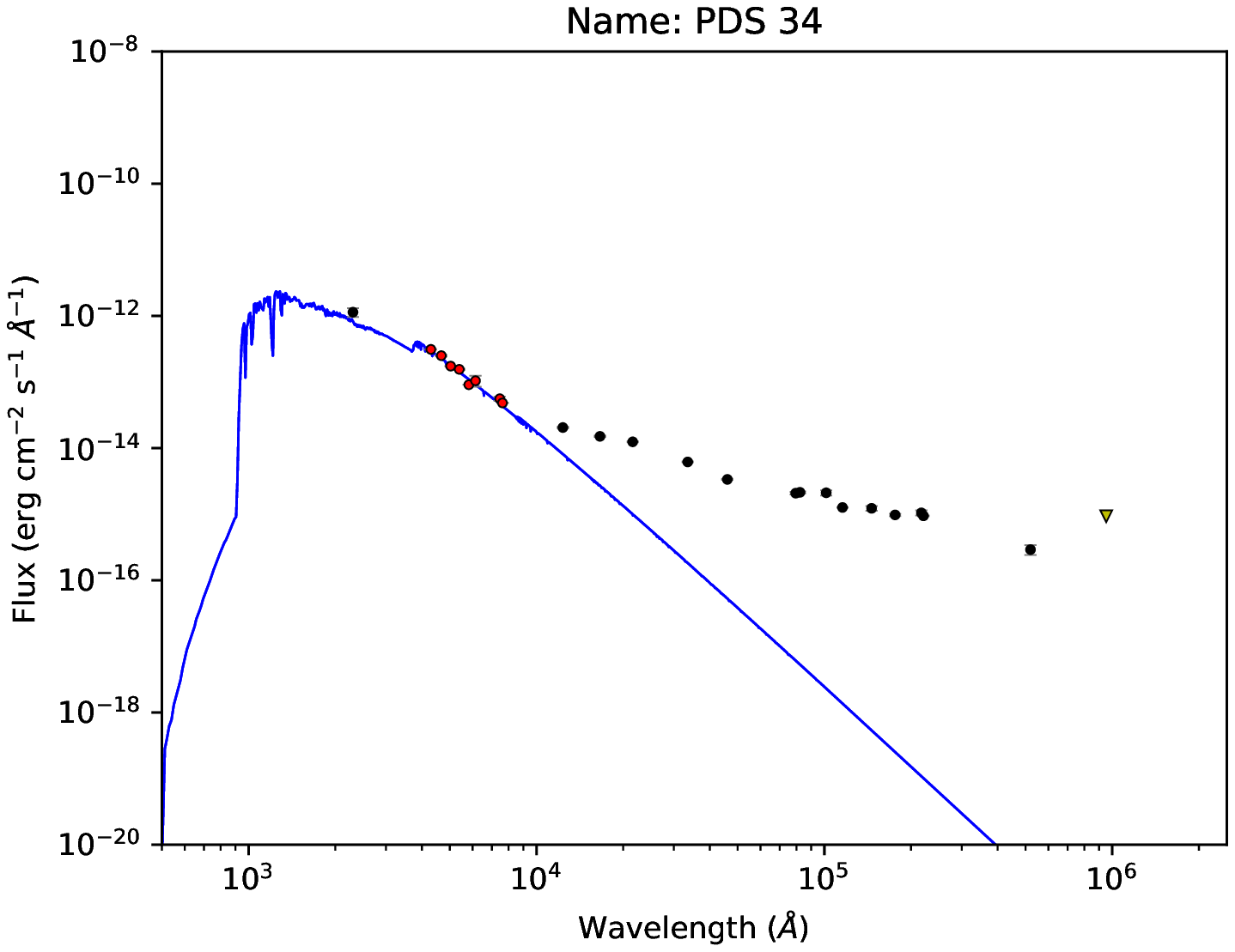}
\end{figure}

\begin{figure} [h]
 \centering
    \includegraphics[width=0.33\textwidth]{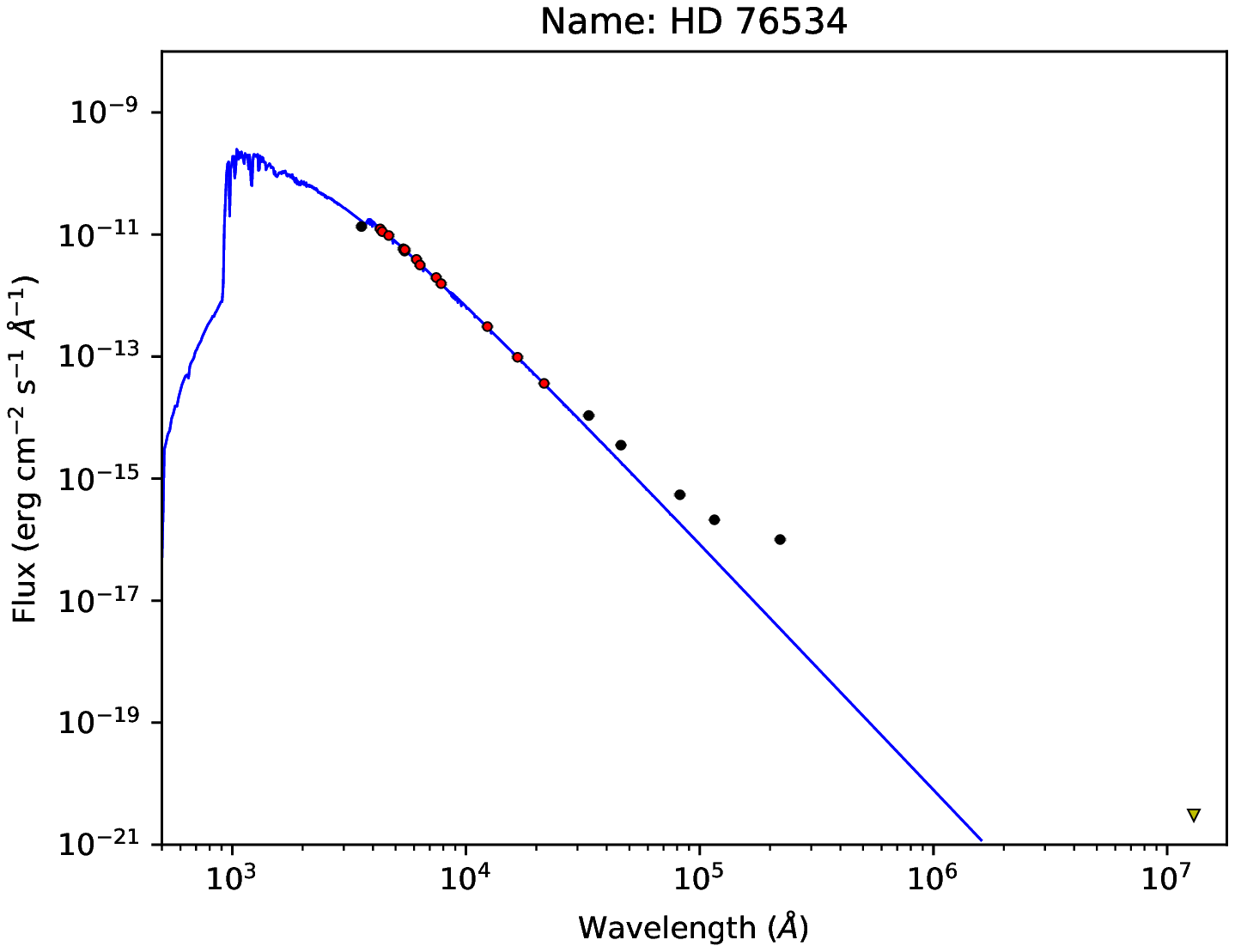}
    \includegraphics[width=0.33\textwidth]{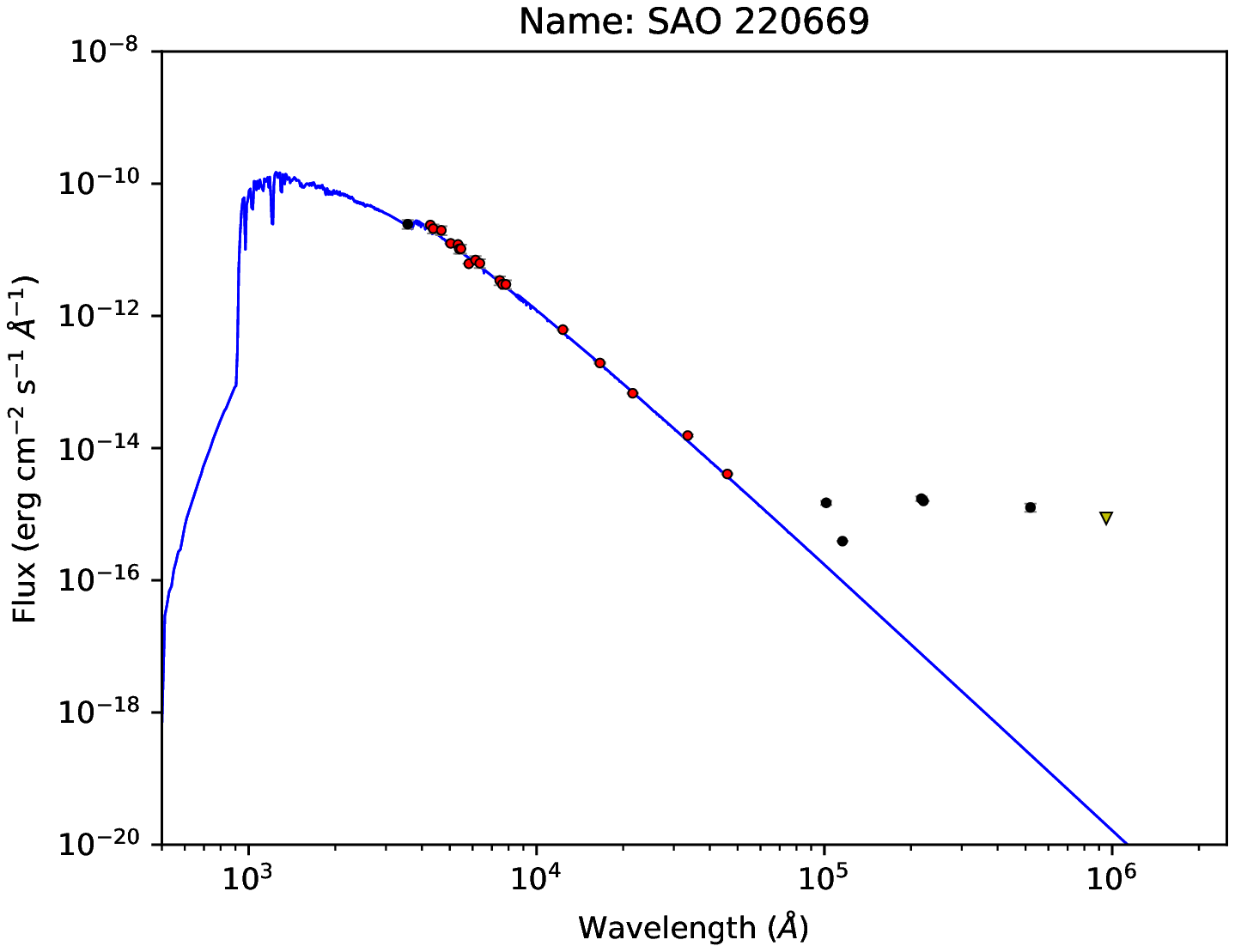}
    \includegraphics[width=0.33\textwidth]{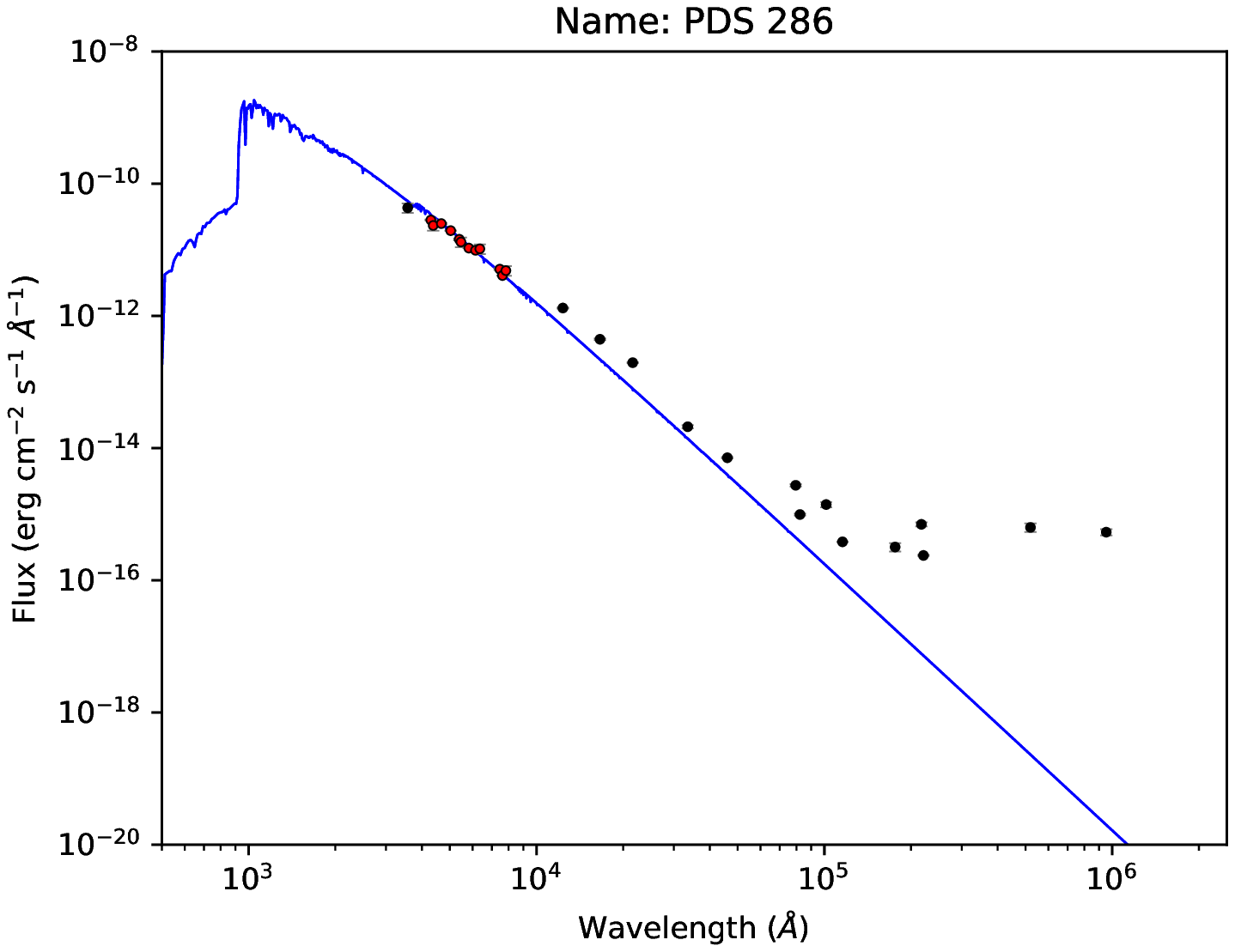}
\end{figure}

\begin{figure} [h]
 \centering
    \includegraphics[width=0.33\textwidth]{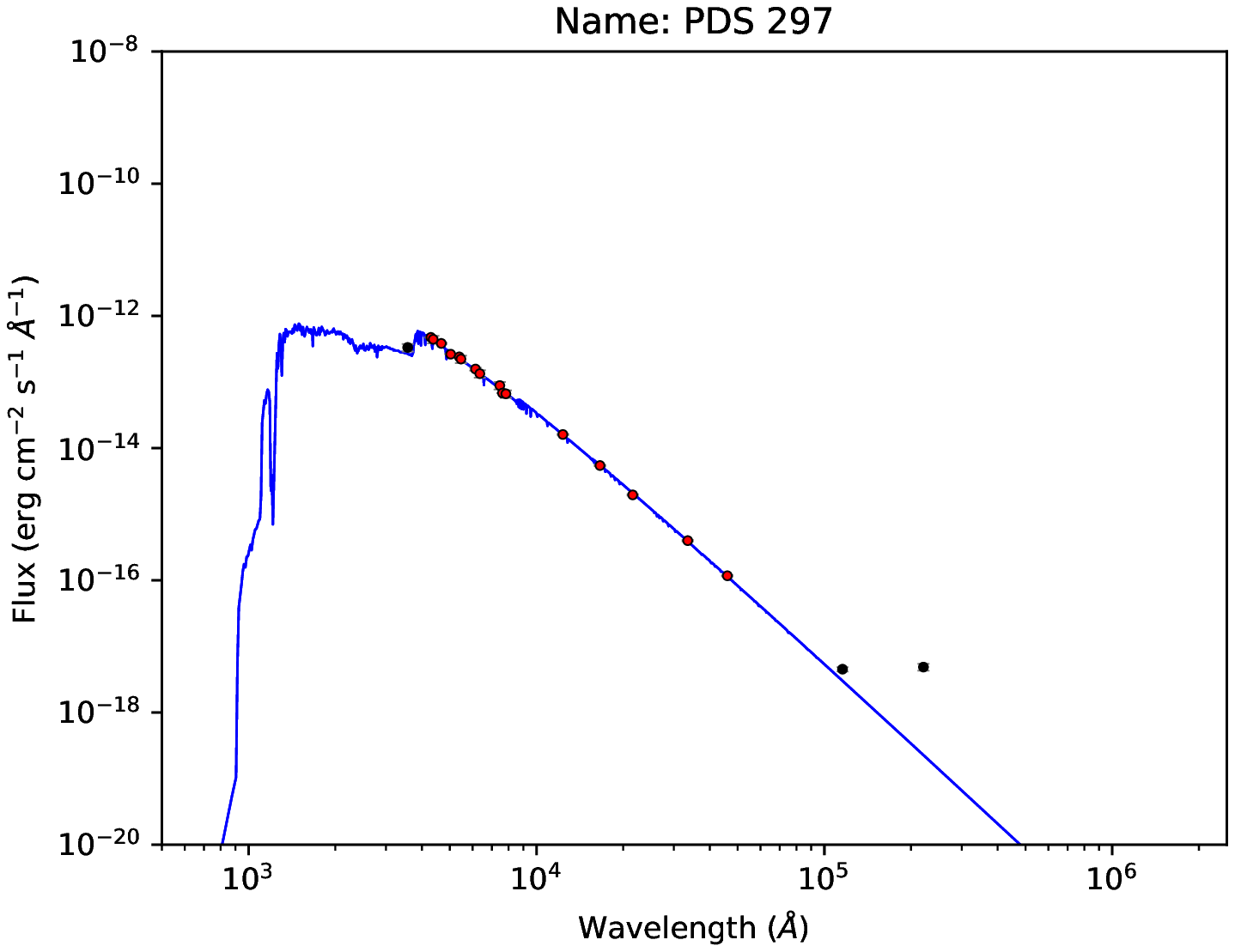}
    \includegraphics[width=0.33\textwidth]{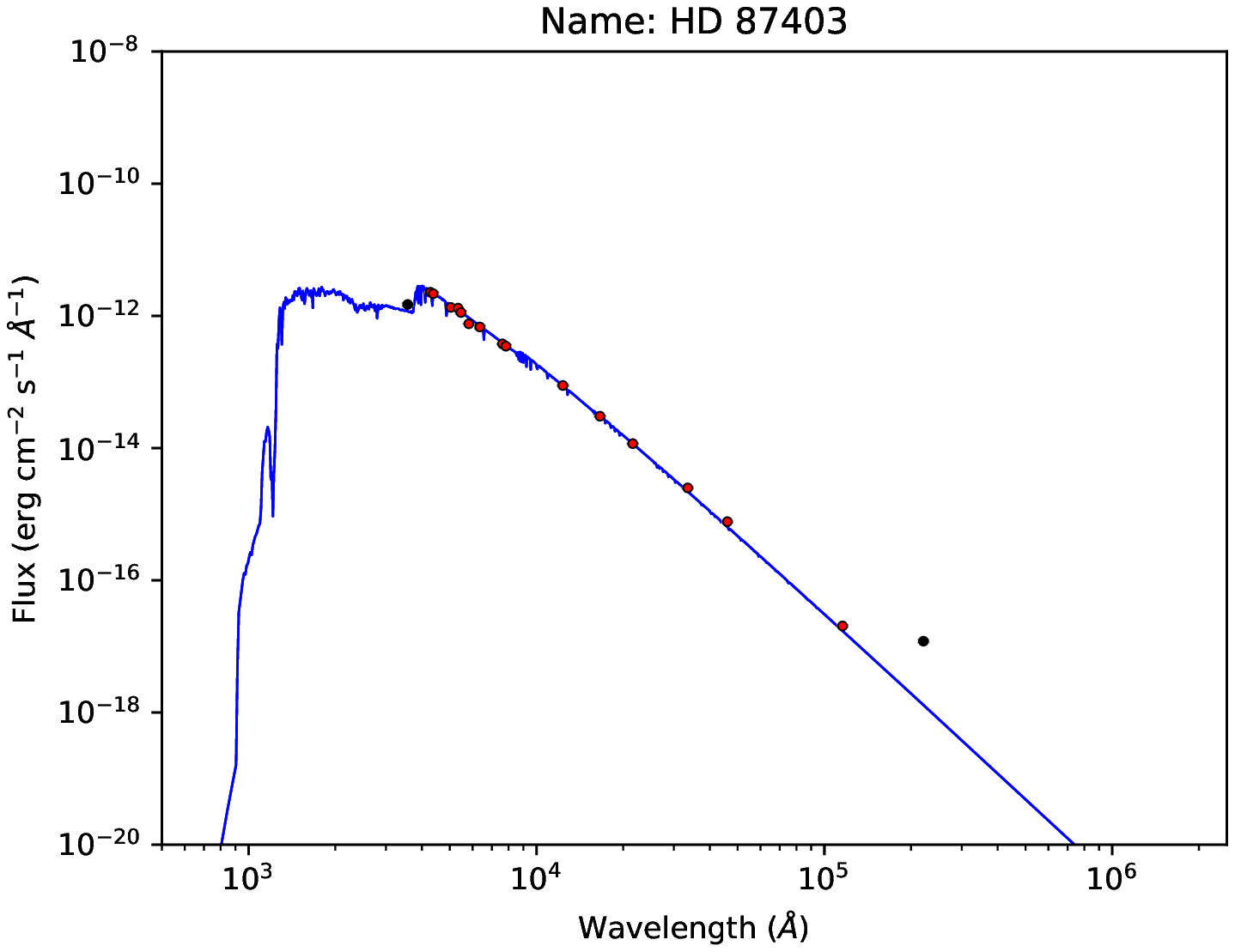}
    \includegraphics[width=0.33\textwidth]{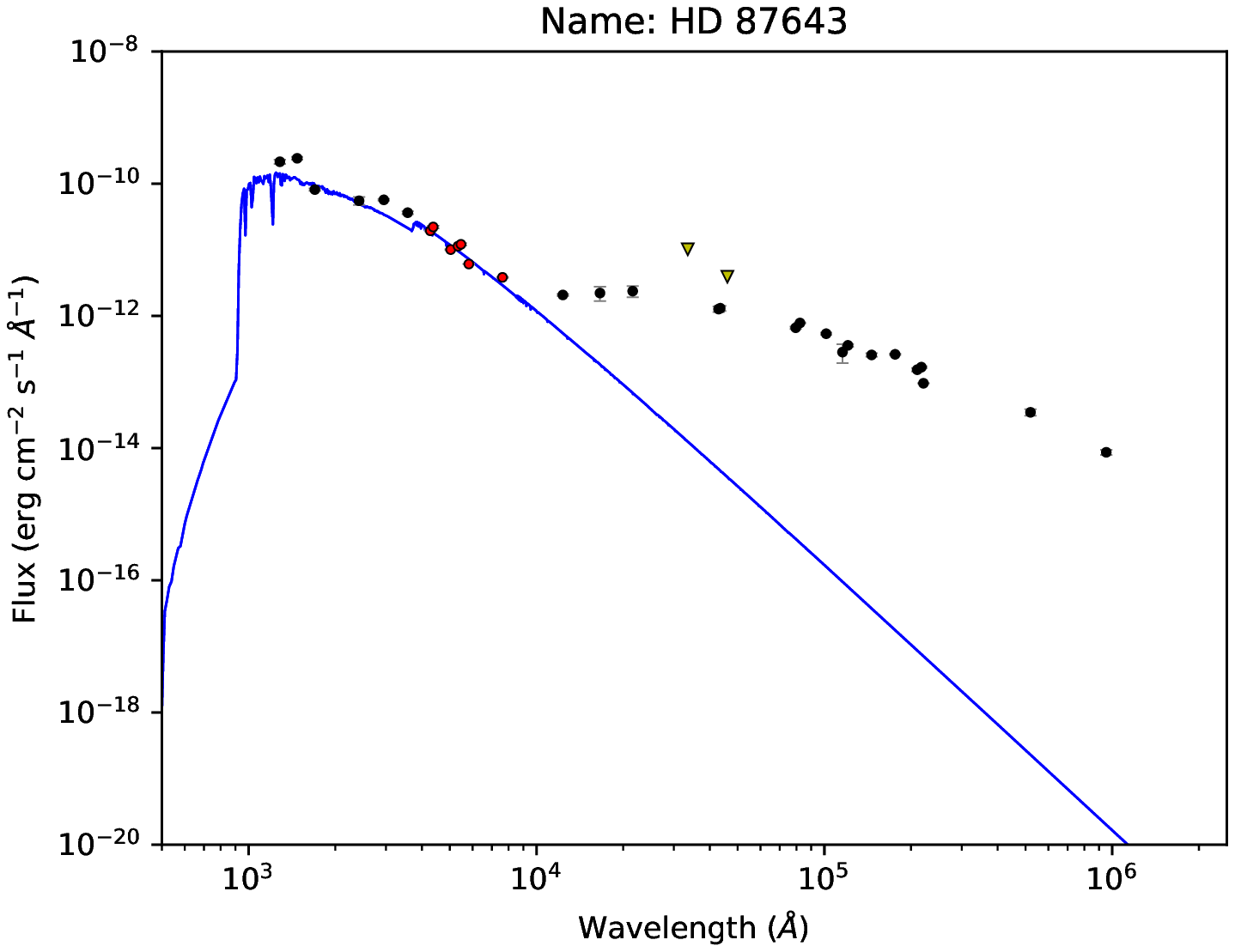}
\end{figure}

\begin{figure} [h]
 \centering
    \includegraphics[width=0.33\textwidth]{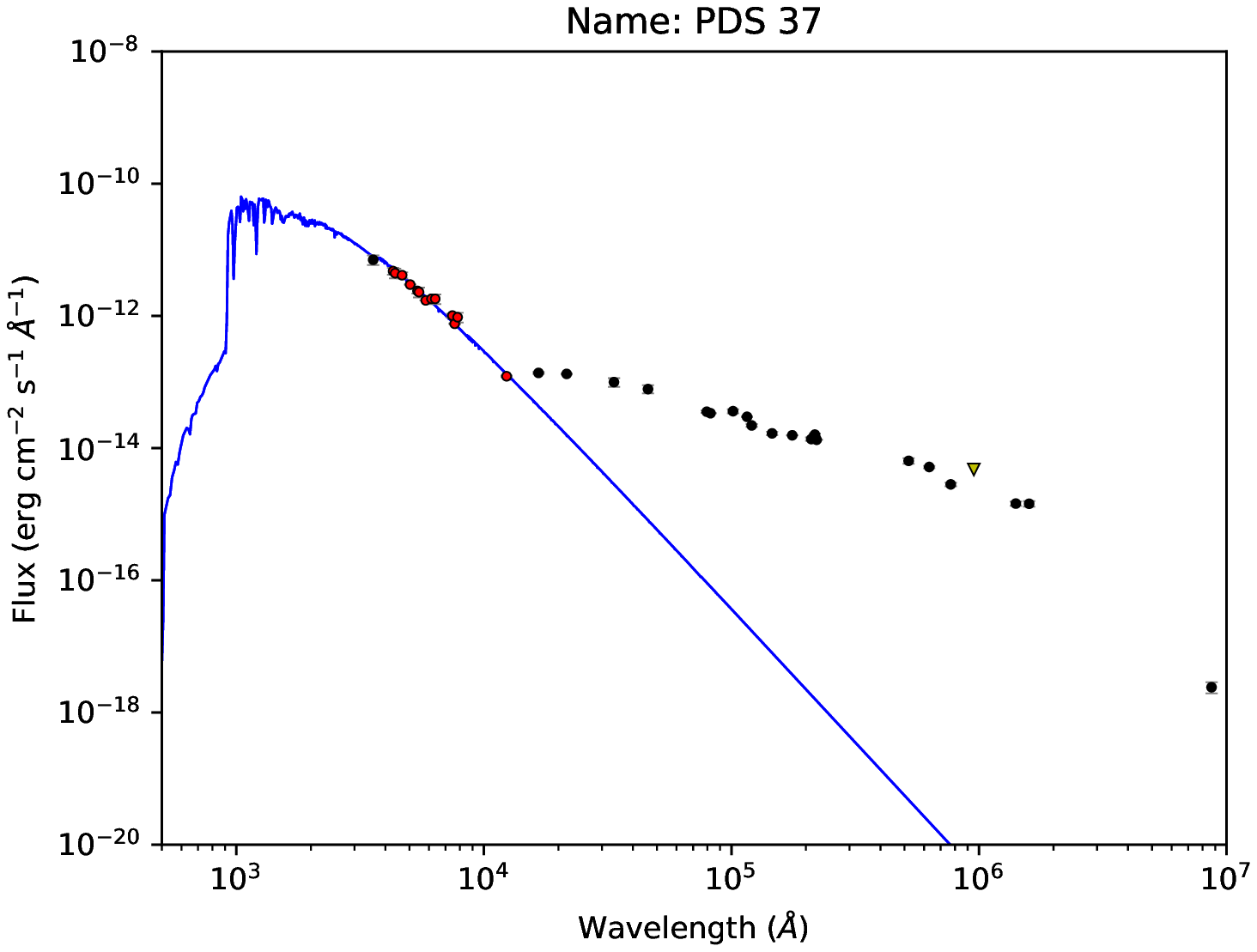}
    \includegraphics[width=0.33\textwidth]{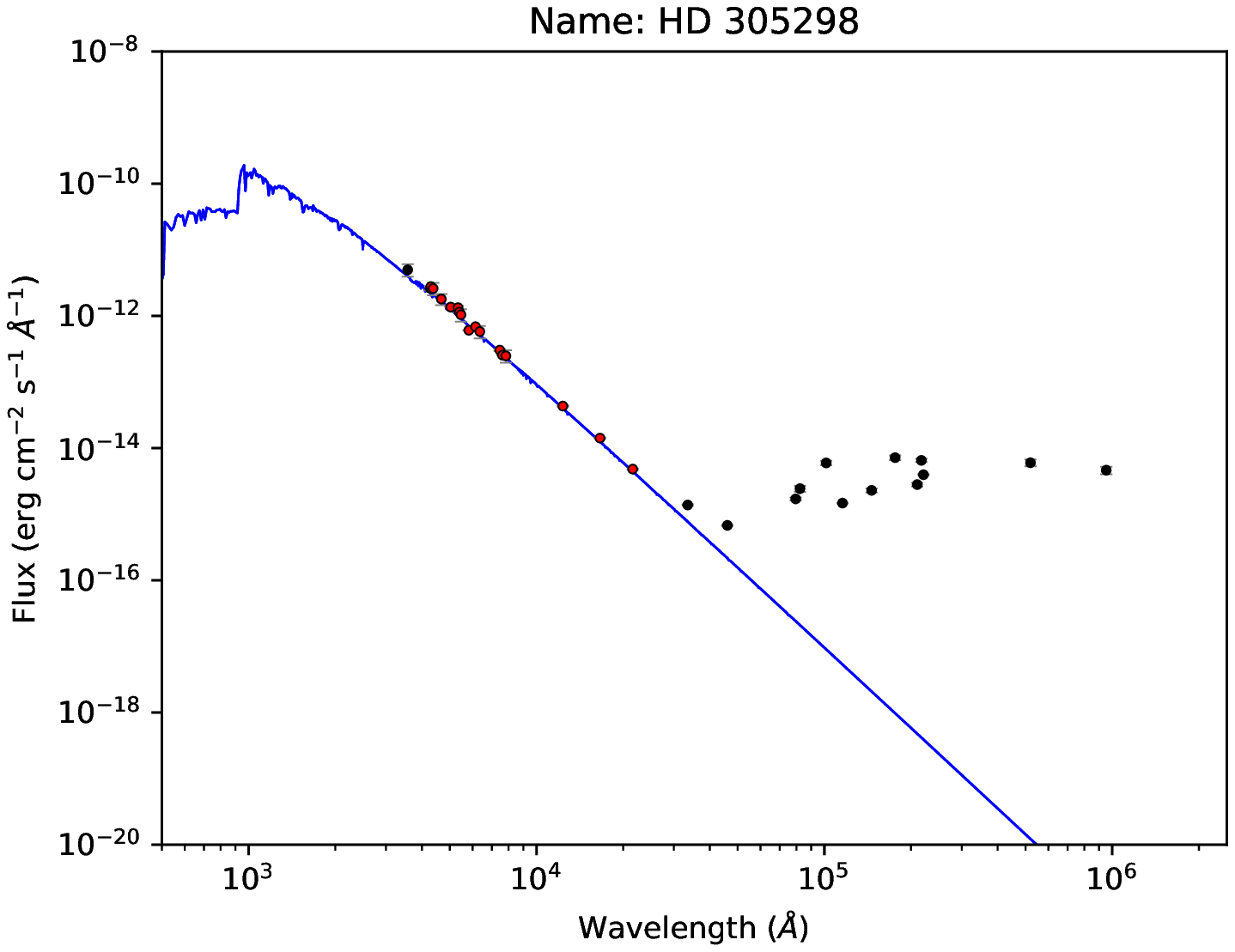}
    \includegraphics[width=0.33\textwidth]{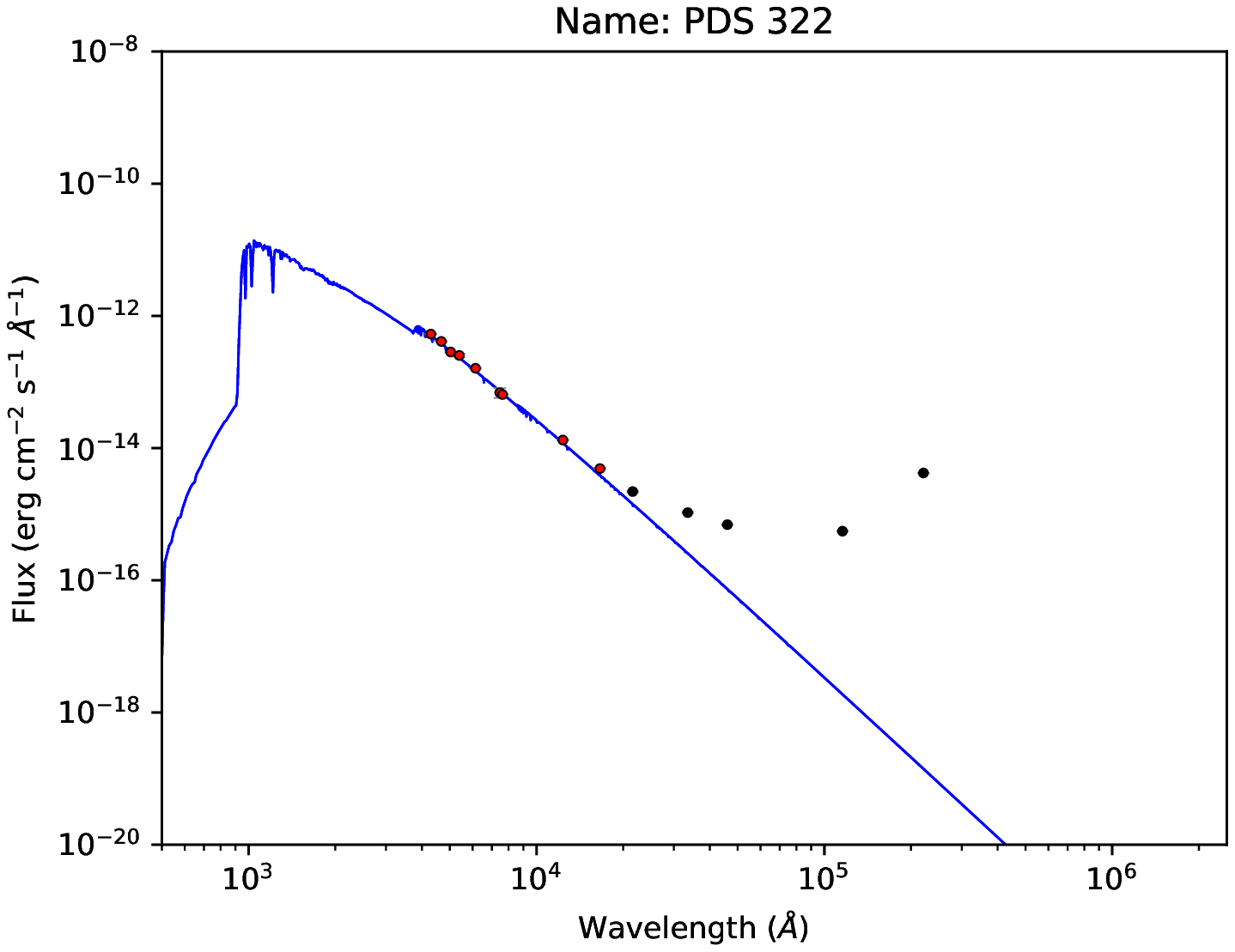}
\end{figure}

\newpage

\onecolumn

\begin{figure} [h]
 \centering
    \includegraphics[width=0.33\textwidth]{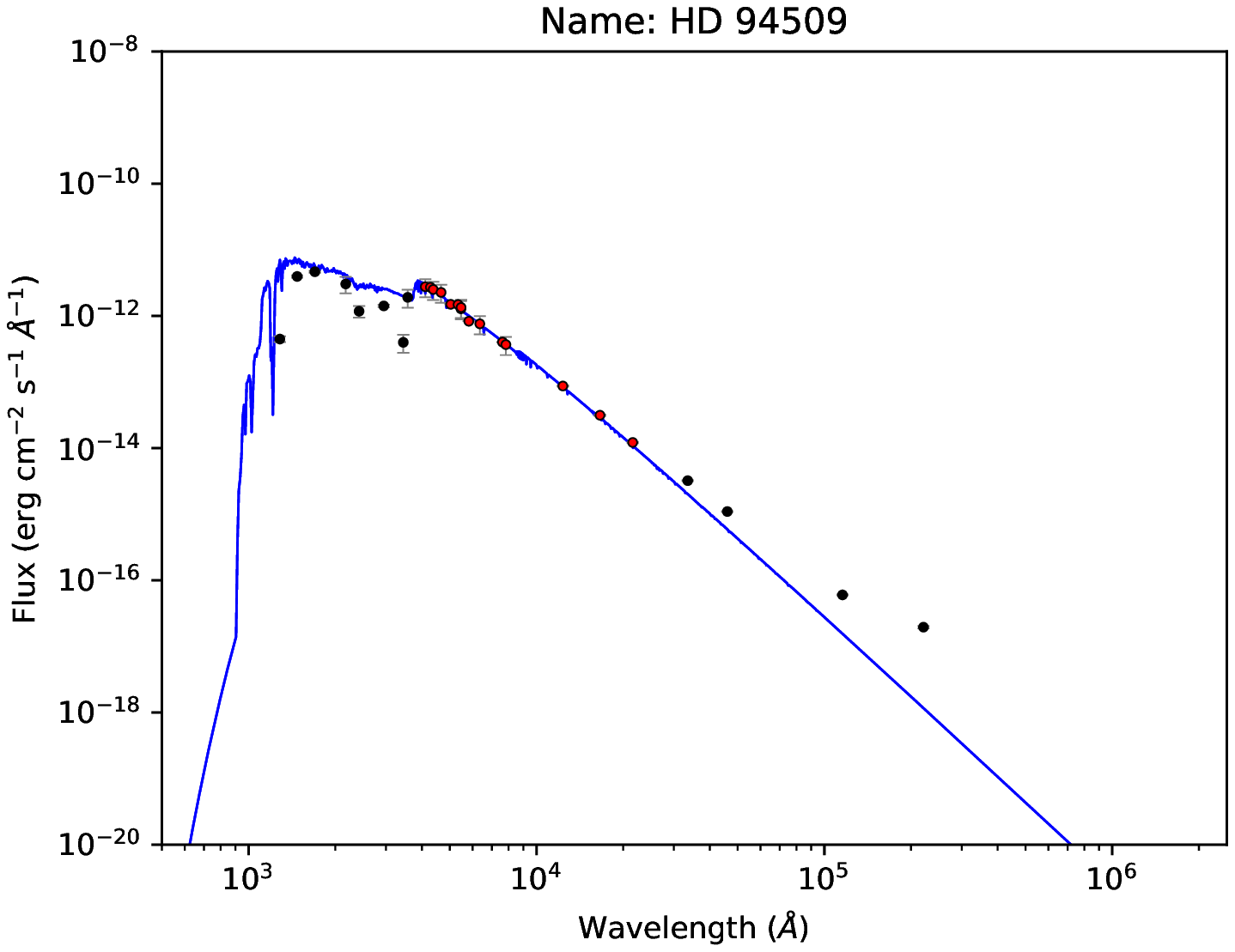}
    \includegraphics[width=0.33\textwidth]{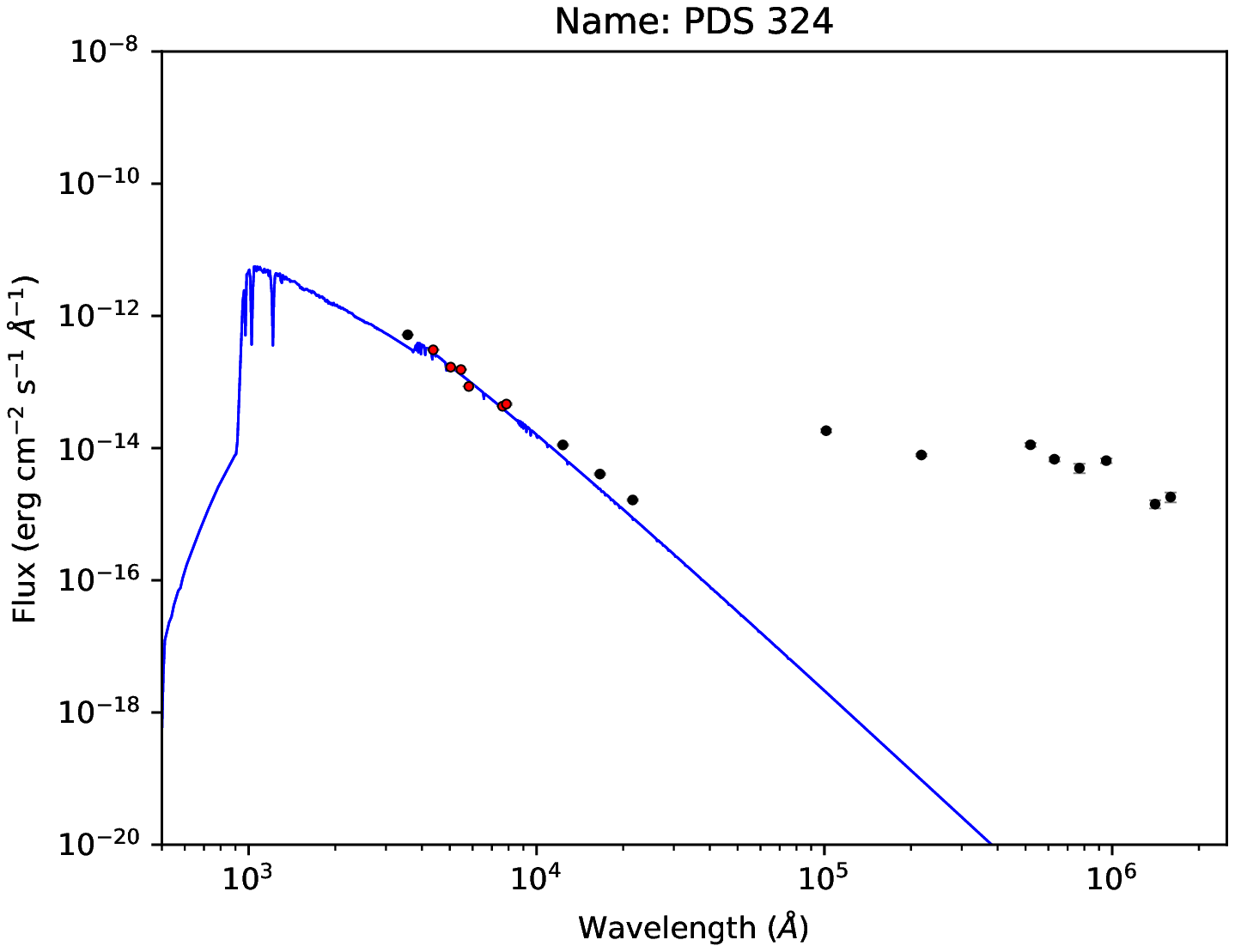}
    \includegraphics[width=0.33\textwidth]{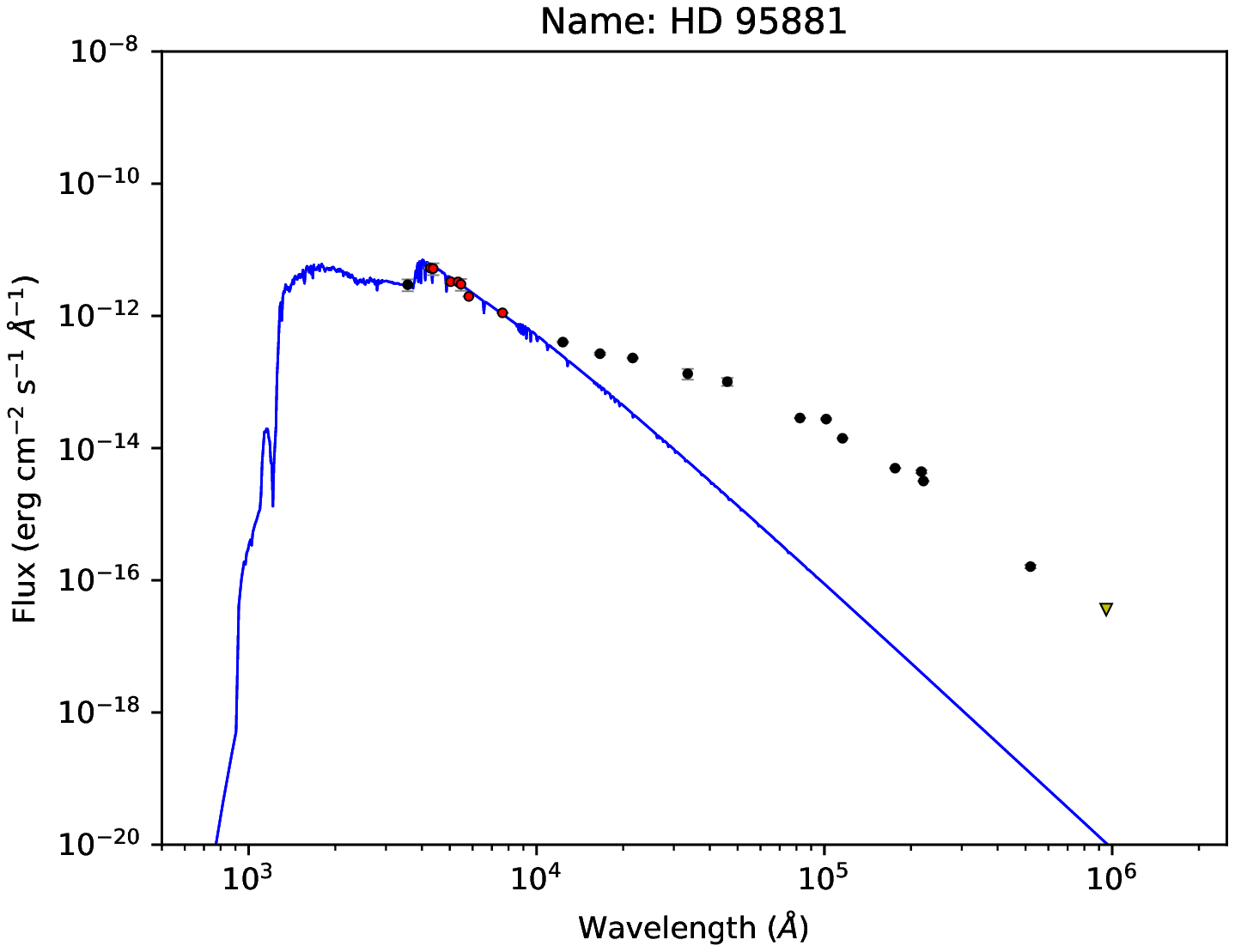}
\end{figure}

\begin{figure} [h]
 \centering
    \includegraphics[width=0.33\textwidth]{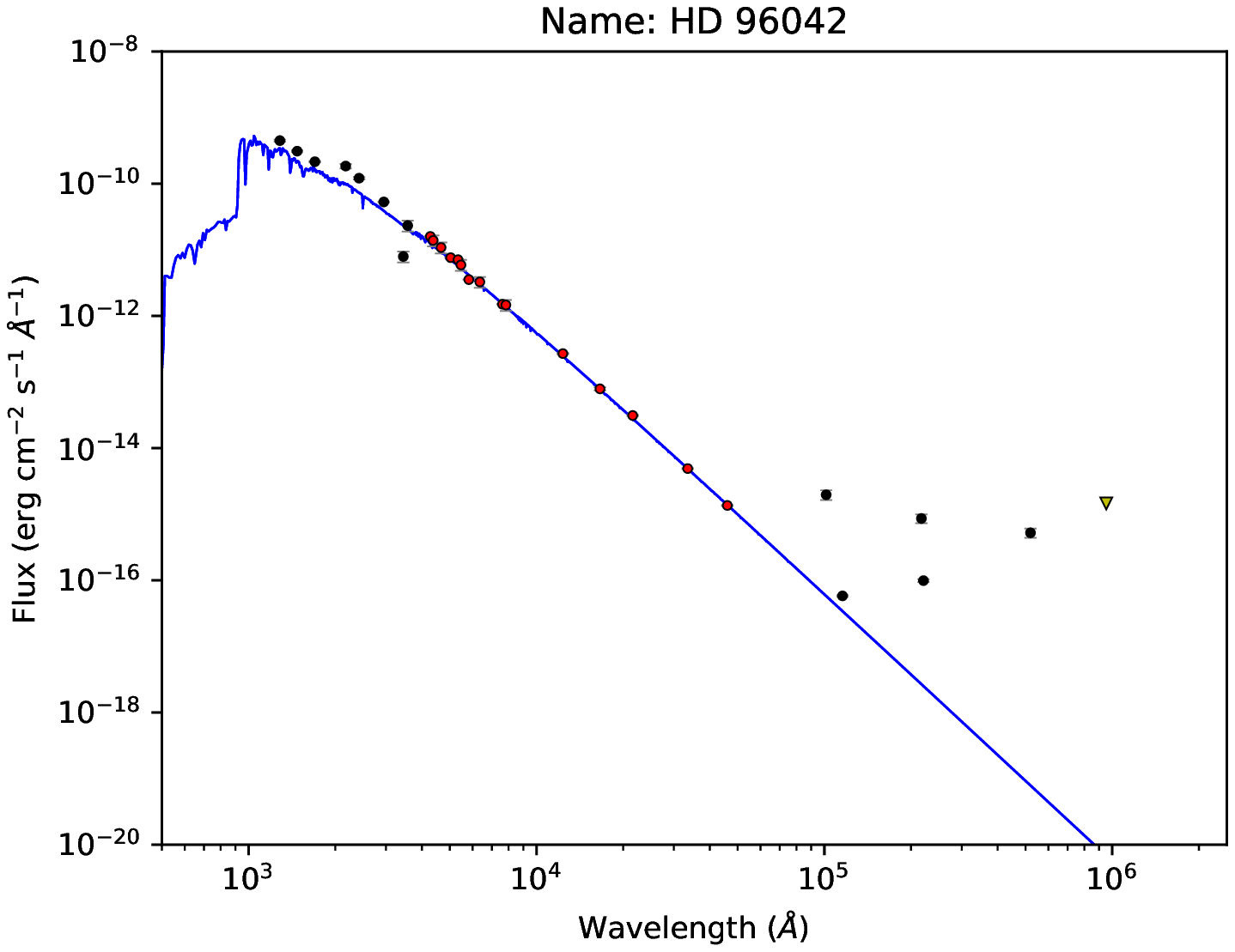}
    \includegraphics[width=0.33\textwidth]{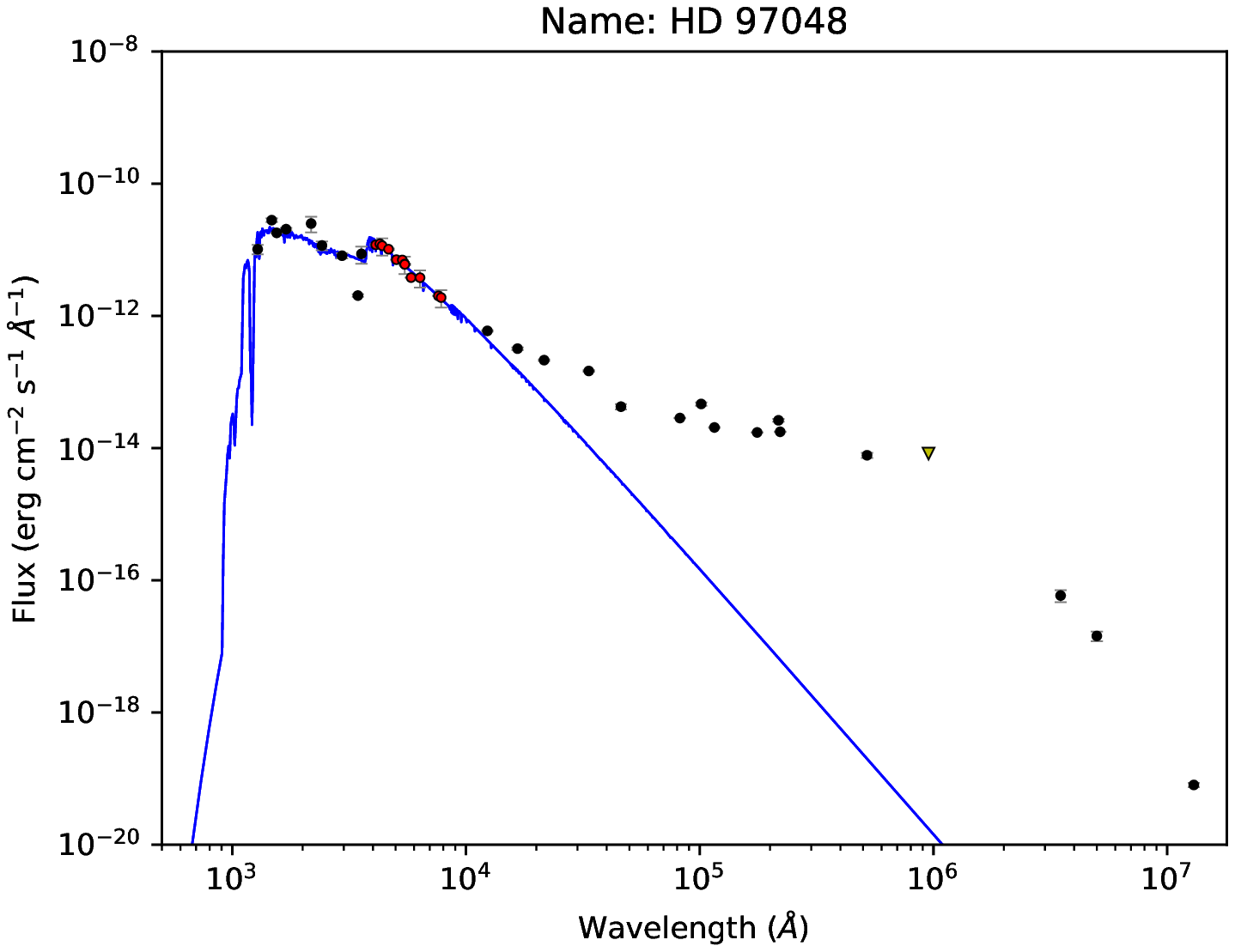}
    \includegraphics[width=0.33\textwidth]{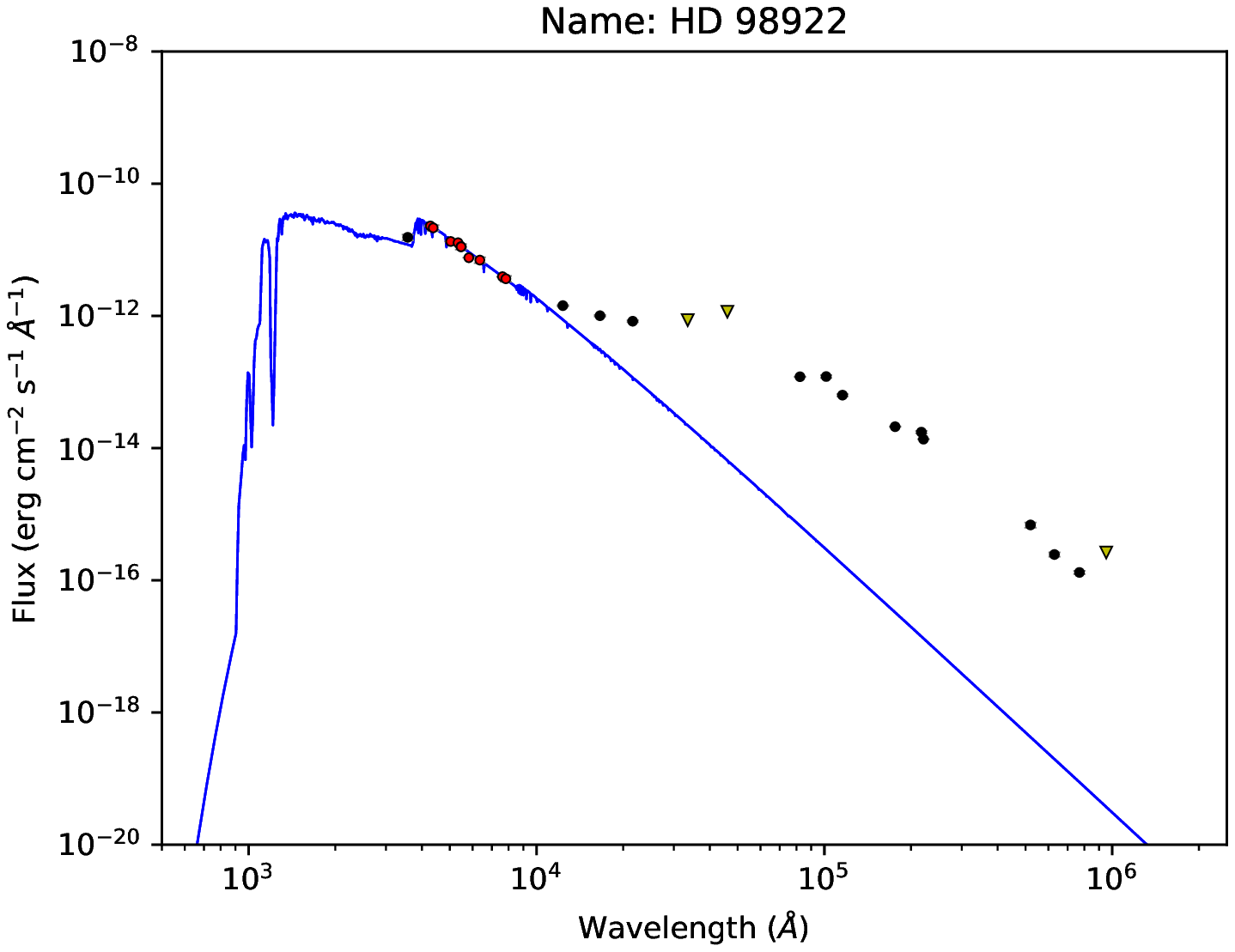}
\end{figure}

\begin{figure} [h]
 \centering
    \includegraphics[width=0.33\textwidth]{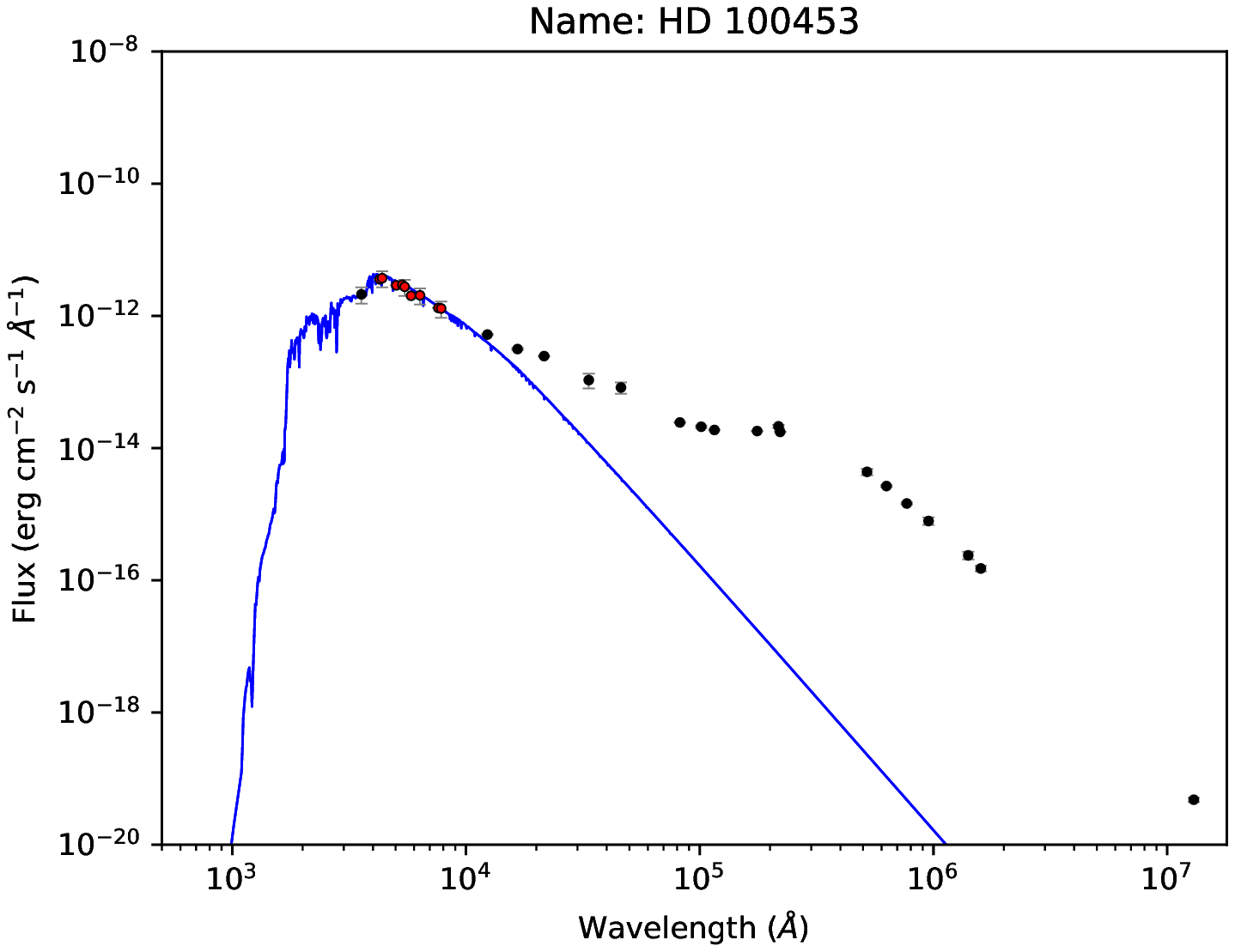}
    \includegraphics[width=0.33\textwidth]{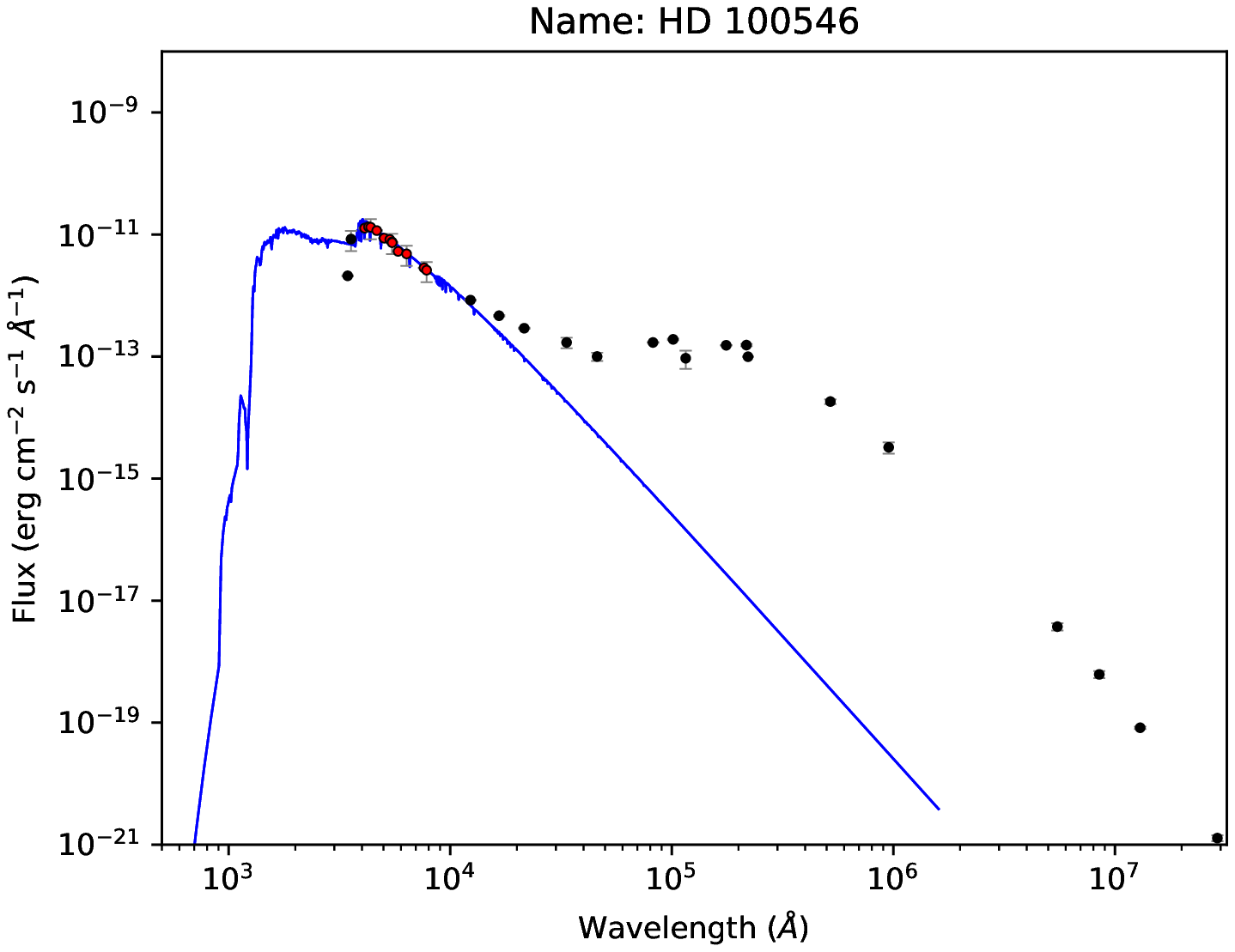}
    \includegraphics[width=0.33\textwidth]{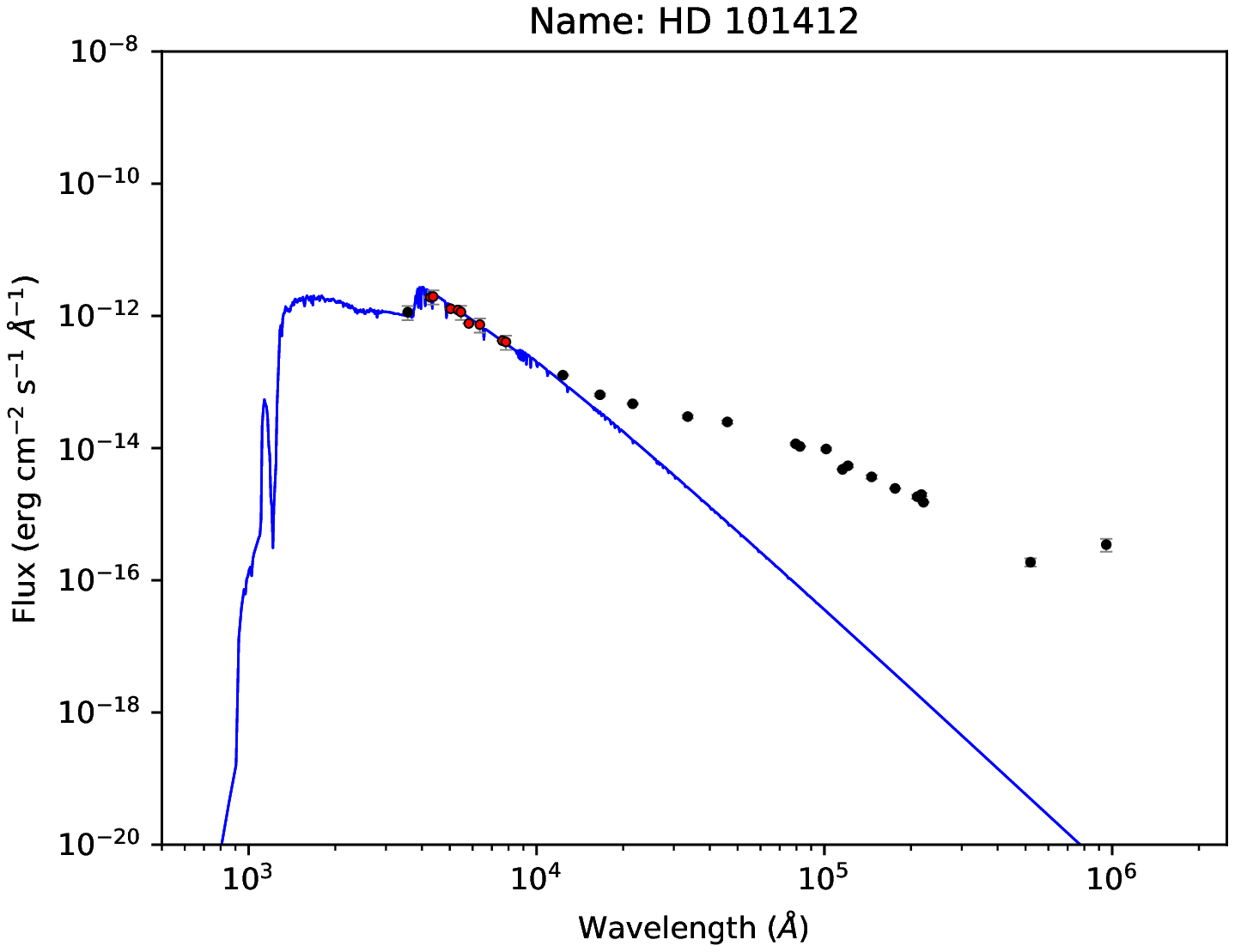}
\end{figure}

\begin{figure} [h]
 \centering
    \includegraphics[width=0.33\textwidth]{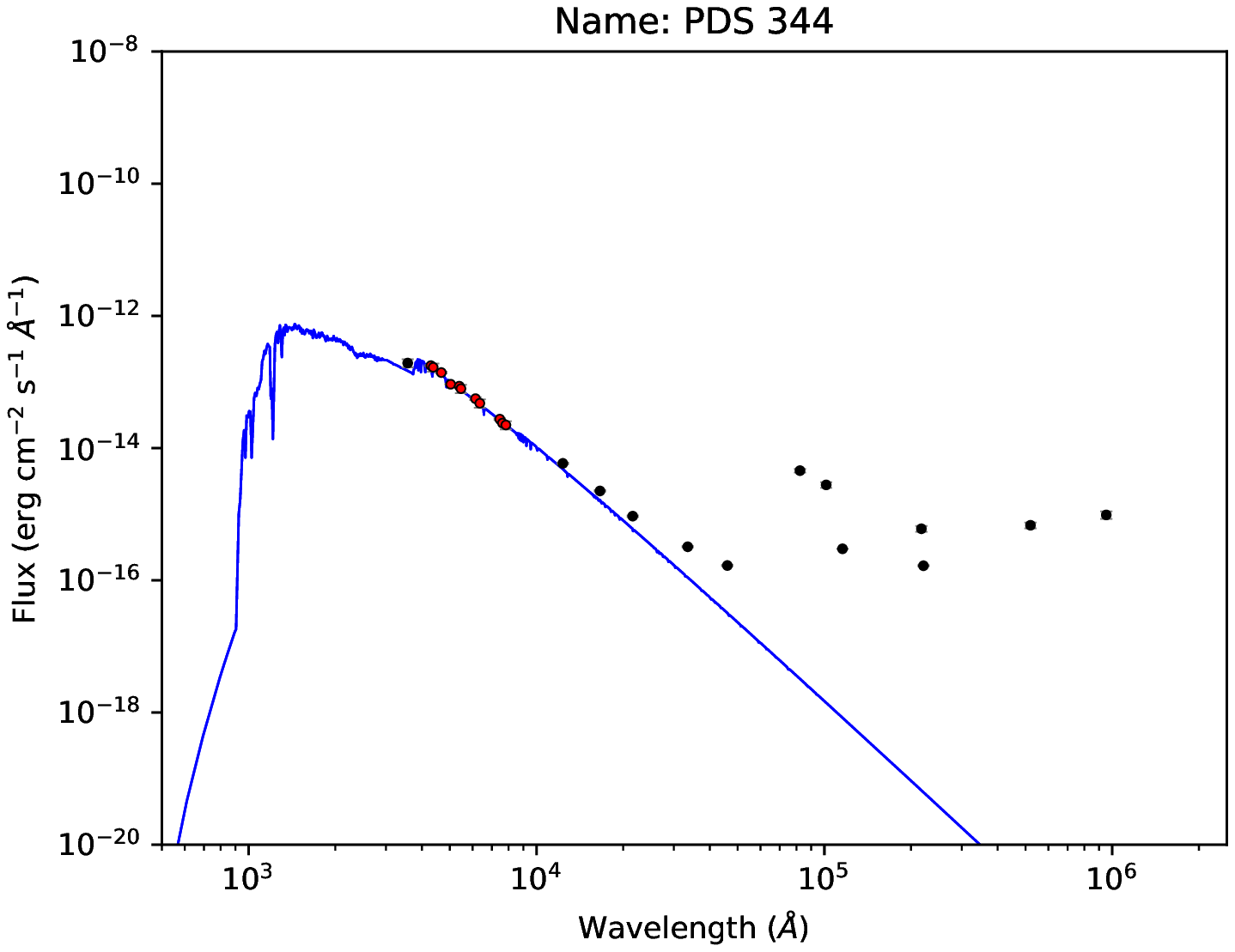}
    \includegraphics[width=0.33\textwidth]{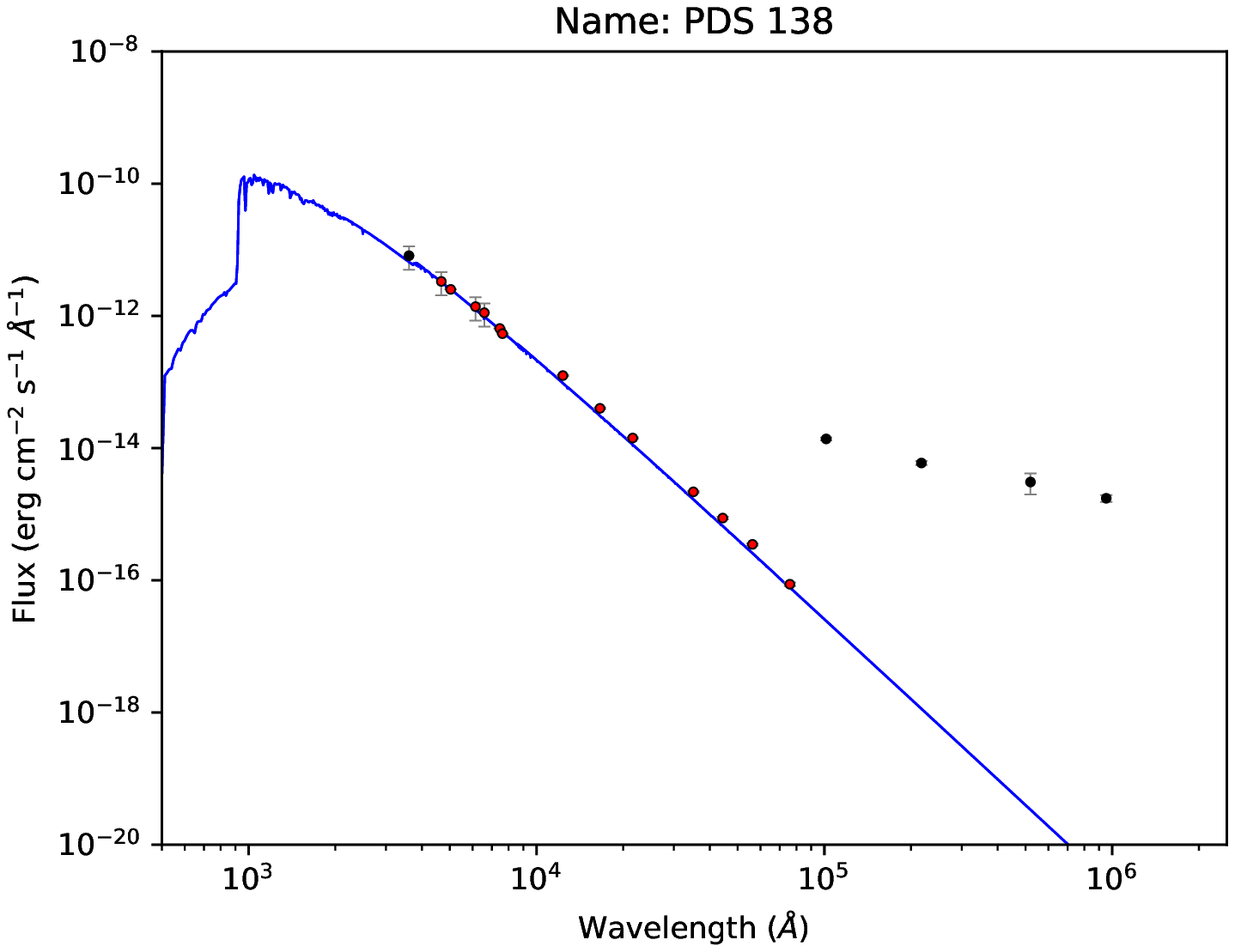}
    \includegraphics[width=0.33\textwidth]{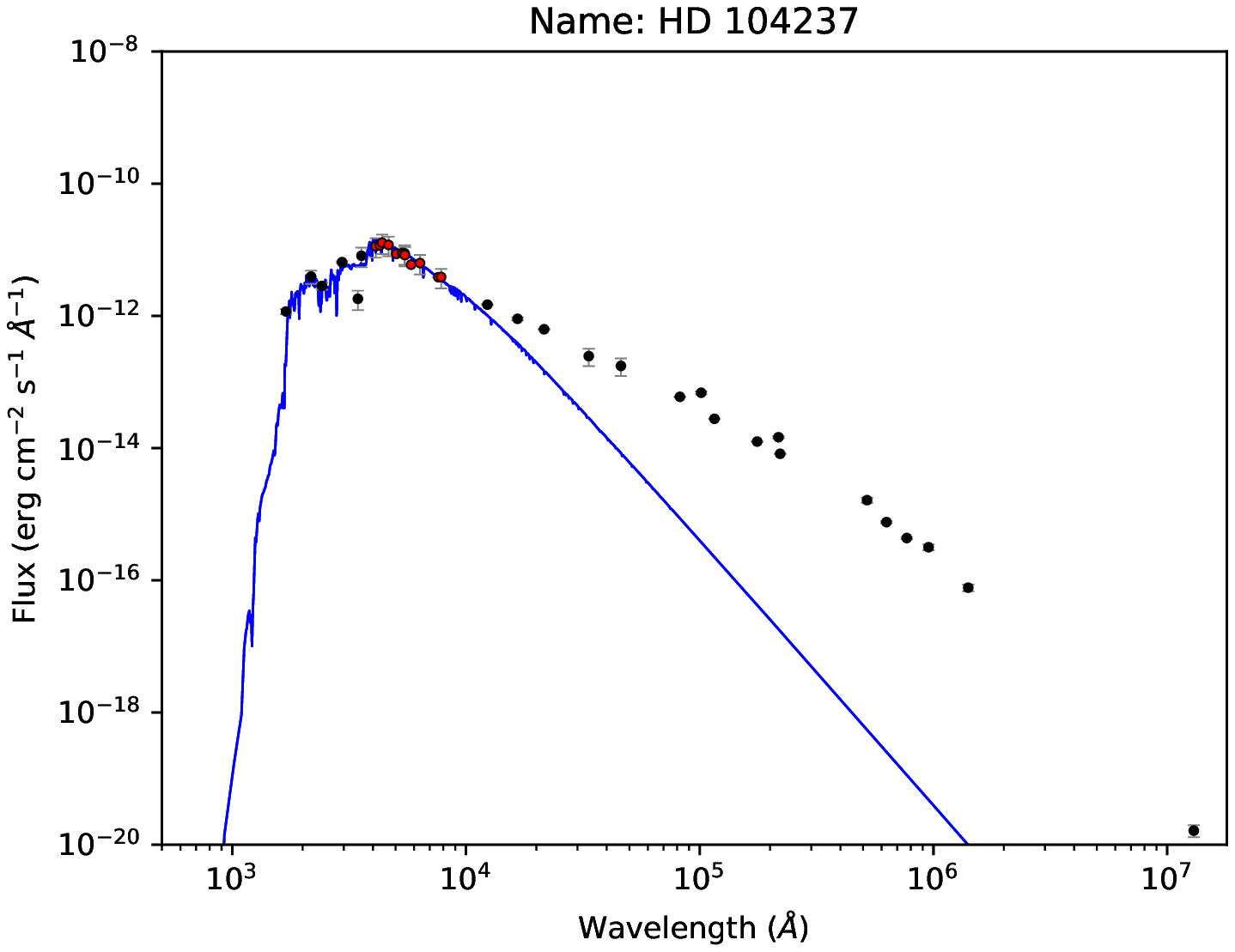}
\end{figure}

\newpage

\onecolumn

\begin{figure} [h]
 \centering
    \includegraphics[width=0.33\textwidth]{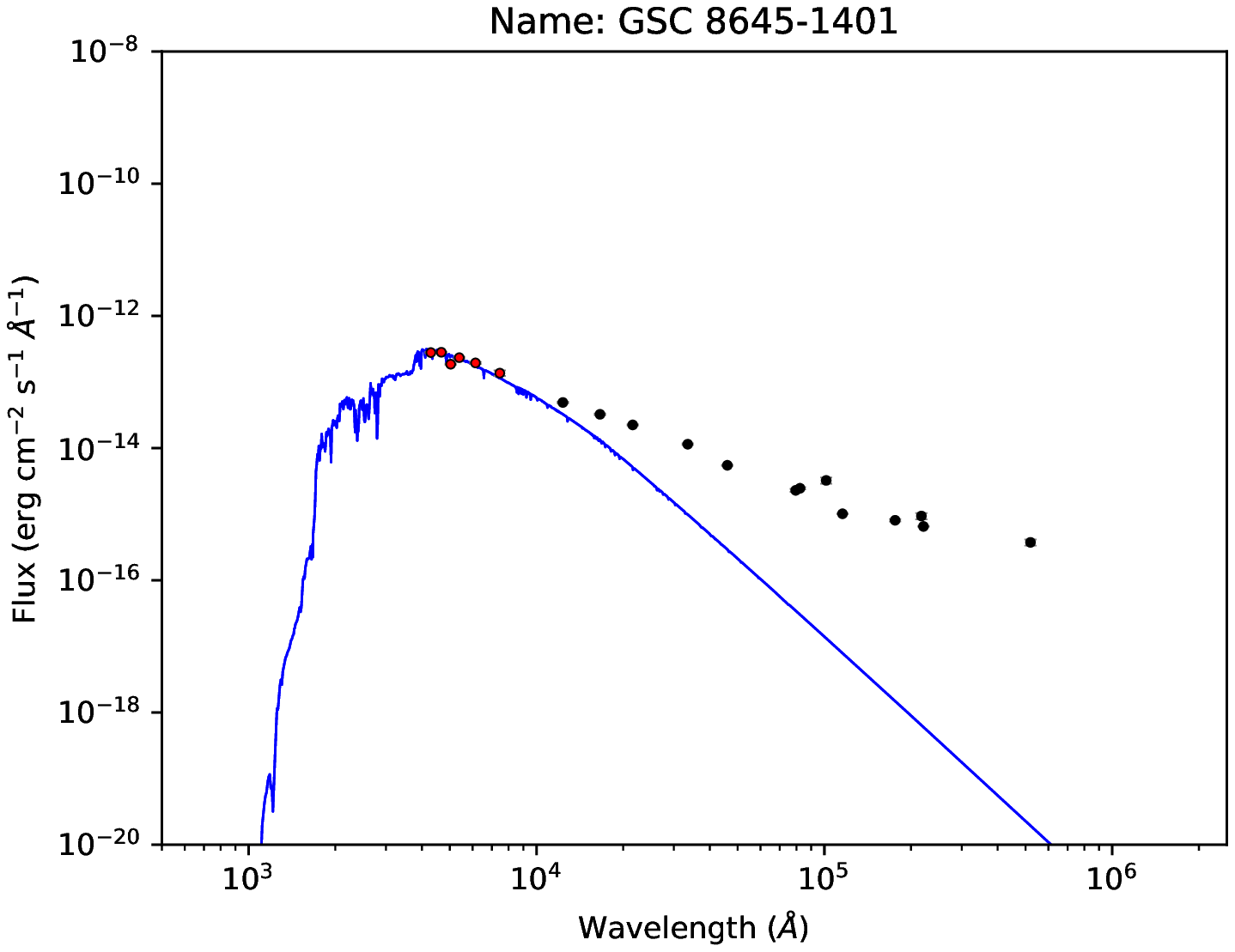}
    \includegraphics[width=0.33\textwidth]{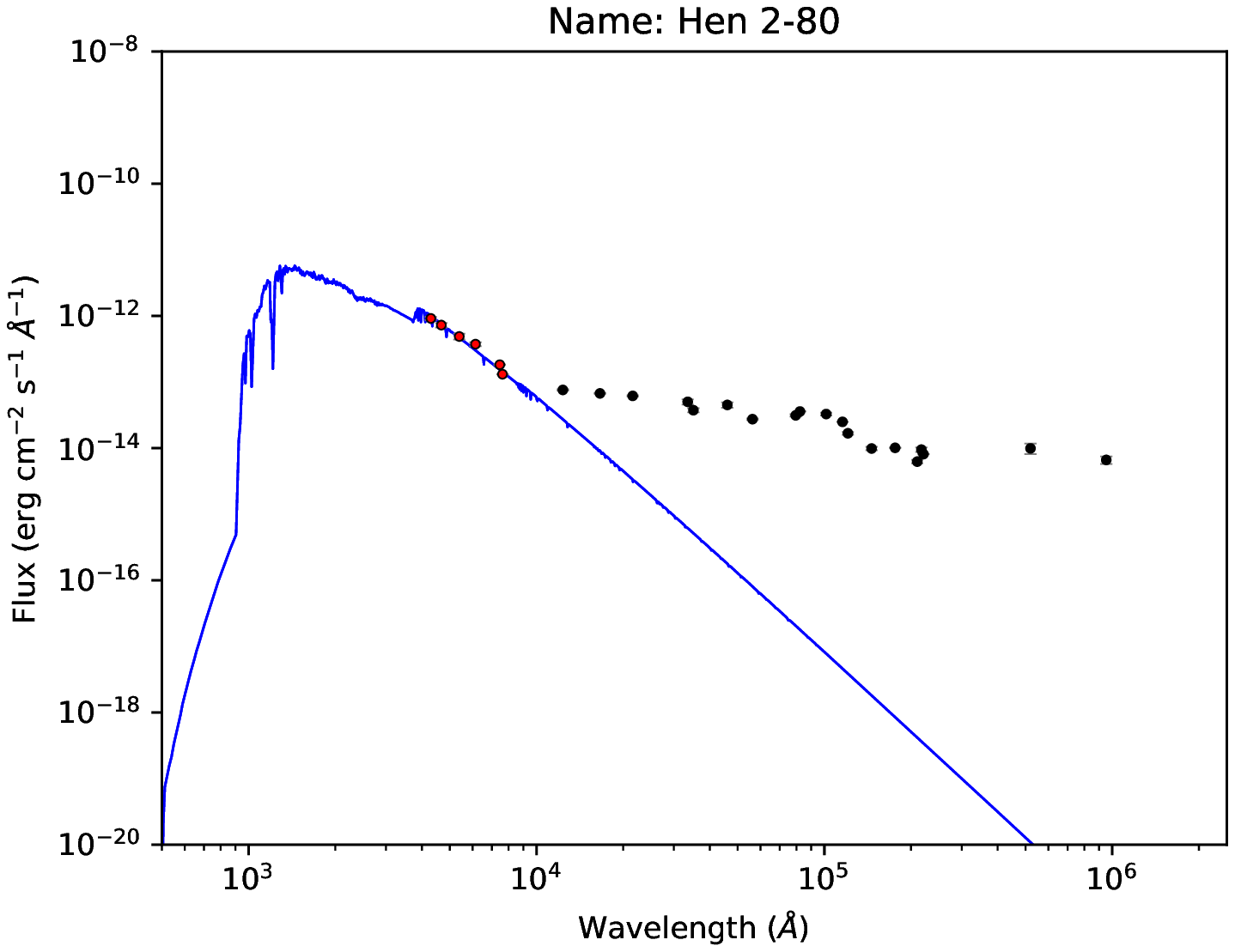}
    \includegraphics[width=0.33\textwidth]{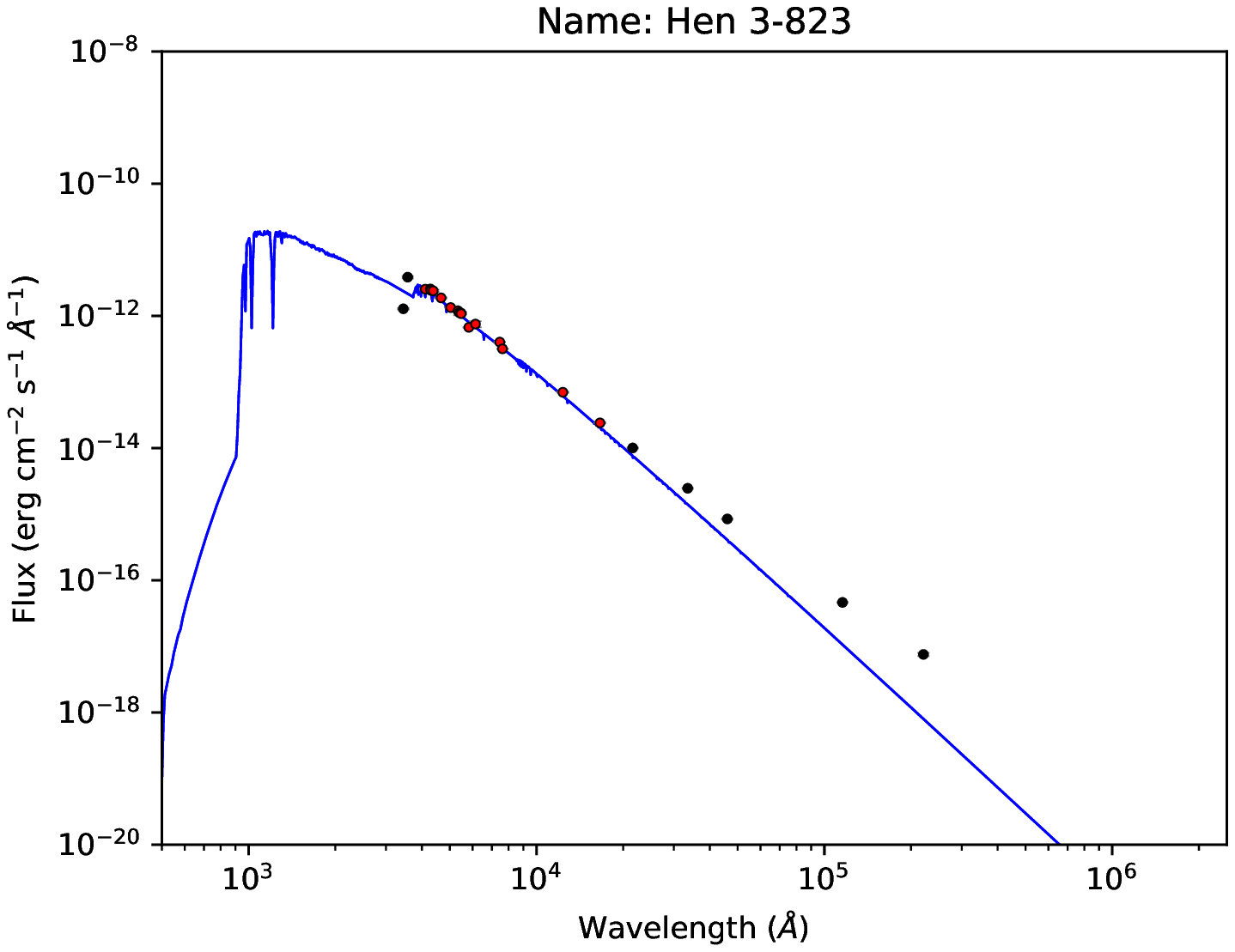}
\end{figure}

\begin{figure} [h]
 \centering
    \includegraphics[width=0.33\textwidth]{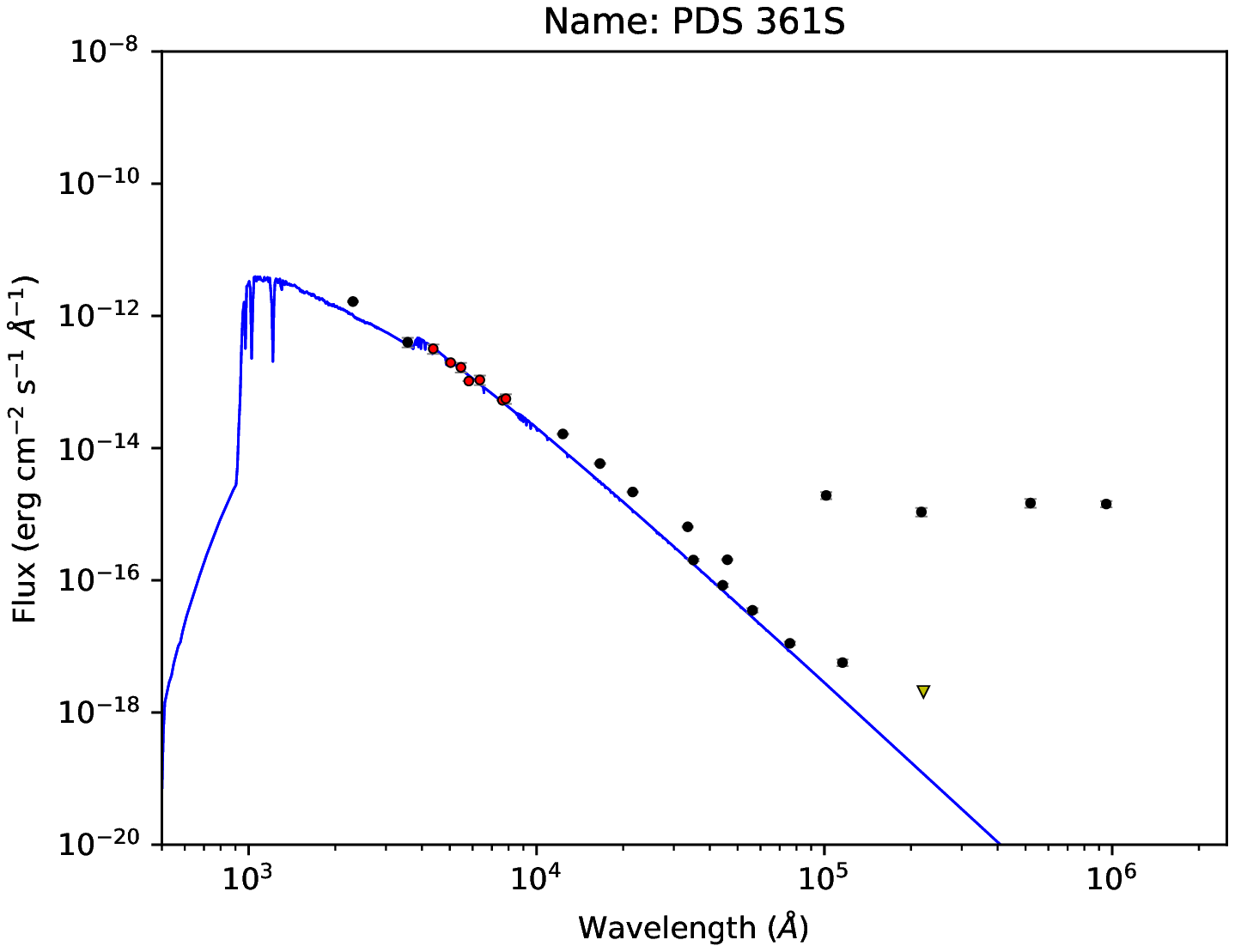}
    \includegraphics[width=0.33\textwidth]{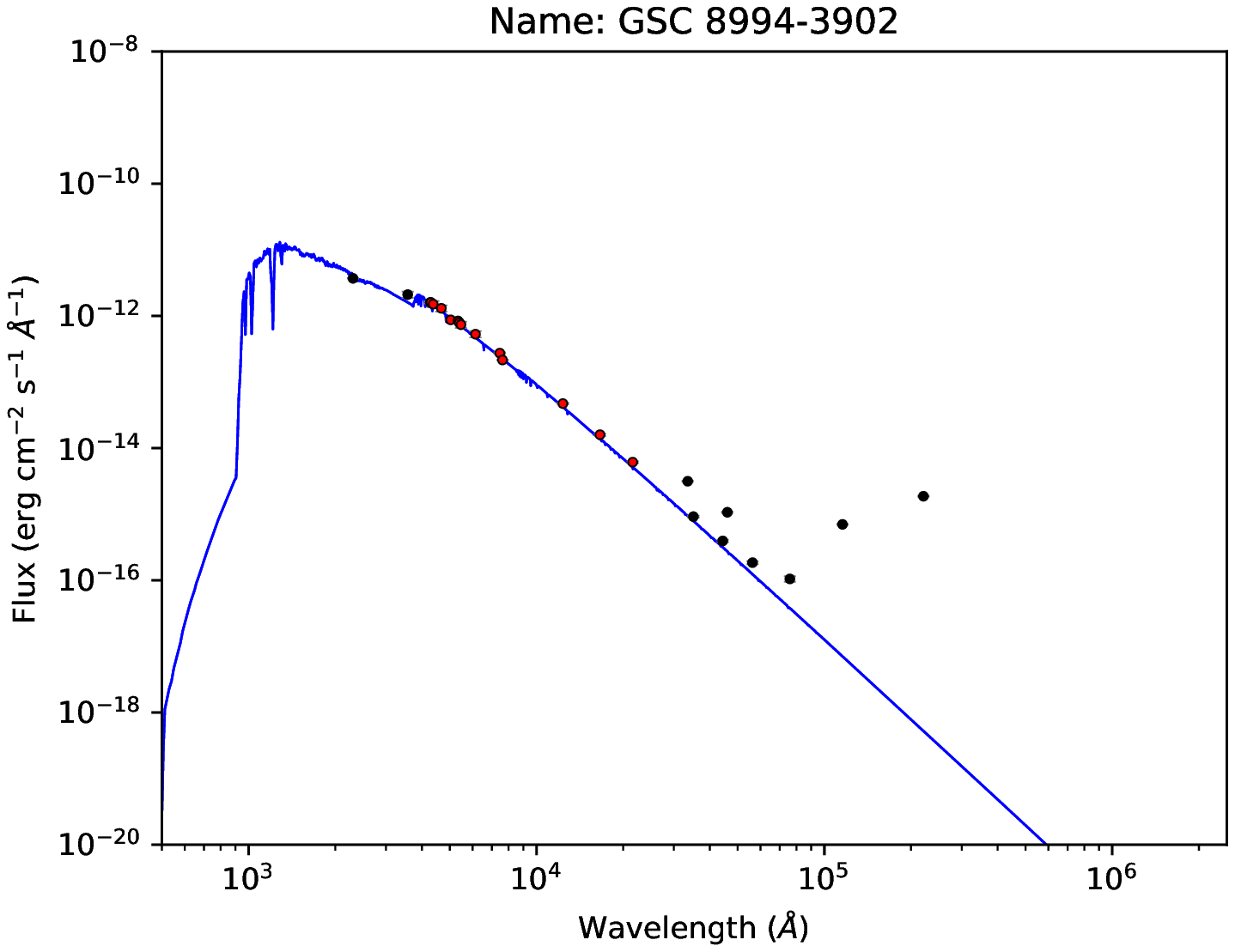}
    \includegraphics[width=0.33\textwidth]{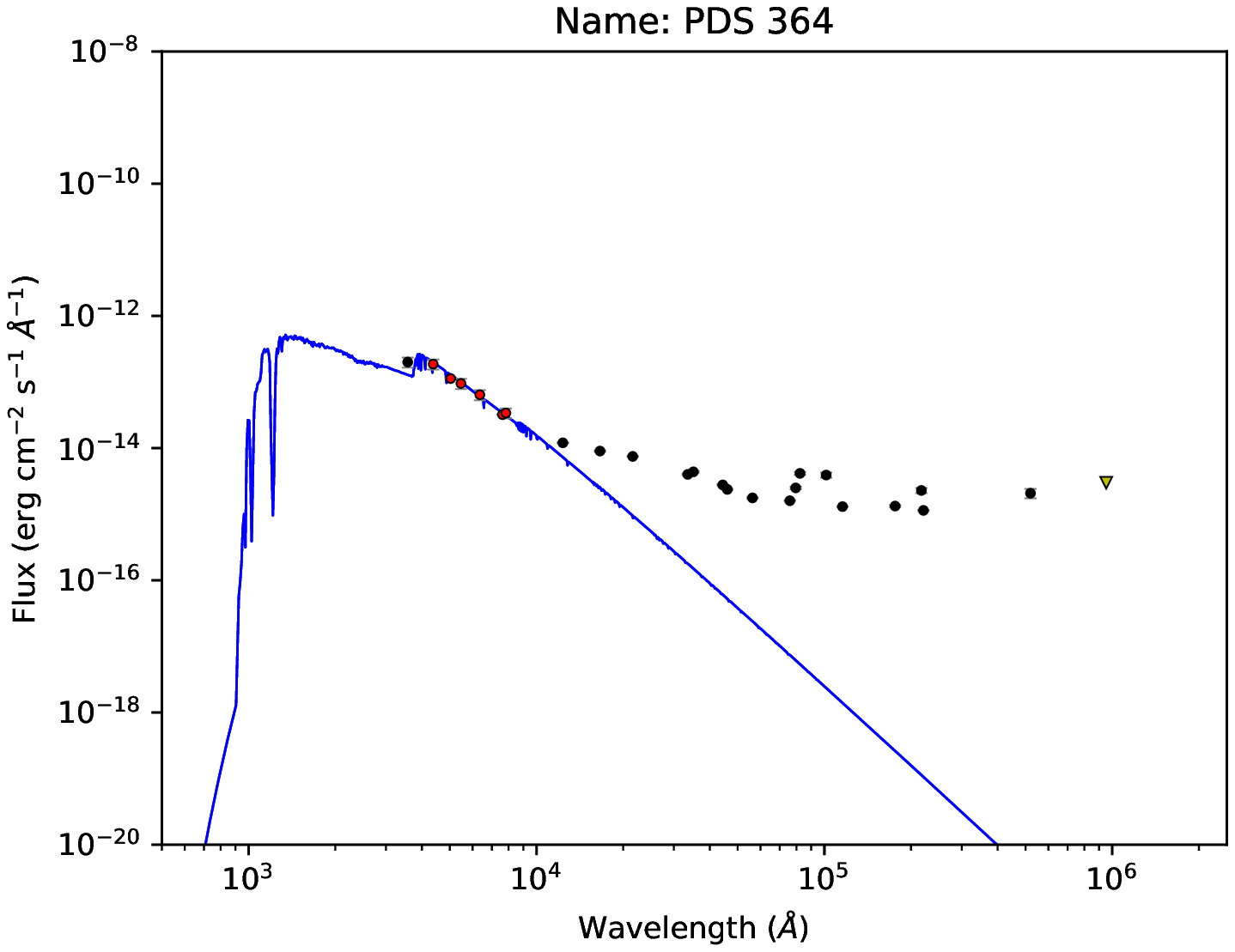}
\end{figure}

\begin{figure} [h]
 \centering
    \includegraphics[width=0.33\textwidth]{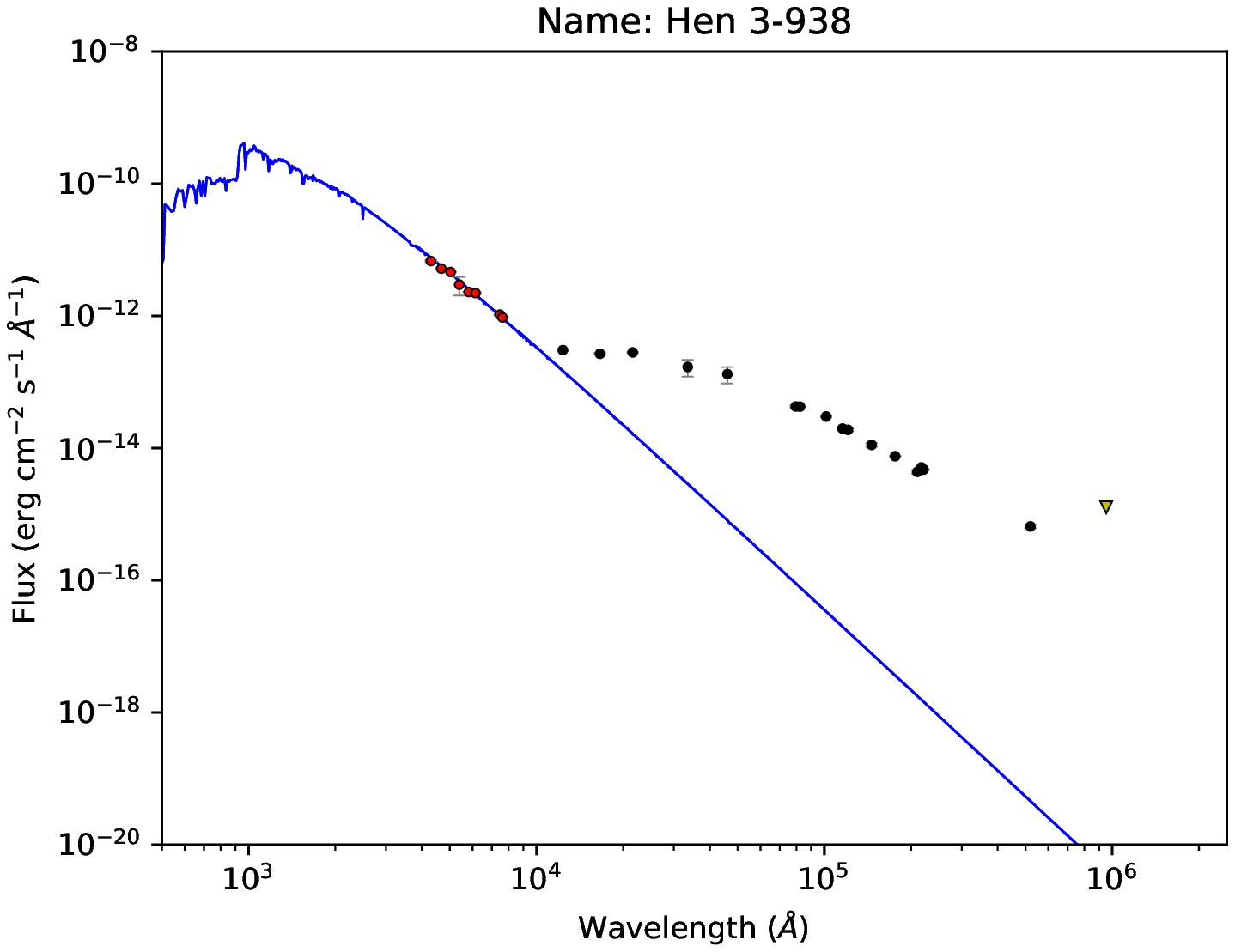}
    \includegraphics[width=0.33\textwidth]{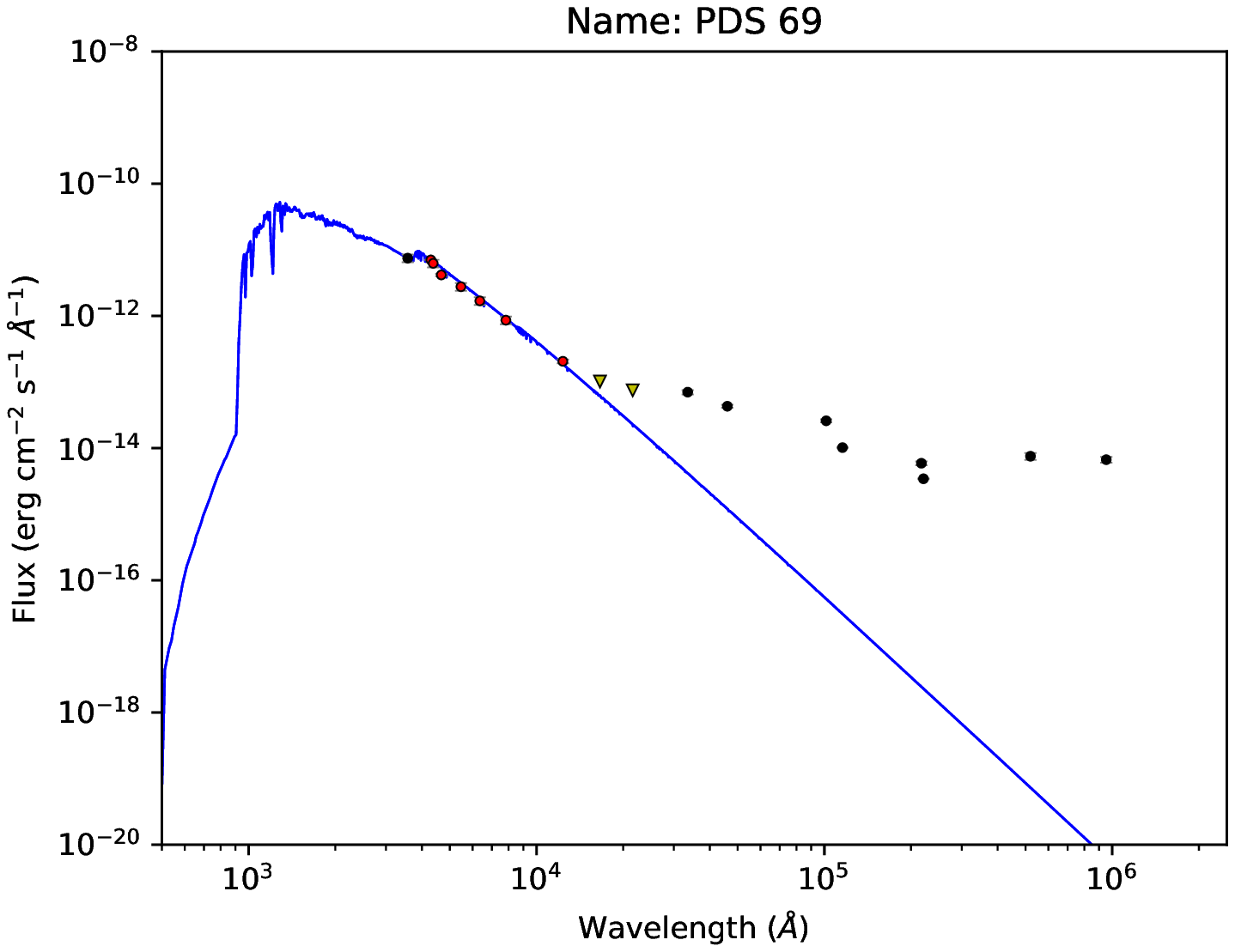}
    \includegraphics[width=0.33\textwidth]{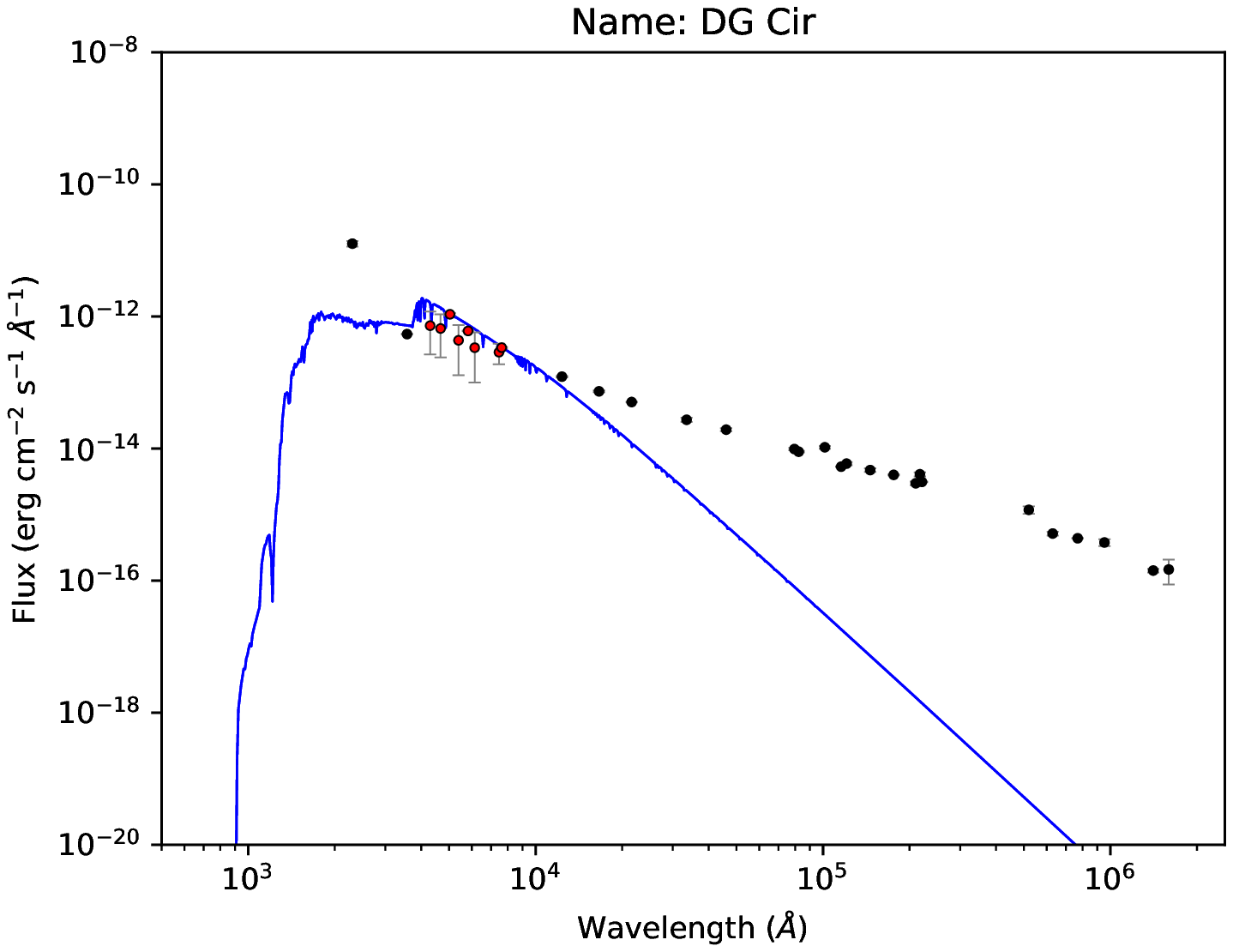}
\end{figure}

\begin{figure} [h]
 \centering
    \includegraphics[width=0.33\textwidth]{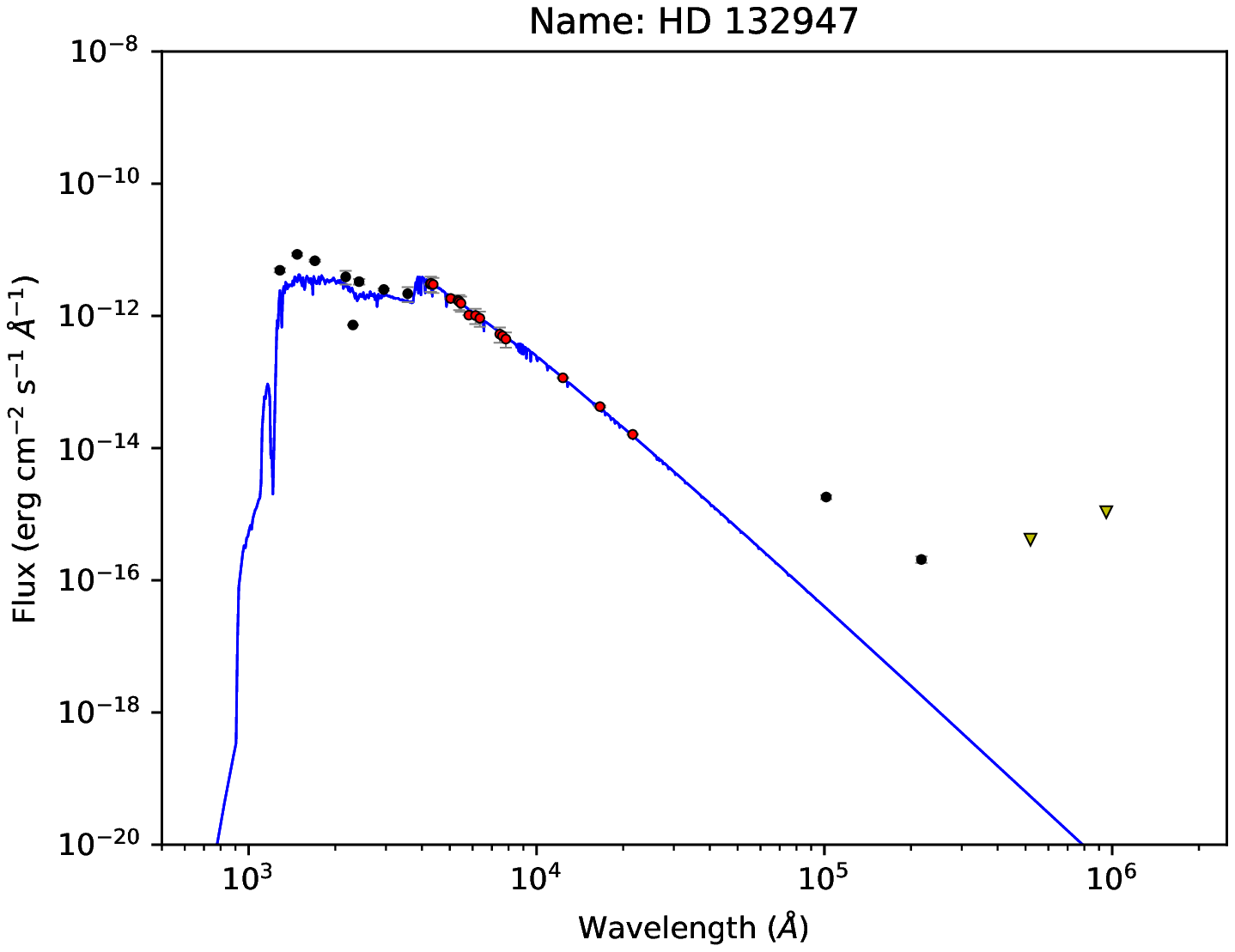}
    \includegraphics[width=0.33\textwidth]{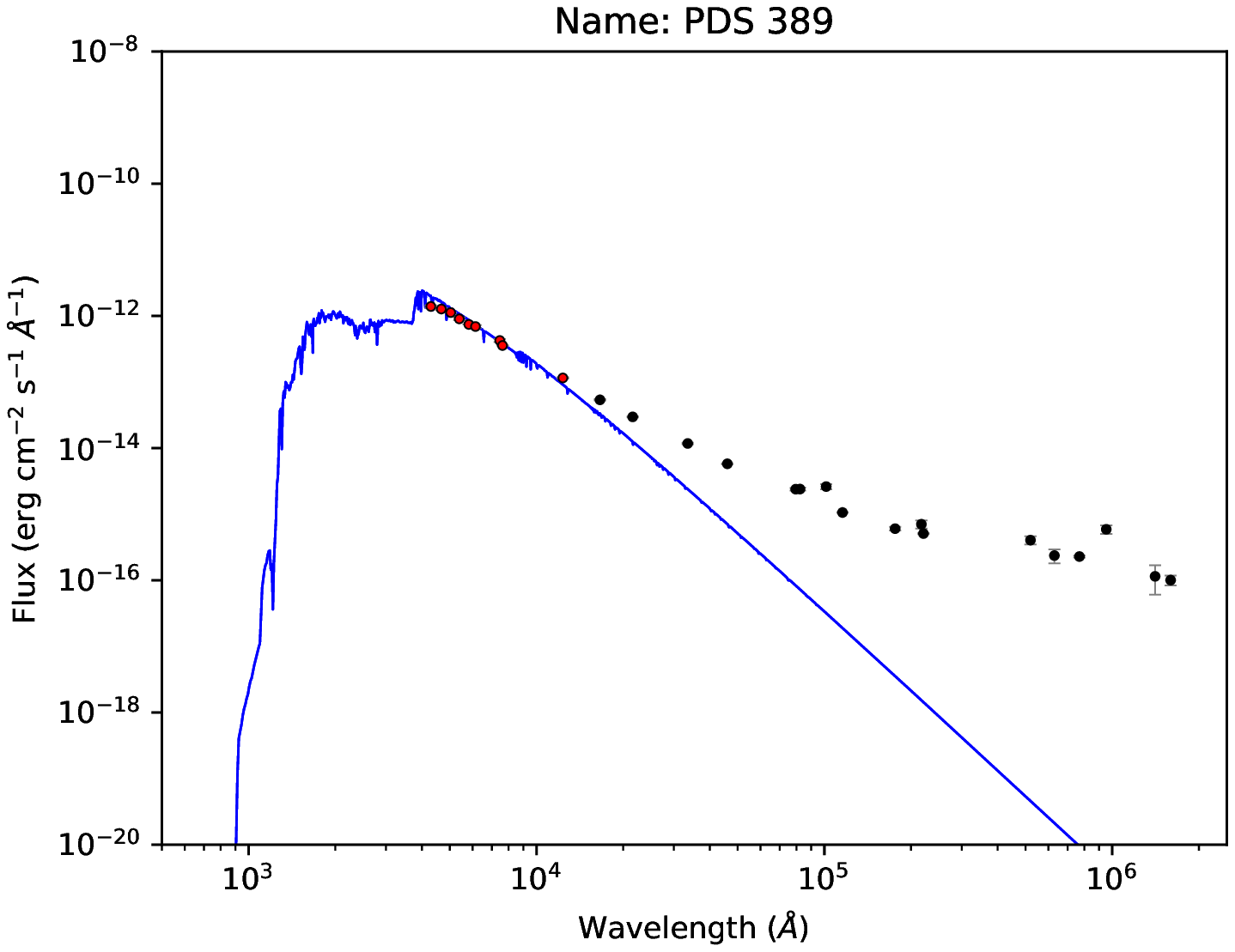}
    \includegraphics[width=0.33\textwidth]{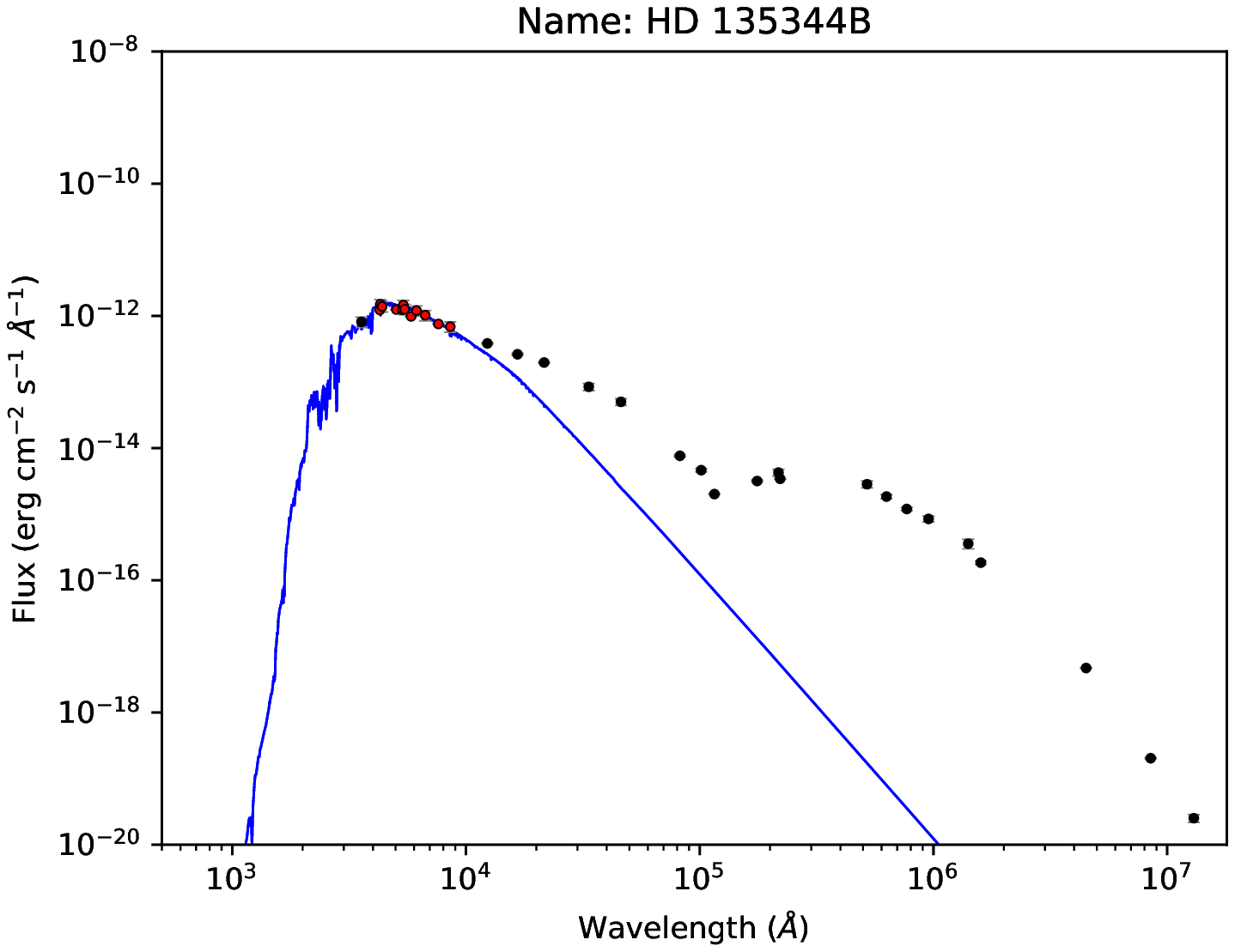}
\end{figure}

\newpage

\onecolumn

\begin{figure} [h]
 \centering
    \includegraphics[width=0.33\textwidth]{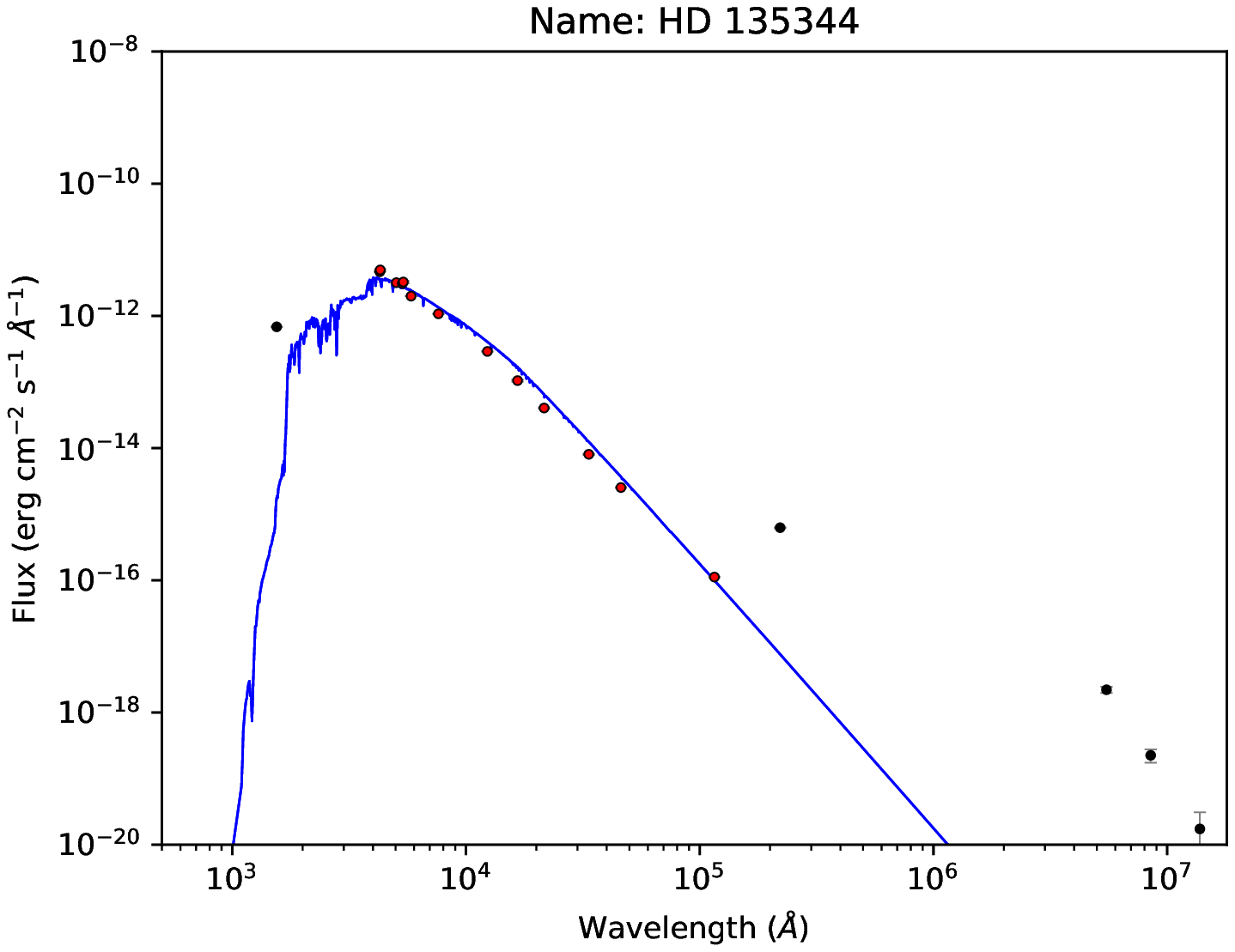}
    \includegraphics[width=0.33\textwidth]{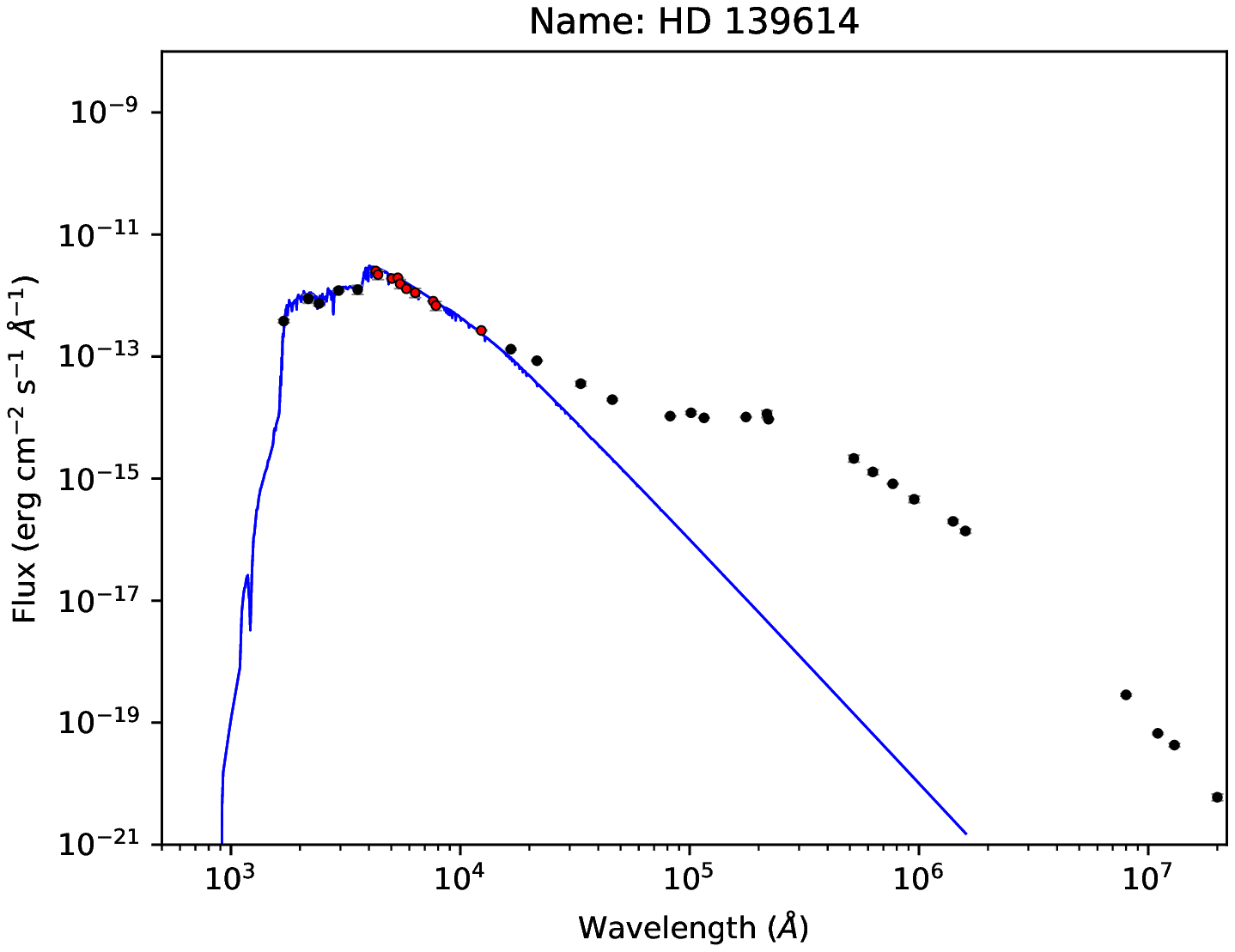}
    \includegraphics[width=0.33\textwidth]{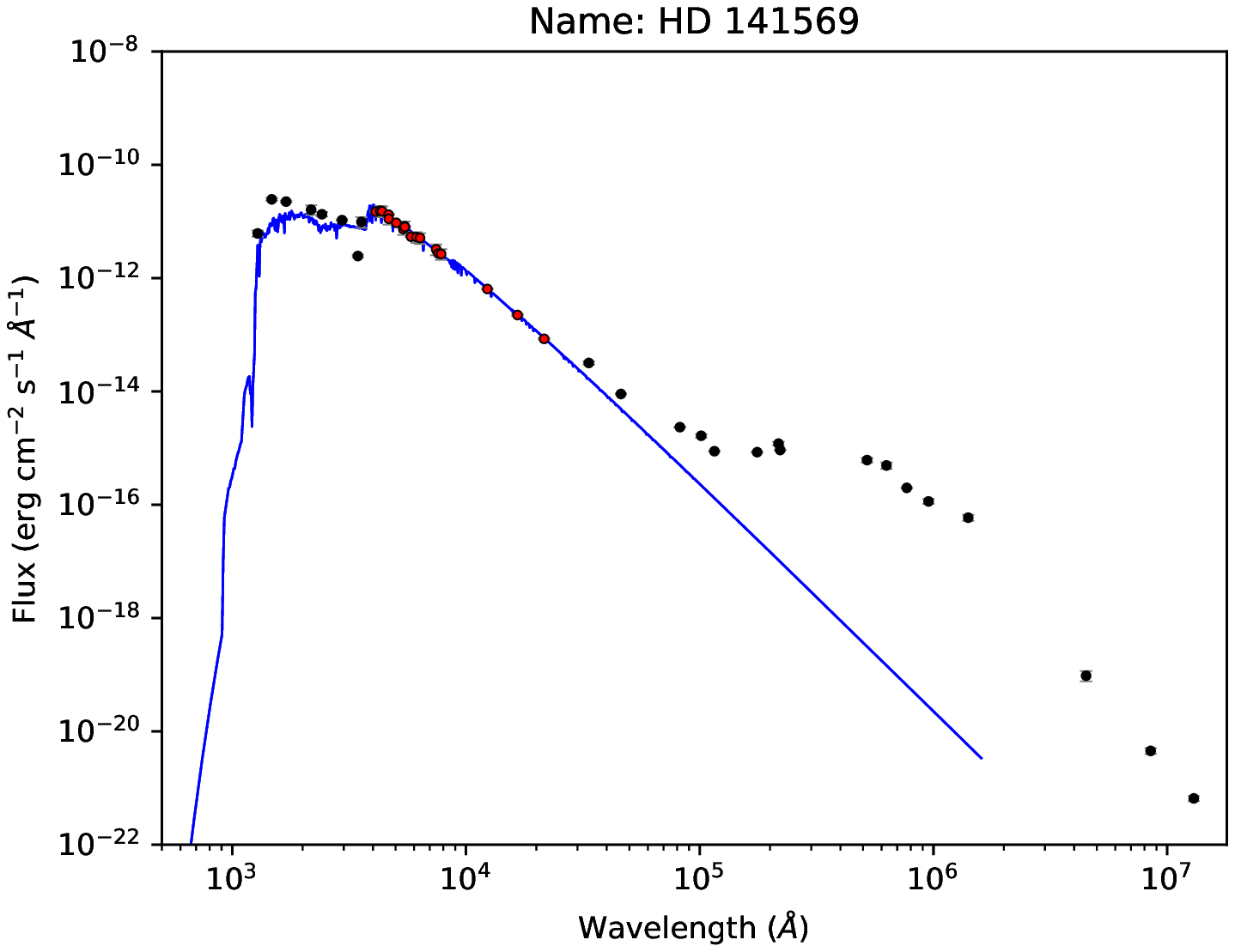}
\end{figure}

\begin{figure} [h]
 \centering
    \includegraphics[width=0.33\textwidth]{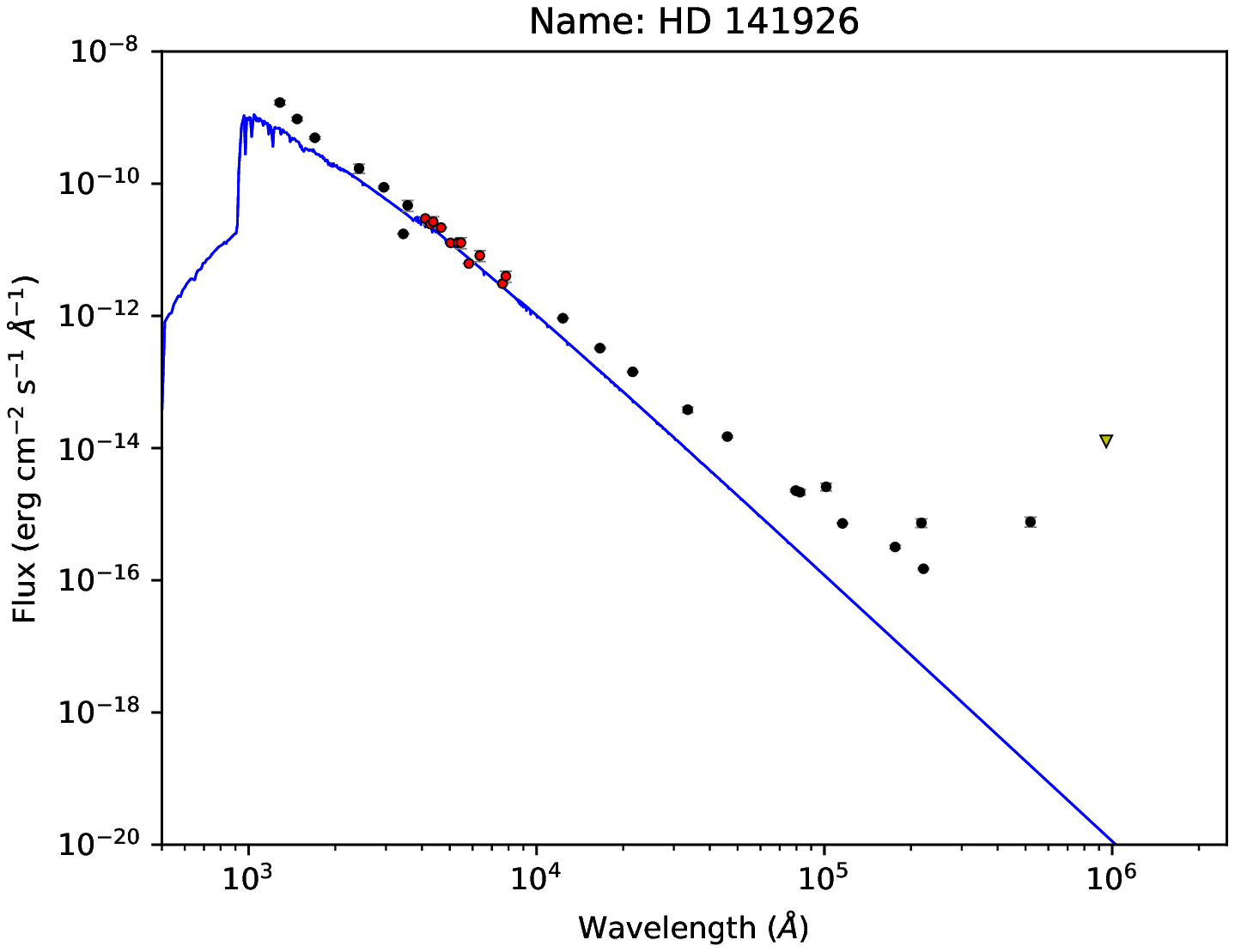}
    \includegraphics[width=0.33\textwidth]{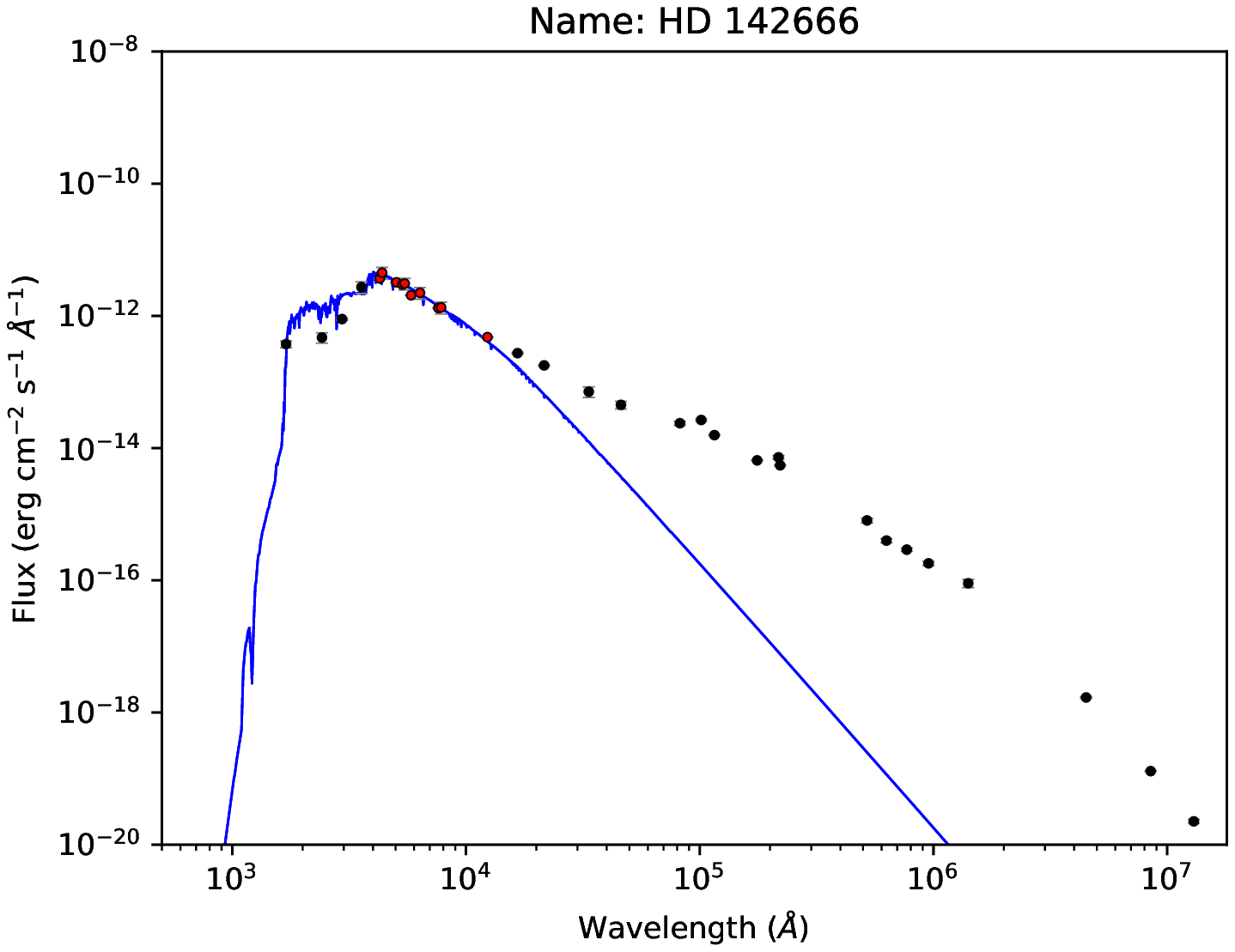}
    \includegraphics[width=0.33\textwidth]{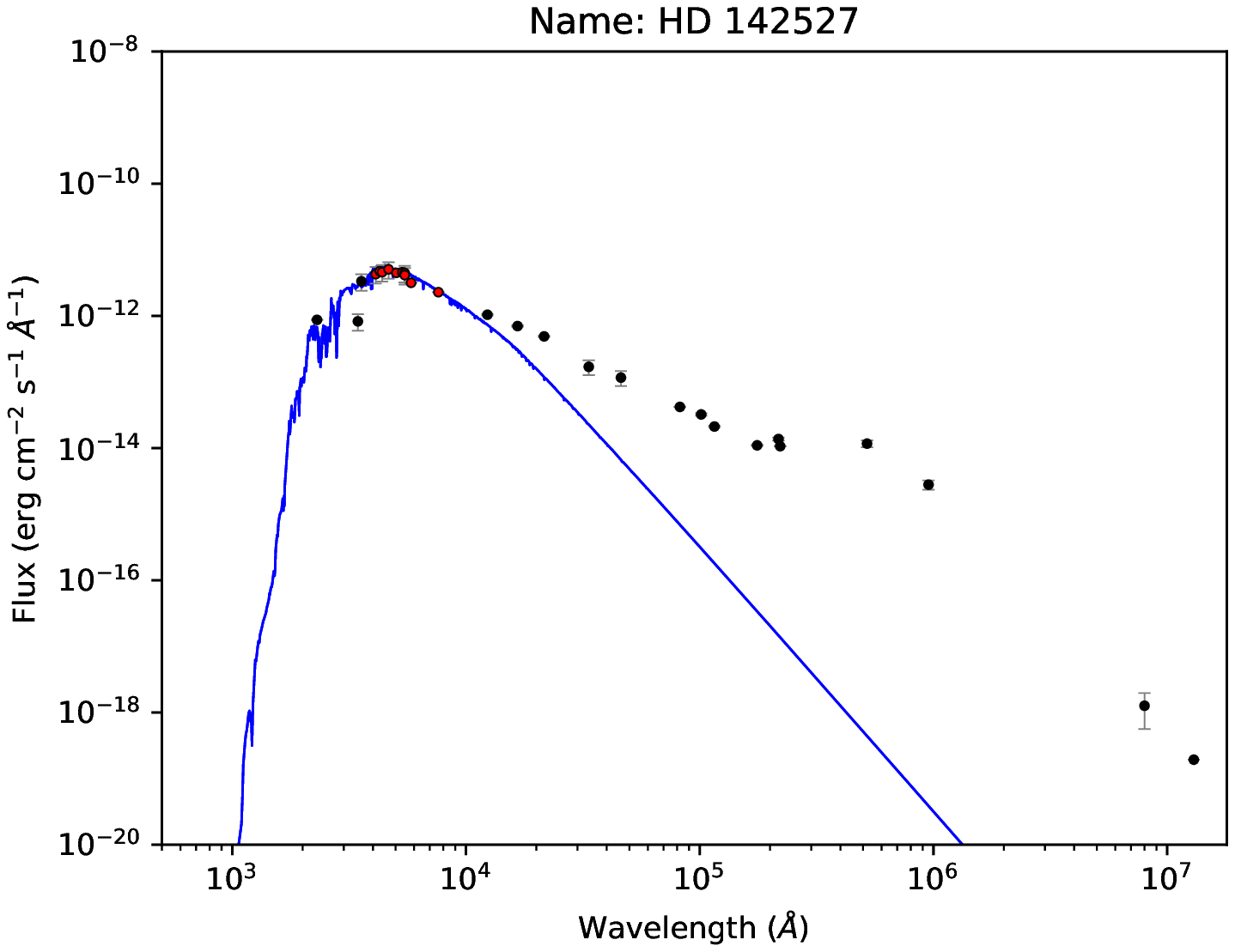}
\end{figure}

\begin{figure} [h]
 \centering
    \includegraphics[width=0.33\textwidth]{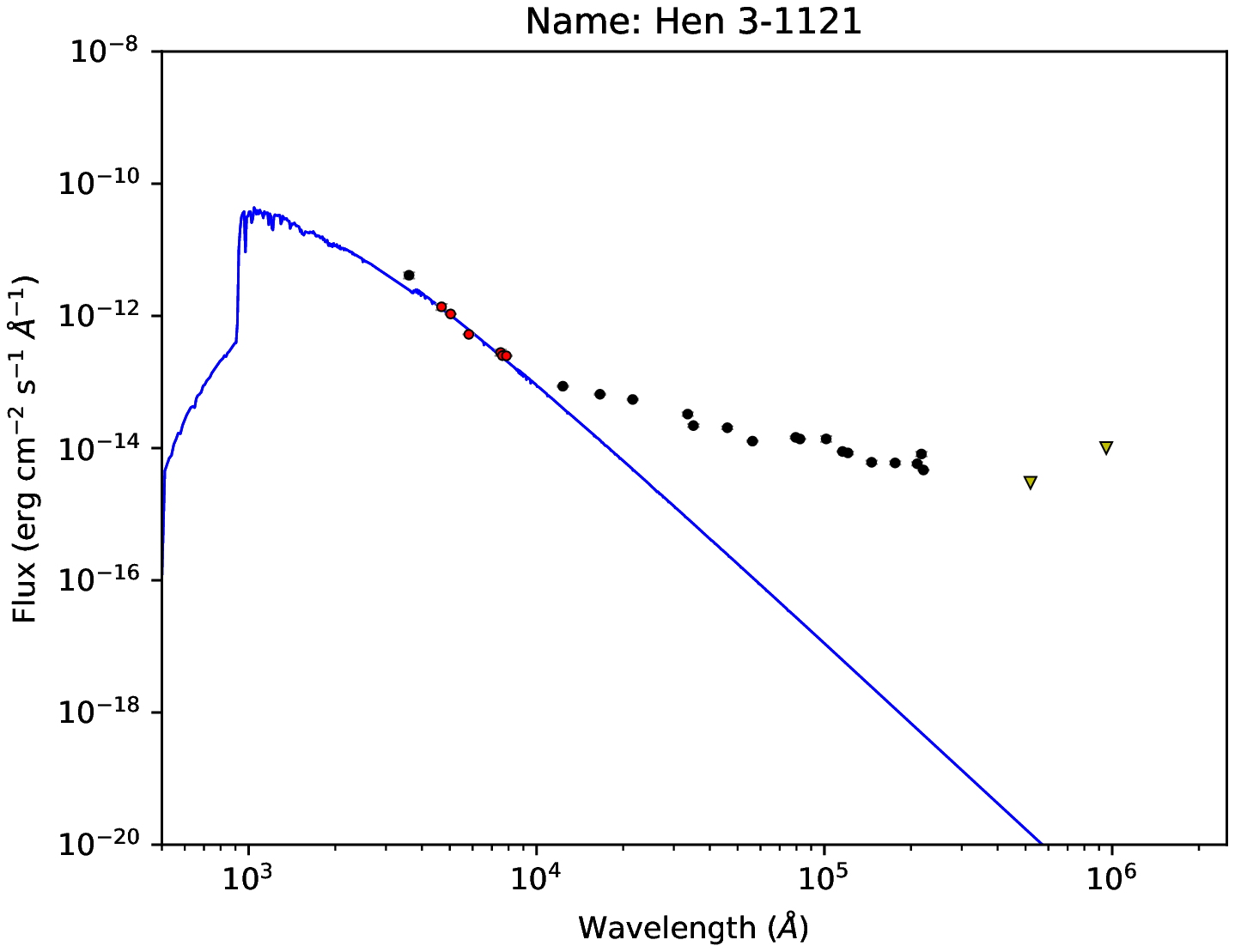}
    \includegraphics[width=0.33\textwidth]{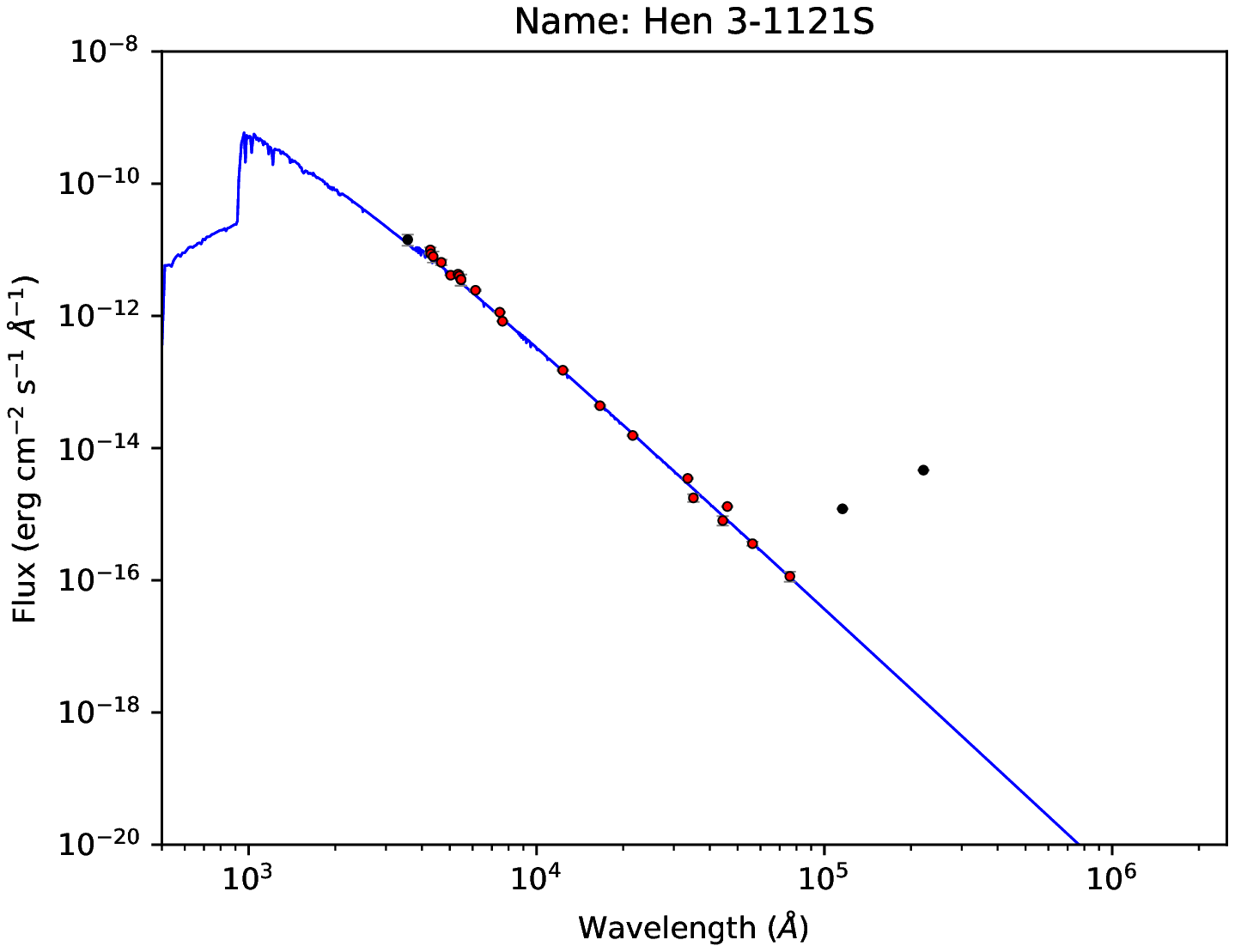}
    \includegraphics[width=0.33\textwidth]{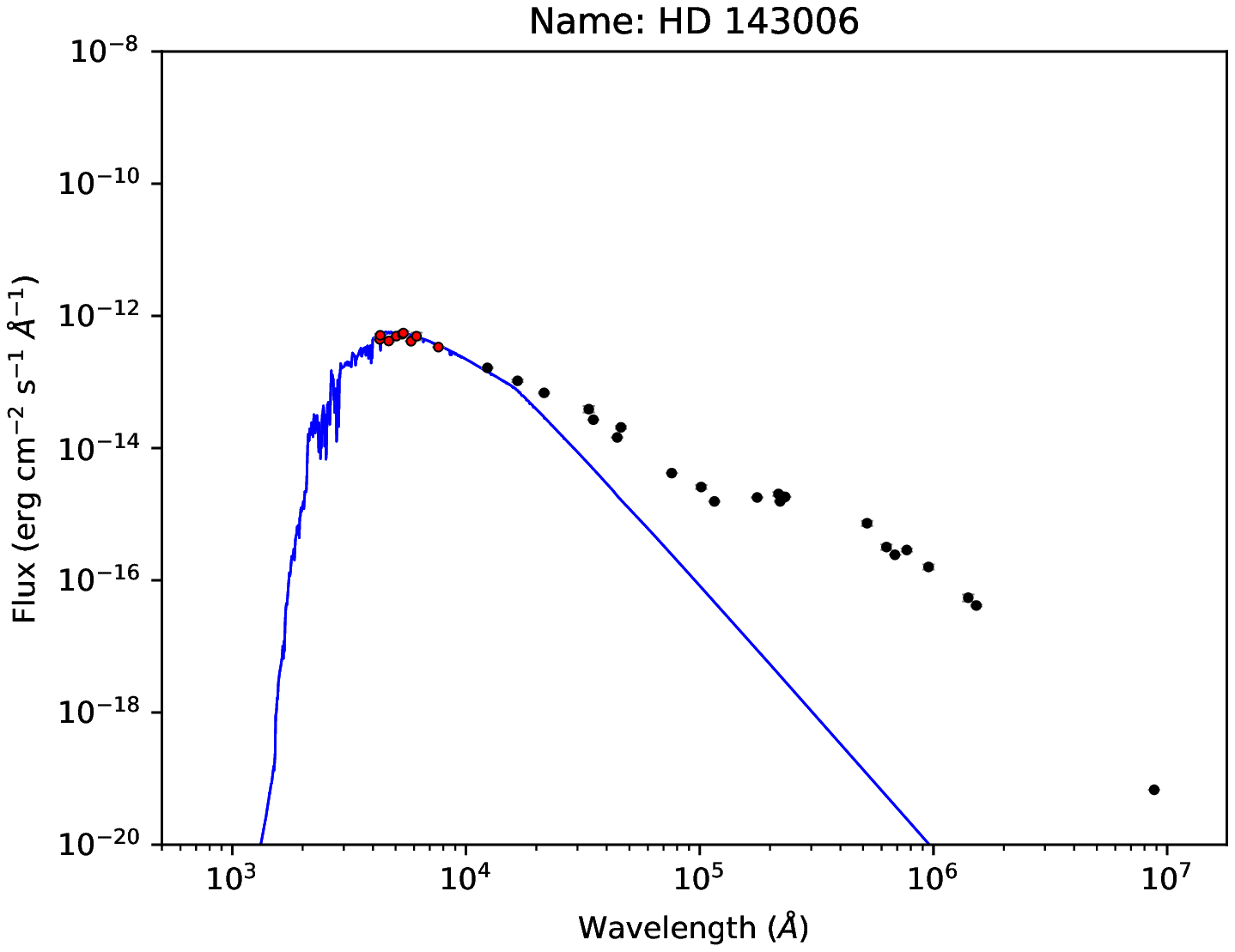}
\end{figure}

\begin{figure} [h]
 \centering
    \includegraphics[width=0.33\textwidth]{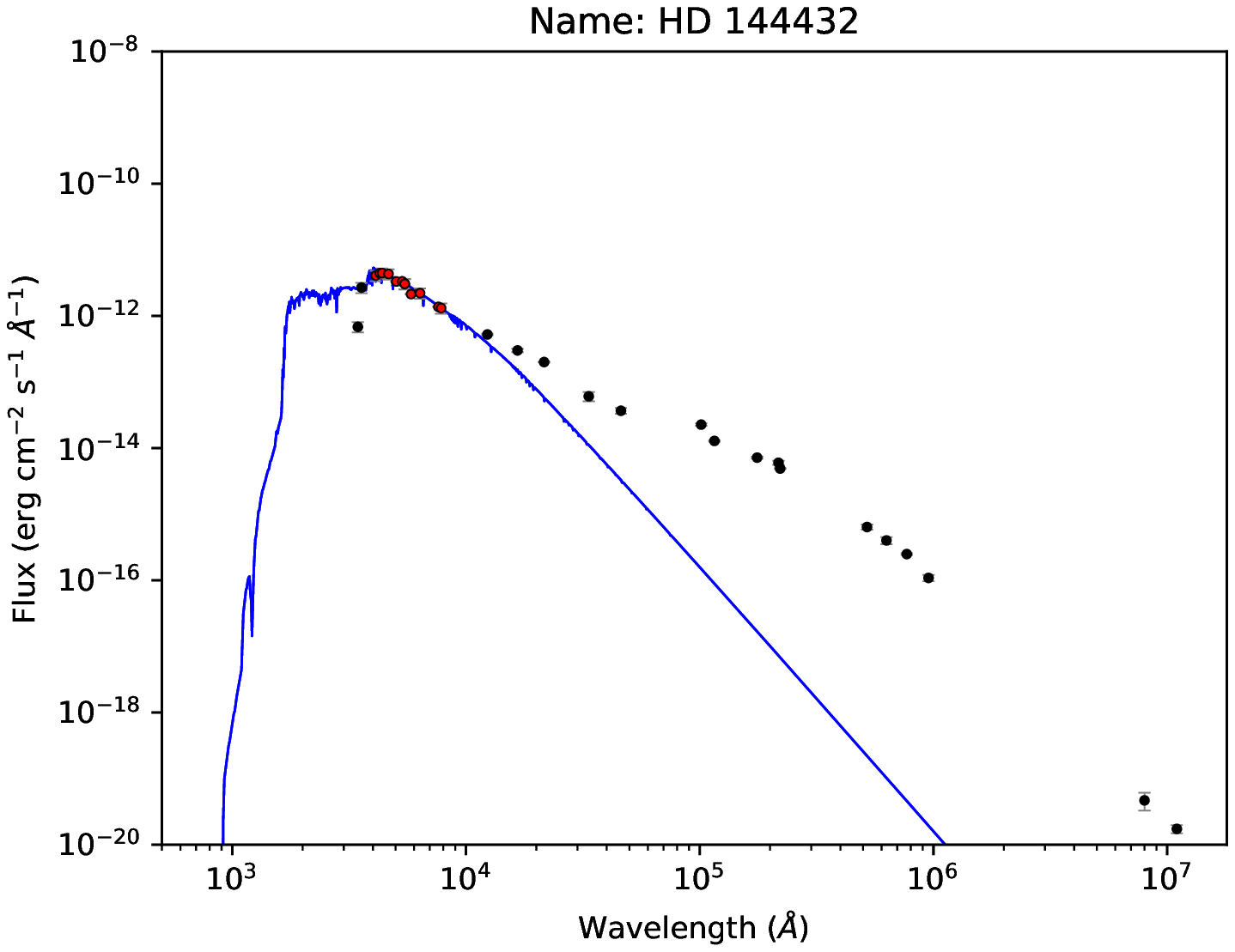}
    \includegraphics[width=0.33\textwidth]{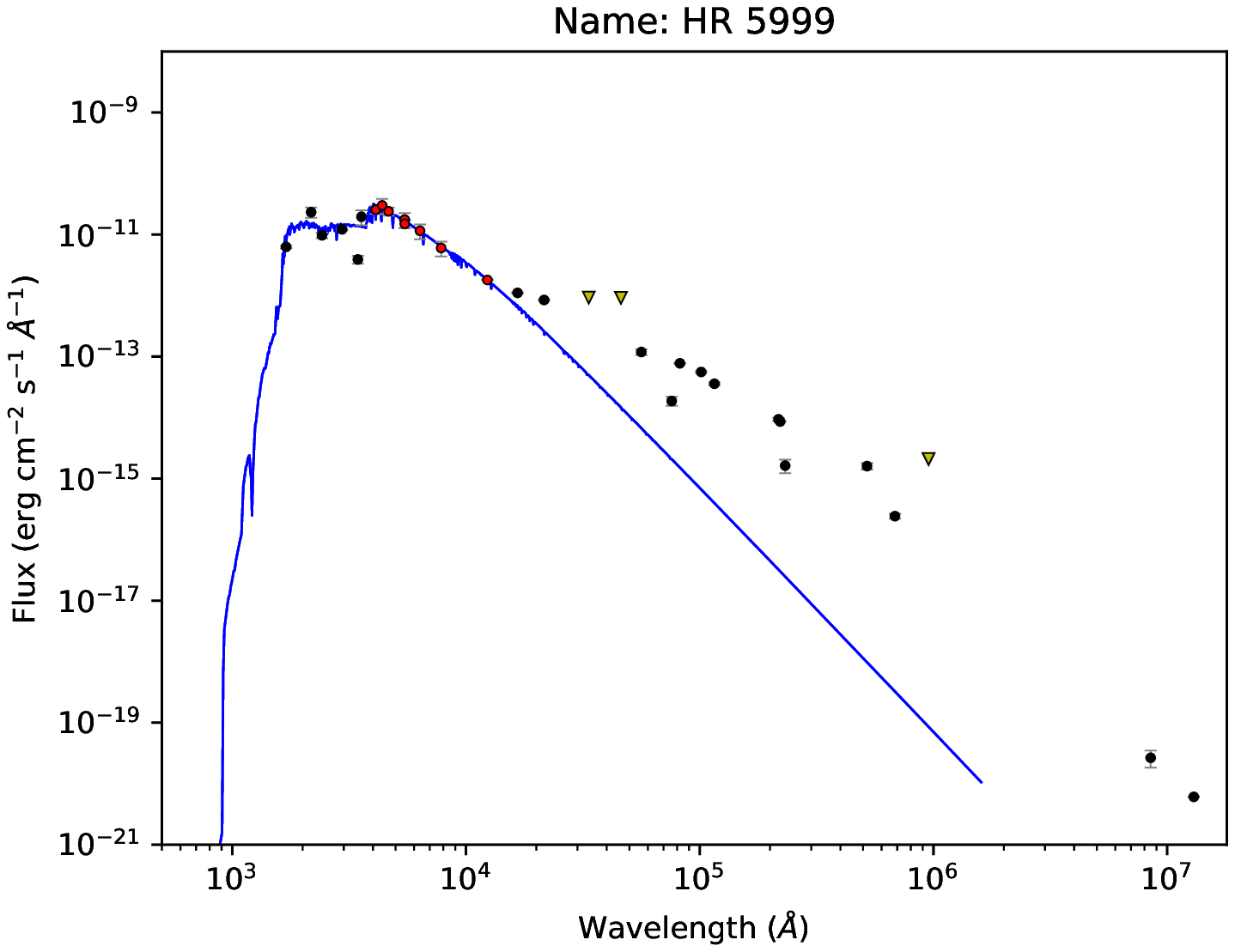}
    \includegraphics[width=0.33\textwidth]{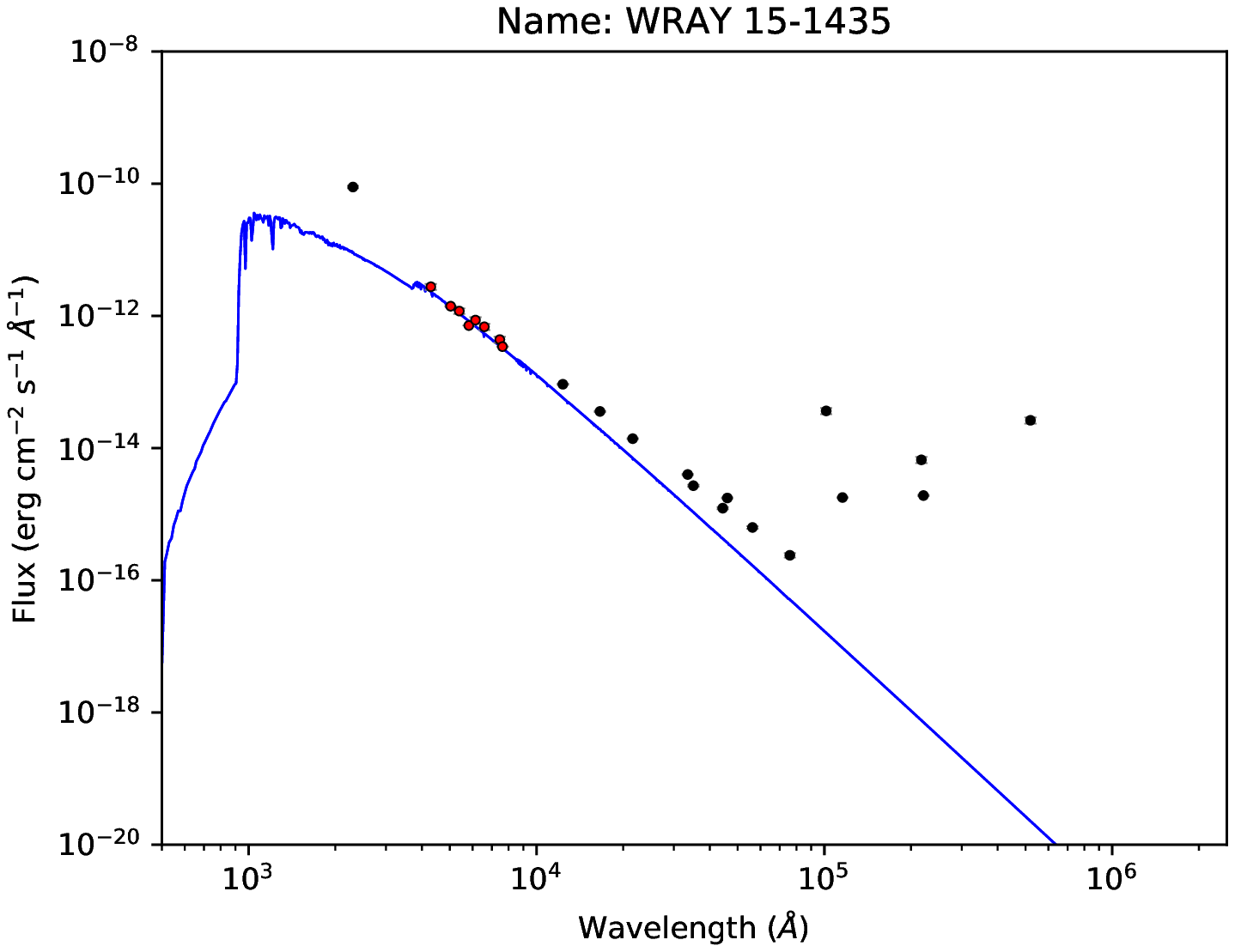}
\end{figure}

\newpage

\onecolumn

\begin{figure} [h]
 \centering
    \includegraphics[width=0.33\textwidth]{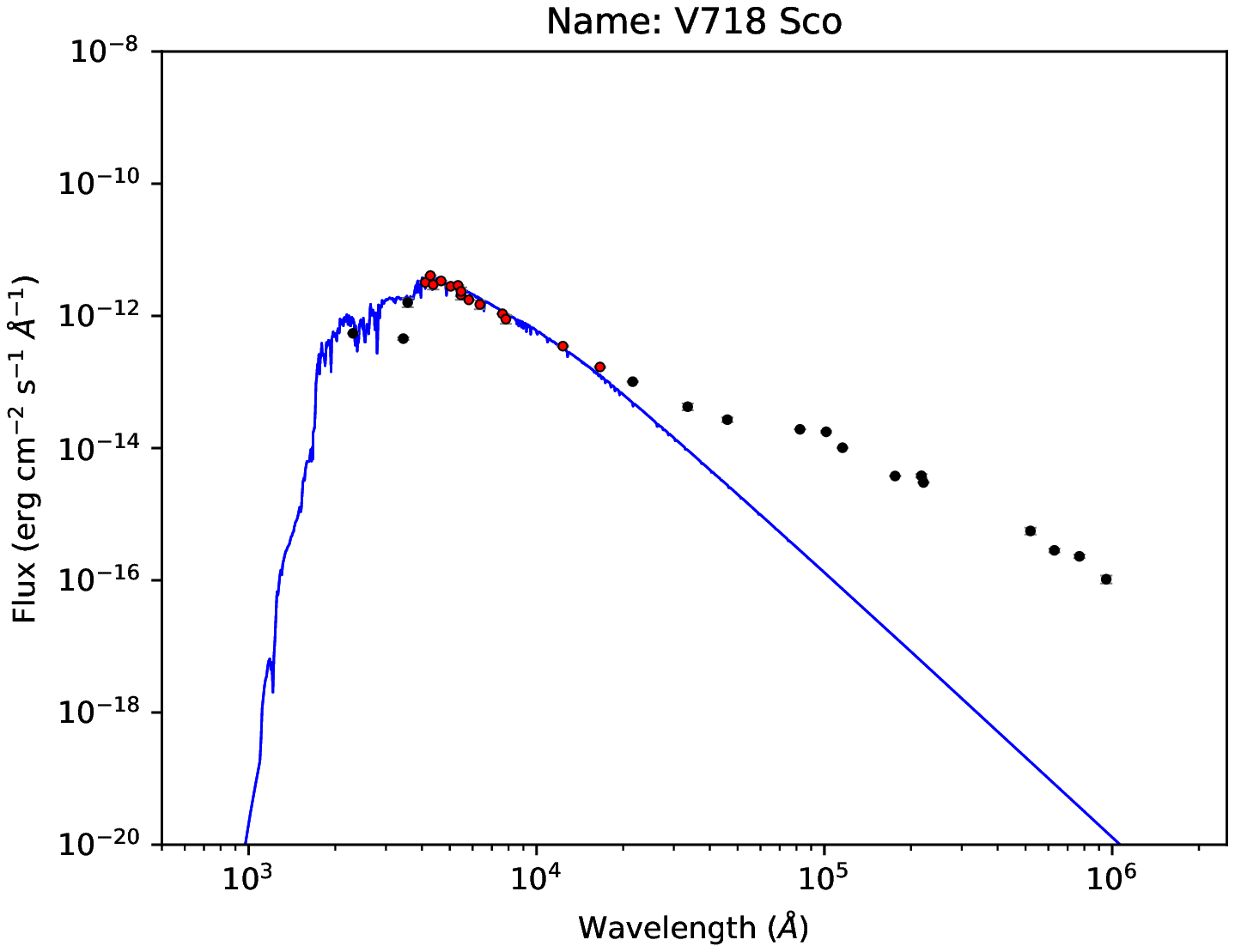}
    \includegraphics[width=0.33\textwidth]{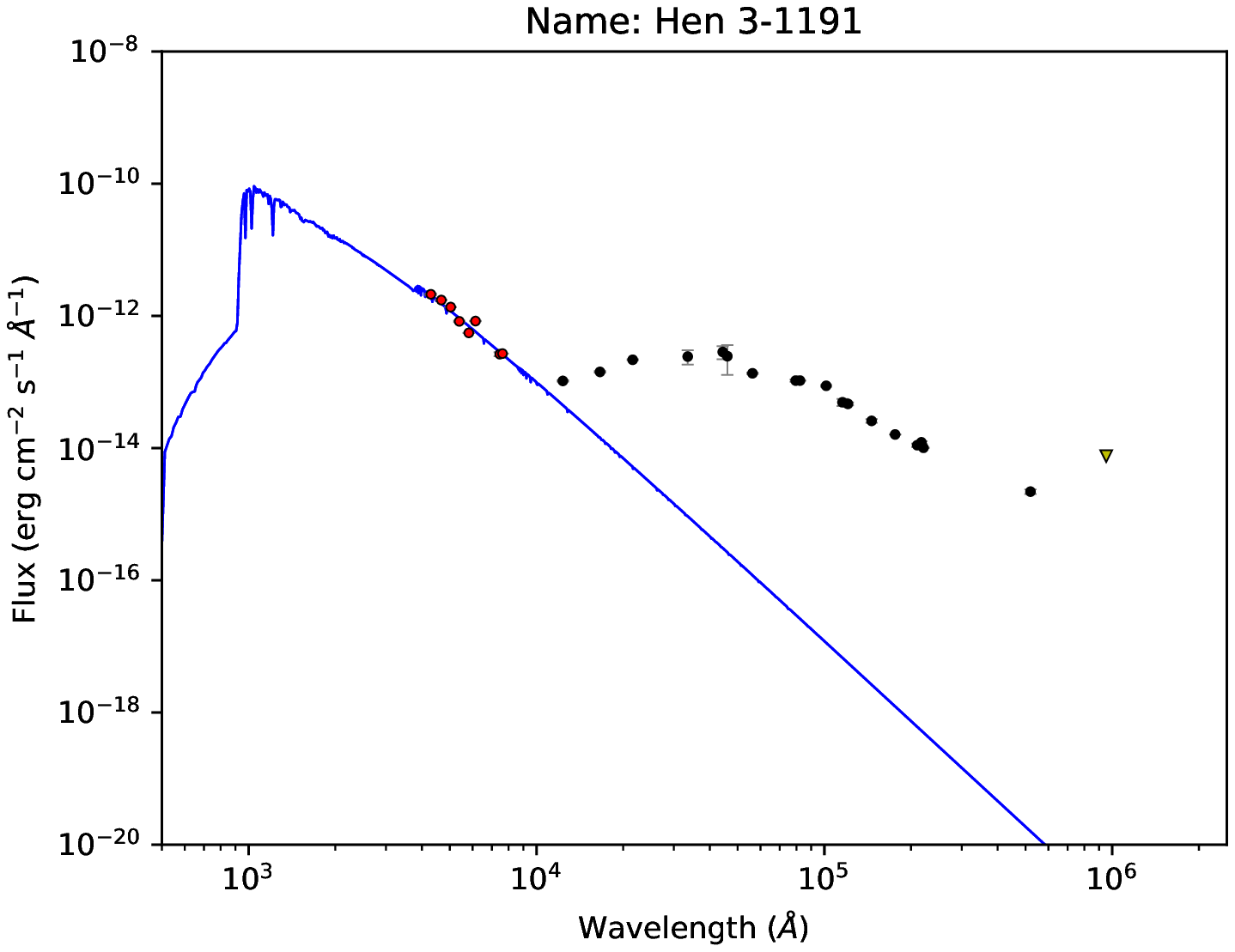}
    \includegraphics[width=0.33\textwidth]{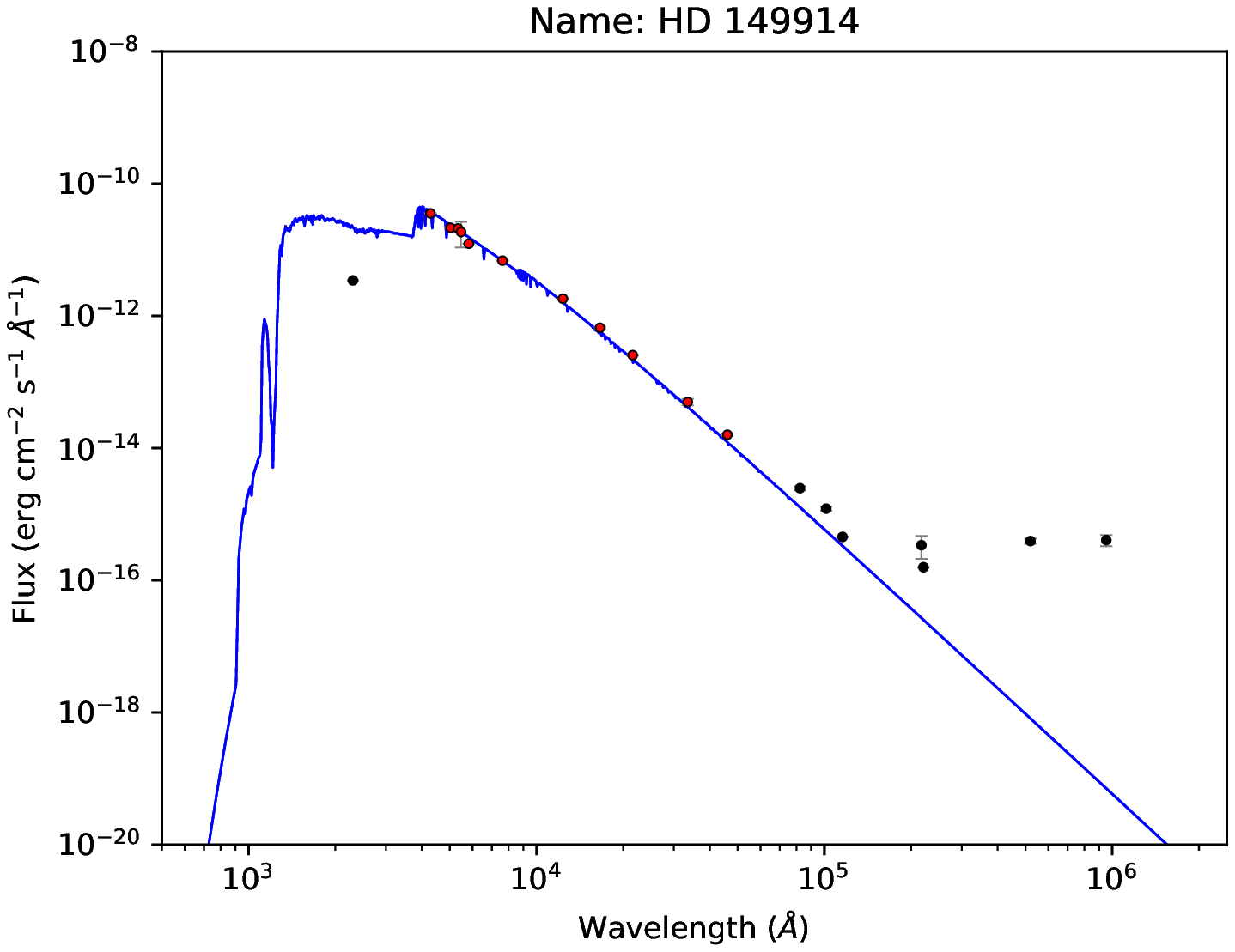}
\end{figure}

\begin{figure} [h]
 \centering 
    \includegraphics[width=0.33\textwidth]{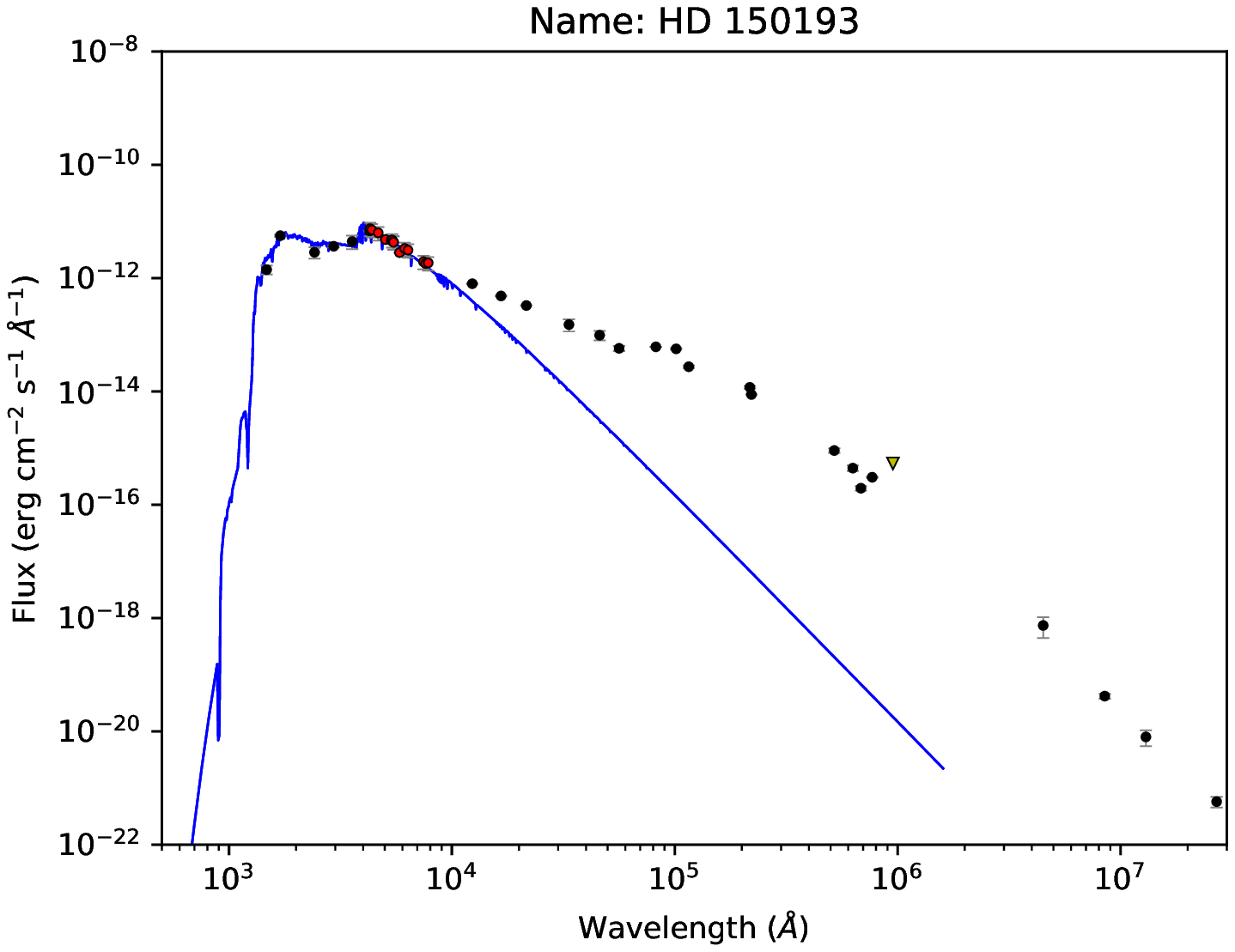}
    \includegraphics[width=0.33\textwidth]{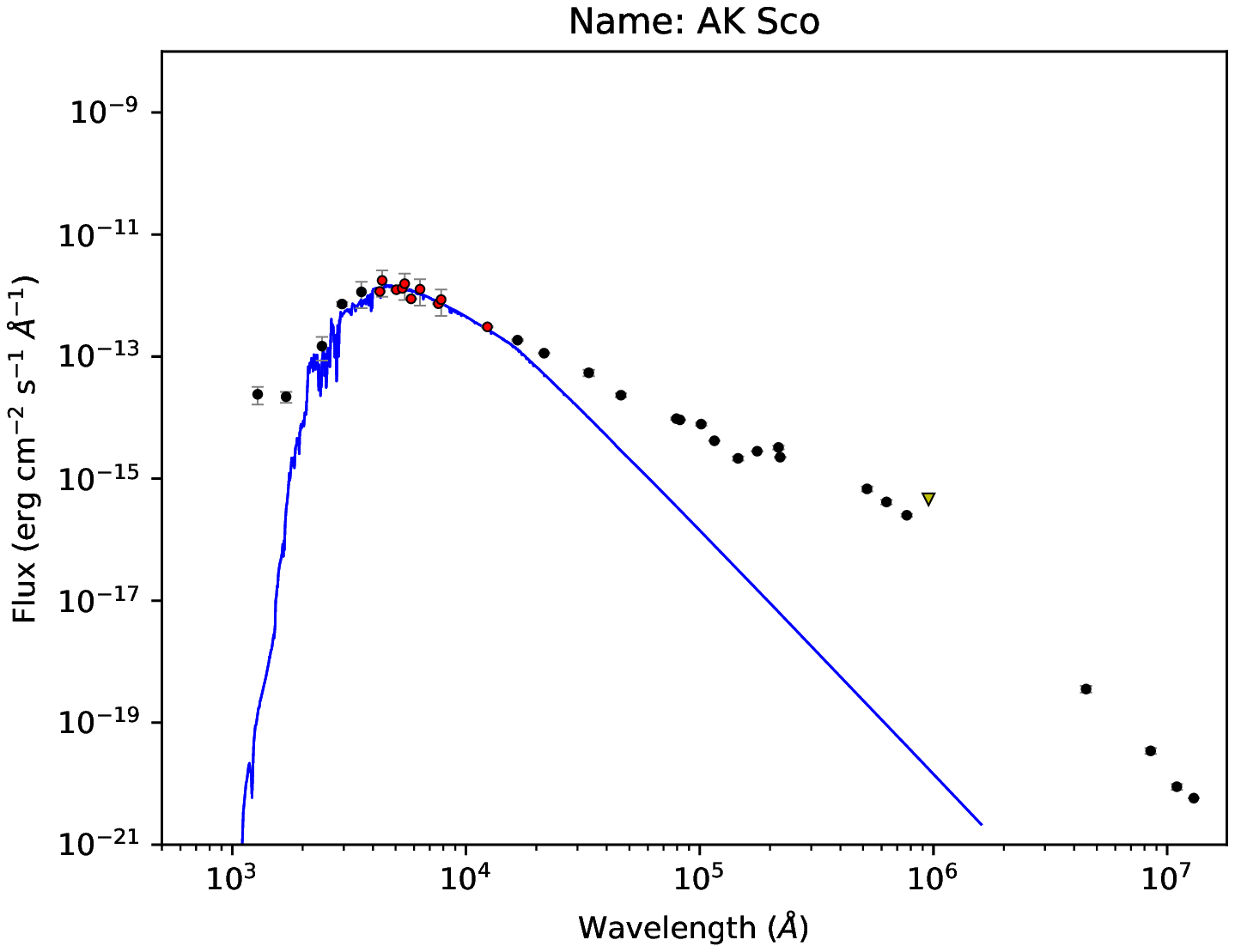}
    \includegraphics[width=0.33\textwidth]{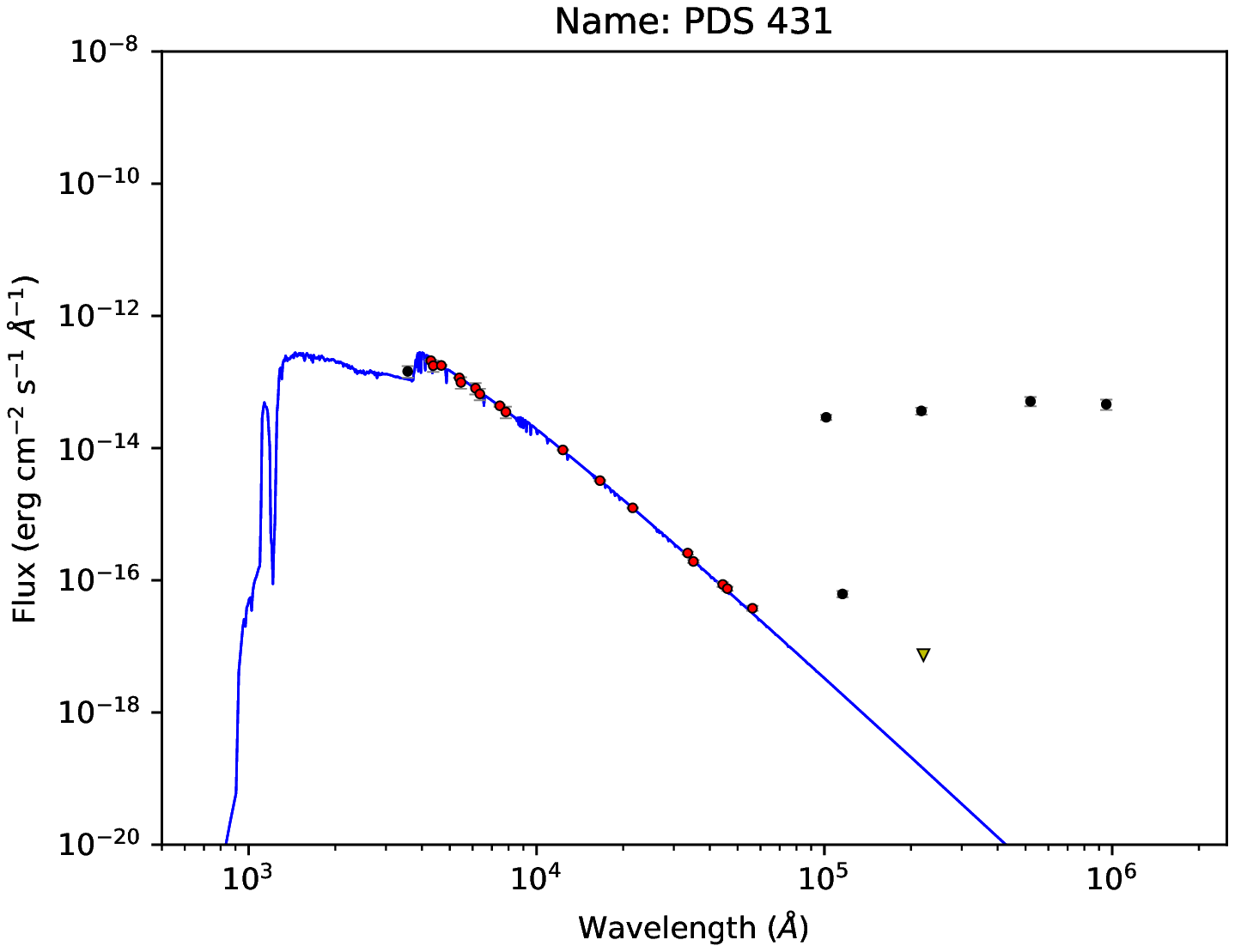}
\end{figure}

\begin{figure} [h]
 \centering
    \includegraphics[width=0.33\textwidth]{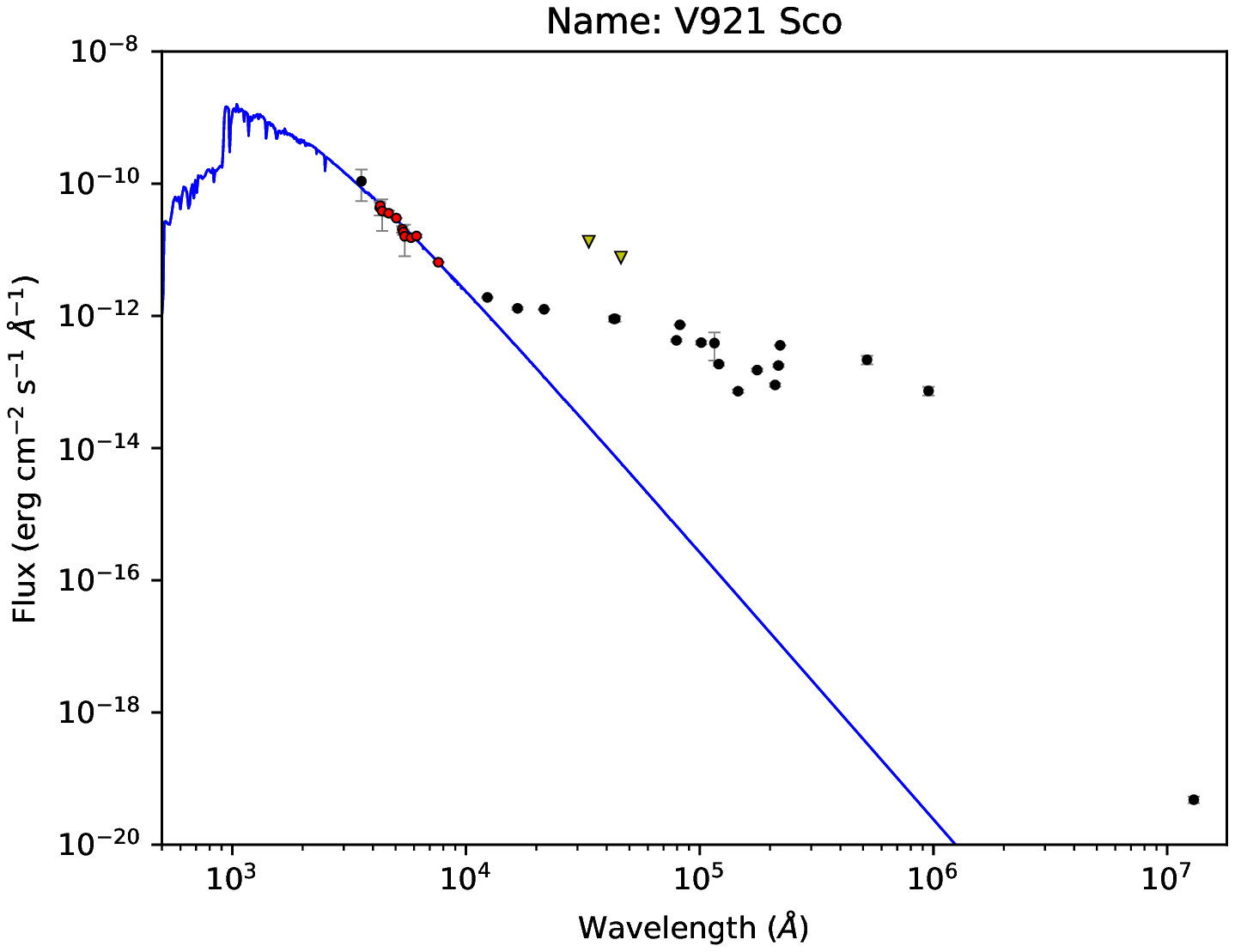}
    \includegraphics[width=0.33\textwidth]{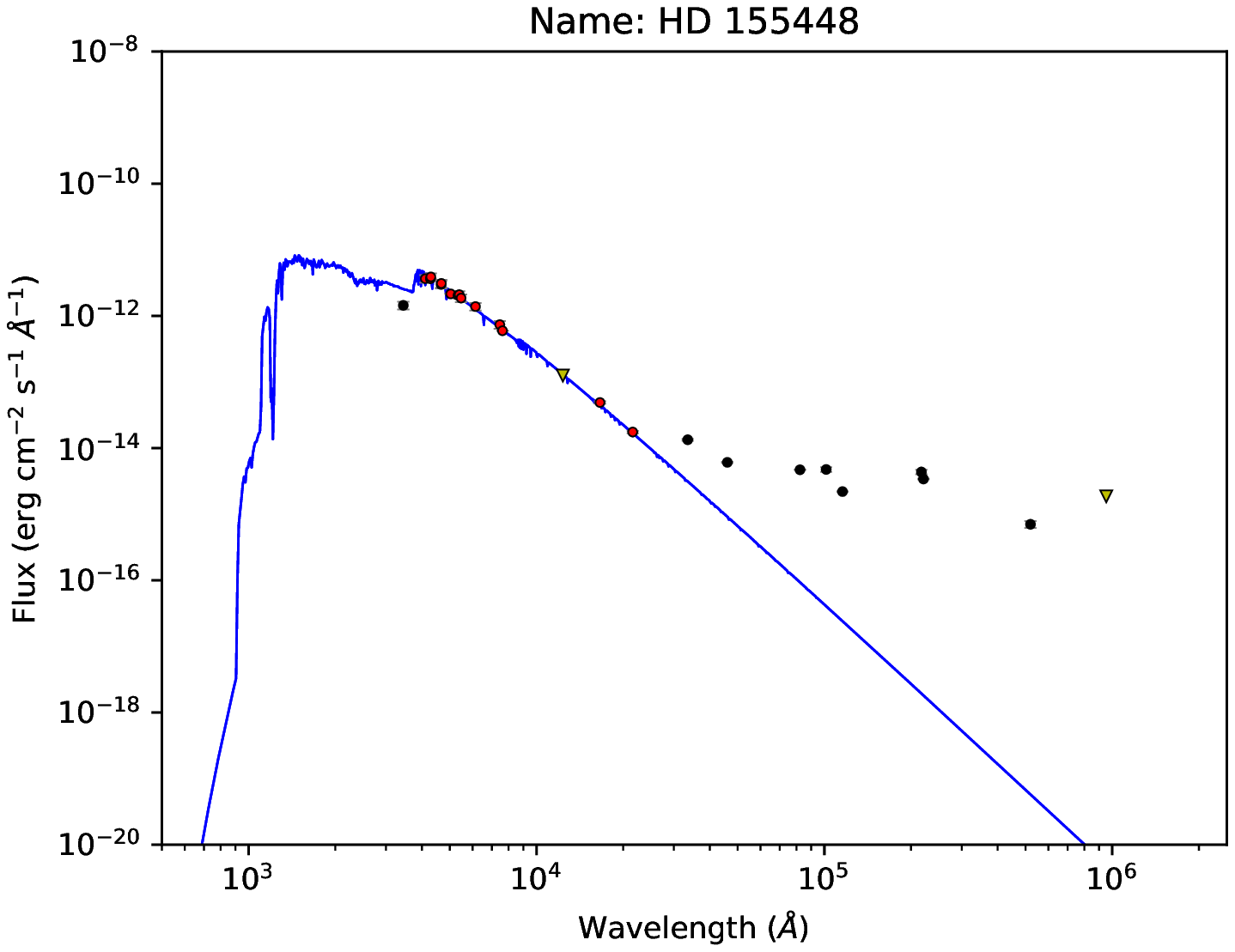}
    \includegraphics[width=0.33\textwidth]{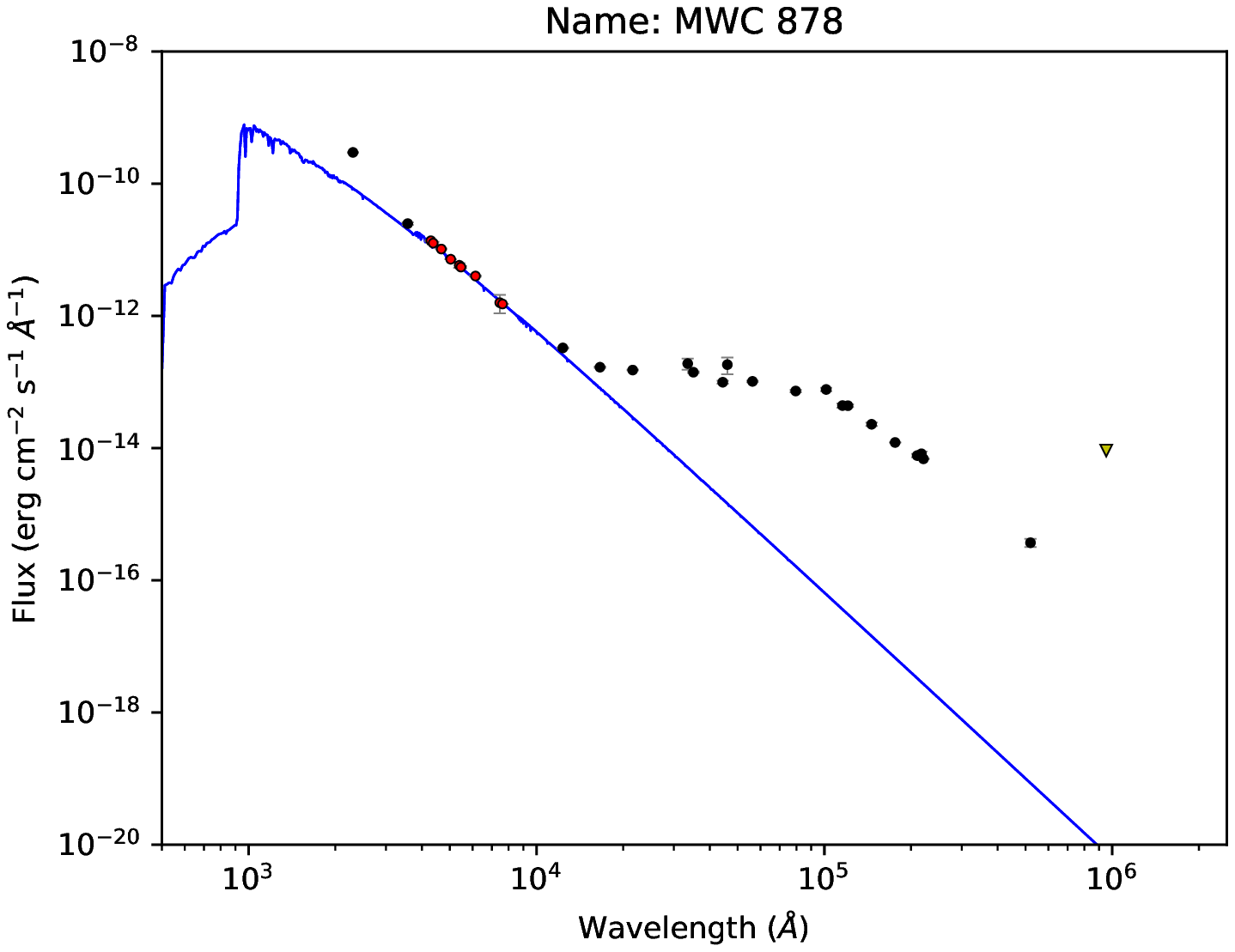}
\end{figure}

\begin{figure} [h]
 \centering
    \includegraphics[width=0.33\textwidth]{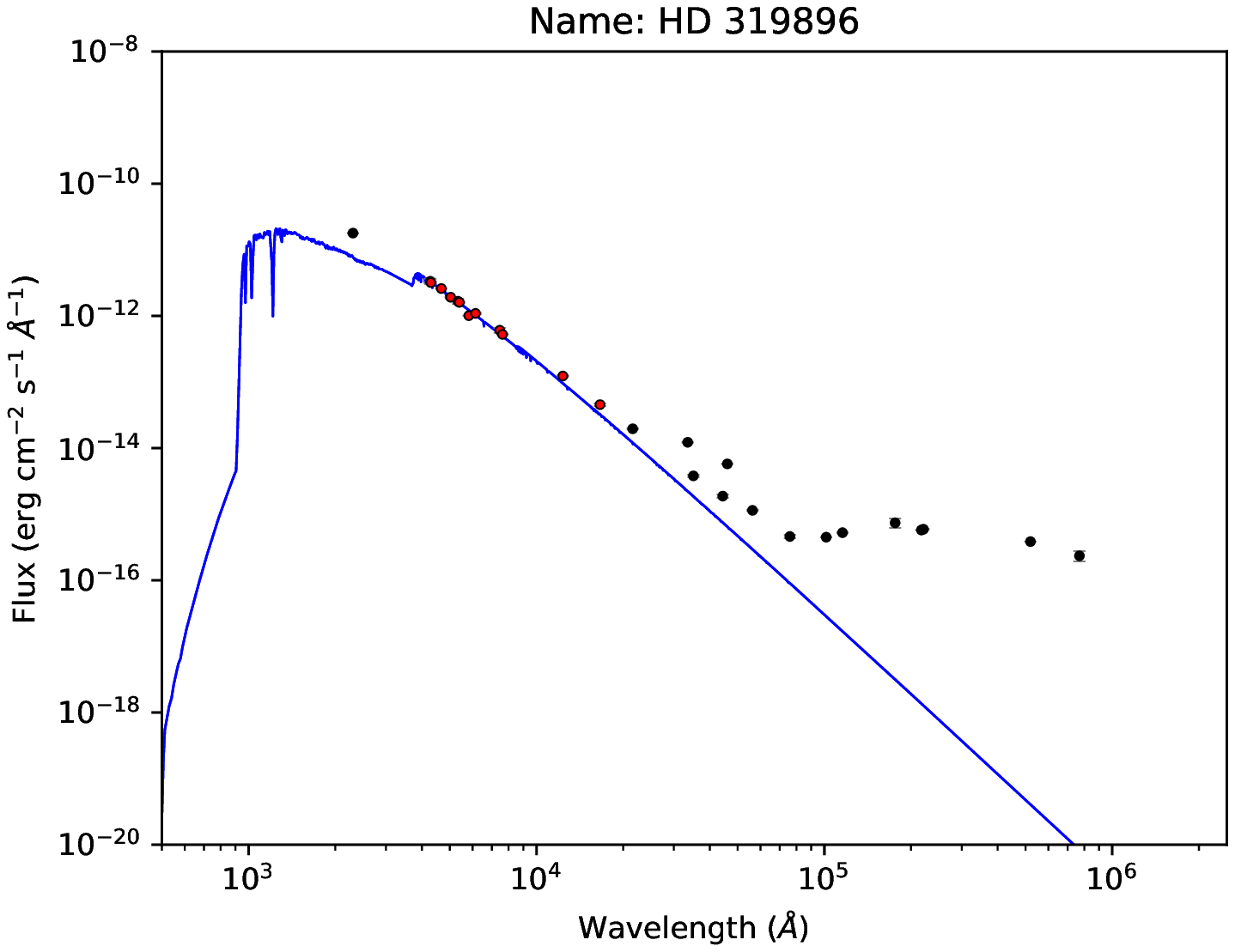}
    \includegraphics[width=0.33\textwidth]{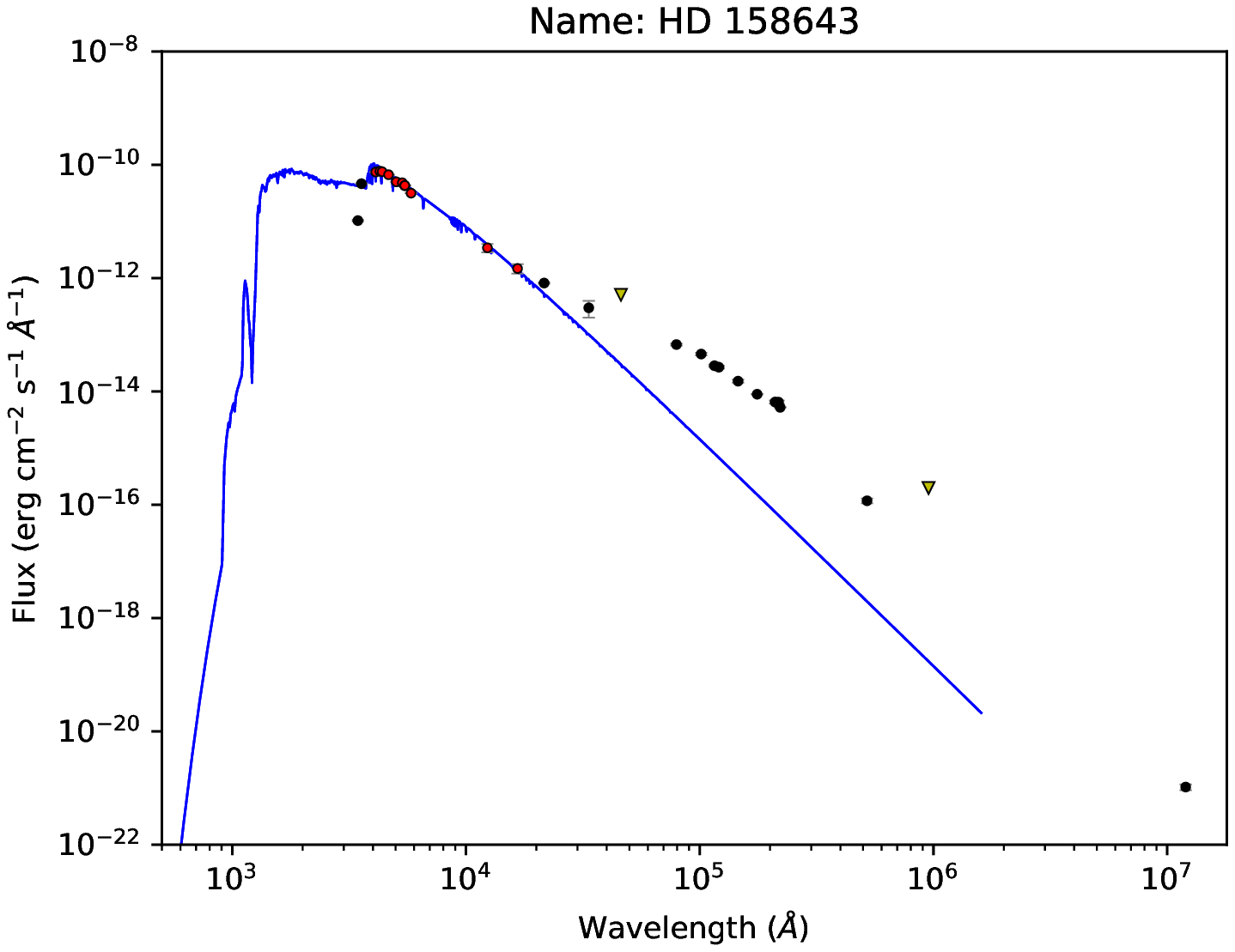}
    \includegraphics[width=0.33\textwidth]{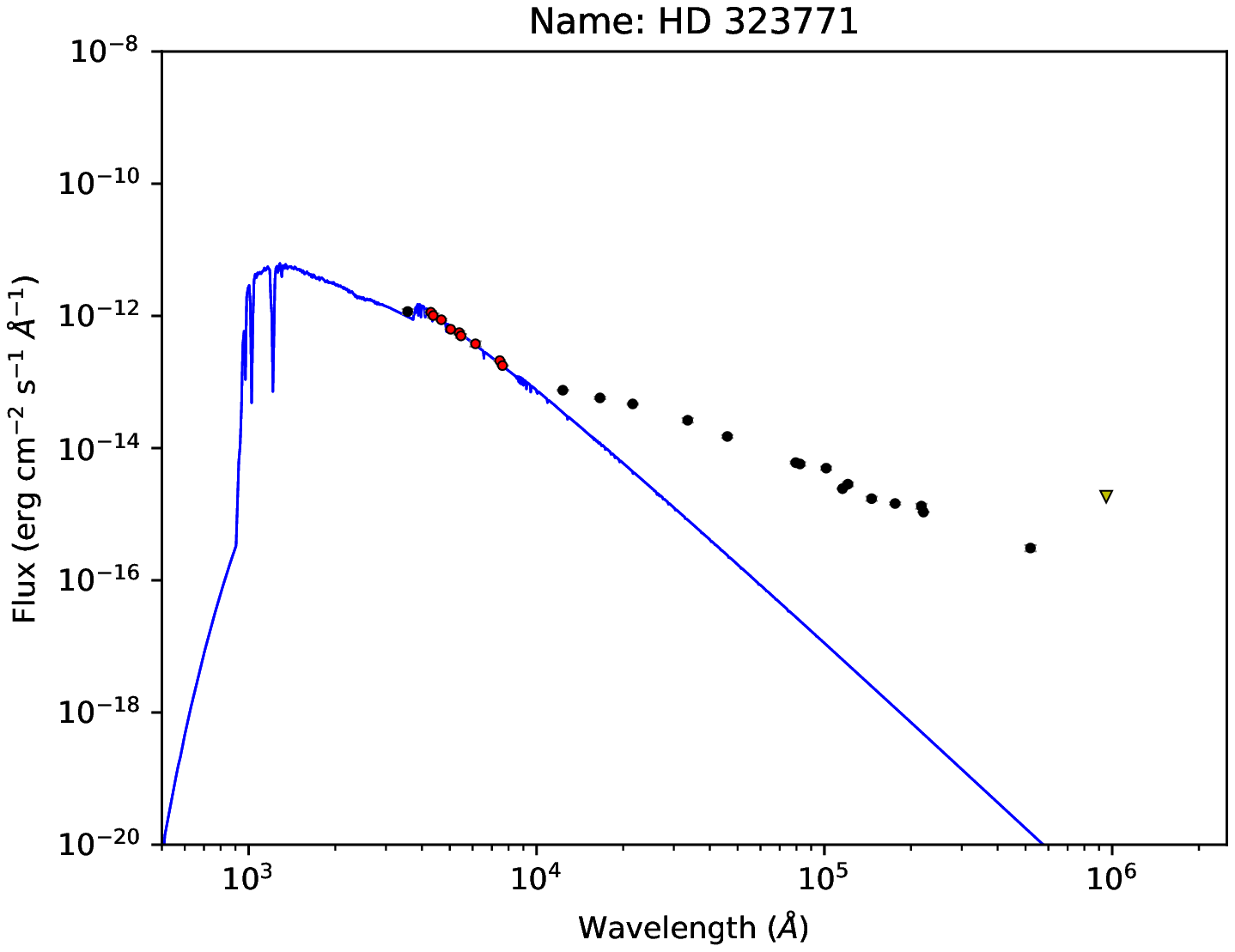}
\end{figure}

\newpage

\onecolumn

\begin{figure} [h]
 \centering
    \includegraphics[width=0.33\textwidth]{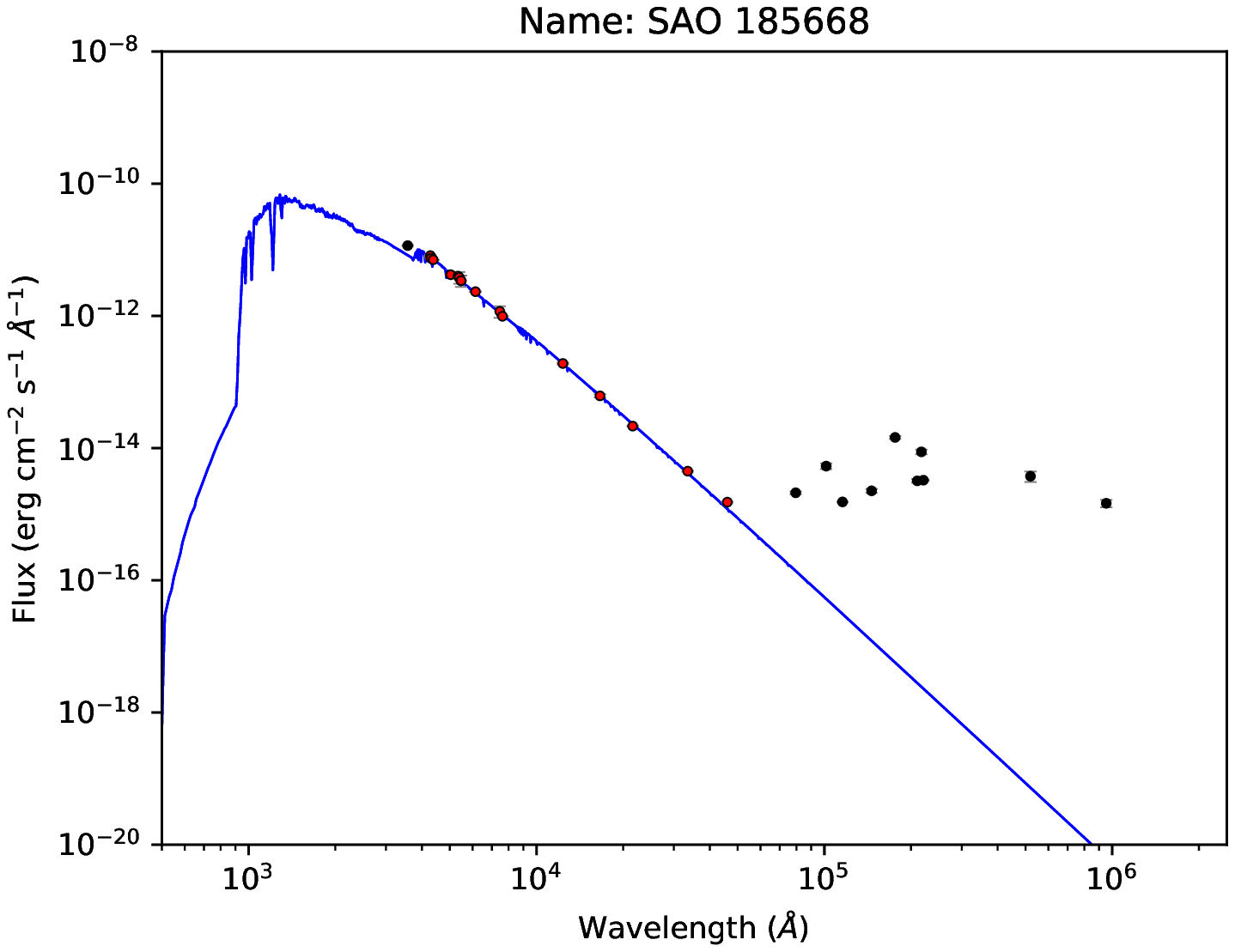}
    \includegraphics[width=0.33\textwidth]{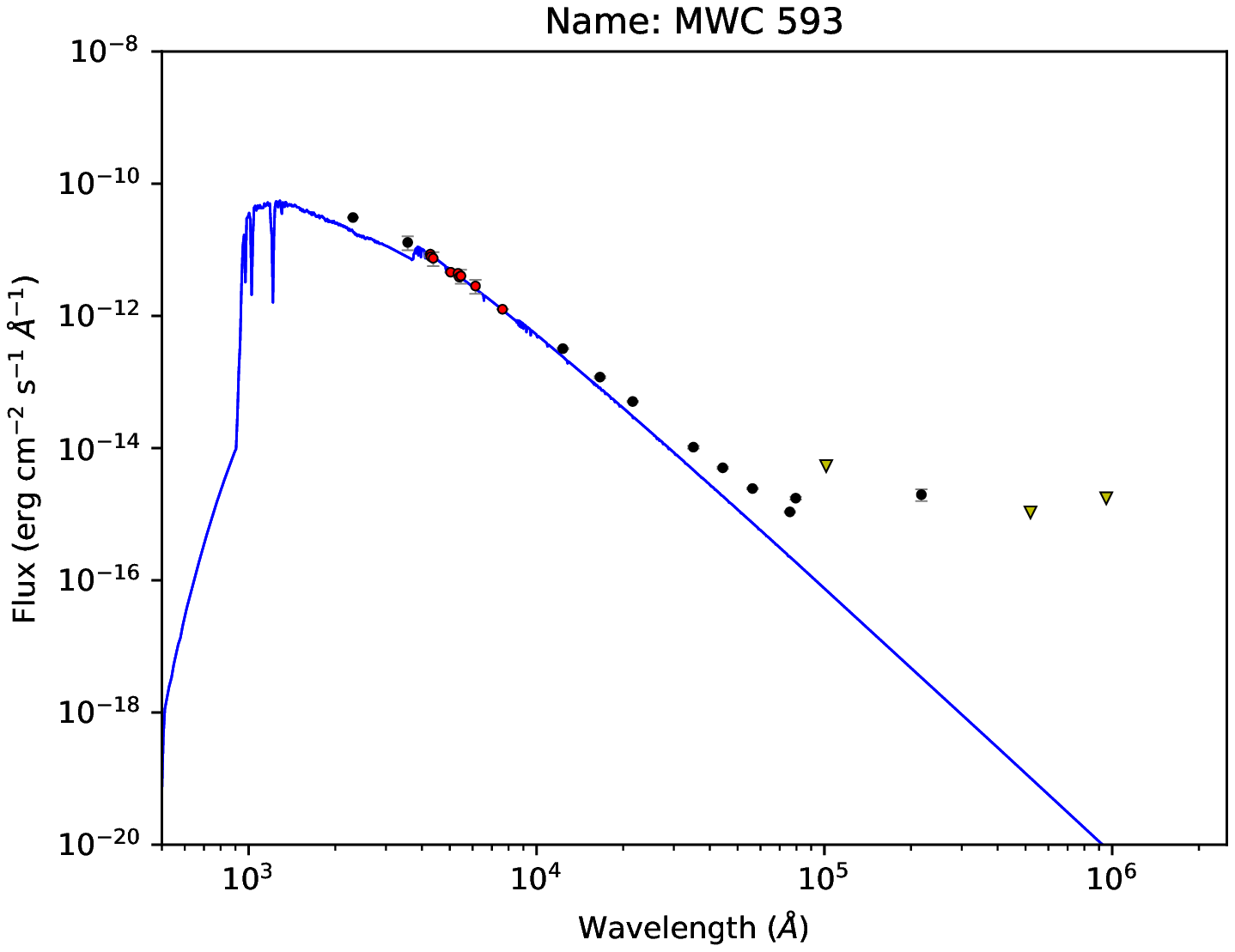}
    \includegraphics[width=0.33\textwidth]{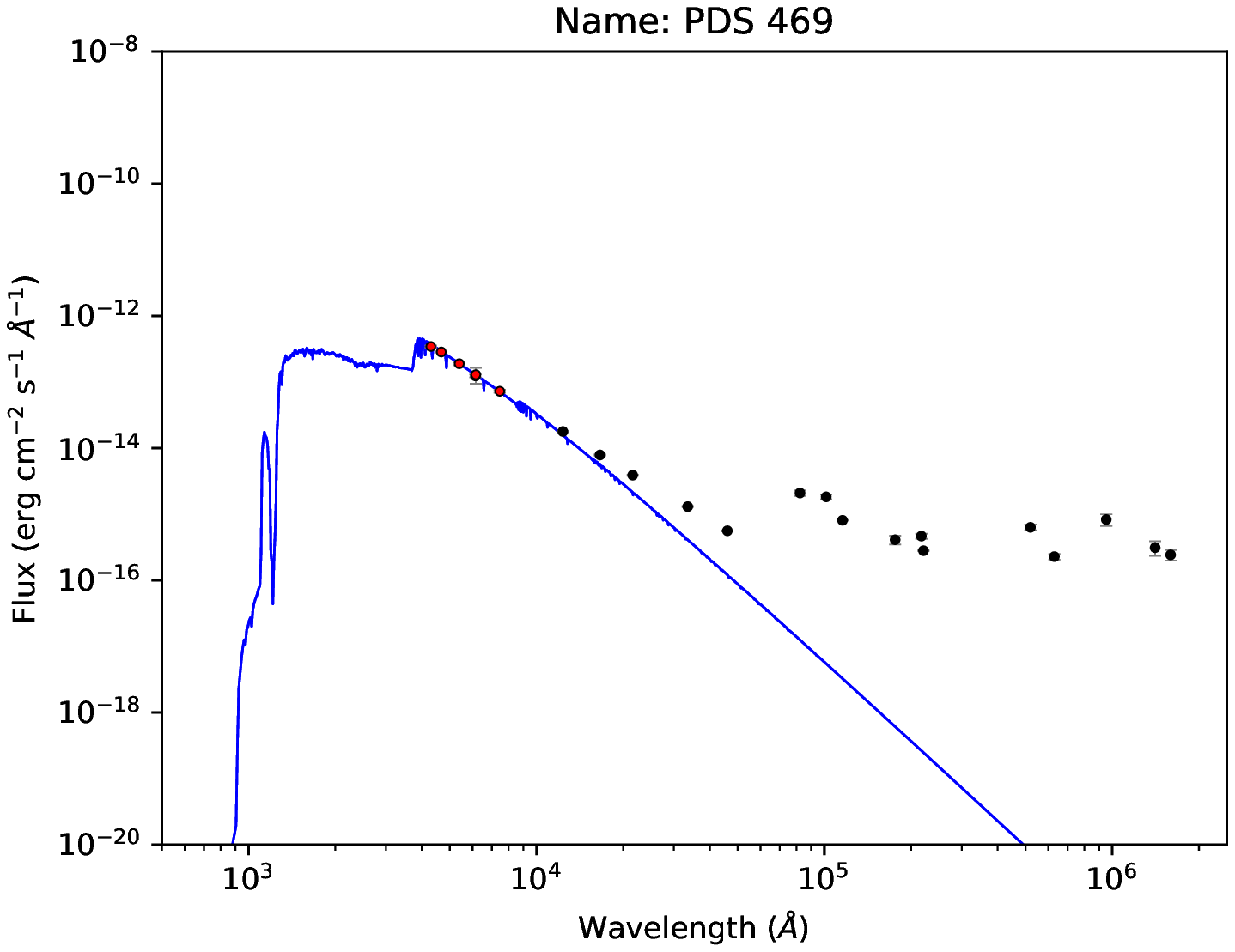}
\end{figure}

\begin{figure} [h]
 \centering
    \includegraphics[width=0.33\textwidth]{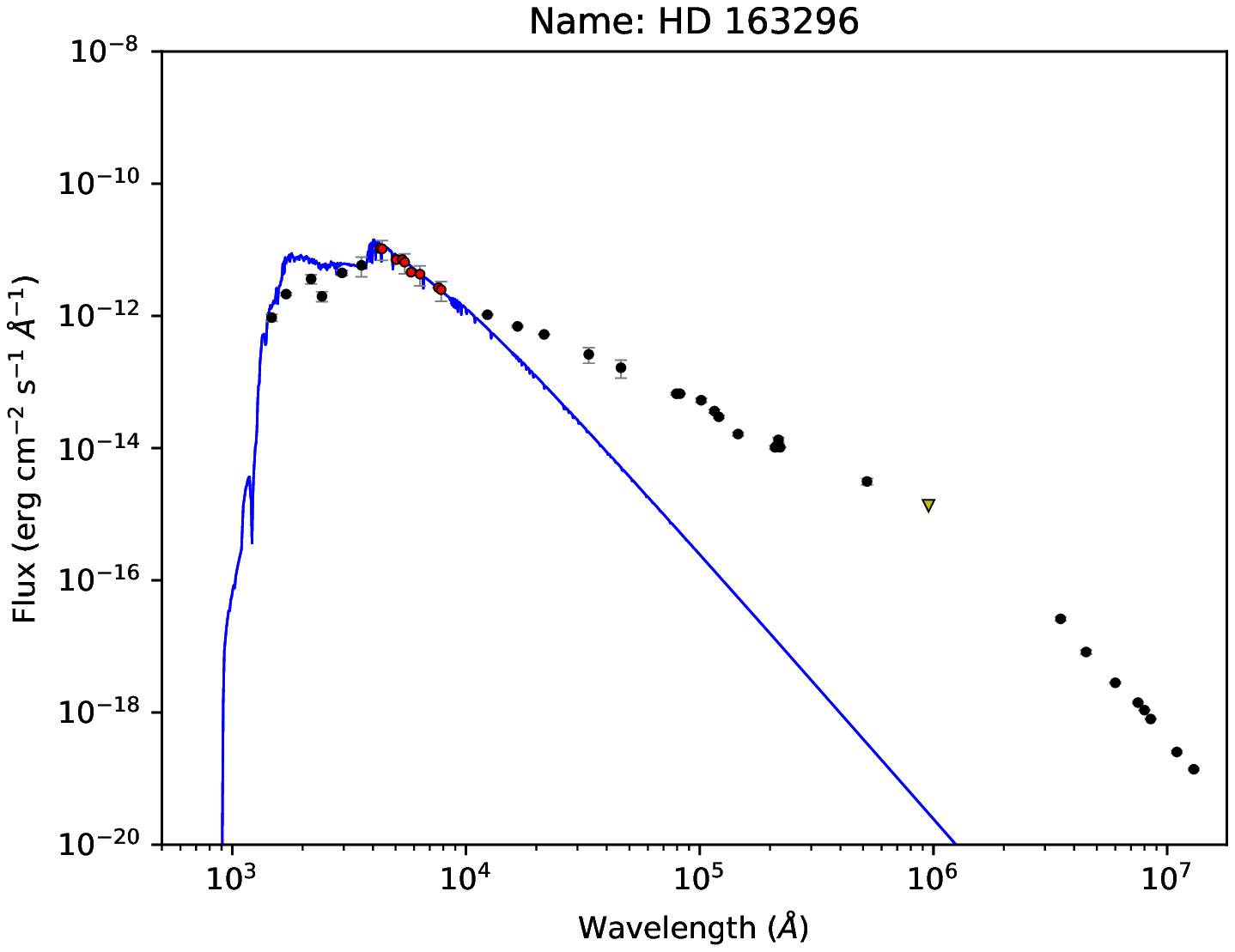}
    \includegraphics[width=0.33\textwidth]{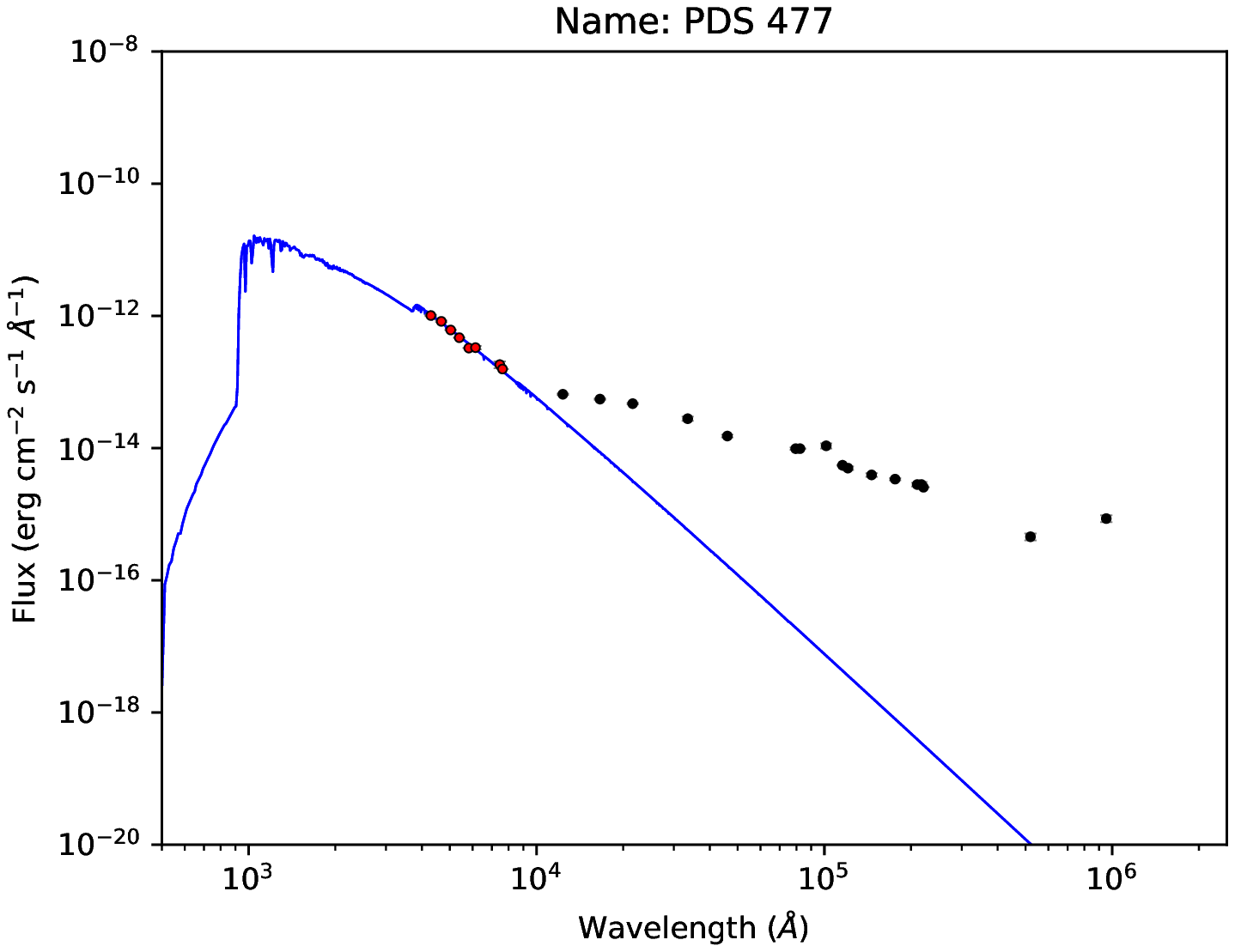}
    \includegraphics[width=0.33\textwidth]{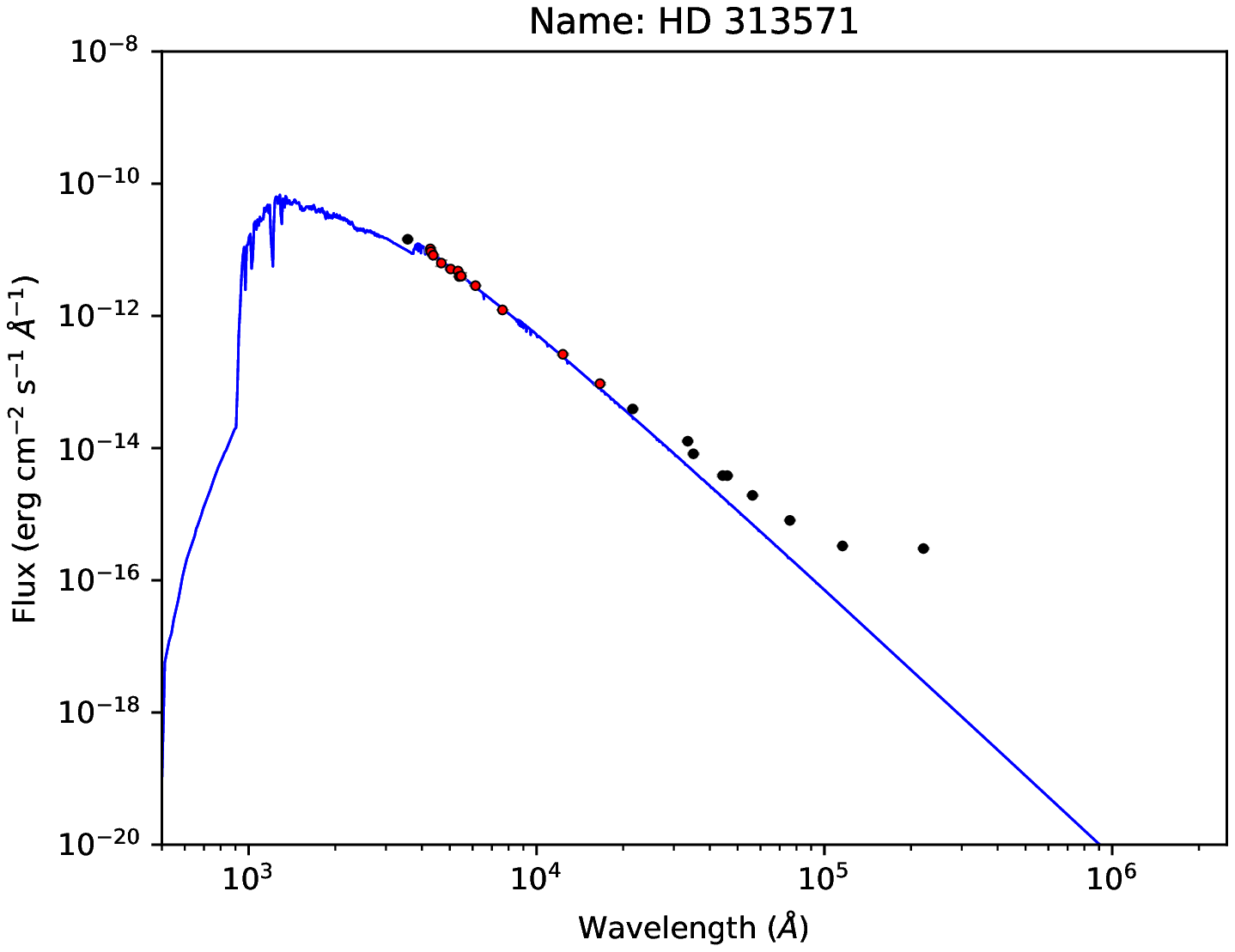}
\end{figure}

\begin{figure} [h]
 \centering
    \includegraphics[width=0.33\textwidth]{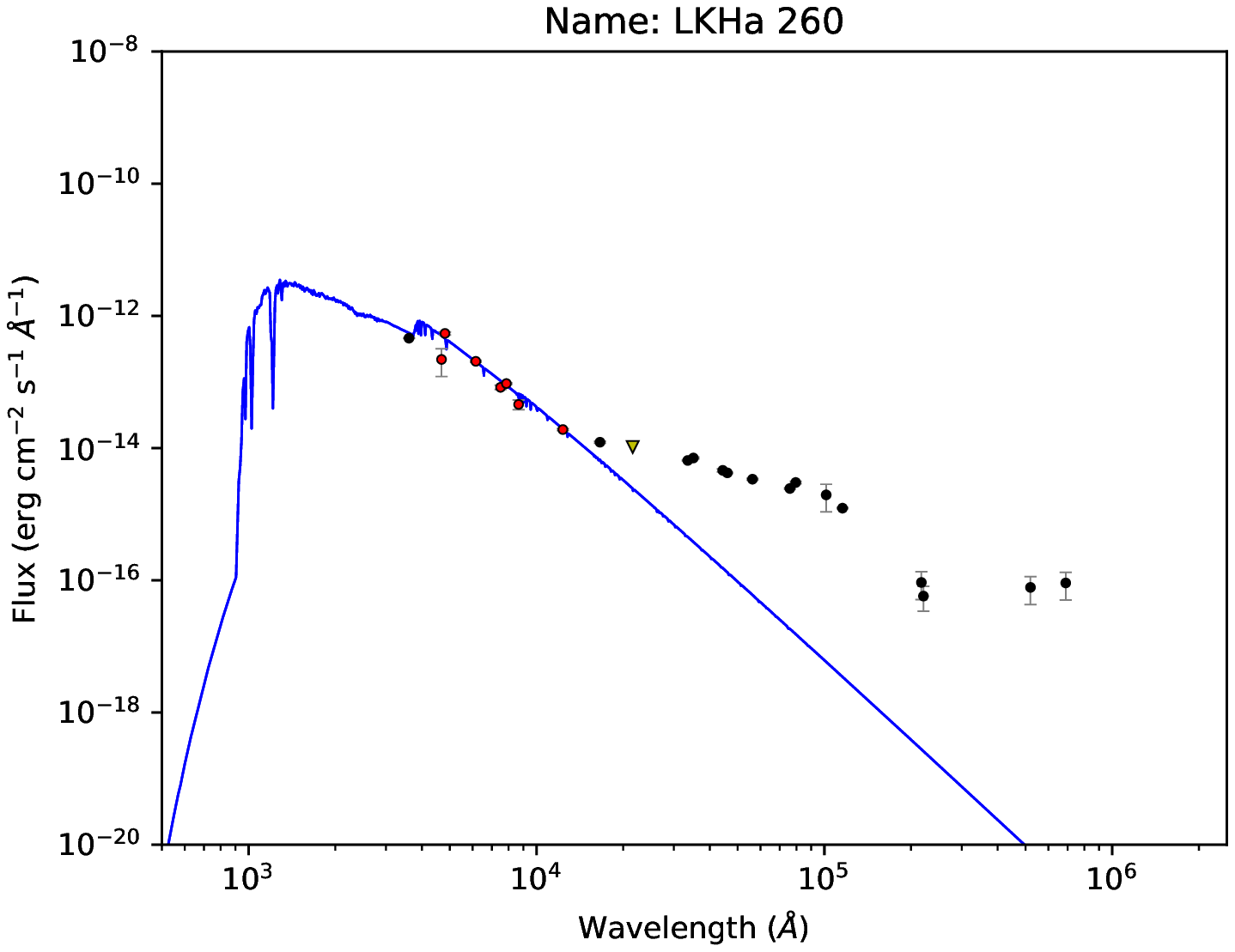}
    \includegraphics[width=0.33\textwidth]{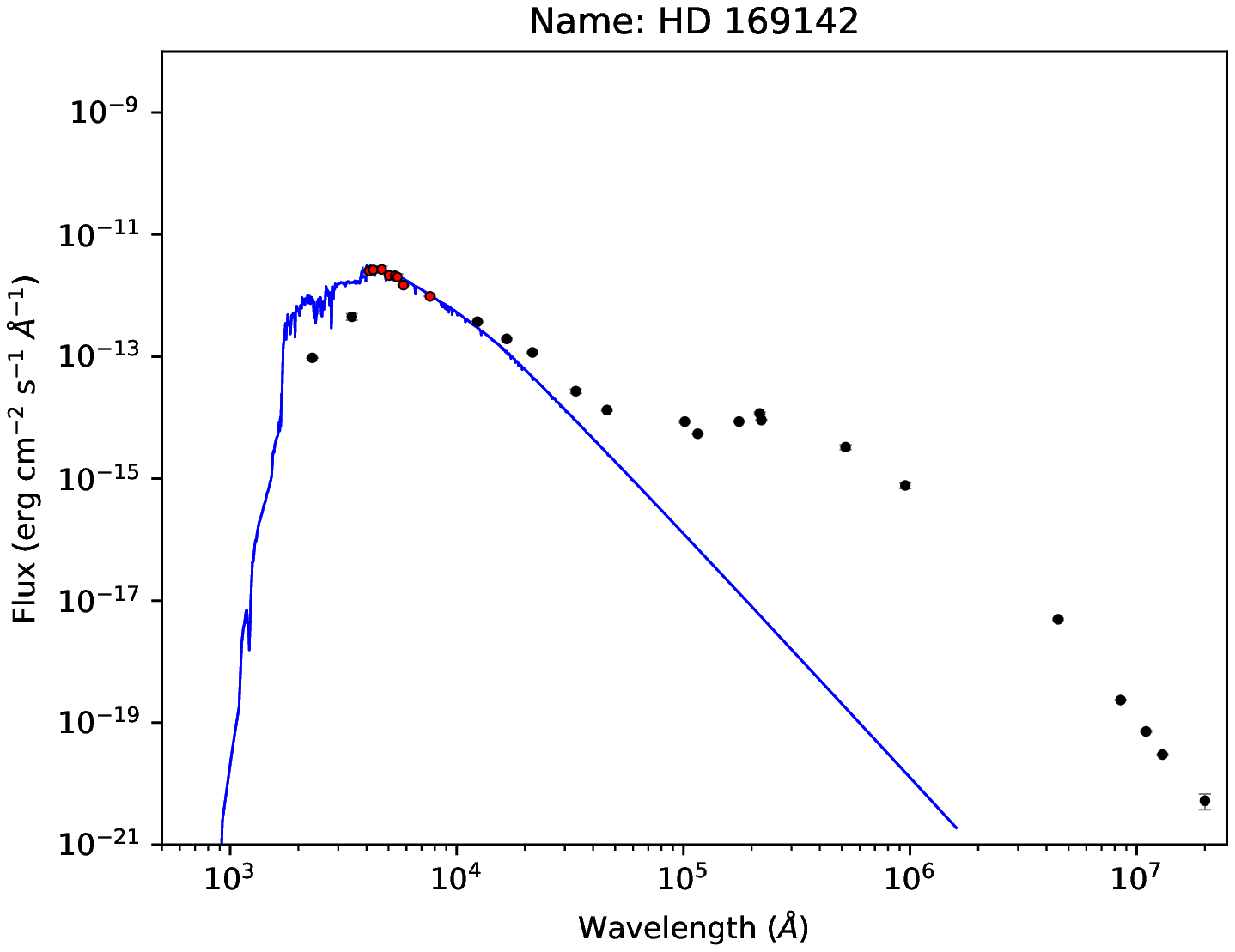}
    \includegraphics[width=0.33\textwidth]{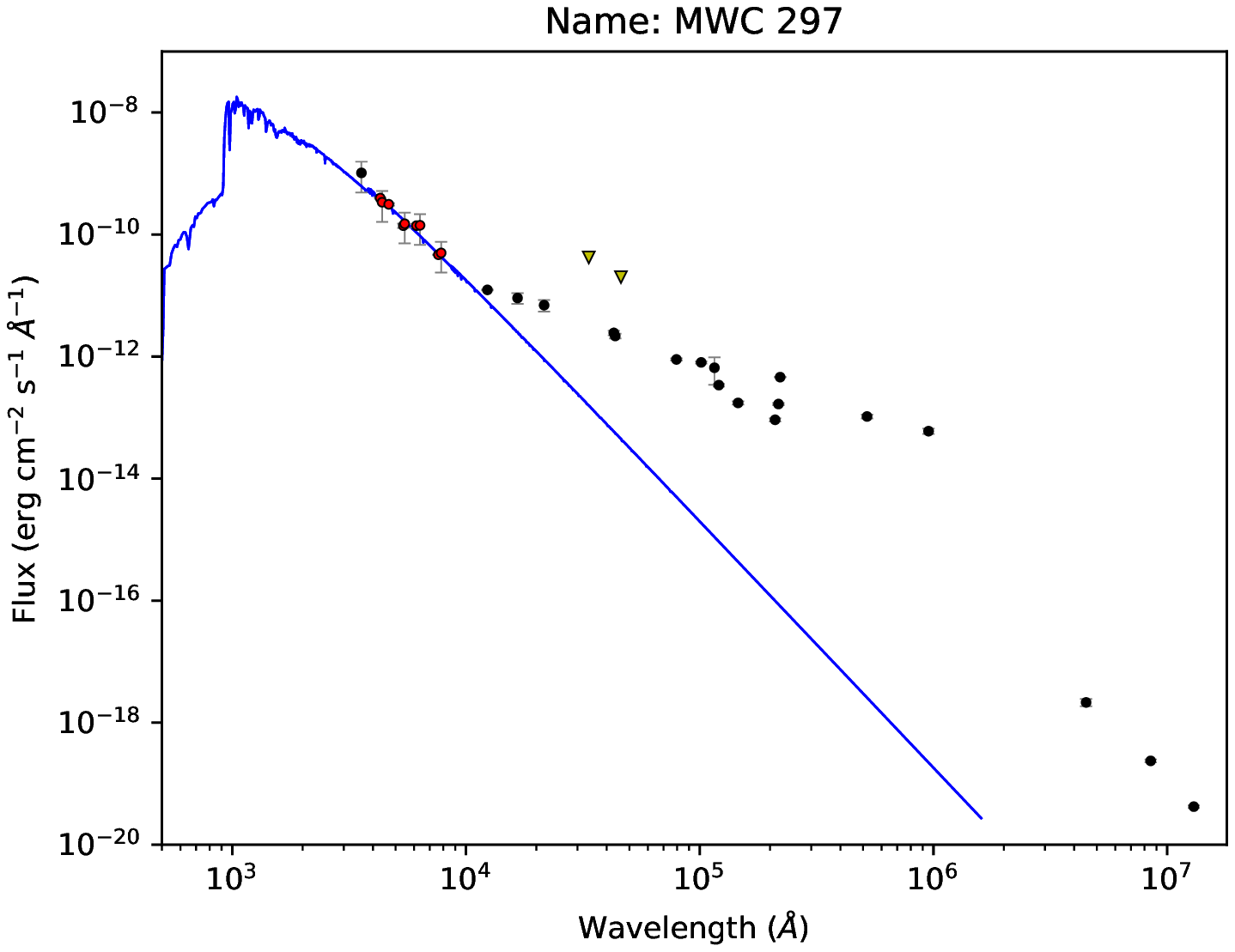}
\end{figure}

\begin{figure} [h]
 \centering
    \includegraphics[width=0.33\textwidth]{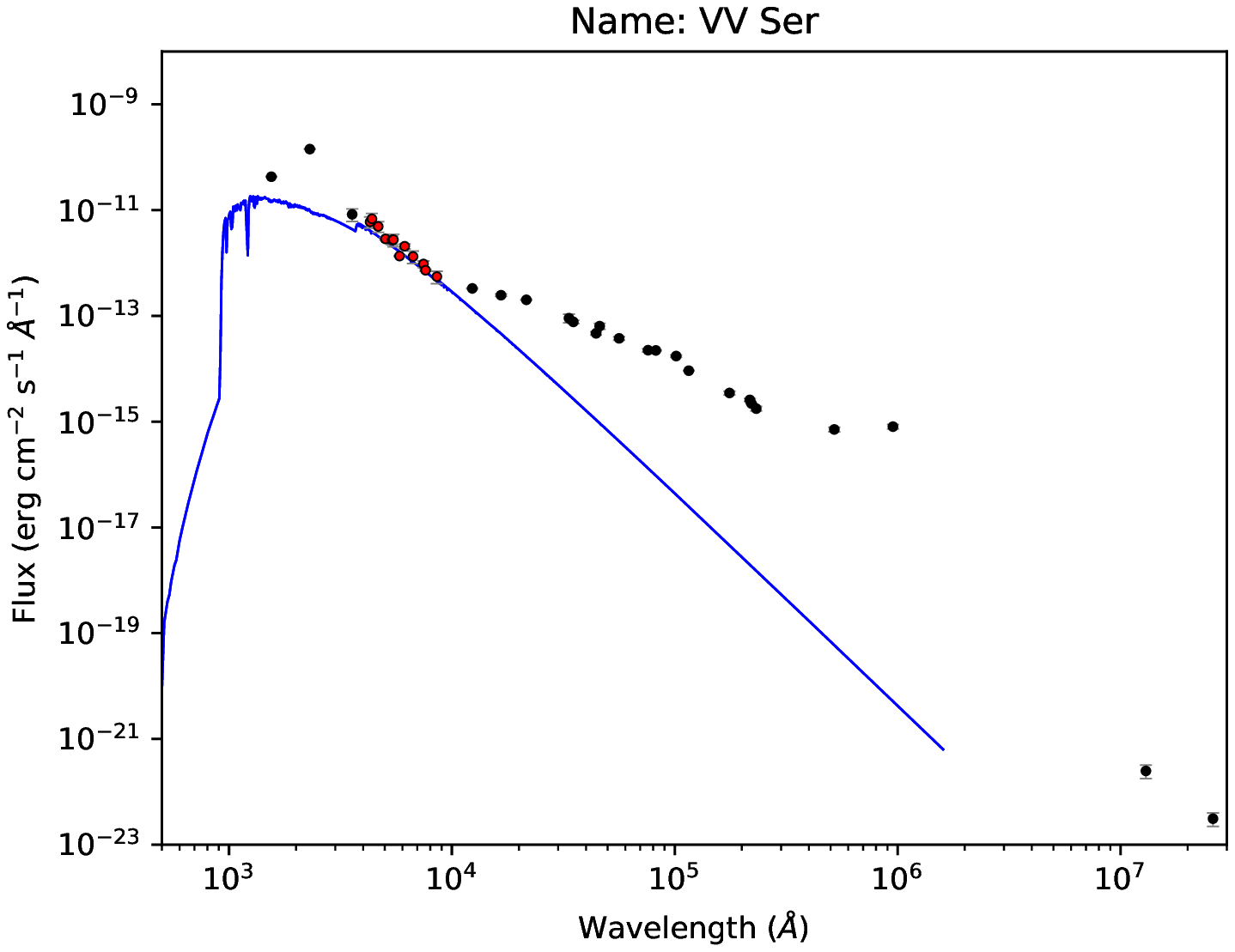}
    \includegraphics[width=0.33\textwidth]{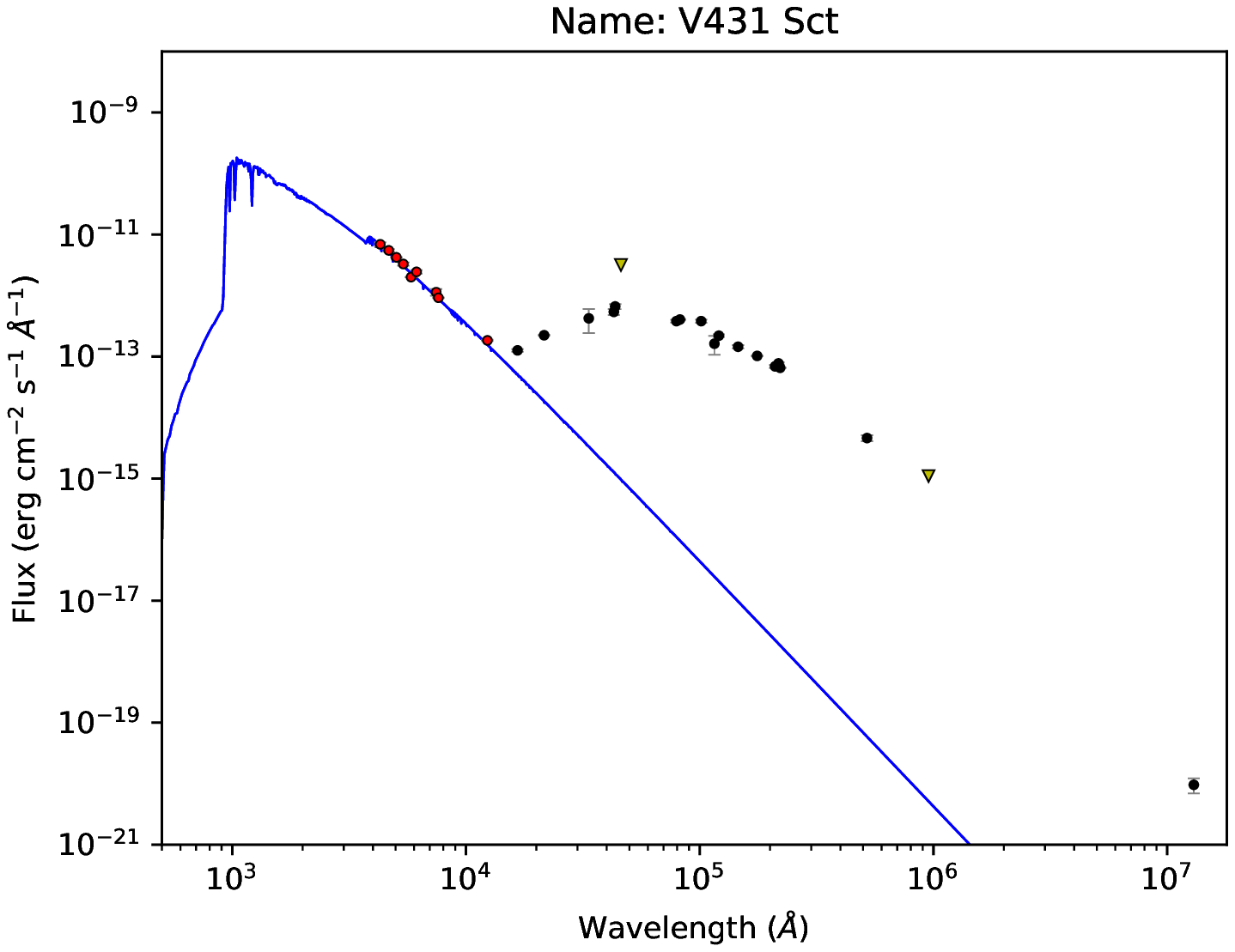}
    \includegraphics[width=0.33\textwidth]{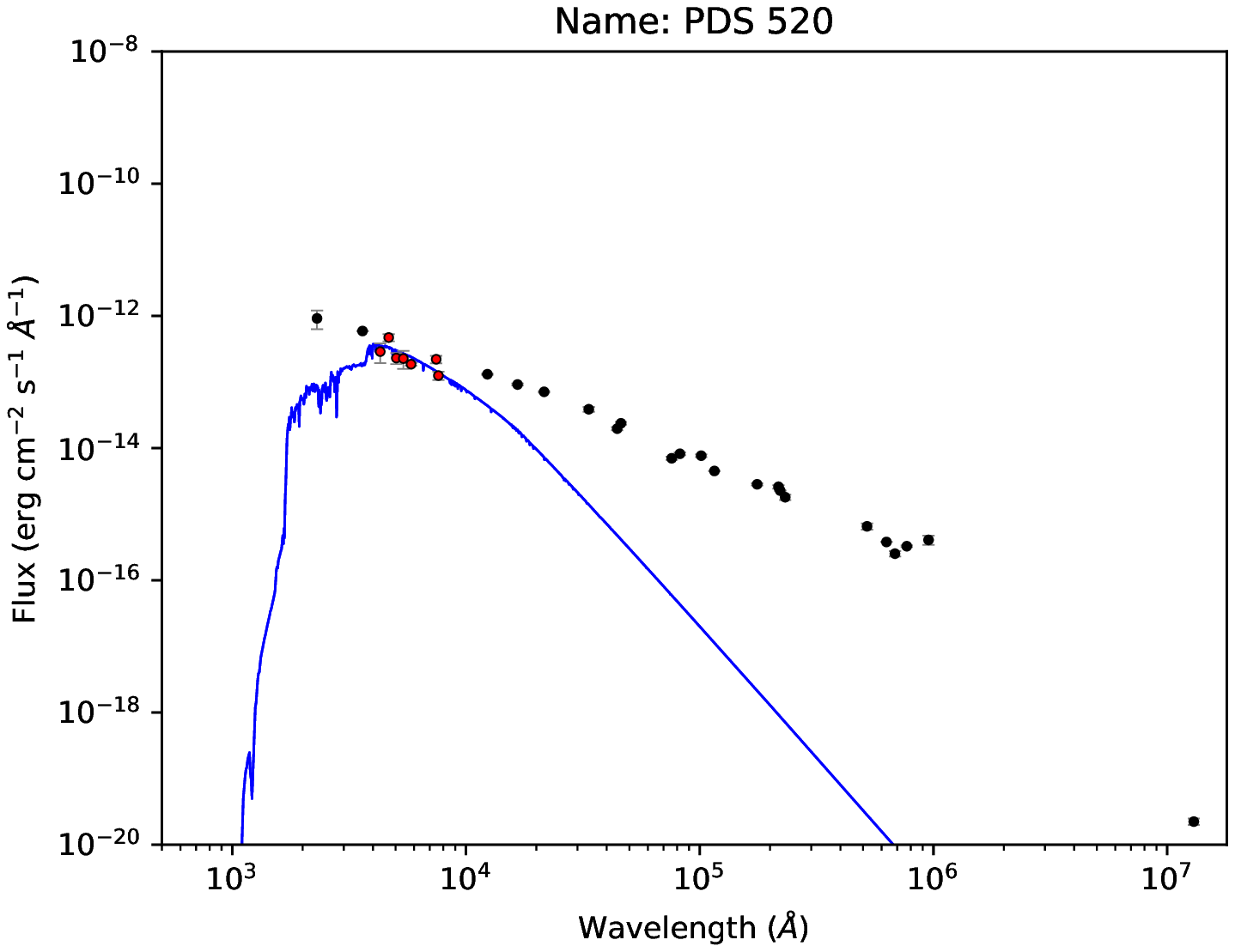}
\end{figure}

\newpage

\onecolumn

\begin{figure} [h]
 \centering
    \includegraphics[width=0.33\textwidth]{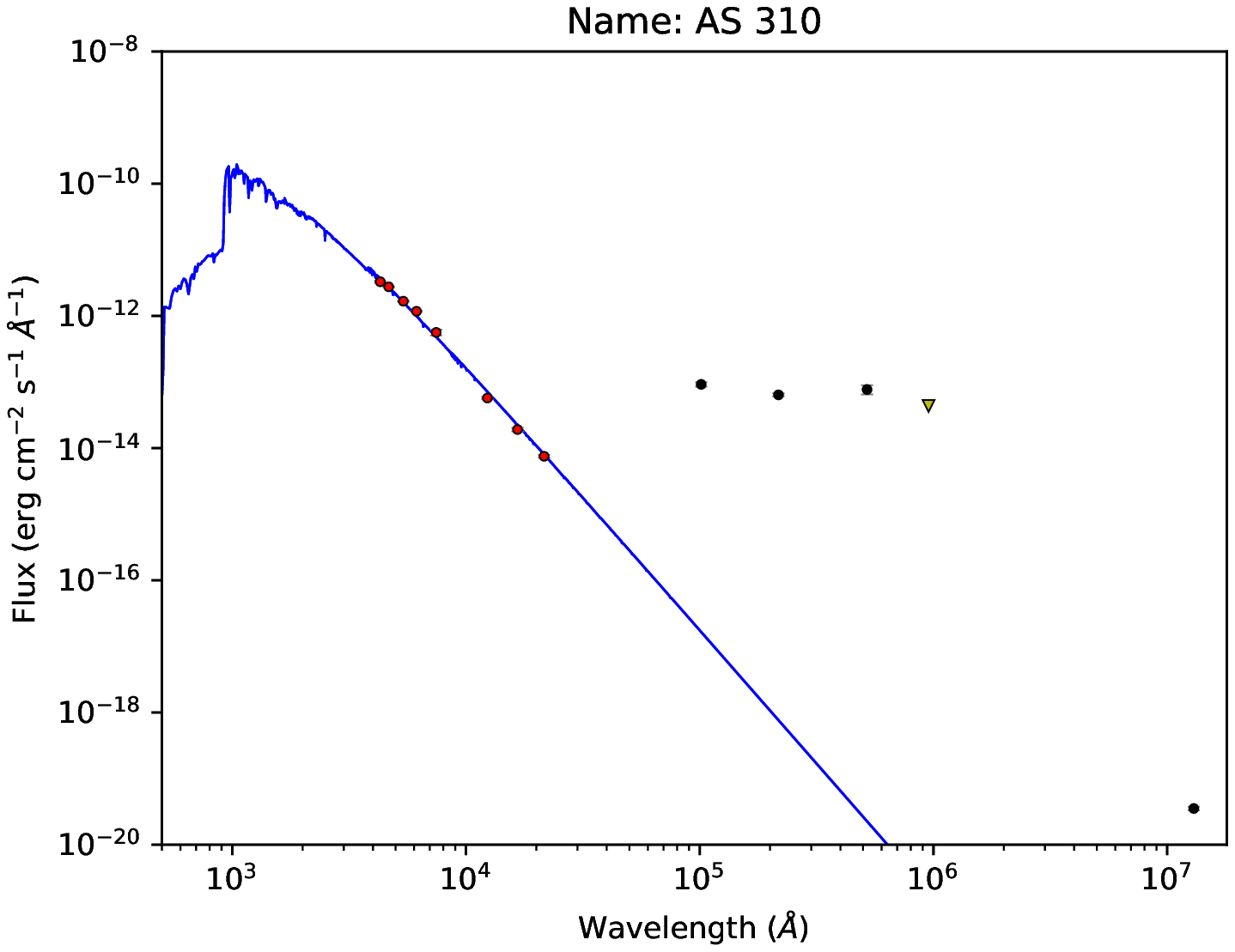}
    \includegraphics[width=0.33\textwidth]{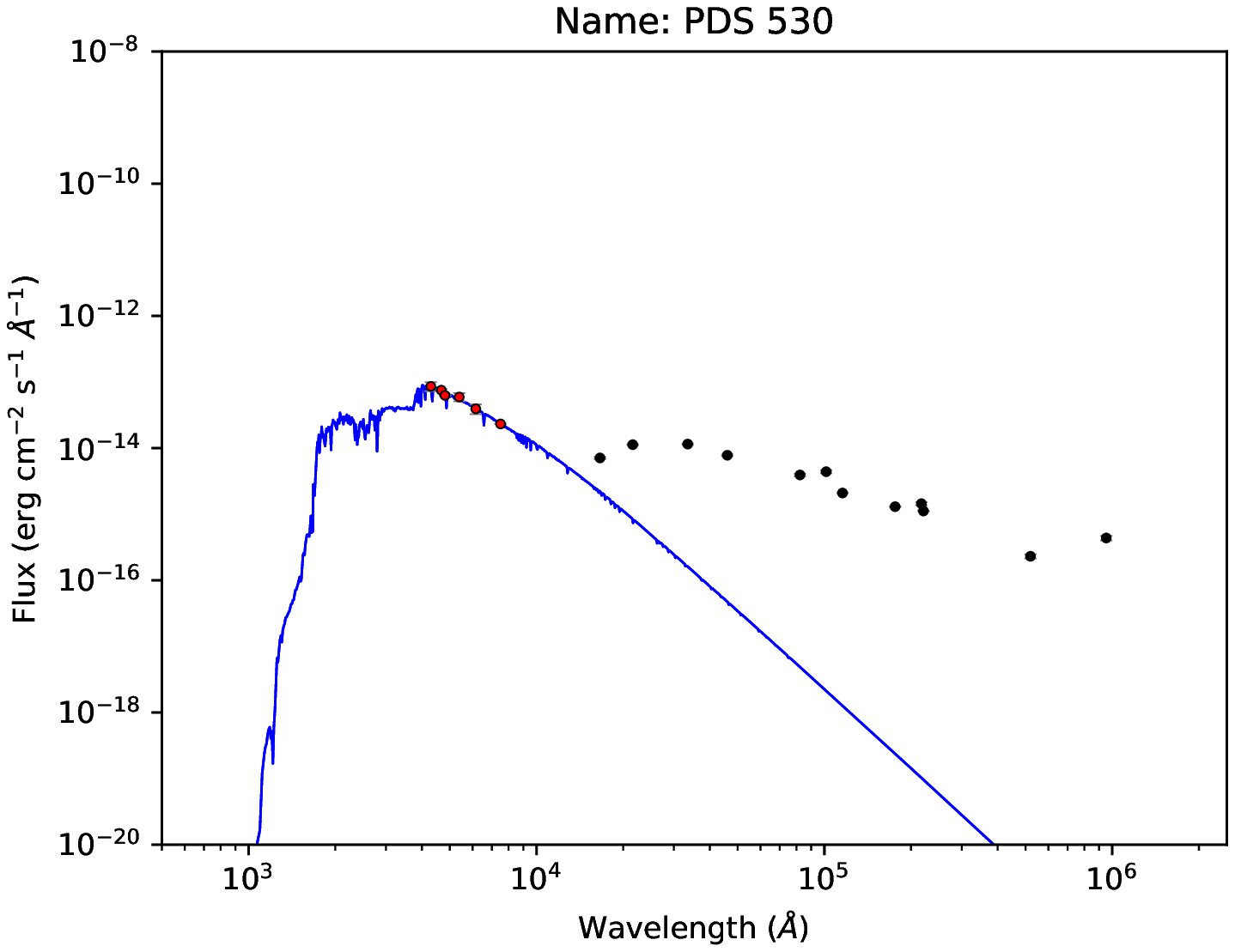}
    \includegraphics[width=0.33\textwidth]{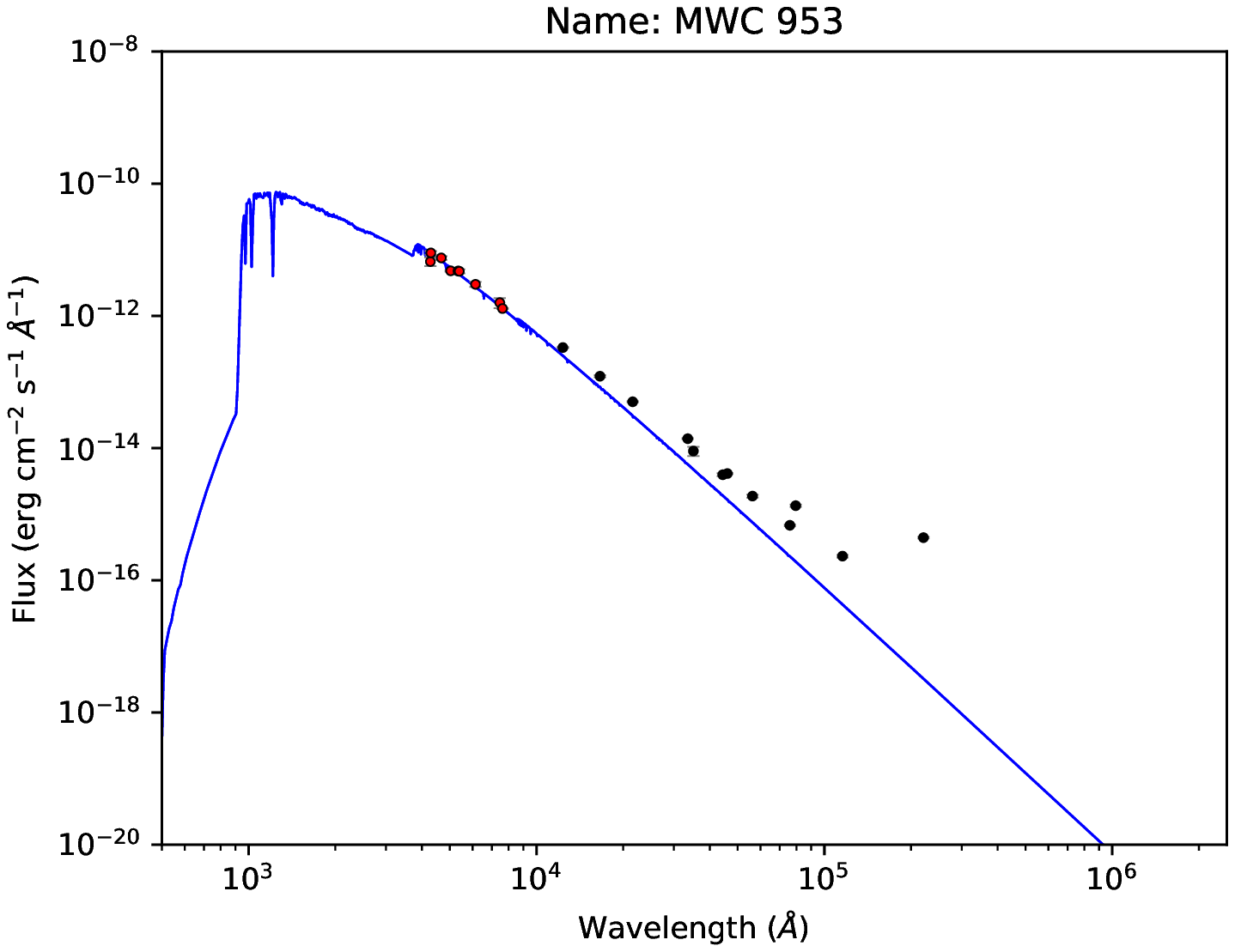}
\end{figure}

\begin{figure} [h]
 \centering
    \includegraphics[width=0.33\textwidth]{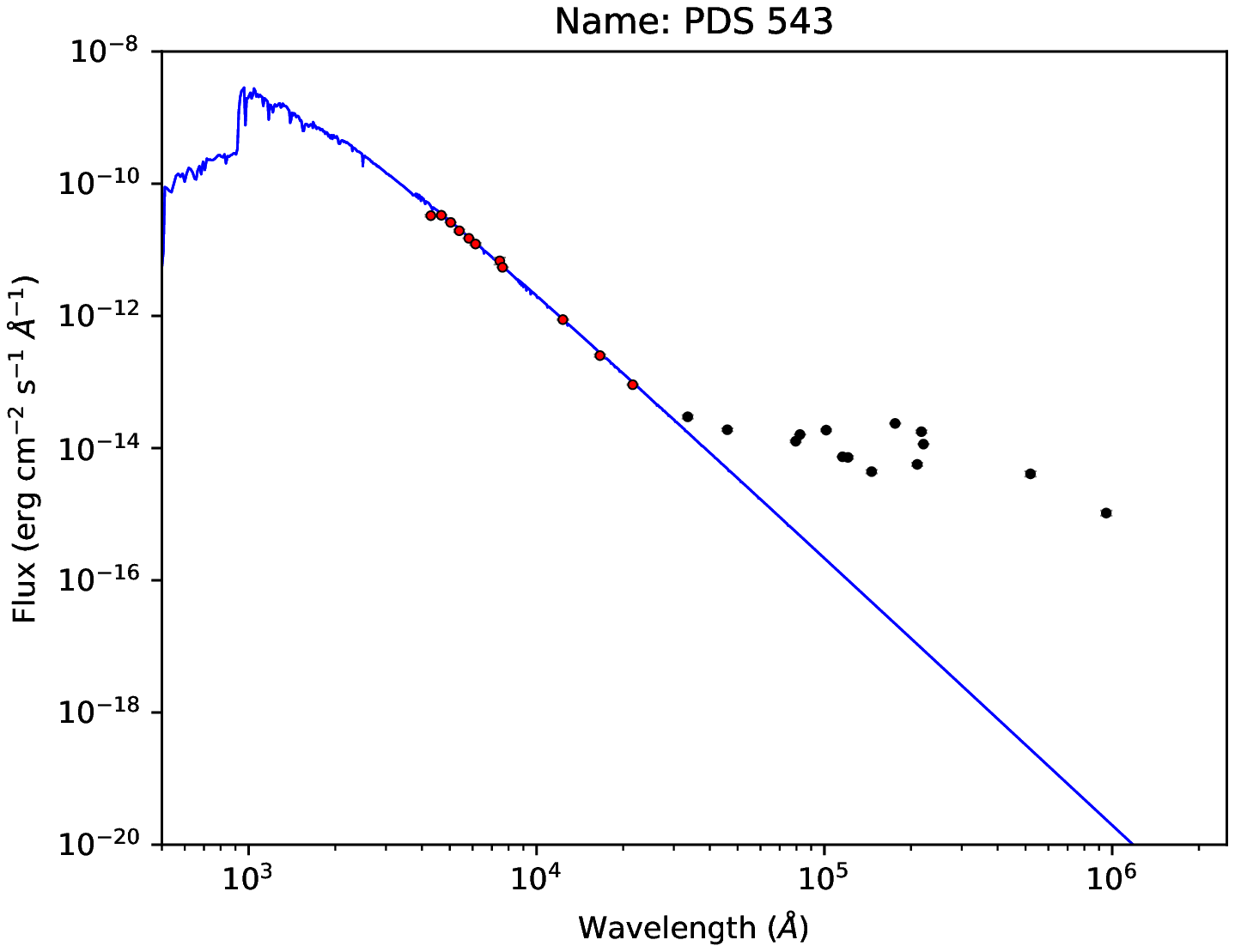}
    \includegraphics[width=0.33\textwidth]{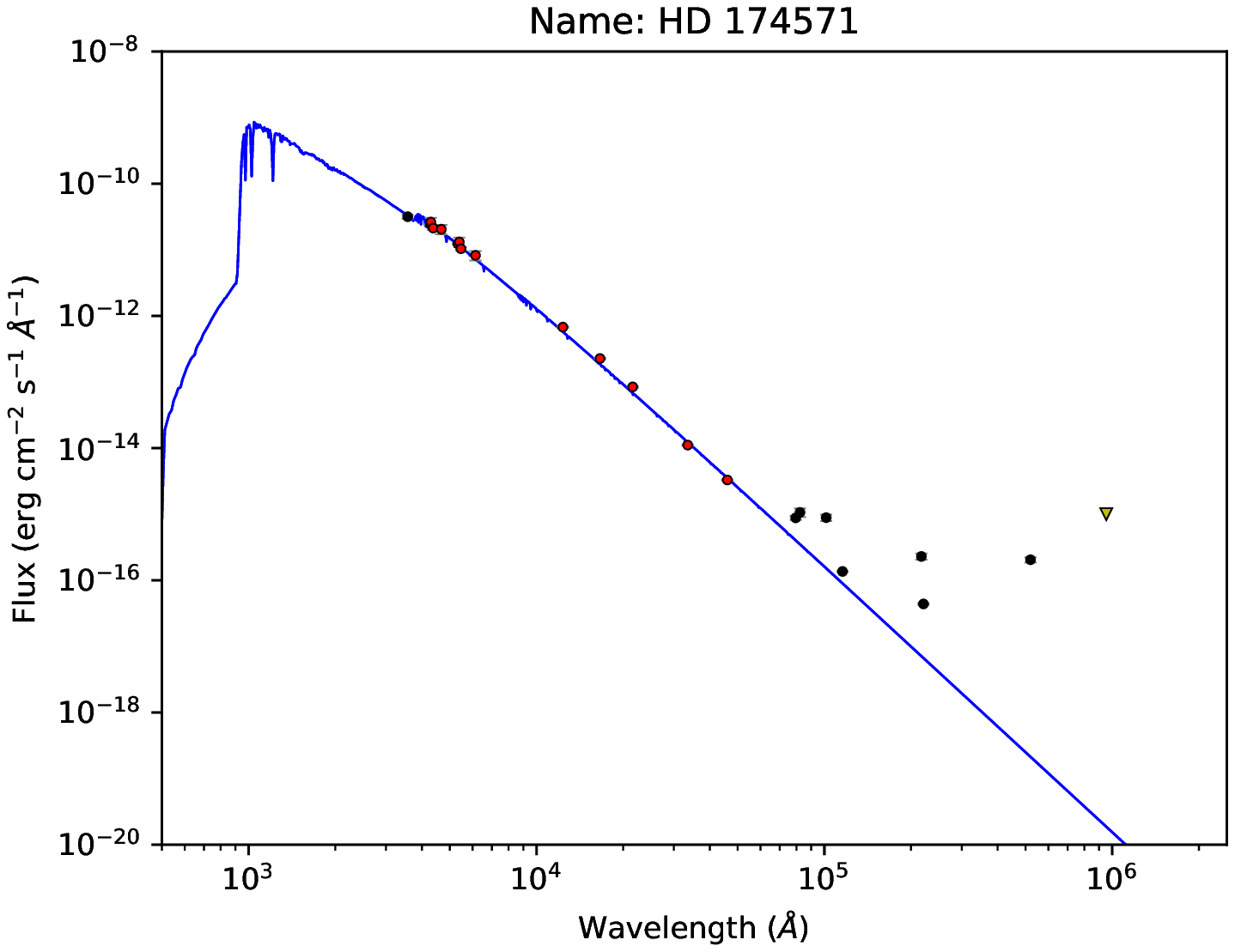}
    \includegraphics[width=0.33\textwidth]{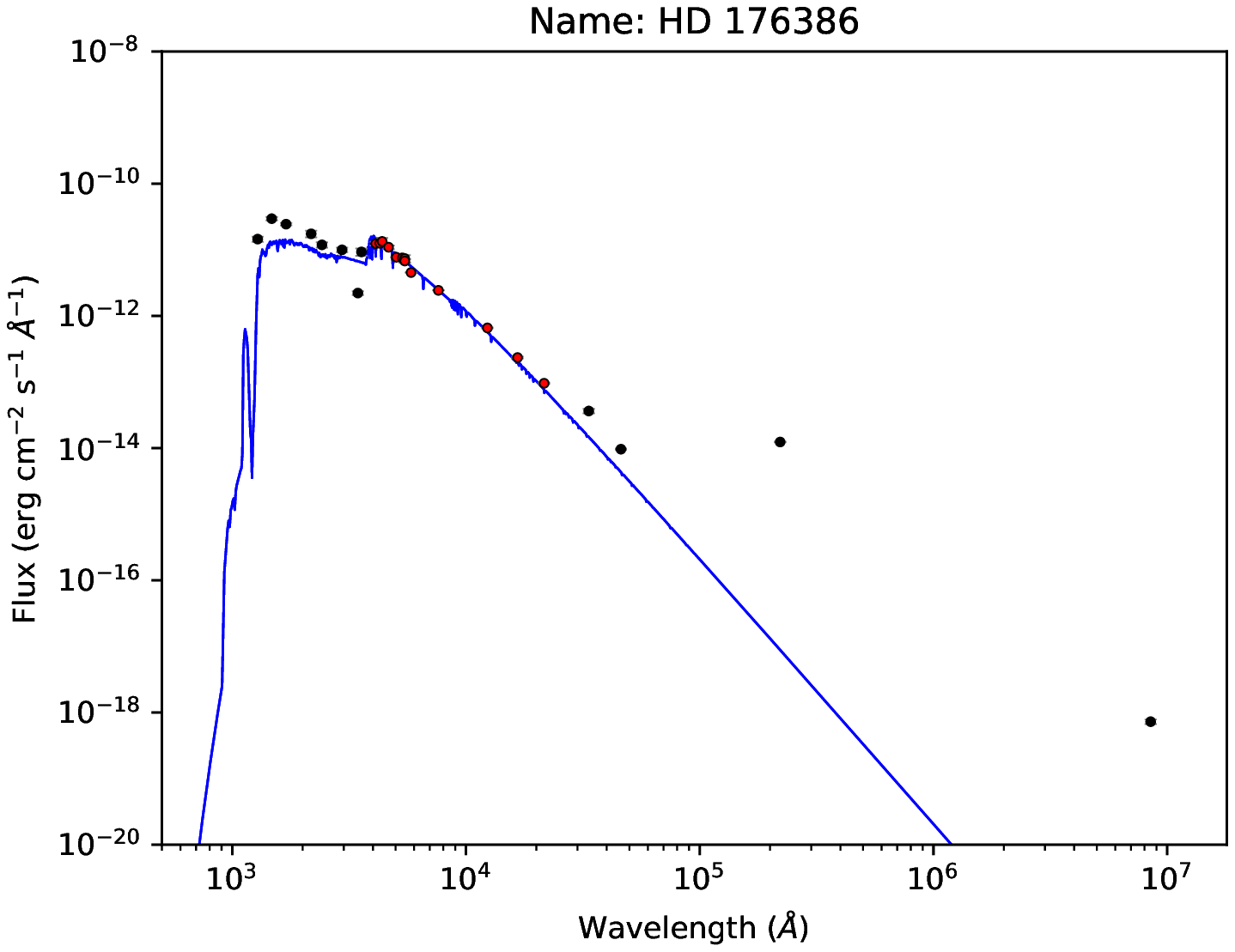}
\end{figure}

\begin{figure} [h]
 \centering
    \includegraphics[width=0.33\textwidth]{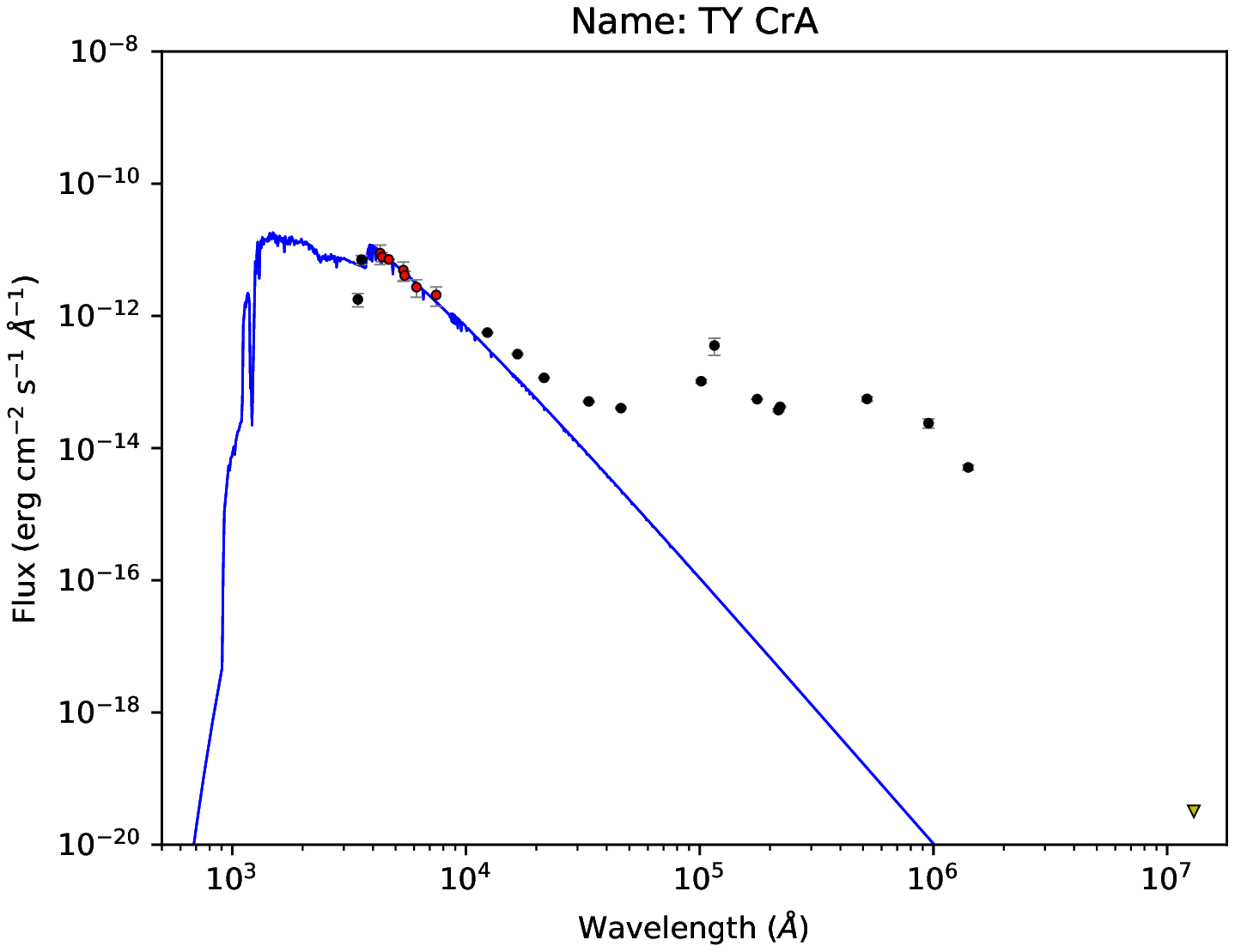}
    \includegraphics[width=0.33\textwidth]{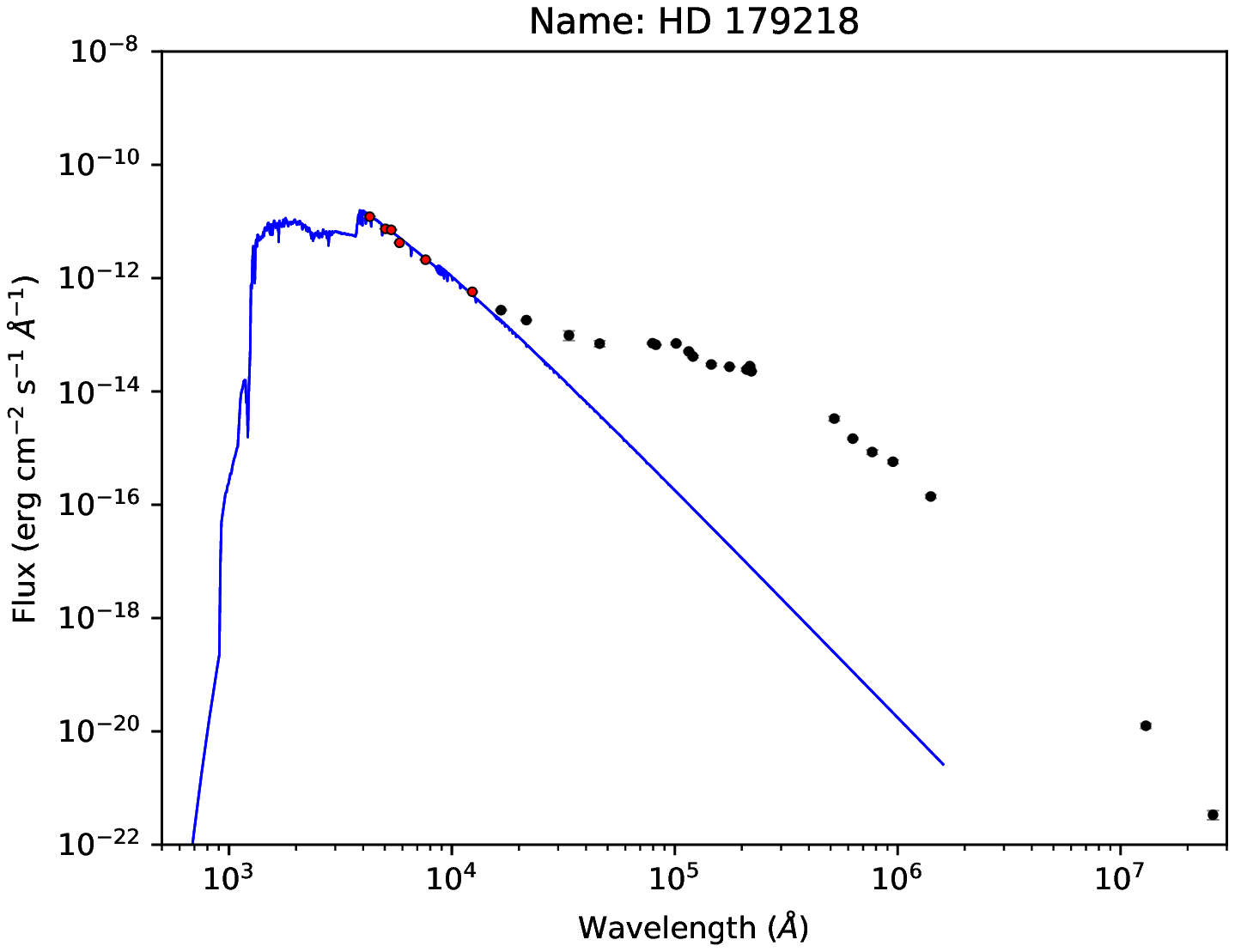}
    \includegraphics[width=0.33\textwidth]{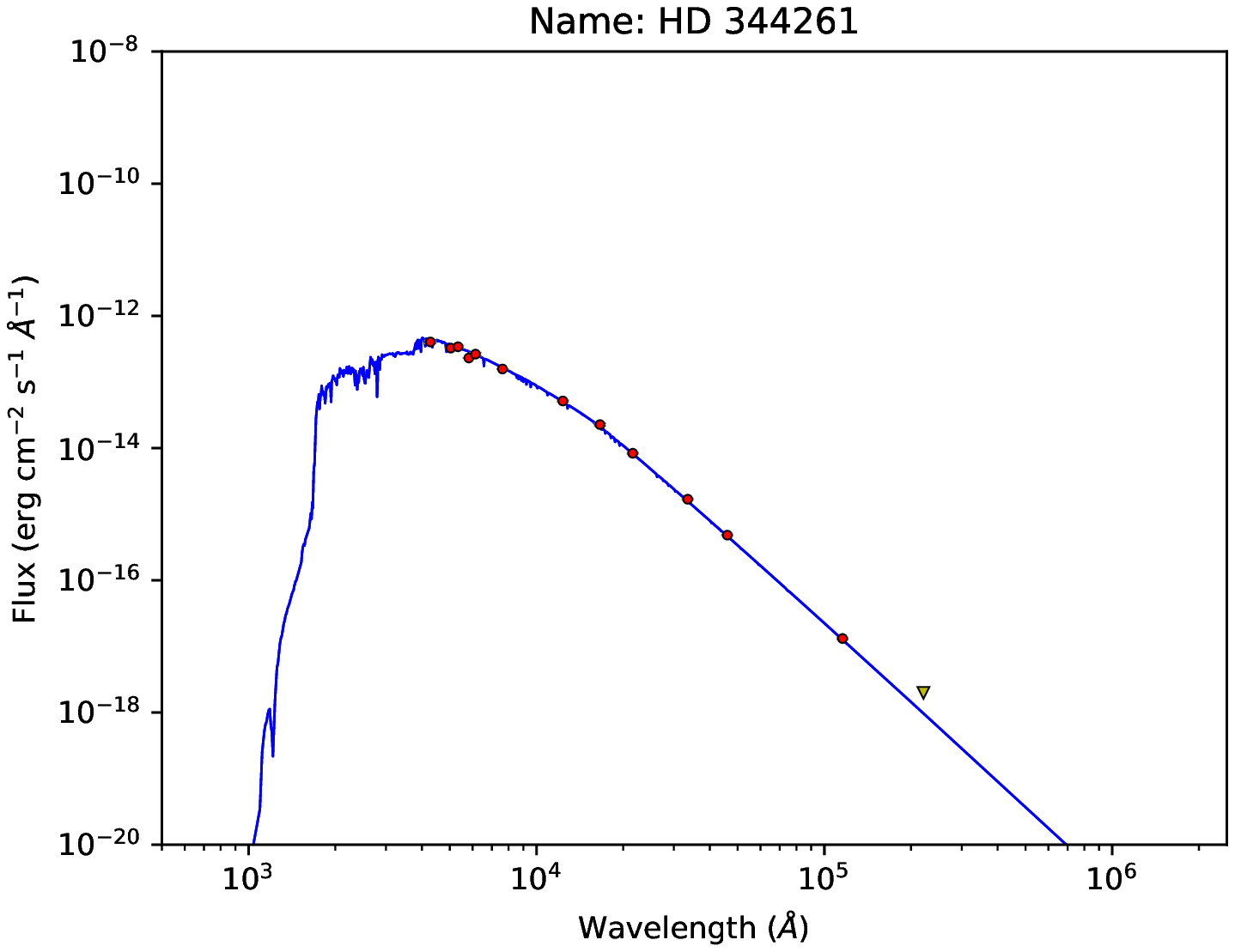}
\end{figure}

\begin{figure} [h]
 \centering
    \includegraphics[width=0.33\textwidth]{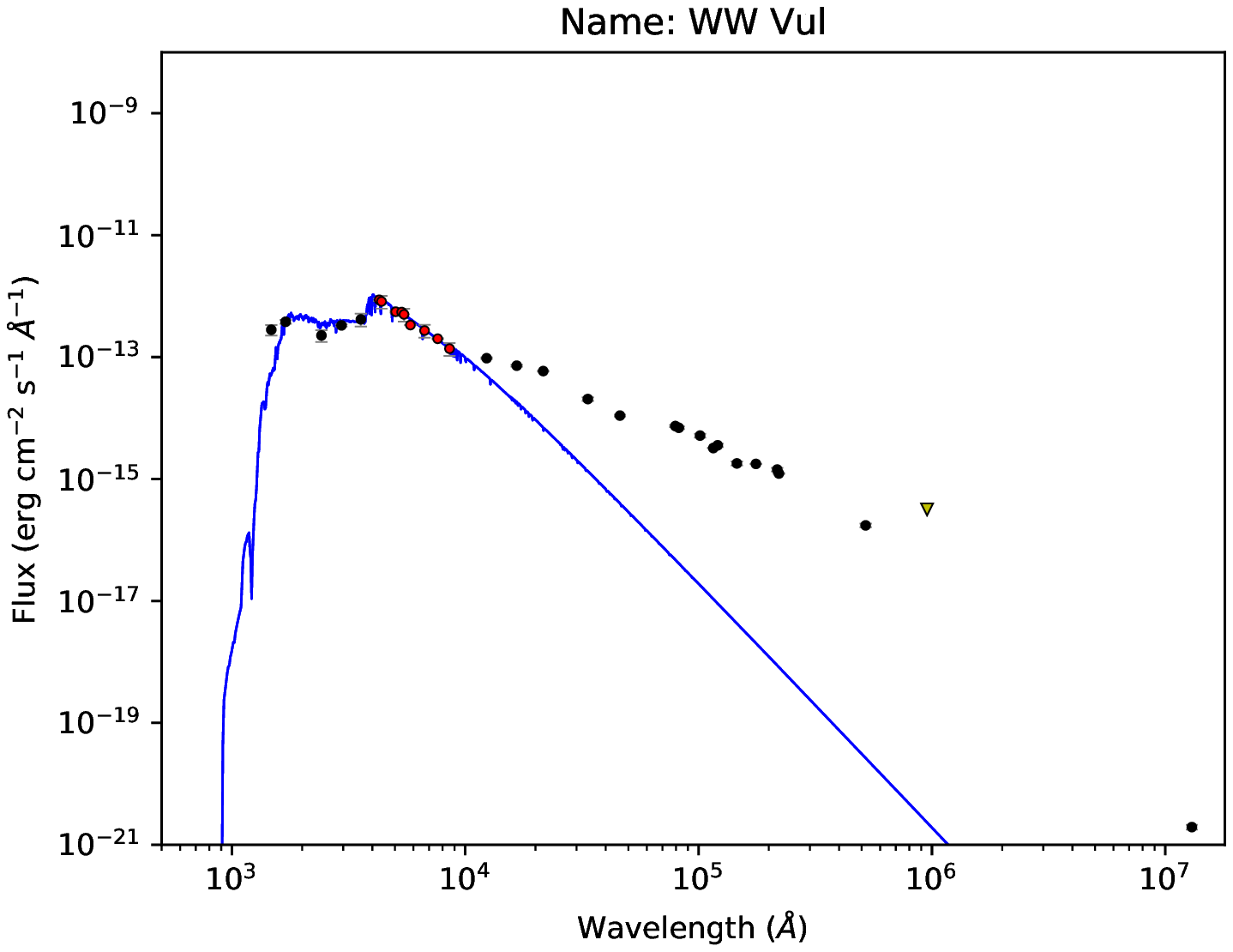}
    \includegraphics[width=0.33\textwidth]{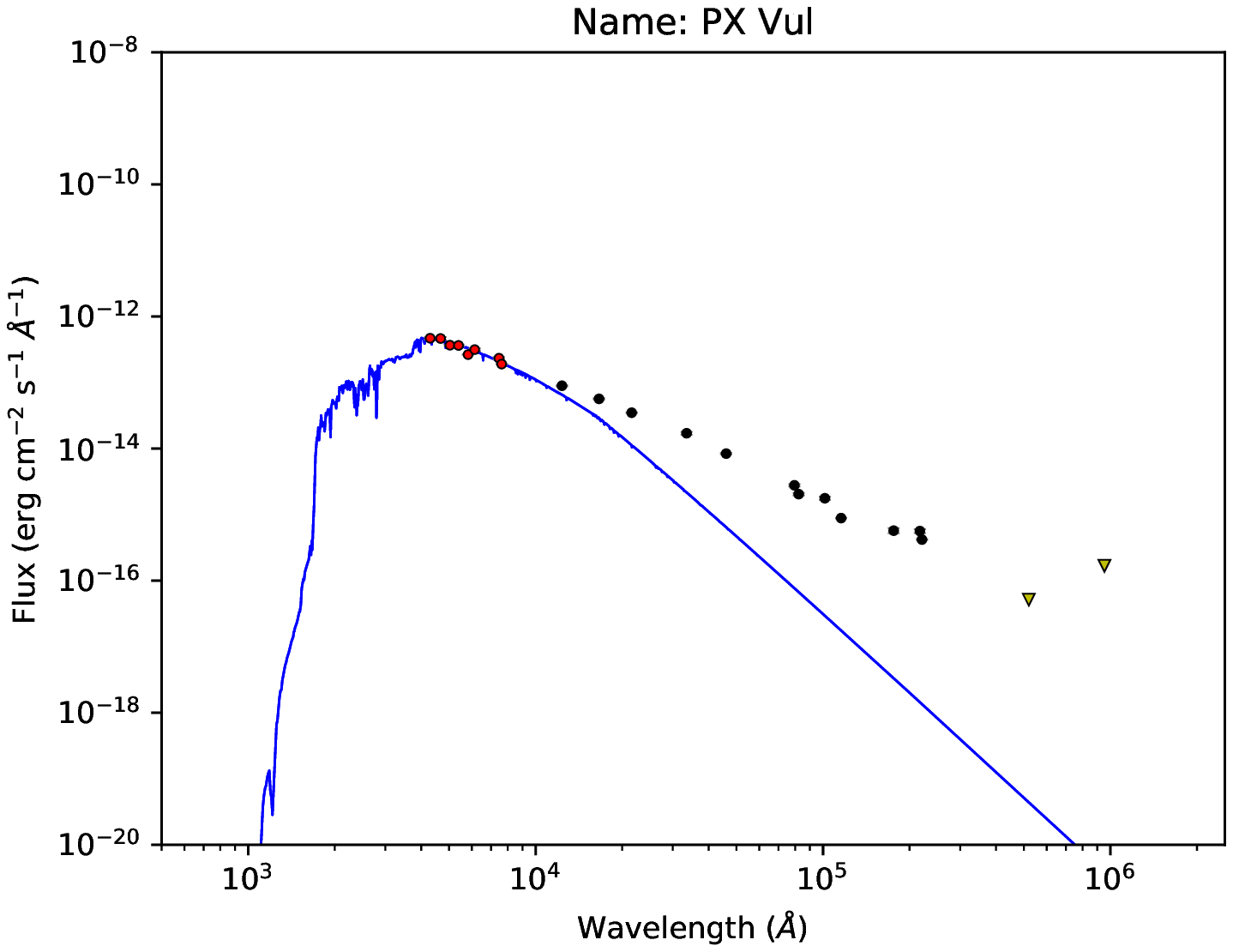}
    \includegraphics[width=0.33\textwidth]{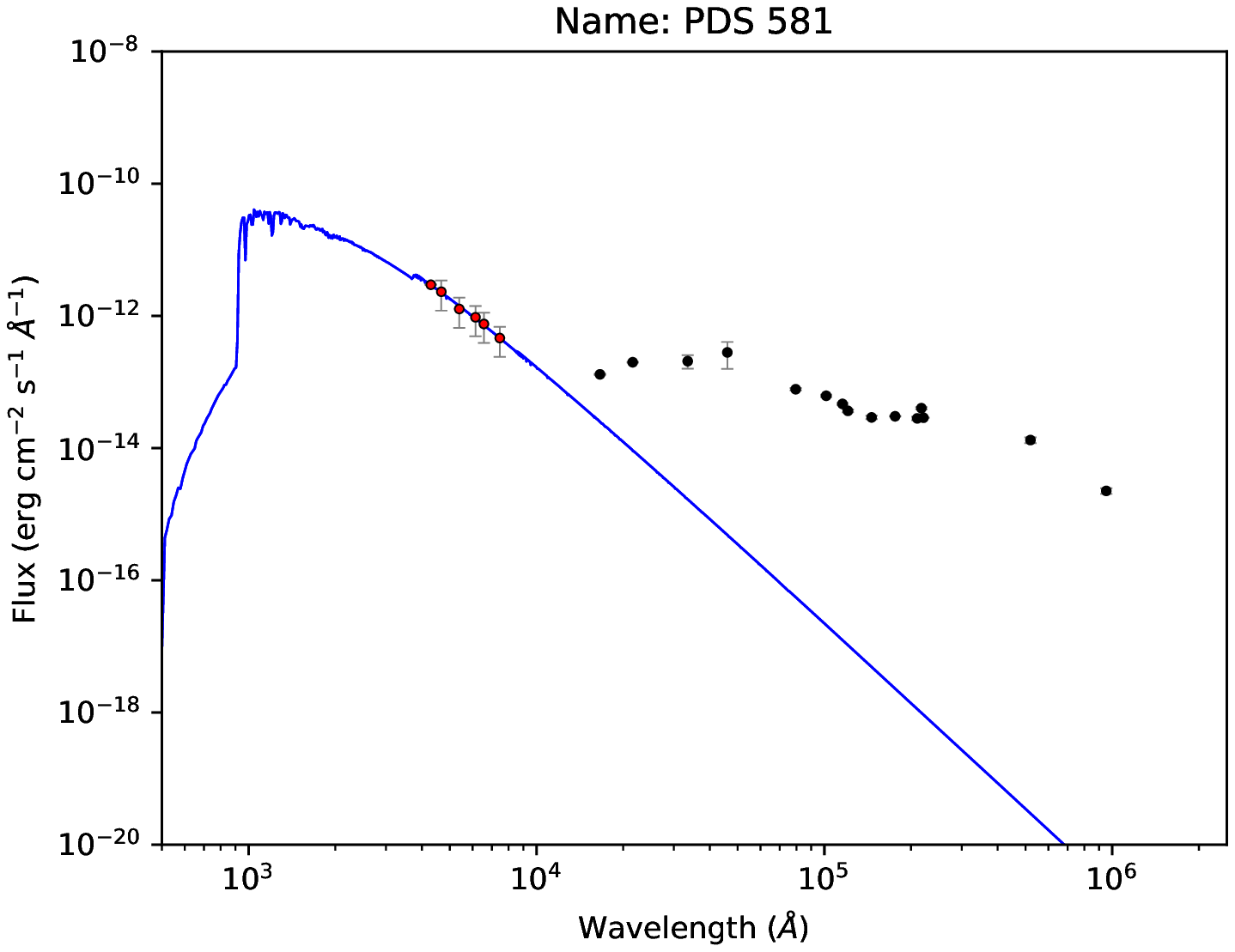}
\end{figure}

\newpage

\onecolumn

\begin{figure} [h]
 \centering
    \includegraphics[width=0.33\textwidth]{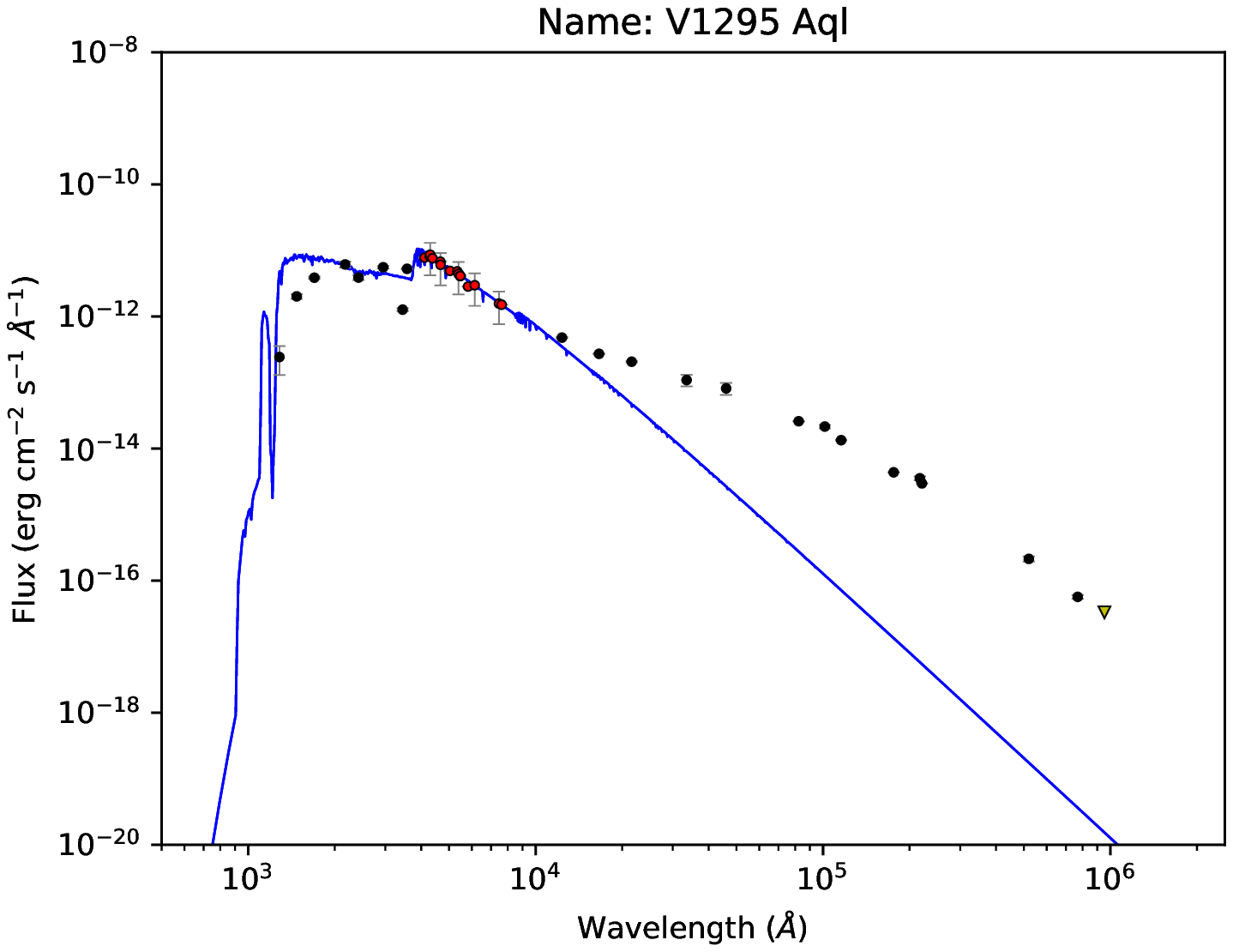}
    \includegraphics[width=0.33\textwidth]{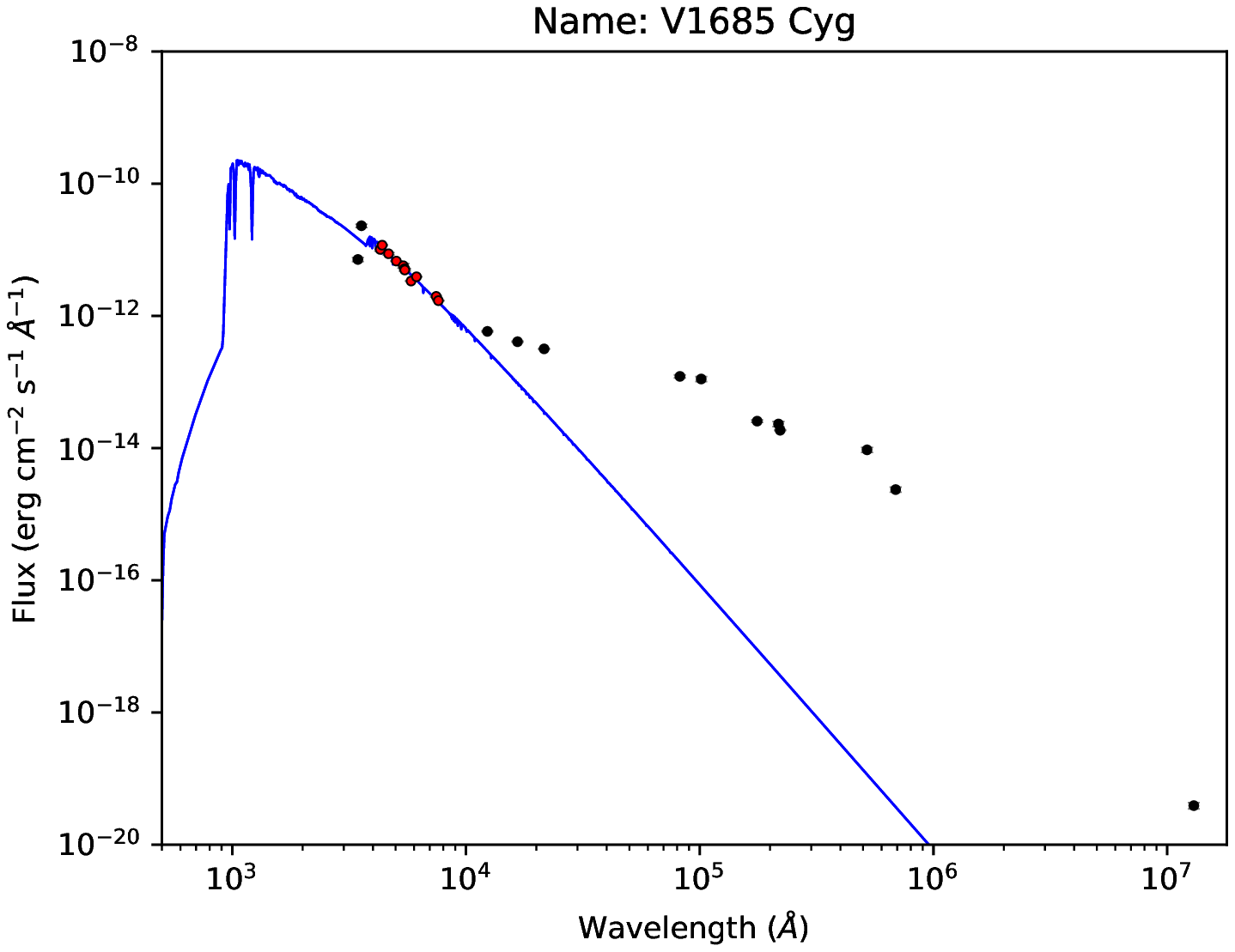}
    \includegraphics[width=0.33\textwidth]{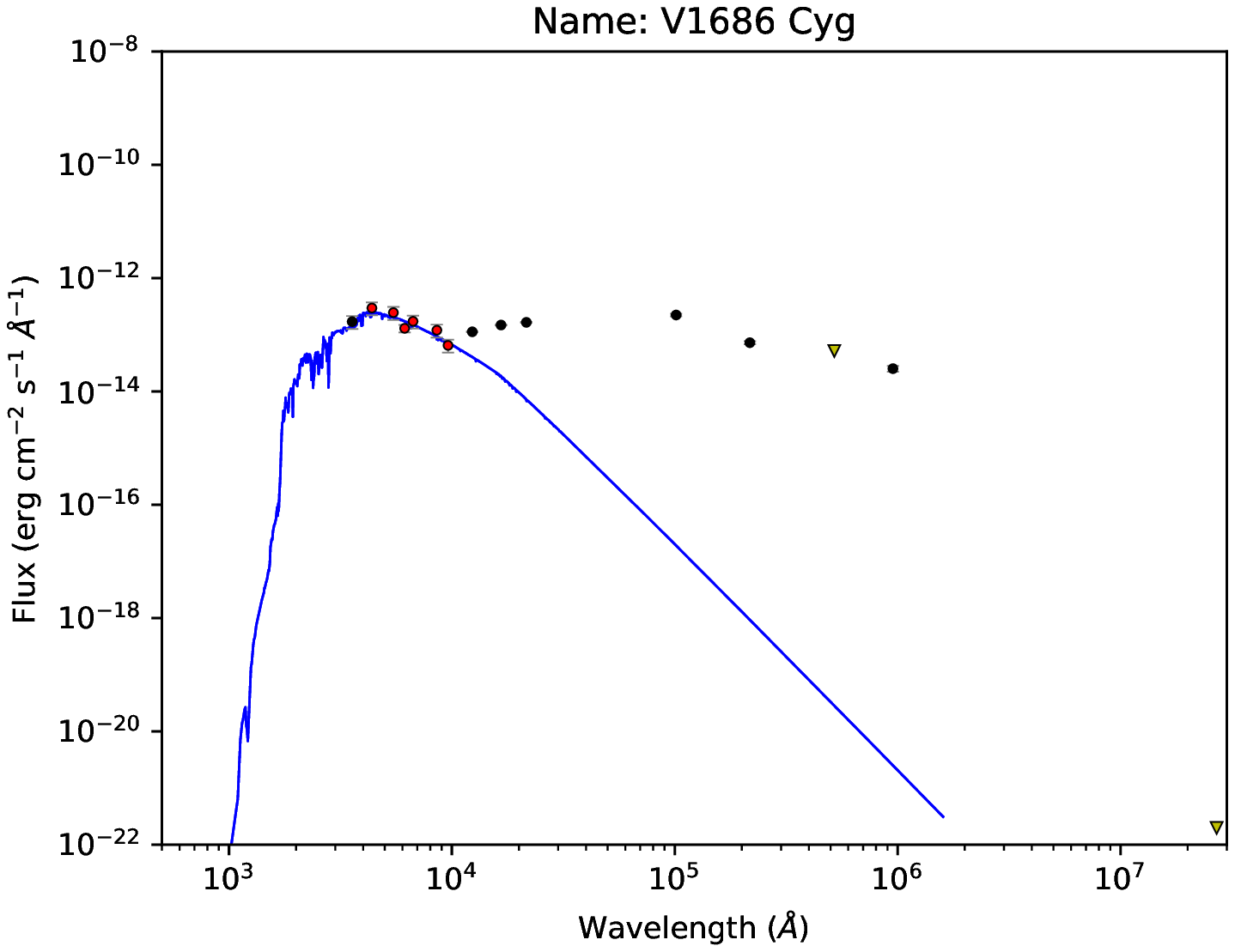}
\end{figure}

\begin{figure} [h]
 \centering
    \includegraphics[width=0.33\textwidth]{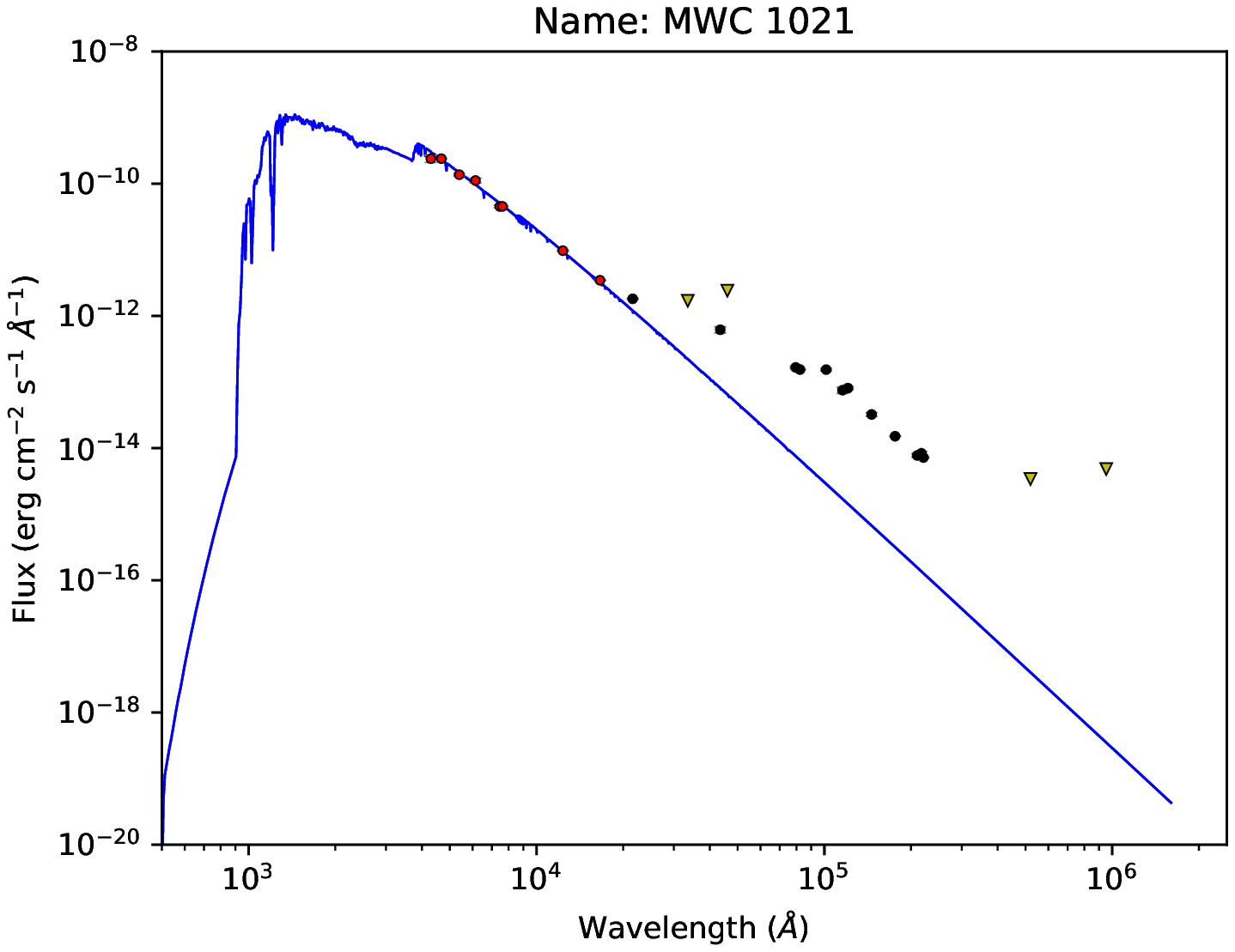}
    \includegraphics[width=0.33\textwidth]{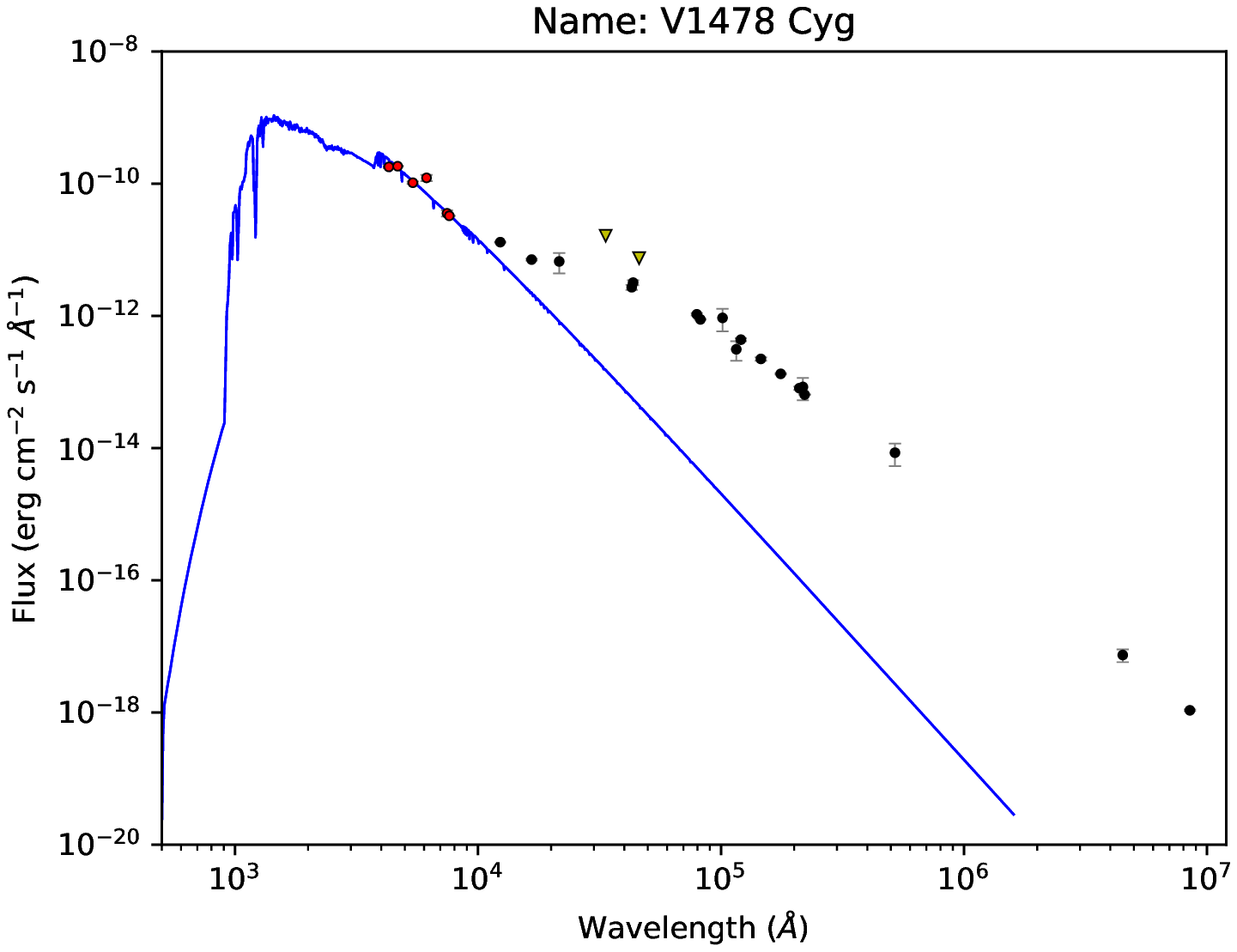}
    \includegraphics[width=0.33\textwidth]{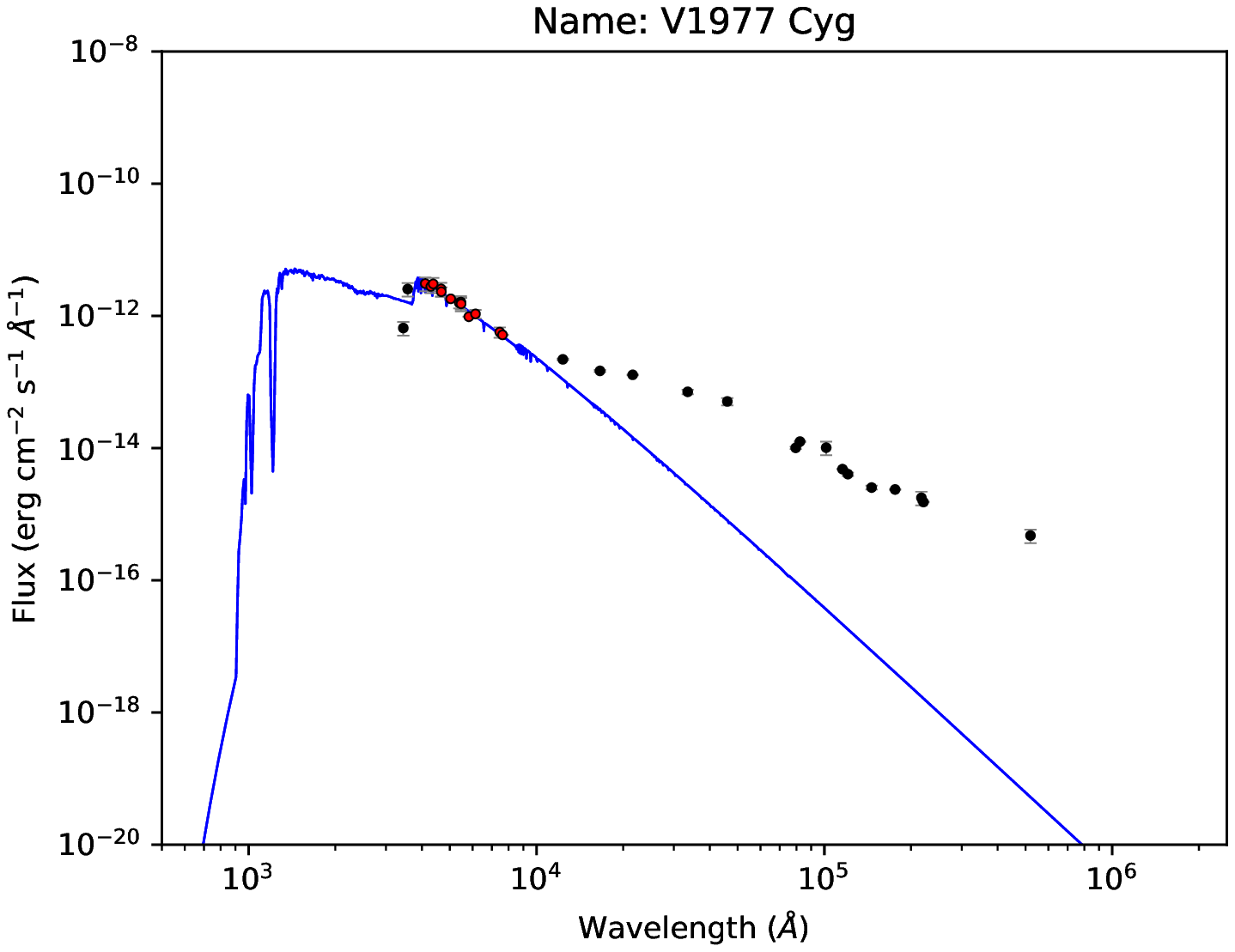}
\end{figure}

\begin{figure} [h]
 \centering
    \includegraphics[width=0.33\textwidth]{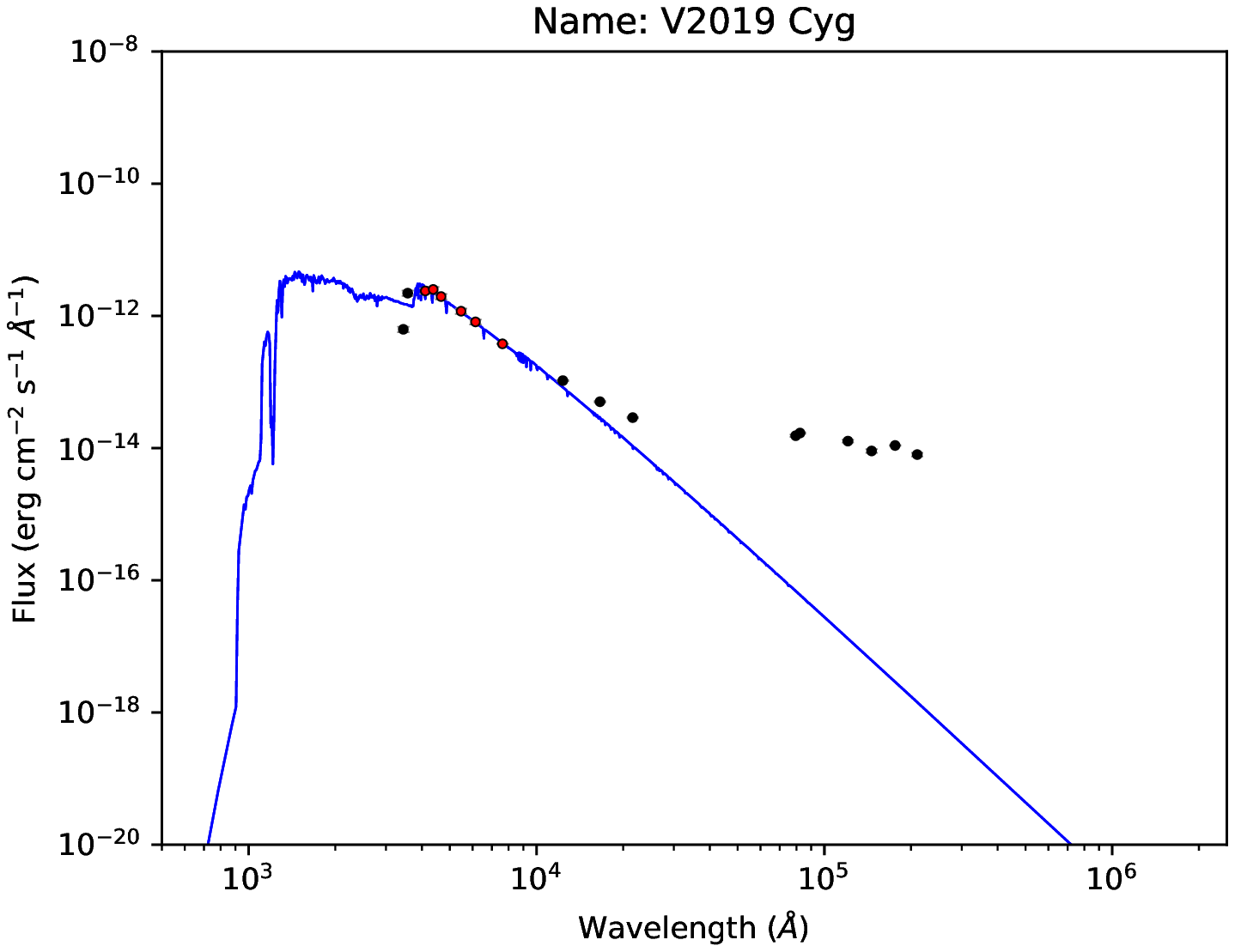}
    \includegraphics[width=0.33\textwidth]{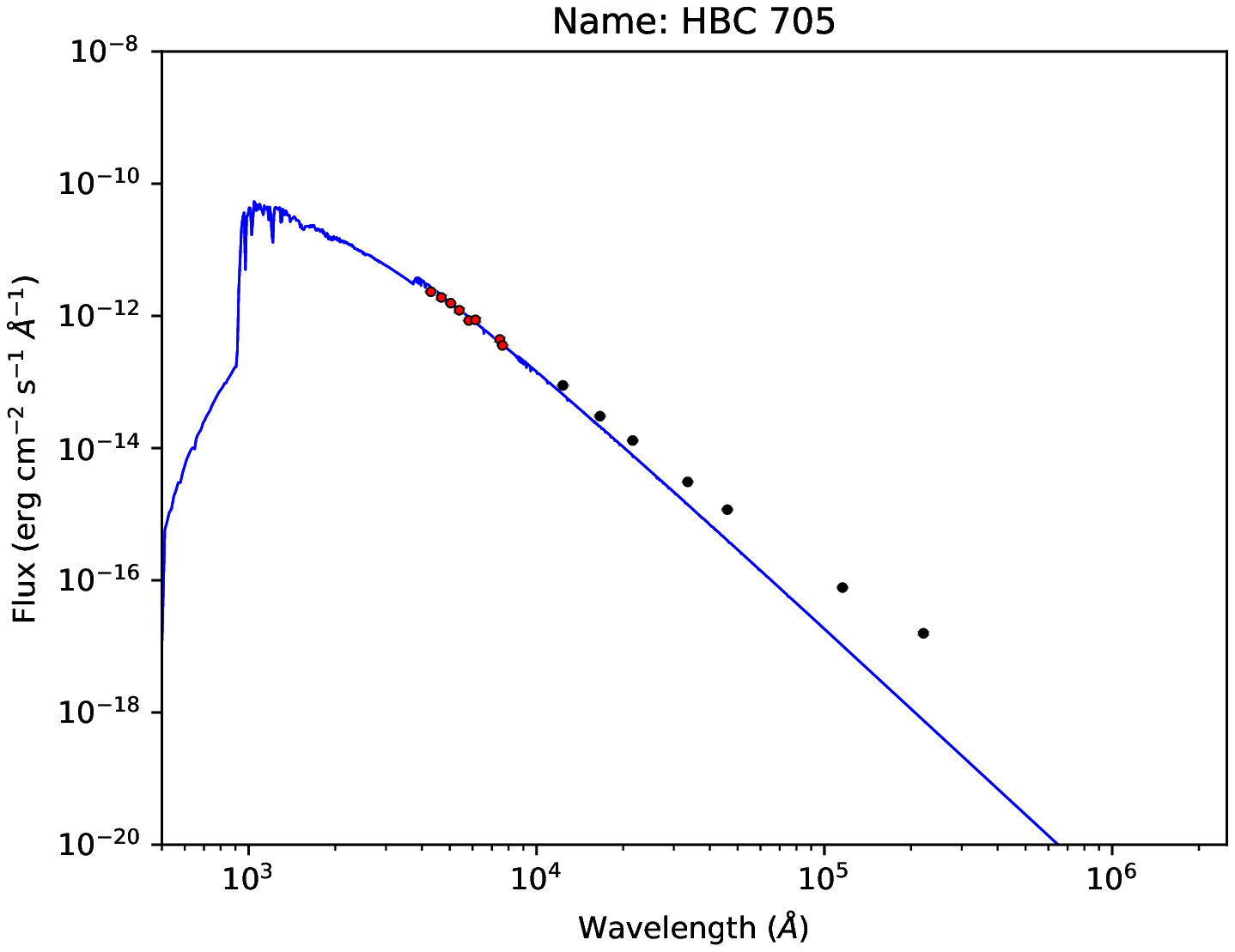}
    \includegraphics[width=0.33\textwidth]{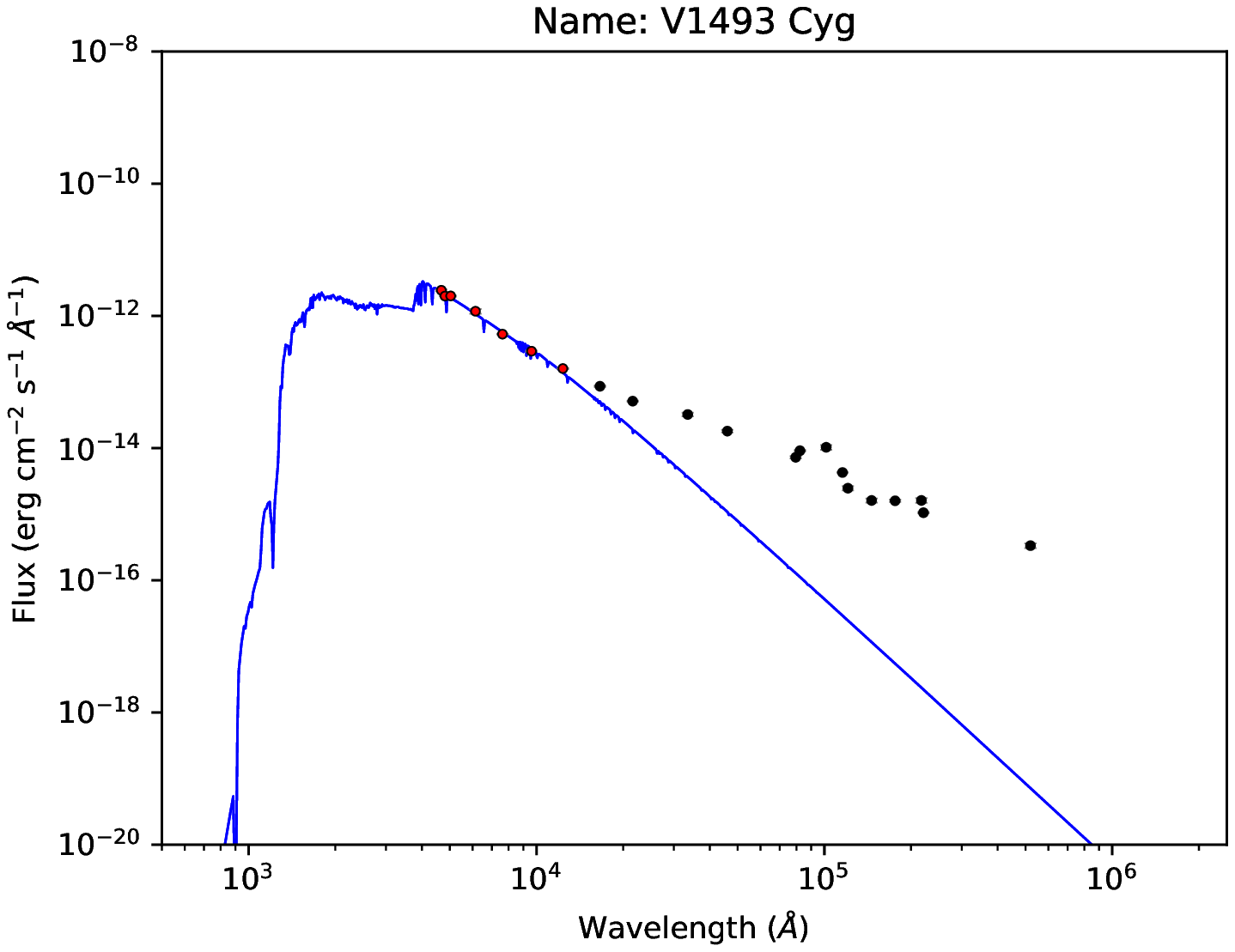}
\end{figure}

\begin{figure} [h]
 \centering
    \includegraphics[width=0.33\textwidth]{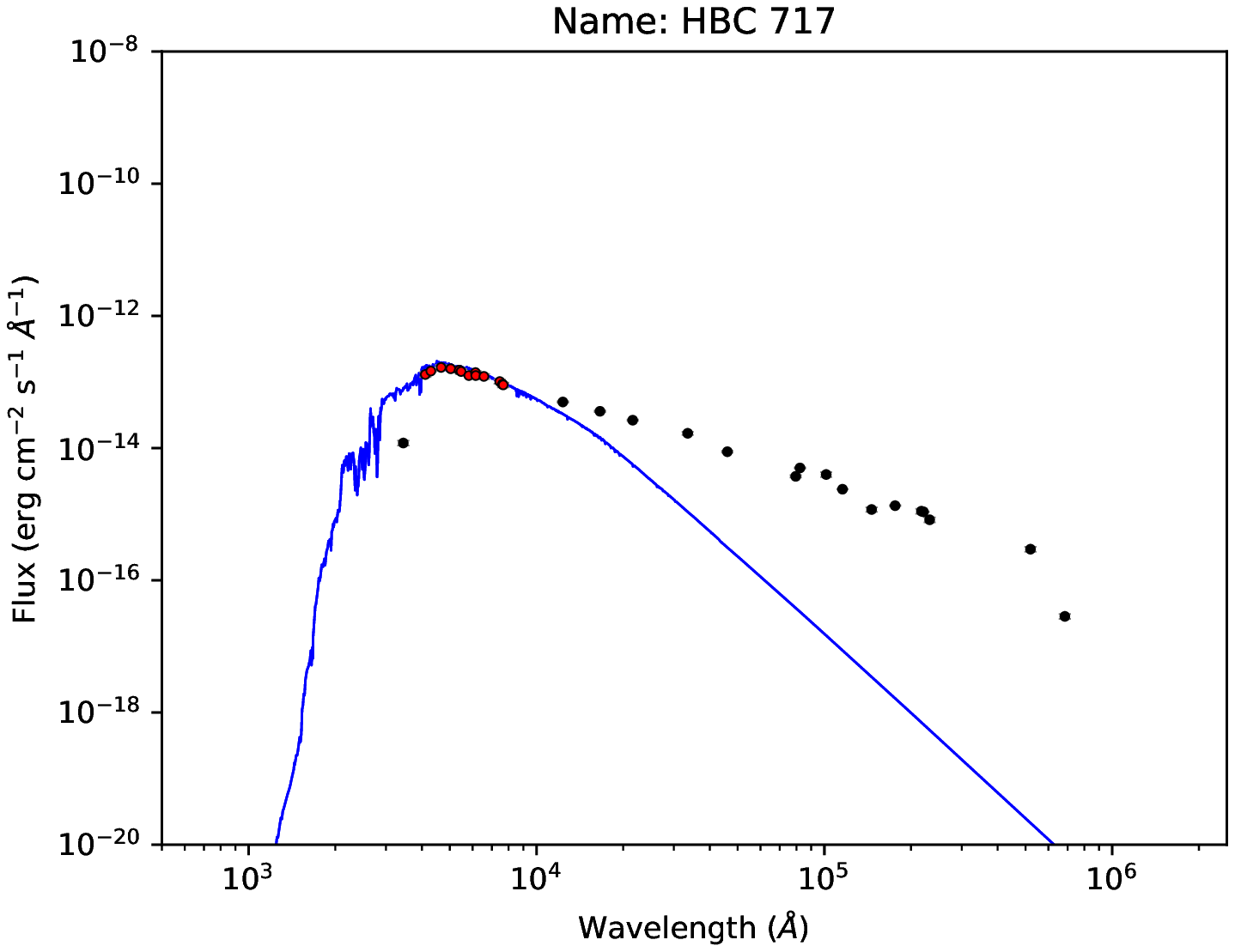}
    \includegraphics[width=0.33\textwidth]{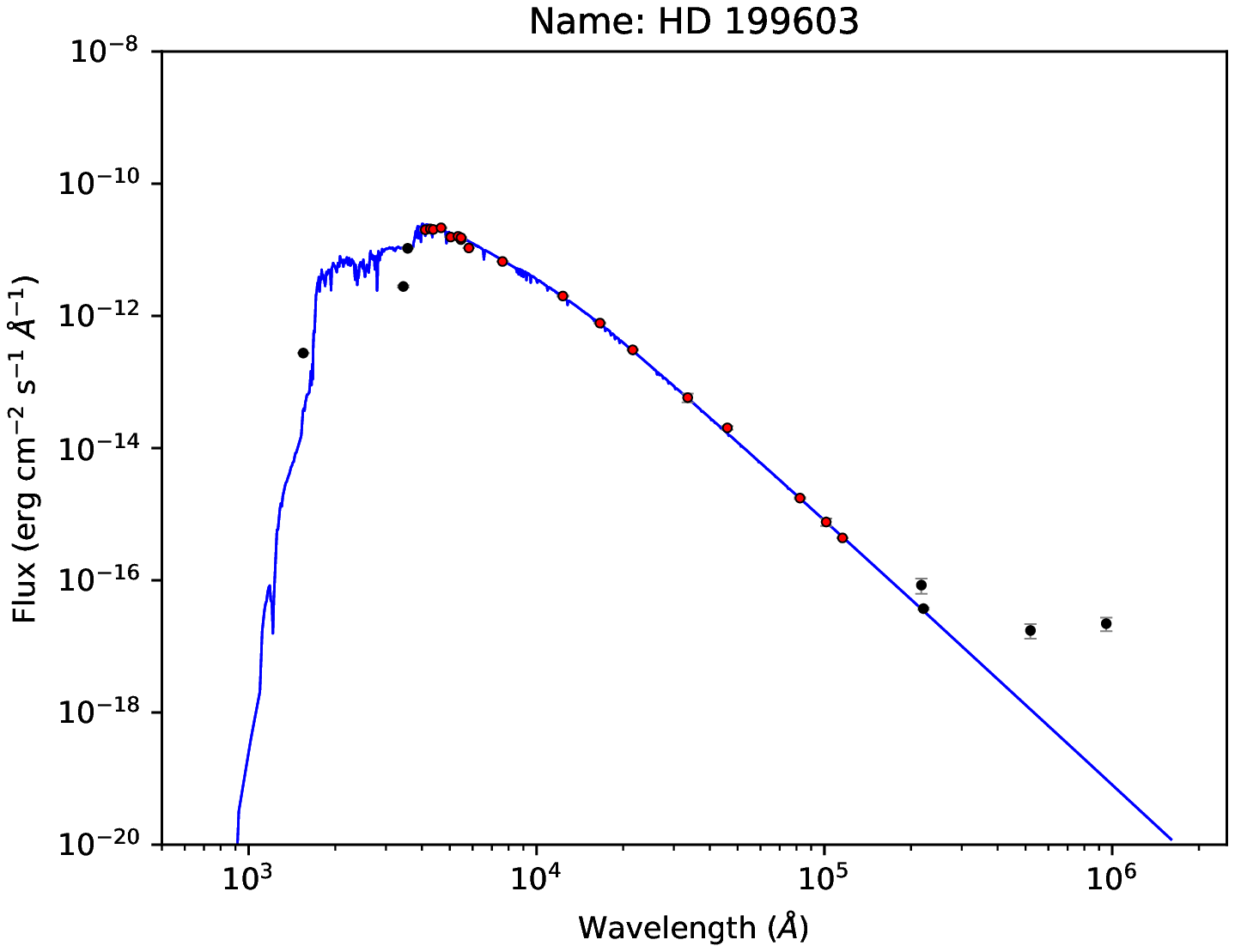}
    \includegraphics[width=0.33\textwidth]{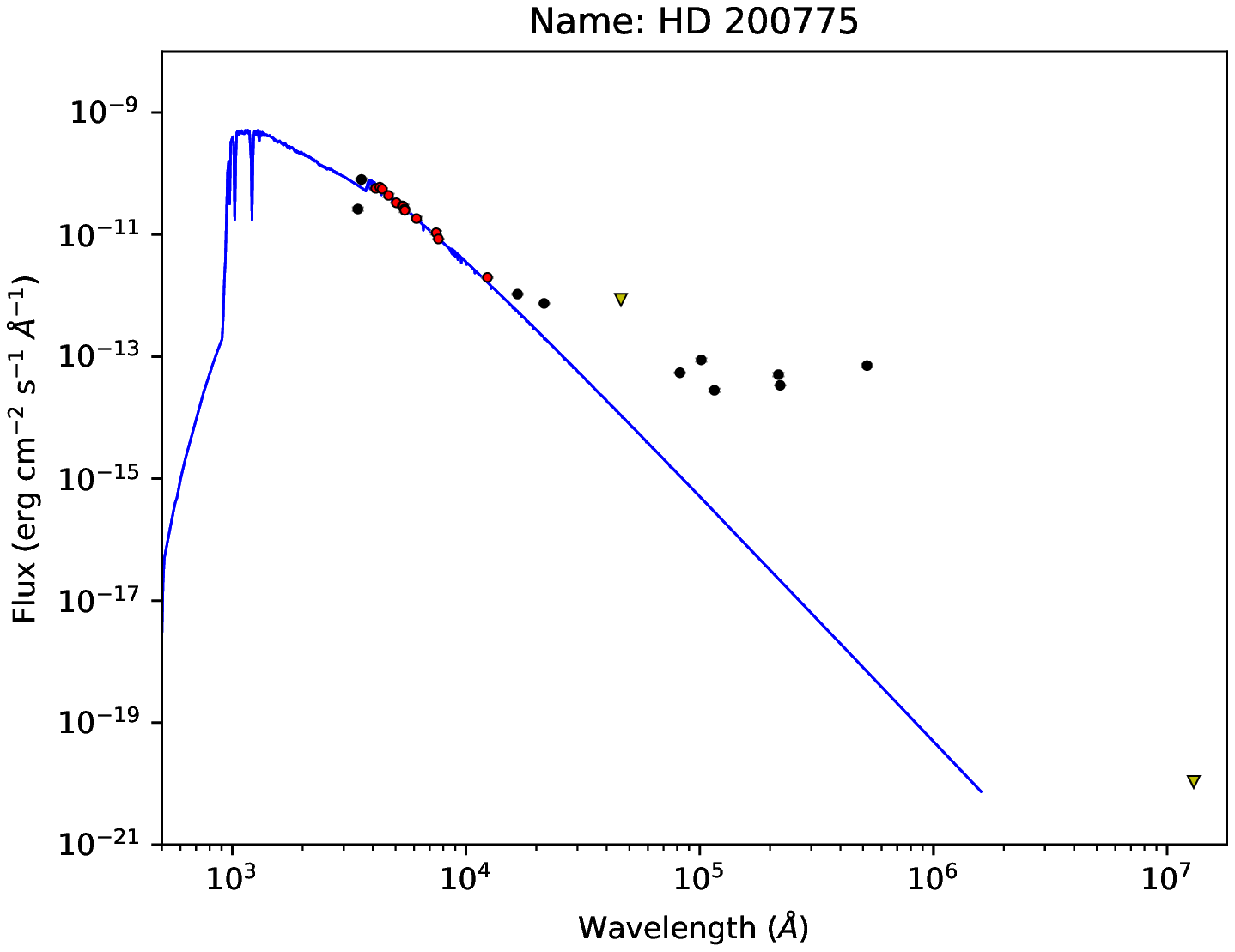}
\end{figure}

\newpage

\onecolumn

\begin{figure} [h]
 \centering
    \includegraphics[width=0.33\textwidth]{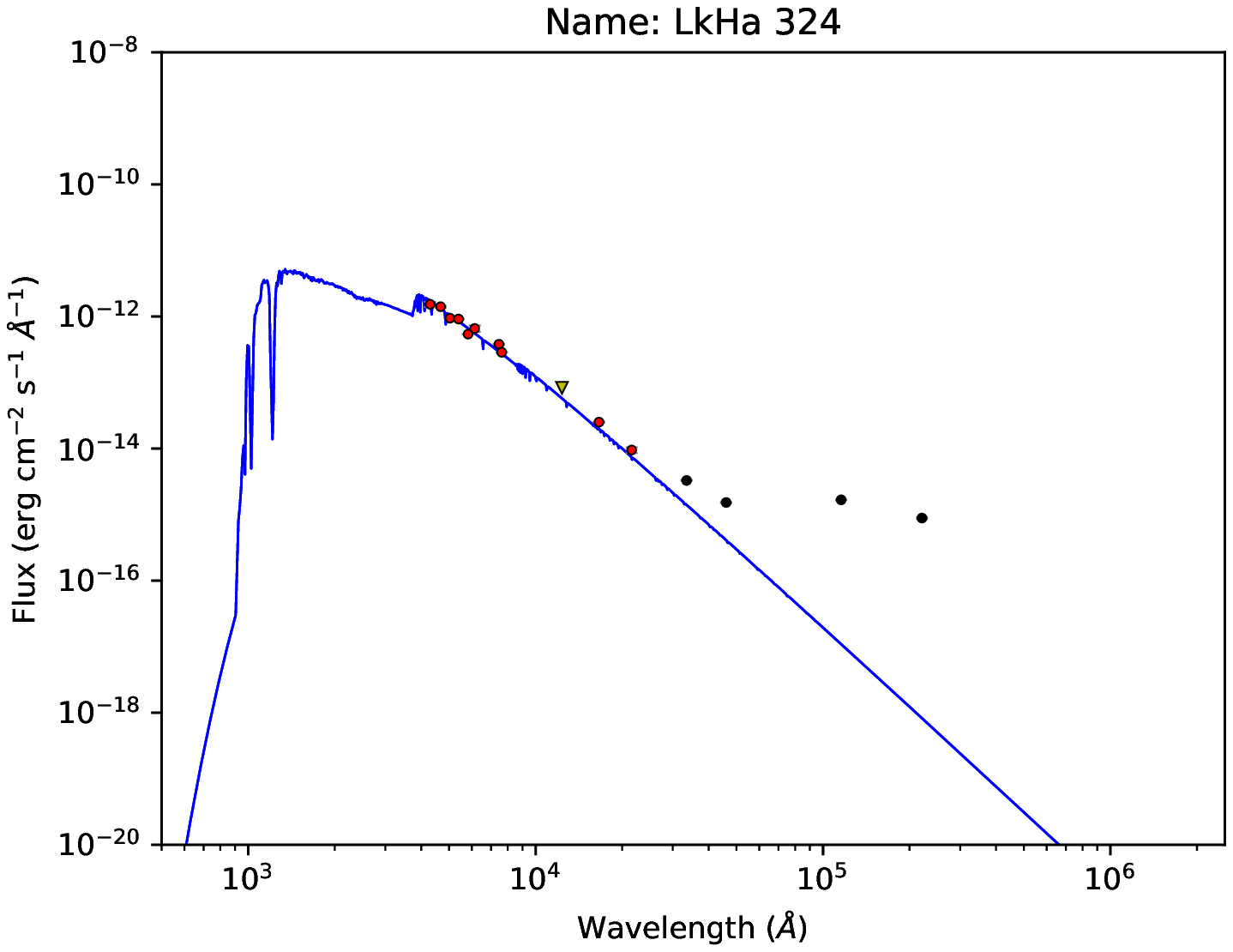}
    \includegraphics[width=0.33\textwidth]{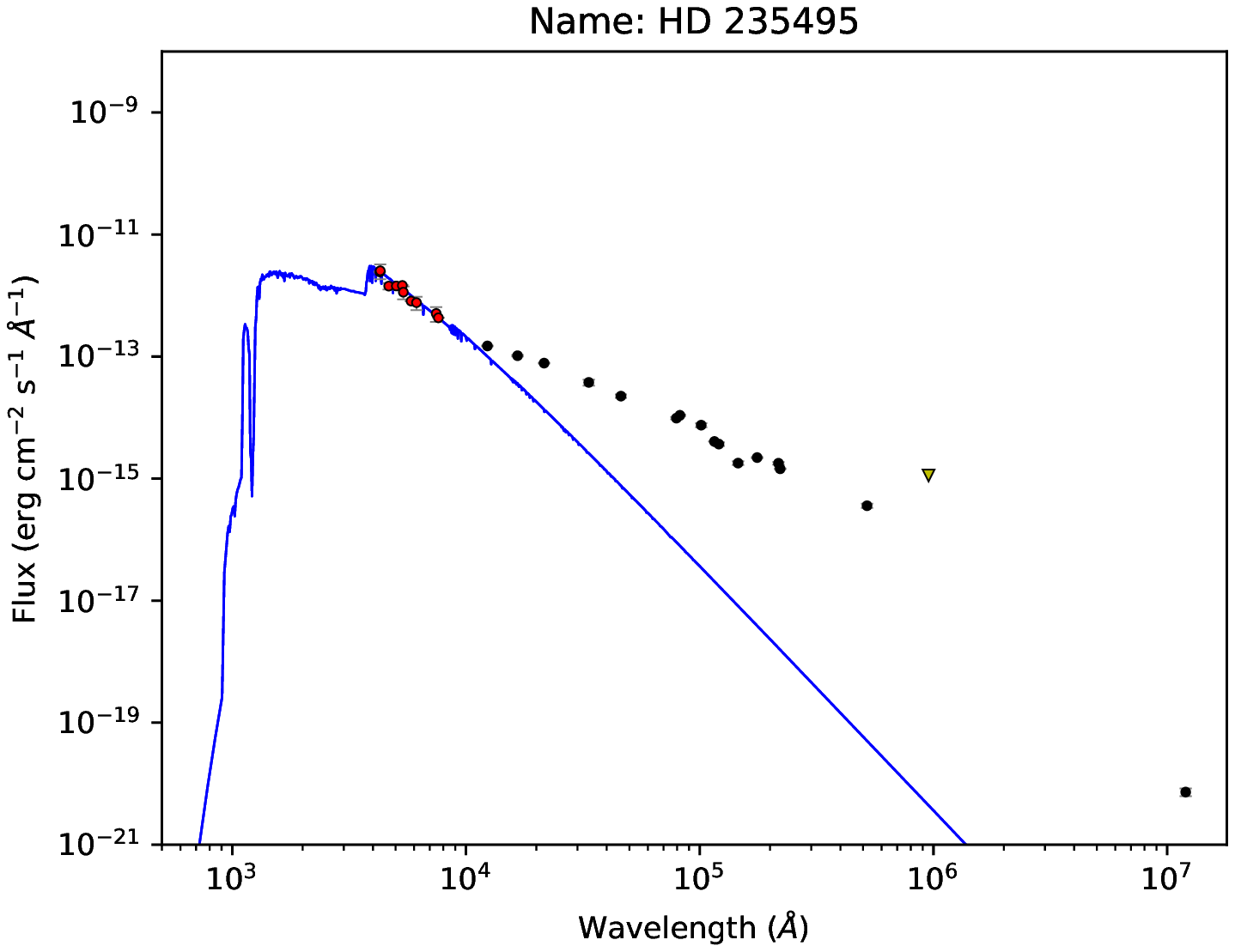}
    \includegraphics[width=0.33\textwidth]{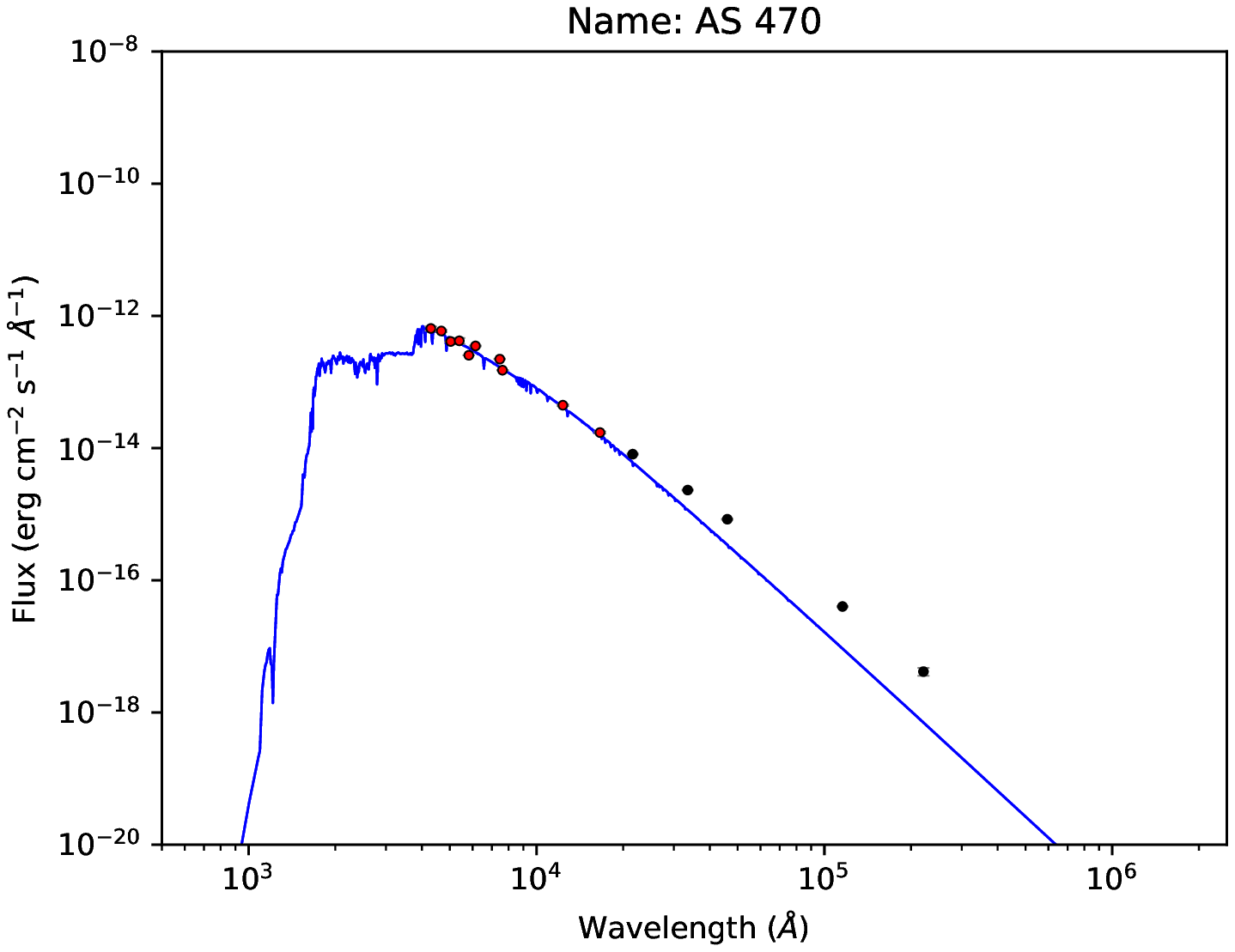}
\end{figure}

\begin{figure} [h]
 \centering
    \includegraphics[width=0.33\textwidth]{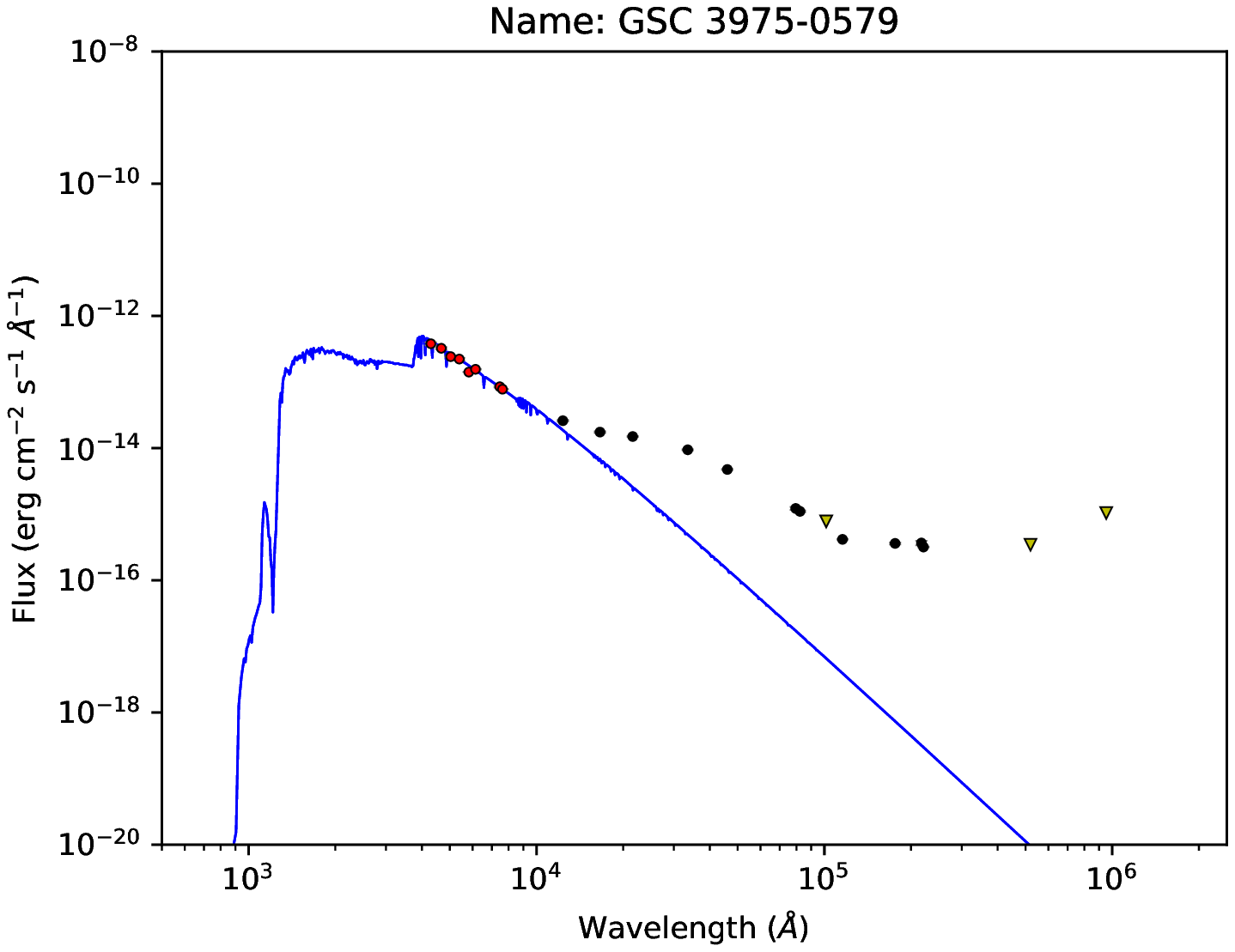}
    \includegraphics[width=0.33\textwidth]{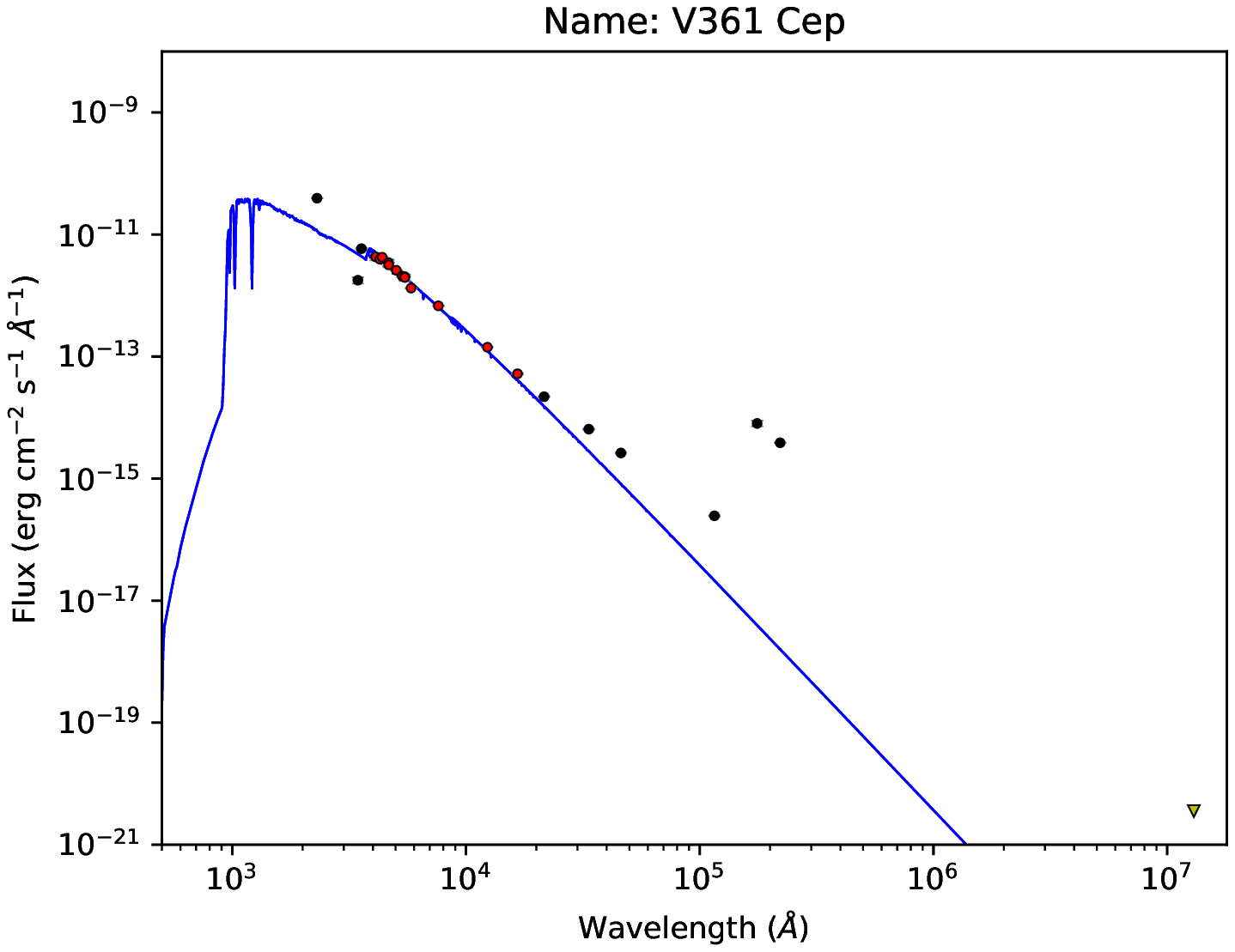}
    \includegraphics[width=0.33\textwidth]{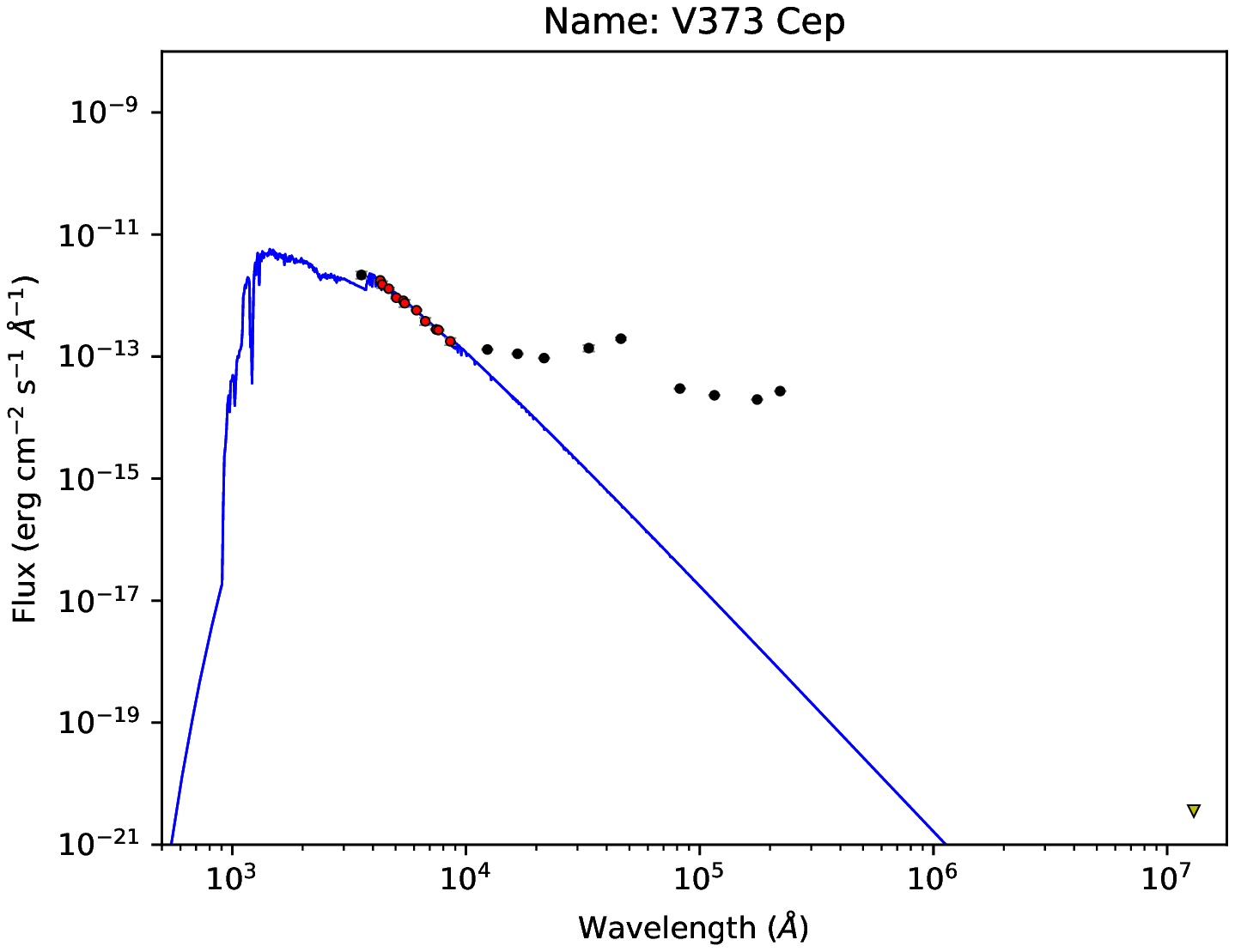}
\end{figure}

\begin{figure} [h]
 \centering
    \includegraphics[width=0.33\textwidth]{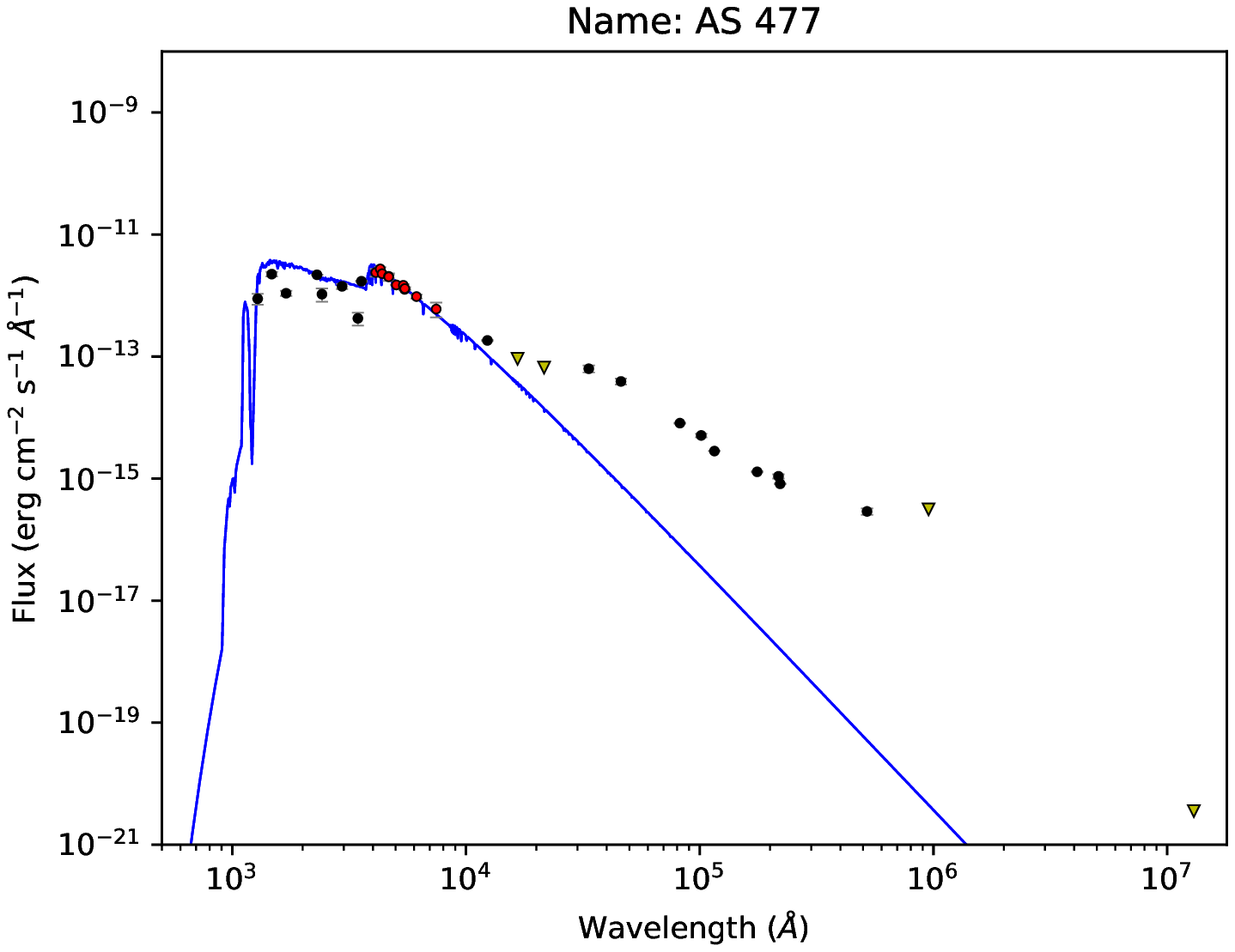}
    \includegraphics[width=0.33\textwidth]{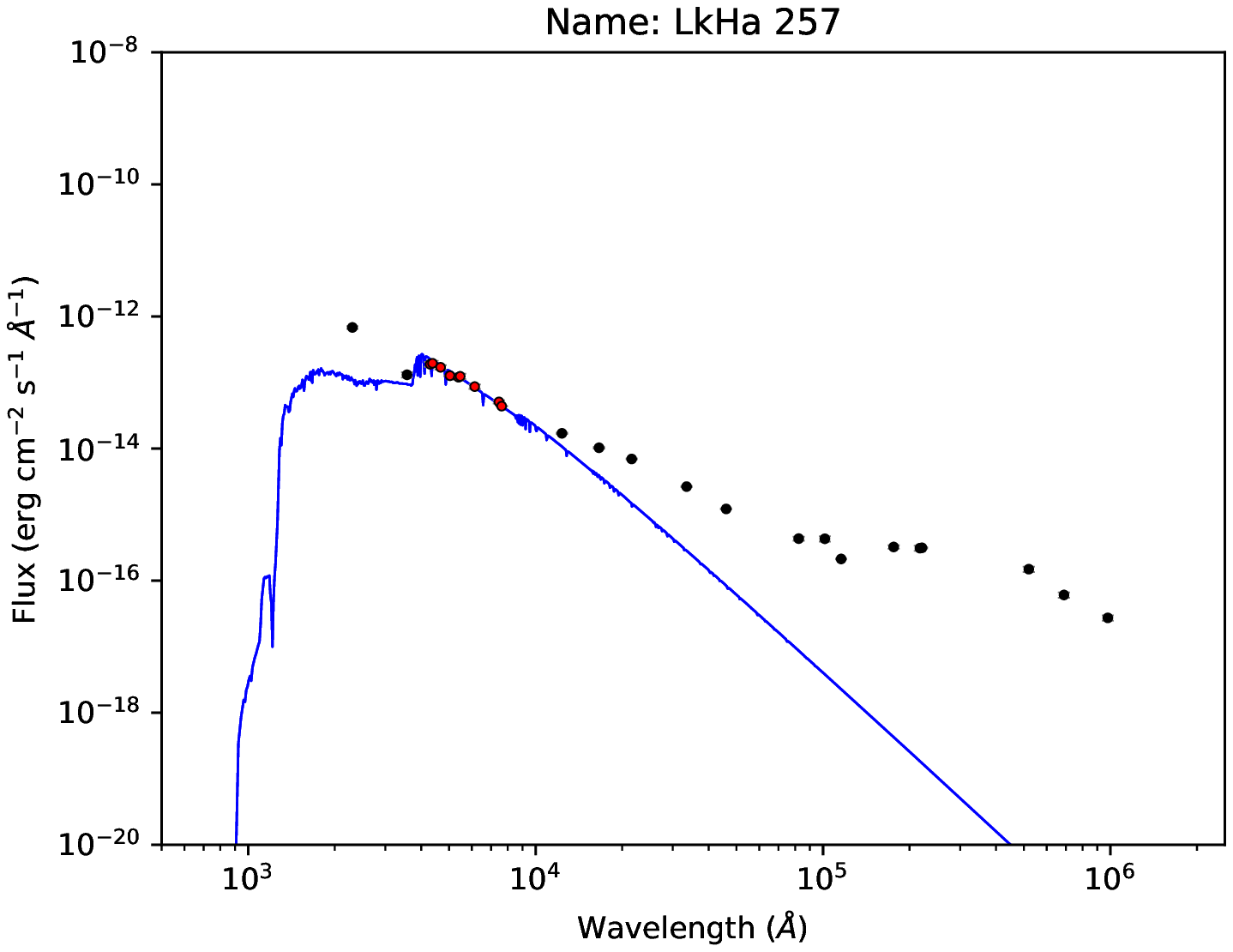}
    \includegraphics[width=0.33\textwidth]{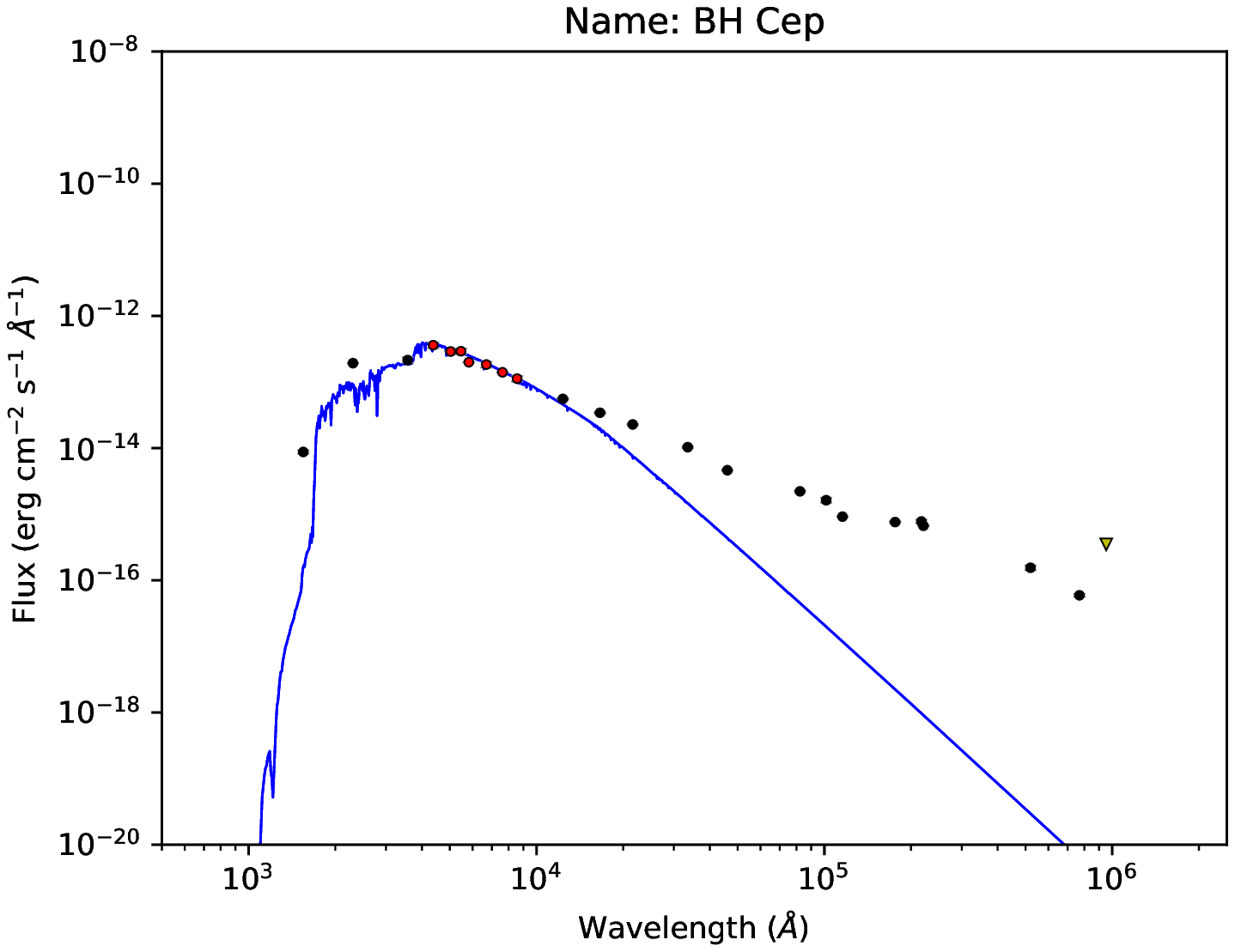}
\end{figure}

\begin{figure} [h]
 \centering
    \includegraphics[width=0.33\textwidth]{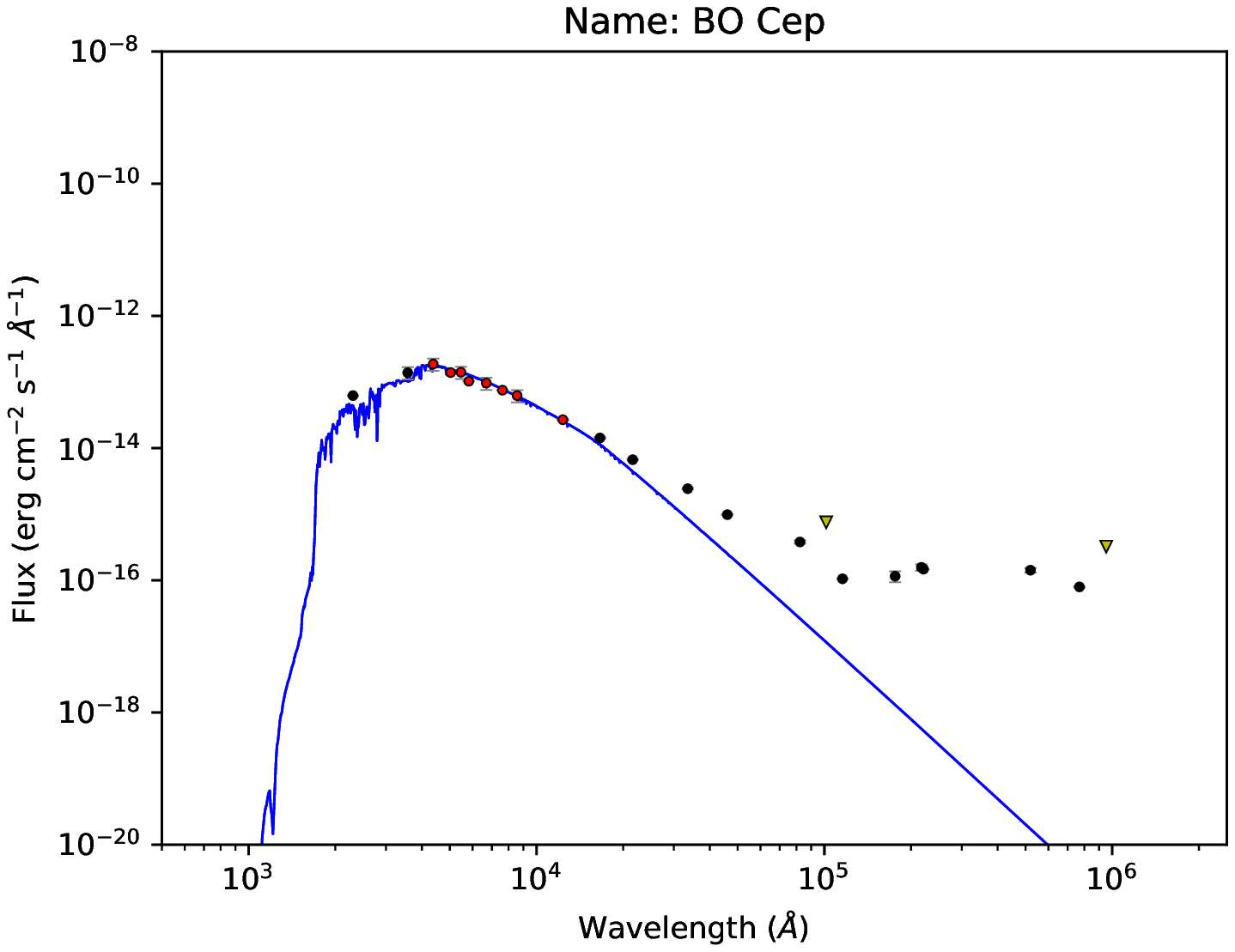}
    \includegraphics[width=0.33\textwidth]{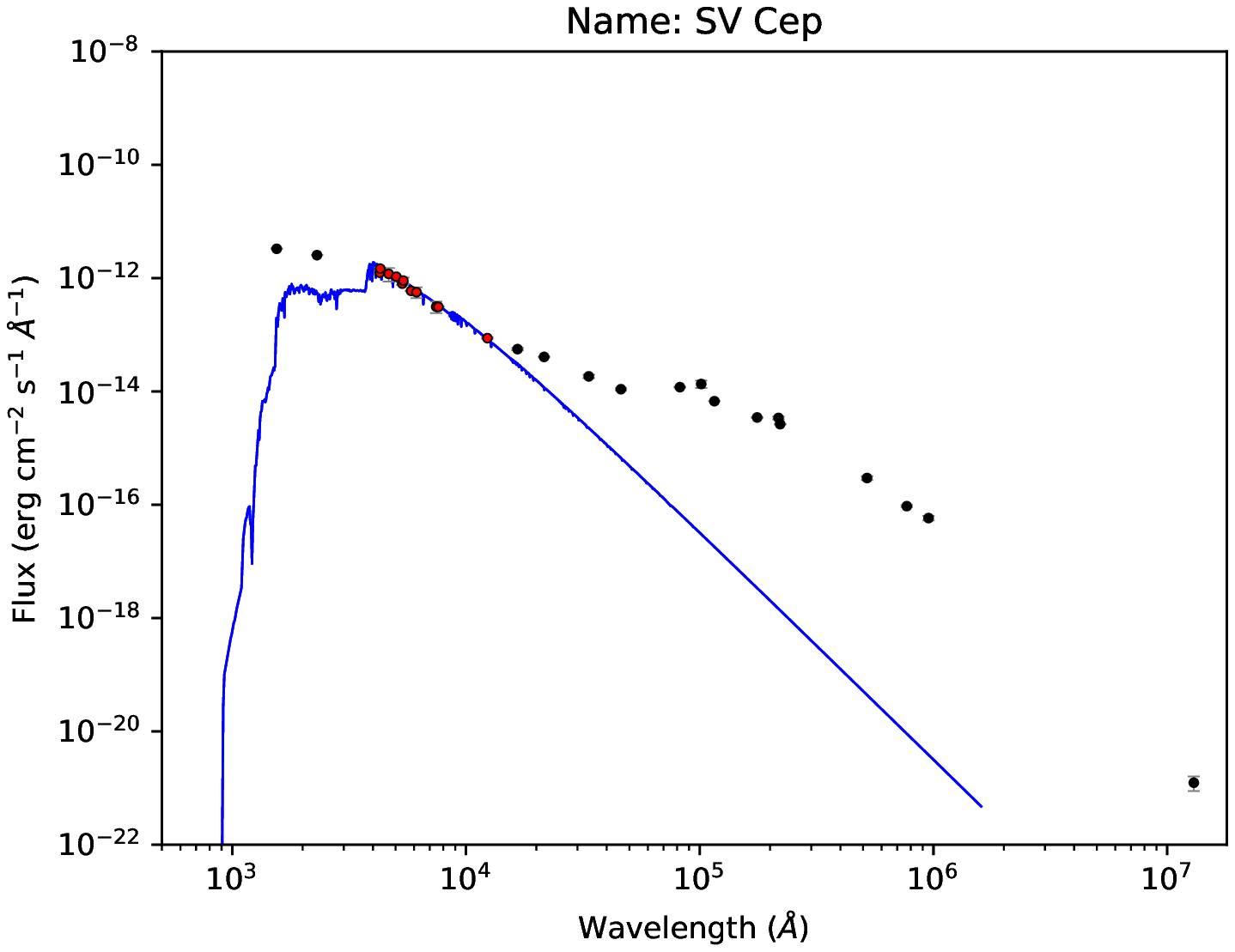}
    \includegraphics[width=0.33\textwidth]{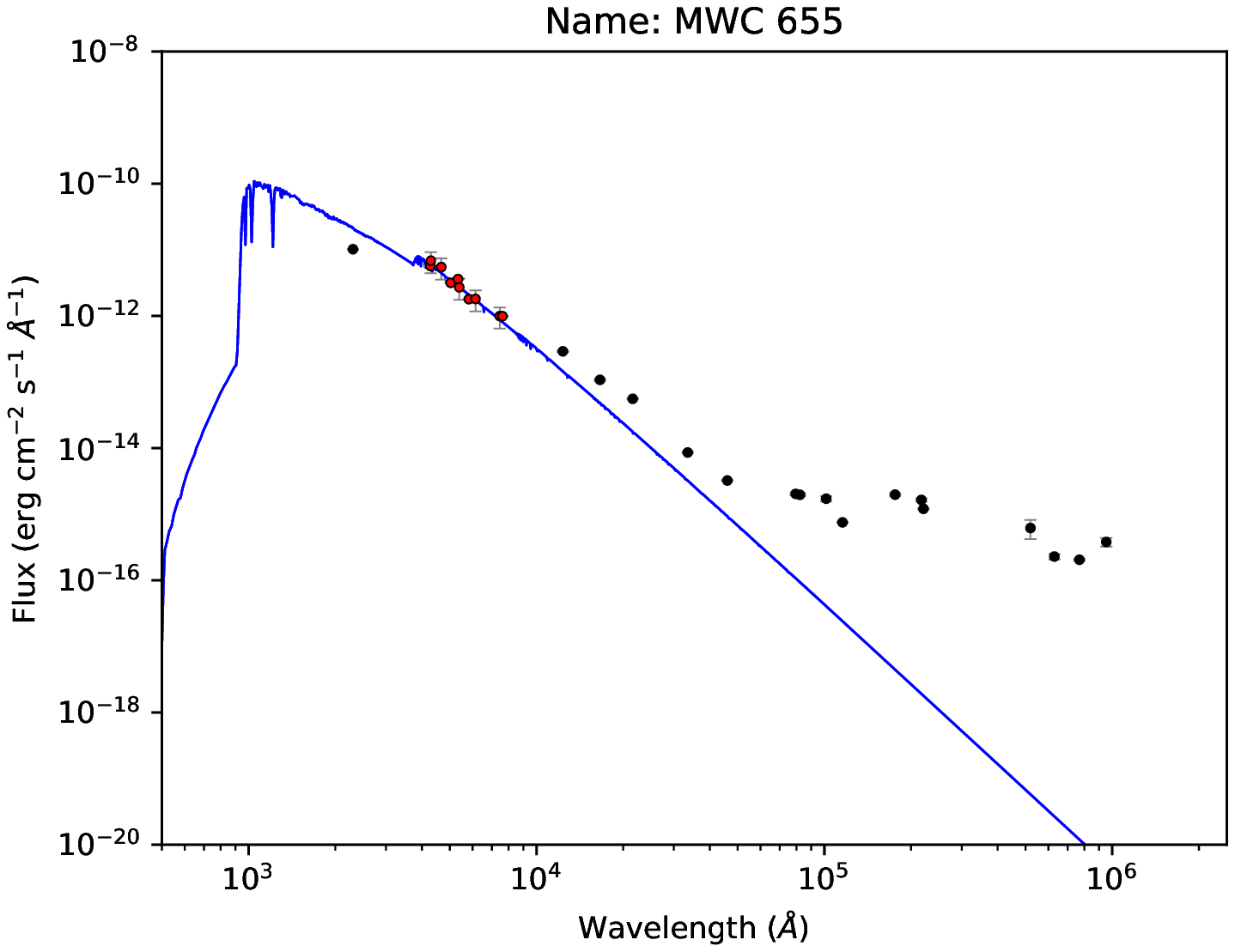}
\end{figure}

\newpage

\onecolumn

\begin{figure} [h]
 \centering
    \includegraphics[width=0.33\textwidth]{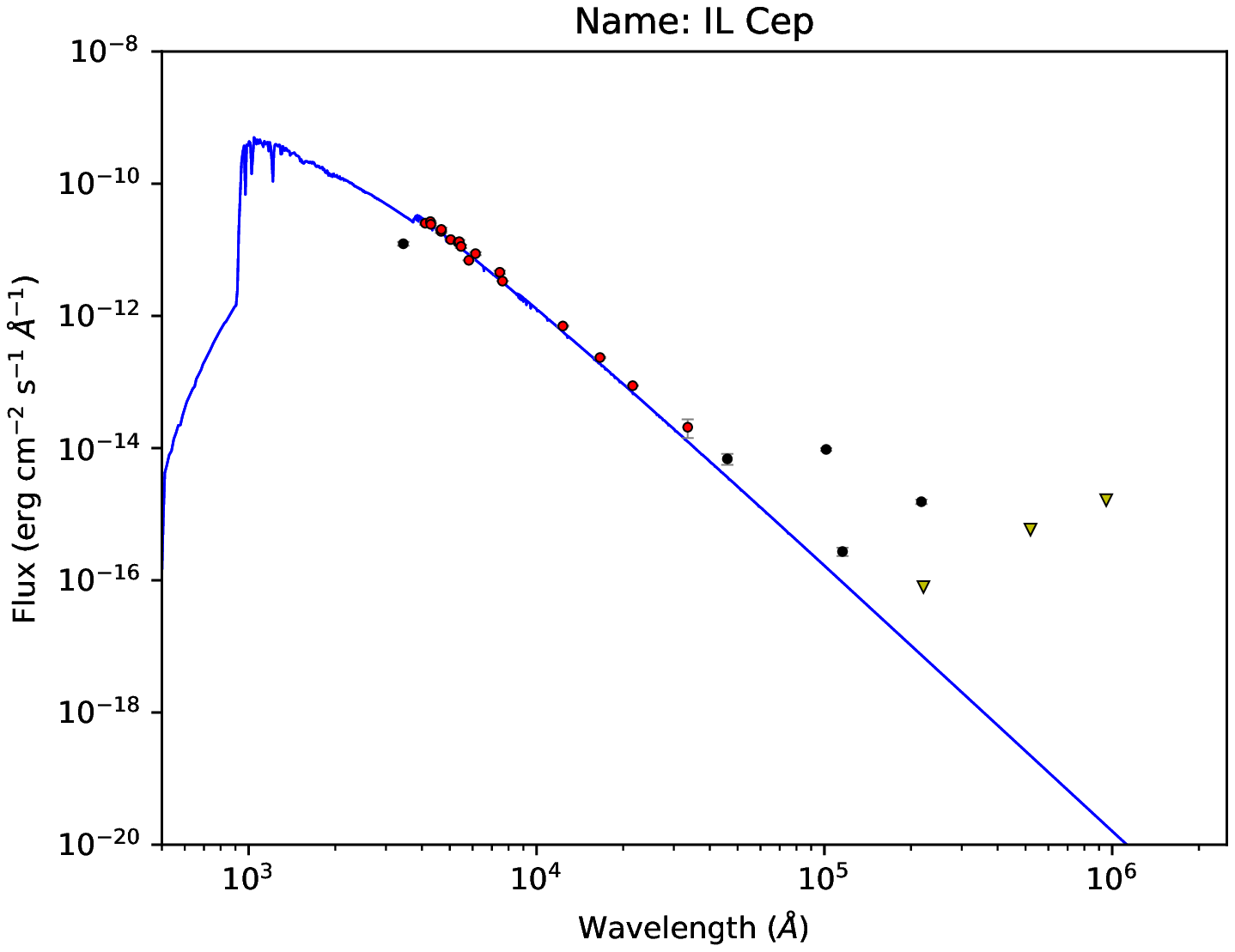}
    \includegraphics[width=0.33\textwidth]{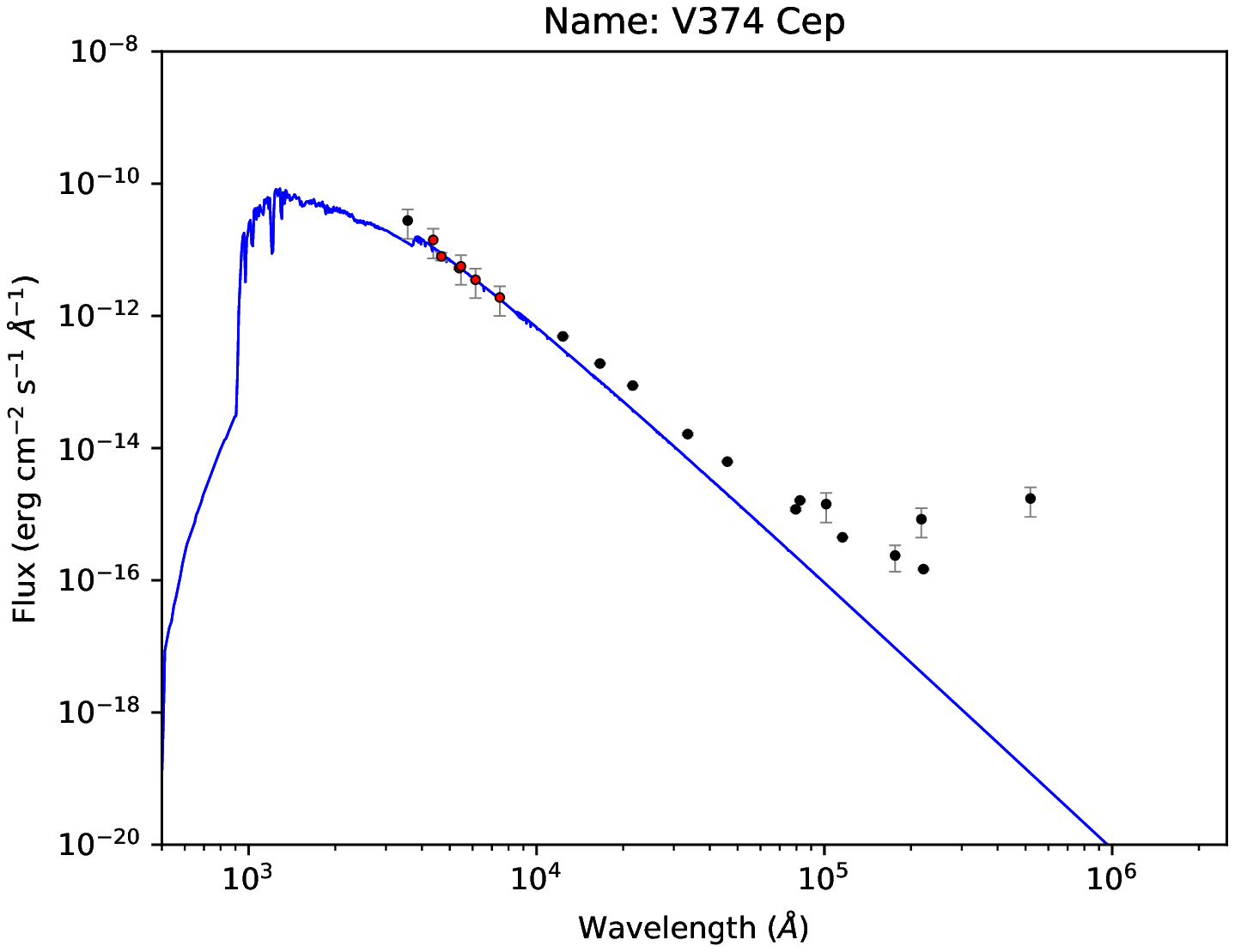}
    \includegraphics[width=0.33\textwidth]{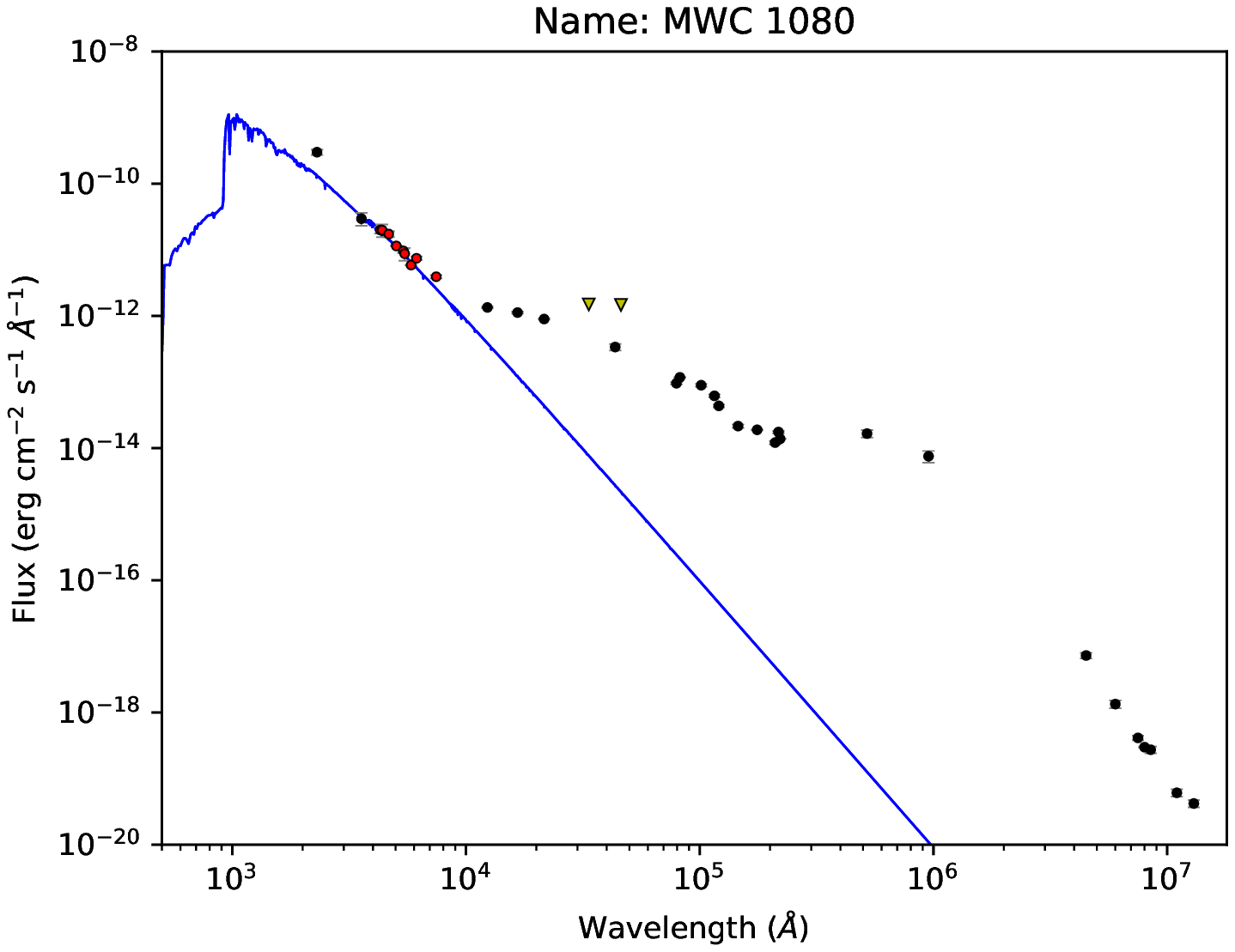}

\end{figure}

\begin{figure} [h]
 \centering
    \includegraphics[width=0.33\textwidth]{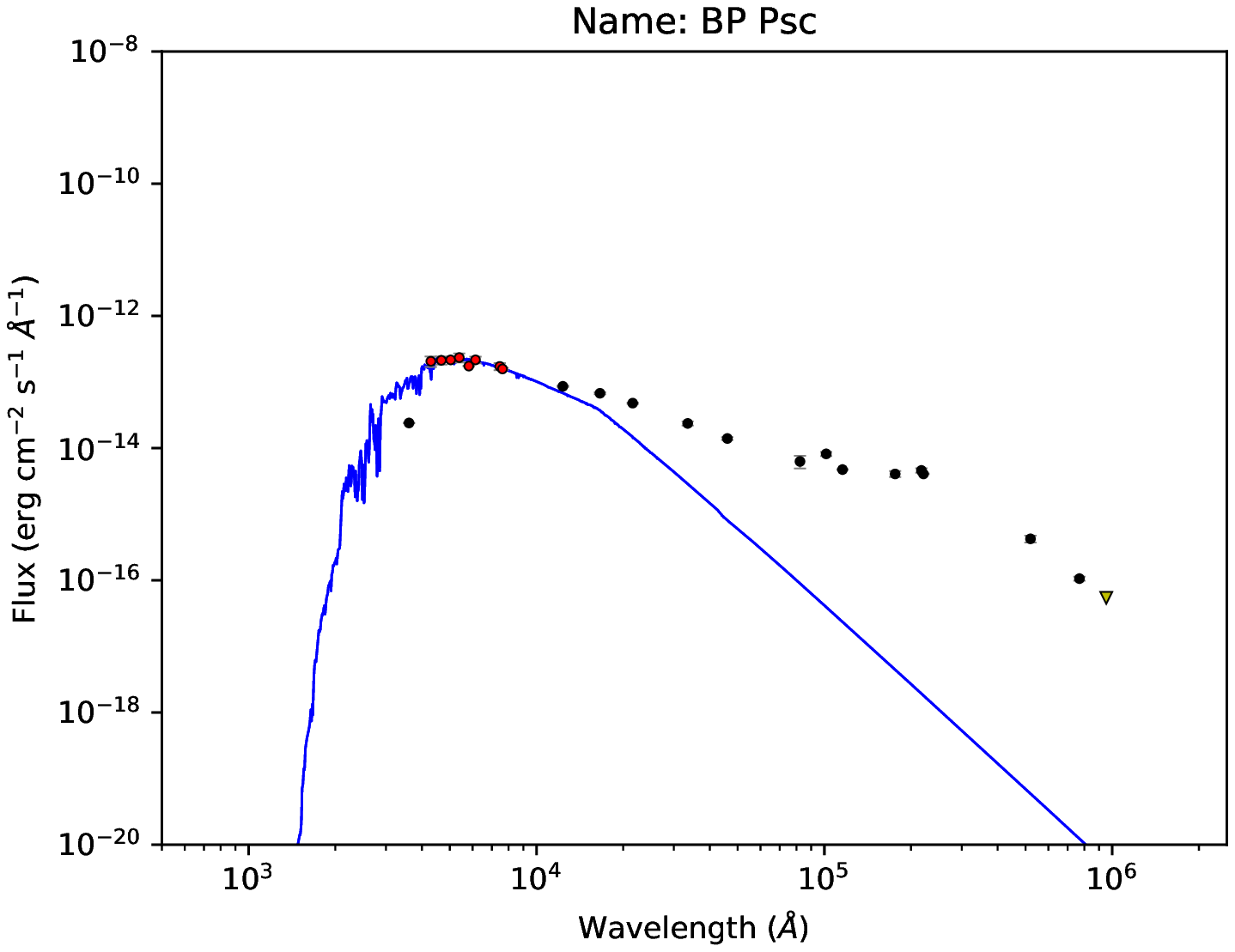}
    \includegraphics[width=0.33\textwidth]{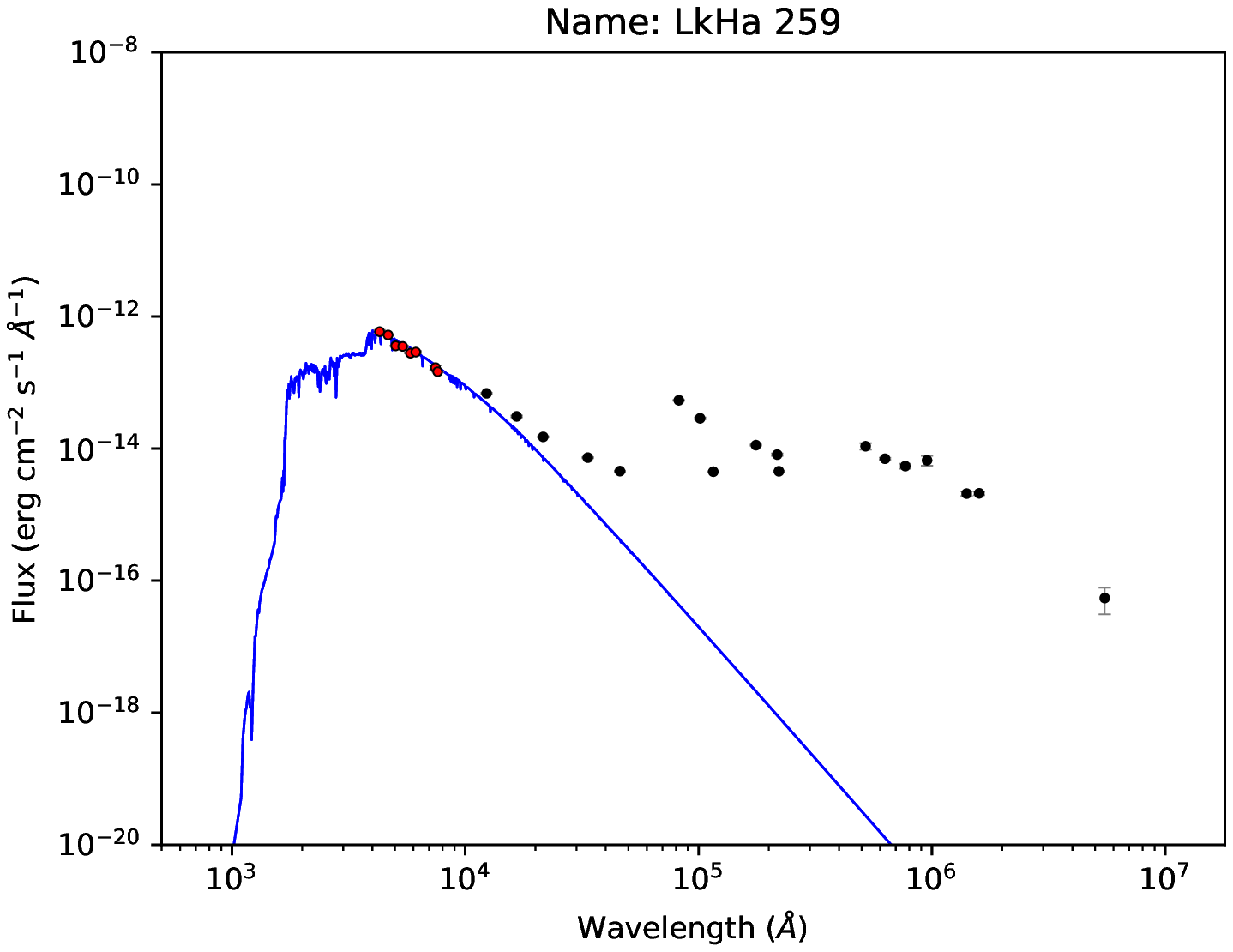}
            \caption{SEDs of the whole sample. The  solid blue line corresponds to the best photospheric model that fits the deredenned optical photometry (red dots). Yellow triangles are upper limits.}
\end{figure}
\end{appendix}

\end{document}